\documentclass[b5paper,10pt,twoside,openright]{book}
\usepackage{lettrine}
\usepackage{etex}
\usepackage[dvipdfm,b5paper,left=25mm,right=18mm,foot=1.5cm,headheight=1.0cm,top=2.1cm,bottom=2.2cm]{geometry}
\usepackage[utf8]{inputenc}
\usepackage[english]{babel}
\usepackage[T1]{fontenc}

\usepackage{tikz}
\usetikzlibrary{arrows,patterns,decorations,decorations.text,decorations.markings,decorations.pathmorphing}

\usepackage{titlesec}

\titleformat
{\chapter} %
[display] %
{\sf\bfseries\Large\itshape} %
{Chapter\ \thechapter} %
{0.5ex} %
{
    \vspace{1ex}
    \huge\sffamily
} %
[
\textcolor{gray}{\rule{\textwidth}{5pt}}\vspace{1em}
] %

\titleformat{\section}[block]%
  {\Large\sffamily}%
  {\thesection}{1em}{\bfseries}

\titleformat*{\paragraph}{\sffamily\bfseries\small}

\titleformat{\subsection}[block]
  {\large\sffamily}
  {\thesubsection}{.5em}{\bfseries}

\titleformat{\subsubsection}[hang]
  {\normalfont\sffamily}
  {}{}{\bfseries}

\usepackage{amssymb,amsmath,amsfonts}
\usepackage{framed}
\usepackage[framed,amsmath,amsthm,thmmarks]{ntheorem}
\usepackage{xcolor}
\definecolor{lightgray}{gray}{0.75}

\theoremstyle{theorem}
\theoremsymbol{}
\theoremprework{\setlength\theorempreskipamount{0pt}\setlength\theorempostskipamount{0pt}}
\newframedtheorem{theorem}{Theorem}[chapter]
\theoremprework{\setlength\theorempreskipamount{0pt}\setlength\theorempostskipamount{0pt}}
\newframedtheorem{lemma}[theorem]{Lemma}
\theoremprework{\setlength\theorempreskipamount{0pt}\setlength\theorempostskipamount{0pt}}
\newframedtheorem{corollary}[theorem]{Corollary}
\theoremprework{\setlength\theorempreskipamount{0pt}\setlength\theorempostskipamount{0pt}}
\newframedtheorem{proposition}[theorem]{Proposition}
\theoremprework{\setlength\theorempreskipamount{0pt}\setlength\theorempostskipamount{0pt}}
\newframedtheorem{algorithm}[theorem]{Algorithm}

\theoremstyle{remark}
\theoremprework{\setlength\theorempreskipamount{0pt}\setlength\theorempostskipamount{0pt}}
\newframedtheorem{remark}[theorem]{Remark}
\theoremprework{\setlength\theorempreskipamount{0pt}\setlength\theorempostskipamount{0pt}}
\newframedtheorem{example}[theorem]{Example}

\theoremstyle{definition}
\theoremsymbol{}
\theoremprework{\setlength\theorempreskipamount{0pt}\setlength\theorempostskipamount{0pt}}
\newframedtheorem{definition}[theorem]{Definition}

\usepackage{textcomp}
\usepackage{marvosym}
\usepackage{pifont}
\usepackage{lmodern}
\usepackage{graphicx}
\usepackage{longtable}
\usepackage{mdwlist}
\usepackage{multirow}
\usepackage{mathdots}
\usepackage{multicol}
\usepackage[nottoc,notlof,notlot]{tocbibind}
\usepackage[subfigure,titles]{tocloft}
\usepackage{rotating}
\usepackage{pdflscape}
\usepackage{scalefnt}
\usepackage{wasysym}
\usepackage{listings}
\usepackage{xspace}
\usepackage{textcomp}
\usepackage{verbatim}
\usepackage{subfig}
\usepackage{array}
\usepackage[square,sort&compress,numbers]{natbib}
\def\@biblabel#1{#1.}

\usepackage{booktabs}
\usepackage{tabularx}
\usepackage{xspace}

\usepackage{lmodern}

\renewcommand\sffamily{\usefont{T1}{lmss}{m}{n}}

\renewcommand{\emph}[1]{{\textsl{#1}}}

\newcommand{\package}[1]{\textsf{#1}\xspace}

\newcommand{\lang}[1]{\textsf{#1}\xspace}

\newcommand{\R}{\lang{R}}

\DeclareMathOperator*{\argmin}{arg\,min}

\DeclareMathOperator*{\intcl}{intcl}

\usepackage[backgroundcolor=gray!10,textsize=scriptsize,shadow]{todonotes}

\newcommand{\indicator}{\text{\bf 1}}%
\renewcommand{\Pr}{\mathrm{P}}%

\newcommand{\vect}[1]{{\mathbf{#1}}}
\newcommand{\func}[1]{{\mathsf{#1}}}

\newcommand{\Naturals}{\mathbb{N}}
\newcommand{\Ival}{\mathbb{I}}
\newcommand{\IvalPow}[1]{\mathbb{I}^{#1}}
\newcommand{\AnyPow}{^{*}}
\newcommand{\IvalAnyPow}{\mathbb{I}\AnyPow}

\usepackage{tikz}
\newcommand*\circled[1]{\tikz[baseline=(char.base)]{
  \node[shape=circle,draw,inner sep=1pt] (char) {#1};}}
\usetikzlibrary{decorations,arrows,shapes,patterns,positioning}

\titleformat{\subsubsection}{\normalsize\sf\bfseries\itshape}{\Alph{subsubsection}. }{0pt}{}
\makeatletter
\renewcommand\p@subsubsection{\thesubsection.}
\makeatother

\newcommand*{\myprime}{^{\prime}\mkern-1.2mu}
\newcommand*{\mydprime}{^{\prime\prime}\mkern-1.2mu}
\newcommand*{\mytrprime}{^{\prime\prime\prime}\mkern-1.2mu}

\author{Marek Gagolewski}
\title{Data fusion. Theory, methods, applications}
\usepackage{ifpdf}
\ifpdf
\RequirePackage[b5paper,final,colorlinks=false,hyperindex,plainpages=false,%
   breaklinks=true,%
   pdfauthor={Marek Gagolewski},%
   pdftitle={Data fusion. Theory, methods, applications}%
]{hyperref}
\else
\RequirePackage[b5paper,final,plainpages=true,breaklinks=true]{hyperref}
\fi

\usepackage{cleveref}
\crefname{theorem}{}{}
\crefrangelabelformat{theorem}{#3#1#4--#5#2#6}

\usepackage{caption}
\addto\captionsenglish{}
\addto\captionsenglish{}

\makeatletter
\providecommand{\bigsqcap}{%
  \mathop{%
    \mathpalette\@updown\bigsqcup
  }%
}
\newcommand*{\@updown}[2]{%
  \rotatebox[origin=c]{180}{$\m@th#1#2$}%
}
\makeatother

\usepackage{fancyhdr}
\fancypagestyle{plain}{%
\fancyhead{}
\fancyfoot{}
}

\addto\captionsenglish{}
\addto\captionsenglish{}

\makeatletter
\def\pagenumbering#1{%
  \gdef\thepage{\csname @#1\endcsname \c@page}}
\makeatother

\usepackage{imakeidx}
\makeindex[intoc]

\begin{document}
\allowdisplaybreaks

\frontmatter
\pagenumbering{arabic}

\author{Marek Gagolewski}
\title{Data fusion: Theory, methods, and applications}

\thispagestyle{empty}

\begin{center}
{\Large MONOGRAPH SERIES}

\smallskip
{\large INFORMATION TECHNOLOGIES:}

RESEARCH AND THEIR INTERDISCIPLINARY APPLICATIONS

\smallskip
{VOL.~7}

\textcolor{gray}{\rule{\textwidth}{2pt}}

\vspace*{5em}

{\LARGE\sf {M}{\large AREK} {G}{\large AGOLEWSKI}}

\vspace*{2em}

\scalebox{1.5}{\Huge\sf\bfseries\scshape {D}{\LARGE ATA} {F}{\LARGE USION}}

\vspace*{0.2cm}

{\huge\sf {T}{\Large HEORY}, {M}{\Large ETHODS}, {\Large AND} {A}{\Large PPLICATIONS}}

\textcolor{gray}{\rule{\textwidth}{5pt}}
\end{center}

\vfill
\begin{center}
\begin{minipage}{2.2cm}
\includegraphics[width=2cm]{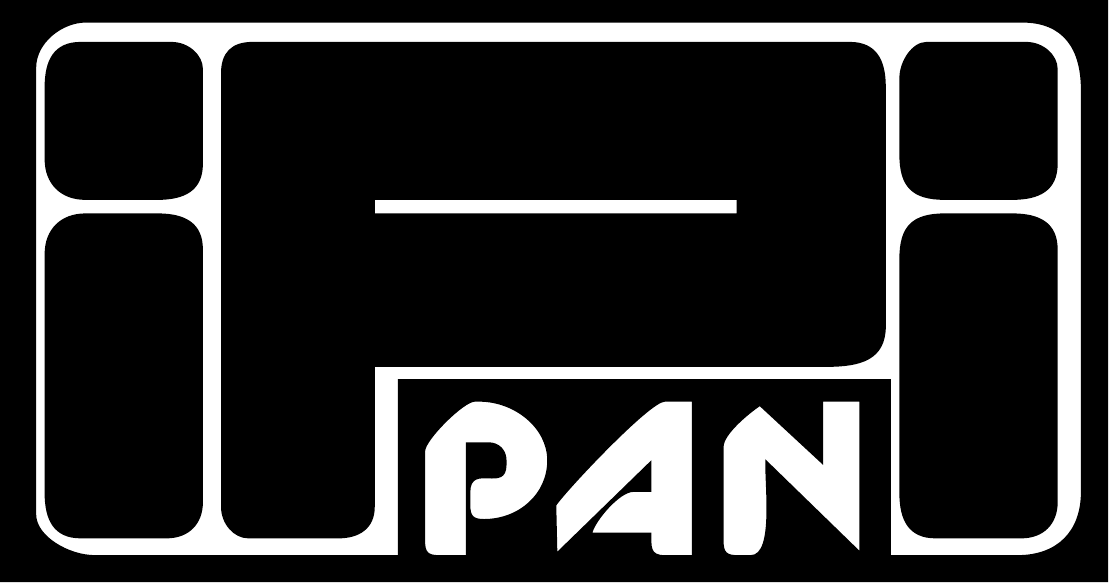}
\end{minipage}
\begin{minipage}{7cm}
INSTITUTE OF COMPUTER SCIENCE

POLISH ACADEMY OF SCIENCES
\end{minipage}

\bigskip
\small WARSAW, 2015
\end{center}

\clearpage

\thispagestyle{empty}

\begin{flushleft}
\textbf{Marek Gagolewski}

\begin{small}
Systems Research Institute

Polish Academy of Sciences

{\sf{https://www.gagolewski.com}}
\end{small}
\end{flushleft}

\bigskip
\begin{small}\noindent
This publication is issued as a part of the project
``Information technologies: \hbox{Research} and their interdisciplinary applications'',
objective 4.1 of the Human \hbox{Capital} Operational Program,
agreement no.~\hbox{UDA-POKL.04.01.01-00-\allowbreak{}051/10-00}. It~is
co-financed by European Union from resources of European Social Fund.

\begin{flushleft}
\textbf{Project leader:}

{\small Institute of Computer Science, Polish Academy of Sciences}

\medskip
\textbf{Project partners:}

{\small Systems Research Institute, Polish Academy of Sciences}

{\small \hbox{Institute of Biocybernetics and Biomedical Engineering, Polish Academy of Sciences}}
\end{flushleft}
\end{small}

\bigskip
\begin{flushleft}
\textbf{Editors-in-chief:}

Olgierd Hryniewicz

Jan Mielniczuk

Wojciech Penczak

Jacek Waniewski

\bigskip
\textbf{Reviewers:}

Gleb Beliakov

Radko Mesiar

\bigskip
\textbf{Typesetting with} L\hspace*{-0.3em}${}^\mathrm{A}$\hspace*{-0.2em}T\hspace*{-0.1em}${}_\mathrm{E}$\hspace*{-0.1em}X\textbf{:}

Marek Gagolewski
\end{flushleft}

\vfill

\begin{flushleft}
\textbf{This material is licensed under the Creative Commons
Attribution-NonCommercial-NoDerivatives 4.0 International License (CC BY-NC-ND 4.0).}

\smallskip\noindent
All trademarks, trade names, and logos mentioned or used in this book
are the property of their respective owners.

\smallskip
Copyright \copyright~2015 Marek Gagolewski\\
All rights reserved.

\bigskip
\hspace{1.8ex}ISBN {978-83-63159-20-7}

\end{flushleft}
\clearpage\normalfont\normalcolor\normalsize

\setcounter{page}{3}

\clearpage{\pagestyle{empty}\cleardoublepage}

\fancyfoot{}
\fancyhf{}
\fancyhead[LO]{\sffamily\footnotesize Introductory exercise\hfill \sffamily\footnotesize\thepage}
\fancyhead[RO]{}
\fancyhead[LE]{}
\fancyhead[RE]{\flushleft\sffamily\footnotesize\thepage\hfill\sffamily\footnotesize Introductory exercise}
\chapter*{Introductory exercise}

\noindent
Please:
\begin{enumerate}
\item Grab two pencils, one in each hand.
\item With one eye closed, try to touch the erasers together.
Was that difficult?
\item Now try it with both eyes open. Was that easier?
\end{enumerate}

\vfill %

Humans are quite unusual in that both their eyes face forward
and that they have overlapping visual fields.
This kind of sensor redundancy
and the nature of the information fusion process applied by the human brain
makes 3D stereo vision possible. Readers who found touching
two pencils much harder with an eye shut already got
an impression of the importance of data fusion
and may proceed to page \pageref{Chapter:onedim}.
Otherwise, in the Preface we shall
explore the role of data fusion in various real-world applications.

\clearpage{\pagestyle{empty}\cleardoublepage}

\fancyhf{}
\fancyhead[LO]{\sffamily\footnotesize Preface\hfill \sffamily\footnotesize\thepage}
\fancyhead[RO]{}
\fancyhead[LE]{}
\fancyhead[RE]{\flushleft\sffamily\footnotesize\thepage\hfill\sffamily\footnotesize Preface}
\chapter{Preface}

\lettrine[lines=3]{A}{ppropriate} fusion of large, complex data sets
is necessary in the information era. Having to deal with just a few records
already forces the human brain to look for patterns in the data and to make
its overall picture instead of conceiving a reality as a set of individual
entities, which are much more difficult to process and analyze.
Quite similarly, the usage of appropriate methods to reduce the information overload
on a computer, may not only increase the quality of the results but also
significantly decrease algorithms' run-time.

It is known that information systems relying on a single information source
(e.g., measurements gathered from one sensor,
opinions of just a single authoritative decision maker,
outputs of one and only one machine learning algorithm,
answers of an individual social survey taker)
are most often neither accurate nor reliable.

The theory of aggregation is a relatively new research field,
even though various particular methods for data fusion were known and
used already by the ancient mathematicians. Since the 1980s, studies of
aggregation functions most often focus on the construction and
formal, mathematical analysis of diverse ways to summarize numerical lists with elements
in some real interval $\Ival=[a,b]$. This covers different kinds of
broadly-conceived means, fuzzy logic connectives (t-norms, fuzzy
implications), as well as copulas. Quite recently, we observe
an increasing interest in aggregation on partially ordered sets --
in particular, on ordinal (linguistic) scales.

Among the seminal monographs on the applied mathematics-oriented
\textit{classical} aggregation theory there are
\textit{Aggregation Functions: A Guide for Practitioners} \cite{BeliakovETAL2007:aggregationpractitioners}
by Beliakov, Pradera, and Calvo and
\textit{Aggregation Functions} \cite{GrabischETAL2009:aggregationfunctions}
by Grabisch, Marichal, Mesiar, and Pap.
We note that the typical mathematical \textit{arsenal}
used by aggregation theoreticians consists of a very creative combination
of approaches known from, among others, algebra, calculus, order and measure theory
(in fact, aggregation theory results strongly contribute to these subfields as well).
What is more, particular subclasses of aggregation functions are studied
in-depth in the following textbooks:
\textit{Triangular norms} \cite{KlementMesiarPap2000:trinorm} authored by Klement, Mesiar, and Pap,
\textit{{Fuzzy implications}}  \cite{BaczynskiJayaram2008:fuzzyimplications}
by Baczy\'{n}ski and Jayaram,
\textit{Handbook of means and their inequalities} \cite{Bullen2003:means} by Bullen,
as well as -- very recently --
\textit{A Practical Guide to Averaging Functions} \cite{BeliakovETAL2015:practicalbook}
by Beliakov, Bustince, and Calvo.
We shall also mention the book by Torra and Narukawa
(\textit{Modeling Decisions: {I}nformation Fusion and Aggregation Operators} \cite{TorraNarukawa2007:moddecis}),
which perhaps is the most computer science-oriented work of the ones listed.
However, in \cite{BeliakovETAL2007:aggregationpractitioners} and \cite{BeliakovETAL2015:practicalbook}
numerous interesting algorithms and computational issues are discussed too.

During the 2013 AGOP -- \textit{International Summer School on Aggregation Opera\-tors} --
conference in Pamplona, Spain, Prof.~Bernard De Baets in his plenary lecture \cite{DeBaets2013:AGOPplenary}
pointed out the need to convey research on the so-called \textit{Aggregation 2.0}.
Of course, Aggregation 2.0 does not aim to replace or in any terms depreciate
the very successful and important \textit{classical} aggregation field, but rather to
attract the investigators' attention to new, more complex domains,
most of which cannot be properly handled without using computational methods.
From this perspective, data fusion tools may be embedded in larger,
more complicated information processing systems and thus studied as
their key components.

A proper complex data fusion has been of interest to many researchers in diverse
fields, including computational statistics, computational geometry,
bioinformatics, machine learning, pattern recognition,
quality management, engineering, statistics, finance, economics, etc.
Let us note that it plays a crucial role in:
\begin{itemize}
   \item a synthetic description of data processes or whole domains,
   \item creation of rule bases for approximate reasoning tasks,
   \item consensus reaching and selection of the optimal
   strategy in decision support systems,
   \item missing values imputations,
   \item data deduplication and consolidation,
   \item record linkage across heterogeneous databases,
   \item automated data segmentation algorithms' construction
   (compare, e.g., the $k$-means and hierarchical clustering algorithms).
\end{itemize}
We observe that many useful machine learning methods
are based on a proper aggregation of information entities.
In particular, the class of ensemble methods for classification
is very successful in practice because of the assumption that no single
``weak'' classifier can perform as well as their whole group.
Interestingly, many of the winning solutions to data mining competitions
on \textit{Kaggle} and similar platforms
base somehow on the random forest and similar algorithms.
What is more, e.g., neural networks -- universal approximators --
and other deep learning tools can be understood as hierarchies of
individual fusion functions. Thus, they can  be
conceived as kinds of aggregation techniques as well.
We should also mention that an appropriate data fusion is crucial to
business enterprises. For numerous reasons, companies are rarely eager
to sell large parts of the data sets they posses to their clients.
Instead, only carefully pre-processed and aggregated data \textit{models}
are delivered to the customers.

This monograph is a
first attempt to integrate the spread-out results from different domains
using the methodology of the well-established \textit{classical}
aggregation framework, introduce
researchers and practitioners to Aggregation 2.0,
as well as to point out the challenges and interesting directions
for further research. It is organized as follows.
\begin{itemize}
   \item In \textit{Chapter \ref{Chapter:onedim}} we review classical aggregation
   results which deal with aggregation of numeric tuples with elements
   in some real interval $\Ival=[a,b]$ or $]a,b[$. We list some interesting properties
   of fusion functions on such a domain which may be crucial in various
   practical applications. Even though the described data model seems to be quite simple
   at a first glance,
   it shall provide us with a deep insight on the nature of more complex
   data fusion processes. In particular, we pay special attention to the notion
   of monotonicity.

   Then we discuss general construction methods that may be used
   to derive new fusion functions from simpler ones. Additionally, we
   present the connection between aggregation functions and monotone (fuzzy)
   measures and integrals as well as introduce the notion of a penalty-based
   and an extended fusion function.

   Further on we present different ways which can aid in an appropriate tool
   selection for diverse tasks. This includes
   characterization theorems, synthetic numerical characteristics,
   as well as algorithms to learn fusion functions from empirical data.

   Moreover, the topic of aggregation of data on an ordinal scale
   and -- more generally -- bounded partially ordered sets, as well
   as on a nominal scale is presented.

   \item \textit{Chapter \ref{Chap:multidim}} deals with aggregation
   of $d$-dimensional data, this time for $d>1$. Our point of departure
   consists of data fusion tools which are studied in fields such as
   computational statistics and computational geometry.
   Among their important properties we find,
   e.g., equivariances to particular geometrical transformations,
   as well as generalizations of some of the properties studied
   in the previous chapter.
   We note that the simplest fusion functions may be constructed by means
   of componentwise extensions of one-dimensional mappings.
   Other ones are based on the concept of data depth or penalty minimizers.

   We are also  interested in aggregation on product lattices and character
   sequences, especially in connection with the Hamming distance.

   \item In \textit{Chapter \ref{Chap:anydim}} we focus on the topic
   of strings' aggregation, that is tuples of not necessarily conforming
   lengths. In this case, various ordering relations may be defined,
   e.g., the lexicographic order. The data types of our interest
   include numeric strings which represent informetric data, as well
   as character strings, like DNA and protein sequences.
   It turns out that the most influential data fusion methods on such a domain
   may be expressed as minimizers of various string distance-based penalties.
   Because of that, we include a comprehensive overview of character string
   metrics. This embraces the notion of a generic edit, $q$-gram, and Dinu rank distance.

   \item \textit{Chapter~\ref{Chap:other}} deals with aggregation of
   much more complex data types: directional data, real intervals,
   fuzzy numbers, random variables, graphs and relations, as well as heterogeneous
   data sets. We shall observe that some of the  key ideas in data fusion
   can be extrapolated to these kinds of data models.

   \item Finally, in \textit{Chapter~\ref{Chap:Characteristics}}, we discuss
   various numerical characteristics of different objects. This topic is inevitably
   connected to data aggregation. In particular, we are interested
   in a synthetic description of probability distributions, spread of
   numeric lists, decision makers' consensus, economical inequity,
   informetric data, fuzzy numbers, and fusion functions themselves.
   We end the chapter with a discussion on the so-called checksums, which --
   as it shall turn out -- require a quite different treatment
   than other measures.

   \item In the \textit{Appendix}, following the excellent approach from
   \cite{BeliakovETAL2007:aggregationpractitioners}, we provide the implementations
   of the most interesting algorithms. For that, we use the \lang{R} \cite{Rproject:home}
   and \lang{C++11} programming language. In the latter case, the
   \package{Rcpp} package classes \cite{EddelbuettelFrancois2013:rcppbook}
   are used as a link between these two languages.

\end{itemize}

Apart from the provision of a global and concise view on fusion functions across different
domains (``Aggregation 2.0''), original contributions in this monograph,
which were not yet published at the time of its writing,
include, but are not limited to:
\begin{itemize}
   \item \textit{Chapter \ref{Chapter:onedim}}:
   The idea of incremental fusion functions
   as a generalization of recursive aggregation tools (Definition~\ref{Def:incrementality});
   new methods for learning aggregation operators from empirical data,
   including the least Chebyshev metric fitting tasks in Section~\ref{Sec:weightfit1d},
   the least squares error fitting with output ranking preservation in Section~\ref{Sec:LSEpreserveoutrank},
   applications of weights' regularization to prevent model overfit,
   fitting weights to quasi-arithmetic means (without variables' linearization);
   some notes on aggregation of elements on nominal scale in Section~\ref{Sec:nominalscale1}.

   \item \textit{Chapter \ref{Chap:multidim}}:
   Extension of results published in \cite{Gagolewski2015:issuesmultidim}
   concerning aggregation of $d$-dimensional real tuples,
   including  Propositions~\ref{Prop:transequiv}, \ref{Prop:uscaleequiv},
   \ref{Prop:rotationinvcomponent}, \ref{Prop:affineequicw}, \ref{Prop:CompositionProps2},
   and \ref{Prop:rotequichull}; a construction of SVD-based similarity transform equivariant
   fusion functions in Sec.~\ref{Sec:equivSimTrans}; proposal of a framework for
   penalty-based multidimensional fusion functions in
   Sec.~\ref{Sec:genpenaltyoptmultidim} and their general properties
   (in particular, Proposition~\ref{Prop:DistanceBasedPenaltyMultidim});
   a new evolutionary algorithm for approximating the Hamming distance-based 1-center
   character sequence.

   \item \textit{Chapter \ref{Chap:anydim}}:
   New results concerning aggregation of informetric data
   (Proposition \ref{Prop:OrderGamma} and \ref{Prop:convexsorted}),
   proposal for a list of desirable properties that such data fusion tools should fulfill,
   new aggregation methods for numeric strings in Section~\ref{Sec:1mediannumstr},
   including the 1-median for informetric data under assumption that $\Ival=[0,\infty]$;
   an exact algorithm to compute a centroid of two character strings
   as well as an evolutionary algorithm for 1-median of arbitrary number or
   character strings with respect to the Levenshtein distance,
   a list of desirable properties of fusion functions for character sequences
   and strings in Section %
   \ref{Sec:nominalscaleany}.

   \item \textit{Chapter~\ref{Chap:other}}:
   Fast approximate set exemplar search algorithm in arbitrary finite
   semimetric spaces in Section \ref{Sec:PseudometricSpace}.

   \item \textit{Chapter~\ref{Chap:Characteristics}}:
   A generalization of a spread relation \cite{Gagolewski2015:spread} in Section~\ref{Sec:SpreadMultidimSample}
   for multidimensional numeric lists and a list of new spread measures'
   construction methods.

\end{itemize}

The author would like to thank Prof.~Gleb Beliakov, Prof.~Radko Mesiar,
and Dr.~Simon James for the useful, in-depth comments on the manuscript
and to Prof.~Olgierd Hryniewicz
who encouraged him to write this book in November 2014. Moreover, he wishes
to thank Prof.~Bernard De~Baets and Prof.~Janusz Kacprzyk for motivating him
to convey research on Aggregation 2.0.
Also, the help of his Ph.D.~students Anna Cena and Maciej Bartoszuk
while dealing with early versions of this work is much appreciated.
He is also indebted to Prof.~Martin \v{S}t\v{e}pni\v{c}ka and other researchers
with the Institute for Research and Applications of Fuzzy Modeling
for their great hospitality during a research visit  at the University of
Ostrava, Czech Republic during which he has written some key parts of this monograph.

The study was cofounded by the European Union from resources of the European
Social Fund, Project PO KL ``Information technologies: Research and
their interdisciplinary applications'', Agreement UDA-POKL.04.01.01-00-051/10-00
as well as by the research task A4.1.2/2015, ``Algorithms for data aggregation and fusion'',
Systems Research Institute, Polish Academy of Sciences
and by the National Science Center, Poland, research project
2014/13/D/HS4/01700 (research in Sections \ref{Sec:aginformetricAny} and \ref{Sec:ImpactFunctions}).

\vspace*{2cm}
\noindent \hspace*{19em} Marek Gagolewski

\noindent \hspace*{19em} Warsaw, December 2015

\clearpage{\pagestyle{empty}\cleardoublepage}

\pdfbookmark{\contentsname}{Contents}
\cleardoublepage
\fancyfoot{}
\fancyhf{}
\fancyhead[LO]{\sffamily\footnotesize Contents\hfill \sffamily\footnotesize\thepage}
\fancyhead[RO]{}
\fancyhead[LE]{}
\fancyhead[RE]{\flushleft\sffamily\footnotesize\thepage\hfill\sffamily\footnotesize Contents}

\renewcommand\cftchapafterpnum{\vskip2pt}
\renewcommand\cftsecafterpnum{\vskip2pt}
\tableofcontents

\clearpage{\pagestyle{empty}\cleardoublepage}
\listoffigures

\clearpage{\pagestyle{empty}\cleardoublepage}
\listoftables

\clearpage{\pagestyle{empty}\cleardoublepage}
\fancyfoot{}
\fancyhf{}
\fancyhead[LO]{\sffamily\footnotesize Notation convention\hfill \sffamily\footnotesize\thepage}
\fancyhead[RO]{}
\fancyhead[LE]{}
\fancyhead[RE]{\flushleft\sffamily\footnotesize\thepage\hfill\sffamily\footnotesize Notation convention}

\chapter{Notation convention and \lang{R} basics}

\lettrine[lines=3]{I}{n} this book we roughly follow the conventions
used in \cite{GrabischETAL2009:aggregationfunctions}, which are
to some degree consistent with the way the \lang{R} \cite{Rproject:home}
environment handles vector and matrix computations. In particular:

\begin{itemize}
\item The set of natural numbers, $\{1,2,\dots\}$, is denoted by $\mathbb{N}$,
by $\mathbb{N}_0$ we mean the set $\mathbb{N}\cup\{0\}$, and the set of all
integers is denoted with $\mathbb{Z}$.
Additionally, $\mathbb{R}$ is the set of reals, $\mathbb{R}_{+}=]0,\infty[$,
and $\mathbb{R}_{0+}=[0,\infty[$. Where it is needed, $\bar{\mathbb{R}}=[-\infty,\infty]$
denotes the set of extended reals. By default, we assume that $+\infty+(-\infty)=-\infty$
and $0\cdot \infty=0$ (unless stated otherwise).

\item The interval closure of the set $S\subseteq \mathbb{R}$,
\index{intcl}i.e.,~the smallest closed interval that contains $S$, is denoted with $\intcl{S}$.
For any $x\in\mathbb{R}$, $\lfloor x \rfloor = \max\{y\in\mathbb{Z}: y\le x\}$
and $\lceil x \rceil = \min\{y\in\mathbb{Z}: y \ge x\}$ denote the floor
and ceiling function, respectively.

\item For any natural number $n$, let $[n]=\{1,2,\dots,n\}$, with convention $[0]=\emptyset$.
Moreover, $[i:j]=\{i,i+1,\dots,j\}$ for any $i\le j$. Thus, $[n] = [1:n]$.
Here is a corresponding \lang{R} code:

\begin{lstlisting}[language=R]
seq_len(0)        # [0]
## integer(0)
seq_len(5)        # [5]
## [1] 1 2 3 4 5
0:4               # [0:4]
## [1] 0 1 2 3 4
\end{lstlisting}

\item Given a set $X$, let $X^*=\bigcup_{n=2}^\infty X^n$ denote the set
of all sequences with elements in $X$ of length at least (if not stated
explicitly otherwise) two.

\item Each sequence is denoted with a bold symbol, e.g., $\vect{x}=(x_1,\dots,x_n)$.
Note that in each case we use $1$-based indexing (like in the \lang{R} programming language).
Having that in mind is crucial when it comes to implementing algorithms to perform
computations on vectors: for instance, languages like \lang{C++}, \lang{Java},
and \lang{Python} use $0$-based indexing.

\begin{lstlisting}[language=R]
x <- c(2, 4, 6, 8)
n <- length(x)     # n == 4
x[1]               # the first element in x
## [1] 2
x[n]               # the last element in x
## [1] 8
\end{lstlisting}

\item Given arbitrary $x\in X$, by $(n\ast x)$ we denote an $n$-tuple (a sequence of length $n$)
$(x,x,\dots,x)\in X^n$.  %
More generally, $(n\ast (x_1,\dots,x_k)) = (x_1,\dots, x_k, \dots, x_1, \dots, x_k)\in X^{nk}$
denotes the fact that $(x_1,\dots,x_k)$ is repeated exactly $n$ times, with
recycling.

\begin{lstlisting}[language=R]
rep(1, 5)          # (5 * 1)
## [1] 1 1 1 1 1
rep(1:2, 3)        # (3 * (1, 2))
## [1] 1 2 1 2 1 2
\end{lstlisting}

\item Given, say, $\vect{x}\in X^n,\vect{y}\in X^m, t\in X$,
$(\vect{x},\vect{y},t)\in X^{n+m+1}$ denotes their concatenation
into a single vector.

\begin{lstlisting}[language=R]
x <- c(1, 2, 3)
y <- c(4, 5)
t <- 6
c(x, y, t)          # (x, y, t)
## [1] 1 2 3 4 5 6
\end{lstlisting}

\item Binary operations like $+,-,\cdot,/,\wedge$ (minimum), and $\vee$ (maximum)
on vectors of equal lengths $n$ are applied elementwise
and thus output a vector of length $n$ too.
On the other hand, if one of the operands is a scalar, then
it is extended to a vector of length $n$ in such a way that $\vect{x}+t=\vect{x}+(n\ast t)$.

\begin{lstlisting}[language=R]
c(-1, 1, -2, 2) * c(1, 2, 3, 4)  # vector * vector
## [1] -1  2 -6  8
2 * c(1, 3, 5)                   # scalar * vector
## [1]  2  6 10
\end{lstlisting}

Note that in fact in \lang{R} there are no separate scalar data types:
single values are represented as vectors of length 1.

\item If $\vect{A}\in\mathbb{R}^{d\times n}$ is a matrix with $d$ rows and $n$ columns
and $\vect{t}\in\mathbb{R}^d$,
then by, e.g., $\vect{A}+\vect{t}$ we mean $\vect{A}+[\vect{t}\ \vect{t}\ \cdots\ \vect{t}]$,
i.e., $\vect{t}$ is treated as a column vector.
Moreover, $\vect{A}+t = \vect{A}+(d\ast t) = \vect{A} + [(d\ast t)\ \cdots\ (d\ast t)]$.

\begin{lstlisting}[language=R]
d <- 2
n <- 3
A <- matrix(byrow=TRUE, nrow=d, ncol=n,
            c(1, 2, 3,
              4, 5, 6))
A
##      [,1] [,2] [,3]
## [1,]    1    2    3
## [2,]    4    5    6
A * c(-1, 1)
##      [,1] [,2] [,3]
## [1,]   -1   -2   -3
## [2,]    4    5    6

A / 2
##      [,1] [,2] [,3]
## [1,]  0.5  1.0  1.5
## [2,]  2.0  2.5  3.0
\end{lstlisting}

\item Regarding evaluation of $n$-argument functions, we interchangeably use notations:
$\func{F}(x_1,\dots,x_n)=\func{F}((x_1,\dots,x_n))=\func{F}(\vect{x})$.
If $\func{F}$ is defined on a domain $X$, then for $Y\subset X$,
$\func{F}|_Y$ denotes the projection of $\func{F}$ onto $Y$
(domain restriction).

\item If a function $\func{F}$ is defined on $X$,
then we implicitly assume that  it may be extended onto $X^n$
by vectorization: $\func{F}(x_1,\dots,x_n)=(\func{F}(x_1), \dots, \func{F}(x_n))$.

\begin{lstlisting}[language=R]
sign(c(-2, 1, 0, 0.5))
## [1] -1  1  0  1
\end{lstlisting}
On a side note, if vectorization is not an \lang{R} function's inherent feature,
we can assure it manually by calling
a functional programming construct called \texttt{sapply()}.

\begin{lstlisting}[language=R]
sapply(c(-2, 1, 0, 0.5), sign)
## [1] -1  1  0  1
\end{lstlisting}

\item $\indicator(p)$ denotes the Boolean indicator function,
$\indicator(p)=1$ whenever a logical statement $p$ is true and $0$ otherwise.
Moreover, the characteristic function is denoted with $\indicator_X(x)=\indicator(x\in X)$ for any set $X$.
Of course, these functions may be vectorized if needed.

\begin{lstlisting}[language=R]
x <- c(-2, 1, 0, 0.5)
as.integer(x > 0)
## [1] 0 1 0 1
\end{lstlisting}

\item For any finite set $X$, $|X|$ denotes its cardinality.
If $\vect{x}$ is a sequence, then the same notion, $|\vect{x}|$,
is used to denote its length.

\item Let $\mathfrak{S}_{Y}$ denote the set of all permutations of a finite set $Y$.
Given $\vect{x}\in X^n$ and $\sigma\in\mathfrak{S}_{[n]}$
let $\vect{x}_\sigma:=(x_{\sigma(1)},\dots,x_{\sigma(n)})$.
\index{order statistic}%
Additionally, let $x_{(i)}$ denote the $i$th \index{order statistic}\emph{order statistic} in a vector $\vect{x}\in X^n$,
i.e., the $i$th smallest value in that vector. The term ``smallest''
is of course relative to some linear order $\le$ on $X$. For instance,
if $X=\mathbb{R}$, we use (if not stated otherwise) standard ordering of reals. Thus, it holds:
\[
   x_{(1)}\le x_{(2)} \le \dots \le x_{(i)} \le \dots \le x_{(n)}
\]
Of course, $x_{(i)}=x_{\sigma(i)}$, where $\sigma$ is a so-called
\index{ordering permutation}\emph{ordering permutation} of $\vect{x}$.
Generally (if there are tied observations in $\vect{x}$) such a permutation
might be ambiguous, so we assume that $\sigma$ is the
\index{stable ordering permutation}\emph{stable ordering permutation}:
for $\vect{x}=(1, 2, 1, 2, 1)$ we always get $\sigma=(1, 3, 5, 2, 4)$.
The linear order $\sqsubseteq$ used here is such that $x_i\sqsubseteq x_j$ whenever
$x_i< x_j \text{ or } (x_i=x_j \text{ and } i\le j)$.

\begin{lstlisting}[language=R]
x <- c(13, 11, 12, 11, 11)
o <- order(x) # a (stable) ordering permutation
o
## [1] 2 4 5 3 1
x[o[1]] # the smallest value in x
## [1] 11
x[o[5]] # the largest value in x
## [1] 13
\end{lstlisting}

\item The uniform distribution on a set $A$ is denoted with $\mathrm{U}A$,
e.g., $\mathrm{U}[0,1]$ or $\mathrm{U}\{-1,1\}$.
The normal distribution with expected value of $\mu$ and standard deviation of $\sigma$
is denoted with $\mathrm{N}(\mu, \sigma)$.
\end{itemize}
Regardless of the differences in vector indexing in the \lang{Python} programming language,
similar code chunks could have been provided for \texttt{ndarray}s
defined in the \package{NumPy} package.

\medskip
Let us also note that how \lang{C++} code can seamlessly  be integrated in \lang{R}
(for instance, to speed up computations, access external libraries, or make use
of lower-level programming concepts, like dynamic data structures).
For that, we use the \package{Rcpp} package \cite{EddelbuettelFrancois2013:rcppbook}.

\lang{C++} source files may be turned into a dynamically linked library
(automatically loaded by \lang{R}) via a call to:

\begin{lstlisting}[language=R,showstringspaces=false]
Rcpp::sourceCpp('filename.cpp')
\end{lstlisting}

For quite simple functions, their \lang{C++} code may be provided inline in the \lang{R} console.
Here is an exemplary function which takes a single numeric argument and
returns a single numeric value:

\begin{lstlisting}[language=R,showstringspaces=false]
Rcpp::cppFunction('
   double square(double x) {
      return x*x;
   }
')
\end{lstlisting}
Equivalently, a complete \lang{C++} source file may be written:
\begin{lstlisting}
#include <Rcpp.h>
// [[Rcpp::plugins("cpp11")]]
using namespace Rcpp;

// [[Rcpp::export]]
double square(double x) {
   return x*x;
}
\end{lstlisting}
Usage in \lang{R}:
\begin{lstlisting}[language=R,showstringspaces=false]
square(2)
## [1] 4
\end{lstlisting}

Moreover, the following function takes a vector as input and returns
a vector of the same size:

\begin{lstlisting}[language=R,showstringspaces=false]
Rcpp::cppFunction('
   NumericVector square_vec(NumericVector x) {
      int n = x.size();
      NumericVector y(n);
      for (int i=0; i<n; ++i)
         y[i] = x[i]*x[i];
      return y;
   }
')

square_vec(c(-1, 2.5, 0))
## [1] 1.00 6.25 0.00
\end{lstlisting}

In this book we use \lang{R} and \lang{C++} to
implement the discussed algorithms. As a \lang{Python} alternative to \package{Rcpp},
we suggest, e.g., \package{Cython} or \package{boost::python}.

\clearpage{\pagestyle{empty}\cleardoublepage}
\mainmatter

\pagestyle{fancy}
\fancyfoot{}

\renewcommand{\chaptermark}[1]{\markboth{{#1}}{}}
\renewcommand{\sectionmark}[1]{\markright{{#1}}{}}

\fancyhf{}
\fancyhead[LO]{\sffamily\footnotesize\thesection.~\rightmark\hfill \sffamily\footnotesize\thepage}
\fancyhead[RO]{}
\fancyhead[LE]{}
\fancyhead[RE]{\flushleft\sffamily\footnotesize\thepage\hfill\sffamily\footnotesize\thechapter.~\leftmark}

\clearpage{\pagestyle{empty}\cleardoublepage}
\chapter{Aggregation of univariate data}\label{Chapter:onedim}

\lettrine[lines=3]{C}{lassically}, the theory of aggregation discusses
methods to summarize \hbox{$n\ge 2$} numeric quantities in some real interval $\Ival=[a,b]$
or $]a,b[$, $a<b$. %
It is assumed that these quantities represent the
results of \textit{measurements} of the same process,
for instance decision makers' preference degrees towards some alternative,
or outputs gathered from sensors of the same kind (thermometers,
traffic speed guns, personality questionnaires in psychology, and so forth).
Of course, further on we shall discuss more complex methods, e.g., aggregating
an arbitrary number of elements (so-called extended fusion functions),
elements on discrete scales (nominal or ordered, like character strings),
more complex objects (like vectors in $\mathbb{R}^d$ for $d>1$ or DNA sequences),
as well as determining numeric characteristics of entities.
Before this happens, our universe of discourse appears to be quite simple
at first glance, both from the mathematical and computational perspective.
However, the kind reader should not be misled by that impression:
the purpose of this introduction is not only to establish basic
notation and key ideas. It shall turn out that even in such an uncomplicated
domain a practitioner is faced with many challenges and interesting issues.

\section{Preliminaries}

To get a general idea of objects that are of our interest
in this  chapter, let us introduce the following definition.

\begin{definition}[\cite{BustinceETAL2014:directionalmonotonicity,BustinceETAL2015:directionalmonotonicity}]
\index{fusion function}\label{Def:FusionFunction}
A \emph{fusion function} is a mapping $\func{F}:\IvalPow{n}\to\Ival$.
\end{definition}

The notion of a fusion function reflects the abstract aim of data fusion:
we take $n$ numbers from some domain and, as a result, get one value
of the same \textit{type}.
For instance, in decision making and fuzzy logic
we often suppose that $\Ival=[0,1]$ or $\Ival=[-1,1]$ and in statistics that $\Ival=]-\infty, \infty[$.
We shall see that the choice of interval $\Ival=[a,b]$ may be crucial;
some of the results presented below may hold only if, e.g., $a=0$
or $b=\infty$ but not otherwise.

\begin{example}\index{Sum@$\mathsf{Sum}$}%
Consider a mapping defined as:
\[\func{Sum}(\vect{x})=\sum_{i=1}^n x_i.\]
It is a fusion function if, e.g., $\Ival=[0,\infty]$
or $[-\infty,\infty]$, but not if $\Ival=[0,1]$ or $[-1,1]$.
\end{example}

\smallskip
Let us review some general cases where fusion functions in $\IvalPow{n}$
are applicable and introduce some well-known data aggregation tools.

\begin{example}\label{Ex:intro:stats0}
Assume that we are given a realization $(x_1,\dots,x_n)$
of a random sample of independent random variables
following a common distribution $D$ with support $\Ival$.
This may denote the results of an IQ test that was taken by a group of students.
Knowing that $D$ is symmetric around some value $t$,
how can we estimate $t$ so that one group of pupils
may be compared to some reference value?
Among examples of fusion functions applicable in this case we find:
\begin{itemize}
   \item \index{AMean@$\mathsf{AMean}$|see {arithmetic mean}}\index{arithmetic mean}%
   $\func{AMean}(\vect{x})=\displaystyle \frac{1}{n} \sum_{i=1}^n x_i$, \hfill(\emph{arithmetic mean})

   \item \index{median}%
   $\func{Median}(\vect{x})=\left\{
   \begin{array}{ll}
      x_{((n+1)/2)} & \text{if $n$ is odd,}\\
      \left(x_{(n/2)}+x_{(n/2+1)}\right)/2 & \text{if $n$ is even.}
   \end{array}
   \right.$ \hfill(\emph{median})
\end{itemize}
\end{example}

Note that the sample median may be written as:
\[\func{Median}(\vect{x})=\frac{x_{\lfloor(n+1)/2\rfloor}+x_{\lceil(n+1)/2\rceil}}{2}\]
and that it is defined using {order statistics}, which are also types of fusion functions.
Namely, for any $k\in[n]$, we may define:
\index{OS@$\mathsf{OS}_k$|see {order statistic}}\index{order statistic}%
\[
\func{OS}_k(\vect{x})=x_{(k)}.
\]
As we shall see in Section~\ref{Sec:AgLatClass}, the two following instances of order statistics
are particularly noteworthy:
\begin{itemize}
   \item \index{Min@$\mathsf{Min}$|see {minimum}}\index{minimum}%
   $\displaystyle \func{Min}(\vect{x})=\func{OS}_1(\vect{x})=\bigwedge_{i=1}^n x_i$, \hfill(\emph{minimum})

   \item \index{Max@$\mathsf{Max}$|see {maximum}}\index{maximum}%
   $\displaystyle \func{Max}(\vect{x})=\func{OS}_n(\vect{x})=\bigvee_{i=1}^n x_i$. \hfill(\emph{maximum})
\end{itemize}

Also, apart from the arithmetic mean, the reader is possibly familiar
with two other types of means:
\begin{itemize}
\item \index{GMean@$\mathsf{GMean}$|see {geometric mean}}\index{geometric mean}%
$\func{GMean}(\vect{x})=\displaystyle\left(\prod_{i=1}^n x_i\right)^{1/n}$, \hfill(\emph{geometric mean})

\item
$\func{HMean}(\vect{x})=\displaystyle\frac{1}{\frac{1}{n}\sum_{i=1}^n\frac{1}{x_i}}$.
\index{HMean@$\mathsf{HMean}$|see {harmonic mean}}\index{harmonic mean}%
\hfill(\emph{harmonic mean})
\end{itemize}

It is well-known that for $a>0$ we have:
\[
   \func{Min}(\vect{x})\le \func{HMean}(\vect{x})\le \func{GMean}(\vect{x})\le \func{AMean}(\vect{x})\le \func{Max}(\vect{x}).
\]
This is in fact one of the first results in the theory of aggregation
-- the Greek mathematicians studied its simplest case $(n=2)$ over 2000 years ago.

\begin{example}\label{Ex:intro:stats}
Let us go back to Example \ref{Ex:intro:stats0}.
Knowing that some of the input observations were contaminated and
that now outliers possibly occur in our data set
(e.g., because the students were not focused enough while performing the tasks),
how can we choose a fusion function $\func{F}$ so that $\func{F}(x_1,x_2,\dots,x_n)$
is a {\normalfont plausible} estimator of $D$'s center point?
Among possible choices we find:
\begin{itemize}
\item \index{TriMean@$\mathsf{TriMean}_k$|see {trimmed mean}}\index{trimmed mean}%
$\func{TriMean}_k(\vect{x}) = \displaystyle \frac{1}{n-2k} \sum_{i=k+1}^{n-k} x_{(i)}$, \hfill(\emph{trimmed mean})
\item \index{WinMean@$\mathsf{WinMean}_k$|see {Winsorized mean}}\index{Winsorized mean}%
$\func{WinMean}_k(\vect{x}) = \displaystyle \frac{1}{n} \sum_{i=k+1}^{n-k} x_{(i)} + \frac{k}{n} x_{(k+1)} + \frac{k}{n} x_{(n-k)}$,

\hfill(\emph{Winsorized mean})
\end{itemize}
for some $k\in \{0,1,\dots,\lfloor n/2\rfloor-1 \}$.
\end{example}

\begin{example}[\cite{BeliakovETAL2007:aggregationpractitioners}]\label{Ex:intro:rbsAND}
Suppose that we have a rule-based system with rules of the form:
\[
\mathtt{IF}\ o_1\ \mathtt{is}\ O_1\ \mathtt{AND}\ o_2\ \mathtt{is}\ O_2\ \mathtt{AND}\ \dots\ \mathtt{AND}\ o_n\ \mathtt{is}\ O_n\ \mathtt{THEN}\dots,
\]
and that $x_i$ denotes the degree of satisfaction of the predicate
``$o_i\ \mathtt{is}\ O_i$'', $x_i\in[0,1]$.
At this point, $0$ may be interpreted as ``no satisfaction'',
$1$ as ``complete satisfaction'', and intermediate values
can depict partial degrees of compliance.
Then the overall degree of satisfaction of all the rules
may be referred to as $\func{F}(x_1,x_2,\dots,x_n)$.
For the sake of this purpose the following fusion functions are sometimes used:
\begin{itemize}
   \item \index{Prod@$\mathsf{Prod}$|see {product}}\index{product}%
   $\func{Prod}(\vect{x})=\displaystyle \prod_{i=1}^n x_i$, \hfill(\emph{product})
   \item \index{minimum}%
   $\func{Min}(\vect{x})$, \hfill(\emph{sample minimum})
   \item \index{TL@$\mathsf{T}_{\mathrm{Ł}}$|see {Łukasiewicz t-norm}}\index{Lukasiewicz t-norm@Łukasiewicz t-norm}%
   $\func{T}_\mathrm{Ł}(\vect{x})=\displaystyle 0\vee \left(\sum_{i=1}^n x_i - n + 1\right)$. \hfill(\emph{Łukasiewicz t-norm})
\end{itemize}
\end{example}

\begin{example}\label{Ex:intro:rbsOR}
Now let us assume that we have a rule-based system with rules of the form:
\[
\mathtt{IF}\ o_1\ \mathtt{is}\ O_1\ \mathtt{OR}\ o_2\ \mathtt{is}\ O_2\ \mathtt{OR}\ \dots\ \mathtt{OR}\ o_n\ \mathtt{is}\ O_n\ \mathtt{THEN}\dots.
\]
Assuming that $x_i\in[0,1]$, as in the previous example, we may take into account
the fusion functions:
\begin{itemize}
   \item \index{maximum}%
   $\func{Max}(\vect{x})$, \hfill(\emph{sample maximum})
   \item \index{SL@$\mathsf{S}_{\mathrm{Ł}}$|see {Łukasiewicz t-conorm}}\index{Lukasiewicz t-conorm@Łukasiewicz t-conorm}%
   $\func{S}_\mathrm{Ł}(\vect{x})=\displaystyle 1\wedge \left(\sum_{i=1}^n x_i\right)$, \hfill(\emph{Łukasiewicz t-conorm}, \emph{bounded sum})
   \item \index{SD@$\mathsf{S}_\mathrm{D}$|see {Drastic t-norm}}\index{drastic t-conorm}%
   $\func{S}_\mathrm{D}(\vect{x})=\displaystyle \left\{
   \begin{array}{ll}
      x_{(n)} &\text{if } x_{(n-1)}=0,\\
      1 & \text{otherwise}.\\
   \end{array}
   \right.$ \hfill(\emph{drastic t-conorm})
\end{itemize}
\end{example}

\begin{example}\label{Ex:intro:weightedbipolar}
Similarly, in group decision making problems,
$x_i$ may designate the degree of preference of the $i$th expert towards an
alternative. Here, $\func{F}$ may be used to combine individual evaluations
to obtain a global score, $\func{F}(x_1,x_2,\dots,x_n)$.
Then, e.g., a bipolar scale (see \cite{DuboisPrade2004:useagopfusion} for discussion),
$\Ival=[-1,1]$, may be used, where $-1$ stands for ``strongly disagree'',
and $1$ for ``strongly agree''.

Additionally, suppose that the experts have different ``esteem'', i.e., some
of them have stronger impact on the final decision than the others
(this is the case of, e.g., stockholders in a company's board).
Assuming that the $i$th expert is assigned weight $w_i\ge 0$,
$\sum_{j=1}^n w_j=1$, $\func{F}$ is often set to be
\index{convex combination}%
a convex combination of input values, that is:
\begin{itemize}
   \item \index{WAMean@$\mathsf{WAMean}$|see {weighted arithmetic mean}}\index{weighted arithmetic mean}%
   $\func{WAMean}_\vect{w}(\vect{x})=\displaystyle \sum_{i=1}^n w_i x_i$. \hfill(\emph{weighted arithmetic mean})
\end{itemize}
\end{example}

\paragraph{FP arithmetic.}
Before going any further let us make a remark concerning
the representation of values in $\Ival\subseteq\mathbb{R}$ on modern computers.

\begin{definition}\label{Def:FP}
\index{floating point numbers}%
For some $s,m\in\mathbb{N}$ and $\mathbb{N}\ni b \ge 2$ let:
\begin{equation}\label{Eq:FPset}
\mathbb{F}_{s,m}^b = \left\{ \pm\sum_{i=0}^{s-1} d_i b^{j-i}:
d_i \in [0:b-1], %
j\in [-m:m]
\right\}\subseteq \mathbb{R}
\end{equation}
denote the set of signed \emph{floating point numbers} with \emph{precision} of $s$ significant digits,
\emph{base} $b$, and \emph{exponent} ranging in $\{-m,\dots,m\}$.
\end{definition}

In particular, if $b=2$, then we have numbers in the binary representation
(e.g., $1.0101_2\cdot 2^4 = 21_{10}$),
and if $b=10$, then we get decimal numbers (e.g., $3.1415_{10}\cdot 10^0$).
Equation~\eqref{Eq:FPset} may be rewritten equivalently as:
\[
\mathbb{F}^b_{s,m} = \left\{ \pm (d_0.d_1d_2\dots d_{{s-1})_b} \cdot b^j, %
d_i \in [0:b-1]: j\in[-m:m]\right\}.
\]
In order to assure that each number $\mathbb{F}_{s,m}^b$ has an unambiguous
representation, we may assume that $d_0\neq 0$ (normalized form).

For some fixed $b,s,m$, let $\bar{\mathbb{F}}^b_{s,m} = \mathbb{F}^b_{s,m}\cup\{\pm\mathtt{Inf}, \mathtt{NaN}\}$,
i.e., the set of extended floating point numbers that also includes
signed infinities (representing values so small or so large that they do not
fit in $\mathbb{F}^b_{s,m}$)
and a not-a-number (an erroneous value, for results of operations like $\sqrt{-1},\log(-1),0/0\not\in\mathbb{R}$).

\begin{definition}
For fixed $b,s,m$, let $\mathsf{fp}: \mathbb{R}\to\bar{\mathbb{F}}^b_{s,m}$ be such that
for arbitrary $x\in\mathbb{R}$ we have:
\begin{equation}
\mathsf{fp}(x)=\left\{
\begin{array}{ll}
\phantom{-}\mathtt{Inf} & \text{if }x > \max\mathbb{F}^b_{s,m},\\
-\mathtt{Inf} & \text{if }x < \min\mathbb{F}^b_{s,m},\\
\arg^*\min_{y\in\mathbb{F}^b_{s,m}} |x-y| & \text{otherwise}.
\end{array}
\right.
\end{equation}
\end{definition}

Thus, if $x\in\mathsf{range}(\mathbb{F}^b_{s,m})$, then
$\mathsf{fp}(x)$ rounds $x$ to the closest value in $\mathbb{F}^b_{s,m}$.
Of course, such a rounding scheme may be ambiguous if $b$ is even,
e.g., in $\mathbb{F}_{3,2}^{10}$ the value $1{,}005_{10}\in\mathbb{R}$
can be represented as $1{,}00_{10}$ and $1{,}01_{10}$.
In order for the function to be well defined,
\index{IEEE-754}\index{round half to even}%
we should introduce some tie-breaking rule (and hence the informal notation $\arg^*\min$).
Here we shall rely on the \package{IEEE-754} standard which
suggests the \textit{round half to even} scheme, that is, $d_{s-1}$
should always be even.

\begin{definition}\index{machine epsilon}%
For fixed $b,s,m$,
the \emph{machine epsilon} is the greatest value $\varepsilon_\mathrm{M}>0$
such that $\mathsf{fp}(1+\varepsilon_\mathrm{M})= 1$.
\end{definition}

\begin{proposition}
In any $\mathbb{F}^b_{s,m}$ it holds that $\varepsilon_\mathrm{M}=b^{-s+1}/2$.
\end{proposition}

For example, in $\mathbb{F}_{3,2}^{10}$
we have $\varepsilon_\mathrm{M}=5\times 10^{-3}=0.005$:
it holds $\mathsf{fp}(1+0.005)=\mathsf{fp}(1.005)=1$
as well as $\mathsf{fp}(1.005+0.0\dots 1)=1.01$.

The machine epsilon gives us an upper bound for the relative rounding error.
This is because for each $0\neq x\in\mathsf{range}(\mathbb{F})$ we have:
\[
   \left|\frac{x-\mathsf{fp}(x)}{x}\right| = \left|1-\frac{\mathsf{fp}(x)}{x}\right| \le \varepsilon_\mathrm{M}.
\]
Also, please note that there always exists $\delta\in[-\varepsilon_\mathrm{M},\varepsilon_\mathrm{M}]$
such that $\mathsf{fp}(x)=x(1+\delta)$.

It turns out that the \texttt{double} type, most often used for floating point computations
\index{double type@\texttt{double} type}%
on modern computers, is roughly equivalent to $\bar{\mathbb{F}}_{52/53,1023}^2$
according to \package{IEEE-754} (we omit issues concerning, among others, subnormal numbers).
In this case, the machine epsilon is equal to $2^{-53}$.

\begin{remark}
${\mathbb{F}^b_{s,m}}$ is not closed with respect to the standard addition operation:
$({\mathbb{F}^b_{s,m}}, +)$ is not a subalgebra of $(\mathbb{R}, +)$.
For instance, in $\mathbb{F}_{3,2}^{10}$ we have $0.001_{10}+1_{10}=1.001_{10}\not\in\mathbb{F}^{10}_{3,2}$.
In other words, even if two values are representable in $\mathbb{F}^b_{s,m}$ exactly,
the result of applying ``$+$'' is might not necessarily be  exact.
\end{remark}

Let $\oplus:\bar{\mathbb{F}}^b_{s,m}\times\bar{\mathbb{F}}^b_{s,m}\to\bar{\mathbb{F}}^b_{s,m}$
be such that for $x,y\in\mathbb{R}$ and
$\mathsf{fp}(\mathsf{fp}(x)+\mathsf{fp}(y)), \mathsf{fp}(x), \mathsf{fp}(y)\in\mathbb{F}^b_{s,m}$,
it holds:
\begin{equation}
\mathsf{fp}(x)\oplus\mathsf{fp}(y) = \mathsf{fp}(\mathsf{fp}(x)+\mathsf{fp}(y)).
\end{equation}
Moreover, let $\mathtt{NaN}\oplus \tilde{z}=\tilde{z}\oplus \mathtt{NaN}=\mathtt{NaN}$
for $\tilde{z}\in\bar{\mathbb{F}}^b_{s,m}$,
$\mathtt{Inf}\oplus \tilde{z} = \tilde{z}\oplus\mathtt{Inf}=\mathtt{Inf}$,
and~$\mathtt{-Inf}\oplus \tilde{z} = \tilde{z}\oplus\mathtt{-Inf}=\mathtt{-Inf}$
for  $\tilde{z}\in\mathbb{F}^b_{s,m}$, as well as
$\mathtt{Inf}\oplus\mathtt{-Inf}=\mathtt{-Inf}\oplus\mathtt{Inf}=\mathtt{NaN}$.
This is a typical redefinition of the ordinary addition operation so that
it acts on elements in $\bar{\mathbb{F}}^b_{s,m}$. Other arithmetic operations,
e.g., $\ominus, \otimes, \oslash$, may be introduced in a similar manner.

\begin{remark}
The $\oplus$ operation is not necessarily associative,
i.e., for $\tilde{x},\tilde{y},\tilde{z}\in\bar{\mathbb{F}}^b_{s,m}$
we may have $(\tilde{x}\oplus\tilde{y})\oplus\tilde{z}\neq\tilde{x}\oplus(\tilde{y}\oplus\tilde{z})$.
For example, in $\bar{\mathbb{F}}_{3,2}^{10}$ it holds that:
\begin{eqnarray*}
(0.005_{10}\oplus 0.025_{10})\oplus 1.00_{10} &=&\\
0.03_{10}\oplus 1.00_{10} &=&\\
1.03_{10}
&\neq& 0.005_{10}\oplus (0.025_{10}\oplus 1.00_{10})\\
&=& 0.005_{10}\oplus 1.02_{10} \\
&=& 1.02_{10}.
\end{eqnarray*}
\end{remark}

\begin{remark}
Let us study the absolute error of the $\oplus$ operation. Let
$\tilde{x}\oplus\tilde{y}\oplus\tilde{z}=(\tilde{x}\oplus\tilde{y})\oplus\tilde{z}$,
where $0<\tilde{x},\tilde{y},\tilde{z}\in\mathbb{F}^b_{s,m}$
and $\tilde{x}\oplus\tilde{y}\oplus\tilde{z}\in\mathbb{F}^b_{s,m}$.
For some $\delta_1,\delta_2\in[-\varepsilon_\mathrm{M},\varepsilon_\mathrm{M}]$ we have:
\begin{eqnarray*}
{\tilde{x}\oplus\tilde{y}\oplus\tilde{z} - (\tilde{x}+\tilde{y}+\tilde{z})}
& = & \mathsf{fp}(\mathsf{fp}(\tilde{x}+\tilde{y})+\tilde{z}) - (\tilde{x}+\tilde{y}+\tilde{z}) \\
& = & \delta_1\tilde{x}+\delta_1\tilde{y}+\delta_2(\tilde{x}+\tilde{y}+\delta_1\tilde{x}+\delta_1\tilde{y}+\tilde{z}) \\
& \le & \varepsilon_\mathrm{M}((2+\varepsilon_\mathrm{M})(\tilde{x}+\tilde{y})+\tilde{z}).
\end{eqnarray*}
The relative error of $\oplus$ is not greater than
$\varepsilon_\mathrm{M}(1+(\tilde{x}+\tilde{y})(1+\varepsilon_\mathrm{M}))/(\tilde{x}+\tilde{y}+\tilde{z})$.
However, we observe that this upper bound depends on the relative magnitude of the inputs:
$\tilde{z}\ge\tilde{x}$ and~$\tilde{z}\ge\tilde{y}$ leads to the largest error.
\index{Sum@$\mathsf{Sum}$}%
From that we may imply that, e.g., the $\mathsf{Sum}$ fusion function imposes
the smallest relative error if we add up nonnegative values in an increasing order.
\end{remark}

We see that even though the fusion functions studied so far seemed to
be very uncomplicated, special care should be taken when they are implemented
on a computer. In such a setting, they of course are
mappings like $\bar{\func{F}}:\bar{\Ival}^{n}\to\bar{\Ival}$,
where $\bar{\Ival}=[a,b]\cap\bar{\mathbb{F}}^b_{s,m}$, $a,b\in\bar{\mathbb{F}}^b_{s,m}$.
For the sole $\func{Sum}$ function there exists a number
of algorithms; one of them is the Kahan (compensated) summation routine \cite{Kahan1965:trunc},
see also \cite{Higham1993:accuracysum}.

\begin{remark}
The $\oplus$ and $\otimes$ operations are not distributive in general:
For example in $\bar{\mathbb{F}}_{3,2}^{10}$ we have:
\begin{eqnarray*}
(0.001_{10}\otimes 0.1_{10})\oplus (0.001_{10}\otimes 1.00_{10}) &=&\\
 0.000_{10}\oplus 0.001_{10} &=&\\
 0.001_{10}
&\neq& 0.001_{10}\otimes (0.1_{10}\oplus 1.00_{10})\\
&=& 0.001_{10}\otimes 0.1_{10} \\
&=& 0.000_{10}.
\end{eqnarray*}
\end{remark}

The reader is suggested to refer, e.g., to \cite[Section~4.2]{Knuth1998:acpvol2},
\cite{Higham}, or \cite{Goldberg1991} for further discussion and issues on the topic.

\begin{remark}
There are a few libraries for performing floating point computations
with higher precision, for instance
\package{MPFR} (Multiple Precision Floating-Point Reliable)
and \package{GMP} (GNU Multiple Precision) libraries.
Unfortunately, the errors in numerical computations are inherent,
they may only be reduced.
This of course comes at a cost of slowing down the computations.

As an alternative, one may consider computer algebra systems performing
\emph{symbolic} computations, e.g., \package{Mathematica}, \package{Maxima},
\package{Maple}, or \package{Sage}.
\end{remark}

\begin{example}
The only discussed so far fusion functions
$\bar{\func{F}}:\bar{\Ival}^{n}\to\bar{\Ival}$ that produce exact values
are $\func{OS}_k$ for arbitrary $k\in[n]$ (but not $\func{Median}$ for arbitrary $n$)
and $\func{S}_\mathrm{D}$.
Generally, among precise operations in $\bar{\mathbb{F}}$
we find $\wedge$ and $\vee$, which are the basis for the class of
weighted lattice polynomial functions discussed in Section~\ref{Sec:LatPolyFun},
see also Equation~\eqref{Eq:WLPFIval}.
\end{example}

\section{Properties of fusion functions}

The definition of a fusion function we presented above is very general.
Thus, we would like to narrow it down and identify some crucial properties
that must always be fulfilled in order to say that $\func{F}$ might
at least be potentially interesting to us. This, however, is relative to
the nature of the practical problem we are faced with.

In the following subsections we therefore explore some noteworthy frameworks,
which include nondecreasingness, symmetry, idempotence, different types of equivariances,
additivity and so forth.

\subsection{Nondecreasingness and preservation of end points}

In Examples \ref{Ex:intro:rbsAND}, \ref{Ex:intro:rbsOR}, and \ref{Ex:intro:weightedbipolar} it seems that
it is reasonable to require that if we increase the degree of satisfaction
of a predicate or the degree of preference stated by the $i$th expert,
then the new overall valuation should not be smaller than the previous one.
Such a property may be formalized as follows.
\index{relation le@relation $\le_n$}%
Let  $\le_n$ be a binary relation on $\IvalPow{n}$ such that
$\vect{x}\le_n\vect{y}$ if for all $i\in[n]$ we have $x_i\le y_i$.

\begin{definition}\index{nondecreasingness}\label{Def:Nondecreasing}
A fusion function $\func{F}:\IvalPow{n}\to\Ival$
is called \emph{nondecreasing} (in each variable),
whenever for all $\vect{x},\vect{y}\in\IvalPow{n}$ it holds that
if $\vect{x}\le_n\vect{y}$, then $\func{F}(\vect{x})\le\func{F}(\vect{y})$.
\end{definition}

\begin{remark}
All the fusion functions reviewed so far are nondecreasing.
\end{remark}

We may also define a \emph{strictly increasing} function
by assuming that $\vect{x}<_n\vect{y}\Rightarrow\func{F}(\vect{x})<\func{F}(\vect{y})$,
where $\vect{x}<_n\vect{y}$ if and only if
\index{relation lt@relation $<_n$}%
$\vect{x}\le_n\vect{y}$ and $\vect{x}\neq\vect{y}$.
Moreover, \emph{unanimous increasingness} (compare \cite{GrabischETAL2009:aggregationfunctions}),
also known as joint strict monotonicity,
can be defined by considering the cases
in which for all $i\in[n]$ it holds $x_i<y_i$.
\index{unanimous increasingness}%

\begin{example}
Among strictly increasing fusion functions we find, e.g.,
$\func{AMean}$. Moreover, $\func{Min}$ and $\func{Max}$
are unanimously increasing.
\end{example}

Moreover, we may require that $\func{F}$ should at least be
normalized in such a way that it preserves the endpoints of $\Ival$.

\begin{definition}\label{Def:EndpointPreserving}
\index{endpoint preservation}%
We say that a fusion function $\func{F}$ is \emph{endpoint-preserving}, %
whenever it holds that $\func{F}(n\ast a)=a$ and $\func{F}(n\ast b)=b$.
\end{definition}

In other words, e.g., in a decision making problem,
if the criteria are not satisfied at all
or each expert finds an alternative totally plausible,
then in such extreme cases the result should be concordant with the inputs.
Note that if $\Ival$ is an open interval, then by, e.g., $\func{F}(n\ast a)$
would of course mean $\func{F}(n\ast a)=\lim_{\vect{x}\to (n\ast a)} \func{F}(\vect{x})$.

\begin{example}
$\func{Prod}$ is an endpoint-preserving fusion function for input elements in $[0,1]$,
but not when $\Ival=[0,0.5]$
or $\Ival=[-1,1]$ and even $n$ is considered.
Moreover, it is not nondecreasing, e.g., in the $[-1,1]$ case.
Hence, we see that some properties indeed depend on the choice of $\Ival=[a,b]$
as well as $n$.

On the other hand, $\func{F}(\vect{x})=b\wedge \func{Prod}(\vect{x})$ is
endpoint-preserving in the case $\Ival=[0,b]$ for any $b\ge 1$. We shall often observe that
some fusion functions may be ``tuned up'': by applying particular transformations
they start to fulfill a given property which is of interest in a particular domain.
\end{example}

Properties given in Definitions \ref{Def:Nondecreasing}
and \ref{Def:EndpointPreserving} lead us to the classical definition of an aggregation function,
as in \cite{GrabischETAL2009:aggregationfunctions,BeliakovETAL2007:aggregationpractitioners,BeliakovETAL2015:practicalbook}.

\begin{definition}
\index{aggregation function (classical)}%
$\func{F}:\IvalPow{n}\to\Ival$ is an \emph{aggregation function}
whenever it is nondecreasing in each variable and it is endpoint-preserving.
\end{definition}

Note that if $\func{F}$ is nondecreasing,
then it is endpoint-preserving if and only if
$\inf_{\vect{x}\in\IvalPow{n}} \func{F}(\vect{x}) =a$ and
$\sup_{\vect{x}\in\IvalPow{n}} \func{F}(\vect{x})= b$.

\subsection{Idempotence and internality}

From another standpoint, fusion functions to be used in application domains
like those mentioned in Examples \ref{Ex:intro:stats0} and \ref{Ex:intro:stats},
might not necessarily be nondecreasing.
For example, already Kolmogorov and Nagumo
\cite{Kolmogorov1930:moyenne,Nagumo1930:mittelwerte},
compare also Aczel's paper \cite{Aczel1948:onmeanvalues},
were interested in discussing various kinds of \emph{means}.
One of the properties they required is the so-called {idempotency}
(unanimity, compensativity), which is well-known from algebra,
where we say that element $x$ is idempotent with respect
to a binary operator $*$, if we have $x*x=x$. The following definition
extends this property to $n$-ary aggregation functions,
see \cite{GrabischETAL2009:aggregationfunctions}.

\begin{definition}\label{Def:idempotency}\index{idempotency}\index{compensativity|see {idempotency}}\index{unanimity|see {idempotency}}%
A fusion function $\func{F}$ is called \emph{idempotent}, whenever:
\begin{equation}
(\forall x\in\Ival)\ \func{F}(n\ast x)=x.
\end{equation}
\end{definition}

Intuitively, if we aggregate $n$ equal inputs, the resulting
value should fully agree with them.
Note that each idempotent fusion function is also endpoint-preserving.
Among idempotent aggregation functions we find
$\func{WAMean}$ and $\func{OS}_k$, but not $\func{T}_\mathrm{Ł}$
and $\func{S}_\mathrm{D}$.

\begin{remark}
Let $n=10$, $\overline{\func{HMean}}(\vect{x})=n\oslash( (1\oslash x_1) \oplus\dots\oplus (1\oslash x_n))$
(the floating point equivalent to the harmonic mean with respect to the \texttt{double} type),
and $x$ be equal to: \[ +1.1101110110000111101111001100111110100011100111101111_2\times 2^0, \]
that is $x \simeq 1.865352440541410805608_{10}$.
Then $\overline{\func{HMean}}(n\ast x)$ is equal to:
\[+1.1101110110000111101111001100111110100011100111110000_2\times 2^0,\]
i.e., $x\neq \overline{\func{HMean}}(n\ast x)\simeq 1.8653524405414110_{10}$.
Even if -- algebraically -- $\func{HMean}$ is idempotent,
its ``computer version'' is not. Therefore, one should be careful when comparing
results of floating point computations, especially with the $=$ operator.
A much more reliable way to do so is to test whether
$|\func{F}(n\ast x)-x|/|x|\le \varepsilon$ for some small $\varepsilon$
\index{machine epsilon}%
being a function of the machine epsilon,
e.g., $\varepsilon=\sqrt{\varepsilon_\mathrm{M}}$.
\end{remark}

Another significant property -- internality
(as named in \cite[Definition 2.53]{GrabischETAL2009:aggregationfunctions}),
also known as compensativity -- requires that a fusion function's output value
must lie ``somewhere in-between'' the input values
(see page~\pageref{Internality2} for an alternative setting).

\begin{definition}%
\index{internality}%
A fusion function $\func{F}$ is \emph{internal}
whenever $(\forall \vect{x}\in\IvalPow{n})$
we have:
\begin{equation}
\func{Min}(\vect{x})\le\func{F}(\vect{x})\le\func{Max}(\vect{x}).
\end{equation}
In other words, $\func{F}(\vect{x})\in\intcl{\vect{x}}$.
\end{definition}

We see that each internal $\func{F}$ is idempotent too.
We often consider fusion functions which are
both idempotent and nondecreasing. Actually,
idempotent aggregation functions are sometimes called
\index{averaging function}%
\emph{averaging functions} in the literature, compare \cite{GrabischETAL2009:aggregationfunctions}.
It turns out that in such a case idempotency and internality coincide,
see \cite[Proposition 2.54]{GrabischETAL2009:aggregationfunctions} for the proof.

\begin{proposition}\label{Proposition:internalVSidempotent}
If $\func{F}$ if nondecreasing and idempotent,
then it is internal.
\end{proposition}

Please observe that nondecreasingness is appealing from the mathematical perspective:
it turns out that many other properties can be simplified owing to
this property. Though, in some applications it may not be fully desirable.

\begin{remark}\label{Remark:MeanCauchy}
\index{mean (Cauchy sense)}%
\index{mean (Gini sense)}%
According to \cite{GrabischETAL2009:aggregationfunctions},
already Cauchy in 1821 considered under a name \emph{mean}
an internal, but not necessarily nondecreasing fusion function.
Similarly, compare \cite{BeliakovETAL2015:practicalbook},
Gini in the 1950s required only this very property when discussing various means.
\end{remark}

\begin{remark}\label{Remark:outliers1d}
A class of idempotent, but not necessarily nondecreasing
fusion functions may be useful in the case of aggregating
data in the presence of outliers: we might sometimes
want to allow that $F(0,0,\dots,0,1) > F(0,0,\dots,0,10^9)$,
just as in Example~\ref{Ex:intro:stats}.
\index{outlier}%
For instance, it is not uncommon to define outliers
(e.g., when building box-and-whisker plots)
as observations $x_i$ such that
$x_i < \func{Q}_{0.25}(\vect{x})-1.5(\func{Q}_{0.75}(\vect{x})-\func{Q}_{0.25}(\vect{x}))$ or
$x_i > \func{Q}_{0.75}(\vect{x})+1.5(\func{Q}_{0.75}(\vect{x})-\func{Q}_{0.25}(\vect{x}))$, where
$\func{Q}_{0.25}$ and $\func{Q}_{0.75}$ stand for the 1st and the 3rd
quartile, compare Example~\ref{Example:quantiles}.
Then, having $\func{F}$ defined as ``the arithmetic mean
of all non-outlying observations'',
we get, e.g., $0.4=\func{F}(-2,-1,0,1,4) \not\le \func{F}(-2,-1,0,1,5)=-0.5$.
In Section~\ref{Sec:OtherMonotonicities} we shall make a review of other types
of monotonicities.
\end{remark}

\begin{remark}\label{Remark:mode}\index{mode}%
The \emph{mode}, well known in exploratory data analysis, is
defined as an observation that appears most often in the input data set.
Of course, such a definition is not strict in the case of multimodal
vectors. What is important here, however, is that it is an idempotent yet
not monotone (at least with respect to $\le_n$) fusion function.
It is because we have, e.g.,
$ 3=\func{Mode}(1,1,2,2,3,3,3) < \func{Mode}(2,2,2,2,3,3,3)=2.$
\end{remark}

\subsection{Conjunctivity and disjunctivity}\label{Sec:ConjDisjDef}

One may well ask if also idempotence or internality
might not be desirable in certain contexts. The answer is of course positive.

\begin{remark}
In an \texttt{AND}-based rule aggregation system from
Example \ref{Ex:intro:rbsAND},
small values of $x_i$ may be treated as ``noise'' and may be cut down
by $\func{F}$ to $0$. What is more, in Example  \ref{Ex:intro:rbsOR}
quite an opposite fusion functions' behavior is expected.
\end{remark}

Actually, Dubois and Prade (see \cite{DuboisPrade1980:fss,DuboisPrade1985:reviewagcon,DuboisPrade2004:useagopfusion})
propose to distinguish the following four main classes of fusion functions:
\begin{itemize}
   \item internal (averaging),
   \item conjunctive (\texttt{AND}-like, e.g., t-norms),
   \item disjunctive (\texttt{OR}-like, e.g., t-conorms),
   \item mixed.
\end{itemize}
\index{minimum}\index{maximum}%
This distinction is based on the relationship between these functions
and $\func{Min}$ or $\func{Max}$.

\begin{definition}%
\index{conjunctivity}%
A fusion function $\func{F}:\IvalPow{n}\to\Ival$ is \emph{conjunctive},
whenever for all $\vect{x}\in\IvalPow{n}$ we have:
\begin{equation}
\func{F}(\vect{x})\le\func{Min}(\vect{x}).
\end{equation}
\end{definition}

\begin{definition}%
\index{disjunctivity}%
A fusion function $\func{F}:\IvalPow{n}\to\Ival$ is called \emph{disjunctive},
whenever for every $\vect{x}\in\IvalPow{n}$
it holds:
\begin{equation}
\func{Max}(\vect{x})\le\func{F}(\vect{x}).
\end{equation}
\end{definition}

Note that $\func{Min}$ and $\func{Max}$ are internal as well as
conjunctive and disjunctive, respectively, at the same time.
On the other hand, mixed fusion functions are neither internal,
conjunctive, nor disjunctive (for all input vectors).
We shall see in Section~\ref{Sec:MixedFunctions}
that this is the case of uninorms (among others).

\begin{example}
Assuming that $\Ival=[0,1]$,  the so-called \index{3Pi@$\func{3\Pi}$ function}%
\emph{3-$\Pi$} function, given by:
\[
   \func{3\Pi}(\vect{x})=\frac{\func{Prod}(\vect{x})}{\func{Prod}(\vect{x})+\func{Prod}(1-\vect{x})},
\]
is an example of a mixed-type fusion function,
with convention $0/0=0$. Yager and Rybalov \cite{YagerRybalov1996:uninorm} showed
that it is conjunctive on $[0,0.5]^n$, disjunctive on $[0.5,1]^n$, and internal otherwise.
\end{example}

Even though the focus of this book is generally on functions that are
idempotent, mappings from other classes are anyway noteworthy,
have influenced, and continue to be a very important part of the theory of aggregation.
For instance, the notion of a copula (a conjunctive -- among others -- fusion function,
e.g., $\func{Min}$ or $\func{T}_\mathrm{Ł}$) will be useful
in Chapter~\ref{Chap:multidim} when we discuss various methods for generating
random observations from $\mathbb{R}^d$ for $d>1$.
Hence, from time to time, we shall refer back to them.

\subsection{Symmetry. Permutations of inputs}

Another useful property is called {symmetry}.
It may be a sine qua non condition in statistics, where all the
observations are treated just as ``points in the real line''.
Moreover, it may be useful in decision making, in a case when
all of the experts are of the same ``esteem'' or all of them are anonymous.

\begin{definition}\index{symmetry}\label{Def:symmetry}%
We say that a fusion function $\func{F}:\IvalPow{n}\to\Ival$ is \emph{symmetric}, if:
\begin{equation}
(\forall \vect{x},\vect{y}\in\IvalPow{n})\
\vect{x}\cong\vect{y}\Longrightarrow\func{F}(\vect{x})=\func{F}(\vect{y}),
\end{equation}
 where  $\vect{x}\cong\vect{y}$ if and only if there exists a permutation
$\sigma$ of $[n]$ such that
$\vect{x}=(y_{\sigma(1)},\dots,y_{\sigma(n)})$.
\end{definition}

In other words, the output value of a symmetric function does not depend
on the ordering of inputs. Of course, each $\func{F}$ that is defined as a
function of $x_{(1)}$, \dots, $x_{(n)}$, i.e., order statistics,
is symmetric by definition, see also Section~\ref{Sec:symmetrization}.

\begin{example}\label{Ex:SkiJumping}
Among instances of symmetric aggregation functions we find the sample median,
all order statistics,
or trimmed and Winsorized means (see Example \ref{Ex:intro:stats}).
Specifically, the $1$-trimmed mean is used in
ski jumping competitions organized by the International Ski Federation, where
each of 5 experts provide scores based on a jumper's balance, body position, and landing style.
In such a case, one lowest and highest score is neglected.
\end{example}

\begin{remark}\index{order statistic}%
For a given $k$, $x_{(k)}$ may be computed in $O(n)$ time by using the
BFPRT (median of medians, \cite{BlumETAL1973:timeboundsselection}) algorithm
without actually sorting the input vector.
Note that often the Quickselect \cite{Hoare1961:quickselect}
or the Floyd-Rivest \cite{FloydRivest1975:select} schemes are preferred,
though; they have $O(n)$ time complexity only on average but, when implemented,
tend to run faster than BFPRT. See also \texttt{std::nth\_element()} function
in the \lang{C++} Standard Library.
\end{remark}

\begin{remark}[\cite{Beliakov2011:fasttrimmedmean}]
As $\func{WinMean}_k(\vect{x})=\frac{1}{n} \sum_{i=1}^n \left((x_i\vee x_{(k+1)})\wedge x_{(n-k)}\right)\allowbreak=
\frac{1}{n} \sum_{i=1}^n \func{Median}(x_{(k+1)},x_i,x_{(n-k)})$,
the computation of a Winsorized mean has $O(n)$ time complexity.
Moreover, the same holds for a trimmed mean, because:
\[\func{TriMean}_k(\vect{x})=\frac{1}{n-2k} \left(n\func{WinMean}_\alpha(\vect{x})-kx_{(n-k)}-kx_{(k+1)}\right).\]
\end{remark}

Another interesting result concerning algorithmic aspects of symmetric fusion functions
is due to J.~Rotman, see \cite[Proposition 2.33]{GrabischETAL2009:aggregationfunctions}.
It states that we need to compute the value of $\func{F}$ only three times
in order to determine if this property holds for a fixed $\vect{x}$.

\begin{proposition}
A fusion function $\func{F}:\IvalPow{n}\to\Ival$ is symmetric if and only if for all $\vect{x}\in\IvalPow{n}$:
\begin{eqnarray*}
   \func{F}(x_1,x_2,x_3,\dots,x_{n-1},x_n) & = & \func{F}(x_2,x_1,x_3,\dots,x_{n-1},x_n) \text{ and}\\
   \func{F}(x_1,x_2,x_3,\dots,x_{n-1},x_n) & = & \func{F}(x_2,x_3,x_4,\dots,x_{n\phantom{-1}},x_1)
\end{eqnarray*}
\end{proposition}

Permutations of input objects play an important role in data fusion theory.
Thus, let us recall an algorithm for generating a random permutation of a
given vector. By ``random'' we of course mean
a situation in which every possible permutation is assigned the same probability
measure, i.e., the distribution is uniform.
Generating a random permutation is not necessarily straightforward: in particular,
a procedure like ``$n$ times swap two randomly selected elements of $\vect{x}$''
does not lead to a uniform distribution. To achieve this goal,
we should rather rely, e.g., on the following algorithm.

\begin{algorithm}\label{Algorithm:randperm}\index{Fischer-Yates shuffle|see {random permutation}}\index{random permutation}%
Generation of a random permutation of elements of a given vector $\vect{x}=(x_1,\dots,x_n)$
with the Fisher-Yates shuffle  \cite{FisherYates1938:tables}
as formulated by Knuth \cite[Algorithm P]{Knuth1998:acpvol2} is done as follows:
\begin{enumerate}
   \item[1.] Let $\sigma$ be such that $\sigma(i) = i$ for all $i\in[n]$;
   \item[2.] For $j=n, n-1,\dots, 2$ do:
   \begin{enumerate}
      \item[2.1.] Let $i$ be a random number uniformly distributed in $\{1, 2, \dots, j\}$;
      \item[2.2.] Swap $\sigma(i) \leftrightarrow \sigma(j)$;
   \end{enumerate}
   \item[3.] Return $\vect{x}_\sigma$.
\end{enumerate}
\end{algorithm}
It is easily seen that the above procedure runs
in $O(n)$ time. Moreover, note that with Algorithm \ref{Algorithm:randperm}
we do not only get a random
rearrangement of elements of $\vect{x}$
but also a random $\sigma\in\mathfrak{S}_{[n]}$ itself.

\medskip
Moreover, later on we  need the notion of comonotonicity.
As it is somehow related to the topics discussed in this subsection,
let us introduce it now.

\begin{definition}\label{Def:comonotonic}\index{comonotonicity}%
According to \cite[Definition~2.123]{GrabischETAL2009:aggregationfunctions},
$\vect{x}, \vect{y}\in\IvalPow{n}$ are \emph{comonotonic},
denoted by $\vect{x} \pitchfork \vect{y}$,
if and only if there exists a permutation $\sigma\in\mathfrak{S}_{[n]}$
such that:
\begin{equation}
x_{\sigma(1)}\le \dots \le x_{\sigma(n)}
\quad\text{and}\quad
y_{\sigma(1)}\le \dots \le y_{\sigma(n)}.
\end{equation}
\end{definition}

Thus, $\sigma$ orders $\vect{x}$ and $\vect{y}$ simultaneously.
It is easily seen that the $\pitchfork$ binary relation
is reflexive and symmetric.%

\begin{remark}
Equivalently, $\vect{x}$ and $\vect{y}$ are comonotonic,
if and only if for every $i,j\in[n]$ it holds that:
\[(x_i-x_j)(y_i-y_j)\ge 0.\]
\end{remark}

It is easily seen that in order to generate two random comonotonic vectors
\index{random comonotonic vectors}%
$\vect{x}, \vect{y}$ we may generate the two vectors independently
(from some desired probability distribution on $\IvalPow{n}$),
sort them separately, generate a random permutation $\sigma$ with
Algorithm~\ref{Algorithm:randperm},
and then return $(\vect{x}_\sigma, \vect{y}_\sigma)$.

If all the elements of $\vect{x}$ are unique,
then to determine if two vectors are comonotonic it is sufficient
\index{ordering permutation}%
to take the (unique) ordering permutation of $\vect{x}$ and
then verify if $\vect{y}_\sigma$ is sorted.
On the other hand, if there are tied observations in $\vect{x}$,
we seek the longest possible sequence
$(x_{\sigma(i)},x_{\sigma(i+1)},\dots,x_{\sigma(i+k)})$
such that $x_{\sigma(i)}=x_{\sigma(i+k)}$,
where $\sigma$ is an ordering permutation of $\vect{x}$.
Then we try to update $\sigma$
so that it also sorts the corresponding observations in $\vect{y}$,
see \cite{Gagolewski2015:normalizedspread} for discussion.
An exemplary implementation is given in Figure~\ref{Fig:comonotonic}.
Here we use a sorting routine from the C++11 Standard Library,
with guaranteed run-time of $O(n\log n)$.
Note that the provided implementation also generates
a common ordering permutation for $\vect{x}$ and $\vect{y}$.
What is more, at some step an ordering permutation of $\vect{x}$ is given.
To guarantee that $\sigma$ is unique and such that for $i<j$ and $x_i = x_j$
we have $\sigma(i)\le \sigma(j)$, one may use \texttt{std::stable\_sort()}
instead of \texttt{std::sort()}.

\subsection{Continuity and convexity}

The notion of continuity is very attractive from the perspective of mathematical analysis.

\begin{definition} $\func{F}:\IvalPow{n}\to\mathbb{R}$
is \index{continuity}\emph{continuous}
if for all $\vect{x}^*\in\IvalPow{n}$ we have:
\begin{equation}
   \lim_{\IvalPow{n}\ni\vect{x}\to\vect{x}^*} \func{F}(\vect{x})=\func{F}(\vect{x}^*).
\end{equation}
\end{definition}

It may be shown that if $\func{F}$ is nondecreasing, then $\func{F}$
is continuous if and only if it is continuous in each variable,
see \cite[Proposition 2.8]{GrabischETAL2009:aggregationfunctions} for a proof.
This corresponds to the so-called \textit{intermediate value property}:
\index{intermediate value property}%
for each $\vect{x},\vect{y}$ with $\vect{x}\le_n\vect{y}$
and $c\in[\func{F}(\vect{x}),\func{F}(\vect{y})]$ there exists $\vect{z}$ such
that $\func{F}(\vect{z})=c$. In fact, $\vect{z}=\alpha\vect{x}+(1-\alpha)\vect{y}$
\index{convex combination}%
for some $\alpha\in[0,1]$, i.e., it is a convex combination of $\vect{x}$ and~$\vect{y}$.

\begin{remark}\label{Remark:MeanKolmogorovNagumoSense}
\index{mean (Kolmogorov-Nagumo sense)}%
A \emph{mean} in the sense of Kolmogorov \cite{Kolmogorov1930:moyenne}
and Nagumo \cite{Nagumo1930:mittelwerte} is a fusion function $\func{F}$
that is nondecreasing, continuous, symmetric, and idempotent.
\end{remark}

Let us recall the definition of a {norm} on an abstract vector space $V$
over a subfield $\mathbb{R}$.

\begin{definition}\label{Def:norm}\index{norm}%
A \emph{norm} on $V$ is a function $\|\cdot\|: V\to[0,\infty]$ such that:
\begin{enumerate}
\item[(a)] $\|\vect{v}\|=0$ if and only if $\vect{v}\equiv\vect{0}$,
\item[(b)] for any $a\in\mathbb{R}$ and $\vect{v}\in V$
\index{homogeneity of the first degree}%
it holds $\|a\vect{v}\|=|a|\,\|\vect{v}\|$ (homogeneity of the first degree), and
\item[(c)] for all $\vect{u},\vect{v}\in V$
we have $\|\vect{u}+\vect{v}\|\le \|\vect{u}\|+\|\vect{v}\|$ (\emph{triangle inequality)}.
\end{enumerate}
Moreover, we call $\|\cdot\|$ a \emph{pseudonorm} if condition (a)
\index{pseudonorm}%
is replaced with:
\begin{enumerate}
   \item[(a')] $\|\vect{v}\|=0$ if $\vect{v}\equiv\vect{0}$.
\end{enumerate}
\end{definition}

Informally, a norm is often used to measure the ``size'' of an object
and is a type of its numeric characteristic, compare Chapter~\ref{Chap:Characteristics}.
Also, we further on recall that norms may be used to generate
various distance metrics, that is functions to measure dissimilarities
between pairs of objects. It shall turn out that such a notion
is meaningful in aggregation theory (and data fusion and mining),
as many fusion functions may be written as minimizers of some penalty function.

\begin{remark}
Here are some notable norms on $\mathbb{R}^n$:
\index{Euclidean norm}\index{Manhattan norm}\index{maximum norm}%
\begin{itemize}
   \item
   \emph{Euclidean norm}, $\displaystyle \|\vect{x}\|_2=\sqrt{\sum_{i=1}^n x_i^2}=\sqrt{\vect{x}^T \vect{x}}$,
   \item
   \emph{Manhattan (Taxicab) norm}, $\displaystyle \|\vect{x}\|_1=\sum_{i=1}^n |x_i|$,
   \item
   \emph{maximum (Chebyshev) norm}, $\displaystyle \|\vect{x}\|_\infty=\bigvee_{i=1}^n |x_i|$,
\end{itemize}
or, more generally:
\index{Minkowski $p$-norm}%
\begin{itemize}
\item \emph{$p$-(Minkowski-)norm}, $p\ge 1$, $\displaystyle \|\vect{x}\|_p = \left( \sum_{i=1}^n |x_i|^p \right)^{1/p}.$
\end{itemize}
\end{remark}

R.~Lipschitz \cite{Lipschitz1864:explicatione} also considered the following
condition.

\begin{definition}%
\index{Lipschitz continuity}%
We say that a fusion function $\func{F}:\IvalPow{n}\to\Ival$ is \emph{Lipschitz continuous}
if for any norm $||\cdot||$ on $\IvalPow{n}$ there exists a finite constant $K \ge 0$ such that
for all $\vect{x},\vect{y}\in\IvalPow{n}$ it holds:
\begin{equation}
   |\func{F}(\vect{x})-\func{F}(\vect{y})|\le K ||\vect{x}-\vect{y}||.
\end{equation}
\end{definition}

Of course, $K$ depends on the choice of the norm. If no norm is mentioned
explicitly, the Manhattan one is assumed. In such a case,
the smallest constant in the above equation such that the Lipschitz
condition is still fulfilled is called the (best) Lipschitz constant.
\index{Lipschitz constant}\index{1-Lipschitz function}%
In particular, we call a function \emph{1-Lipschitz} if:
\[ |\func{F}(\vect{x})-\func{F}(\vect{y})|\le \sum_{i=1}^n |x_i-y_i|. \]
For example, this is the case of the arithmetic mean,
with best $K$ being equal to $1/n$ in the case of the 1-norm.

Generally, if the best $K$ is not greater than 1 we call $\func{F}$
\emph{non-expansive}, and if $K<1$ then $\func{F}$ is a
\emph{contraction}.

It might be shown that if $\func{F}$ is a Lipschitz function, then it is continuous.
On the other hand, $\func{GMean}$ is an example of a continuous aggregation
function on $[0,\infty[^n$ which is not Lipschitz.

Further on we shall see that, e.g., copulas, widely used in probability
and mathematical modeling (hydrology, finance, risk, etc.)~are Lipschitz functions.
In this regard, note that a continuous fusion function $\func{F}$
acting on a list with elements in a real closed interval fulfills
the property that for each $\varepsilon$,
there exists $\delta>0$ such that
$|\func{F}(\vect{x})-\func{F}(\vect{y})|\le\varepsilon$ if $\vect{x},\vect{y}$
are such that $\|\vect{x}-\vect{y}\|\le\delta$.

\begin{remark}
As the set of floating point numbers $\mathbb{F}$ is countable, the notion
of continuity is rather of theoretical interest.
The Lipschitz condition is even stronger: it guarantees that an
arbitrarily small change of input elements does not
lead to uncontrolled behavior of the output. Nevertheless, its milder version is
\index{numerical stability}useful from the computational perspective, where it is called \emph{numerical stability}
and concerns ``small'' perturbations of values in $\vect{x}$.
\end{remark}

The next property is important, e.g., when dealing with optimization tasks.

\begin{definition}\index{convexity}%
We say that a fusion function $\func{F}$ is \emph{convex}
whenever for all $\vect{x},\vect{y}\in\IvalPow{n}$ and $\lambda\in[0,1]$
it holds that:
\begin{equation}
   \func{F}(\lambda\vect{x}+(1-\lambda)\vect{y})\le \lambda\func{F}(\vect{x})+(1-\lambda)\func{F}(\vect{y}).
\end{equation}
\index{concativity}%
Moreover, $\func{F}$ is called \emph{concave}, if $(b+a)-\func{F}$ is convex.
\end{definition}

\begin{example}
$\func{Max}$ is an example of a convex function
and $\func{GMean}$ is a concave one if $a>0$. $\func{WAMean}$ is both convex and concave at the same time.
Moreover, by definition, every norm on $\mathbb{R}^n$ is convex.
\end{example}

Note that if $\func{F}$ is continuous and twice differentiable, then it is
convex if and only if its Hessian is positive semidefinite.
Also, if $\func{F}$ and $\func{G}$ are convex fusion functions,
then all of their convex combinations ($c\func{F}+d\func{G}$ for any $c,d\ge 0$, $c+d=1$) are also convex.

\subsection{Equivariance to translation and scaling}

In the practice of data analysis, transformations of input variables such as \emph{standardization}:
\[
   \vect{x}\mapsto \frac{\vect{x}-\func{AMean}(\vect{x})}{\func{SD}(\vect{x})},
\]
\index{standardization}%
\emph{robust standardization}:
\[
   \vect{x}\mapsto \frac{\vect{x}-\func{Median}(\vect{x})}{\func{MAD}(\vect{x})},
\]
\index{robust standardization}%
or \emph{normalization}:
\[
   \vect{x}\mapsto \frac{\vect{x}-\func{Min}(\vect{x})}{\func{Range}(\vect{x})},
\]
\index{normalization}%
where $0/0=0$, are often applied.
Here $\func{SD}(\vect{x})=\sqrt{\frac{1}{n-1} \sum_{i=1}^n (x_i-\func{AMean}(\vect{x}))^2}$ is the sample
\index{standard deviation}\index{SD@$\mathsf{SD}$|see {standard deviation}}%
\emph{standard deviation},
$\func{MAD}(\vect{x})=1.4826\func{Median}(|\vect{x}-\func{Median}(\vect{x})|)$ is the
\index{median absolute deviation}\index{MAD@$\mathsf{MAD}$|see {median absolute deviation}}%
\emph{median absolute deviation}, and
$\func{Range}(\vect{x})=\func{Max}(\vect{x})-\func{Min}(\vect{x})$ is the
\index{range}%
\emph{range},
see also Section~\ref{Sec:SpreadMeasures}.

The classic standardization implies that the transformed vector
is of mean~0 and standard deviation of 1 and normalization
assures that the output values are in $[0,1]$. The three transformations retain
relative distances between the observations.

Additionally, it is not that uncommon to convert the measurement units (e.g., Fahrenheit to Celsius,
feet to meters, etc.). Note that standardization and normalization result in \textit{unitless} values.

Taking the above into account, sometimes it would be useful to assure that a fusion function is equivariant
to translation (shifting) and/or scaling.

\begin{definition}\index{translation equivariance}%
We say that a fusion function $\func{F}$ is \emph{translation (shift, difference scale) equivariant}
if for all $t\in\mathbb{R}$ and $\vect{x}\in\IvalPow{n}$ such that $t+\vect{x}\in\IvalPow{n}$ it holds:
\begin{equation}
\func{F}(t+\vect{x})=t+\func{F}(\vect{x}).
\end{equation}
\end{definition}

Note that \index{translation invariance}\emph{translation \textit{in}variance}
would imply that for all $t$ and $\vect{x}$
it held $\func{F}(t+\vect{x})=\func{F}(\vect{x}).$

\begin{remark}\index{mean (Bullen sense)}%
Notably, Bullen in his seminal monograph \cite{Bullen2003:means}
defines a \emph{mean} as a nondecreasing,
symmetric, idempotent, and translation equivariant fusion function.
Interestingly, he assumes that means are most often computed on elements
in the interval $\Ival=[0,\infty[$.
\end{remark}

\begin{definition}\index{scale equivariance}%
A fusion function $\func{F}$ is called \emph{(ratio) scale equivariant} %
if for all $s> 0$ and $\vect{x}\in\IvalPow{n}$ such that $s\vect{x}\in\IvalPow{n}$ it holds that:
\begin{equation}
\func{F}(s\vect{x})=s\func{F}(\vect{x}).
\end{equation}
\end{definition}

Similarly to the translation or scale equivariance,
$\wedge$- and \emph{$\vee$-equivariance} may be defined:
it suffices to replace the addition or multiplication operation with
the minimum and maximum, respectively.
\index{0vee-equivariance@$\vee$-equivariance}%
\index{0wedge-equivariance@$\wedge$-equivariance}%
In this regard, translation and scale equivariance may be combined as follows.

\begin{definition}\label{Def:IntervalScaleEquivariant}\index{interval scale equivariance}%
A fusion function $\func{F}$ is \emph{interval scale equivariant}
if for all $s> 0$, $t\in\mathbb{R}$, $\vect{x}\in\IvalPow{n}$ with $t+s\vect{x}\in\IvalPow{n}$
we have:
\begin{equation}
\func{F}(t+s\vect{x})=t+s\func{F}(\vect{x}).
\end{equation}
\end{definition}

\begin{remark}
Note that if $0\in\Ival$ and $\func{F}$ is at least scale equivariant, then for every $s$ we have that
$s\func{F}(\vect{0})=\func{F}(s\vect{0})=\func{F}(\vect{0})$. Thus, $\func{F}(\vect{0})=0$.
If $\func{F}$ is additionally translation equivariant, then for any $t$ we have
that $\func{F}(n\ast t)=\func{F}(t\vect{1})=t\func{F}(\vect{0}+1)=t$. Thus,
$\func{F}$ is idempotent.
\end{remark}

\begin{remark}\label{Remark:MeanPitman}\index{mean (Pitman sense)}%
It is worth noting that Pitman in 1939
\cite{Pitman1939:estlocscale}
considered \emph{estimators} of
a location parameter $l\in\mathbb{R}$
under the transformation:
\[ f(x) \mapsto \frac{1}{s} f\left( \frac{x-l}{s} \right) \]
of a density function $f$ with $s>0$.
We see that it is a simple translate-scale model.
He posed that $\func{A}:\mathbb{R}^n\to\mathbb{R}$, being the estimator of $l$,
should fulfill:
\begin{displaymath}
\func{A}\left(\frac{x_1+\lambda}{\mu}, \dots, \frac{x_n+\lambda}{\mu}\right)
=\frac{\func{A}(x_1,\dots,x_n)+\lambda}{\mu}
\end{displaymath}
for all $\lambda\in\mathbb{R}$ and $\mu>0$,
and be independent of $l$.
The definition of the $\func{A}$ function is a very appealing,
early approach to aggregation
as we know today: ``any function of this type will be called an estimate of $l$'',
see \cite[page~409]{Pitman1939:estlocscale}.
He also wrote on page~420: ``any function of the sample values whose value
may be used as an estimate of an unknown parameter is called
an estimator of that parameter''.
Moreover, he pointed out that there are many estimators, each of which may fulfill
different properties (e.g., one that minimizes the minimum mean absolute error
or the minimum mean square error).
\end{remark}

Sometimes we might be interested in the following, much stronger version
of interval scale equivariance (compare Proposition~\ref{Proposition:agfuntrans}):

\begin{definition}\index{ordinal scale equivariance}%
A fusion function $\func{F}$ is said to be \emph{ordinal scale equivariant}
if for all increasing bijections $\varphi:\Ival\to\Ival$ and every $\vect{x}\in\IvalPow{n}$ it holds that:
\begin{equation}
\func{F}(\varphi(x_1),\dots,\varphi(x_n))=\varphi(\func{F}(\vect{x})).
\end{equation}
\end{definition}

\subsection{Additivity}

Recall that the ``$+$'', ``$\wedge$'', and ``$\vee$''
operations on vectors are applied elementwise.

\begin{definition}\index{additivity}%
A fusion function $\func{F}$ is said to be \emph{additive},
whenever:
\begin{equation}
\func{F}(\vect{x}+\vect{y})=\func{F}(\vect{x})+\func{F}(\vect{y}),
\end{equation}
for all $\vect{x}, \vect{y}\in\IvalPow{n}$ such that $\vect{x}+\vect{y}\in\IvalPow{n}$.
\end{definition}

It is easily seen that each idempotent and additive fusion function is also
translation equivariant.

\begin{definition}\index{modularity}%
A fusion function $\func{F}$ is said to be \emph{modular},
whenever for all $\vect{x}, \vect{y}\in\IvalPow{n}$:
\begin{equation}
\func{F}(\vect{x}\wedge\vect{y})+\func{F}(\vect{x}\vee\vect{y})=\func{F}(\vect{x})+\func{F}(\vect{y}).
\end{equation}
\end{definition}

Due to the fact that $\vect{x}\wedge\vect{y}+\vect{x}\vee\vect{y} = \vect{x}+\vect{y}$,
each additive function is necessarily also modular.

\begin{definition}\index{maxitivity}%
A fusion function $\func{F}$ is said to be \emph{maxitive},
whenever for all $\vect{x}, \vect{y}\in\IvalPow{n}$:
\begin{equation}
\func{F}(\vect{x}\vee\vect{y})=\func{F}(\vect{x})\vee\func{F}(\vect{y}).
\end{equation}
\end{definition}

\begin{definition}\index{minitivity}%
A fusion function $\func{F}$ is said to be \emph{minitive},
whenever for all $\vect{x}, \vect{y}\in\IvalPow{n}$:
\begin{equation}
\func{F}(\vect{x}\wedge\vect{y})=\func{F}(\vect{x})\wedge\func{F}(\vect{y}).
\end{equation}
\end{definition}

\subsection{Other types of monotonicity}\label{Sec:OtherMonotonicities}

Over history, there have been different approaches to define the concept of
a mean. In the Pitman sense (see Remark~\ref{Remark:MeanPitman}), a mean is meant to be
translation and scale equivariant, in the Cauchy or Gini sense (see Remark~\ref{Remark:MeanCauchy})
it is just an internal fusion function.
Classical aggregation theory focuses on fusion functions that are monotone with respect
to all their arguments. However, it is known that some classes of broadly
conceived means are nonmonotone. One example of such a fusion function is the \emph{mode},
defined as an observation that occurs most often in a data sample
(in the case of unimodal data sets), see Remark~\ref{Remark:mode}.
Some other examples of nonmonotone fusion functions
may be found in the class of \emph{Bajraktarević means} (see \cite{Bullen2003:means}
and Equation~\eqref{Eq:Bajraktarevic})
or density-based fusion functions (see \cite{AngelovYager2013:densitybasedavg}
as well as \cite{BeliakovWilkin2014:weightedavgvarw}).

As Beliakov, Calvo, and Wilkin in \cite{BeliakovCalvoWilkin2014:3typesmono} note,
unexceptional monotonicity with respect to $\le_n$ also might not be desirable
in certain contexts. For example, it can reduce the robustness of an averaging
method in the case of outliers (compare Remark~\ref{Remark:outliers1d}).
Moreover, as we shall see in further chapters,
there are indeed many issues in regard to defining order preserving transformations
in more complex domains than $\IvalPow{n}$.

Due to the fact that a kind of monotonicity in~the $\IvalPow{n}$ space
is nevertheless very appealing,
quite recently, some researchers in aggregation theory
\index{relation le@relation $\le_n$}%
introduced mappings that preserve  orders other than $\le_n$.
Therefore, in this section we review  a few of them.

\medskip
The concept of weak monotonicity has been introduced by Wilkin and Beliakov
in \cite{WilkinBeliakov2015:weaklymonotone}, see also \cite{Wilkin2014:phd}.
It requires that the output of an aggregation function surely does not decrease
whenever we increase all the input values by the same amount.

\begin{definition}\label{Def:weakmonotonicity}\index{weak monotonicity}%
A fusion function $\func{F}:\IvalPow{n}\to\Ival$ is \emph{weakly monotone}
whenever $\func{F}(\vect{x}+t)\ge \func{F}(\vect{x})$
for any $t\ge 0$ and $\vect{x}\in\IvalPow{n}$ such that
$\vect{x}+t\in\IvalPow{n}$.
\end{definition}

Of course, each fusion function that is nondecreasing, is also weakly monotone.
The same is true for any translation equivariant mapping.

\medskip
In \cite{BeliakovCalvoWilkin2014:3typesmono} it is noted that
the standard nondecreasingness and weak monotonicity
are two extremes of a more general situation called
\textit{monotonicity with respect to coalitions} or \emph{quantiles}
($\alpha$-monotonicity).

\begin{definition}\index{alpha-monotonicity@$\alpha$-monotonicity}%
A fusion function $\func{F}:\IvalPow{n}\to\Ival$
is \emph{monotone with respect to the $\alpha$-quantile of the inputs},
$\alpha\in[0,1[$,
whenever $\func{F}(\vect{x}+t\vect{u})\ge \func{F}(\vect{x})$
for any $t\ge 0$, $\vect{u}\in\{0,1\}^n$ such that $\{i: u_i=1\}\ge \lfloor \alpha n + 1\rfloor$,
and $\vect{x}\in\IvalPow{n}$ such that $\vect{x}+t\vect{u}\in\IvalPow{n}$.
\end{definition}

\medskip
What is more, Bustince, Fernandez, Koles\'{a}rov\'{a}, and Mesiar
introduced in \cite{BustinceETAL2014:directionalmonotonicity,BustinceETAL2015:directionalmonotonicity}
another concept -- directional monotonicity.

\begin{definition}\index{directional monotonicity}%
For a given $n$-dimensional vector $\vec{r}\neq \vect{0}$ a fusion function $\func{F}$
is called $\vec{r}$-nondecreasing, whenever for all $t>0$ such that
$\vect{x}+t\vec{r}\in\IvalPow{n}$ it holds:
\begin{equation}
\func{F}(\vect{x})\le\func{F}(\vect{x}+t\vec{r}).
\end{equation}
\end{definition}

Clearly, $(n\ast 1)$-nondecreasing fusion functions are weakly monotone and vice versa.
This concept is interesting if one wants to study in which \textit{directions} a function is monotone:
please notice that $\le_n$-monotonic fusion functions are also $\vec{r}$-nondecreasing
for all $\vec{r}\ge_n \vect{0}$.
Lucca et~al.~in~\cite{LuccaETAL2015:preaggregation} called a
\index{pre-aggregation function}fusion function $\func{F}$ a \emph{pre-aggregation mapping},
whenever it is $\vec{r}$-nondecreasing for some $\vec{r}$ and endpoint-preserving.

\section{Construction methods}

Let us discuss a few notable fusion function construction methods.
   Firstly, we focus on functions that are created by a fusion
   (composition) or modification of other, perhaps simpler mappings.
   Due to that we may try to obtain
   data aggregation tools that start to fulfill  originally missing properties or behavior.
   Further on we note that many interesting fusion functions are related to
   universal integrals with respect to monotone measures, tools known from
   -- among others -- decision making. An appropriate choice of a monotone measure
   and/or integral provides us with new ways to aggregate data.

   Finally, we study the concept of
   fusion functions which can be expressed as minimizers of some penalty.

\subsection{Compositions and transforms of fusion functions}

New fusion functions may be obtained
by a proper composition of simpler ones.
It turns out that under certain circumstances some of the properties
of the underlying mappings may be preserved.

\begin{proposition}\label{Prop:CompositionProps}
Let $\func{F}:\IvalPow{k}\to\Ival$, $\func{G}_1,\dots,\func{G}_k:\IvalPow{n}\to\Ival$,
and $\func{H}:\IvalPow{n}\to\Ival$ be given by $\func{H}(\vect{x})=\func{F}(\func{G}_1(\vect{x}),\dots,\func{G}_k(\vect{x}))$
for $\vect{x}\in\IvalPow{n}$.
\begin{itemize}
   \item If $\func{F}$ is $\le_k$-nondecreasing and $\func{G}_1,\dots,\func{G}_k$ are $\le_n$-nondecreasing (respectively,
   idempotent, internal, translation equivariant, scale equivariant),
   then $\func{H}$ is $\le_n$-nondecreasing (respectively,
   idempotent, internal, and so forth).

   \item If $\func{F}$ is $\le_k$-nondecreasing and idempotent and $\func{G}_1,\dots,\func{G}_k$
   are $\le_n$-nondecreasing and conjunctive (disjunctive)
   then $\func{H}$ is also $\le_n$-nondecreasing and conjunctive (respectively,
   disjunctive).

   \item
   If $\func{F}$ is $\le_k$-nondecreasing and
   $\func{G}_1,\dots,\func{G}_k$ are $\vec{r}$-nondecreasing, then
   $\func{H}$ is $\vec{r}$-nondecreasing, \cite{BustinceETAL2015:directionalmonotonicity}.

   \item If $\func{F}$ is weakly monotone and
   $\func{G}_1,\dots,\func{G}_k$ are translation equivariant, then
   $\func{H}$ is weakly monotone, \cite{WilkinBeliakov2015:weaklymonotone}.
\end{itemize}
\end{proposition}

\begin{example}
The $\func{Median}$ function for even $n$ is defined as an arithmetic mean
(nondecreasing, idempotent, internal, translation, and scale equivariant)
of two order statistics (which also fulfill these properties).
\end{example}

In particular, in Section~\ref{Sec:Hierarchies} we study an exemplary
\textit{hierarchy} of fusion functions, which leads us to the concept
of an artificial neural network.

\bigskip
In certain contexts, it may be desirable to apply a fusion function
on transformed inputs or to remap the produced outputs.
For instance, we may note that nondecreasingness is a very mild condition.
Because of this, we have what follows.

\begin{proposition}\label{Proposition:agfuntrans}
If $\varphi:\Ival\to\Ival$
is a nondecreasing univariate function with $\varphi(a)=a$ and $\varphi(b)=b$,
then for each aggregation function $\func{F}$, $\func{G}=\varphi\circ\func{F}$, i.e.:
\[
   \func{G}(x_1,\dots,x_n)=\varphi\left(\func{F}(x_1,\dots,x_n)\right)
\]
is an aggregation function too.
A similar result holds for a function given by:
\[\func{H}(x_1,\dots,x_n)=\func{F}(\varphi(x_1),\dots,\varphi(x_n)).\]
\end{proposition}

In Section~\ref{Sec:QuasiArithmeticMeans} we study the notion of
a $\varphi$-isomorphism of a given fusion function. This shall lead us to the
class of quasi-arithmetic means.

\medskip
Sometimes it is also possible to transform a fusion function in such a way
that its modified version starts to fulfill a desired property which was
missing in the original setting.

Let $\delta_\func{F}:\Ival\to\Ival$ denote the so-called
\index{diagonal section}\emph{diagonal section} of a fusion function $\func{F}$,
that is $\delta_\func{F}(x)=\func{F}(n\ast x)$.
The following result allows us to generate an idempotent fusion function $\func{G}$
having been given $\func{F}$ whose diagonal section is strictly increasing
and such that $\mathsf{range}(\delta_\func{F}) = \mathsf{range}(\func{F})$.
Such a process is called \index{idempotization}\emph{idempotization}.

\begin{proposition}[\cite{CalvoETAL2002:aggoppsChapter1}]
If $\func{F}$ is such that $\delta_\func{F}$ is strictly increasing
and there exists a fusion function $\func{G}$ such that $\func{F}=\delta_\func{F}\circ\func{G}$,
then $\func{G}$ is idempotent.
\end{proposition}
For instance, the arithmetic mean and the geometric mean are results of idempotentization
of the sum and the product, respectively.

We may also assure internality in the following way.
\index{internalization}%
Let $\func{F}$ be a fusion function. Then $\func{G}$ given for example by:
\begin{itemize}
\item cut-off: \[
\func{G}(\vect{x})=\func{Min}(\vect{x})\vee(\func{F}(\vect{x})\wedge\func{Max}(\vect{x})),
\] or
\item normalization (by, e.g., \cite[Proposition 2.55]{GrabischETAL2009:aggregationfunctions}):
\[
   \func{G}(\vect{x}) =
      \func{Min}(\vect{x}) + (\func{Max}(\vect{x})-\func{Min}(\vect{x}))\psi{(\func{F}(\vect{x}))}, %
\]
where $\psi:\Ival\to[0,1]$ is some strictly increasing mapping,
e.g., $\psi(x)=(x-a)/(b-a)$ in the case of a bounded $\Ival=[a,b]$,
\end{itemize}
is internal (recall that $\Ival=[a,b]$). Note that in both cases if $\func{F}$ is
nondecreasing, $\func{G}$ is nondecreasing too.

Additionally, in Section~\ref{Sec:symmetrization} we shall illustrate the
concept of symmetrization.

\medskip
What is more, in some applications it is useful to assume that
not all the input observations have the same \textit{impact} on the resulting value.
In order to take this into account, in Section~\ref{Sec:Weighting} we introduce the concept of
fusion functions' weighting.

\bigskip
Bullen \cite[page~60]{Bullen2003:means} notes:
\begin{quote}\it\small
[The arithmetic mean] is the simplest mean and by far the most common;
in fact for a non-mathematician this is probably the only concept for averaging
a set of numbers. The arithmetic mean of two numbers $a$ and $b$,
$(a+b)/2$, was known and used by the Babylonians in 7000 B.C.,
and occurs in several contexts in the works of the Pythagorean school,
sixth-fifth century B.C. [\dots] Aristotle,
[\dots] used the arithmetic mean but did not give it
this name. [\dots] The idea of arithmetic mean is also found in the concept
of centroid used by Heron, and earlier by Archimedes in the third century B.C. [\dots]
\end{quote}

\index{arithmetic mean}%
In the sequel we consecutively modify $\func{AMean}$ so that we approach more
and more complex (and thus interesting) fusion functions. Despite its
first-glance simplicity, we shall notice that the arithmetic mean is in fact
a ``sleeping beauty''.

\subsubsection{$\varphi$-isomorphisms: Quasi-arithmetic means}\label{Sec:QuasiArithmeticMeans}

Let us first introduce the notion of a $\varphi$-isomorphism.

\begin{definition}\label{Def:isomorphism}
\index{Phi isomorphism@$\varphi$-isomorphism}%
Let $\Ival=[a,b],\mathbb{J}=[a',b']$, and $\varphi:\Ival\to\mathbb{J}$ be a strictly monotone bijection.
Then the $\varphi$-isomorphism of a fusion function $\func{F}:\mathbb{J}^n\to\mathbb{J}$
is a fusion function $\func{F}_{[\varphi]}:\IvalPow{n}\to\Ival$ defined as:
\begin{equation}
\func{F}_{[\varphi]}(x_1,\dots,x_n)=\varphi^{-1}\left(\func{F}(\varphi(x_1),\dots,\varphi(x_n))\right).
\end{equation}
\end{definition}

For instance, on $\Ival=\mathbb{J}=[a,b]$, we have $\func{Max}(\vect{x})=b+a-\func{Min}(b+a-\vect{x})$.
Thus, $\func{Max}$ is a $(x\mapsto b+a-x)$-isomorphism of $\func{Min}$.

We have the following result, compare also Proposition~\ref{Proposition:agfuntrans}.

\begin{proposition}\label{Prop:isomorphism}
If $\varphi:\Ival\to\mathbb{J}$ is a strictly monotone bijection
and $\func{F}:\mathbb{J}^n\to\mathbb{J}$ is an idempotent aggregation function, then
$\func{F}_{[\varphi]}:\IvalPow{n}\to\Ival$ is an idempotent aggregation function too.
Moreover, in the case of a weakly monotone fusion function $\func{F}$ the same is true whenever $\varphi$
is linear (but not in general), see \cite{WilkinBeliakov2015:weaklymonotone}.
\end{proposition}

This serves as a basis for the definition of quasi arithmetic means, which
have already been studied  in the 1930s \cite{Kolmogorov1930:moyenne,Nagumo1930:mittelwerte}
by, e.g., Kolmogorov and Nagumo.

\begin{definition}\index{QAMean@$\mathsf{QAMean}_\varphi$|see {quasi-arithmetic mean}}\index{quasi-arithmetic mean}%
Let $\varphi:\Ival\to\bar{\mathbb{R}}$ be a continuous and strictly monotonic function.
Then a \emph{quasi-arithmetic mean} generated by $\varphi$ is a fusion function
$\func{QAMean}_\varphi:\IvalPow{n}\to\Ival$ given by:
\begin{equation}
\func{QAMean}_\varphi(\vect{x}) = \varphi^{-1}\left( \frac{1}{n} \sum_{i=1}^n \varphi(x_i)\right).
\end{equation}
\end{definition}
In other words, a quasi arithmetic mean is a $\varphi$-isomorphism of
(the nondecreasing and idempotent) $\func{AMean}:\bar{\mathbb{R}}^n\to\bar{\mathbb{R}}$.
We have $\func{QAMean}_\varphi=\func{AMean}_{[\varphi]}$.

\begin{table}[thb!]
\caption[Examples of quasi-arithmetic means.]{\label{Tab:QMeans} Examples of quasi-arithmetic means under~the~assumption~that~$\Ival=[0,b]$~for~some~$b>0$.}
\centering
\begin{tabularx}{1.0\linewidth}{llX}
\toprule
\small\bf $\varphi(x)$ & \small\bf name & \small\bf $\func{QAMean}_\varphi(\vect{x})$  \\
\midrule\index{arithmetic mean}
$x$   & arithmetic mean & $\displaystyle \func{AMean}(\vect{x})=\frac{1}{n} \sum_{i=1}^n x_i$ \\
\midrule\index{QMean@$\mathsf{QMean}$|see {quadratic mean}}\index{quadratic mean}
$x^2$ & quadratic mean  & $\displaystyle \func{QMean}(\vect{x})=\sqrt{\frac{1}{n} \sum_{i=1}^n x_i^2}$ \\
\midrule\index{harmonic mean}
$1/x$ & harmonic mean & $\displaystyle \func{HMean}(\vect{x})=\frac{1}{\frac{1}{n}\sum_{i=1}^n\frac{1}{x_i}}$\\
\midrule\index{PMean@$\mathsf{PMean}_r$|see {power mean}}\index{power mean}
$x^r$, $r\neq 0$ & power mean & $\displaystyle \func{PMean}_r(\vect{x})=\left( \frac{1}{n} \sum_{i=1}^n x_i^r \right)^{1/r}$ \\
\midrule\index{geometric mean}
$\log x$ & geometric mean & $\displaystyle \func{GMean}(\vect{x})=\left(\prod_{i=1}^n x_i\right)^{1/n}$ \\
\midrule\index{EMean@$\mathsf{EMean}_\gamma$|see {exponential mean}}\index{exponential mean}
$e^{\gamma x}$, $\gamma\neq 0$ & exponential mean & $\displaystyle \func{EMean}_\gamma(\vect{x})=\frac{1}{\gamma} \log\left(\frac{1}{n} \sum_{i=1}^n e^{\gamma x_i}\right)$\\
\bottomrule
\end{tabularx}
\end{table}

Table~\ref{Tab:QMeans} lists notable instances of quasi-arithmetic means.
Like in \cite{Bullen2003:means,GrabischETAL2009:aggregationfunctions},
we assume that $a=0$, i.e., $\Ival=[0, b]$ for some $b>0$. Note that among power means we have the arithmetic,
quadratic, and harmonic means and that power means for $r\ge 1$ are actually norms.

\begin{example}
Suppose that a driver uses a cruise control device while driving a freeway.
He/she always drives with the same speed at a fixed distance.
Assuming that the consecutive speeds are $x_1,\dots,x_n$,
the average speed is equal to $\func{HMean}(\vect{x})$.
\end{example}

\begin{example}
   The exponential mean with $\gamma=1$ (the so-called
\index{LogSumExp@$\func{LogSumExp}$}$\func{LogSumExp}$ function) is used
in certain optimization tasks (e.g., in some machine learning algorithms)
as a smooth, strictly increasing, and convex approximation
to the $\func{Max}$ function. It is because for any $\vect{x}\in\IvalPow{n}$
it holds $\func{Max}(\vect{x})\le \func{EMean}_1(\vect{x})\le\func{Max}(\vect{x})+\log n$.
\end{example}

\begin{remark}
Each quasi-arithmetic mean is, among others,
nondecreasing,
continuous,
idempotent, and
symmetric,
see, e.g., \cite{Aczel1948:onmeanvalues}.
Moreover, the arithmetic mean and all the exponential means are translation equivariant
and the geometric mean as well as all the power means are scale equivariant,
compare Theorems \ref{Thm:QAMeanTranslation} and \ref{Thm:QAMeanScale}.
\end{remark}

\subsubsection{Weighting: Weighted quasi-arithmetic means}\label{Sec:Weighting}

It is not unusual for the observations in an input vector
to have a non-equal impact on data fusion results.
For instance, in a decision making context, the opinions of some agents
may be of greater importance than of the other ones,
just as in Example \ref{Ex:intro:weightedbipolar}.
Also in physics, when there is a need to calculate the center of mass
of a system of particles, we may need to take into account
different ``amounts of matter'' constituting the objects of concern.

To quantify the degrees of importance of the aggregated entities,
we may associate with each observation $x_i$ its weight, $w_i$.
Most commonly, a weighting vector, which must be of the same length as $\vect{x}$,
is assumed to satisfy the following conditions.

\begin{definition}\index{weighting vector}%
We call $\vect{w}=(w_1,\dots,w_n)$ a \emph{weighting vector}
if for all $i$ it holds $w_i\ge 0$ and $\sum_{j=1}^n w_j = 1$.
\end{definition}

\begin{remark}
Of course, if we are given nonnegative degrees of importance $d_1,\dots,d_n$
that do not sum up to 1, we may always create a weighting vector as
$\vect{w}=\vect{d}/\sum_{i=1}^n d_i$, under the assumption that
$(\forall i)$ $d_i=0 \Longrightarrow w_i=1/n$.
\end{remark}

A weighted version of quasi-arithmetic means (also known as quasi-linear means)
was introduced by Kitagawa in \cite{Kitagawa1934:weightedmeans}.

\begin{definition}\index{WQAMean@$\mathsf{WQAMean}_\varphi$|see {weighted quasi-arithmetic mean}}\index{weighted quasi-arithmetic mean}%
Let $\varphi:\Ival\to\bar{\mathbb{R}}$ be a continuous and strictly monotonic  function
and $\vect{w}$ be a weighting vector.
Then a \emph{weighted quasi-arithmetic mean} generated by $\varphi$ and $\vect{w}$
is a fusion function $\func{WQAMean}_{\varphi,\vect{w}}:\IvalPow{n}\to\Ival$ given by:
\begin{equation}\label{Eq:wqmean}
\func{WQAMean}_{\varphi,\vect{w}}(\vect{x}) = \varphi^{-1}\left( \sum_{i=1}^n w_i\varphi(x_i)\right) = \varphi^{-1}\left( \vect{w}^T \varphi(\vect{x}) \right).
\end{equation}
\end{definition}

Clearly, if for all $i$ it holds $w_i=1/n$, then a weighted quasi-arithmetic mean
reduces to a quasi-arithmetic mean.
Among examples of such fusion functions we have, e.g.:
\begin{itemize}
   \item \index{weighted arithmetic mean}%
   $\displaystyle \func{WAMean}_\vect{w}(\vect{x})=\sum_{i=1}^n w_i x_i=\vect{w}^T \vect{x}$,\par\hfill (weighted arithmetic mean, convex combination of inputs)

   \item \index{WHMean@$\mathsf{WHMean}$|see {weighted harmonic mean}}\index{weighted harmonic mean}%
   $\displaystyle \func{WHMean}(\vect{x})=\frac{1}{\sum_{i=1}^n {w_i/x_i}}$, \hfill(weighted harmonic mean)

   \item \index{WGMean@$\mathsf{WGMean}$|see {weighted geometric mean}}\index{weighted geometric mean}%
   $\displaystyle \func{WGMean}(\vect{x})=\prod_{i=1}^n x_i^{w_i}$, \hfill(weighted geometric mean)
\end{itemize}
and so forth. Note that $\mathsf{WQAMean}_\varphi$ is a $\varphi$-isomorphism of the fusion function $\func{WAMean}:\bar{\mathbb{R}}^n\to\bar{\mathbb{R}}$.

\begin{remark}
If $\varphi^{-1}$ is convex, then by the Jensen inequality we have that
for all weighting vectors $\vect{w}$:
\[
   \func{WQAMean}_{\varphi,\vect{w}}(\vect{x}) \le \sum_{i=1}^n w_i x_i = \func{WAMean}(\vect{x}).
\]
\end{remark}

Weights may also be dependent on the order of magnitude of inputs.
An intuitively appealing generalization of weighted quasi-arithmetic means
(and other weighted fusion functions) may be obtained by replacing
a weighting vector in Equation~\eqref{Eq:wqmean} with a vector of \textit{weighting functions}.
This leads to the concept of \emph{Bajraktarević means} (compare \cite{Bullen2003:means}):
\index{BajMean@$\mathsf{BajMean}$|see {Bajraktarević mean}}\index{Bajraktarević mean}%
\begin{equation}\label{Eq:Bajraktarevic}
   \func{BajMean}_{\varphi,\vect{w}}(\vect{x}) = \varphi^{-1}\left(
      \frac{\displaystyle \sum_{i=1}^n \func{w}_i(x_i) \varphi(x_i)}{\displaystyle \sum_{i=1}^n \func{w}_i(x_i)}
   \right),
\end{equation}
where $\vect{w}=(\func{w}_1,\dots,\func{w}_n)$ is a vector of weighting
functions, $\func{w}_i:\Ival\to[0,\infty[$ for all $i\in[n]$, and $\vect{\varphi}:\Ival\to\bar{\mathbb{R}}$
is a strictly monotone bijection.

\begin{remark}
The case $\func{\varphi}(x)=x$ and $(\forall i\in[n])$ $\func{w}_i=\func{w}$ for some function $\func{w}$
generates the so-called \emph{mixture operator}.
\index{mixture operator}%
Also note that if $\vect{w}_i$ are constant functions for all $i$, then a Bajraktarević
mean reduces to a weighted arithmetic mean.
Other particular cases may be formed by, e.g., setting $\func{w}_i$ to be power functions.
In such a way we get the \emph{Gini means}:
\index{Gini mean}%
\[
   \func{GiniMean}^{p,q}_\vect{w}(\vect{x}) = \left\{
   \begin{array}{ll}
   \left(
   \frac{\sum_{i=1}^n w_i x_i^p}{\sum_{i=1}^n w_i x_i^q}
   \right)^{1/(p-q)}
   & \text{if }p\neq q,\\
   \left(
   \prod_{i=1}^n x_i^{w_i x_i^p}
   \right)^{1/\sum_{i=1}^n w_i x_i^p}
   & \text{if }p= q,\\
   \end{array}
   \right.
\]
where $\vect{w}$ is a weighting vector and $p,q\in\mathbb{R}$.
The case $p=q-1$ generates the so-called \emph{Lehmer means}.
Note that if $q=0$, then a Gini mean reduces to a nondecreasing power mean.
\end{remark}

All the Bajraktarević means are of course idempotent.
On the other hand, it is quite easy to find many examples of Bajraktarević
means that are not nondecreasing.
More generally, Beliakov, Wilkin, and Calvo in \cite{BeliakovCalvoWilkin2015:weakmonogini,WilkinBeliakovCalvo2014:weakmono}
studied sufficient conditions for Gini means and some other
Bajraktarević means to be weakly monotone.

\subsubsection{Symmetrization: OWA operators}\label{Sec:symmetrization}

Note that if there exists $i\neq j$ such that $w_i\neq w_j$,
then a weighted quasi-arithmetic mean is no longer symmetric.
However, it turns out that each symmetric fusion function $\func{F}$
may be generated by using another function $\func{G}$ applied to
an input vector's consecutive order statistics.

\begin{proposition}[\cite{GrabischETAL2009:aggregationfunctions}]%
\label{Prop:symmetrization}%
$\func{F}:\IvalPow{n}\to\Ival$ is symmetric if and only if there exists
a function $\func{G}:\IvalPow{n}\to\Ival$ such that:
\[\func{F}(x_{1},\dots,x_{n})=\func{G}(x_{(1)},\dots,x_{(n)}).\]
\end{proposition}
\index{symmetrization}%
Technically, note that in fact the domain of $\func{G}$ might be set to
$\{\vect{x}\in\IvalPow{n}: x_1\le\dots\le x_n\}\subseteq\IvalPow{n}$ here.
In other words, each fusion function may be \emph{symmetrized} by replacing all $x_i$'s
with $x_{(i)}$'s, i.e., $i$th order statistics, in its definition.

For instance, a symmetrized version of a weighted arithmetic mean
is called in decision making an OWA operator:
\index{OWA operator}%
\begin{equation}
\func{OWA}_\vect{w}(\vect{x})=\sum_{i=1}^n w_i x_{(i)}.
\end{equation}
Its name -- ordered weighted averaging -- is due to Yager \cite{Yager1988:owa},
see also \cite{YagerKacprzyk1997:owaops,YagerETAL2011:oward}.

\begin{example}\label{Ex:multiset}
Weighting and symmetrization naturally occurs in a case when we aggregate
elements of a multiset: identical values may occur multiple times in an input
data set and we do not pay attention to their order.
Let us consider a multiset $\{ (1, 3), (2, 1), (3,4) \}$ over $\mathbb{N}$,
i.e., such that we have 3 ones, 1 two, and 4 threes.  Then the corresponding
weighting vector may be created according to the number of occurrences of elements:

\begin{center}
\begin{tabularx}{1.0\linewidth}{XrrrX}
\toprule
&\small\bf value & \small\bf \#occurrences & \small\bf weight & \\
\midrule
&1 & 3 & 0.375& \\
&2 & 1 & 0.125 &\\
&3 & 4 & 0.5\phantom{00} &\\
\midrule
&$\Sigma$ & 8 & 1.0\phantom{00} &\\
\bottomrule
\end{tabularx}
\end{center}
\end{example}

\begin{example}
$\func{Median}$, $\func{WinMean}$, and $\func{TriMean}$ are in fact
OWA operators -- they are used as \textit{robust}, i.e., less sensitive
to the presence of a few outliers, estimators of an underlying probability distribution
location parameters.
\end{example}

\begin{example}\label{Example:quantiles}
Let us also recall the notion of a \emph{sample quantile} of order $\alpha\in[0,1]$.
\index{Q@$\mathsf{Q}_\alpha$|see {quantile}}\index{quantile}%
Although there are many various definitions in the literature
and implementations in statistical software packages,
see~\cite{HyndmanFan1996:samplequantiles} for a review,
it is generally accepted that this kind of an aggregation
function is an OWA operator given by:
\[
\func{Q}_\alpha(\vect{x})=
\left\{
\begin{array}{ll}
\func{Min}(\vect{x}) & \text{for }\alpha=0,\\
\func{Median}(\vect{x}) & \text{for }\alpha=0.5,\\
\func{Max}(\vect{x}) & \text{for }\alpha=1,\\
\gamma\, x_{(k)} + (1-\gamma)\,x_{(k+1)} & \text{otherwise},
\end{array}
\right.
\]
for some $\gamma=\gamma(\alpha,k)\in]0,1]$ and
$k\in\left\{\lfloor n\alpha-1\rfloor, \lfloor n\alpha+1\rfloor\right\}$
such that for each fixed $\vect{x}$ it is a~nondecreasing function of~$\alpha$.

More precisely, Hyndman and Fan in \cite{HyndmanFan1996:samplequantiles}
list nine quantile function types, see Table~\ref{Tab:quantiles1}.
The first three types are discontinuous functions of $\alpha$.
The other types (IV-IX) define continuous quantile functions.
Each of the types may exhibit different properties,
either algebraic or probabilistic.
For instance, type VIII is approximately median-unbiased regardless of the
distribution of input data (in an i.i.d.~model).
\R by default uses type VII.
Types I, III, and IV are not nondecreasing functions of $\alpha$,
therefore and not appropriate from our perspective.
\end{example}

\begin{table}[p!]
\caption{\label{Tab:quantiles1} Different quantile functions
listed in \cite{HyndmanFan1996:samplequantiles}, see Example~\ref{Example:quantiles}.}
\centering
\begin{tabularx}{1.0\linewidth}{llX}
\toprule
\small\bf method & \multicolumn{2}{l}{\small\bf parameters} \\
\midrule
I*   & $k=\lfloor n\alpha \rfloor$    & $\gamma=\left\{\begin{array}{ll}
                         1 & \text{if }k=n\alpha\\
                         0 & \text{otherwise}
                      \end{array}\right.$ \\
\midrule
II  & $k=\lfloor n\alpha \rfloor$  & $\gamma=\left\{\begin{array}{ll}
                         0.5 & \text{if }k=n\alpha\\
                         0 & \text{otherwise}
                      \end{array}\right.$ \\
\midrule
III* & $k=\lfloor n\alpha-0.5 \rfloor$ & $\gamma=\left\{\begin{array}{ll}
                         1 & \text{if }k=n\alpha-0.5\text{ and $k$ is even}\\
                         0 & \text{otherwise}
                      \end{array}\right.$ \\
\midrule
IV*   & $k=\lfloor n\alpha\rfloor$ & $\gamma=k+1-\alpha n$ \\
\midrule
V  & $k=\lfloor n\alpha+0.5\rfloor$ & $\gamma=k+1-\alpha n-0.5$ \\
\midrule
VI & $k=\lfloor n\alpha+\alpha\rfloor$ & $\gamma=k+1-\alpha n-\alpha$ \\
\midrule
VII   & $k=\lfloor n\alpha+1-\alpha\rfloor$ & $\gamma=k-\alpha n+\alpha$ \\
\midrule
VIII  & $k=\lfloor n\alpha+\frac{p+1}{3}\rfloor$ & $\gamma=k+1-\alpha n-\frac{p+1}{3}$ \\
\midrule
IX & $k=\lfloor n\alpha+\frac{p}{4}+\frac{3}{8}\rfloor$ & $\gamma=k+\frac{5}{8}-\alpha n-\frac{p}{4}$ \\
\bottomrule
\end{tabularx}
\end{table}

\subsubsection{Hierarchies of fusion functions}\label{Sec:Hierarchies}

\begin{figure}[p!]
\begin{center}
\scalebox{0.9}{
\begin{tikzpicture}
\draw[fill=red!2] (1,-5.25) rectangle (11.5,0.25);
\node at (11, -4.75) {\LARGE $\func{F}$};

\node[draw,minimum size=8mm,fill=yellow!2] (x1) at (0,-1) {$x_1$};
\node[draw,minimum size=8mm,fill=yellow!2] (x2) at (0,-2) {$x_2$};
\node (xi) at (0,-3) {$\vdots$};
\node[draw,minimum size=8mm,fill=yellow!2] (xn) at (0,-4) {$x_n$};
\node[text width=1.5cm,align=center,anchor=north] (l0) at (0,-5.5) {\it inputs};

\node[draw,circle,inner sep=0pt,minimum size=12mm,fill=white] (f11) at (2.5,-0.5) {$\func{F}_{1}^{(1)}$};
\node[draw,circle,inner sep=0pt,minimum size=12mm,fill=white] (f12) at (2.5,-2) {$\func{F}_{2}^{(1)}$};
\node (f1i) at (2.5,-3.25) {$\vdots$};
\node[draw,circle,inner sep=0pt,minimum size=12mm,fill=white] (f1m) at (2.5,-4.5) {$\func{F}_{m_1}^{(1)}$};
\node[text width=1.5cm,align=center,anchor=north] (l1) at (2.5,-5.5) {\it layer~$1$};

\draw[->] (x1)--(f11);
\draw[->] (x1)--(f12);
\draw[->] (x1)--(f1m);
\draw[->] (x2)--(f11);
\draw[->] (x2)--(f12);
\draw[->] (x2)--(f1m);
\draw[->] (xn)--(f11);
\draw[->] (xn)--(f12);
\draw[->] (xn)--(f1m);

\node[draw,circle,inner sep=0pt,minimum size=18mm,fill=white] (fl11) at (8,-1.0) {$\func{F}_{1}^{(l-1)}$};
\node (fl1i) at (8,-2.5) {$\vdots$};
\node[draw,circle,inner sep=0pt,minimum size=18mm,fill=white] (fl1m) at (8,-4) {$\func{F}_{m_{l-1}}^{(l-1)}$};

\node (hid1) at (5.25, -1) {};
\node (hid2) at (5.25, -2) {};
\node (hid3) at (5.25, -3) {};
\node (hid5) at (5.25, -2.5) {};
\node (hid4) at (5.25, -4) {};
\draw[->] (f11)--(hid1);
\draw[->] (f12)--(hid1);
\draw[->] (f1m)--(hid1);
\draw[->] (f11)--(hid2);
\draw[->] (f12)--(hid2);
\draw[->] (f1m)--(hid2);
\draw[->] (f11)--(hid3);
\draw[->] (f12)--(hid3);
\draw[->] (f1m)--(hid3);
\draw[->] (f11)--(hid4);
\draw[->] (f12)--(hid4);
\draw[->] (f1m)--(hid4);

\draw[->] (hid1)--(fl11);
\draw[->] (hid5)--(fl1m);
\draw[->] (hid4)--(fl11);
\draw[->] (hid1)--(fl1m);
\draw[->] (hid5)--(fl11);
\draw[->] (hid4)--(fl1m);
\draw[fill=blue!5] (4.25,0) rectangle (6.25,-5);
\node at (5.25,-2.5) {$\dots$};

\node[text width=1.5cm,align=center,anchor=north] (ll1) at (5.5,-5.5) {\dots};
\node[text width=1.5cm,align=center,anchor=north] (ll1) at (8,-5.5) {\it layer~$l$-$1$};

\node[draw,circle,inner sep=0pt,minimum size=14mm,fill=white] (fl1) at (10.5,-2.5) {$\func{F}_{l}^{(1)}$};
\node[draw,minimum size=10mm,fill=green!2] (y) at (12.5,-2.5) {$y$};
\draw[->] (fl11)--(fl1);
\draw[->] (fl1m)--(fl1);
\draw[->] (fl1)--(y);

\node[text width=1.5cm,align=center,anchor=north] (ll) at (10.5,-5.5) {\it layer~$l$};
\node[text width=1.5cm,align=center,anchor=north] (ly) at (12.5,-5.5) {\it output};

\end{tikzpicture}
}
\end{center}

\caption{\label{Fig:Hierarchy}A hierarchy of fusion functions.}
\end{figure}
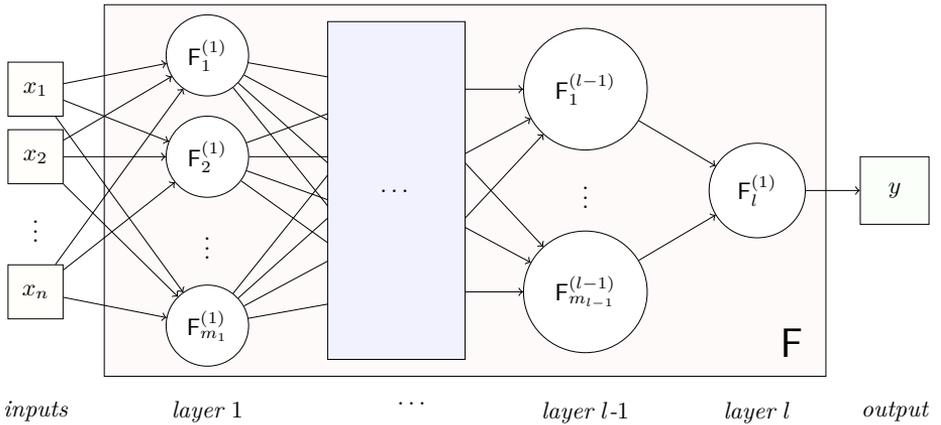

Being inspired by Torra's \cite{Torra1999:neuralnet} paper,
let us consider the concept of a general fusion function hierarchy,
see Figure~\ref{Fig:Hierarchy}.

\begin{definition}\index{hierarchy of fusion functions}%
A hierarchy of fusion functions is a tuple
$\mathcal{F}=(l, \vect{m}, {\func{\vect{F}}})$,
where $l\in\mathbb{N}$ denotes the number of layers,
$\vect{m}=(m_0, m_1,\dots,m_l)\in\mathbb{N}^l$, where $m_i$ gives the number
of fusion functions in layer $i\in[l]$, $m_l=1$ and $m_0=n$ is the number of inputs,
and $\func{\vect{F}}=(\func{F}_{j}^{(i)})_{i\in[l], j\in[m_i]}$ is a sequence
of fusion functions like $\func{F}_j^{(i)}:\IvalPow{m_{i-1}}\to\Ival$.
\end{definition}

A hierarchy of fusion functions $\mathcal{F}$ determines in fact a new fusion function, $\func{F}$,
whose output is determined as follows.

\begin{algorithm}
To determine the output of a hierarchy of fusion functions $\mathcal{F}=(l, \vect{m}, {\func{\vect{F}}})$,
do:
\begin{enumerate}
   \item[1.] Let $y_j^{(0)} := x_j$ for $j\in[n]$;
   \item[2.] For $i=1,2,\dots,l$ do:
   \begin{enumerate}
      \item[2.1.] For $j=1,2,\dots,m_i$ do:
      \begin{enumerate}
      \item[2.1.1.] Let $y_{j}^{(i)} := \func{F}_{j}^{(i)}(y_{1}^{(i-1)},\dots, y_{m_{i-1}}^{(i-1)})$;
      \end{enumerate}
   \end{enumerate}
   \item[3.] Return $y_{1}^{(l)}$ as result;
\end{enumerate}
\end{algorithm}

Surely, $l=1$ gives the case of an ordinary, ``single fusion function'' setting.
Please note that if $\func{F}_j^{(i)}$ consequently fulfills certain properties,
then by recursively applying Proposition~\ref{Prop:CompositionProps}
we may deduce the implied properties of the outcoming $\func{F}$.

\begin{example}\label{Ex:FFNN}\index{feedforward neural network}%
Let $\Ival=[-1,1]$.
A feedforward neural network $(l, \vect{m}, \vect{p})$, see \cite{HastieTibshiraniFriedman2009:esl},
is a particular hierarchy of fusion functions $(l, \vect{m}, {\func{\vect{F}}})$ with:
\[
\func{F}_{j}^{(i)}(x_1,\dots,x_{m_{i-1}}) = \func{f}\left(
\sum_{k=1}^{m_{i-1}} p_{k,j}^{(i)} x_k + p_{0,j}^{(i)}1
\right),
\]
where $\func{f}:\mathbb{R}\to\Ival$ is the so-called activation function,
typically:
\[
   \func{f}(x) = \frac{1}{1+\exp(-x)},
\]
i.e., the sigmoidal function,
and $p_{k,j}^{(i)}\in\mathbb{R}$ are arbitrary coefficients,
$i\in[l]$, $j\in[m_i]$, $k\in[0:m_{i-1}]$.
Here, $y_j^{(i)}$ are  called \index{neuron}\emph{neurons}. Note that
$p_{0,j}^{(i)}$ may be treated as a coefficient standing near a so-called
bias neuron, whose value is fixed at $1$.

Artificial neural networks are widely used in (deep) machine learning
for automated data classification.
\end{example}

\begin{example}\index{quasi-arithmetic mean}%
Torra in \cite{Torra1999:neuralnet} showed that
a feedforward neural network is isomorphic to a hierarchy of quasi-arithmetic means.
Here is a sketch of its possible construction.
First of all, every input element is copied with sign changed so that only nonnegative
coefficients may from now on be taken into account.
Then, another artificial neuron is added and the coefficients are accordingly
normalized -- now they are indeed weights (thus, they sum up to $1$).
Further on, by an appropriate choice of the generator function $\varphi$, closely related
to the activation mapping, we may note that in fact only functions of quasi-arithmetic means
are used.
\end{example}

\subsection{Monotone measures and integrals}\label{Sec:Integrals}

It turns out that some fusion functions are tightly related to monotone measures
and respective integrals -- tools known from decision making, social choice theory,
as well as engineering. Here we shall present the notion of a universal integral,
which gives a common framework to the famous Choquet \cite{Choquet1954:theorycapa},
Sugeno \cite{Sugeno1974:PhD}, and Shilkret \cite{Shilkret1971:maxitivemeasure} integrals.
Due to this,  we may not only explore new interpretations of already introduced
data fusion tools, but also generate new ones.

\bigskip
First we shall review some basic definitions and concepts,
see also, e.g., \cite[Chapter 5]{GrabischETAL2009:aggregationfunctions} or
\cite[Chapter 4]{BeliakovETAL2015:practicalbook}.
Let $(\Omega,\mathcal{F})$ be a measurable space,
i.e., a nonempty set $\Omega$ equipped with a $\sigma$-algebra.

\begin{definition}
\index{capacity|see {monotone measure}}\index{monotone measure}%
We call $\mu:\mathcal{F}\to[0,\infty]$ a \emph{monotone measure} (a capacity)
on $(\Omega,\mathcal{F})$,
if:
\begin{itemize}
   \item[(a)] $\mu(\emptyset)=0$,
   \item[(b)] $\mu(\Omega)>0$, and
   \item[(c)] $\mu(U)\le \mu(V)$ for $U\subseteq V$, $U,V\in\mathcal{F}$.
\end{itemize}
\end{definition}

Note that a monotone measure is not necessarily ($\sigma$-)additive.
A normalized monotone measure, i.e., one which has $\mu(\Omega)=1$
from now on shall be called a \index{fuzzy measure}\emph{fuzzy measure}.

Denoting by $\mathcal{B}([0,\infty])$ the $\sigma$-algebra
of Borel subsets of $[0,\infty]$,
we say  that a function $\func{X}: \Omega\to[0,\infty]$ is \index{measurable function}%
\textit{$\mathcal{F}$-measurable}, if for each $T\in\mathcal{B}([0,\infty])$
its inverse image $\func{X}^{-1}(T)$ is an element of $\mathcal{F}$.

Let $\mathcal{M}^{(\Omega,\mathcal{F})}$ denote the set of all monotone measures on $(\Omega,\mathcal{F})$
and $\mathcal{R}^{(\Omega,\mathcal{F})}$ designate the set of all $\mathcal{F}$-measurable
functions $\func{X}: \Omega\to[0,\infty]$.

\begin{remark}
Please note that for both $\mathcal{M}^{(\Omega,\mathcal{F})}$ and
$\mathcal{R}^{(\Omega,\mathcal{F})}$ natural partial orders $\preceq_\mathcal{M}$
and $\preceq_\mathcal{R}$, respectively, may be constructed.
This is because we have, e.g., $\func{X}\preceq_\mathcal{R}\func{Y}$
if and only if for all $\omega\in \Omega$ it holds $\func{X}(\omega)\le\func{Y}(\omega)$.
Moreover, the spaces $(\mathcal{M}^{(\Omega,\mathcal{F})},\preceq_\mathcal{M})$ and
$(\mathcal{R}^{(\Omega,\mathcal{F})},\preceq_\mathcal{R})$ are lattices
(see Section~\ref{Sec:ProdLat}).
\end{remark}

\smallskip
For further discussion we shall also need the notion
of a pseudomultiplication operation.

\begin{definition}
A bivariate fusion function $\otimes:[0,\infty]^2\to[0,\infty]$ is called
\index{pseudomultiplication}%
a~\textit{pseudomultiplication} operation, whenever:
\begin{enumerate}
   \item[(a)] it is nondecreasing in each variable,
   i.e., for $0\le x_1\le x_2$ and $0\le y_1\le y_2$,
   we have $x_1\otimes y_1\le x_2\otimes y_2$,
   \item[(b)] it has $0$ as the annihilator element,
   i.e., for all $x\in[0,\infty]$, $x\otimes 0=0\otimes x=0$,
   \item[(c)] it has a neutral element $e>0$,
   i.e., for all $x\in[0,\infty]$, $x\otimes e=e\otimes x=x$.
\end{enumerate}
\end{definition}

Note that $\otimes$ is neither necessarily associative nor commutative.
Standard multiplication $\cdot$ ($e=1$) and
minimum $\wedge$ ($e=\infty$) are particular examples
of pseudomultiplication operations. On the other hand, e.g., maximum $\vee$
does not annihilate at $0$, thus does not fall into this class.

\smallskip
What is more, let $\{\omega\in\Omega: \func{X}(\omega) \ge t\}\in\mathcal{F}$ be the so-called
\index{t-level set}\emph{$t$-level set} of $\func{X}$, $t\in[0,\infty]$.

\begin{example}\label{Example:MeasureVector}
Let $(\Omega,\mathcal{F})=([n],2^{[n]})$
and take any $\vect{x}\in\IvalPow{n}$, $\Ival=[0,b]$.
By setting $\func{X}(i)=x_{i}$ we have that for any $t \ge 0$ the $t$-level
set of $\func{X}$ fulfills
$\{\omega: \func{X}(\omega) \ge t\} = \{i: x_i \ge t\}$,
i.e., there is a one-to-one correspondence between $\vect{x}$ and $\func{X}$.
\end{example}

It is easily seen that $\{\omega: \func{X}(\omega) \ge t\}_{t\in[0,\infty]}$ forms a
left-continuous, nonincreasing chain (with respect to~$t$).
Thus,
\begin{equation}
\func{S}^{(\mu,\func{X})}(t):=\mu(\{\omega\in\Omega: \func{X}(\omega) \ge t\})
\end{equation}
is a nonincreasing function of $t$.

\begin{example}\label{Example:ProbSpace}
   Let $(\Omega, \mathcal{F}, P)$ be a probability space,
   \index{probability measure}%
   i.e., a measurable space equipped with a
   \index{probability measure}\emph{probability measure}
   (a $\sigma$-additive fuzzy measure) $P$, see \cite{Billingsley1979:probmeasure}.
   In this setting, $\Omega$ is called a \index{sample space}\emph{sample space},
   any $\func{X}\in \mathcal{R}^{(\Omega,\mathcal{F})}$ is named a (nonnegative real-valued)
   \index{random variable}\emph{random variable},
   and $\func{S}^{(\mu,\func{X})}(t)=P(\{\omega\in \Omega: \func{X}(\omega)\ge t\})$
   is often shortened as $P(\func{X}\ge t)$ and called a \emph{survival function}.
   \index{survival function}\index{c.d.f.|see {cumulative distribution function}}\index{cumulative distribution function}%
   It might also be observed that a \emph{cumulative distribution function} is tightly connected to it:
   we have $\func{F}^{(\mu,\func{X})}(t) = P(\func{X}\le t) = 1-\func{S}^{(\mu,\func{X})}(t)+P(\{\omega\in \Omega: \func{X}(\omega)= t\})$.

   Note that in probability theory it is customary to say just
   ``let $X$ be a random variable with c.d.f.~$F$'' -- to some degree the definitions
   of all the underlying objects may be inferred implicitly.
\end{example}

\begin{figure}[tb!]
\centering

\includegraphics[width=8.25cm]{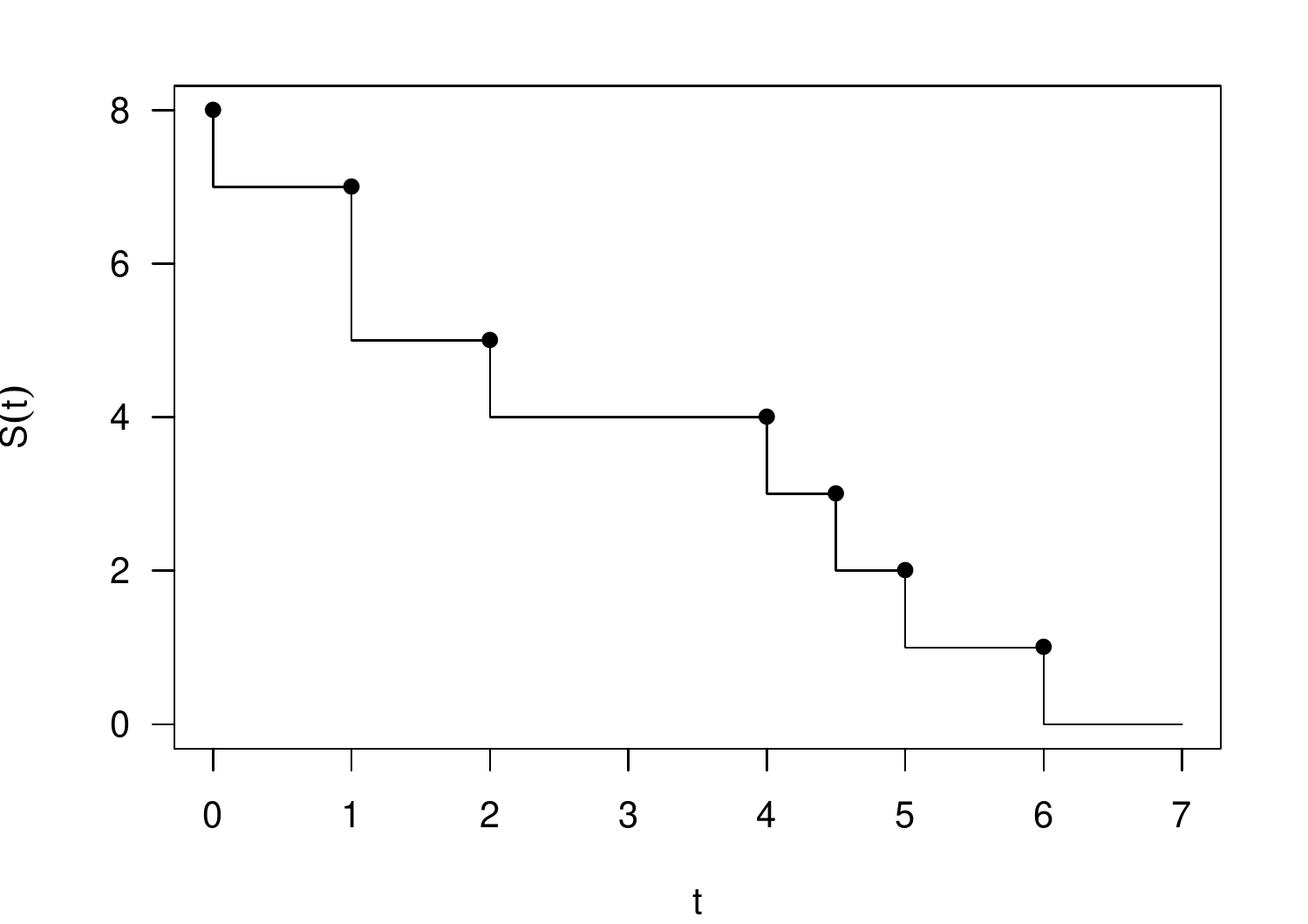}

\caption[A graphical representation of a numeric list.]{\label{Fig:countingmeasure} A graphical representation
of an ordered version of the numeric list $(1, 6, 4.5, 1, 0, 5, 2, 4)$.}
\end{figure}

\begin{example}\label{Example:MeasureVector2}
In Example~\ref{Example:MeasureVector},
\index{counting measure}%
if $\mu(U)=|U|$ for $U\in\mathcal{F}$ is the \emph{counting measure},
$\func{S}^{(\mu,\func{X})}(t)$ gives us an appealing graphical representation
of $\vect{x}_\sigma$ (in the form of a step function), where $\sigma$ is a permutation that orders
observations in $\vect{x}$ nonincreasingly, see Figure~\ref{Fig:countingmeasure}.
Here, a choice of a different \emph{symmetric monotone measure}, i.e., one such that
\index{symmetric monotone measure}
$\mu(U)=\varphi(|U|)$ for some nondecreasing $\varphi$, $\varphi(0)=0$, $\varphi(n)>0$,
corresponds to some transformation of the plot's  $y$ axis.
Also, please refer, e.g., to \cite{Grabisch1997:additivediscrfm} for a
review of basic classes of discrete fuzzy measures.
\end{example}

As noted in \cite{KlementMesiarPap2010:universalintegral},
which function shall be called an integral of
$\func{X}\in\mathcal{R}^{(\Omega,\mathcal{F})}$ is still
a disputable issue.
Generally, it is agreed that an integral:
\begin{itemize}
   \item should map the space $\mathcal{M}^{(\Omega,\mathcal{F})}\times\mathcal{R}^{(\Omega,\mathcal{F})}$ into $[0,\infty]$,
   \item should be at least nondecreasing with respect to each coordinate, and
   \item for $\func{X}\equiv 0$ it should return the value 0.
\end{itemize}

In this book, we rely on the notion of a universal integral, introduced
by Klement, Mesiar, and Pap. The following characterization (for the purpose of this book, we use it as a definition)
was provided for it in \cite[Proposition 2.7]{KlementMesiarPap2010:universalintegral},
see also \cite{GrecoMesiarRindone2014:twocharactunivine} for an alternative
setting in the discrete (thus, particular) case.

\begin{definition}\label{Def:universalintegral}\index{universal integral}%
A \textit{universal integral} corresponding to a pseudomultiplication operation $\otimes$
is a function $\mathcal{I}:\mathcal{M}^{(\Omega,\mathcal{F})}\times\mathcal{R}^{(\Omega,\mathcal{F})}\to[0,\infty]$
given by:
\begin{equation}
\mathcal{I}\left(\mu,\func{X}\right)=\mathcal{J}\left(\func{S}^{(\mu,\func{X})}\right),
\end{equation}
where
$\mathcal{J}: \mathcal{R}^{([0,\infty],\mathcal{B}([0,\infty]))} \to [0,\infty]$ is nondecreasing
and such that for each $c,d\in[0,\infty]$ we have $\mathcal{J}(d\indicator_{(0,c]})=c\otimes d$.
\end{definition}

Please note that $\mathcal{I}\left(\mu,e\indicator_{U}\right)=\mu(U)$
for all $U\in\mathcal{F}$, where $e$ is the neutral element of $\otimes$.
Given a measurable space $(\Omega,\mathcal{F})$,
below are a few well-known examples of universal integrals
of $\func{X}\in\mathcal{R}^{(\Omega,\mathcal{F})}$
with respect to~a monotone measure $\mu\in\mathcal{M}^{(\Omega,\mathcal{F})}$:

\begin{definition}
\index{Choquet integral}The \emph{Choquet integral} \cite{Choquet1954:theorycapa}
is given by:
\begin{equation}\label{Eq:IntChoquet}
\mathrm{Ch}(\mu,\func{X}) = \int_{[0,\infty]} \func{S}^{(\mu,\func{X})}(t)\,dt.
\end{equation}
\end{definition}

Here we have $\otimes=\cdot$ (standard multiplication).
Note that this integral is defined in the same way as the one by Lebesgue,
but with respect to an arbitrary monotone measure.
In this regard, for brevity, we often write $\int_{[0,\infty]} \func{S}^{(\mu,\func{X})}(t)\,dt. = \int \func{X}\,d\mu$.

\begin{example}
Referring back to the setting from Example~\ref{Example:ProbSpace}
(a probability space), the Choquet integral corresponds to the expected value
of a nonnegative random variable $\func{X}$. This is because
$\mathbb{E}\,\func{X}=\int_0^\infty P(\func{X}\ge t)\,dt=\int \func{X}\,dP$.
\end{example}

\begin{example}
In example \ref{Example:MeasureVector},
if $\sigma\in\mathfrak{S}_{[n]}$ is an ordering permutation of $\vect{x}$,
assuming that $x_{\sigma(0)}=0$, it holds:
\begin{eqnarray*}
&&\mathrm{Ch}(\mu,\func{X}) = \int_{[0,\infty]} \func{S}^{(\mu,\func{X})}(t)\,dt = \\
&=& \sum_{i=1}^n \big(x_{\sigma(i)}-x_{\sigma(i-1)}\big) \mu(\{\sigma(i),\dots,\sigma(n)\})\\
&=& \sum_{i=1}^n x_{\sigma(i)} \big(\mu(\{\sigma(i),\dots,\sigma(n)\})-\mu(\{\sigma(i+1),\dots,\sigma(n)\})\big).
\end{eqnarray*}
Thus, if $\mu$ is a symmetric fuzzy measure, then the Choquet integral corresponds
to some \index{OWA operator}OWA operator
-- here a monotone measure in fact generates a weighting vector.
Moreover, if $\mu$ is an additive fuzzy measure,
\index{weighted arithmetic mean}%
then we get the case of a weighted arithmetic mean, $\func{WAMean}$, see, e.g.,
\cite{Marichal2000:axiomaticdiscrchoquet}.
\end{example}

\begin{definition}
\index{Shilkret integral}The \emph{Shilkret integral} \cite{Shilkret1971:maxitivemeasure} is given by:
\begin{equation}\label{Eq:IntShilkret}
\mathrm{Sh}(\mu,\func{X}) = \sup_{t\in{[0,\infty]}} \{ t\cdot \func{S}^{(\mu,\func{X})}(t) \},
\end{equation}
with convention $0\cdot\infty=0$.
\end{definition}

In this case we have $\otimes=\cdot$ as well.
Following is a very intuitive example of the usefulness of the introduced concepts
which we attribute to Mesiar \cite{Mesiar2014:slidesWarszawa}.

\begin{example}[\cite{Mesiar2014:slidesWarszawa}]
Suppose that $\Omega=\{a,b,c\}$ represents the set of three blue-collar workers
and $\mathcal{F}=2^\Omega$.
Let $\mu:\mathcal{F}\to[0,\infty]$ give their per-hour overall performance
when they work either alone or in teams.
\begin{center}
\begin{tabularx}{1.0\linewidth}{lXXXXXXXX}
\toprule
$U\in\mathcal{F}$ &  $\emptyset$ & $\{ a\}$  & $\{ b\}$ & $\{ c\}$ & $\{ a,b\}$ & $\{ a,c\}$ & $\{ b,c\}$ & $\{ a,b,c\}$ \\
\midrule
$\mu(U)$          &   $0$          &  $2$        &   $3$      &    $4$     & $7$          & $4$          & $5$          & $8$ \\
\bottomrule
\end{tabularx}
\end{center}
We see that, due to various reasons, working together on the same task does not necessarily
increase their performance additively. Hence, $\mu$ is not a measure in the classical sense.

Now let $\func{X}: \Omega\to[0,\infty]$ be a function denoting each worker's
availability -- how many hours they can work in a certain day:
\begin{center}
\begin{tabularx}{1.0\linewidth}{Xrrr}
\toprule
$\omega\in\Omega$                & $a$ & $b$ & $c$ \\
\midrule
$\func{X}(\omega)$               &$ 5$ & $4$ & $3$ \\
\bottomrule
\end{tabularx}
\end{center}
The corresponding $t$-level sets and $\func{X}^{(\mu,\func{X})}(t)$ are as follows:
\begin{center}
\begin{tabularx}{1.0\linewidth}{lXXXX}
\toprule
$T$                                                 & $[0,3]$ & $]3,4]$ & $]4,5]$ & $]5,\infty]$ \\
\midrule
$\{\omega\in \Omega: \func{X}(\omega) \ge t\}$, $t\in T$           & $\{a,b,c\}$ & $\{a,b\}$ & $\{a\}$ & $\emptyset$ \\
\midrule
$\func{S}^{(\mu,\func{X})}(t)$, $t\in T$            & $8$ & $7$ & $2$ & $0$ \\
\bottomrule
\end{tabularx}
\end{center}
For instance, only $a$ and $b$ may work for no less than $3.5$ hours that day.

Here the Shilkret integral yields the result equal to $4\cdot 7 = 28$ --
this is the best total performance under the constraint that only one group may work this day.
On the other hand, the Choquet integral gives $3\cdot 8+1\cdot 7+1\cdot 2 = 33$ --
all the workers start their work at the beginning of the time period,
and then once one of them stops, he/she goes home and does not continue that day.
\end{example}

\begin{example}
In Example~\ref{Example:MeasureVector2} the two integrals have an appealing
graphical interpretation: the Choquet integral corresponds to the area
below the step function representing a vector $\vect{x}$, and the Shilkret integral
is the area of the largest rectangle that can be fitted under such a function.
\end{example}

Let us consider an example of a universal integral that uses a different
pseudomultiplication operation, $\otimes=\wedge$. Hence, its value
has a quite different interpretation. Even if its present form is due to Sugeno
-- as noted in \cite{GrabischETAL2009:aggregationfunctions} --
some of its aspects were already studied by Ky Fan in the 1940s \cite{Fan1943:metric}.

\begin{definition}
\index{Sugeno integral}The \emph{Sugeno integral} \cite{Sugeno1974:PhD}
may be expressed as:
\begin{eqnarray}
\mathrm{Su}(\mu,\func{X}) &=&\label{Eq:IntSugeno1} \sup_{t\in{[0,\infty]}} \{ t \wedge \func{S}^{(\mu,\func{X})}(t) \}.
\end{eqnarray}
\end{definition}

\begin{example}
In the setting established in Examples~\ref{Example:MeasureVector} and \ref{Example:MeasureVector2}
the graphical interpretation of the discrete Sugeno integral is as follows:
it is the side of the largest square that can be fitted under the step function.
Here, this universal integral may be expressed as:
\[
   \bigvee_{i=1}^n x_{\sigma(i)} \wedge \mu(\{ \sigma(i),\dots,\sigma(n) \}),
\]
where $\sigma\in\mathfrak{S}_{[n]}$ is the ordering permutation of a given vector $\vect{x}$,
see, e.g., \cite{Marichal2000:sugenointagfun}.

Let $\vect{v}\in\IvalPow{n}$ be such that $\bigvee_{i=1}^n v_i = b = \sup\Ival$,
and $\mathcal{A}=\{A_j\}_{j\in[k]}$, $\emptyset\neq A_j\subseteq[n]$ for some $k$.
This integral generalizes all \index{order statistic}order statistics
as well as the following fusion functions:
\begin{itemize}
   \item \index{WMax@$\mathsf{WMax}$|see {weighted maximum}}\index{weighted maximum}%
   $\func{WMax}_\vect{v}(\vect{x})=\displaystyle \bigvee_{i=1}^n v_i\wedge x_i$,
   \hfill(\emph{weighted maximum}, see \cite{DuboisETAL1988:wfpm})

   \item \index{WMin@$\mathsf{WMin}$|see {weighted minimum}}\index{weighted minimum}%
   $\func{WMin}_\vect{v}(\vect{x})=\displaystyle \bigwedge_{i=1}^n (b-v_i)\vee x_i$,
   \hfill(\emph{weighted minimum})

   \item \index{OWMax@$\mathsf{OWMax}$|see {ordered weighted maximum}}\index{ordered weighted maximum}%
   $\func{OWMax}_\vect{v}(\vect{x})=\displaystyle \bigvee_{i=1}^n v_i\wedge x_{(i)}$,
   where $v_1\ge\dots\ge v_n$,\par
   \hfill(\emph{ordered~weighted~maximum, see \cite{DuboisPrade1996:owmin}})

   \item \index{OWMin@$\mathsf{OWMin}$|see {ordered weighted minimum}}\index{ordered weighted minimum}%
   $\func{OWMin}_\vect{v}(\vect{x})=\displaystyle \bigwedge_{i=1}^n (b-v_i)\vee x_{(i)}$,
   where $v_1\le\dots\le v_n$,\par
   \hfill(\emph{ordered~weighted~minimum})
   \item \index{LPF@$\mathsf{LPF}$|see {lattice polynomial function}}\index{lattice polynomial function}%
   $\func{LPF}_\mathcal{A}(\vect{x}) = \bigvee_{j=1}^k \bigwedge_{i\in A_j} x_i$.
   \hfill(\emph{lattice polynomial function})
\end{itemize}
Note that the class of OWMax and OWMin fusion functions coincide
and for each $\func{LPF}_\mathcal{A}$ there exists
$\mathcal{B}=\{B_j\}_{j\in[l]}$, $\emptyset\neq B_j\subseteq[n]$ for some $l$
such that $\func{LPF}_\mathcal{A}(\vect{x}) = \bigwedge_{j=1}^l \bigvee_{i\in B_j} x_i$,
see \cite[Proposition 5.55]{GrabischETAL2009:aggregationfunctions}.

Due to the fact that this integral is defined only using $\wedge$ and $\vee$ operations,
it can be applied on purely ordinal data -- we will refer back to it
in Section~\ref{Sec:LatPolyFun}. As a matter of fact, the discrete Sugeno integral
may also be written as, see \cite[Proposition 5.63]{GrabischETAL2009:aggregationfunctions}:
\[
   \mathrm{Su}(\mu,\vect{X}) = \bigvee_{A\subseteq[n]}\left(
   \bigwedge_{i\in A} x_i\wedge \mu(A)
   \right).
\]
Hence, it is a special case of the so-called
\index{WLPF@$\mathsf{WLPF}$|see {weighted lattice polynomial function}}%
\index{weighted lattice polynomial function}%
\emph{weighted lattice polynomial functions}, given by:
\begin{equation}\label{Eq:WLPFIval}
   \func{WLPF}_{\vect{v}, \mathcal{A}}(\vect{x}) = \bigvee_{j=1}^k \left(\bigwedge_{i\in A_j} x_i \wedge v_j\right),
\end{equation}
for some $k$, $\mathcal{A}=\{A_j\}_{j\in[k]}$, $\emptyset\neq A_j\subseteq[n]$, and $\vect{v}\in\IvalPow{k}$.
\end{example}

\begin{remark}
Many interesting applications of discrete Sugeno integrals have been reported in decision making,
please refer, e.g., to \cite{TorraNarukawa2006:interpretationfuzzyint}.
Moreover, it is also used in the problem of multiple significance testing in statistics,
as a measure of false discovery rate, see \cite{BenjaminiHochberg1995:FDR}
and the issue of measuring performance of scientists, see Section~\ref{Sec:ImpactFunctions}
and \cite{GagolewskiMesiar2014:integrals,TorraNarukawa2008:h2fuzzyintegrals}.
\end{remark}

\medskip
Note that not all the classes of integrals known in the literature
are universal integrals. For example,
\index{decomposition integral}decomposition integrals
introduced by Even and Lehrer \cite{EvenLehrer2014:decompositionintegral},
see also \cite{MesiarStupnanova2013:decompositionintegrals},
include the non-universal Yang's PAN \cite{Yang1985:PanIntegral} and
Lehrer's concave \cite{Lehrer2009:concaveintegral} integral
as well as the discussed above Choquet and Shilkret integral.

\subsection{Penalty-based aggregation functions}

Firstly, we shall recall the notion of a metric and a pseudometric.

\begin{definition}
A \index{metric}\emph{metric} on a set $Z$ is a function $\mathfrak{d}: Z\times Z\to[0,\infty]$
such that for any $\vect{x}, \vect{y}, \vect{z}\in Z$:
\begin{enumerate}
   \item[(a)] $\mathfrak{d}$ fulfills the triangle inequality $\mathfrak{d}(\vect{x}, \vect{y}) \leq \mathfrak{d}(\vect{x}, \vect{z}) + \mathfrak{d}(\vect{z}, \vect{y})$,

   \item[(b)] $\mathfrak{d}$ is symmetric, i.e., $\mathfrak{d}(\vect{x}, \vect{y}) = \mathfrak{d}(\vect{y}, \vect{x})$,

   \item[(c)] it holds $\mathfrak{d}(\vect{x}, \vect{y}) = 0$ if and only if $\vect{x}=\vect{y}$.
\end{enumerate}
Moreover, a \index{pseudometric}\emph{pseudometric} is a function
$\mathfrak{d}': Z\times Z\to[0,\infty]$ that fulfills the triangle inequality, is symmetric,
and such that for $\vect{x}=\vect{y}$ we have $\mathfrak{d}'(\vect{x}, \vect{y}) = 0$.
\end{definition}

If $\mathfrak{d}(\vect{x}, \vect{y})=d$, then it is customary
to say that ``the \emph{distance} between $\vect{x}$ and $\vect{y}$ is $d$''.

Notably, metrics themselves may be aggregated: if $\func{F}:[0,\infty]^k\to[0,\infty]$
is nondecreasing, subadditive, and such that $\func{F}(k\ast 0)=0$, %
then given arbitrary metrics $\mathfrak{d}^{(1)},\dots,\mathfrak{d}^{(k)}$ we have that
$\mathfrak{d}(\vect{x},\vect{y})=\func{F}(\mathfrak{d}^{(1)}(\vect{x},\vect{y}),\dots,\mathfrak{d}^{(k)}(\vect{x},\vect{y}))$ is a metric too,
see, e.g., \cite{BorsikDobos1981:productmetric,MartinMayorValero2011:fixedpointdist}.
Moreover, if $\|\cdot\|$ is a norm on a vector space $Z$,
then $\mathfrak{d}(\vect{x},\vect{y})=\|\vect{x}-\vect{y}\|$ is a metric.
In particular, for given $p$ from now on we denote with $\mathfrak{d}_p$
the $p$-norm-based metric ($L_p$ metric).
Note that all $L_p$ metrics coincide on the vector space $\mathbb{R}$,
and thus on any $Z=\Ival\subseteq \mathbb{R}$: they may be expressed
as $\mathfrak{d}(x, y)=|x-y|$.

\medskip
The fusion functions studied in this section may be expressed as minimizers of
some kind of penalty or dissimilarity measure between
the observations in an input sample and the resulting value. Intuitively,
this represents the idea behind widely conceived \textit{averaging}: we seek the
$y$ that is a good ``compromise'', on the whole being not ``far away'' from the inputs.

In this section we deal with idempotent aggregation functions.
To measure the overall (aggregated) dissimilarity, here we rely on the concept
of a penalty function which was introduced by Yager and Rybalov in \cite{YagerRybalov1997:medianfusion}
and then extended in the works of Calvo and others, see, e.g., \cite{CalvoMesiarYager2004:quantweight,CalvoBeliakov2010:penalties}.
The general idea behind them is well explained, e.g., in \cite{BeliakovETAL2011:penaltyconsensus}:
if we have equal values on input, then the output $y$ is the same value,
we have a unanimous vote. On the other hand, if some input
$x_i\neq y$, then we impose a kind of ``penalty'' for such a disagreement.
The larger the disagreement, then the more the inputs disagree with the output
and the larger the penalty.

\begin{definition}[\cite{CalvoBeliakov2010:penalties}]\label{Def:PenaltyFunction}
\index{penalty function}%
The function $P:\Ival\times\IvalPow{n}\to[0,\infty]$
is a \emph{penalty function}, whenever:
\begin{enumerate}
   \item[(a)] $P(y;\vect{x})=0$ if $\vect{x}=(n\ast y)$,
   \item[(b)] for every fixed $\vect{x}$, the set of minimizers of $P(y;\vect{x})$ is either
   a singleton or an interval.
\end{enumerate}
\end{definition}

\begin{definition}[\cite{CalvoBeliakov2010:penalties}]
\index{penalty-based function}%
Given a penalty function $P$, a $P$-based function is  defined as:
\begin{equation}
\func{F}(\vect{x})=\argmin_y P(y;\vect{x})
\end{equation}
if $y$ is the unique minimizer of $P(y;\vect{x})$,
and $y=(u+v)/2$ if the set of minimizers is an (open or closed) interval $]u, v[$.
\end{definition}

Based on the fact that we may always take $P(y;\vect{x})=(\func{F}(\vect{x})-y)^2$,
we have the following simple yet appealing result,
which states that every idempotent fusion function is a penalty-based one.

\begin{theorem}[\cite{CalvoBeliakov2010:penalties}]
   Let $\func{F}:\IvalPow{n}\to\Ival$ be an idempotent function.
   Then there exists a penalty function $P$ such that
   $ \func{F}(\vect{x})=\argmin_y P(y;\vect{x})$ for all $\vect{x}$.
\end{theorem}

As a particular class of penalty functions, we may consider, e.g., one that
consists of mappings given by:
\begin{equation}\label{Eq:faithfulpenaltyfunction}
   P(y; \vect{x}) = \sum_{i=1}^n w_i p(y, x_i),
\end{equation}
where $\vect{w}$ is a weighting vector and
$p:\Ival\times\Ival\to[0,\infty]$ is a dissimilarity function
that fulfills:
\begin{itemize}
   \item $p(y,x)=0$ if and only if $x=y$,
   \item $p(y,x)\ge p(y,x')$ whenever $x\ge x'\ge y$ or $x\le x'\le y$.
\end{itemize}
\index{faithful penalty function}%
\emph{Faithful penalty functions} \cite{CalvoMesiarYager2004:quantweight}
are defined via $p(y,x)=K(h(y),h(x))$ where $h$ is continuous monotone
and $K$ is convex.

Among faithful penalty-based aggregation functions
we have, e.g., the \index{weighted arithmetic mean}weighted arithmetic mean:
\[
   \func{WAMean}(\vect{x})=\argmin_y \sum_{i=1}^n w_i p(y, x_i)=\argmin_y \sum_{i=1}^n w_i (x_i-y)^2
\]
and \index{median}median (again note that the minimizer might not be unique):
\[
   \func{Median}(\vect{x})=\argmin_y \sum_{i=1}^n |x_i-y|.
\]
According to \cite{BeliakovETAL2011:penaltyconsensus}, these two results were already known to Laplace.

On the other hand, if, e.g., $p(y,x)=(\varphi(x)-\varphi(y))^2$, then we obtain
\index{weighted quasi-arithmetic mean}a weighted quasi-arithmetic mean with generator $\varphi$,
and if we use $p(y,x_{(i)})$ instead of $p(y,x_i)$ in Equation~\ref{Eq:faithfulpenaltyfunction},
then we obtain a symmetric function which, unfortunately,
might not always be monotonic and well-defined, see \cite{BeliakovETAL2011:penaltyconsensus}.
Yet, in this way it is possible to obtain, e.g., \index{OWA operator}OWA operators.
Other classes of (non necessarily faithful) penalty-based aggregation functions
include, e.g., deviation and entropic means, see \cite{BeliakovETAL2011:penaltyconsensus}
and functions generated by so-called restricted \cite{BustinceETAL2008:reldisfun,BustinceETAL2011:restrictedpenalty}
dissimilarity functions, see also \cite{Mesiar2007:fsutility}.

\bigskip
Viewing idempotent fusion functions as minimizers of some penalty function
is a very inspiring concept, especially when we shall deal with aggregation
of more complex objects in the following chapters.
In particular, soon we are going to consider the concept of a centroid
(minimizer of the sum of squared distances),
1-median (minimizer of sums of distances),
and 1-center (minimizer of maximums of distances), among others.

\section{Extended aggregation functions}

Sometimes we do not know in advance the value of $n$ (an input vector's length)
or we just would like to be ``prepared'' to aggregate any number of observations.
Here is the definition of a data fusion tool that reflects this need.

\begin{definition}\index{extended fusion function}%
An \emph{extended fusion function} is a mapping $\func{F}^*:\Ival^*\to \Ival$.
\end{definition}

Recall that if $X$ is a set, then $X^*=\bigcup_{n=2}^\infty X^n$ designates the family
of all the vectors with elements in $X$ of length at least 2.
This is because aggregation of a single value is not particularly interesting,
we usually set $\func{F}(x)=x$ if it is indeed necessary.

Thus, an extended fusion function may be treated as a family of $2,3,\dots$-ary fusion functions,
each acting on a vector of fixed arity. This may be written as:
\[
   \func{F}^* = \left(\func{F}^{(2)}, \func{F}^{(3)}, \func{F}^{(4)}, \dots \right),
\]
where $\func{F}^{(n)}=\func{F}^*|_{\IvalPow{n}}$, i.e., a projection
of $\func{F}^*$ onto $\IvalPow{n}$.
According to \cite{CalvoMayor1999:remarks2eaf},
the concept of extended aggregation functions
has been introduced by Mayor and Calvo in~\cite{MayorCalvo1997:eaf}.

\begin{example}
Let us go back to the definition of the arithmetic mean.
Up to now, we assumed that $n$ is fixed. Thus, formally, we have introduced:
\[
\func{AMean}^{(n)}(\vect{x})=\frac{1}{n} \sum_{i=1}^n x_i.
\]
However, this definition may naturally be extended so that input vectors of
any length are accepted:
\[
\func{AMean}^*(\vect{x})=\frac{1}{|\vect{x}|} \sum_{i=1}^{|\vect{x}|} x_i,
\]
which we may simply write as $\func{AMean}^*(x_1,\dots,x_n)=\sum_{i=1}^n x_i/n$
(but now keeping in mind that we may provide a vector of any length $n$ on input).
Note that this indeed may be expressed as a family of aggregation functions,
\[
\func{AMean}^*=\left(
\begin{array}{lcl}
   (x_1,x_2) & \mapsto & \frac{1}{2} (x_1+x_2),\\
   (x_1,x_2,x_3)& \mapsto & \frac{1}{3}(x_1+x_2+x_3),\\
   \dots
\end{array}
\dots\right).
\]
\end{example}

\subsection{Weighting}\label{Sec:WeightingTriangles}

Now let us go back to the definition of a weighted arithmetic mean,
$\func{WAMean}^{(n)}_\vect{w}(\vect{x})\allowbreak=\sum_{i=1}^n w_i x_i$,
where $\vect{w}$ is a weighting vector of length $n$. The
question in this very context is of course how to extend it
to the domain of tuples of arbitrary length? For that
we need the following definition.

\begin{definition}\index{weighting triangle}%
A \emph{weighting triangle}
(see \cite{MayorCalvo1997:eaf,CarbonellMasMayor1997:monotonicextendedowa})
is a sequence
$\triangle=(w_{i,n}\in[0,1]: i\in[n], n\in\{2,3,\dots\})$ with
$\sum_{i=1}^n w_{i,n}=1$ for all $n\ge 2$.
\end{definition}

A weighting triangle can be represented graphically as:
\begin{equation}
\triangle=\left(
\begin{array}{ccccccccc}
&&&w_{1,2} && w_{2,2}&& \\
&&w_{1,3} &&w_{2,3}&& w_{3,3}&&\\
&w_{1,4} &&w_{2,4}&& w_{3,4}&& w_{4,4}\\
\iddots &  & & &  \dots &  & &   & \ddots\\
\end{array}\right)
\end{equation}

Based on the notion of a weighting triangle, we are now able to define,
e.g., an extended weighting arithmetic mean:
\index{weighted arithmetic mean}%
\[
   \func{WAMean}_\triangle^*(x_1,\dots,x_n) = \sum_{i=1}^n w_{i,n} x_{i},
\]
and extended OWA (see \cite{MayorCalvo1997:eaf,CarbonellMasMayor1997:monotonicextendedowa})
operators:
\index{OWA operator}%
\[
   \func{OWA}_\triangle^*(x_1,\dots,x_n) = \sum_{i=1}^n w_{i,n} x_{(i)}.
\]

\begin{example}
A weighting triangle which corresponds to the (extended) sample median
\index{median}%
generated by an OWA operator is given by:
\begin{equation*}\small
\triangle=\left(
\begin{array}{ccccccccccc}
&&&&0.5 && 0.5&&& \\
&&&0 &&1&& 0&&&\\
&&0 &&0.5&& 0.5&& 0&&\\
&0&&0 &&1&& 0&& 0&\\
\iddots &  & & &&  \dots &  & &&   & \ddots\\
\end{array}\right).
\end{equation*}
Another example is the normalized Pascal triangle
with $w_{i,n}={n-1 \choose i-1}/2^{n-1}$, see~\cite{BeliakovETAL2007:aggregationpractitioners}:
\begin{equation*}\small
\triangle=\left(
\begin{array}{ccccccccccc}
&&&&1/2 && 1/2&&& \\
&&&1/4 &&2/4&& 2/4&&&\\
&&1/8 &&3/8&& 3/8&& 1/8&&\\
&1/16&&4/16 &&6/16&& 4/16&& 1/16&\\
\iddots &  & & &&  \dots &  & &&   & \ddots\\
\end{array}\right).
\end{equation*}
\end{example}

Generally, there are a few possible schemes to generate weighting triangles
like $\triangle=(w_{i,n}\in[0,1]: i\in[n], n\in\{2,3,\dots\})$.
\begin{itemize}
   \item Let $\vect{c}=(c_1,c_2,\dots)$ with $c_i\ge 0$ for $i=1,2,\dots$
   and $c_1+c_2>0$. Then we may set (see, e.g., \cite{JamisonETAL1965:convwave}):
   \[
      w_{i,n} = \frac{c_i}{\sum_{j=1}^n c_j}.
   \]
   \item Let $\func{w}:[0,1]\to[0,1]$ be a nondecreasing function
   with $\func{w}(0)=0$ and $\func{w}(1)=1$. In such a case we can set
   (see, e.g., \cite{Borovskikh1981:Lstatistics,Yager1991:connectives}):
   \[
      w_{i,n} = \func{w}\left(\frac{i}{n}\right)-\func{w}\left(\frac{i-1}{n}\right).
   \]
\end{itemize}

More generally, triangles of coefficients like $\triangle=(w_{i,n}: i\in[n], n\in\{2,3,\dots\})$
(with different constraints on $w_{i,n}$) may be considered when extending,
e.g., $\func{WMax}$ or $\func{OWMax}$ operators.
In a similar way we may define a triangle weighting of functions
like $\triangle=(\func{w}_{i,n}:\Ival\to[0,\infty[: i\in[n], n\in\{2,3,\dots\})$
for the purpose of defining extended Bajraktarević means.

We shall refer back to these concepts when considering the so-called
$\alpha$- and $\beta$-orderings in Section~\ref{Sec:AlphaBetaOrdering}
and when studying asymptotic properties of fusion functions applied
on random data in Section~\ref{Sec:RandomVariables}.

\subsection{Arity-dependent vs arity-free properties}

Formally, an abstract property $P$ of a fusion function is a kind of logical predicate:
the statement ``$\func{F}$ fulfills $P$'' might be true or false.
There is a semantic equivalence (one to one correspondence) between such a predicate and
the class of fusion functions that fulfill it.
By defining:
\[
\mathcal{P}=\left\{
\text{fusion function }\func{F}: \func{F}\text{ fulfills } P
\right\},
\]
the statements ``$\func{F}$ fulfills $P$'' (e.g., symmetry)
and ``$\func{F}\in\mathcal{P}$'' (e.g., the class of all symmetric fusion functions) coincide.

Having said that, we may introduce the following classification
of extended fusion functions' properties, see \cite{GagolewskiGrzegorzewski2010:ipmu}.
A property $\mathcal{P}$ may either be:
\begin{itemize}
\item an \index{arity-free property}\emph{arity-free (weak) property},
if it deals only with $n$-ary mappings.
More precisely, it is such that
for all $n,m,n\neq m$ and some $\func{G}^{(n)}\in\mathcal{P}|_{\IvalPow{n}}$
it holds:
\begin{equation*}
\Big\{\func{F}^*|_{\IvalPow{m}}:
        \func{F}^*\in \mathcal{P},\
        \func{F}^*|_{\IvalPow{n}}=\func{G}^{(n)}
\Big\}
    =
\Big\{
\func{F}^*|_{\IvalPow{m}}: \func{F}^*\in \mathcal{P}
\Big\},
\end{equation*}
equivalently:
\[
\left(\forall \func{F}^{(2)}\in\mathcal{P}|_{\IvalPow{2}}, \func{F}^{(3)}\in\mathcal{P}|_{\IvalPow{3}}, \dots\right)
\quad
\left(\func{F}^{(2)}, \func{F}^{(3)}, \dots\right)\in\mathcal{P},
\]

\item or an \emph{arity-dependent (strong) property} otherwise.
\index{arity-dependent property}%
\end{itemize}

\begin{remark}
All the properties we considered up to now are arity-free.
This concerns: nondecreasingness, symmetry, translation and scale equivariance,
continuity, idempotence, etc.
\end{remark}

\subsection{Some arity-dependent properties}\label{Sec:ArityDependentProperties}

Let us make a review of a few interesting arity-dependent properties.
We shall begin with a stronger version of idempotency.

\begin{definition}\index{strong idempotence}%
An extended fusion function $\func{F}^*$ is said to be
\emph{strongly idempotent}, whenever:
for all $\vect{x}\in\bigcup_{n=1}^\infty \IvalPow{n}$ and $k>1$
it holds:
\begin{equation}
\func{F}^*(k\ast\vect{x})=\func{F}^*(\vect{x}).
\end{equation}
\end{definition}

Each strongly idempotent extended fusion function is of course idempotent.
Also note that Ghiselli-Ricci \cite{GhiselliRicci2009:asidempotent,GhiselliRicci2004:nonidempotent}
studied the concept of asymptotic idempotency.

The following property is well known from algebra,
see also \cite[Definition 2.63]{GrabischETAL2009:aggregationfunctions}.

\begin{definition}\label{Def:associativity}\index{associativity}%
We say that $\func{F}^*$ is \emph{associative},
if and for any $\vect{x},\vect{y}\in\bigcup_{n=1}^\infty \Ival^n$ it holds:
\begin{equation}
\func{F}^*(\vect{x},\vect{y})=\func{F}^*\big(\func{F}^*(\vect{x}),\func{F}^*(\vect{y})\big),
\end{equation}
with assumption $\func{F}^*(x)=x$.
\end{definition}

\begin{remark}
In order to define an associative fusion function,
it is sufficient only to provide a formula/algorithm that deals with an input
vector of length~$2$.
The following recursive formula may be used to compute the value
of an associative function:
\[
\func{F}^*(x_1,\dots,x_{n})=\func{F}^*(\func{F}^*(x_1,\dots,x_{n-1}), x_{n}).
\]
In other words, to compute $\func{F}^*(x_1,\dots,x_n)$, we may use the following algorithm:
\begin{enumerate}
   \item[1.] Let $y := x_1$;
   \item[2.] For $i=2,3,\dots, n$:
   \begin{enumerate}
      \item[2.1.] Set $y := \func{F}^*(y, x_i)$;
   \end{enumerate}
   \item[3.] Return $y$ as result;
\end{enumerate}
\end{remark}

\begin{example}
In functional programming, the above scheme is called fold, reduce,
or accumulate. Among associative fusion functions we find, e.g.,
$\func{Prod}$, $\func{Min}$, $\func{Sum}$, and $\func{T}_{Ł}$
-- it appears as a sine qua non condition in the definitions
of t-norms and t-conorms in Section~\ref{Sec:tnorm}.
Below we compute the two latter functions in \R{}
by using a call to the \texttt{Reduce()} built-in.
\index{Sum@$\mathsf{Sum}$}\index{product}\index{minimum}\index{Lukasiewicz t-norm@Łukasiewicz t-norm}%

\begin{lstlisting}[language=R]
Reduce("+", c(1, 2, 3, 4)) # the same as sum(c(1,2,3,4))
## [1] 10
x <- c(0.6, 0.8, 0.7, 1)
Reduce(function(x, y) # Lukasiewicz t-norm
       max(0, x+y-1), x) # i.e., max(0, sum(x)-length(x)+1)
## [1] 0.1
\end{lstlisting}
\end{example}

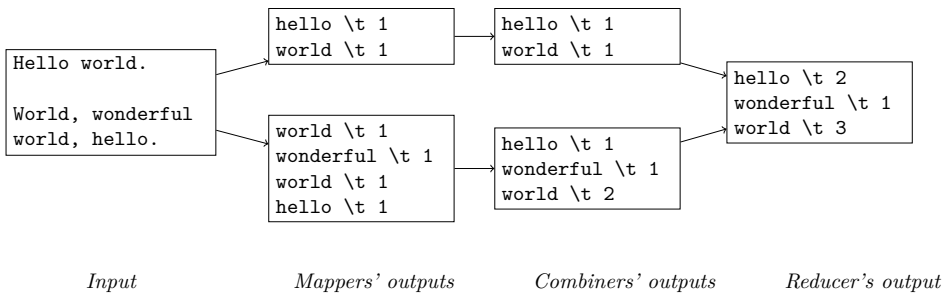
\begin{figure}[b!]
\centering

\scalebox{0.8}{
\begin{tikzpicture}

\node[rectangle,draw] (A) at (0,0) {\parbox{3.2cm}{\texttt{%
Hello world.\newline
\hrulefill\newline
World, wonderful\newline
world, hello.}}};

\node[rectangle,draw] (B1) at (4.1,1.1) {\parbox{2.8cm}{\texttt{%
hello \textbackslash{}t 1\newline
world \textbackslash{}t 1}}};

\node[rectangle,draw] (B2) at (4.1,-1.1) {\parbox{2.8cm}{\texttt{%
world \textbackslash{}t 1\newline
wonderful \textbackslash{}t 1\newline
world \textbackslash{}t 1\newline
hello \textbackslash{}t 1
}}};

\node[rectangle,draw] (C1) at (7.8,1.1) {\parbox{2.8cm}{\texttt{%
hello \textbackslash{}t 1\newline
world \textbackslash{}t 1}}};

\node[rectangle,draw] (C2) at (7.8,-1.1) {\parbox{2.8cm}{\texttt{%
hello \textbackslash{}t 1\newline
wonderful \textbackslash{}t 1\newline
world \textbackslash{}t 2
}}};

\node[rectangle,draw] (D) at (11.6,0) {\parbox{2.8cm}{\texttt{%
hello \textbackslash{}t 2\newline
wonderful \textbackslash{}t 1\newline
world \textbackslash{}t 3
}}};

\node at(0, -3) {\it Input};

\node at(4.3, -3) {\it Mappers' outputs};

\node at(8.4, -3) {\it Combiners' outputs};

\node at(12.3, -3) {\it Reducer's output};

\draw[->] (A)--(B1);
\draw[->] (A)--(B2);

\draw[->] (B1)--(C1);
\draw[->] (B2)--(C2);

\draw[->] (C1)--(D);
\draw[->] (C2)--(D);

\end{tikzpicture}
}

\caption{\label{Fig:MapCombinerReduce} An exemplary Map-Combine-Reduce
word count procedure.}
\end{figure}

\begin{example}
\index{Hadoop Map-Reduce}%
Apache Hadoop Streaming API (at least as far as version 2.6 is concerned)
allows to run scalable Map-Reduce (see \cite{DeanGhemawat2008:googlemr})
jobs with any programs acting as the mapper and/or the reducer.

By default, an input file is processed line-by-line.
Mapper programs receive appropriate chunks of the input file and their aim
is to convert them to key-value pairs. Such pairs should be output
to \texttt{stdout} using a form like:

\begin{lstlisting}
key1 \t val1
key2 \t val2
\end{lstlisting}

Then the mappers' outputs are sorted and merged so that
the reducer receives a sequence of key-value pairs
(on \texttt{stdin}) which are sorted with respect to keys.
Thus, a Hadoop Streaming job acts like a scalable version of:

\begin{lstlisting}
cat input | mapper | sort | reducer > output
\end{lstlisting}

In particular, a simple word count job (a Hadoop ``hello world''-like program)
may be implemented as follows.
\begin{itemize}
   \item Mapper:
   \begin{enumerate}
      \item[1.] For each text line \texttt{l} read from \texttt{stdin}:
          \begin{enumerate}
            \item[1.1.] Split \texttt{l} into separate words;
            \item[1.2.] For each word \texttt{w}:
            \begin{enumerate}
               \item [1.2.1.] Write \texttt{"{}w \textbackslash{}t 1 \textbackslash{}n"{}} to \texttt{stdout};
            \end{enumerate}
         \end{enumerate}
     \item[]\textit{Exemplary input (\texttt{stdin}):}
\begin{lstlisting}
Hello world.
World, wonderful world, hello.
\end{lstlisting}
     \item[]\textit{Desired output (\texttt{stdout}):}
\begin{lstlisting}
hello \t 1
world \t 1
world \t 1
wonderful \t 1
world \t 1
hello \t 1
\end{lstlisting}
   \end{enumerate}

   \item Reducer:
      \begin{enumerate}
         \item[1.] Count the number \texttt{c} of consecutive key-value pairs
         with the same key~\texttt{w};
         \item[2.] Write \texttt{"{}w \textbackslash{}t c \textbackslash{}n"{}} to \texttt{stdout};
      \end{enumerate}
     \item[]\textit{Exemplary input (\texttt{stdin}):}
\begin{lstlisting}
<stdout of the above exemplary mapper job>
\end{lstlisting}
     \item[]\textit{Desired output (\texttt{stdout}):}
\begin{lstlisting}
hello \t 2
wonderful \t 1
world \t 3
\end{lstlisting}
\end{itemize}
By default, the number of mappers is set to be a function of the input
file's size -- the mapper jobs are executed in parallel on each
available cluster node. On the other hand, most often
only a single reducer job is run to collect the output of all the mappers,
which often creates a performance bottleneck.

However, if an aggregation function computed by the reducer
is symmetric and associative, then we may set up an additional job
called \emph{combiner}, which is performed directly
on the outputs generated by mappers. Its aim is to pre-aggregate
chunks of data so that the single-threaded reducer has less
work to do (in our word count example the combiner is exactly the
same program as the reducer). It may be observed that in such a way
some significant speed ups may be obtained.
An exemplary work flow is graphically depicted in Figure~\ref{Fig:MapCombinerReduce}
\end{example}

\begin{remark}
Assume that we have a bivariate fusion function $\func{F}:X^2\to X$, where
$X=\{a_1, \dots, a_k\}$ is a finite set. To check whether its extension
is associative, we may compute a matrix which stores the results of
$\func{F}(a_i, a_j)$, $i\neq j$, and then apply Light's associativity
test algorithm, see, e.g., \cite{Bednarek1968:LightAssociativityTest}.
Moreover, e.g., Rajagopalan and Schulman in
\cite{RajagopalanSchulman2000:RandomizedAssociativityTest}
give an approximate randomized algorithm which runs in $O(k^2 \log(1/p))$
with error probability $p$.
\end{remark}

\medskip
A generalization of the associativity property
is due to Montero and others \cite{DelamoMonteroMolina2001:recursive}
(compare also the notion of a quasi-associative fusion function
-- a function of $k$ associative mappings \cite{Yager1987:quasiassociative}).

\begin{definition}\index{recursivity}%
An extended  fusion function $\func{F}^*$ is (left)-\emph{recursive},
whenever there exist binary functions $\psi_2,\psi_3,\dots: \Ival\times\Ival\to\Ival$
such that for all $\vect{x}\in\bigcup_{n=2}^\infty \IvalPow{n}$ it holds:
\begin{equation}
\func{F}^*(x_1,\dots,x_n) = \psi_n(\func{F}^*(x_1,\dots,x_{n-1}), x_n),
\end{equation}
with assumption $\func{F}^*(x)=x$.
\end{definition}

In other words, we have:
\begin{eqnarray*}
   \func{F}^*(x_1)&=&x_1, \\
   \func{F}^*(x_1, x_2)&=&\psi_2(\func{F}^*(x_1),x_2), \\
   \func{F}^*(x_1, x_2, x_3)&=&\psi_3(\func{F}^*(x_1, x_2),x_3), \\
   \vdots & \vdots & \vdots\\
   \func{F}^*(x_1, x_2\dots,x_n)&=&\psi_{n}(\func{F}^*(x_1,x_2,\dots,x_{n-1}),x_n).
\end{eqnarray*}
Note that an associative fusion function is recursive.
It is generated by setting $\psi_2=\psi_3=\dots=\func{F}^*|_{\IvalPow{2}}$.

Recursive fusion functions are suitable for
on-line processing of input data streams. It is only necessary
to read an input stream sequentially, without the need to have
it in its entirety available from the very beginning (hence, there are memory savings).

\begin{example}
The arithmetic mean is a recursive fusion function. It is because we have:
\[\func{AMean}^*(x_1,\dots,x_n) = \frac{1}{n} \Big( (n-1)\func{AMean}^*(x_1,\dots,x_{n-1}) + x_n \Big). \]
Hence, in this case the generating functions are of the form:
\[
\psi_n(x, y) = \frac{(n-1)x+y}{n}.
\]
\end{example}

In a similar manner we may define a class of functions
that need to have access only to the $k$ last elements of the input sequence
and/or intermediate aggregation results, for some fixed $k$.
This leads, e.g., to the notion of aggregation of ``bags'' of data,
as discussed by Koles\'{a}rov\'{a}, Mesiar, and Montero in \cite{KolesarovaMesiarMontero2015:bags}.
Here, we assume that data come in groups of a few observations.

This idea may be elaborated even further as follows.
We may consider functions which require only a constant number of auxiliary variables
and consecutive observations from an input data stream and thus operate in $O(1)$ memory.

\begin{definition}\index{incrementality}\label{Def:incrementality}
We say that a function $\func{F}^*:\mathbb{R}^*\to\mathbb{R}$ is $k$-\emph{incremental},
$k\in\mathbb{N}$, if there exists $\vect{p}^{(0)}\in\mathbb{R}^k$
and a function $\psi: \mathbb{R}^k\times\Ival\to\mathbb{R}^k\times\Ival$
such that $\func{F}^*(\vect{x})$ for any $\vect{x}\in\IvalAnyPow$ may be computed as follows:
\begin{enumerate}
   \item[1.] Let $\vect{p} = \vect{p}^{(0)}\in\mathbb{R}^k$; \textit{(initialize auxiliary variables)}
   \item[2.] For $i=1,2,\dots, |\vect{x}|$:
   \begin{enumerate}
      \item[2.1.] Set $(\vect{p}, y) := \psi(\vect{p}, x_i)$;
   \end{enumerate}
   \item[3.] Return $y$ as result;
\end{enumerate}
\end{definition}

Of course, a $k$-incremental fusion function is also $k'$-incremental
for $k'>k$. Every associative fusion function is $1$-incremental (store previous $y$),
and each recursive fusion function is $2$-incremental (store previous $y$ and $n$).

\begin{example}
Here are some exemplary $k$-incremental functions (some of them are not
necessarily fusion functions according to Definition~\ref{Def:FusionFunction}):
\begin{itemize}
   \item $\func{Sum}^*$, $\func{Prod}^*$, $\func{Max}^*$, $\func{Min}^*$ are $1$-incremental,
   \item $\func{AMean}^*$ is $2$-incremental,
   \item sample variance and standard deviation is $3$-incremental,
   \item $\func{OS}_k$ is $k$-incremental.
\end{itemize}
\end{example}

Moreover, in practice we may  also be interested in fusion functions which may
be computed using \index{online algorithm}\emph{online algorithms},
i.e., ones for which $\func{F}^*|_{\bigcup_{i=2}^n \IvalPow{i}}$ is $O(n)$-incremental.
Such functions need to examine each observation only once -- this is the case
of the $\func{Median}$ function, among others.
The \lang{C++} Boost \package{accumulators} library includes a set of such
tools.

\bigskip
A concept somehow related to associativity is called \textit{decomposability}
and was already studied in the 1930s by Kolmogorov \cite{Kolmogorov1930:moyenne}
and Nagumo \cite{Nagumo1930:mittelwerte}.

\begin{definition}\label{Def:decomposability}\index{decomposability}%
We call an extended fusion function $\func{F}$ \emph{decomposable} if
for all $\vect{x}\in\Ival^*$ and $k\in[0:|\vect{x}|]$
it holds:
\begin{eqnarray}
   && \func{F}^*(x_1,\dots,x_k,x_{k+1}\dots,x_n)\nonumber\\
   &=&\func{F}^*(k\ast\func{F}^*(x_1,\dots,x_k),(n-k)\ast\func{F}^{*}(x_{k+1},\dots,x_n)),
\end{eqnarray}
with assumption $\func{F}^*(x)=x$.
\end{definition}

\begin{example}
It is known (see, e.g., \cite{Aczel1948:onmeanvalues}) that quasi-arithmetic means
are decomposable. However, as it is noted in \cite[Remark 2.70]{GrabischETAL2009:aggregationfunctions},
decomposability (unlike associativity) does not determine the relationship
between the result of aggregation of $n-1$ elements and $n$ elements.
\end{example}

Sometimes we may also be interested in a property called (strong) \emph{bisymmetry},
see \cite{MarichalMesiar2004:agfinordscal}.

\begin{example}[\cite{BeliakovETAL2007:aggregationpractitioners}]
Suppose that there are $n$ decision makers that express their opinions
on $m$ criteria. Let $x_{ij}$ represent the score given
by the $i$th expert to the $j$th attribute.
Let us suppose that we would like to compute the global score.
How should we do that?
There are at least three possibilities:
\begin{itemize}
   \item Use an $(nm)$-ary fusion function.
   \item For each expert, aggregate his/her opinions on all the criteria,
   and then aggregate $n$ intermediate results to obtain the global score.
   \item Aggregate the experts' opinions separately for each criterion.
   Then use another fusion function to combine such values to get a single
   number.
\end{itemize}
\end{example}

\begin{definition}\label{Def:bisymmetry}\index{bisymmetry}%
We call an extended fusion function \emph{strongly bisymmetric}, if
for all $n,m$, and $\vect{x}=(x_{i,j})\in\IvalPow{n\times m}$
it holds:
\begin{eqnarray}
&&\func{F}^*(\vect{x})=\\
&&\begin{array}{rcccl}
=\func{F}^*\bigg( & \func{F}^*(x_{1,1},\dots,x_{1,m}),  & \dots, & \func{F}^*(x_{n,1},\dots,x_{n,m}) & \bigg)\\
=\func{F}^*\bigg( & \func{F}^*(x_{1,1},\dots,x_{n,1}),  & \dots, & \func{F}^*(x_{1,m},\dots,x_{n,m}) & \bigg).
\end{array}\nonumber
\end{eqnarray}
\end{definition}

This may be represented graphically as:

\smallskip
\[
\begin{array}{|cccc|c|c|}
\cline{1-4}\cline{6-6}
x_{1,1} & x_{1,2} & \dots & x_{1,m} & \rightarrow & f_{1,\cdot} \\
x_{2,1} & x_{2,2} & \dots & x_{2,m} & \rightarrow & f_{2,\cdot} \\
\vdots & \vdots & \ddots &\vdots &  & \vdots \\
x_{n,1} & x_{n,2} & \dots & x_{n,m} & \rightarrow & f_{n,\cdot} \\
\cline{1-4}\cline{6-6}
\multicolumn{1}{c}{\downarrow} & \multicolumn{1}{c}{\downarrow} && \multicolumn{1}{c}{\downarrow} & \multicolumn{1}{c}{} & \multicolumn{1}{c}{\downarrow}\\
\cline{1-4}\cline{6-6}
f_{\cdot, 1} & f_{\cdot, 2} & \dots & f_{\cdot, m} & \rightarrow & f_{\cdot, \cdot}\\
\cline{1-4}\cline{6-6}
\end{array}
\]

\smallskip
Moreover, we may introduce \emph{weak bisymmetry} of an $n$-ary fusion function $\func{F}^{(n)}$
(an arity-free property) by considering only the condition that for all
$\vect{x}=(x_{ij})\in\IvalPow{n\times n}$ it holds:
\begin{equation}
\begin{array}{rcccl}
\func{F}^{(n)}\bigg(  & \func{F}^{(n)}(x_{1,1},\dots,x_{1,n}),  & \dots, & \func{F}^{(n)}(x_{n,1},\dots,x_{n,n}) & \bigg)\\
=\func{F}^{(n)}\bigg( & \func{F}^{(n)}(x_{1,1},\dots,x_{n,1}),  & \dots, & \func{F}^{(n)}(x_{1,n},\dots,x_{n,n}) & \bigg).\\
\end{array}
\end{equation}

\section[Choosing an aggregation method (I): Desired properties]%
{Choosing an aggregation method (I):\newline Desired properties}%
\label{Sec:ChooseAg1Characterize}

Contrary to popular belief, aggregation is ``not only about applying the arithmetic mean''.
We already explored (uncountably) many interesting fusion functions.
One of the crucial questions is thus of course: Which one shall be chosen to
reflect one's needs arising in a particular application domain?

In this section we briefly indicate a few general selection methods --
each of them is based on an expected functions' behavioral specification (set up a priori).
One of the possible schemes relies on known \emph{characterization theorems},
which aim to provide a concrete definition of a class of fusion functions
that fulfill a given set of properties.
For instance, if we need a mapping which is symmetric,
strictly increasing, continuous, idempotent, and decomposable,
then by the famous Kolmogorov-Nagumo theorem (\ref{Theorem:KolmogorovNagumo})
we shall conclude that we are in fact interested in one of the quasi-arithmetic means.
On the other hand, if a conjunctive and -- at the same time -- disjunctive
function is desired, then -- unfortunately -- it turns out that our
needs are contradictory.

We shall also discuss a few notable subclasses of fusion functions,
especially useful in approximate reasoning and statistics, e.g.,
t-norms, copulas, and fuzzy implications. These do not indicate concrete
aggregation tools, but at least narrow the ``search space'' down.

Moreover, we sketch some numerical characteristics of
fusion functions, which can also aid in the selection process.
Their aim is to quantify the degree to which a function characterizes itself with
a kind of behavior, what is its ``typical'' outcome, etc.
Please keep in mind that this topic shall be significantly extended in
Chapter~\ref{Chap:Characteristics}.

\bigskip
Note that in this section we  make use of a few ``classical'' assumptions
in aggregation theory, namely that all the considered
(extended) fusion functions are:
\begin{itemize}
   \item defined for $\Ival=[0,1]$,
   \item nondecreasing, and
   \item endpoint-preserving.
\end{itemize}
In other words, we  focus on (extended) aggregation functions.

\newcommand{\pt}{{\LARGE\textbullet}}
\newcommand{\pf}{{\LARGE\textopenbullet}}
\newcommand{\pb}{{\small\Yinyang}}%

\begin{table}[bth!]
\centering
\caption[Exemplary fusion functions and some basic properties they fulfill.]{\label{Tab:extransonedim}
Exemplary fusion functions ($\Ival=[0,1]$) and some basic properties they fulfill:
ND -- nondecreasingness,
SM -- symmetry,
ID -- idempotency,
CJ -- conjunctivity,
DJ -- disjunctivity,
TE -- translation equivariance,
SE -- scale equivariance,
OE -- ordinal scale equivariance.%
}

\newcommand{\U}{?}%
\newcolumntype{H}{>{\setbox0=\hbox\bgroup}c<{\egroup}@{}}%
\begin{tabularx}{1.0\linewidth}{XcccccHcccH}
\toprule
\bf\small function                & \bf\small ND & \bf\small SM & \bf\small ID & \bf\small CJ
& \bf\small DJ
& \bf\small CT
& \bf\small TE & \bf\small SE & \bf\small OE
& \bf\small AD
\\
\midrule                               %
$\func{AMean}$                         & \pt & \pt & \pt & \pf & \pf & \U & \pt & \pt & \pf & \U \\
$\func{QMean}$                         & \pt & \pt & \pt & \pf & \pf & \U & \pf & \pt & \pf & \U \\
$\func{HMean}$                         & \pt & \pt & \pt & \pf & \pf & \U & \pf & \pt & \pf & \U \\
$\func{GMean}$                         & \pt & \pt & \pt & \pf & \pf & \U & \pf & \pt & \pf & \U \\
$\func{Median}$                        & \pt & \pt & \pt & \pf & \pf & \U & \pt & \pt & \pb & \U \\
$\func{Max}$                           & \pt & \pt & \pt & \pf & \pt & \U & \pt & \pt & \pt & \U \\
$\func{Min}$                           & \pt & \pt & \pt & \pt & \pf & \U & \pt & \pt & \pt & \U \\
$\func{T}_\mathrm{Ł}$                  & \pt & \pt & \pf & \pt & \pf & \U & \pf & \pf & \pf & \U \\
$\func{S}_\mathrm{Ł}$                  & \pt & \pt & \pf & \pf & \pt & \U & \pf & \pf & \pf & \U \\
$\func{3\Pi}$                          & \pt & \pt & \pf & \pf & \pf & \U & \pf & \pf & \pf & \U \\
\midrule
$\func{WMean}_\vect{w}$                & \pt & \pb & \pt & \pf & \pf & \U & \pt & \pt & \pf & \U \\
$\func{OWA}_\vect{w}$                  & \pt & \pt & \pt & \pb & \pb & \U & \pt & \pt & \pb & \U \\
$\func{WMax}_\vect{v}$                 & \pt & \pb & \pb & \pf & \pb & \U & \pb & \pb & \pb & \U \\
$\func{WMin}_\vect{v}$                 & \pt & \pb & \pb & \pb & \pf & \U & \pb & \pb & \pb & \U \\
$\func{OWMax}_\vect{v}$                & \pt & \pt & \pb & \pb & \pb & \U & \pb & \pb & \pb & \U \\
$\func{QAMean}_\varphi$                & \pt & \pt & \pt & \pf & \pf & \U & \pb & \pb & \pf & \U \\
$\func{BajMean}_{\varphi,\vect{w}}$    & \pb & \pb & \pt & \pf & \pf & \U & \pb & \pb & \pf & \U \\
\bottomrule
\end{tabularx}
\end{table}

\begin{example}
Table~\ref{Tab:extransonedim} summarizes some of the fusion functions and their classes
discussed so far along with the properties they fulfill (marked with ``\textbullet'').
Wherever ``{\pb}'' appears in the table, there are cases in which
a behavior is present as well as cases where the opposite is true.
This is exactly a situation in which characterization theorems are useful.
\end{example}

To complete the discussion, in Section~\ref{Sec:ChooseAg2Fit} we introduce methods
for fitting fusion functions (from some classes which can be
established by applying results presented in this part of the book)
to empirical data.

\subsection{Internal functions}

First let us explore a few noteworthy results that concern internal
(idempotent) aggregation functions. The following characterization
of quasi-arithmetic means was obtained
independently by Kolmogorov and Nagumo in 1930.

\begin{theorem}[\cite{Kolmogorov1930:moyenne,Nagumo1930:mittelwerte}]\label{Theorem:KolmogorovNagumo}
An extended fusion function $\func{F}^*$ is symmetric, strictly increasing,
continuous, idempotent, and decomposable if and only if
there exists a continuous strictly monotonic function
$\varphi:\Ival\to\mathbb{R}$ such that $\func{F}^*$
is an extended quasi-arithmetic mean generated by $\varphi$.
\end{theorem}

Here is a theorem by Aczel in which weak bisymmetry
is substituted for decomposability.

\begin{theorem}[\cite{Aczel1948:onmeanvalues}]
 An $n$-ary fusion function $\func{F}^{(n)}$ is strictly increasing,
 continuous, idempotent, and weakly bisymmetric if and only if
there exists a continuous strictly monotonic function
$\varphi:\Ival\to\mathbb{R}$ and a weighting vector $\vect{w}>0$
such that $\func{F}^{(n)}$ is a weighted quasi-arithmetic mean generated by $\varphi$
and $\vect{w}$.
\end{theorem}

\begin{corollary}[\cite{Aczel1948:onmeanvalues}]
 An $n$-ary fusion function $\func{F}^{(n)}$ is symmetric, strictly increasing,
 continuous, idempotent, and weakly bisymmetric if and only if
there exists a continuous strictly monotonic function
$\varphi:\Ival\to\mathbb{R}$ such that $\func{F}^{(n)}$
is an quasi-arithmetic mean generated by $\varphi$.
\end{corollary}

According to \cite{GrabischETAL2009:aggregationfunctions},
 here is how Nagumo \cite{Nagumo1930:mittelwerte}
 characterized all the quasi-arithmetic means that fulfill
 translation and scale equivariance.

\begin{theorem}\label{Thm:QAMeanTranslation}
A $n$-ary quasi-arithmetic mean $\func{QAMean}^{(n)}$ is translation equivariant
if and only if it is either the arithmetic mean
or it is an exponential mean.
\index{exponential mean}\index{arithmetic mean}%
\end{theorem}

\begin{theorem}\label{Thm:QAMeanScale}
A $n$-ary quasi-arithmetic mean $\func{QAMean}^{(n)}$ is scale equivariant
if and only if it is either the geometric mean
or it is a power mean.
\index{power mean}\index{geometric mean}%
\end{theorem}

Recall that among power means we have the arithmetic, quadratic, and harmonic means.
Taking the two above results into account we imply that
 the only quasi-arithmetic mean that is interval scale equivariant is the arithmetic mean.

\bigskip
Let us now discuss additivity and related concepts.

\begin{theorem}\label{Thm:CharacterizationAdditive}\index{weighted arithmetic mean}%
\index{additivity}%
An $n$-ary fusion function $\func{F}^{(n)}$
is additive, nondecreasing, and idempotent if and only if
$\func{F}^{(n)}$ is a weighted arithmetic mean.
\end{theorem}

See \cite[Proposition 4.21]{GrabischETAL2009:aggregationfunctions} for a proof.
As a corollary, we have that an $n$-ary fusion function $\func{F}^{(n)}$
is additive, nondecreasing, idempotent, and symmetric
if and only if it is the arithmetic mean.
Moreover, please note that nondecreasingness can be replaced with continuity
in this theorem.

\begin{theorem}[\cite{MesiarMesiarova2011:OMA}]\index{modularity}%
An $n$-ary fusion function $\func{F}^{(n)}$
is modular, nondecreasing, and idempotent if and only if:
\[
   \func{F}^{(n)}(\vect{x})=\sum_{i=1}^n \func{f}_i(x_i)
\]
for any nondecreasing $\func{f}_1,\dots,\func{f}_n:\Ival\to\Ival$
such that $(\forall x\in\Ival)$ $\sum_{i=1}^n \func{f}_i(x)=x$.
\end{theorem}

Let us consider two variants of additivity.
\index{additivity}The first one assumes that
the vectors on which the addition operation is applied on comonotonic vectors
(see Definition~\ref{Def:comonotonic}).

\begin{definition}\label{Def:ComonotonicAdditive}\index{comonotonic additivity}%
An $n$-ary fusion function $\func{F}^{(n)}$ is said to be \emph{comonotonic additive},
whenever:
\[
\func{F}^{(n)}(\vect{x}+\vect{y})=\func{F}^{(n)}(\vect{x})+\func{F}^{(n)}(\vect{y}),
\]
for all comonotonic $\vect{x}, \vect{y}\in\IvalPow{n}$ such that $\vect{x}+\vect{y}\in\IvalPow{n}$.
\end{definition}

\begin{theorem}
An $n$-ary fusion function $\func{F}^{(n)}$ is comonotonic additive,
nondecreasing, and idempotent if and only if $\func{F}^{(n)}$ is a discrete Choquet integral
with respect to a fuzzy measure.
\end{theorem}

We may also deal with a symmetrized version of the additivity property.

\begin{definition}\label{Def:SymmetricAdditive}\index{symmetric additivity}%
An $n$-ary fusion function $\func{F}^{(n)}$ is said to be \emph{symmetric additive},
whenever:
\[
\func{F}^{(n)}(\vect{x}\stackrel{S}{+}\vect{y})=\func{F}^{(n)}(\vect{x})+\func{F}^{(n)}(\vect{y}),
\]
for all $\vect{x}, \vect{y}\in\IvalPow{n}$ such that $\vect{x}+\vect{y}\in\IvalPow{n}$,
where $\vect{x}\stackrel{S}{+}\vect{y}=(x_{(1)}+y_{(1)},\dots,x_{(n)}+y_{(n)})$.
\end{definition}

Clearly, each {symmetric additive} fusion function is necessarily symmetric.

\begin{theorem}\index{OWA operator}%
An $n$-ary fusion function $\func{F}^{(n)}$
is symmetric additive, nondecreasing, and idempotent if and only if
$\func{F}^{(n)}$ is an OWA operator.
\end{theorem}

For a different characterization of OWA operators, see, e.g., \cite{FodorMarichaRoubens1995:OWAchar}.
Let us now present a characterization concerning associativity.

\begin{theorem}\cite{Marichal2000:associativity,Fodor1996:fungfu} %
An extended fusion function $\func{F}^*$ is nondecreasing, continuous, idempotent,
and associative if and only if there exist $\alpha,\beta\in\Ival$ such that:
\[
\func{F}|_{\IvalPow{2}}(x_1,x_2) = (\alpha\wedge x_1)\vee(\beta\wedge x_2)\vee(x_1\wedge x_2).
\]
\end{theorem}

Together with symmetry the above result restricts itself to the so-called
\index{alpha-median@$\alpha$-median}\emph{$\alpha$-median},
$\func{F}|_{\IvalPow{2}}(x_1,x_2)=\func{Median}(x_1,\alpha,x_2)$
for some $\alpha\in\Ival$,
see \cite{DuboisPrade1985:reviewagcon}.
Moreover, Czogała and Drewniak in \cite{CzogalaDrewniak1984:assomono}
presented one of the possible characterizations of the $\func{Min}$ and $\func{Max}$ functions.

\subsection{Conjunctive and disjunctive functions}\label{Sec:tnorm}

Another set of tools in which aggregation theory is interested in
consists of fuzzy logic connectives (useful in, e.g., approximate
reasoning, preference modeling, etc.) and copulas (very important in probability
and statistics, compare Remark~\ref{Remark:CopulasMultiDim}),
see \cite{KlirYuan1995:fuzzybook,BaczynskiJayaram2008:fuzzyimplications,Nelsen1999:Copulas}.
Most of them are considered as binary operations on members of $\Ival=[0,1]$,
but they may be extended to $\IvalAnyPow$ easily.

The two properties provided below are well-known from algebra.

\begin{definition}\label{Def:annihilatorelement}\index{annihilator element}%
We say that $\func{F}^{(n)}$ has an \emph{annihilator element} $h\in\Ival$, whenever
for all $\vect{x}\in\IvalPow{n}$ and $i\in[n]$ we have:
\begin{equation}
\func{F}^{(n)}(x_1,x_2,\dots,x_{i-1}, h,x_{i+1},\dots,x_n)=h.
\end{equation}
\end{definition}

\begin{definition}\label{Def:neutralelement}\index{neutral element}%
$\func{F}^{(n)}$ has a \emph{neutral element} $e\in\Ival$, if
for all $\vect{x}\in\IvalPow{n}$ and $i\in[n]$  it holds:
\begin{equation}
\func{F}^{(n)}(e,e,\dots, e, x_i,e,\dots,e)=x_i.
\end{equation}
\end{definition}

This property may be extended as follows.

\begin{definition}\index{strong neutral element}%
Given an extended fusion function $\func{F}^*$, we call $e\in\Ival$ its
\emph{strong neutral element}, whenever for all $n$ and $\vect{x}$ it holds:
\begin{equation}
\func{F}^{(n+1)}(x_1,x_2,\dots,x_{i-1}, e,x_{i},\dots,x_n)=\func{F}^{(n)}(\vect{x}).
\end{equation}
\end{definition}

\paragraph{T-norms.}
Triangular norms were first introduced by Schweizer and Sklar
in the context of probabilistic metric spaces (see \cite{SchweizerSklar1983:probmetricspace})
and are used, among others, for defining intersections of fuzzy sets
and modeling the conjunction operation in fuzzy logic.

\begin{definition}\label{Def:tnorm}\index{triangular norm|see {t-norm}}\index{t-norm}%
An aggregation function $\func{T}^{(2)}: [0,1]\times[0,1]\to[0,1]$ is a \emph{t-norm}
if for all $x,y,z\in[0,1]$ it holds:
\begin{enumerate}
\item[(a)] if $y \le z$, then $\func{T}^{(2)}(x, y)\le \func{T}^{(2)}(x, z)$, \hfill(nondecreasingness)
\item[(b)] $\func{T}^{(2)}(x, y)=\func{T}^{(2)}(y, x)$, \hfill(symmetry)
\item[(c)] $\func{T}^{(2)}(x, \func{T}^{(2)}(y, z))=\func{T}^{(2)}(\func{T}^{(2)}(x, y), z)$, \hfill(associativity)
\item[(d)] $\func{T}^{(2)}(x, 1)=x$. \hfill(neutral element $1$)
\end{enumerate}
\end{definition}
Thus, a t-norm is a symmetric conjunctive aggregation function on $[0,1]^2$.
It is easily seen that the restriction of any t-norm to $\{0,1\}^2$ gives us the conjunction
operation known from  classical Boolean logic.
Moreover, each t-norm has $0$ as its annihilator element,
i.e., $\func{T}^{(2)}(x,0)=\func{T}^{(2)}(0,x)=0$ for all $x$.

Table \ref{Tab:tnorms} lists some seminal t-norms.
For any t-norm $\func{T}^{(2)}$ and all $x,y$ it holds $\func{T}^{(2)}_\mathrm{D}(x,y)\le \func{T}^{(2)}(x,y)\le \func{Min}^{(2)}(x, y)$.
Moreover, we have $\func{T}^{(2)}_\mathrm{Ł}(x,y)\le \func{Prod}^{(2)}(x,y)$.

Recall that in Proposition~\ref{Prop:isomorphism}
we stated that for every nondecreasing fusion function $\func{F}^{(n)}$,
its $\varphi$-isomorphism is also nondecreasing.

\begin{proposition}
For any strictly increasing and continuous function $\varphi:\Ival\to\Ival$,
if $\func{F}^{(n)}$ is conjunctive, then $\func{F}^{(n)}_{[\varphi]}$ is conjunctive too.
Moreover, if $\func{F}^{(n)}$ is a t-norm, then $\func{F}^{(n)}_{[\varphi]}$ is also a t-norm.
\end{proposition}

\paragraph{T-conorms.}
First of all, let us note what follows.

\begin{proposition}
For any strictly decreasing and continuous function $\varphi:\Ival\to\Ival$,
$\func{F}^{(n)}$ is disjunctive if and only if $\func{F}^{(n)}_{[\varphi]}$ is conjunctive.
\end{proposition}

Triangular conorms generalize the notion of the classical Boolean logic alternative operator.
They are defined as $(x\mapsto 1-x)$-isomorphisms of t-norms.

\begin{definition}\label{Def:cotnorm}\index{triangular conorm|see {t-conorm}}\index{t-conorm}%
A function $\func{S}^{(2)}: [0,1]\times[0,1]\to[0,1]$ is a \emph{t-conorm}
if for all $x,y,z\in[0,1]$ it holds:
\begin{enumerate}
\item[(a)] if $y \le z$, then $\func{S}^{(2)}(x, y)\le \func{S}^{(2)}(x, z)$, \hfill(nondecreasingness)
\item[(b)] $\func{S}^{(2)}(x, y)=\func{S}^{(2)}(y, x)$, \hfill(symmetry)
\item[(c)] $\func{S}^{(2)}(x, \func{S}^{(2)}(y, z))=\func{S}^{(2)}(\func{S}^{(2)}(x, y), z)$, \hfill(associativity)
\item[(d)] $\func{S}^{(2)}(x, 0)=x$. \hfill(neutral element $0$)
\end{enumerate}
\end{definition}
It is evident that all t-conorms are {disjunctive}.
Table \ref{Tab:tconorms} lists a few noteworthy t-conorms -- all of them
are dual to respective t-norms in Table~\ref{Tab:tnorms}.
For any t-conorm $\func{S}^{(2)}$ and all $x,y$ it holds $\func{Max}^{(2)}(x,y)\le \func{S}^{(2)}(x,y)\le \func{S}^{(2)}_\mathrm{D}(x, y)$.
Moreover, we have $\func{S}^{(2)}_\mathrm{P}(x,y)\le \func{S}^{(2)}_\mathrm{Ł}(x,y)$.

\bigskip
Please refer to the seminal monograph of Klement, Mesiar, and Pap
\cite{KlementMesiarPap2000:trinorm} and their so-called
position papers \cite{KlementMesiarPap2004:tnormposition1,KlementMesiarPap2004:tnormposition2,
KlementMesiarPap2004:tnormposition3} as well as to \cite[Chapter~3]{GrabischETAL2009:aggregationfunctions}
for more details on t-norms and t-conorms.

\paragraph{Copulas.}
Copulas form another group of interesting and useful aggregation functions.
They may be used in probability and statistics to model
dependencies between random variables, see, e.g., \cite{Nelsen1999:Copulas}
and also Remark~\ref{Remark:CopulasMultiDim}.

For given $n$, each $n$-copula $\func{C}^{(n)}:[0,1]^n\to[0,1]$ is a cumulative distribution
function  of an $n$-dimensional random variable
having uniform margins. In particular, for $n=2$ we have what follows.

\begin{definition}\label{Def:copula}\index{copula}%
A function $\func{C}^{(2)}: [0,1]\times[0,1]\to[0,1]$ is a \emph{2-copula}
if for all $x,y,x',y'\in[0,1]$ it holds:
\begin{enumerate}
\item[(a)] if $x\le x'$ and $y\le y'$, then: \hfill\index{2-increasingness}(2-increasingness)
\[\func{C}^{(2)}(x,y)+\func{C}^{(2)}(x',y')-\func{C}^{(2)}(x,y')-\func{C}^{(2)}(x',y)\ge 0,\]
\item[(b)] $\func{C}^{(2)}(x, 0)=\func{C}^{(2)}(0,x)=0$, \hfill(annihilator element)
\item[(c)] $\func{C}^{(2)}(x, 1)=x$. \hfill(neutral element)
\end{enumerate}
\end{definition}
Note that each t-norm fulfills conditions (b) and (c).
Moreover, each 2-copula is nondecreasing and 1-Lipschitz. There are 2-copulas
that are not t-norms and vice versa (see \cite{KlementMesiarPap2000:trinorm}).
However, e.g., associative copulas are exactly 1-Lipschitz t-norms.

$\func{T}_\mathrm{Ł}$ and $\func{Min}$ are particular examples of such fusion functions.
By the famous Fr\'{e}chet-Hoeffding theorem (compare \cite{Nelsen1999:Copulas}),
these are the smallest and the largest copulas, respectively. Hence, copulas are conjunctive.

An important class of associative copulas consists
of \index{Archimedean copula}Archimedean ones.
Let $\varphi:[0,1]\to[0,\infty[$ be a continuous, convex, and decreasing function with
$\varphi(1)=0$. Then we may define:
\begin{equation}
\func{C}^{(2)}_\varphi(x, y) = \varphi^{-1}\left(\varphi(x)+\varphi(y)\right),
\end{equation}
where $\varphi^{-1}$, $\varphi^{-1}(y)=\inf\{x\in[0,1]:\varphi(x)\ge y\}$, denotes the pseudoinverse of $\varphi$.
Table \ref{Tab:archcopula} lists a few particular subfamilies of Archimedean copulas.
Note that the Gumbel copula with $\theta = 1$ is equivalent to the
$\func{Prod}$ fusion function, which models the case of independent
\index{independence copula|see {product}}random variables.
What is more, $\func{C}_{\mathrm{C},-1}^{(2)}\equiv\func{T}_\mathrm{Ł}^{(2)}$.

Another noteworthy class consists of \index{Gaussian copula}Gaussian copulas.
If $\Phi$ denotes the standard normal cumulative distribution function
(note that no analytical closed-form expression exists for it)
and $\Phi_\vect{V}$ denotes the joint cumulative distribution function
of the bivariate normal distribution with expectation $\mathbf{0}$ and covariance matrix $\vect{V}$,
then:
\begin{equation}
\func{C}_{\mathrm{Gauss}, \vect{V}}^{(2)}(x, y) = \Phi_\vect{V}\left( \Phi^{-1}(x), \Phi^{-1}(y) \right).
\end{equation}

\begin{table}[p!]
\caption{\label{Tab:tnorms} Exemplary t-norms.}
\centering
\begin{tabularx}{1.0\linewidth}{p{3.5cm}X}
\toprule
\small\bf name & \small\bf definition \\
\midrule
minimum  \index{minimum}%
& $\func{Min}^{(2)}(x, y)=x\wedge y$ \\
\midrule
product \index{product}%
& $\func{Prod}^{(2)}(x, y)=xy$ \\
\midrule
Łukasiewicz \index{Lukasiewicz t-norm@Łukasiewicz t-norm}%
& $\func{T}^{(2)}_\mathrm{Ł}(x, y)=(x+y-1)\vee 0$ \\
\midrule
drastic \index{TD@$\mathsf{T}_{\mathrm{D}}$|see {drastic t-norm}}\index{drastic t-norm}%
& $\func{T}^{(2)}_\mathrm{D}(x, y)=\left\{
\begin{array}{ll}
0 & \text{if }x, y\in[0,1[\\
x\wedge y & \text{if }x=1\text{ or }y=1\\
\end{array}
\right.$ \\
\midrule
Fodor  \index{TF@$\mathsf{T}_{\mathrm{F}}$|see {Fodor t-norm}}\index{Fodor t-norm}%
& $\func{T}^{(2)}_\mathrm{F}(x, y)=\left\{
\begin{array}{ll}
0 & \text{if }x+y\le 1\\
x\wedge y & \text{if }x+y>1\\
\end{array}
\right.$ \\
\bottomrule
\end{tabularx}
\end{table}

\begin{table}[p!]
\caption{\label{Tab:tconorms} Exemplary t-conorms.}
\centering
\begin{tabularx}{1.0\linewidth}{p{3.5cm}X}
\toprule
\small\bf name & \small\bf definition \\
\midrule
maximum \index{maximum}%
& $\func{Max}^{(2)}(x, y)=x\vee y$ \\
\midrule
product \index{SP@$\mathsf{S}_{\mathrm{P}}$|see {product t-conorm}}\index{product t-conorm}%
& $\func{S}^{(2)}_\mathrm{P}(x, y)=x+y-xy$ \\
\midrule
Łukasiewicz \index{Lukasiewicz t-conorm@Łukasiewicz t-conorm}%
& $\func{S}^{(2)}_\mathrm{Ł}(x, y)=(x+y)\wedge 1$ \\
\midrule
drastic   \index{drastic t-conorm}%
& $\func{S}^{(2)}_\mathrm{D}(x, y)=\left\{
\begin{array}{ll}
1 & \text{if }x, y\in]0,1]\\
x\vee y & \text{if }x=0\text{ or }y=0\\
\end{array}
\right.$ \\
\midrule
Fodor \index{SF@$\mathsf{S}_{\mathrm{F}}$|see {Fodor t-conorm}}\index{Fodor t-conorm}%
& $\func{S}^{(2)}_\mathrm{F}(x, y)=\left\{
\begin{array}{ll}
1 & \text{if }x+y\ge 1\\
x\vee y & \text{if }x+y<1\\
\end{array}
\right.$ \\
\bottomrule
\end{tabularx}
\end{table}

\begin{table}[p!]
\caption{\label{Tab:archcopula} Exemplary Archimedean 2-copulas.}
\centering
\begin{tabularx}{1.0\linewidth}{p{3.5cm}X}
\toprule
\small\bf name,\newline parameter & \small\bf definition,\newline generator  \\
\midrule
Clayton,\index{Clayton copula}%
\newline $\theta\ge -1, \theta\neq 0$
& $\func{C}_{\mathrm{C},\theta}^{(2)}(x, y)=\left( (x^{-\theta}+y^{-\theta}-1)\vee 0 \right)^{-1/\theta}$, \newline
$\varphi(t)=(t^{-\theta}-1)/\theta$\\
\midrule
Gumbel,\index{Gumbel copula}%
\newline $\theta\ge 1$
& $\func{C}_{\mathrm{G},\theta}^{(2)}(x, y)=\exp\left( -\left( (\log 1/x)^\theta + (\log 1/y)^\theta \right)^{1/\theta} \right)$, \newline
$\varphi(t)=(\log 1/t)^\theta$\\
\midrule
Frank,\index{Frank copula}%
\newline $\theta\neq 0$
& $\func{C}_{\mathrm{F},\theta}^{(2)}(x, y)=-\frac{1}{\theta} \log\left( 1-\frac{(1-\exp(-\theta x))(1-\exp(-\theta y))}{1-\exp(-\theta)} \right)$, \newline
$\varphi(t)=-\log\left( \frac{1-\exp(-\theta t)}{1-\exp(-\theta)} \right)$\\
\bottomrule
\end{tabularx}
\end{table}

\subsection{Mixed, non-aggregation, and other functions}\label{Sec:MixedFunctions}

\index{comonotonic maxitivity}%
In a quite similar manner to comonotonic additivity
(compare Definition~\ref{Def:ComonotonicAdditive}),
we may introduce the comonotonic maxitivity (among others).

\begin{theorem}[see {\cite[Theorem 5.81]{GrabischETAL2009:aggregationfunctions}}]
An $n$-ary fusion function $\func{F}^{(n)}$ is comonotonic maxitive,
$\wedge$-equivariant, and such that $\func{F}^{(n)}(n\ast 1)=1$
if and only if $\func{F}^{(n)}$ is a discrete Sugeno integral
with respect to a fuzzy measure.
\end{theorem}

Please observe that a different characterization (using nondecreasingness,
$\wedge$- and $\vee$-equivariance)
\index{0vee-equivariance@$\vee$-equivariance}%
\index{0wedge-equivariance@$\wedge$-equivariance}%
of the discrete Sugeno integral
was proposed by Marichal in \cite{Marichal2000:sugenointagfun}.

\medskip
On the other hand, we may also introduce symmetrized versions of modularity,
maxitivity, and minitivity (compare also Definition~\ref{Def:SymmetricAdditive}).
Each of them implies nondecreasingness and symmetry,
at least in the case $\Ival=[0,1]$ (which is fixed in this section).
\index{symmetric modularity}%
\index{symmetric minitivity}%
\index{symmetric maxitivity}%

\begin{theorem}[\cite{Gagolewski2013:om3}, see also \cite{MesiarMesiarova2011:OMA}]%
An $n$-ary fusion function $\func{F}^{(n)}$
is symmetric modular if and only if:
\[
   \func{F}^{(n)}(\vect{x})=\sum_{i=1}^n \func{f}_i(x_{(i)})
\]
for any nondecreasing $\func{f}_1,\dots,\func{f}_n:[0,1]\to[0,1]$
such that $(\forall x\in[0,1])$ $\sum_{i=1}^n \func{f}_i(x)\le 1$.
\end{theorem}

Idempotent symmetric modular aggregation functions are called
\index{OMA operator} \emph{OMA operators} (ordered modular averages)
in the Mesiar and Mesiarov{\'a}-Zem{\'a}n\-kov{\'a} paper \cite{MesiarMesiarova2011:OMA}.

\begin{theorem}[\cite{Gagolewski2013:om3}, see also \cite{GrabischETAL2009:aggregationfunctions}]
An $n$-ary fusion function $\func{F}^{(n)}$
is symmetric minitive if and only if:
\[
   \func{F}^{(n)}(\vect{x})=\bigwedge_{i=1}^n \func{f}_i(x_{(i)})
\]
for any nondecreasing $\func{f}_1,\dots,\func{f}_n:[0,1]\to[0,1]$.
\end{theorem}

A particular subclass of minitive fusion functions,
so-called effort dominating operators, see \cite{Gagolewski2012:ipmu},
shall be referred to in Section~\ref{Sec:ImpactFunctions}.

\begin{theorem}[\cite{Gagolewski2013:om3}, see also \cite{GrabischETAL2009:aggregationfunctions}]
An $n$-ary fusion function $\func{F}^{(n)}$
is symmetric maxitive if and only if:
\[
   \func{F}^{(n)}(\vect{x})=\bigvee_{i=1}^n \func{f}_i(x_{(i)})
\]
for any nondecreasing $\func{f}_1,\dots,\func{f}_n:[0,1]\to[0,1]$.
\end{theorem}

The following result is due to Gagolewski \cite{Gagolewski2013:om3}.

\begin{theorem}[\cite{Gagolewski2013:om3}]
For an $n$-ary fusion function $\func{F}^{(n)}$ the following conditions are equivalent:
\begin{itemize}
   \item $\func{F}^{(n)}$ is both symmetric minitive and symmetric maxitive,
   \item $\func{F}^{(n)}$ is both symmetric minitive and symmetric modular,
   \item $\func{F}^{(n)}$ is both symmetric modular and symmetric maxitive,
   \item $\func{F}^{(n)}$ is given by:
   \[
      \func{F}^{(n)}(\vect{x}) =
      \bigvee_{i=1}^n \func{f}(x_{(i)})\wedge v_i,
   \]
   for some nondecreasing $\func{f}:[0,1]\to[0,1]$
   and $\vect{v}\in[0,1]^{n}$ such that
   $0\le \func{f}(0) \le v_n \le \dots \le v_1 \le 1$.
\end{itemize}
\end{theorem}

As a corollary, the only idempotent as well as symmetric modular,
minitive, and maxitive fusion function is an ordered weighted maximum
($\func{OWMax}$)  operator.\index{ordered weighted maximum}%

\medskip
We already considered some characterizations which
takes translation, scale, interval scale, $\wedge$-, or $\vee$-equivariance
into account. Let us mention the remaining property of this kind.

\begin{theorem}[\cite{MarichalRoubens1993:charstable}]
A fusion function $\func{F}$ is nondecreasing and ordinal scale equivariant
if and only if $\func{F}$ is a lattice polynomial function.
\index{lattice polynomial function}%
\end{theorem}

Under ordinal scale equivariance, nondecreasingness and continuity coincide,
see, e.g., \cite[Proposition~8.13]{GrabischETAL2009:aggregationfunctions}.
Note that, as showed by Marichal in \cite{Marichal2002:orderinvsynth},
\index{order statistic}%
the only symmetric lattice polynomial functions are exactly the order statistics,
$\func{OS}_k$, $k\in[n]$.

\medskip
Here is a whole family of functions which falls into the class of ``mixed'' type
aggregation.

\paragraph{Uninorms.}
Recall that a t-norm is a symmetric and associative aggregation function
with neutral element $1$. A t-conorm, on the other hand,
has the neutral element $0$. Here is a class of fusion functions
which have a neutral element, but such that it is neither equal to $0$
nor to $1$.

\begin{definition}\label{Def:uninorm}\index{uninorm}%
A fusion function $\func{U}^{(2)}: [0,1]\times[0,1]\to[0,1]$ is a \emph{uninorm}
if for all $x,y,z\in[0,1]$ it holds:
\begin{enumerate}
\item[(a)] if $y \le z$, then $\func{U}^{(2)}(x, y)\le \func{U}^{(2)}(x, z)$, \hfill(nondecreasingness)
\item[(b)] $\func{U}^{(2)}(x, y)=\func{U}^{(2)}(y, x)$, \hfill(symmetry)
\item[(c)] $\func{U}^{(2)}(x, \func{U}^{(2)}(y, z))=\func{U}^{(2)}(\func{U}^{(2)}(x, y), z)$, \hfill(associativity)
\item[(d)] for some $e\in]0,1[$ it holds $\func{U}^{(2)}(x, e)=x$.\par
\hfill(neutral element $e\not\in\{0,1\}$)
\end{enumerate}
\end{definition}

Here is an important result on representation of uninorms,
see \cite[Proposition 3.95]{GrabischETAL2009:aggregationfunctions}.

\begin{proposition}
Let $\func{U}^{(2)}: [0,1]\times[0,1]\to[0,1]$ be a {uninorm}
with neutral element $e$. Then there exists a t-norm $\func{T}^{(2)}$,
a t-conorm $\func{S}^{(2)}$, and a symmetric, idempotent aggregation function $\func{A}^{(2)}$
such that for any $\vect{x}\in\IvalPow{2}$ it holds:
\[
   \func{U}^{(2)}(\vect{x})=\left\{
   \begin{array}{ll}
      \func{T}^{(2)}(\vect{x}) & \text{if }\vect{x}\in[0,e]^2,\\
      \func{S}^{(2)}(\vect{x}) & \text{if }\vect{x}\in[e,1]^2,\\
      \func{A}^{(2)}(\vect{x}) & \text{otherwise}.
   \end{array}
   \right.
\]
\end{proposition}
Thus, a uninorm is neither internal, conjunctive, nor disjunctive.
The $3\Pi$ function \index{3Pi@$\func{3\Pi}$ function}%
is an exemplary uninorm.

\paragraph{Fuzzy implications.}
As it was noted earlier, even if the nondecreasingness property
is very influential in aggregation theory
(many of the results presented so far would not be possible to obtain
without such an assumption), it should not be treated dogmatically
(compare the notion of weak monotonicity, among others).
Here is a useful class of functions that generalizes the concept
of the Boolean logic implication operator.

\begin{table}[t!]
\caption{\label{Tab:fimplications} Exemplary fuzzy implications.}
\centering
\begin{tabularx}{1.0\linewidth}{p{3.5cm}X}
\toprule
\bf\small name & \bf\small definition \\
\midrule
minimal  & $\func{I}_0^{(2)}(x,y)=\left\{
\begin{array}{ll}
1 & \text{if }x=0\text{ or }y=1\\
0 & \text{otherwise}\\
\end{array}
\right.$ \\
\midrule
maximal & $\func{I}_1^{(2)}(x,y)=\left\{
\begin{array}{ll}
0 & \text{if }x=1\text{ and }y=0\\
1 & \text{otherwise}\\
\end{array}
\right.$ \\
\midrule
Kleene-Dienes & $\func{I}_\mathrm{KD}^{(2)}(x,y)=(1-x)\vee y$ \\
\midrule
Łukasiewicz&  $\func{I}_\mathrm{Ł}^{(2)}(x,y)=(1-x+y)\wedge 1$ \\
\midrule
Reichenbach& $\func{I}_\mathrm{RB}^{(2)}(x,y)=1-x+xy$ \\
\midrule
Fodor & $\func{I}_\mathrm{F}^{(2)}(x,y)=\left\{
\begin{array}{ll}
1 & \text{if }x\le y\\
(1-x)\vee y & \text{if }x> y\\
\end{array}
\right.$ \\
\midrule
Goguen & $\func{I}_\mathrm{GG}^{(2)}(x,y)=\left\{
\begin{array}{ll}
1  & \text{if }x\le y\\
y/x  & \text{if }x> y\\
\end{array}
\right.$ \\
\midrule
G\"{o}del & $\func{I}_\mathrm{GD}^{(2)}(x,y)=\left\{
\begin{array}{ll}
1  & \text{if }x\le y\\
y  & \text{if }x> y\\
\end{array}
\right.$ \\
\midrule
Rescher & $\func{I}_\mathrm{RS}^{(2)}(x,y)=\left\{
\begin{array}{ll}
1  & \text{if }x\le y\\
0  & \text{if }x> y\\
\end{array}
\right.$ \\
\midrule
Weber & $\func{I}_\mathrm{W}^{(2)}(x,y)=\left\{
\begin{array}{ll}
1 & \text{if }x<1\\
y & \text{if }x=1\\
\end{array}
\right.$ \\
\midrule
Yager & $\func{I}_\mathrm{Y}^{(2)}(x,y)=\left\{
\begin{array}{ll}
1 & \text{if }x=0\text{ and }y=0\\
y^x & \text{otherwise}\\
\end{array}
\right.$ \\
\bottomrule
\end{tabularx}
\end{table}

\begin{definition}\label{Def:fimpl}\index{fuzzy implication}%
A function $\func{I}^{(2)}: [0,1]\times[0,1]\to[0,1]$ is a \emph{fuzzy implication}
if for all $x,y,x',y'\in[0,1]$ it holds:
\begin{enumerate}
\item[(a)] if $x \le x'$, then $\func{I}^{(2)}(x, y)\ge \func{I}^{(2)}(x', y)$, \hfill(nonincreasingness w.r.t.~$x$)
\item[(b)] if $y \le y'$, then $\func{I}^{(2)}(x, y)\le \func{I}^{(2)}(x, y')$, \hfill(nondecreasingness w.r.t.~$y$)
\item[(c)] $\func{I}^{(2)}(1,1)=1$,
\item[(d)] $\func{I}^{(2)}(0,0)=1$,
\item[(e)] $\func{I}^{(2)}(1,0)=0$.
\end{enumerate}
\end{definition}
It is easily seen that $\func{I}^{(2)}(x,1)=1$ and $\func{I}^{(2)}(0,y)=1$ for all $x,y$.
Table~\ref{Tab:fimplications} lists some exemplary fuzzy implications.
The reader is referred to
the monograph by Baczyński and Jayaram \cite{BaczynskiJayaram2008:fuzzyimplications}
and, e.g., to \cite{Reiser2013:fuzzimplagg,BaczynskiJayaram2008:snrsurvey}
for a comprehensive overview of this class of fusion functions as well as
its relation to t-norms, t-conorms, and other aggregation tools.

\subsection{Andness, orness, and other numerical characteristics}

In Section~\ref{Sec:FusionFunctionsCharacteristics} we shall discuss
methods for ``measuring'' the degree to which a fusion function obeys some
particular behavior. This may be used to aid in the aggregation tool
selection process too.

To get a general intuition standing behind these
numerical characteristics, let us at least list a few of them here.
\begin{itemize}
\item Let $\func{F}^{(n)}$ be an averaging aggregation function on $[0,1]^n$.
\index{orness}Its \emph{orness}  \cite{Dujmovic1974:orness} is given by:
\[
\mathrm{orness}(\func{F}^{(n)}) =
   \frac{\int_{[0,1]^n} \func{F}^{(n)}(\vect{x})\,d\vect{x} - \int_{[0,1]^n} \func{Min}^{(n)}(\vect{x})\,d\vect{x}}%
   {\int_{[0,1]^n} \func{Max}^{(n)}(\vect{x})\,d\vect{x} - \int_{[0,1]^n} \func{Min}^{(n)}(\vect{x})\,d\vect{x}}.
\]
Of course, $\mathrm{orness}(\func{Min}^{(n)})=0$ and $\mathrm{orness}(\func{Max}^{(n)})=1$.
In a dual manner, \emph{andness} may be defined.

\item The \index{average orness}\emph{average orness} \cite{FernandezMurakami2003:aveorness} of $\func{F}^{(n)}$ is given by:
\[
\mathrm{aveorness}(\func{F}^{(n)}) = \displaystyle\int_{[0,1]^n}
\frac{\func{F}^{(n)}(\vect{x})-\func{Min}^{(n)}(\vect{x})}{\func{Max}^{(n)}(\vect{x})-\func{Min}^{(n)}(\vect{x})}\,d\vect{x},
\]
where we assume that $0/0=0$.

\item For the arithmetic mean, it suffices to contaminate a single point and
set it to $\pm\infty$ to obtain an infinite value.
Yet, it is known that, e.g., the sample median serves as a robust estimate
for the center of an empirical distribution -- it needs
up to roughly 50\% of the data to be contaminated to change its output value drastically.
\index{breakdown value}The so-called \emph{breakdown value} measures a fusion
function's sensitivity to the presence of outliers, compare  \cite{Donoho1982:phd}.
\end{itemize}

\section[Choosing an aggregation method (II):  Fitting to data]%
{Choosing an aggregation method (II):\newline  Fitting fusion functions to data}
\label{Sec:ChooseAg2Fit}

Let us presume that we have established our favorite class of fusion functions
(e.g., by stating  a desired set of properties that must be fulfilled and
then by choosing it according to one of the characterization theorems from
the previous section). For simplicity, first we are going to assume that
a fusion function of our interest, $\func{F}_\vect{w}$, is
parametrized via a weighting vector (or, more generally,
a vector of some parameters) $\vect{w}$. For instance, it may be
a weighted quasi-arithmetic mean with a fixed generator function $\varphi$
(further on we shall discuss methods for automated $\varphi$ selection as well).
Our main concern in this  section is how to choose $\vect{w}$.

Of course, one may rely on experts' knowledge at this point.
This was the case of the aggregation method used in Ski jumping
competitions, see Example~\ref{Ex:SkiJumping}.
However, if the experts are unavailable, another common option
is based on a methodology widely used in data mining/machine learning
(see, e.g., \cite{Torra2005:agopmodel}).
Namely, we may obtain  an (empirical) data set of input points somehow and then:
\begin{itemize}
\item if we have access to desired output values for corresponding input cases
provided, we may rely on supervised learning-like algorithms;
the weight fitting methods discussed in this section
assure consistency of the obtained fusion function's outputs
with prototypes at hand;
\item if we do not have initial preferences towards desired output
data, unsupervised learning-like techniques may be used,
see, e.g., \cite{Kojadinovic2004:unsupervised,Kojadinovic2008:unsupervised};
note that this task is much more vague than the previous one.
\end{itemize}
Note also that  other approaches may be useful, for example reinforcement
learning-based ones. Nevertheless, in this monograph we are
interested in examining the first scenario.

\bigskip
More formally, we would like to fit a fusion function $\func{F}_\vect{w}$
parametrized via a vector $\vect{w}$ to empirical data, see, e.g., \cite{Beliakov2003:buildagopdata}.
We observe $m\ge n$ input vectors $\vect{x}^{(1)},\dots,\vect{x}^{(m)}\in\IvalPow{n}$
together with $m$ desired output values $y^{(1)},\dots,\allowbreak y^{(m)}\in\Ival$.
Our task is to compute the weighting vector $\vect{w}$ that best ``fits'' the
given data set.
Assuming that desired input and output data are represented as
matrices $\vect{X}\in\IvalPow{n\times m}, \vect{Y}\in\IvalPow{1\times m}$,
and that $\func{F}_\vect{w}(\vect{X})=[\func{F}_\vect{w}(\vect{x}^{(1)})\ \cdots\ \func{F}_\vect{w}(\vect{x}^{(m)})]\in\IvalPow{1\times m}$,
we are faced with a constrained optimization problem:
\[
   \mathrm{minimize}\ E\left(\func{F}_\vect{w}(\vect{X}), \vect{Y}\right)
   \quad \text{w.r.t.~}\vect{w}
\]
subject to some conditions on $\vect{w}$ that guarantee monotonicity,
idempotency, or any other valuable property,
where $E: \IvalPow{m}\times\IvalPow{m}\to[0,\infty]$ is some
\emph{loss function} (typically a function of some metric)
that we shall use as a goodness-of-fit measure.

\begin{remark}
If there exists a fusion function that \emph{interpolates}
a set of prototypical observations provided ($\func{F}_\vect{w}(\vect{X})=\vect{Y}$), algorithms
like those in \cite{Beliakov2007:constragopinterpolation,Beliakov2005:monopreservapprox},
where very general Lipschitz aggregation functions are fit to data, may be used.
In our case, we presume that there is a kind of ``noise'' in the data set
and it may not always be possible to find a function that goes through
all the observations. In other words, we are faced with a discrete approximation task.
\end{remark}

\subsection{Fitting weighted arithmetic means}\label{Sec:weightfit1d}

Let us start by examining a quite simple case of weighted arithmetic means.
At this point, only some simple linear algebra and mathematical programming
tools are involved in the computations.
As it shall turn out below, optimization problems utilizing the most
common goodness of fit measures:
squared Euclidean (least squares error, LSE), Manhattan (least absolute deviation, LAD),
and Chebyshev (least maximal absolute deviation, LMD) metrics reduce themselves
to quadratic and linear programming tasks (see, e.g., \cite{NocedalWright}).

\begin{remark}
   The discussed algorithms may also be easily modified to fit
   OWA operators' weights (by ordering elements in $\vect{X}$ appropriately).
   Also note that fitting WAM weights to data is a more difficult
   problem than performing linear regression, as in our case weights
   must fulfill some additional constraints.
\end{remark}

\subsubsection{Least squares fitting}

\index{LSE|see {least squares error}}\index{least squares error}%
Most often, we would like to find the least squares error (LSE) solution
to a weight fit problem:
\begin{equation}\label{Eq:qp2}
   \mathrm{minimize}\  \sum_{j=1}^m \left(
   \sum_{i=1}^n w_i x_i^{(j)} - y^{(j)}
   \right)^2
      \quad \text{w.r.t.~}\vect{w}
\end{equation}
subject to $\vect{w}\ge_n \vect{0}$ and $\vect{1}^T \vect{w}=1$.
This task is a quadratic programming (QP) problem,
see, e.g., \cite[Chapter 5]{BeliakovETAL2007:aggregationpractitioners}
or \cite{Torra2002:learningweightsqwam}.

\begin{definition}\index{QP task|see {quadratic programming task}}\index{quadratic programming task}%
A \emph{quadratic programming problem} may be expressed as:
\[
   \mathrm{minimize}\ 0.5\,\vect{v}^T \vect{D} \vect{v} + \vect{c}^T \vect{v} + c_0
   \quad \text{w.r.t.~}\vect{v}=(v_1,\dots,v_n)
\]
subject to:
\begin{eqnarray*}
   \vect{A}\vect{v} & \gtreqless_n & \vect{b}, \\
   \vect{v}         & \le_n & \vect{u}, \\
   \vect{v}         & \ge_n & \vect{l},
\end{eqnarray*}
where $\vect{D}\in\mathbb{R}^{n\times n}$ is symmetric and positive semidefinite,
$\vect{c}\in\mathbb{R}^n$,
$c_0\in\mathbb{R}$,
$\vect{l}\in\bar{\mathbb{R}}^n$,
$\vect{u}\in\bar{\mathbb{R}}^n$,
$\vect{l}\le_n\vect{u}$,
and
$\vect{A}\in\mathbb{R}^{k\times n},
\vect{b}\in\mathbb{R}^k$
for some $k\ge 0$.
\end{definition}

\begin{remark}\label{Remark:CGAL}
Figure \ref{Fig:CGAL_quadprog1} and \ref{Fig:CGAL_quadprog2}
gives the source code of an \R{} language interface to the quadratic
programming solver from the open source \package{CGAL} \cite{cgal:eb-14b}
library. The implemented algorithm is based on a generalized
simplex method, see also \cite{GartnerSchonherr2000:qpball,Schoenherr2002:phd}.
This solver has a particularly good performance for tasks with a small number of variables
but large number of constraints or a large number of variables and
small number of constraints.
Other \R{} QP solvers (e.g., the \texttt{solve.QP()} function
from the \package{quadprog} package) either assume that $\vect{D}$ is
(strictly) positive definite or require additional commercial software
installed, e.g., \package{CPLEX}, \package{MOSEK}, or \package{LocalSolver}.
\end{remark}

The optimization problem given by Equation~\eqref{Eq:qp2}
may be rewritten in terms of a QP task as follows:
\begin{equation}\label{Eq:qp2mod}
   \mathrm{minimize}\  0.5\, \vect{w}^T \vect{X} \vect{X}^T \vect{w} - (\vect{X}\vect{Y}^T)^T \vect{w}
   \quad \text{w.r.t.~}\vect{w}
\end{equation}
with 1 linear equality constraint under the assumption that $\vect{w}\ge_n\vect{0}$,
see Figure~\ref{Fig:fit_wam_L2_quadprog} for an exemplary \R{} implementation.
Note that $\vect{X} \vect{X}^T$ is surely at least positive semidefinite,
see also \cite{Torra2002:learningweightsqwam} for discussion
on linearly dependent rows in~$\mathbf{X}$.

\subsubsection{Least absolute deviation fitting}

Beliakov in \cite{Beliakov2009:linearprogramming}
(see also \cite[Chapter 5]{BeliakovETAL2007:aggregationpractitioners})
considered methods for fitting aggregation operators to
observed input data using the least absolute deviation
(LAD, i.e., $L_1$ metric) criterion, which is less sensitive
to outliers than the least squares error.
Nevertheless, we shall note that in this setting the solutions may be ambiguous and unstable.
\index{LAD|see {least absolute deviation}}\index{least absolute deviation}%

We aim to find a weighting vector $\vect{w}$ that
is a solution to the optimization problem:
\begin{equation}\label{Eq:lp1}
    \mathrm{minimize}\ \sum_{j=1}^m \left|\sum_{i=1}^n w_i x_i^{(j)}-y^{(j)} \right|
    \quad \text{w.r.t.~}\vect{w}
\end{equation}
subject to $\vect{w}\ge_n \vect{0}$ and $\vect{1}^T \vect{w}=1$.

It turns out that our LAD minimization task may be translated
to a linear programming (LP) problem.

\begin{definition}\index{LP task|see {linear programming task}}\index{linear programming task}%
A \emph{linear programming problem}  may be expressed as:
\[
   \mathrm{minimize}\ \vect{c}^T\vect{v} + c_0
   \quad \text{w.r.t.~}\vect{v}=(v_1,\dots,v_n)
\]
subject to:
\begin{eqnarray*}
   \vect{A}\vect{v} & \gtreqless_n & \vect{b}, \\
   \vect{v}         & \le_n & \vect{u}, \\
   \vect{v}         & \ge_n & \vect{l},
\end{eqnarray*}
where $\vect{c}\in\mathbb{R}^n$,
$c_0\in\mathbb{R}$,
$\vect{l}\in\bar{\mathbb{R}}^n$,
$\vect{u}\in\bar{\mathbb{R}}^n$,
$\vect{l}\le_n\vect{u}$,
and
$\vect{A}\in\mathbb{R}^{k\times n},
\vect{b}\in\mathbb{R}^k$
for some $k\ge 0$.
\end{definition}

\begin{remark}
The simplex or interior-point methods, among others, may be used to solve LP tasks.
Note that in some LP software, like \package{lp\_solve},
the condition $\vect{v} \ge_n \vect{0}$ is always implicitly assumed.
Interestingly, LP tasks may also be computed by using the mentioned-above
\package{CGAL} library QP solver by simply assuming that $\vect{D}=\mathbf{0}$.
\end{remark}

Let us introduce $2m$ auxiliary variables $r_j^+, r_j^-$, $j=1,\dots,m$,
such that $r_j^+ - r_j^-=\sum_{i=1}^n w_i x_i^{(j)}-y^{(j)}$
and $r_j^+, r_j^-\ge 0$.
With this, the optimization problem given by Equation~\eqref{Eq:lp1} may be rewritten,
see \cite[Chapter 6]{BloomfieldSteiger1983:lad},
\cite[Chapter 6]{ConvexOptimization},
and \cite[Chapter 6]{AIMMSoptimizmod2012}, as:
\begin{equation}\label{Eq:lp1mod}
      \mathrm{minimize}\ \sum_{j=1}^m \left( r_j^+ + r_j^- \right)
      \quad \text{w.r.t.~}\vect{w}, \vect{r}^+, \vect{r}^-
\end{equation}
subject to:
\begin{eqnarray*}
\sum_{i=1}^n w_i x_i^{(j)} - r_j^+ + r_j^- &=& y^{(j)}, \quad j=1,\dots,m\\
\sum_{i=1}^n w_i &=& 1,\\
(\vect{w}, \vect{r}^+, \vect{r}^-) & \ge_{n+2m} & \vect{0}.
\end{eqnarray*}
Figure~\ref{Fig:fit_wam_L1_linprog} presents an \R{} implementation
of this LP task setup, which again is based on the \package{CGAL} QP solver.

\subsubsection{Least Chebyshev metric fitting}

Let us now suppose that we would like to find the least maximum absolute deviation (LMD)
solution to a weight fitting problem, i.e., one that minimizes
the Chebyshev $L_\infty$ metric:
\begin{equation}\label{Eq:lpInf}
    \mathrm{minimize}\ \bigvee_{j=1}^m \left|\sum_{i=1}^n w_i x_i^{(j)}-y^{(j)} \right|
    \quad \text{w.r.t.~}\vect{w}
\end{equation}
subject to $\vect{w}\ge_n \vect{0}$ and $\vect{1}^T \vect{w}=1$.
It turns out that, see \cite[Chapter 6]{ConvexOptimization} or
\cite[Chapter 6]{AIMMSoptimizmod2012}, the Chebyshev metric minimization task
may also be represented as an LP problem. Thus, by rewriting Equation~\eqref{Eq:lpInf},
we get what follows:
\begin{equation*}\label{Eq:lpInfLinProg}
    \mathrm{minimize}\ t
    \quad \text{w.r.t.~}\vect{w}, t
\end{equation*}
with linear constraints of the form:
\begin{eqnarray*}
\sum_{i=1}^n w_i x_i^{(j)} - t &\le &  y^{(j)}, \quad j=1,\dots,m\\
\sum_{i=1}^n w_i x_i^{(j)} + t &\ge &  y^{(j)}, \quad j=1,\dots,m\\
\sum_{i=1}^n w_i &=& 1,\\
(\vect{w}, t) &\ge_{n+1}& 0.\\
\end{eqnarray*}
Figure \ref{Fig:fit_wam_LInf_linprog} gives an exemplary \R{} language implementation
for least Chebyshev metric fitting.

\begin{example}\label{Ex:WAMfit1}
Suppose that $n=5$ and we are given $m=9$ toy data points as follows:
\begin{center}
\begin{tabularx}{1.0\linewidth}{Xrrrrrrrrr}
\toprule
\bf\small $j$ & \bf\small 1 & \bf\small 2 & \bf\small 3 & \bf\small 4 & \bf\small 5 & \bf\small 6 & \bf\small 7 & \bf\small 8 & \bf\small 9  \\
\midrule
\bf\small $x_1^{(j)}$ & 0.12 & 0.48 & 0.65 & 0.07 & 0.37 & 0.22 & 0.29 & 0.57 & 0.84 \\
\bf\small $x_2^{(j)}$ & 0.73 & 0.41 & 0.45 & 0.79 & 0.92 & 0.23 & 0.90 & 0.40 & 0.57 \\
\bf\small $x_3^{(j)}$ & 0.43 & 0.84 & 0.70 & 0.96 & 0.81 & 0.86 & 0.72 & 0.53 & 0.42 \\
\bf\small $x_4^{(j)}$ & 0.52 & 0.75 & 0.48 & 0.40 & 0.62 & 0.28 & 0.80 & 0.92 & 0.79 \\
\bf\small $x_5^{(j)}$ & 0.69 & 0.70 & 0.24 & 0.22 & 0.92 & 0.34 & 0.15 & 0.50 & 0.50 \\
\midrule
\bf\small $y^{(j)}$   & 0.58 & 0.56 & 0.70 & 0.40 & 0.78 & 0.50 & 0.64 & 0.62 & 0.73 \\
\bottomrule
\end{tabularx}
\end{center}

$\vect{Y}$ was generated in such a way that firstly $\vect{w}=(0.33, 0.43, 0.10, 0.08, 0.06)$ was assumed and then some
random white noise was added ($\sigma=0.05$). Here are the results of applying the above-presented
algorithms (weights and corresponding errors).

\begin{center}\small
\begin{tabularx}{1.0\linewidth}{Xrrrrrrrr}
\toprule
\bf\small $E$ & \bf\small $w_1$ & \bf\small $w_2$ & \bf\small $w_3$ & \bf\small $w_4$ & \bf\small $w_5$ & \bf\small $\mathfrak{d}_1$ & \bf\small $\mathfrak{d}_2$ & \bf\small $\mathfrak{d}_\infty$ \\
\midrule
 LAD       & 0.1131 & 0.3324 & 0.0000 & 0.3460 & 0.2085 & \underline{0.6764} & 0.3618 & 0.2608  \\
 LSE       & 0.2349 & 0.2026 & 0.2235 & 0.2500 & 0.0890 & 0.7654 & \underline{0.2882} & 0.1583 \\
 LMD      & 0.1747 & 0.0996 & 0.2719 & 0.4538 & 0.0000 & 0.9276 & 0.3243 & \underline{0.1335} \\
 ---       & 0.3300 & 0.4300 & 0.1000 & 0.0800 & 0.0600 & 0.8773 & 0.3360 & 0.1997 \\
\bottomrule
\end{tabularx}
\end{center}
\end{example}

\begin{remark}
If given exemplars have different degrees of importance, weighted
goodness of fit measures can straightforwardly be incorporated
into the three above optimization tasks.
\end{remark}

\subsection{Preservation of output rankings}

Beliakov~et~al.~in \cite{BeliakovETAL2007:aggregationpractitioners},
see also, e.g., \cite{Beliakov2009:linearprogramming}, point out that
sometimes a decision modeler may be interested in preserving the ranking
of outputs. To do so, we find a permutation $\sigma\in\mathfrak{S}_{[m]}$
such that $y^{(\sigma(1))}\le\dots\le y^{(\sigma(m))}$. With that,
we introduce additional constrains into our optimization task:
\[
   \func{F}_\vect{w}(\vect{x}^{(\sigma(j+1))})-\func{F}_\vect{w}(\vect{x}^{(\sigma(j))}) \ge 0
   \quad\text{ for } j=1,\dots,m-1.
\]
Let $\vect{X}^{(-k)} = (x_{i,j})_{i\in[n], j\in[m], j\neq k}$ denote
the $\vect{X}$ matrix with the $k$th column omitted.
If $\func{F}_\vect{w}$ is again a weighted arithmetic mean, we
get further linear inequalities of the form:
\[
   \vect{w}^T \left(\vect{X}^{(-1)}-\vect{X}^{(-m)}\right) \ge 0.
\]

However, let us note that some input data may lead
to optimization problems that are inconsistent, i.e., that have no feasible solutions.
In order to overcome this limitation we may try to incorporate an additional term
into our goodness of fit measure which acts as a penalty for violating
the desired output ranking:
\[
   \mathrm{minimize}\ E\left(\vect{w}^T\vect{X}, \vect{Y}\right) +
   P\left(\vect{w}^T (\vect{X}^{(-m)}-\vect{X}^{(-1)})\vee 0\right)
   \quad \text{w.r.t.~}\vect{w}
\]
Typically, we set $P(\vect{z})=p\sum_{i=1}^{m-1} z_i^2$
or $P(\vect{z})=p\sum_{i=1}^{m-1} |z_i|$ for some tuning parameter $p>0$
that must be set up empirically, e.g., by further numeric experiments.
For instance, we may try to seek the smallest $p$
such that the Kendall correlation coefficient between $\vect{Y}$
and $\vect{w}_p^T\vect{X}$ is as large as possible.

\subsubsection{LAD fit with $P$ being the $L_1$ norm}

In the case that $E$ is the $L_1$ metric and $P(\vect{z})=p\sum_{i=1}^{m-1} |z_i|$
we get an LP problem, see \cite[page~267]{BeliakovETAL2007:aggregationpractitioners},
which is a version of Equation~\eqref{Eq:lp1mod} with $m-1$ additional ($n+3m-1$ in total) variables
and exactly $n+5m-1$ constraints:
\[
   \mathrm{minimize}\ \sum_{j=1}^m \left( r_j^+ + r_j^- \right) + p \sum_{j=1}^{m-1} q_j
   \quad \text{w.r.t.~}\vect{w}, \vect{r}^+, \vect{r}^-, \vect{q}
\]
subject to:
\begin{eqnarray*}
\sum_{i=1}^n w_i x_i^{(j)} - r_j^+ + r_j^- &=& y^{(j)}, \quad j=1,\dots,m\\
\sum_{i=1}^n w_i &=& 1,\\
(\vect{w}, \vect{r}^+, \vect{r}^-, \vect{q}) & \ge_{n+3m-1} & \vect{0},\\
\sum_{i=1}^n w_i \left(x_i^{(\sigma(j+1))}-x_i^{(\sigma(j))}\right) + q_j &\ge& 0, \quad j=1,\dots,m-1
\end{eqnarray*}
where $\sigma$ is an ordering permutation of $(y^{(1)},\dots,y^{(m)})$.

\subsubsection{LSE fit with $P$ being the squared $L_2$ norm}\label{Sec:LSEpreserveoutrank}

It turns out (in \cite{BeliakovETAL2007:aggregationpractitioners}
only the case of $P$ being the $L_1$ norm is considered) that
the case of least squared error fitting with $P(\vect{z})=p\sum_{i=1}^{m-1} z_i^2$
is quite similar to the previous one.
We may incorporate $m-1$ additional variables into the quadratic programming
task given by Equation~\eqref{Eq:qp2mod} and approach the following optimization problem:
\[
   \mathrm{minimize}\  0.5\, \vect{v}^T \vect{D} \vect{v} + \vect{c}^T \vect{v}
   \quad \text{w.r.t.~}\vect{v}=(\vect{w}, \vect{q})
\]
subject to:
\begin{eqnarray*}
\sum_{i=1}^n w_i &=& 1,\\
(\vect{w}, \vect{q}) & \ge_{2n-1} & \vect{0},\\
\sum_{i=1}^n w_i \left(x_i^{(\sigma(j+1))}-x_i^{(\sigma(j))}\right) + q_j &\ge& 0, \quad j=1,\dots,m-1
\end{eqnarray*}
where:
\[
   \vect{D}=\left[
   \begin{array}{ccc|ccc}
      &                      &   &   &             &   \\
      & \vect{X} \vect{X}^T  &   &   &  \vect{0}   &   \\
      &                      &   &   &             &   \\
      \hline
      &                      &   & p & \cdots      & 0 \\
      &            \vect{0}  &   & 0 & \ddots      & 0 \\
      &                      &   & 0 & \cdots      & p \\
   \end{array}
   \right], \quad
   \vect{c}=\left[
   \begin{array}{c}
      \\
      -\vect{X}\vect{Y}^T\\
      \\
      \hline
      \\
      \vect{0}\\
      \\
   \end{array}
   \right].
\]

\begin{example}
Let us go back to the data set studied in Example~\ref{Ex:WAMfit1}.
Below are the results of finding the best fitting WAM weights,
together with Kendall's $\tau$ correlation coefficient
between $\vect{Y}$ and the output generated by the computed model.
Parameters $p$ were selected so that $\tau$ is maximized
and then the error metric of interest is minimized.
\begin{center}
\begin{tabularx}{1.0\linewidth}{Xrrrrr}
\toprule
\bf\small $E$ & $P(\vect{z})$ & \bf\small $\mathfrak{d}_1$ & \bf\small $\mathfrak{d}_2$ & \bf\small $\mathfrak{d}_\infty$ & $\tau$\\
\midrule
LAD         & 0                    & \underline{0.6764} & 0.3618 & 0.2608 & 0.28 \\
LSE         & 0                    & 0.7654 & \underline{0.2882} & 0.1583 & 0.56 \\
LMD         & 0                    & 0.9276 & 0.3243 & \underline{0.1335} & 0.33\\
LAD         & $1.2\sum_i |z_i|$    & 0.8059 & 0.3775 & 0.2575 & \underline{0.72} \\
LSE         & $2.8\sum_i z_i^2$    & 0.8914 & 0.3339 & 0.2063 & \underline{0.72} \\
\bottomrule
\end{tabularx}
\end{center}
We see that we were able to match the output ranking quite well,
however, at the cost of increasing the minimized goodness-of-fit measure.
Also please keep in mind that there are data sets
for which we cannot increase the initial $\tau$.
\end{example}

\subsection{Regularization}

A well-known fact from machine learning is that
even if we establish ``good'' weights on a given input sample,
we do not necessarily obtain a model which exhibits satisfactory behavior on other
data that come from the same source. For instance, an estimated fusion
function may be overfitted.
For this reason, in regression analysis the concept of
\index{regularization}\emph{parameter regularization}
is sometimes used. It has a form of an additional penalty term
dependent on some norm (or its function) of the vector of parameters.
And so, e.g., ridge regression aims to minimize the squared
prediction error plus a properly scaled, squared $L_2$ norm
of the variables.

In our case we may consider, for some $\lambda$, an optimization task:
\[
   \mathrm{minimize}\ E\left(\func{F}_\vect{w}(\vect{X}), \vect{Y}\right) + \lambda\|\vect{w}\|
   \quad \text{w.r.t.~}\vect{w}
\]
subject to $\vect{1}^T\vect{w}=1$ and $\vect{w}\ge_n\vect{0}$,
where $\|\cdot\|$ is some norm (or its function), typically squared $L_2$.
Note that due to the usual constraints on $\vect{w}$, the use of the $L_1$ norm
(like, e.g., in Lasso regression) does not make much sense at this point.

Incorporating the penalty term $\|\cdot\|_2^2$ in optimization
tasks discussed above is relatively easy, therefore it is left to the kind reader.

\begin{remark}
Regularization in the case of WAM weights estimation works quite well
if $n$ or $m$ is relatively small. If this is not the case,
we often do not observe positive effects of introducing the mentioned penalty.
Unlike in regression problems, where we always presuppose that $\lambda\ge 0$,
in our framework we are bounded with the constraint $\sum_i w_i=1$ which,
for large $\lambda$, tends to generate weighting vectors
such that $w_i\to 1/n$. On the other hand, in the current framework the case of $\lambda<0$
may also lead to useful outcomes. Yet, we should note that for
$\lambda\to -\infty$ we observe that $w_j\to 1$ for some $j\in[n]$.
\end{remark}

\begin{figure}[b!]
\centering

\includegraphics[width=8.25cm]{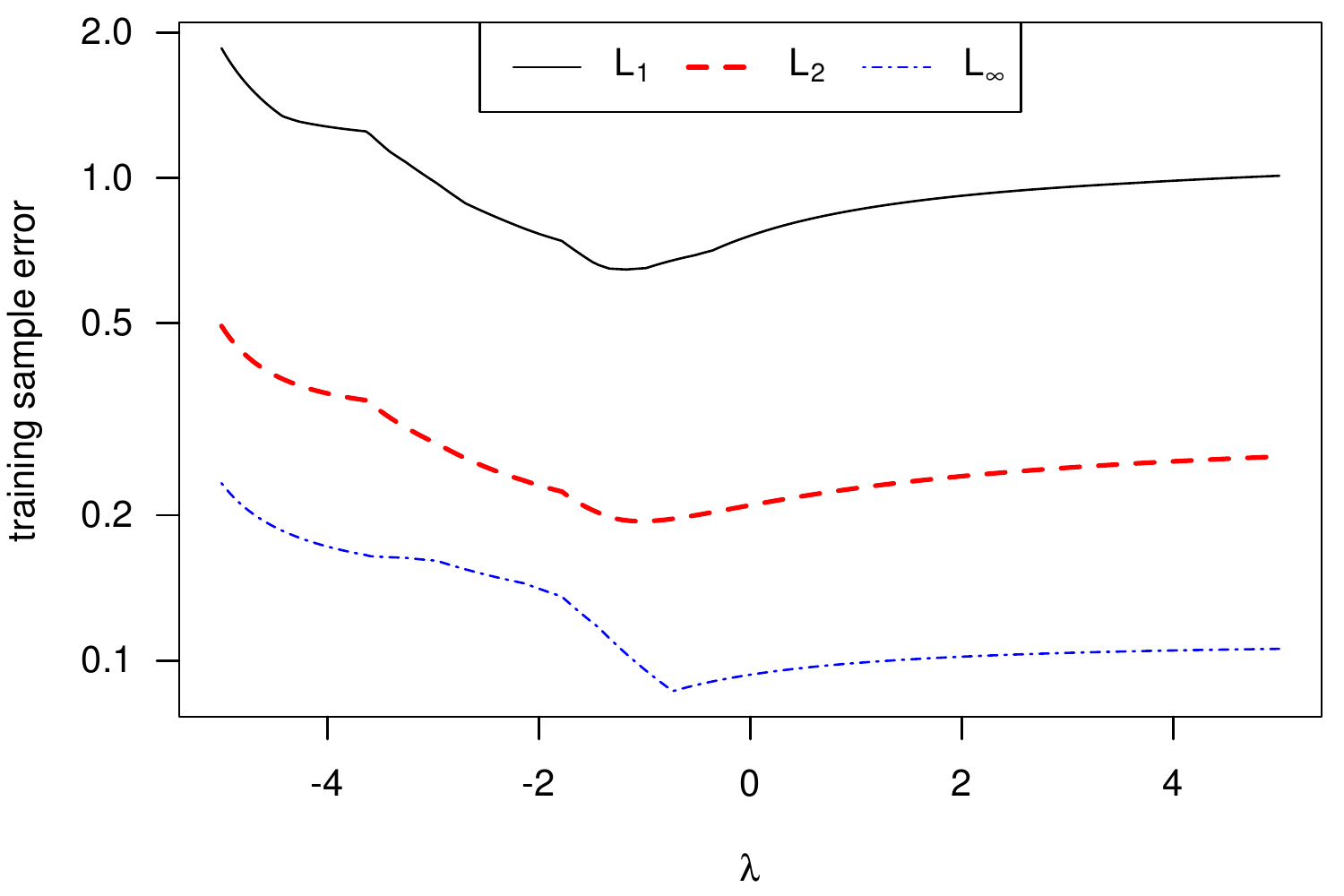}

\caption{\label{Fig:wam_regularization_L2} Three error measures on a test data set
from Example~\ref{Ex:regularization} as a function of
regularization penalty $\lambda$.}
\end{figure}

\begin{example}\label{Ex:regularization}
Let us consider a data set generated randomly with \R{} as follows:
\begin{lstlisting}[language=R]
set.seed(321)
n <- 10
m <- 100
realw <- runif(n)
realw <- realw/sum(realw)
X <- t(round(matrix(runif(n*m, 0, 1), nrow=m), 2))
Y <- t(realw) %

train <- sample(1:m, m*0.8)
X_test <- X[,-train,drop=FALSE] # test sample
Y_test <- Y[,-train,drop=FALSE]
X <- X[,train,drop=FALSE]       # training sample
Y <- Y[,train,drop=FALSE]
\end{lstlisting}
The set is divided into two parts: a training sample (80\% of the observations,
used to compute the weights) and a test sample (20\%, used to estimate the error).
Here we consider a QP task:
\begin{equation}
   \mathrm{minimize}\  0.5\, \vect{w}^T (\vect{X} \vect{X}^T+\lambda\vect{I}) \vect{w} - (\vect{X}\vect{Y}^T)^T \vect{w}
   \quad \text{w.r.t.~}\vect{w}
\end{equation}
subject to $\vect{w}\ge_n\vect{0}$, $\vect{1}^T\vect{w}=1$,
which optimizes the squared error plus a $\lambda\|\vect{w}\|_2^2$ penalty term.
Figure~\ref{Fig:wam_regularization_L2} depicts three goodness of fit measures
as a function of~$\lambda$.  We see that in this example we are able to
improve the least squares error measure (which was minimized in this case).
\begin{center}
\begin{tabularx}{1.0\linewidth}{Xrrrr}
\toprule
\bf\small $E$    &  \bf\small $\lambda$    &   \bf\small $\mathfrak{d}_1$ & \bf\small $\mathfrak{d}_2$ & \bf\small $\mathfrak{d}_\infty$ \\
\midrule
--- (using \texttt{realw})  & $0$       & 0.8113 & 0.2256 & 0.1009 \\
LAD            & $0$       & 0.8525 & 0.2315 & 0.0956 \\
LSE            & $0$       & 0.7589 & 0.2098 & \underline{0.0935} \\
LMD       & $0$       & 0.9137 & 0.2384 & 0.0975 \\
LSE          & $-1.03$   & \underline{0.6492} & \underline{0.1944} & 0.0970 \\
\bottomrule
\end{tabularx}
\end{center}
\end{example}

\subsection{Fitting weights of weighted quasi-arithmetic means}

Let us now consider the case of $\func{F}_\vect{w}=\varphi^{-1}(\vect{w}^T\varphi(\vect{x}))$,
i.e., weighted quasi-arithmetic means, for an
arbitrary but known and fixed continuous,
strictly increasing generator function $\varphi$.
Note that the case of fitting $\varphi$ to empirical data is discussed
later on.

Torra in \cite{Torra2002:learningweightsqwam,Torra2004:owadatreindentification} discussed weighted
quasi-arithmetic mean fitting using the $L_2$-metric minimization criterion.
Yet, he noted that the problem is difficult in general,
so he assumed that the exemplars are not subject to errors.
In such a case, noting that $\varphi$ is surely invertible, we have for all $j$:
\[
   \sum_{i=1}^n w_i \varphi(x_i^{(j)}) = \varphi(y^{(j)}).
\]
Using this assumption, instead of minimizing:
\[
   \| \varphi^{-1}\left( \vect{w}^T \varphi(\vect{X}) \right) - \vect{Y} \|_2
\]
one can minimize a quite different (in general) goodness of fit measure:
\[
   \| \vect{w}^T \varphi(\vect{X}) - \varphi(\vect{Y}) \|_2.
\]
A similar approach, this time concerning the $L_1$ metric, was
utilized by Beliakov et al.~in, e.g., \cite{Beliakov2009:linearprogramming,%
BeliakovETAL2007:aggregationpractitioners,BeliakovJames2012:linprogbonferroni}.
For this task, exactly the methods presented in the previous subsection can be applied,
but this time on appropriately transformed $\vect{X}$ and $\vect{Y}$.
Such an approach is often called \emph{linearization} of inputs.%
\label{Page:Linearization}\index{linearization}

Let us suppose, however, that we would like to solve the original weight fit
problem and not the simplified one. This leads (in general)
to a nonlinear optimization task.

\begin{example}
Let $n=5, m=9$, and $\vect{x}^{(1)},\dots,\vect{x}^{(m)}$ be the same
as in Example~\ref{Ex:WAMfit1}. This time, however, $\varphi(x)=x^2$
and $y_{(1)},\dots, y_{(m)}$ is as follows:
\begin{center}
\begin{tabularx}{1.0\linewidth}{Xrrrrrrrrr}
\toprule
\bf\small $j$ & \bf\small 1 & \bf\small 2 & \bf\small 3 & \bf\small 4 & \bf\small 5 & \bf\small 6 & \bf\small 7 & \bf\small 8 & \bf\small 9  \\
\midrule
\bf\small $y^{(j)}$   & 0.65 & 0.58 & 0.70 & 0.51 & 0.82 & 0.56 &  0.70 & 0.64 & 0.75 \\
\bottomrule
\end{tabularx}
\end{center}

\noindent
Here are the true $\mathfrak{d}_1$ and $\mathfrak{d}_2$ errors in the case of
linearized and optimal goodness-of-fit measure minimization tasks.
The differences are quite small, but not negligible. Yet, we may observe
that often the linearized and ``exact'' $E$ minimization tasks lead to solutions
which are very close to each other.

\begin{center}\small
\begin{tabularx}{1.0\linewidth}{Xrrr}
\toprule
\bf\small $E$ & \bf\small $\mathfrak{d}_1$ & \bf\small $\mathfrak{d}_2$ & \bf\small $\mathfrak{d}_\infty$ \\
\midrule
\small LAD -- linearization & 0.7385 & 0.4120 & 0.2798  \\
\small LSE -- linearization & 0.7423 & 0.2859 & 0.1626 \\
\small LAD -- optimal       & \underline{0.7157} & 0.3170 & 0.2044  \\
\small LSE -- optimal       & 0.7587 & \underline{0.2817} & \underline{0.1501} \\
\bottomrule
\end{tabularx}
\end{center}
\end{example}

\subsubsection{LSE fit of WQAMean weights}

We aim to:
\begin{equation}\label{Eq:LSEWQAM}
   \mathrm{minimize}\  \sum_{j=1}^m \left(
   \varphi^{-1}\left(\sum_{i=1}^n w_i \varphi\left(x_i^{(j)}\right)\right) - y^{(j)}
   \right)^2
      \quad \text{w.r.t.~}\vect{w}
\end{equation}
subject to $\vect{w}\ge_n \vect{0}$ and $\vect{1}^T \vect{w}=1$.
By homogeneity and triangle inequality of $\|\cdot\|_2$ we have that
this is a convex optimization problem.
To drop the constraints on $\vect{w}$,
let us use an approach considered by Filev and Yager \cite{FilevYager1998:issueobtainowaweights},
see also \cite{Torra2004:owadatreindentification}
(a barrier function could also be used for that, among others).
We take a different parameter space, $\boldsymbol\lambda\in\mathbb{R}^n$,
such that:
\[
w_i = \frac{\exp(\lambda_i)}{\sum_{k=1}^n \exp(\lambda_k)}.
\]
Assuming that $\varphi^{-1}$ is differentiable,
let us determine the gradient $\nabla E(\boldsymbol\lambda)$. For any $k\in[n]$ it holds:
\begin{eqnarray*}
\frac{\partial}{\partial \lambda_k} E(\boldsymbol\lambda)
&=& 2\frac{\exp(\lambda_k)}{\sum_{i=1}^n \exp(\lambda_i)} \sum_{j=1}^m \left(\varphi^{-1}\left(\frac{\sum_{i=1}^n \exp(\lambda_i) \varphi\left(x_i^{(j)}\right)}{\sum_{i=1}^n \exp(\lambda_i)}\right) - y^{(j)} \right)\\
&\cdot&  (\varphi^{-1})'\left(\frac{\sum_{i=1}^n \exp(\lambda_i) \varphi\left(x_i^{(j)}\right)}{\sum_{i=1}^n \exp(\lambda_i)} \right)\\
&\cdot&
      \left(
         \varphi\left(x_k^{(j)}\right) - \frac{\sum_{i=1}^n \exp(\lambda_i) \varphi\left(x_i^{(j)}\right)}{\sum_{i=1}^n \exp(\lambda_i)}
      \right).
\end{eqnarray*}
Assuming that $\vect{Z}=\vect{w}^T\varphi(\vect{X})$ and $\vect{w}=\exp(\boldsymbol\lambda)/\vect{1}^T\exp(\boldsymbol\lambda)$, we have:
\begin{eqnarray*}
\nabla E(\boldsymbol\lambda) &=&
2\cdot \vect{w}\cdot \Bigg(
   \left(\left(\phi^{-1}(\vect{Z})-Y\right)\cdot(\varphi^{-1})'\left(\vect{Z}\right)\right)\times\left(\varphi(\vect{X})^T-\vect{Z}\right)
\Bigg),
\end{eqnarray*}
where $\cdot, -$ stand for elementwise vectorized multiplication and subtraction, respectively,
$\times$ denotes matrix multiplication,
and $\varphi(\vect{X})^T-\vect{Z}$ means that we subtract $\vect{Z}$ from
each column in $\varphi(\vect{X})^T$ (this is in fact how matrix and vector
arithmetic operations are vectorized in \R{}).
Figure~\ref{Fig:fit_wqam_L2_optim} gives an \R{} implementation of a weight
fitting procedure which is based on a quasi-Newton nonlinear optimization method
by Broyden, Fletcher, Goldfarb and Shanno (the BFGS algorithm, see \cite{NocedalWright}).
Please note that while using the mentioned reparametrization,
the BFGS algorithm may occasionally fail to converge.

\subsubsection{LAD fit of WQAMean weights}

Now let us:
\begin{equation}
   \mathrm{minimize}\  \sum_{j=1}^m \left|
   \varphi^{-1}\left(\sum_{i=1}^n w_i \varphi\left(x_i^{(j)}\right)\right) - y^{(j)}
   \right|
      \quad \text{w.r.t.~}\vect{w}
\end{equation}
subject to $\vect{w}\ge_n \vect{0}$ and $\vect{1}^T \vect{w}=1$.
This case is problematic to nonlinear solvers, as our goodness-of-fit measure
is not differentiable at 0 and we observe that methods like BFGS
(using numeric  finite-difference approximation of the gradient)
may return results that are not close enough to the optimum.

In order to overcome this limitation, we propose the following heuristic.
Instead of minimizing $\sum_{j=1}^m |z_j|$, we may
consider $\sum_{j=1}^m \sqrt{z_j^2+\varepsilon^2}$ for some $\varepsilon>0$,
typically $\varepsilon=10^{-12}$.
This is because $|x|\le\sqrt{x^2+\varepsilon^2}$ and
$\sqrt{x^2+\varepsilon^2}\to_{\varepsilon\to 0} |x|$ for all $x$.
Thus, our task is now to:
\begin{equation}%
   \mathrm{minimize}\  \sum_{j=1}^m \sqrt{\left(
   \varphi^{-1}\left(\frac{\sum_{i=1}^n \exp(\lambda_i)\varphi\left(x_i^{(j)}\right)}{\sum_{i=1}^n  \exp(\lambda_k)}\right) - y^{(j)}
   \right)^2 + \varepsilon^2}
      \quad \text{w.r.t.~}\boldsymbol\lambda
\end{equation}
where again we use the reparametrization $w_i = \frac{\exp(\lambda_i)}{\sum_{k=1}^n \exp(\lambda_k)}$,
which enables us to drop any additional constraints.
In such a case we have:
\begin{eqnarray*}\small
\frac{\partial}{\partial \lambda_k} E(\boldsymbol\lambda)
&=& \frac{\exp(\lambda_k)}{\sum_{i=1}^n \exp(\lambda_i)} \sum_{j=1}^m
\frac{\left(\varphi^{-1}\left(\frac{\sum_{i=1}^n \exp(\lambda_i) \varphi\left(x_i^{(j)}\right)}{\sum_{i=1}^n \exp(\lambda_i)}\right) - y^{(j)} \right)}{ \sqrt{\left(
   \varphi^{-1}\left(\frac{\sum_{i=1}^n \exp(\lambda_i)\varphi\left(x_i^{(j)}\right)}{\sum_{i=1}^n  \exp(\lambda_k)}\right) - y^{(j)}
   \right)^2 + \varepsilon^2}
}\\
&\cdot&  (\varphi^{-1})'\left(\frac{\sum_{i=1}^n \exp(\lambda_i) \varphi\left(x_i^{(j)}\right)}{\sum_{i=1}^n \exp(\lambda_i)} \right)\\
&\cdot&
      \left(
         \varphi\left(x_k^{(j)}\right) - \frac{\sum_{i=1}^n \exp(\lambda_i) \varphi\left(x_i^{(j)}\right)}{\sum_{i=1}^n \exp(\lambda_i)}
      \right).
\end{eqnarray*}
Assuming that $\vect{Z}=\vect{w}^T\varphi(\vect{X})$ and $\vect{w}=\exp(\boldsymbol\lambda)/(\vect{1}^T\exp(\boldsymbol\lambda))$, we have:
\begin{eqnarray*}
\nabla E(\boldsymbol\lambda) &=&
\vect{w}\cdot \Bigg(
\frac{
   \left(\phi^{-1}(\vect{Z})-Y\right)\cdot(\varphi^{-1})'\left(\vect{Z}\right)
}{
   \sqrt{(\varphi^{-1}(\vect{Z})-\vect{Y})\cdot(\varphi^{-1}(\vect{Z})-\vect{Y}) + \varepsilon^2}
}\times
\left(\varphi(\vect{X})^T-\vect{Z}\right)
\Bigg).
\end{eqnarray*}

\begin{remark}
In Figure~\ref{Fig:fit_wqam_L1_optim} we provide an implementation of
the aforementioned weight fitting procedure. It is based on the BFGS algorithm
available via the \texttt{optim()} function in \R{}. For testing purposes,
we set up convergence criteria to be
$\mathtt{reltol=1e-16}$, $\mathtt{maxiter=10000}$.

It is well-known that LAD optimization using nonlinear solvers
does not guarantee that the output result is the global optimum:
the BFGS algorithm may sometimes get stuck in a suboptimal
solution or fail to converge in a predefined number of iterations.

For instance, suppose that
$\varphi(x)=x^2$, $n=5$, $m=25$, $\varepsilon=10^{-12}$
and that $\vect{X}$ and $\vect{Y}$ are generated randomly
like in Example~\ref{Ex:regularization}.
The presented procedure gives median relative $L_1$ error
(as compared to the optimal solution determined
by the routine in Figure~\ref{Fig:fit_wam_L1_linprog}) of $1\times\simeq 10^{-12}$
($M=10000$ MC iterations).
On the other hand, the BFGS algorithm applied directly on an
absolute value-based error function gives median relative
error of $\simeq 6\times 10^{-4}$.
The 99\%-quantiles are, respectively, around $3\times 10^{-7}$ and
$0.5$. Thus, the suggested approximation works far better than
the direct approach.

Sometimes it may be advisable to run the optimization
routine a few times, starting each time from a different initial point
and then choose the best (in terms of $L_1$ error) solution.
For instance, in the current experiment setting, using 10 trials
reduces the median error of the ``exact method''
to $\simeq 6\times 10^{-5}$, i.e., by a factor of 10.
However, in the case of the approximate method, we did not get
any significant improvement in terms of the median error,
which already is close to the accuracy limits of computers' floating
point arithmetic.
Yet, the 99\% quantile is now 10 times lower
and we detected only 1 outlier case (instead of 11 -- out of 10000) in which the
relative error is greater than 1\%.
The 10-fold procedure failed to converge 76 (instead of 825) times
within the presumed \texttt{reltol} and \texttt{maxit} settings -- in such
circumstances one may try to rerun the BFGS algorithm
from different random initial points until convergence criteria are satisfied.
\end{remark}

\subsection{Fitting weighted power means}

Up to now we studied a case of weighted power means
where $\varphi$ was fixed (e.g., to an identity function
which lead to the weighted arithmetic means).
Let us now assume that we have a suspicion that $\varphi$ might be
an instance of some parametrized class of functions, e.g.,
$\varphi(x)=x^p, p>1$. In other words, we are interested
in fitting weighted power means to data.

In the case of the squared error, our task now becomes a bi-level optimization
problem:
\[
   \mathrm{minimize}\ E(p) \quad \text{w.r.t.~}p
\]
subject to $p\in[p_\mathrm{min}, p_\mathrm{max}]$
where $E(p)$ is a solution to:
\[
   \mathrm{minimize}\  \sum_{j=1}^m \left(
   \sqrt[p]{\sum_{i=1}^n w_i \left(x_i^{(j)}\right)^p} - y^{(j)}
   \right)^2
      \quad \text{w.r.t.~}\vect{w}
\]
subject to $\vect{w}\ge \vect{0}$, $\vect{1}^T \vect{w}=1$.
On a side note, Beliakov in \cite{Beliakov2005:learnweightgowa}
and Torra \cite{Torra2002:learningweightsqwam} consider a similar
problem, however using the linearization technique.
The $L_1$ error may be incorporated accordingly.
Note that most often we observe that $E$ is a quite well-behaving,
unimodal function, therefore one-dimensional nonlinear solvers
(like the Brent method \cite{Brent1973:minim}) may be utilized.

\begin{figure}[htb!]
\centering

\includegraphics[width=8.25cm]{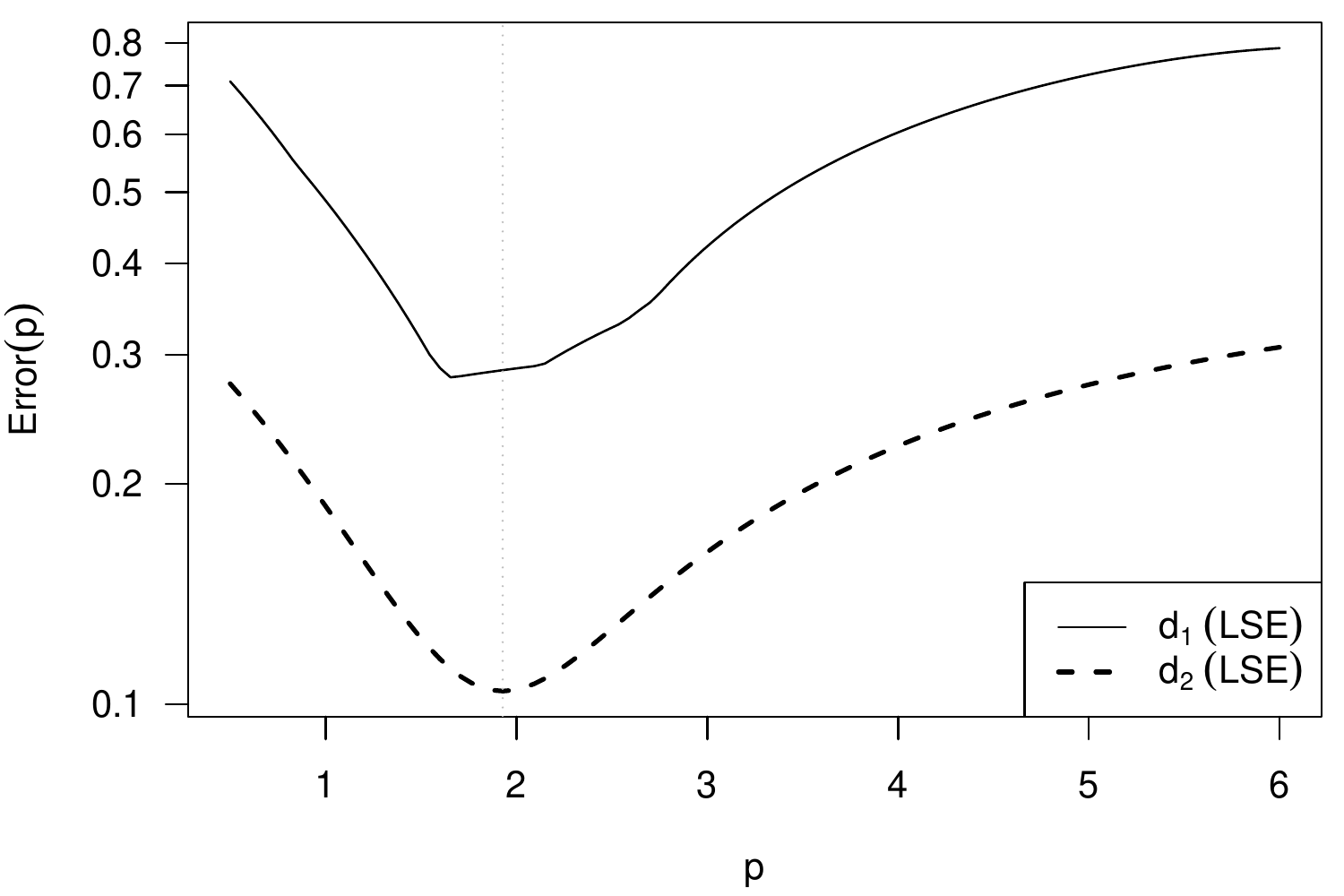}

\caption{\label{Fig:fit_powmean_error_p} Approximation error ($L_1$ and $L_2$)
as a function of $p$, see Example~\ref{Ex:fit_powmean_error_p}.}
\end{figure}

\begin{example}\label{Ex:fit_powmean_error_p}
Let us consider the data set generated as follows:
\begin{lstlisting}[language=R]
set.seed(132)
n <- 2
m <- 9
X <- t(matrix(runif(n*m, 0, 1), nrow=m))
p <- 2
realw <- runif(n)
realw <- realw/sum(realw)
Y <- matrix(as.numeric((t(X^p) %
     rt(m, 5)*0.05, ncol=m)
\end{lstlisting}

For least squares fitting we use the function in the mentioned Figure~\ref{Fig:fit_powmean_L2_optim}.
which relies on the already discussed solver for an optimization task
given by Equation~\eqref{Eq:LSEWQAM}.
The obtained $L_1$ and $L_2$ errors as a function of $p$
are depicted in Figure~\ref{Fig:fit_powmean_error_p}.
Here, the minimum was obtained for $p^* = 1.928388$,
giving the total $L_2$ error of $0.1041785$.
\end{example}

\subsection{Determining generator functions of quasi-arithmetic means}

What happens, however, if we would like to fit a weighted quasi-arithmetic
mean to empirical data but we have no knowledge on how a $\varphi$ generating
function might be defined?
In such a case, Beliakov et al.~suggest to rely on the notion of spline functions, see, e.g.,
\cite{BeliakovWarren2001:appropriatechoiceagop,%
BeliakovETAL2007:aggregationpractitioners,Beliakov2005:learnweightgowa,Beliakov2002:monotapproxlsqspline}.
Namely, we are now interested in a method that uses B-splines to construct the $\varphi$
generator functions that are the basis of weighted quasi-arithmetic means.

Suppose that  $p\ge 1$, $\Ival=[a,b]$ and let $\mathbf{t}=(t_1,\dots,t_k)$ be an
increasingly ordered \index{knot vector}\emph{knot vector} of length $k$ for some $k \ge 0$
such that $a<t_i<t_{i+1}<b$ for all $i\in[k]$.
For brevity of notation we assume that $t_i = a$ for $i<1$
and $t_i=b$ whenever $i>k$. Let us define \emph{B-spline basis functions} for $j=0,\dots,p$
and $\theta\in[a,b]$ recursively as:\index{B-spline basis functions}%
\begin{eqnarray*}
N_{i,j}^\vect{t}(\theta) & = &
\left\{
\begin{array}{ll}
   1 & \text{if } \theta\in [t_{i-1}, t_{i}[,\\
   0 & \text{otherwise},
\end{array}
\right.\quad(j=0)
\\
N_{i,j}^\vect{t}(\theta) & = &
\frac{\theta-t_{i-1}}{t_{i+j-1}-t_{i-1}} N_{i,j-1}^\vect{t}(\theta) +
\frac{t_{i+j}-\theta}{t_{i+j}-t_{i}} N_{i+1,j-1}^\vect{t}(\theta),\quad(j>0)
\end{eqnarray*}
with convention $\cdot/0=0$.

\begin{figure}[t!]
   \centering
   \includegraphics[width=5.5cm]{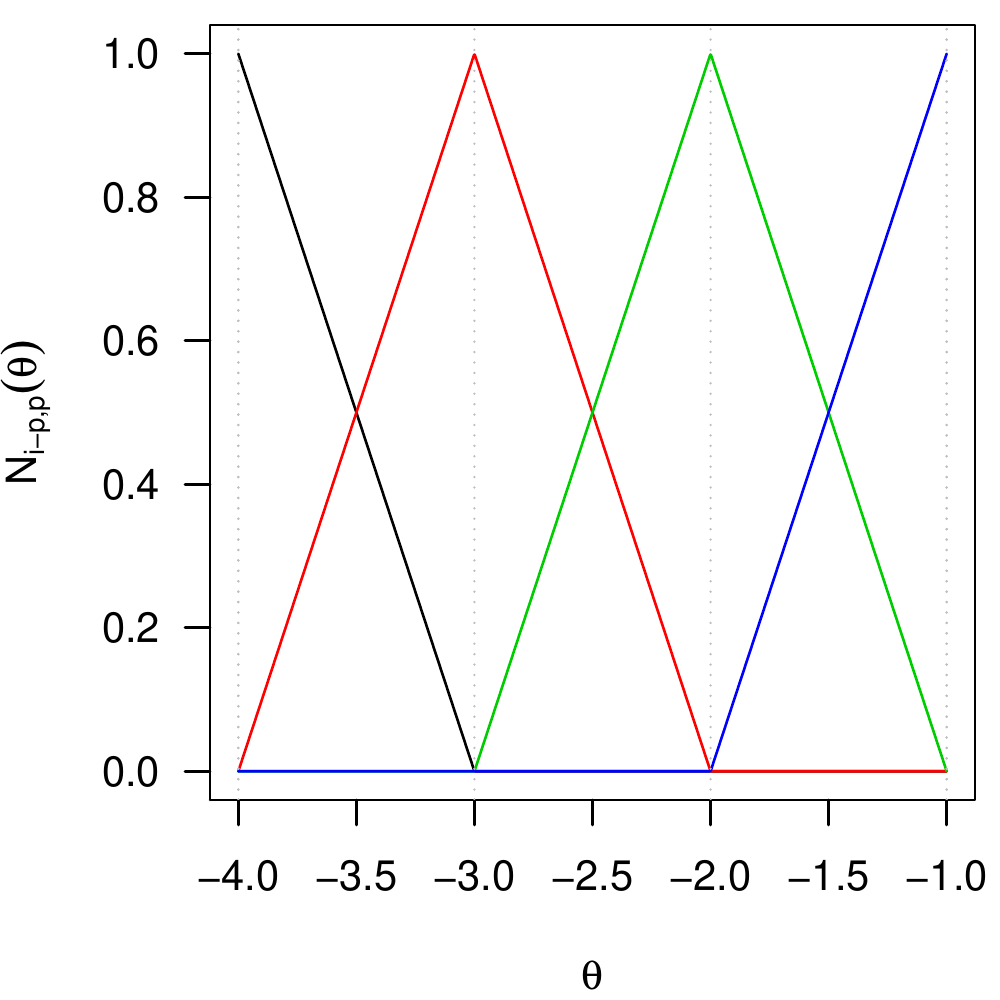}
   \includegraphics[width=5.5cm]{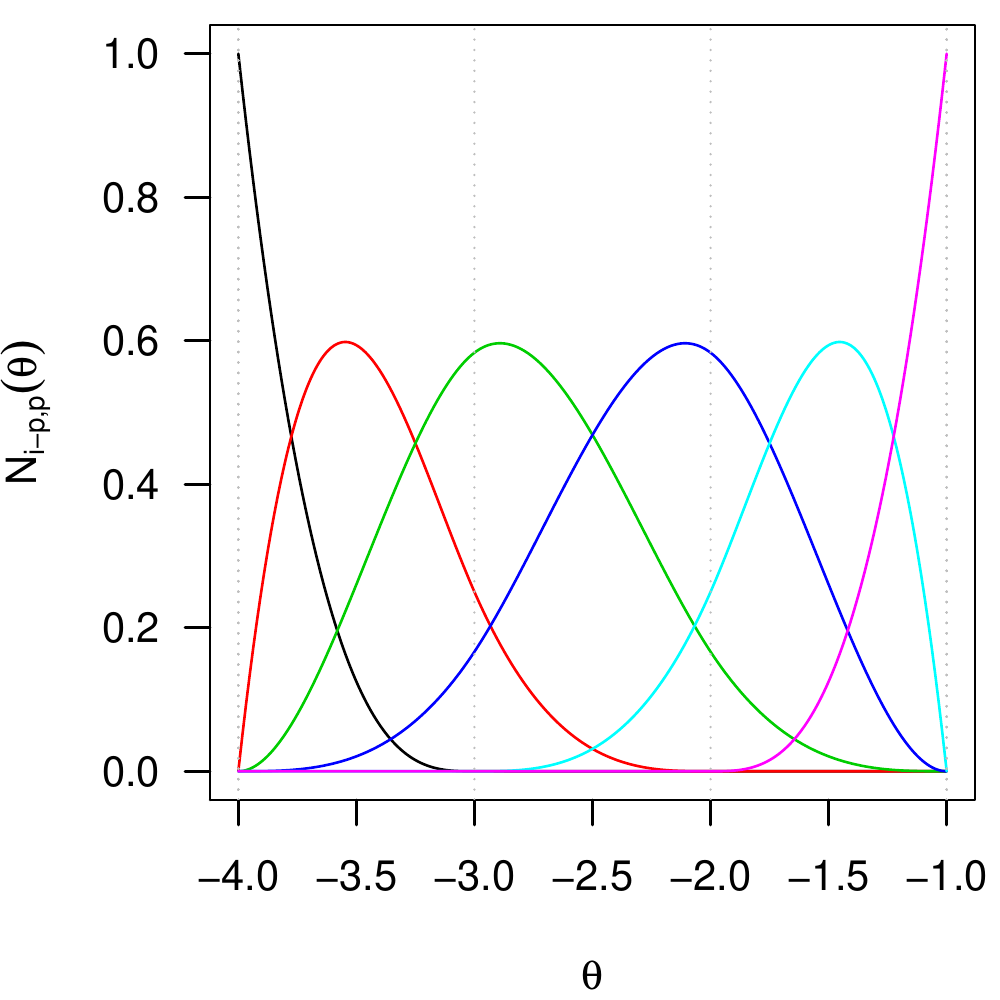}

   \caption[B-spline basis functions.]{\label{Fig:bspline_basis_2}  B-spline basis functions
   in the case of $k=2$ equidistant internal knots and $p=1$ (left) as well as $p=3$ (right).}
\end{figure}

\begin{example}
   Figure \ref{Fig:bspline_basis_2} depicts B-spline basis functions
   $N_{i-p,p}^\vect{t}$ for $i=1,\dots,p+k+1$
   in the case of $k=2$ equidistant internal knots and $p=1$ as well as $p=3$
   with $\Ival=[0,1]$.
   Note that for all $\theta\in\Ival$ it holds
   $\sum_{i=1}^{p+k+1} N_{i-p,p}^\vect{t}(\theta) = 1$.
\end{example}

Let $\vect{v}\in\IvalPow{\eta}$ be a vector of \index{nonperiodic B-spline}\emph{control points},
where $\eta=p+k+1$. Then ${B}_\vect{v}^\vect{t}:\Ival\to\Ival$ given by:
\begin{eqnarray}\label{Eq:Bspline}
B_\vect{v}^\vect{t}(\theta)=\sum_{i=1}^\eta v_i N_{i-p,p}^\vect{t}(\theta)
\end{eqnarray}
is a \emph{nonperiodic B-spline of degree $p$} based on a knot vector $\vect{t}$,
see, e.g., \cite{Schumaker2007:spline}.
In particular, for $p=1$ we get a piecewise linear function
interpolating $(a,v_1),(t_1,v_2),\dots,\allowbreak(t_k,v_{\eta-1}),\allowbreak(b,v_\eta)$.
On the other hand, for $p=3$ we get a cubic B-spline.

\begin{remark}
De Boor's algorithm (see, e.g., \cite{Schumaker2007:spline,Lee1982:bspline})
may be used to compute B-splines. In \R{}, this may be done using, for instance,
the \texttt{splineDesign()} function from the \package{splines} package.
\end{remark}

\begin{figure}[t!]
   \centering
   \includegraphics[width=5.5cm]{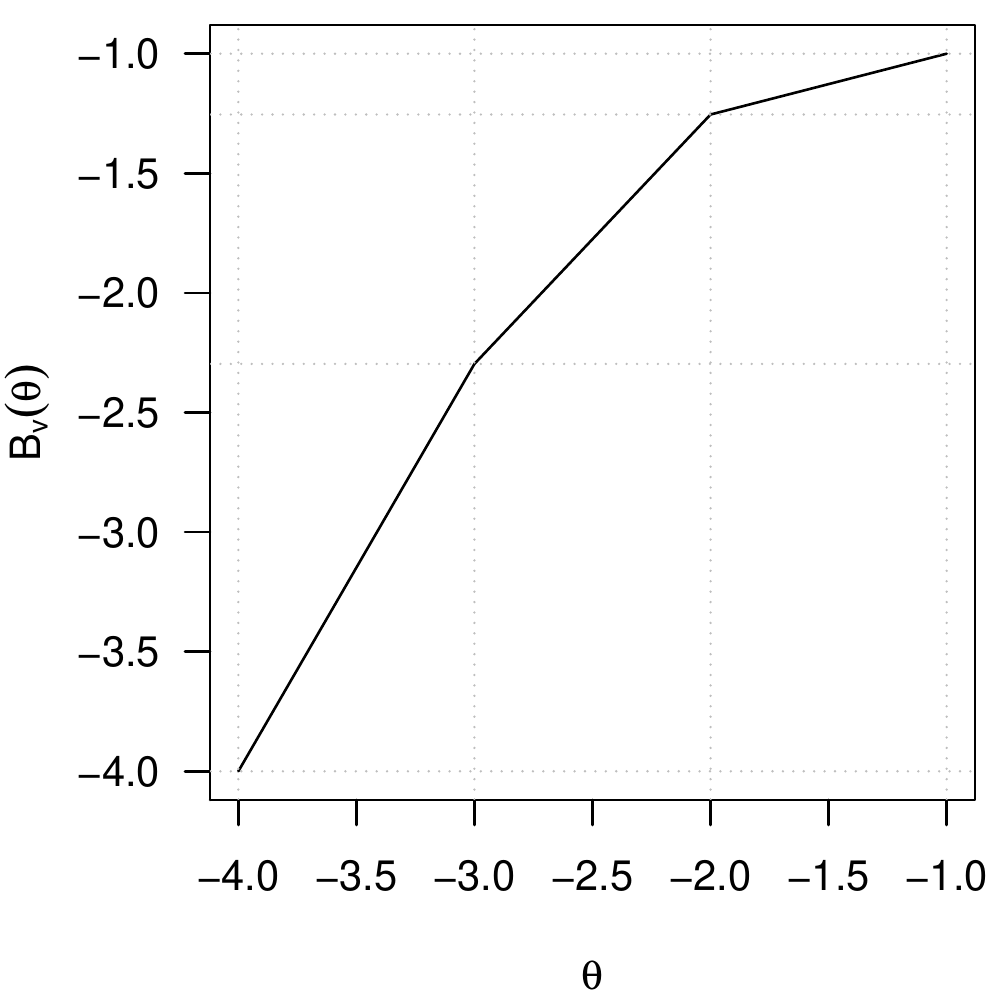}
   \includegraphics[width=5.5cm]{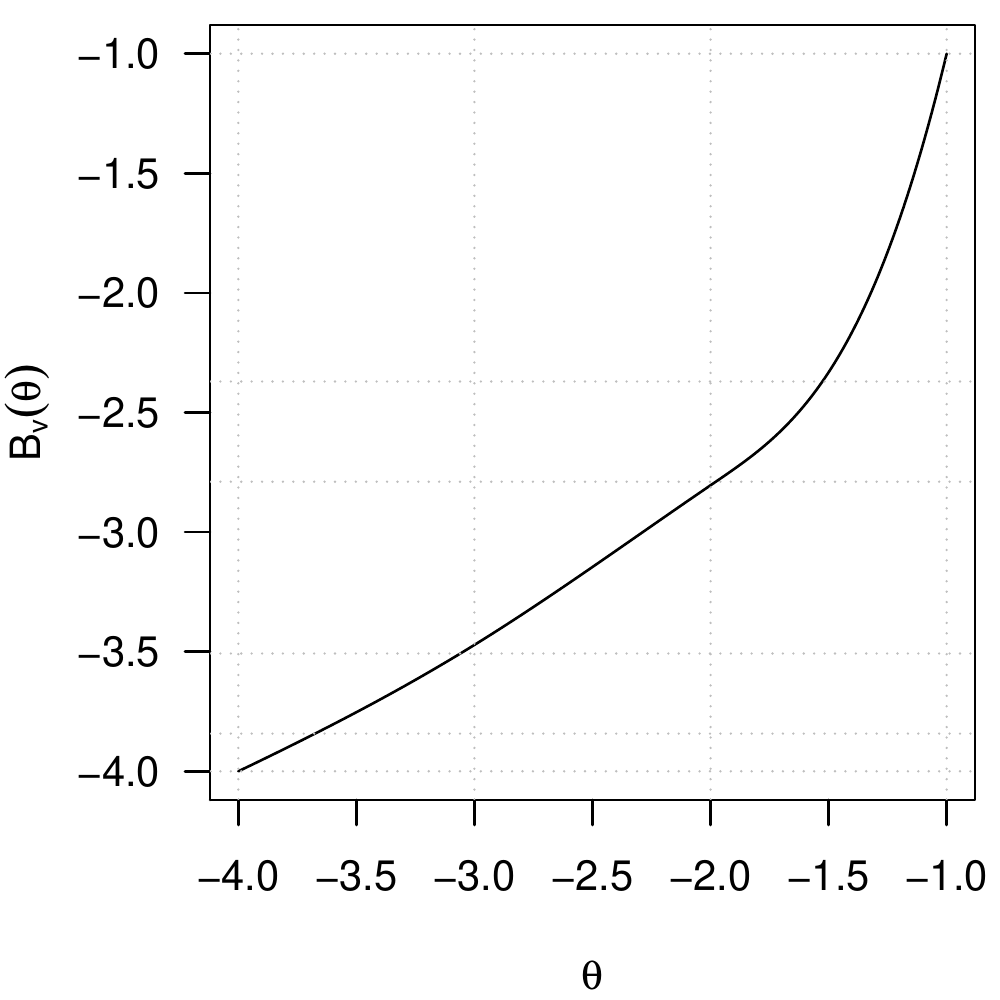}

   \caption[Exemplary B-splines.]%
   {\label{Fig:bspline_example_2} B-splines in the case of $k=2$ equidistant
   internal knots and $p=1$ (left, $\vect{v}=(0, 0.25, 0.8, 1)$)
   as well as $p=3$ (right, $\vect{v}=(0, 0.1, 0.15, 0.2, 0.95, 1)$).}
\end{figure}

\begin{example}
Figure \ref{Fig:bspline_example_2} depicts two exemplary
B-splines: a piecewise linear one and a cubic one; we assume $\Ival=[0,1]$.
\end{example}

\begin{remark}
The derivative of a B-spline of degree $p$ is itself a B-spline
of degree $p-1$. It might be easily shown that if $\vect{v}$ is ordered
increasingly, then its corresponding B-spline is strictly increasing.
It is worth noting that if $v_1=a$ and $v_\eta=b$, then $B_\vect{v}^\vect{t}$
is a function onto $\Ival$.
Of course, if $p=1$, then the inverse of an increasing B-spline
is a B-spline of degree $1$ (piecewise linear spline) as well.
However, to the best of our knowledge, for $p>1$ there are no analytic methods
to determine $(B_\vect{v}^\vect{t})^{-1}$. Yet, the inverse
may easily be computed numerically using, e.g., a root finding algorithm.
Also, it may be approximated with other B-splines.
\end{remark}

Assume that $\vect{t}$ is fixed (see, e.g., \cite{HeShi1998:monotonebspline}
and references therein for a discussion on knot selection) and that
$\varphi(x) = B_\vect{v}^\vect{t}(x)=\sum_{i=1}^\eta v_i N_{i-p,p}^\vect{t}(x)$
for some increasing $\vect{v}\in\IvalPow{\eta}$ such that $v_1=a$ and $v_\eta=b$.
If $\vect{w}$ is given a priori and we rely on the linearization technique
(see page~\pageref{Page:Linearization}), our $\func{WQAMean}$ fitting procedure
may be expressed as:
\[
   \textrm{minimize}\ \sum_{j=1}^m \left(
      \sum_{k=1}^\eta v_k u_{j,k}
   \right)^2
   \quad \text{w.r.t.~}(v_2,\dots,v_{\eta-1})
\]
in the case of the squared error, or:
\[
   \textrm{minimize}\ \sum_{j=1}^m \left|
      \sum_{k=1}^\eta v_k u_{j,k}
   \right|
   \quad \text{w.r.t.~}(v_2,\dots,v_{\eta-1})
\]
in the case of the absolute error,
subject to:
\begin{eqnarray*}
   v_2 &>& a\\
   v_i - v_{i-1} &>& 0 \text{ for } i=3,\dots,\eta-1\\
   v_{\eta-1} & < & b,\\
\end{eqnarray*}
where:
\[
   u_{j,k} = \sum_{i=1}^n w_i  N_{k-p,p}^\vect{t}(x_{i}^{(j)})
      - N_{k-p,p}^\vect{t}(y^{(j)}),
\]
see, e.g., \cite{BeliakovETAL2007:aggregationpractitioners,%
Beliakov2000:shapepreserspline,Beliakov2009:linearprogramming}.
If $\vect{w}$ is also unknown, then a two-stage optimization
procedure may be used, see, e.g., \cite{Beliakov2009:linearprogramming,Beliakov2005:learnweightgowa}.
Alternatively, one may rely on a ``global'' optimization routine
like CMA-ES \cite{Hansen2006:cmaes}. Note that assuring
that $v_1<v_2<\dots<v_\eta$ may be done via reparametrization:
one may use variables like $v_i'$ with boundary constraints on $v_i' > 0$
for $i=2,\dots,\eta$, where $v_i = \sum_{j=1}^n v_i'$.
In is also worth noting that Beliakov and James in \cite{BeliakovJames2012:linprogbonferroni}
also considered B-splines fitting in the case of a LAD task and Bonferroni means.

\subsection{A note on hierarchies of quasi-arithmetic means}

Recall that in Example~\ref{Ex:FFNN} we considered the case
of feedforward neural networks, which were isomorphic to a hierarchy
of quasi-arithmetic means.

It is well known that a neural network serves as a universal approximator:
for instance, many successful applications of such machine learning algorithms
were reported in classification problems.
To \textit{train} a neural network,
the Widrow-Hoff ``backpropagation'' (backward error propagation) algorithm, see, e.g., \cite{WidrowWinter1998:nnet}
may be used (among others) -- it is based on stochastic gradient descent techniques;
the updating algorithm is applied until weights no longer change
significantly under the mean square error  minimization criterion.

\section{Aggregation on bounded posets}\label{Sec:ProdLat}

It turns out that in some intelligent systems and other
applications, elements we aggregate
are non-numeric or although they are represented as numbers,
albeit cannot be treated as being defined on the so-called \textit{interval scale}.
In such a context operations like $+,-,\cdot,/$, as well as
$\sqrt{\cdot}$, $\exp\cdot$, $\sin\cdot$ may not be meaningful at all.

In this section, we relax our (strong up to now) assumptions on the input domain
and suppose that the aggregated elements may only be somehow ordered.
This is the case of, for example, linguistic information:
values of some attributes may be represented as labels like ``low'',
``medium'', ``high'' or ``bad'', ``good'', ``excellent'', etc.,
compare also the Zadeh \textit{computing with words} methodology
\cite{Zadeh1996:computingwithwords}.
It is clear that here statements like ``3$\cdot$bad'' or ``warm+10'' make
no sense. This implies that most of the previously defined data fusion techniques,
e.g., OWA and weighted averaging, must be replaced with some more elaborated
solutions.

\subsection{Basic order theory concepts}

Assume that elements we aggregate come from a set $P$ (possibly uncountable)
and a preordering relation has  been established.
Recall that a \index{preorder}\emph{preorder} over $P$ is a binary relation $\sqsubseteq\, \subseteq P\times P$
which is:
\begin{enumerate}
   \item[(a)] \index{reflexivity}\emph{reflexive}, i.e., $(\forall p\in P)$ it holds $p\sqsubseteq p$,
   \item[(b)] \index{transitivity}\emph{transitive}, i.e., $(\forall p,q,r\in P)$
   $p\sqsubseteq q$ and $q\sqsubseteq r$ $\Longrightarrow$   $p\sqsubseteq r$.
\end{enumerate}
A set $P$ equipped with a preorder $\sqsubseteq$, i.e., $(P, \sqsubseteq)$,
is called a \emph{preordered set}.

Moreover, any \index{antisymmetry}\emph{antisymmetric} preorder $\sqsubseteq$, that is, a binary relation
such that $(\forall p,q\in P)$ if $p\sqsubseteq q$ and $q \sqsubseteq p$,
then $p=q$, is called a \index{partial order}\emph{partial order}
and then $(P,\sqsubseteq)$ is called a \index{poset}\emph{poset} (partially ordered set).
In such a case, we  sometimes write $p\sqsubset q$ to indicate the
fact that $p\sqsubseteq q$ and $p\neq q$.

\begin{example}\label{Ex:brfw}
Let $P=\{ \text{beautiful}, \text{rich}, \text{famous}, \text{wise} \}$.
A decision maker introduces the following partial order $\sqsubseteq$ over $P$,
expressing his/her ``life desires'':
\[
   \sqsubseteq\,=\left\{\begin{array}{l}
      (\text{beautiful}, \text{beautiful}),
      (\text{rich}, \text{rich}),
      (\text{famous}, \text{famous}),
      (\text{wise}, \text{wise}),\\
      (\text{beautiful}, \text{rich}),
      (\text{beautiful}, \text{famous}),
      (\text{rich}, \text{wise}),
      (\text{famous}, \text{wise}),\\
      (\text{beautiful}, \text{wise}).
   \end{array}
   \right\},
\]
Note that, actually, the pairs in the second row above are the most
``informative''. The elements in the first row are implied
by reflexivity and in the third row -- by transitivity.
Also please notice that rich and famous are not comparable
with~$\sqsubseteq$.
\end{example}

\begin{figure}[b!]
\centering
\begin{tikzpicture}
\node (A) at (0,0) { wise };
\node (B) at (-2,-2) { rich };
\node (C) at (2,-2) { famous };
\node (D) at (0,-4) { beautiful };
\draw[->] (D)--(C);
\draw[->] (D)--(B);
\draw[->] (B)--(A);
\draw[->] (C)--(A);
\end{tikzpicture}

\caption{\label{Fig:HasseBeautiful} An illustration for Example \ref{Ex:HasseBeautiful}.}
\end{figure}
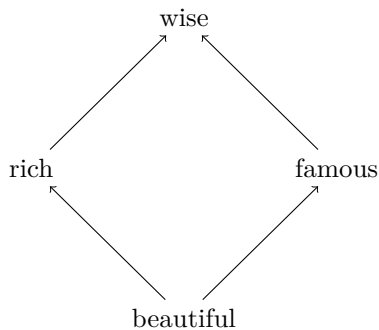

\begin{remark}\label{Ex:HasseBeautiful}
If $P$ is finite, then
from the formal (syntactic) perspective
each preordered set may be represented as a directed graph
(there is a one-to-one correspondence between directed graphs and binary relations).
Thus, instead of writing $p\sqsubseteq q$ we may presume that there is an edge
from $p$ to $q$, where $p,q\in P$.
A simplified version of the poset (directed graph) from
Example \ref{Ex:brfw} may be depicted as in Figure~\ref{Fig:HasseBeautiful}.

What we see there is a \emph{Hasse diagram}.
An arrow (edge) from $p$ to $q$, $p,q\in P$, means that
$p\sqsubseteq q$. Loops, i.e., edges from each $p$ to $p$ itself,
are not included in the diagram for readability.
Moreover, please notice that edges implied by transitivity
are also hidden. In other words, an ordering relation
may be obtained from a Hasse diagram by calculating its
reflexive and transitive closure.
Also please observe that, e.g., the Warshall algorithm \cite{Warshall1962:transitiveclosure}
may be used to find a transitive closure of a graph represented as a 0-1 matrix
in $O(|P|^3)$ time.
The opposite operation, transitive reduction,
may be obtained by a method by Aho, Garey, and Ullman \cite{AhoGareyUllman1972:transitivereduction},
who additionally showed that this problem is of the same computational
complexity as that of finding the corresponding closure.
It might be shown that both tasks may be efficiently solved via
binary matrix multiplication, i.e., in at most $O(|P|^{2.3728639})$-time \cite{LeGall2014:matrixmul}.
\end{remark}

Additionally, a \index{total order}\emph{total} partial order $\sqsubseteq$, i.e.,
such that $(\forall p,q\in P)$ it holds $p\sqsubseteq q$ or $q\sqsubseteq p$,
is called a \index{linear order}\emph{linear order}.

\begin{example}\label{Ex:tinyhuge}
Let $P=\{ \text{tiny}, \text{small}, \text{normal}, \text{large}, \text{huge} \}$
denote the set of T\hspace*{-0.1em}${}_\mathrm{E}$\hspace*{-0.1em}X font sizes.
We may establish a linear order $\sqsubseteq$ over $P$ with
the Hasse diagram below:
\[
   \text{\tiny{tiny}} \longrightarrow
   \text{\small{small}} \longrightarrow
   \text{{normal}} \longrightarrow
   \text{\large{large}} \longrightarrow
   \text{\huge{huge}}.
\]
By transitivity, we of course have
$\text{\small{small}} \sqsubset   \text{\large{large}}$, etc.
\end{example}

\begin{remark}
If $(P,\sqsubseteq)$ is a finite chain, then
it may be represented as a real interval $I$ (with standard ordering of reals)
by means of an order-preserving \emph{utility function}
$f:P\to I$, which is defined up to a strictly increasing bijection
$\varphi: I\to I$, see, e.g., \cite{Marichal2002:orderinvsynth,MarichalMesiar2004:agfinordscal} for discussion.
For instance, in Example \ref{Ex:tinyhuge}
$f$ may be such that
$f(\mathrm{tiny})=1$,
$f(\mathrm{small})=2$,
$f(\mathrm{normal})=3$,
$f(\mathrm{large})=4$,
$f(\mathrm{huge})=5$.
\end{remark}

Given a poset $(P, \sqsubseteq)$,
if there exists $\underline{0}\in P$ for which $(\forall p\in P)$
it holds $\underline{0}\sqsubseteq p$, then we call such
$\underline{0}$ \index{least element}\emph{the least element} of $P$.
Similarly, \index{greatest element}\emph{the greatest element} of $P$
is defined as $\overline{1}\in P$ such that $(\forall p\in P)$
we have $p\sqsubseteq\overline{1}$ (if it exists).
$(P, \sqsubseteq, \underline{0}, \overline{1})$ is called a \index{bounded poset}\emph{bounded poset},
if the poset $(P, \sqsubseteq)$ has the least element $\underline{0}$
and the greatest element $\overline{1}$.

\begin{example}
In Example \ref{Ex:brfw} we presented a bounded poset
with $\underline{0}=\text{beautiful}$ and $\overline{1}=\text{wise}$.
\end{example}

A \index{lattice}\emph{lattice} $(P,\sqsubseteq,\sqcap,\sqcup)$
is a poset in which every pair of elements
has a unique \index{meet}infimum (meet, $\sqcap$,
the greatest element of common lower bounds)
and supremum \index{join}(join, $\sqcup$,
the smallest element of common upper bounds).
If $\sqsubseteq$ is a linear order, then a lattice
is called a \index{chain}\emph{chain}.

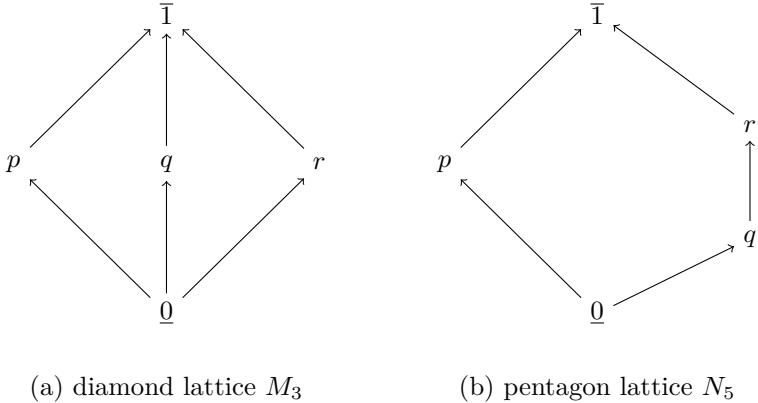
\begin{figure}[htb!]
\centering
\begin{tikzpicture}
\node (A) at (0,0) { $\overline{1}$ };
\node (B) at (-2,-2) { $p$ };
\node (C) at (0,-2) { $q$ };
\node (D) at (2,-2) { $r$ };
\node (E) at (0,-4) { $\underline{0}$ };
\draw[->] (E)--(B);
\draw[->] (E)--(C);
\draw[->] (E)--(D);
\draw[->] (B)--(A);
\draw[->] (C)--(A);
\draw[->] (D)--(A);

\node at (0,-5) { (a) diamond lattice $M_3$ };
\end{tikzpicture}
\hspace*{1cm}
\begin{tikzpicture}
\node (A) at (0,0) { $\overline{1}$ };
\node (B) at (-2,-2) { $p$ };
\node (C) at (2,-3) { $q$ };
\node (D) at (2,-1.5) { $r$ };
\node (E) at (0,-4) { $\underline{0}$ };
\draw[->] (E)--(B);
\draw[->] (E)--(C);
\draw[->] (C)--(D);
\draw[->] (B)--(A);
\draw[->] (D)--(A);

\node at (0,-5) { (b) pentagon lattice $N_5$ };
\end{tikzpicture}

\caption{\label{Fig:NonDistributiveLattice} The two simplest non-distributive lattices.}
\end{figure}

A lattice $(P,\sqsubseteq,\sqcap,\sqcup)$ is called \index{distributive lattice}\emph{distributive}
whenever for all $p,q,r\in P$
\begin{equation}
p\sqcup(q\sqcap r) = (p\sqcup q)\sqcap(p\sqcup r)
\end{equation}
or, equivalently,
\begin{equation}
p\sqcap(q\sqcup r) = (p\sqcap q)\sqcup(p\sqcap r),
\end{equation}
which is exactly the same as requiring:
\begin{equation}
p\sqcap r= q\sqcap r \text{ and } p\sqcup r= q\sqcup r \Longrightarrow p = q.
\end{equation}
Additionally, it may be shown that a lattice is distributive if and only if
none of its sublattices is isomorphic to any of the two simplest
non-distributive lattices depicted in Figure~\ref{Fig:NonDistributiveLattice}.

Moreover, we call a lattice $(P,\sqsubseteq,\sqcap,\sqcup)$
\index{complete lattice}\emph{complete}, whenever every subset $P'\subseteq P$
has a unique supremum (denoted with $\bigsqcup P'=\bigsqcup_{p'\in P'} p'$) and infimum ($\bigsqcap P'$).
Clearly, every complete lattice is bounded.

\subsection{Aggregation functions on bounded posets}

We have established the three most common scenarios,
from the most to the least general:
\begin{enumerate}
   \item bounded posets,
   \item bounded lattices,
   \item chains.
\end{enumerate}

\index{fusion function}%
By the term \emph{fusion function} we now mean any mapping $\func{F}^{(n)}:P^n\to P$.
In order to reintroduce the concept of an aggregation function,
this time in a bounded poset setting, we follow the definition
given, e.g., in \cite{Demirci2006:agposets}.
\begin{definition}\label{Def:AgFunLattice}
Let $(P,\sqsubseteq,\underline{0},\overline{1})$ be a bounded poset.
\index{aggregation function (on bounded posets)}
A mapping $\func{F}^{(n)}: P^n\to P$, is called an \emph{aggregation function} if:
\begin{enumerate}
   \item[(a)] $(\forall \mathbf{x},\mathbf{y}\in P^n)$
   if $(\forall i\in[n])$ $x_i\sqsubseteq y_i$, then
   $\func{F}^{(n)}(\mathbf{x})\sqsubseteq \func{F}^{(n)}(\mathbf{y})$,
   \item[(b)] $\func{F}^{(n)}(n\ast\underline{0})=\underline{0}$, \hfill(lower boundary condition)
   \item[(c)] $\func{F}^{(n)}(n\ast\overline{1})=\overline{1}$. \hfill(upper boundary condition)
\end{enumerate}
\end{definition}

\begin{remark}
Using the introduced notion, we may define
extended aggregation functions like $\func{F}^*: P^*\to P$
by assuming that for all $n$ the restriction $\func{F}^{*}|_{P^n}$ is an aggregation function.
\end{remark}

\begin{remark}%
Let $\Ival=[a,b]$ and $\le$ denote the standard ordering of reals.
In the case of the bounded chain $(\Ival, \le, a, b, \wedge, \vee)$,
the above definition coincides with the classical
one as given in \cite{GrabischETAL2009:aggregationfunctions}.
An example of such an aggregation function is the sample minimum.
On the other hand, the  arithmetic mean cannot be
given as an instance of this class, as in its definition some
``illegal'' arithmetic operations occur.
\end{remark}

Komorn\'{i}kov\'{a} and Mesiar note in \cite{KomornikovaMesiar2011:agbndposet}
that some properties of ``ordinary'' aggregation functions
may be straightforwardly transformed to the case
of fusion functions on bounded posets. This happens, e.g., in the case of:
\begin{itemize}
   \item symmetry (see \index{symmetry}Definition~\ref{Def:symmetry}),
   \item idempotency (\index{idempotency}Definition~\ref{Def:idempotency}),
   \item associativity (\index{associativity}Definition~\ref{Def:associativity}),
   \item decomposability (\index{decomposability}Definition~\ref{Def:decomposability}),
   \item bisymmetry (\index{bisymmetry}Definition~\ref{Def:bisymmetry}),
   \item annihilator (Definition~\ref{Def:annihilatorelement})
   and neutral (Definition~\ref{Def:neutralelement}) element.
\end{itemize}
\index{internality} Internality is sometimes defined as
(see \cite{KomornikovaMesiar2011:agbndposet,Ovchinnikov1996:meanordset,Ovchinnikov1998:invfunsmpord}):
\begin{equation}
\func{F}^{(n)}(x_1,\dots,x_n)\in\{x_1,\dots,x_n\}
\end{equation}
or, if we act on a complete lattice, alternatively as  (see \cite{Ovchinnikov1996:meanordset})\label{Internality2}:
\begin{equation}
\bigsqcap_{i=1}^n x_i \sqsubseteq \func{F}^{(n)}(x_1,\dots,x_n) \sqsubseteq
\bigsqcup_{i=1}^n x_i.
\end{equation}
In this regard, the former one naturally arises when we require that a fusion
function is comparison meaningful (preserves relative output order under any inputs' order automorphism)
and the latter one stands for a basis of \emph{means in the Ovchinnikov sense}.

Both cases may lead to undesired consequences
if the aggregated elements are incomparable.
Thus, in the following section we  review an appealing proposal
on how to solve this issue.
As a side effect, we  also present a categorization
of aggregation functions on bounded posets.

\subsection{Classes of fusion functions}\label{Sec:AgLatClass}

Recall that we distinguished four main classes of nondecreasing fusion functions
(see \cite{DuboisPrade2004:useagopfusion}):
\begin{itemize}
   \item internal (averaging),
   \item conjunctive (\texttt{AND}-like, e.g., t-norms),
   \item disjunctive (\texttt{OR}-like, e.g., t-conorms),
   \item mixed.
\end{itemize}
On the interval scale, the distinction was based on their relationship
to $\func{Min}$ and $\func{Max}$. If we act on chains,
we may replace $\func{Min}$ and $\func{Max}$ with $\inf$ and $\sup$,
respectively. Yet, on more general (bounded) posets the situation is
somehow more complicated.

At this point, let us follow the classification proposed
by Komorn\'{i}kov\'{a} and Mesiar
in \cite{KomornikovaMesiar2011:agbndposet},
which was inspired by the notion of $k$-intolerance introduced in \cite{Marichal2007:kintolerance}.
Given an aggregation function $\func{F}^{(n)}: P^n\to P$,
let:
\begin{eqnarray}
\gamma(\func{F}^{(n)}) &=& \inf\left\{
\left|i: \func{F}^{(n)}(\vect{x})\sqsubseteq x_i\right|:
\vect{x}\in P^n
\right\},\\
\sigma(\func{F}^{(n)}) &=& \inf\left\{
\left|i: x_i \sqsubseteq \func{F}^{(n)}(\vect{x}) \right|:
\vect{x}\in P^n
\right\}.
\end{eqnarray}

\begin{definition}[\cite{KomornikovaMesiar2011:agbndposet}]%
\index{strong conjunctivity}%
\index{strong disjunctivity}%
We call $\func{F}^{(n)}: P^n\to P$ \emph{strongly conjunctive},
whenever $\func{F}^{(n)}\in\gamma^{-1}(n)$.
What is more, $\func{F}^{(n)}: P^n\to P$ is \emph{strongly disjunctive},
if $\func{F}^{(n)}\in\sigma^{-1}(n)$.
\end{definition}

\begin{remark}
In other words, $\func{F}^{(n)}$ is strongly conjunctive,
if for all $\vect{x}\in P^n$ and all $i\in[n]$ it holds
$\func{F}^{(n)}(\vect{x})\sqsubseteq x_i$.
Recall that on an interval scale
we called $\func{F}^{(n)}$ conjunctive, whenever
for all $\vect{x}$ it held $\func{F}^{(n)}(\vect{x})\le\func{Min}(\vect{x})$,
or equivalently $\func{F}^{(n)}(\vect{x})\le x_i$
for all $i\in[n]$.
\end{remark}

\begin{definition}[\cite{KomornikovaMesiar2011:agbndposet}]%
\index{weak conjunctivity}%
\index{weak disjunctivity}%
We call $\func{F}^{(n)}: P^n\to P$ \emph{weakly conjunctive},
whenever $\func{F}^{(n)}\in\bigcup_{i=1}^{n-1}\gamma^{-1}(i)$.
Moreover, we say that $\func{F}^{(n)}: P^n\to P$  is \emph{weakly disjunctive},
if $\func{F}^{(n)}\in\bigcup_{i=1}^{n-1}\sigma^{-1}(i)$.
\end{definition}

Based on the above notion, we may introduce the concept
of an averaging aggregation function.

\begin{definition}[\cite{KomornikovaMesiar2011:agbndposet}]%
\index{strongly averaging function}%
\index{weakly averaging function}%
We call $\func{F}^{(n)}: P^n\to P$ \emph{weakly averaging},
whenever it is weakly conjunctive or weakly disjunctive.
Moreover, it is \emph{strongly averaging}
if it is both weakly conjunctive and weakly disjunctive.
\end{definition}

Intuitively, a weakly averaging function outputs values that are
greater than or less than some elements we aggregate, but definitely
not greater than or less than all such elements in every possible case.
Each weakly/strongly averaging function is idempotent.
Moreover, if $P$ is a bounded lattice, the only strongly conjunctive
(disjunctive) and idempotent aggregation function is the $\func{Min}$
(respectively, $\func{Max}$).

\medskip
Having said that, aggregation functions on bounded posets may be classified~as:
\begin{enumerate}
   \item[(a)] weakly averaging ($\bigcup_{i=1}^{n-1}\gamma^{-1}(i)\cup\bigcup_{i=1}^{n-1}\sigma^{-1}(i)$),
   \item[(b)] strongly conjunctive ($\gamma^{-1}(n)$),
   \item[(c)] strongly disjunctive ($\sigma^{-1}(n)$),
   \item[(d)] mixed ($\gamma^{-1}(0)\cap\sigma^{-1}(0)$).
\end{enumerate}
This is what was called in \cite{KomornikovaMesiar2011:agbndposet}
a weak classification. Its strong version assumes that
the class of weakly averaging functions may additionally be
considered as consisting of aggregation functions that are either:
\begin{enumerate}
   \item[(a$\myprime$)] strongly averaging ($\bigcup_{i=1}^{n-1}\gamma^{-1}(i)\cap\bigcup_{i=1}^{n-1}\sigma^{-1}(i)$),
   \item[(a$\mydprime$)] weakly conjunctive but not weakly disjunctive\newline
   ($\bigcup_{i=1}^{n-1}\gamma^{-1}(i)\setminus\bigcup_{i=1}^{n-1}\sigma^{-1}(i)$),
   \item[(a$\mytrprime$)] weakly disjunctive but not weakly conjunctive\newline
   ($\bigcup_{i=1}^{n-1}\sigma^{-1}(i)\setminus\bigcup_{i=1}^{n-1}\gamma^{-1}(i)$).
\end{enumerate}
Both classification schemes are \textit{complete}
in the sense that any function falls exactly into one category.

\begin{remark}
If we act on a bounded chain, then the weak and strong
classification in the
Komorn\'{i}kov\'{a}-Mesiar  \cite{KomornikovaMesiar2011:agbndposet} sense
and the ``classical'' Dubois-Prade classification \cite{DuboisPrade2004:useagopfusion}
are equivalent.
\end{remark}

\subsection{Idempotent fusion functions}

\index{t-norm}\index{t-conorm}\index{uninorm}%
Please note that triangular norms and conorms on a bounded poset $(P, \sqsubseteq,\allowbreak \underline{0}, \overline{1})$
may be defined via a straightforward generalization of the case presented
in Definitions~\ref{Def:tnorm} and~\ref{Def:cotnorm},
see \cite{DeCoomanKerre1994:ordnorm,DeBaetsMesiar1999:trinormprodlat}.
This is because none of their sine qua non properties
are  specific to $[0,1]$ and a natural linear order $\le$.
Moreover, e.g., Kara\c{c}al and Mesiar \cite{KaracalMesiar2015:uninormbndlat}
studied uninorms on bounded lattices.

Nevertheless, our main focus in this book is on the study of
fusion functions that are at least idempotent. Various authors also
translate some well-known classical averaging aggregation functions to the
framework of aggregation on posets. Here are a few examples.

\begin{example}
Let $P=\{p_1,\dots,p_k\}$ be a finite set equipped with a total ordering relation $\sqsubseteq$
and assume that $p_1\sqsubseteq\dots\sqsubseteq p_k$. Moreover, let $\vect{w}$ be a weighting
vector of length $n$ and suppose that $w_i$ is a weight corresponding to $p_i$.
Then the Yager \cite{Yager1998:ordinffusion} \emph{weighted median}
is defined as $p_l$ with the smallest possible $l$ such that
$\sum_{i=1}^l w_i\ge 0.5$.
Input data of this kind naturally occur when elements come from
a multiset over a totally ordered set (see Example~\ref{Ex:multiset}).
It is easily seen that the input median is idempotent,
averaging, symmetric, and monotone. Noteworthily,
an iterative algorithm for weights fitting was also provided
in this case, see \cite[Section~4]{Yager1998:ordinffusion}.
\end{example}

\begin{example}
Let $P=\{p_0,p_1,\dots,p_k\}$ be a finite set equipped with
a total ordering relation $\sqsubseteq$ and suppose $p_0\sqsubseteq\dots\sqsubseteq p_n$.
The \emph{linguistic OWA operator} introduced by Herrera,
Herrera-Viedma, and Verdegay in \cite{HerreraETAL1996:gdmlingowa}
is generated by a weighting vector $\vect{w}$
and is defined for a given $\vect{x}\in P^n$, $n\ge 2$ as follows.
Assume that $\sigma\in\mathfrak{S}_{[n]}$ is such that
$x_{\sigma(1)}\sqsubseteq\dots\sqsubseteq x_{\sigma(n)}$. Then:
\[
   \func{LOWA}_\vect{w}(\vect{x})=C^n(\vect{w},\vect{x}),
\]
where the ``convex combination'' of elements in $P$ operator \cite{DelgadoETAL1993:agoplinglab}
$C^n$ is defined for $n>2$ recursively as:
\begin{eqnarray*}
   &&C^n(\vect{w},\vect{x})=
   C^2\Bigg(
      (1-w_n,w_n),\\
      &&
      \left(C^{n-1}\left(
         \left(\frac{w_1}{1-w_n},\dots,\frac{w_{n-1}}{1-w_n}\right),
         \left(x_{\sigma(1)},\dots,x_{\sigma(n-1)}\right)
      \right), x_{\sigma(n)}\right)
   \Bigg),
\end{eqnarray*}
and for $n=2$ -- under assumption $k\ge j\ge i\ge 0$ and $w+w'=1$ -- as:
\[
   C^2((w, w'), (p_j, p_i))=p_{k\wedge\left(
      i+\mathrm{round}\left(w(j-i)\right)
   \right)}.
\]
We see that in fact we map elements in $P$ to the set of nonnegative integers.
It can be shown that the linguistic OWA operator is
monotonic, averaging, idempotent, and symmetric.
The above idea was enhanced by Godo and Torra
\cite{GodoTorra2000:agopord} who introduced the so-called qualitative
OWA operators. Such operators utilize the notion of the t-norm instead
of the $C$ function. Moreover, Koles\'{a}rov\'{a}, Mayor, and Mesiar
in \cite{KolesarovaETAL2007:weighordmean}
study a different approach for constructing weighted ordinal means
based on divisible discrete t-norms.
\end{example}

\begin{example}
Lizasoain and Moreno \cite{LizasoainMoreno2013:owacompletelattices}
note that the original OWA operator for $\vect{x}\in[0,1]^n$:
\[
   \func{OWA}_\vect{w}(\vect{x})=\sum_{i=1}^n w_i x_{(i)}
\]
generated by a weighting vector $\vect{w}$  may be rewritten as:
\[
   \func{OWA}_\vect{w}(\vect{x})=\func{S}_\mathrm{Ł}\left(
      \func{T}_\mathrm{P}(w_1, x_{(1)}),\dots,\func{T}_\mathrm{P}(w_n, x_{(n)})
   \right)
\]
with assumption $\func{S}_\mathrm{Ł}(w_1,\dots,w_n)=1$,
where $\func{T}_\mathrm{P}$ and $\func{S}_\mathrm{Ł}$ denote the product t-norm
and Łukasiewicz t-conorm, respectively.
Assuming that we act on a complete lattice
and substituting arbitrary t-norms and t-conorms valid there
for $\func{T}_\mathrm{P}$ and $\func{S}_\mathrm{Ł}$,
we may introduce OWA-like lattice operators as long as
we are able to order the input observations. Of course, if we are on
a chain, this task is trivial.
In other cases, the authors propose to follow the approach of, e.g.,
Ovchinnikov \cite{Ovchinnikov1996:meanordset}, and compute an OWA operator on inputs
like $y_i$ (instead of $x_{(i)}$), where:
\begin{eqnarray*}
y_1 & = & x_1\sqcap\dots\sqcap x_n, \\
&\vdots&\\
y_i & = & \mathop{\bigsqcup}_{\{j_1,\dots,j_{n-i+1}\}\subseteq[n] } x_{j_1}\sqcap\dots\sqcap x_{j_{n-i+1}}, \\
&\vdots&\\
y_{n-1} & = & \mathop{\bigsqcup}_{\{j_1, j_2\}\subseteq[n]} x_{j_1}\sqcap x_{j_2},\\
y_n & = & x_1\sqcup\dots\sqcup x_n,
\end{eqnarray*}
which fulfill:
\[
   y_1 \sqsubseteq y_2 \sqsubseteq \dots \sqsubseteq y_n.
\]
It can be noted that if we are on a chain, then $y_i=x_{(i)}$.
All OWA-like lattice operators are idempotent.
\end{example}

\subsection{Lattice polynomial functions}\label{Sec:LatPolyFun}

Let us generalize the notion of a (weighted) lattice polynomial
function, see Equation~\eqref{Eq:WLPFIval}, to the case of
a complete distributive lattice $(P,\sqsubseteq,\allowbreak\sqcap,\sqcup,\underline{0},\overline{1})$.
Assume that
$\mathop{\bigsqcup}_{x\in\emptyset} x = \underline{0}$
and
$\mathop{\bigsqcap}_{x\in\emptyset} x = \overline{1}$.
Lattice polynomial functions are formed as expressions that consist of
variables in $P$ which are linked by the $\sqcap$, $\sqcup$ lattice operations
applied in any order, see \cite{Birkhoff1967:latticetheory}.

\begin{example}
Here is an exemplary lattice polynomial function of four variables:
$\func{F}^{(n)}(x_1, x_2, x_3, x_4)=(x_1\sqcap x_2)\sqcup (x_3\sqcap x_4).$
\end{example}

\begin{definition}\index{lattice polynomial function}%
The class of $n$-argument \emph{lattice polynomial functions} ($n$-LPF)
from $P^n$ to $P$ is defined by applying the following rules
finitely many times:
\begin{enumerate}
   \item[(a)] $\func{F}^{(n)}(x_1,\dots,x_n)=x_i$ is an $n$-LPF
   for any $i\in[n]$,
   \item[(b)] If $\func{F}^{(n)}$ and $\func{G}^{(n)}$ are $n$-LPFs,
   then $\func{F}^{(n)}\sqcap\func{G}^{(n)}$ and $\func{F}^{(n)}\sqcup\func{G}^{(n)}$
   are $n$-LPFs.
\end{enumerate}
\end{definition}

\begin{remark}
Each lattice polynomial function is nondecreasing with respect to $\sqsubseteq$.
\end{remark}

\begin{example}
A ternary median on a bounded distributive lattice is given by:
\begin{eqnarray*}
   \func{Median}^{(3)}(x_1,x_2,x_3) &=&
      (x_1\sqcap x_2)\sqcup(x_2\sqcap x_3)\sqcup(x_3\sqcap x_1)\\
   &=&(x_1\sqcup x_2)\sqcap(x_2\sqcup x_3)\sqcap(x_3\sqcup x_1).
\end{eqnarray*}
\end{example}

It turns out that each $n$-LPF may be written in a simpler form.
We have what follows, see \cite{Birkhoff1967:latticetheory}.

\begin{proposition}\label{Prop:lpfform}
Let $\func{F}^{(n)}$ be an $n$-LPF. Then there exist $k,l\ge 1$
and families $A_1,\dots,A_k,B_1,\dots,B_l$ of nonempty subsets of $[n]$
such that:
\begin{equation}
   \func{F}^{(n)}(x_1,\dots,x_n)=
   \mathop{\bigsqcup}_{j=1}^k \mathop{\bigsqcap}_{i\in A_j} x_i
  =\mathop{\bigsqcap}_{j=1}^l \mathop{\bigsqcup}_{i\in B_j} x_i.
\end{equation}
\end{proposition}

\begin{example}
Fix $t\in[n]$.
Let $k= {n\choose t}$
and $\mathcal{A}=\{A_1,\dots,A_k\}=\{ \{i_1,\dots,i_t\}\subseteq[n] \}$.
If we are on a chain, then for any $\vect{x}\in P^n$ it holds that
$
   x_{(n-t+1)}=\mathop{\bigsqcup}_{j=1}^k \mathop{\bigsqcap}_{i\in A_j} x_i,
$
i.e., the $t$th order statistic,  see \cite{Ovchinnikov1996:meanordset,Ovchinnikov1998:invfunsmpord}.
In fact, see \cite{Marichal2002:orderinvsynth}, any symmetric $n$-LPF on a chain
is an order statistic.
\end{example}

The class of weighted lattice polynomial functions
has been generalized by Marichal in \cite{Marichal2009:weightedlatpolynom}.

\begin{definition}\index{weighted lattice polynomial function}%
The class of $n$-argument \emph{weighted lattice polynomial functions} ($n$-WLPF)
from $P^n$ to $P$ is defined by applying the following rules
finitely many times:
\begin{enumerate}
   \item[(a)] $\func{F}^{(n)}(x_1,\dots,x_n)=x_i$ is an $n$-WLPF
   for any $i\in[n]$,
   \item[(b)] $\func{F}^{(n)}(x_1,\dots,x_n)=p$ is an $n$-WLPF
   for any $p\in P$,
   \item[(c)] If $\func{F}^{(n)}$ and $\func{G}^{(n)}$ are $n$-WLPFs,
   then $\func{F}^{(n)}\sqcap\func{G}^{(n)}$ and $\func{F}^{(n)}\sqcup\func{G}^{(n)}$
   are $n$-WLPFs.
\end{enumerate}
\end{definition}

As an analogue of Proposition~\ref{Prop:lpfform}, we have the following result.

\begin{proposition}
Let $\func{F}^{(n)}$ be an $n$-WLPF. Then there exist $k,l\ge 1$,
constants $a_1,\dots,a_k,b_1,\dots,b_l\in P$,
and families $A_1,\dots,A_k,B_1,\dots,B_l$ of nonempty subsets of $[n]$
such that:
\begin{equation}
   \func{F}^{(n)}(x_1,\dots,x_n)=
   \mathop{\bigsqcup}_{j=1}^k \left(a_j\sqcap\mathop{\bigsqcap}_{i\in A_j} x_i\right)
  =\mathop{\bigsqcap}_{j=1}^l \left(b_j\sqcup\mathop{\bigsqcup}_{i\in B_j} x_i\right).
\end{equation}
\end{proposition}

Interestingly, it turns out that $n$-ary weighted lattice polynomial functions may
also be represented as below.

\begin{proposition}\label{Prop:dcnfwlpf}
Let $\func{F}^{(n)}$ be an $n$-WLPF. Then there exist
set functions $\alpha,\beta:2^{[n]}\to P$ such that:
\begin{equation}
   \func{F}^{(n)}(x_1,\dots,x_n)=
   \mathop{\bigsqcup}_{S\subseteq[n]} \left(\alpha(S) \sqcap\mathop{\bigsqcap}_{i\in S} x_i\right)
  =\mathop{\bigsqcap}_{S\subseteq[n]} \left(\beta(S)  \sqcup\mathop{\bigsqcup}_{i\in S} x_i\right).
\end{equation}
\end{proposition}

It can be shown that, e.g., $\alpha(S)=\beta([n]\setminus S)$ in the above equation.
An $n$-WLPF formulated as above is said to be either in
disjunctive (left) or conjunctive (right) normal form.

\begin{example}\index{Sugeno integral}%
   By  \cite[Corollary 13]{Marichal2009:weightedlatpolynom},
   see also \cite{Marichal2000:sugenointagfun},
   $\func{F}^{(n)}$ is a Sugeno integral if and only if
   $\func{F}^{(n)}$ is an idempotent $n$-WLPF.
   And this happens if and only if
   $\func{F}^{(n)}(n\ast\underline{0})=\underline{0}$
   and $\func{F}^{(n)}(n\ast\overline{1})=\overline{1}$,
   i.e., it is endpoint preserving.
\end{example}

\begin{proposition}[\cite{CouceiroMarichal2010:charsugeno}]
$\func{F}^{(n)}$ is a symmetric $n$-WLPF
if and only if it can be represented
in a disjunctive or conjunctive normal form
(see Proposition~\ref{Prop:dcnfwlpf})
with $\alpha(S)$ and $\beta(S)$ being cardinality-based,
i.e., solely functions of~$|S|$.
\end{proposition}

\section{Aggregation on a nominal scale}\label{Sec:nominalscale1}

Having been given a space of objects on which an ordering
relation is defined is a quite comfortable situation. It is even more pleasant,
if we can rely on this assumption in such a way that we may require
that a fusion function must preserve such an order.
Unfortunately, in some practical applications we do not have as much as that.

Let us assume that the elements to be aggregated
are defined on a nominal scale. That is,
\index{alphabet}\index{character}%
there is a finite set, $\Sigma=\{\mathtt{a}_1,\dots,\mathtt{a}_k\}$, called an  \emph{alphabet},
on which only an equivalence relation, $=$, is defined.
Each element of $\Sigma$ is called a \emph{character}.

\begin{example}\label{Ex:NominalACGT}
In molecular biology and bioinformatics (among others),
we may assume $\Sigma=\{\mathtt{A},\mathtt{C},\mathtt{G},\mathtt{T}\}$,
i.e., a set consisting of
the primary nucleobases: adenine, cytosine, guanine, and thymine,
respectively.
Here, we may also be interested in the protein alphabet,
which is of cardinality 20.
\end{example}

\begin{example}\label{Ex:NominalASCII}
$\Sigma$ may also be the set of code points covered by the Unicode
standard. The Universal Coded Character Set defines more than
110,000 characters (letters, numbers, symbols, etc.)
from most languages, scripts, and locales.
Alternatively, it may be the set of characters covered by the ASCII
(see Table~\ref{Tab:ASCII}) or ISO-8859-1 standard.
Note that even though encoding standards define mappings between
sets of characters and integers (on which a natural linear order exists),
it does not mean that we obtain anything more than just a nominal scale here.
\end{example}

\begin{remark}
In the \R{} programming language, there is a special data type
to store information on a nominal scale called \texttt{factor}.
Such objects are represented as integer vectors
with a special attribute, \textit{levels},
which is used to decode the numeric indices into string labels.

\begin{lstlisting}[language=R]
x <- factor(c("a", "g", "c", "a", "t", "g"))
print(x)
## [1] a g c a t g
## Levels: a c g t
table(x)
## a c g t
## 2 1 2 1
unclass(x)       # internal representation
## [1] 1 3 2 1 4 3       # integer indices
## attr(,"levels")
## [1] "a" "c" "g" "t"   # decoding scheme
\end{lstlisting}
\end{remark}

\begin{example}
We may also assume that $\Sigma=\{\mathtt{0},\mathtt{1}\}$
is a set of bits, i.e., \textit{b}inary dig\textit{it}s.
\end{example}

It turns out that fusion functions defined on objects on a nominal scale,
although useful in practical applications, are not too
``mathematically interesting''.
Perhaps the only sensible
family of metrics we may define in the current setting is given by:
\begin{equation}
\mathfrak{d}_c(\mathtt{a}, \mathtt{b})=c\indicator(\mathtt{a}\neq \mathtt{b}),
\end{equation}
where $\mathtt{a},\mathtt{b}\in\Sigma$,
which for $c=1$ is in fact the Hamming distance on $\Sigma^1$,
see also Section~\ref{Sec:nominalscaled}.

Given $\vect{x}\in\Sigma^n$, $\mathtt{a}\in\Sigma$ such that:
\[
a=\argmin_{\mathtt{a}\in\Sigma} \sum_{i\in[n]}
\mathfrak{d}_c(x_i, \mathtt{a}).
\]
is equivalent to the \emph{mode} of $\vect{x}$, i.e., the most frequently occurring observation
\index{mode}%
in $\vect{x}$, see also Remark~\ref{Remark:mode}. Note that the solution to the above equation may be non-unique.
Nevertheless, assuming that $\Sigma=\{ \mathtt{a}_1,\dots,\mathtt{a}_k \}$, we may introduce
a fusion function, e.g.,
like:
\[
   \func{Median}_{\mathfrak{d}_c}^{(n)}(x_1,\dots,x_n) = \mathtt{a}_j,
\]
where $j = \min\{ j: \mathtt{a}_j = \argmin_{\mathtt{a}\in\Sigma} \sum_{i\in[n]}
\mathfrak{d}_c(x_i, \mathtt{a}) \}$, which now is well-defined.

\begin{remark}\label{Algorithm:mode}
Assuming that $\Sigma=\{1,2,\dots,k\}$, there are a few possible
approaches to determine a mode:
   \begin{itemize}
      \item a bucket-sort like algorithm requires $O(k+n)$ time,
      \item the elements may be sorted
      with the radix sort algorithm, which requires $O(n\log k)$ time,
      \item a hash-table-based procedure requires amortized $O(n)$ time,
   \end{itemize}
   and so on.
\end{remark}

The introduced fusion function is:
\begin{itemize}
   \item symmetric,
   \item idempotent,
   \item such that $\func{Median}_{\mathfrak{d}_c}^{(n)}(\vect{x})=\func{Median}_{\mathfrak{d}_c}^{(n)}(\vect{y})$
   where $y_i\in\{x_i, \func{Median}_{\mathfrak{d}_c}^{(n)}(\vect{x})\}$,
   \item stable, see \cite{Rojas2012:consistency,GagolewskiGrzegorzewski2010:ipmu}, i.e.,
   \index{stability}%
   \[\func{Median}_{\mathfrak{d}_c}^{(n+1)}(x_1,\dots,x_n, \func{Median}_{\mathfrak{d}_c}^{(n)}(x_1,\allowbreak\dots,x_n))\allowbreak
   =\allowbreak\func{Median}_{\mathfrak{d}_c}^{(n)}(x_1,\dots,x_n).\]
\end{itemize}

\begin{example}
A \index{weighted mode}\emph{weighted mode}
is a fusion function which minimizes:
\[
\mathtt{a}=\argmin_{\mathtt{a}\in\Sigma} \sum_{i\in[n]}
w_i \mathfrak{d}_c(x_i, \mathtt{a}).
\]
for some weighting vector $\vect{w}$.
This tool is used in a class of machine learning algorithms
for classification called \index{ensemble methods}\emph{ensemble methods}.
For instance, in the so-called \index{bagging}\emph{bagging} (bootstrap averaging),
see, e.g., \cite{Breiman2001:randomforest},
$w_i$ is defined as $\alpha_i/\sum_j \alpha_j$, where $\alpha_i$ is the
$i$th classifier's accuracy. The famous
\index{random forest}\emph{random forest}
algorithm is based on the very same idea, compare \cite{Breiman2001:randomforest}.
\end{example}

Please note that the case of aggregating observations on a nominal scale becomes
much more challenging when we shall consider $d$- or arbitrary-dimensional
data.

\clearpage{\pagestyle{empty}\cleardoublepage}
\chapter{Aggregation of multivariate data}\label{Chap:multidim}

\lettrine[lines=3]{L}{et} us focus on the task dealing with aggregation of $n$ objects in
a ${d}$-dimensional space $X^d$, where this time $d>1$. This is a case
of, e.g., real vectors in $\mathbb{R}^d$,
Cartesian products of $d$ identical bounded posets, as well as $d$-digits
binary or nucleobase sequences.

For fixed $d$, consider a fusion function
$\func{F}:(X^d)^n\to X^d$ that aims to aggregate
a set of $n$ objects $\vect{x}^{(1)},\dots,\vect{x}^{(n)}\in X^d$.
By using this mapping we obtain a single object from the set $X^d$.
In other words, $\func{F}$ is such that:

\begin{equation}
\func{F}\left(
\left[
\begin{array}{c}
x_1^{(1)} \\
x_2^{(1)} \\
\vdots \\
x_d^{(1)} \\
\end{array}
\right],
\dots,
\left[
\begin{array}{c}
x_1^{(n)} \\
x_2^{(n)} \\
\vdots \\
x_d^{(n)} \\
\end{array}
\right]
\right)
=
\left[
\begin{array}{c}
y_1\\
y_2\\
\vdots \\
y_d
\end{array}
\right].
\end{equation}
Equivalently, we may conceive $\func{F}$ as a function
acting on a $d\times n$  matrix:
\[\vect{X}=[\vect{x}^{(1)}\ \vect{x}^{(2)}\ \cdots\ \vect{x}^{(n)}].\]
From now on we assume that all vectors are column vectors.
Note that in data analysis, $\vect{x}^{(i)}$ is often called an \emph{observation} --
it designates an object or experimental unit.
On the other hand, $x_j^{(i)}$ denotes the result of measuring
the $j$th variable or feature (such as temperature, weight, velocity, etc.)~of
the $i$th observation (e.g., a person, autonomous vehicle, spatial location).

\medskip
First we shall review the task of real vectors' fusion from the perspective of
aggregation theory. In the consecutive sections, we significantly extend
the results presented in \cite{Gagolewski2015:issuesmultidim}.

\section{Aggregation of real vectors}

Most of the aggregation methods reviewed in this section
come from areas like computational statistics and geometry.
Therefore, here we shall assume that $X=\mathbb{R}$.

\begin{example}
Let us take any three non-colinear points in $\mathbb{R}^2$.
Even in such a simple case there are many useful ways
to aggregate a triad, see the triangle center problem
\cite{Johnson1929:moderngeometry,Kimberlng1998:trianglecenters,Kimberlng1994:trianglecenters}.
Most often this issue is conceptualized by using the so-called
\index{triangle center function}%
\emph{triangle center function}, see \cite{Bottema1981:trianglefunction},
which is a homogeneous real-valued function of
a triangle's side lengths. Thus, when rewritten in terms of vertex coordinates,
this leads us to a fusion function which is -- among others --
rotation and scale equivariant (see below).
Among the most well-known triangle centers we find the centroid, in-, circum-,
and ortho-center. What is interesting, C.~Kimberling's
\textit{Encyclopedia of Triangle Centers} (available online
at \url{http://faculty.evansville.edu/ck6/encyclopedia/ETC.html})
as of December 10, 2015 lists, names, and characterizes over 8781 such aggregation methods.
\end{example}

As we know from Chapter~\ref{Chapter:onedim}, in classical aggregation
theory, we mostly focus on the $d=1$ case. Recall that the notion of a mean
(internal aggregation function) $\func{F}:\mathbb{R}^n\to\mathbb{R}$,
may be used to determine the ``most typical observation'' among a given set of values.
We know that identifying the sine qua non conditions that $\func{F}$ should
fulfill in order to be useful in particular applications is very important,
as the class of all fusion functions is of course too broad.
Following the axiomatic framework by Kolmogorov and Nagumo, see, e.g.,
\cite{Bullen2003:means,Kolmogorov1930:moyenne,Nagumo1930:mittelwerte} and
Remark \ref{Remark:MeanKolmogorovNagumoSense},
\index{mean (Kolmogorov-Nagumo sense)}%
we could require the fulfillment of at least the three following properties:
\begin{itemize}
   \item symmetry,
   \item nondecreasingness, and
   \item internality.
\end{itemize}
Let us extend them in such a way that they are valid for any $d$.

\paragraph{Symmetry.}\index{symmetry}%
The first property is the least problematic one.
We may simply assume that for any $\sigma\in\mathfrak{S}_{[n]}$ it holds:
\begin{equation}
\func{F}(\vect{x}^{(1)},\dots,\vect{x}^{(n)})=\func{F}(\vect{x}^{(\sigma(1))},\dots,\vect{x}^{(\sigma(n))})
\end{equation}

\medskip
It turns out that the easiest and perhaps the most natural approach to
extend the other two properties is to apply them in a \emph{componentwise manner}.

\paragraph{Nondecreasingness.}\index{nondecreasingness}%
First of all, note that the ordering
structure on $\mathbb{R}$ may easily be extended to $\mathbb{R}^d$
by determining the so-called \emph{product order}.
The partial order $\le_d$ is defined in such a way that for any
$\mathbf{x}, \mathbf{y}\in\mathbb{R}^d$ we have:
\begin{equation}
\mathbf{x}\le_d\mathbf{y}\text{ if and only if }(\forall i\in[d])\ x_i\le y_i.
\end{equation}
This leads us to the concept of $\le_d$ (componentwise)-nondecreasingness.
Such an approach is often used when the topic
of aggregation on posets is explored,
see, e.g., \cite{DeBaetsMesiar1999:trinormprodlat,BustinceETAL2014:cartprodlat,%
KomornikovaMesiar2011:agbndposet},
and also Section~\ref{Sec:ProdLatAg}.

\begin{definition}
A fusion function $\func{F}:(\mathbb{R}^d)^n\to \mathbb{R}^d$
is \emph{$\le_d$-nondecreasing} whenever for all
$\vect{x}^{(1)},\dots,\vect{x}^{(n)},\vect{y}^{(1)},\dots,\vect{y}^{(n)}\in\mathbb{R}^d$
such that $\vect{x}^{(i)}\le_d\vect{y}^{(i)}$ for all $i=1,\dots,n$
it holds
$\func{F}(\vect{x}^{(1)},\dots,\vect{x}^{(n)})\le_d\func{F}(\vect{y}^{(1)},\dots,\vect{y}^{(n)})$.
\end{definition}

\paragraph{Internality.}
On the other hand, componentwise internality may be defined
as follows.\index{internality}%

\begin{definition}
   A fusion function $\func{F}:(\mathbb{R}^d)^n\to \mathbb{R}^d$
   is \emph{componentwise internal}
   if for all $\vect{x}^{(1)},\dots,\vect{x}^{(n)}\in\mathbb{R}^d$
it holds:
\begin{equation}
\func{F}(\vect{x}^{(1)},\dots,\vect{x}^{(n)})\in
\Bigg[\bigwedge_{i=1}^n x^{(i)}_1, \bigvee_{i=1}^n x^{(i)}_1\Bigg]
\times\dots\times
\Bigg[\bigwedge_{i=1}^n x^{(i)}_d, \bigvee_{i=1}^n x^{(i)}_d\Bigg].
\end{equation}
\end{definition}
Basically, above we deal with the bounding (hyper)rectangle
of a given set of input points.

\bigskip
Here are two exemplary fusion functions that fulfill
symmetry as well as componentwise monotonicity and internality.

\begin{definition}
The componentwise extension of the arithmetic mean is given~by:
\index{CwAMean@$\mathsf{CwAMean}$|see {componentwise arithmetic mean}}%
\index{componentwise arithmetic mean}%
\[
\func{CwAMean}(\vect{x}^{(1)},\dots,\vect{x}^{(n)})=
\left[
\begin{array}{c}
\frac{1}{n}\sum_{i=1}^n x_1^{(i)}\\
\vdots\\
\frac{1}{n}\sum_{i=1}^n x_d^{(i)}\\
\end{array}
\right].
\]
\end{definition}

\index{centroid}%
This fusion function is also called the \emph{centroid} (barycenter, geometric center) of a set of points.
This notion is crucial, e.g., in the definition of the $k$-means \cite{MacQueen1967:kmeans}
clustering algorithm.

On the other hand, the following mapping
is sometimes used, see \cite{Small1990:medianssurv}, as a robust
estimate of a multidimensional probability distribution's median.

\begin{definition}
\index{CwMedian@$\mathsf{CwMedian}$|see {componentwise median}}%
\index{componentwise median}%
The componentwise extension of the sample median  is defined~as:
\[
\func{CwMedian}(\vect{x}^{(1)},\dots,\vect{x}^{(n)})=
\left[
\begin{array}{c}
\func{Median}\left(x_1^{(1)},\dots,x_1^{(n)}\right)\\
\vdots\\
\func{Median}\left(x_d^{(1)},\dots,x_d^{(n)}\right)\\
\end{array}
\right].
\]
\end{definition}

Both functions are examples of componentwise extensions
of an internal aggregation function $\func{G}:\mathbb{R}^n\to\mathbb{R}$.
The induced fusion function $\func{CwG}:(\mathbb{R}^d)^n\to\mathbb{R}^d$
combines each data dimension independently. Thus, we have:
\begin{equation}
\func{CwG}\left(
\left[
\begin{array}{c}
x_1^{(1)} \\
x_2^{(1)} \\
\vdots \\
x_d^{(1)} \\
\end{array}
\right],
\dots,
\left[
\begin{array}{c}
x_1^{(n)} \\
x_2^{(n)} \\
\vdots \\
x_d^{(n)} \\
\end{array}
\right]
\right)
=
\left[
\begin{array}{c}
y_1\\
y_2\\
\vdots \\
y_d
\end{array}
\right]
=
\left[
\begin{array}{c}
\func{G}(x^{(1)}_1, \dots, x^{(n)}_1) \\
\func{G}(x^{(1)}_2, \dots, x^{(n)}_2) \\
\vdots \\
\func{G}(x^{(1)}_d, \dots, x^{(n)}_d),
\end{array}
\right].
\end{equation}

It is easily seen that if $\func{G}$ is nondecreasing in each variable,
then for $\vect{x}^{(1)}\le_d \vect{y}^{(1)},\dots,\vect{x}^{(n)}\le_d \vect{y}^{(n)}$,
we get $\func{CwG}(\vect{x}^{(1)},\dots,\vect{x}^{(n)})\le_d\func{CwG}(\vect{y}^{(1)},\dots,\allowbreak\vect{y}^{(n)})$.
Thus, $\func{CwG}$ is $\le_d$-nondecreasing.

Even more generally, we may of course consider the class of \emph{decomposable}
(as named, e.g., in \cite{KomornikovaMesiar2011:agbndposet})
\index{decomposability}%
fusion functions:
\begin{equation}
\func{F}_{\func{G}_1,\dots,\func{G}_d}\left(
\left[
\begin{array}{c}
x_1^{(1)} \\
x_2^{(1)} \\
\vdots \\
x_d^{(1)} \\
\end{array}
\right],
\dots,
\left[
\begin{array}{c}
x_1^{(n)} \\
x_2^{(n)} \\
\vdots \\
x_d^{(n)} \\
\end{array}
\right]
\right)
=
\left[
\begin{array}{c}
\func{G}_1(x^{(1)}_1, \dots, x^{(n)}_1) \\
\func{G}_2(x^{(1)}_2, \dots, x^{(n)}_2) \\
\vdots \\
\func{G}_d(x^{(1)}_d, \dots, x^{(n)}_d),
\end{array}
\right],
\end{equation}
where $\func{G}_i: \mathbb{R}^{n}\to\mathbb{R}$, $i=1,\dots,d$.
However, we shall note that in the case of such a class of fusion functions,
no interactions between different dimensions are taken into account explicitly.

Thus, in practice more intricate fusion functions are used.
Let us note that, see \cite{Gagolewski2015:issuesmultidim},
the following data aggregation tools -- well known in data analysis --
do not fulfill the componentwise monotonicity.
We will inspect them in much greater detail later on,
so now let us only provide their basic definitions.

\begin{definition}\index{1median@$\mathsf{1median}$|see {Euclidean 1-median}}%
\index{Euclidean 1-median}%
The (Euclidean) \emph{1-median} is a point $\vect{y}$ such that:
\begin{equation}\label{Eq:1median}
 \func{1median}_{\mathfrak{d}_2}(\vect{x}^{(1)},\dots,\vect{x}^{(n)})
= \argmin_{\vect{y}\in\mathbb{R}^{d}} \frac{1}{n} \sum_{i=1}^n \mathfrak{d}_2(\vect{x}^{(i)}, \vect{y}),
\end{equation}
where $\mathfrak{d}_2$ is again the Euclidean distance.
\end{definition}

\begin{example}\label{Ex:1mediannotmonotone}
Also 1-median is not componentwise monotone.
Take $d=2$, $n=3$, and $\vect{x}^{(1)}=[0,0]^T$, $\vect{x}^{(2)}=[1,-5]^T$,
$\vect{x}^{(3)}=[20,1]^T$. We have $\func{1median}_{\mathfrak{d}_2}(\vect{x}^{(1)}, \vect{x}^{(2)}, \vect{x}^{(3)})\simeq[1.961,-2.305]^T$.
However, when we take $\vect{x}'^{(3)}=\vect{x}^{(3)}+[1980,1]^T$,
then we get $\func{1median}_{\mathfrak{d}_2}(\vect{x}^{(1)}, \vect{x}^{(2)}, \vect{x}'^{(3)})\simeq[1.946,-3.351]^T<_2[1.961,-2.305]^T$.
\end{example}

\begin{definition}\index{1center@$\mathsf{1center}$|see {Euclidean 1-center}}%
\index{Euclidean 1-center}%
The \emph{Euclidean 1-center} (smallest enclosing ball radius)
is given~by:
\begin{equation}
\func{1center}_{\mathfrak{d}_2}(\vect{x}^{(1)},\dots,\vect{x}^{(n)})
=\argmin_{\vect{y}\in\mathbb{R}^{d}} \bigvee_{i=1}^n \mathfrak{d}_2(\vect{x}^{(i)}, \vect{y}),
\end{equation}
where $\mathfrak{d}_2$ is the Euclidean metric.
\end{definition}

\begin{example}
Euclidean 1-center is not componentwise monotone.
Consider $n=3$ and $d=2$ with $\vect{x}^{(1)}=[1,-1]^T$,
$\vect{x}^{(2)}=[-1,1]^T$, $\vect{x}^{(3)}=[-\sqrt{2},0]^T$.
We have $\func{1center}_{\mathfrak{d}_2}(\vect{x}^{(1)}, \vect{x}^{(2)}, \vect{x}^{(3)})=[0,0]^T$.
Letting $\vect{x}'^{(1)}=\vect{x}^{(1)}+[3,0]^T$
we get $\func{1center}_{\mathfrak{d}_2}(\vect{x}'^{(1)}, \vect{x}^{(2)}, \vect{x}^{(3)})\approx[1.3, -0.5]^T\not\ge_2[0,0]^T$.
\end{example}

Moreover, Tukey \cite{Tukey1974:mathpict} introduced
\index{halfplane location depth|see {Tukey depth}}%
\index{tdepth@$\mathsf{tdepth}$|see {Tukey depth}}%
\index{Tukey depth}%
the concept of \emph{the halfplane location depth} of $\vect{y}$ relative to
a given set of points in $\mathbb{R}^d$.
It is the smallest number of points contained in any closed halfhyperplane
with boundary line through~$\vect{y}$. In other words: %
\begin{equation}
\func{tdepth}_d(\vect{y}; \vect{x}^{(1)},\dots,\vect{x}^{(n)})
=\min_{\vect{u}\in\mathbb{R}^d, |\vect{u}|=1} |\{i: \vect{u}^T\vect{x}^{(i)}\ge \vect{u}^T\vect{y}\}|.
\end{equation}
Observe that the deepest point in $d=1$ generalizes the concept of the median,
at least for odd $n$. Therefore, a deepest value in higher dimensions can be
thought of as a multidimensional median.

\begin{definition}
The center of gravity of the deepest halfplane location depth region
is called the
\index{TkMedian@$\mathsf{TkMedian}$|see {Tukey median}}%
\index{Tukey median}\emph{Tukey median}, $\func{TkMedian}$.
\end{definition}

\begin{example}
Tukey median is not componentwise monotone.
Consider $n=4$ and $d=2$ with $\vect{x}^{(1)}=[0,0]^T$,
$\vect{x}^{(2)}=[1,0]^T$, $\vect{x}^{(3)}=[1,1]^T$, and $\vect{x}^{(4)}=[0,1]^T$.
We have $\func{TkMedian}(\vect{x}^{(1)}, \vect{x}^{(2)}, \vect{x}^{(3)}, \vect{x}^{(4)})=[0.5,0.5]^T$.
Letting $\vect{x}'^{(4)}=\vect{x}^{(4)}+[1,0]^T$
we get $\func{TkMedian}(\vect{x}^{(1)}, \vect{x}^{(2)}, \vect{x}^{(3)}, \vect{x}'^{(4)})=[2/3, 1/3]^T\not\ge_2[0.5,0.5]^T$.
\end{example}

\begin{figure}[t!]
\centering
\begin{tabular}{ccc}
\includegraphics[width=3.8cm]{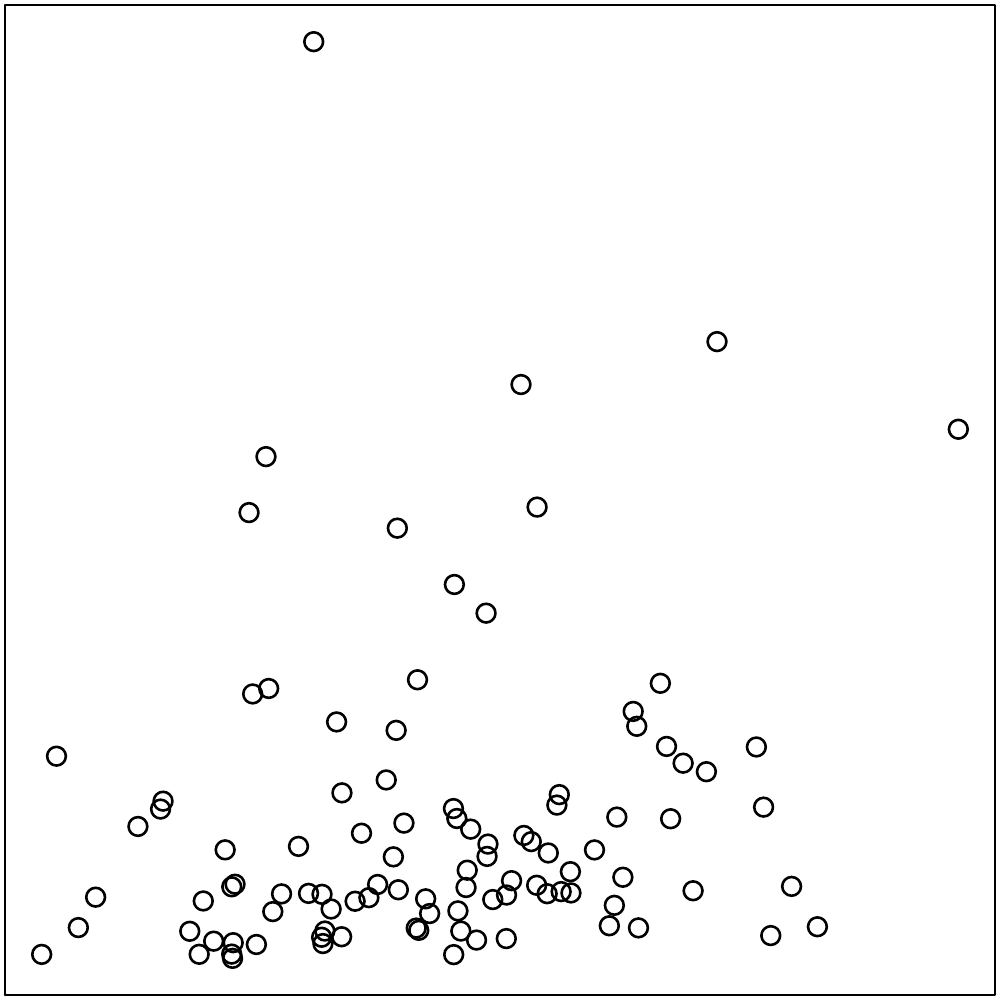}&
\includegraphics[width=3.8cm]{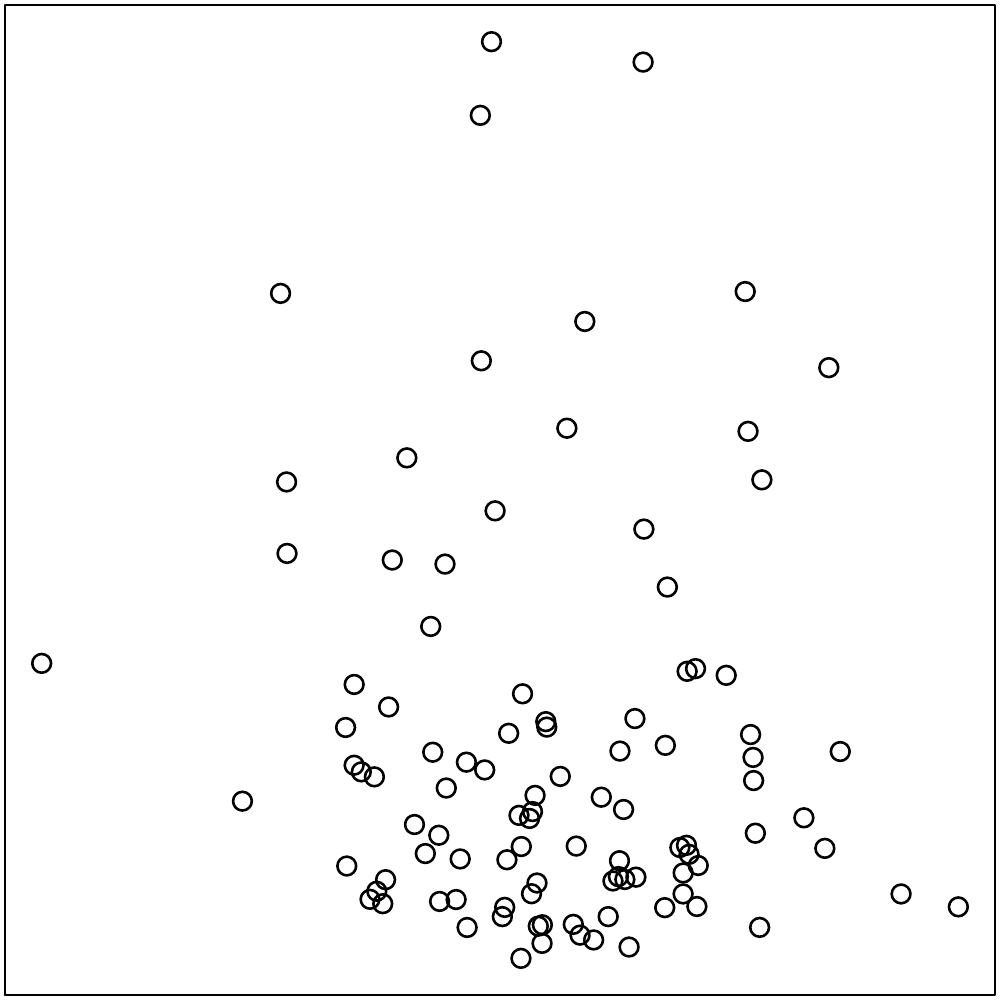}&
\includegraphics[width=3.8cm]{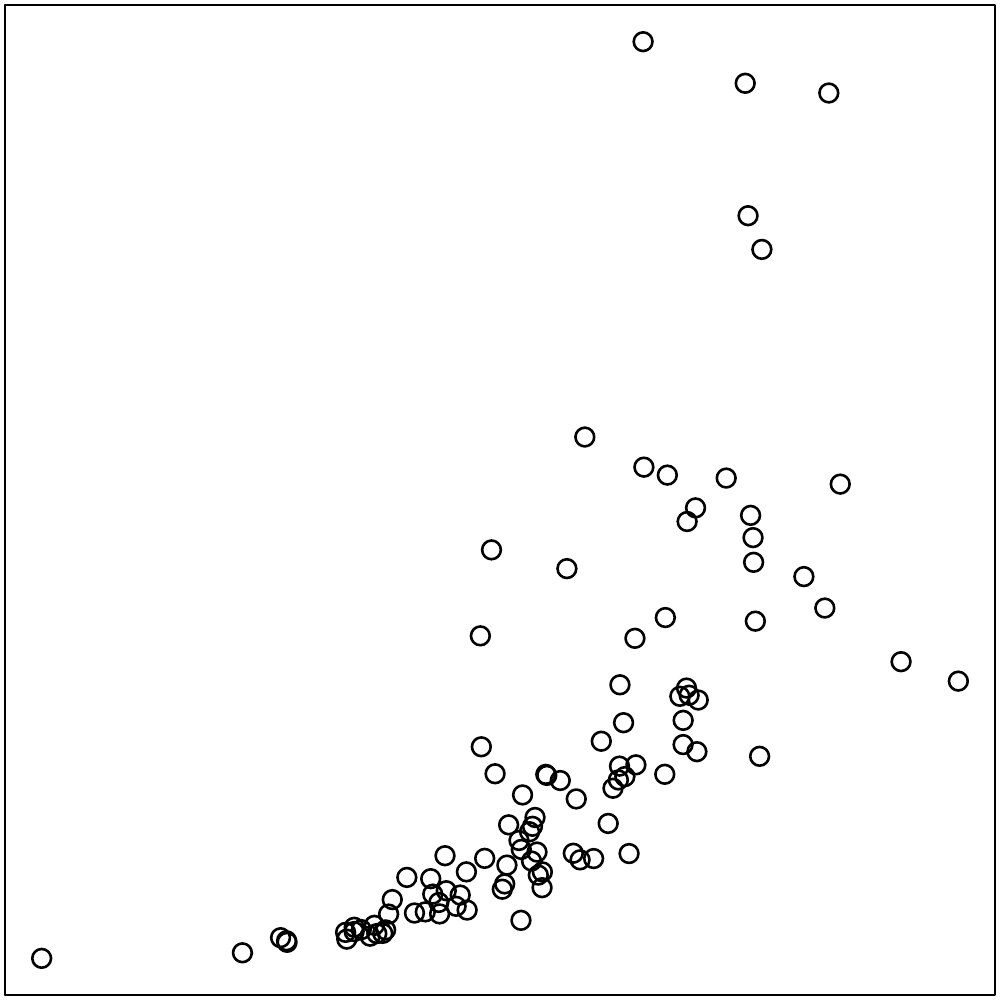}\\
\small (a) Gaussian ($\varrho=0.4$) & \small (b) Product & \small (c) Clayton ($\theta=4$)\\
\end{tabular}
\caption[Effects of choosing different copulas.]%
{\label{Fig:Copulas} Effects of choosing different copulas
if marginal cumulative distribution functions are  $F_1=\mathrm{N}(0,1), F_2=\mathrm{Exp}(0.1)$.}
\end{figure}

\begin{remark}\label{Remark:CopulasMultiDim}\index{copula}%
To model a $d$-dimensional data set $\vect{X}$ we may make
use of marginal cumulative distribution functions $F_1,\dots,F_d$
and a $d$-dimensional copula $\func{C}$ (see Definition~\ref{Def:copula}),
which describes the interdependence between
individual data dimensions. This is because, according to the famous
Sklar theorem \cite{Sklar1959:copulas}, see also \cite{Nelsen1999:Copulas},
whatever  the joint cumulative distribution function $H$ of $(X_1,\dots,X_d)$ is,
i.e.:
\[
   H(x_1,\dots,x_d)=\Pr(X_1\le x_1,\dots,X_d\le x_d),
\]
there always exists (unique if $H$ is continuous) $\func{C}, F_1,\dots,F_d$ such that:
\[
   H(x_1,\dots,x_d)=\func{C}(F_1(x_1),\dots,F_d(x_d)).
\]
Such a description is also useful if random variates generation is needed.
A~procedure for obtaining a single random vector in $\mathbb{R}^d$ may thus look as follows:
\begin{enumerate}
   \item[1.] Generate $(Y_1,\dots,Y_d)\sim \func{C}$ (note that copula $\func{C}$ is in fact
   a cumulative distribution function on the unit hypercube);
   \item[2.] Return $(F_1^{-1}(Y_1),\dots,F_d^{-1}(Y_d))\sim H$ as result.
\end{enumerate}
Here is an exemplary \R{} code that uses the \package{copula}
\cite{Yan2007:copula} package to generate a sample of $n=100$ $(d=2)$-dimensional
points using the Clayton copula with parameter $\theta=4$,
$\func{C}(u,v)=\left(0\vee u^{-\theta}+v^{-\theta}-1\right)^{-1/\theta}$,
and $F_1=\mathrm{N}(0,1)$ (standard normal), $F_2=\mathrm{Exp}(0.1)$
(an exponential distribution).

\begin{lstlisting}[language=R]
n <- 100
d <- 2
C <- copula::claytonCopula(dim=d, param=4)
Finv <- list( # marginal c.d.f.s (inverses)
   function(y) qnorm(y, 0, 1), function(y) qexp(y, 0.1)
)

X <- t(copula::rCopula(n, C))
for (i in 1:d) X[i,] <- Finv[[i]](X[i,])
\end{lstlisting}
Refer to Figure~\ref{Fig:Copulas} for an illustration of effects of choosing
different copulas.

\end{remark}

\section{Equivariance to geometric transforms}

Instead of focusing on monotonicity and internality,
researchers in such fields as computational statistics and geometry  most
often consider equivariances with respect to specific
classes of geometrical transformations.
This is in line with the aforementioned fact that
the necessity of the notion of monotonicity is being put into question
in the classical framework too, see, e.g.,
\cite{BeliakovCalvoWilkin2014:3typesmono,BustinceETAL2014:directionalmonotonicity}.
In the $d=1$ case this property seems quite natural and moreover it simplifies the way
the analytic results are derived. However, the situation is much different
in higher dimensions.

Namely, one might be interested in finding a fusion function $\func{F}$
which fulfills for all input vectors:
\begin{itemize}
   \item \index{translation equivariance}\emph{translation equivariance}:
   for all $\vect{t}\in\mathbb{R}^d$,
   \[\func{F}(\vect{x}^{(1)}+\vect{t}, \dots, \vect{x}^{(n)}+\vect{t})
   =\func{F}(\vect{x}^{(1)}, \dots, \vect{x}^{(n)})+\vect{t},\]
   \item \index{uniform scale equivariance}\emph{uniform scale equivariance}:
   for all $s>0$,
   \[\func{F}(s\vect{x}^{(1)}, \dots, s\vect{x}^{(n)})
   =s\func{F}(\vect{x}^{(1)}, \dots, \vect{x}^{(n)}),\]
   \item \index{d-scale equivariance@$d$-scale equivariance}%
   \emph{$d$-scale equivariance}: for all $\mathbb{R}^d\ni\vect{s}>\vect{0}$,
   \[\func{F}(\vect{s}\vect{x}^{(1)}, \dots, \vect{s}\vect{x}^{(n)})
   =\vect{s}\func{F}(\vect{x}^{(1)}, \dots, \vect{x}^{(n)}).\]
   \item \index{orthogonal equivariance}\emph{orthogonal equivariance}:
   for all orthogonal matrices $\vect{A}\in\mathbb{R}^{d\times d}$,
   \[\func{F}(\vect{A}\vect{x}^{(1)}, \dots, \vect{A}\vect{x}^{(n)})
   =\vect{A}\func{F}(\vect{x}^{(1)}, \dots, \vect{x}^{(n)}),\]
   and/or
   \item \index{affine equivariance}\emph{affine equivariance}:
   for all matrices $\vect{A}\in\mathbb{R}^{d\times d}$ of full rank
   and all $\vect{t}\in\mathbb{R}^d$,
   \[\func{F}(\vect{A}\vect{x}^{(1)}+\vect{t}, \dots, \vect{A}\vect{x}^{(n)}+\vect{t})
   =\vect{A}\func{F}(\vect{x}^{(1)}, \dots, \vect{x}^{(n)})+\vect{t}.\]
\end{itemize}
Affine equivariance implies translation, uniform scale, $d$-scale, and orthogonal equivariance.
Recall that with the notation convention used throughout this book, e.g.,
affine equivariance may be written as $\func{F}(\vect{A}\vect{X}+\vect{t})=\vect{A}\func{F}(\vect{X})+\vect{t}$.
Thus, an affine equivariant fusion function is independent of the chosen coordinate system.
It is a very strong property, so let us start our discussion
with simpler transformations. Also, we cover equivariance to similarity transforms,
which includes the translation, uniform scale, and orthogonal equivariance.

We are interested in exploring basic facts about different types of equivariances,
as well as different ways to modify a given mapping (especially one that is
a componentwise extension of a classical aggregation function) so that it obeys
the most important properties.

\subsection{Translation and scale equivariance}

It turns out that, given any fusion function $\func{F}$,
it is quite easy to transform it in such a way that it becomes translation
and uniform scale equivariant.

\begin{proposition}\label{Prop:transequiv}
Let $\func{F},\func{G}:(\mathbb{R}^d)^n\to\mathbb{R}^d$ be two fusion functions and
assume that $\func{G}$ is translation equivariant. Then $\func{F}'$ given by:
\[
   \func{F}'(\vect{X}) = \func{F}(\vect{X}-\func{G}(\vect{X}))+\func{G}(\vect{X})
\]
is translation equivariant.
\end{proposition}

Note that $\func{G}$ is often set to be the componentwise mean.

\begin{proposition}\label{Prop:uscaleequiv}
Let $\func{g}:(\mathbb{R}^d)^n\to\mathbb{R}$ be a function such that
$\func{g}(s\vect{X})=s\func{g}(\vect{X})$ with $\func{g}(\vect{X})\neq 0$
for all nondegenerate $\vect{X}$. Assuming that
$\func{F}:(\mathbb{R}^d)^n\to\mathbb{R}^d$ is a fusion function, we have that
$\func{F}'$ given by:
\[
   \func{F}'(\vect{X}) = \func{g}(\vect{X})\func{F}\left(\frac{1}{\func{g}(\vect{X})}\vect{X}\right)
\]
is uniform scale equivariant for all nondegenerate $\vect{X}$.
\end{proposition}

In practice, we may set, e.g., $\vect{g}(\vect{X})=\sqrt{\sum_{i=1}^n \sum_{j=1}^n \mathfrak{d}_2^2(\vect{x}^{(i)}, \vect{x}^{(j)})}$,
which may be thought of as a multivariate extension of the sample standard deviation,
see Section~\ref{Sec:SpreadMeasures}.

A quite similar result may be provided for the $d$-scale equivariance.
Notably, each $d$-scale equivariant fusion function is also uniform scale equivariant.

\begin{remark}
Translation and $d$-scale equivariance
is highly useful in the practice of data analysis,
as one often standardizes the input variables: \[
x_{j}^{(i)} \mapsto \frac{x_{j}^{(i)} - \func{AMean}(x_j^{(1)},\dots,x_j^{(n)})}{\func{SD}(x_j^{(1)},\dots,x_j^{(n)})},
\]
where $\func{AMean}$ and $\func{SD}$ stand for the arithmetic mean
and standard deviation, respectively, which are applied on the $j$th coordinate,
$j=1,\dots,d$.
\end{remark}

Here is a result concerning componentwise extensions of interval scale equivariant
univariate fusion functions, see Definition~\ref{Def:IntervalScaleEquivariant}.
On a side note, recall that we stated in Section~\ref{Sec:ChooseAg1Characterize}
that  the only quasi-arithmetic mean that is interval scale equivariant is
the arithmetic mean.

\begin{proposition}
If $\func{G}:\mathbb{R}^n\to\mathbb{R}$ is such that
$\func{G}(s\vect{x}+t)=s\func{G}(\vect{x})+t$ for all $\vect{x}\in\mathbb{R}^n, t\in\mathbb{R}, s>0$,
then its componentwise extension $\func{CwG}$ is translation and $d$-scale equivariant.
\end{proposition}

\subsection{Orthogonal equivariance}

Some machine learning algorithms (such as principal component analysis,
see Remark~\ref{Remark:PCA})
assume that the data points may freely be rotated.
Orthogonal equivariance implies equivariance to all possible rotations
of input points, reflections against the axes, and their combinations.
The discussed equivariance type -- especially together with translation equivariance --
may be important if we do not wish to be dependent on the choice of a coordinate system.

Recall that $\mathbf{A}$ is an \index{orthogonal matrix}\emph{orthogonal matrix}
whenever it holds $\mathbf{A}\mathbf{A}^T=\mathbf{A}^T\mathbf{A}=\mathbf{I}$
or, equivalently,  $\mathbf{A}^T=\mathbf{A}^{-1}$.

\begin{remark}
If $\vect{A}$ is orthogonal, then $|\mathrm{det}\,\mathbf{A}|=1$
and columns of $\mathbf{A}$ are orthogonal unit vectors --
they form an orthonormal basis of the Euclidean space $\mathbb{R}^d$.
An $\mathbf{A}$-based transformation is \index{unitary transformation}\emph{unitary},
i.e., it preserves the dot product of vectors.
Thus, it preserves the Euclidean distance between two points
(it is an isometry of the Euclidean space).
\end{remark}

\paragraph{Generating random orthogonal matrices.}
Methods for random generation of orthogonal matrices
may be used, e.g., for empirically testing whether a fusion
function is orthogonal equivariant. Let $\mathcal{O}(d)$
denote the group of orthogonal $d\times d$  matrices.

Following \cite{DiaconisShahshahani1987:subgroupalg},
we may be interested in a uniform distribution on $\mathcal{O}(d)$ with respect
to the Haar measure, see \cite{Halmos1950:measuretheory}.
In other words, a random matrix $\vect{A}$ is uniformly
distributed if $\Pr(\vect{A}\in \mathcal{A})=\Pr(\vect{A}\in \boldsymbol{\Gamma}\mathcal{A})$,
for any $\mathcal{A}\subset\mathcal{O}(d)$ and $\boldsymbol{\Gamma}\in\mathcal{O}(d)$.

For $d=2$, a random matrix may be generated by considering
$\vartheta\sim\mathrm{U}[0,2\pi[$ and $b\sim\mathrm{U}\{-1, 1\}$
and then taking:
\begin{equation}\label{Eq:RandomOrthogonal2}
\vect{A}=\left[
\begin{array}{cc}
\cos\vartheta & \sin\vartheta \\
-b\sin\vartheta & b\cos\vartheta
\end{array}
\right].
\end{equation}

One way to generate a random orthogonal matrix for $d>2$ is to produce
a $d\times d$ matrix with i.i.d.~elements following a standard normal distribution.
Then, by applying the Gram-Schmidt orthogonalization algorithm on such a matrix,
we get a desired  object, see \cite[page~234]{Eaton1983:multivarstat} for a proof.
This gives an $O(d^3)$ algorithm, but in practice its implementation
characterizes itself with  slow performance.
For this reason, we may rather want to use the following procedure.

\begin{algorithm}{\cite[Section~3]{DiaconisShahshahani1987:subgroupalg}}\label{Arg:rortho}
To generate a random orthogonal $d\times d$ matrix for given $d > 2$ proceed as follows:
\begin{enumerate}
   \item[1.] Generate a random orthogonal $2\times 2$ matrix $\mathbf{A}^{(2)}$, see Equation~\eqref{Eq:RandomOrthogonal2}.
   \item[2.] For $i=3,4,\dots,d$ do:
   \begin{enumerate}
      \item[2.1.] Let $\vect{v}\in\mathbb{R}^i$ be a randomly generated vector distributed uniformly
      on a unit $i$-sphere; for that we may generate $\vect{z}=(z_1,\dots,z_i)$ i.i.d.~$\mathrm{N}(0,1)$
      and set $\vect{v}:=\vect{z}/\|\vect{z}\|_2$, see \cite{Marsaglia1972:pointsphere};
      \item[2.2.] Let $\vect{x}:=(\vect{e}^{(i)}-\vect{v})/\|\vect{e}^{(i)}-\vect{v}\|_2$,
      where $\vect{e}^{(i)}=(1,0,0,\dots,0)\in\mathbb{R}^i$;
      \item[2.3.] Set:
      \[
         \vect{A}^{(i)} := \left(\vect{I}^{(i)} - 2\vect{x}\vect{x}^T\right)\left[
         \begin{array}{cccc}
         1 & 0 & \dots & 0 \\
         0 &   &       &   \\
         \vdots & & \vect{A}^{(i-1)} & \\
         0 &   &       &   \\
         \end{array}
         \right],
      \]
      where $\vect{I}^{(i)}$ is the diagonal $i\times i$ matrix.
   \end{enumerate}
   \item[3.] Return $\vect{A}^{(d)}$ as result.
\end{enumerate}
\end{algorithm}
By carefully setting vector/matrix multiplication order in Step 2.3.
we may get $O(d^3)$ time complexity,
see Figure~\ref{Fig:rortho} for an exemplary C++ implementation.

\paragraph{Orthogonal equivariant componentwise fusion functions.}
A special class of orthogonal projections consists of
a kind of rotation combined with reflection.
We take $\mathbf{A}=\mathbf{I}_\sigma$ for some $\sigma\in\mathfrak{S}_{[d]}$,
i.e., an identity matrix with permuted rows. The equivariance with respect to
such a transformation is the same as requiring that for each $\vect{X}\in(\mathbb{R}^d)^n$ it holds:
\begin{equation}
\func{F}\left(
\left[
\begin{array}{c}
x_1^{(1)} \\
x_2^{(1)} \\
\vdots \\
x_d^{(1)} \\
\end{array}
\right],
\dots,
\left[
\begin{array}{c}
x_1^{(n)} \\
x_2^{(n)} \\
\vdots \\
x_d^{(n)} \\
\end{array}
\right]
\right)
=
\func{F}\left(
\left[
\begin{array}{c}
x_{\sigma(1)}^{(1)} \\
x_{\sigma(2)}^{(1)} \\
\vdots \\
x_{\sigma(d)}^{(1)} \\
\end{array}
\right],
\dots,
\left[
\begin{array}{c}
x_{\sigma(1)}^{(n)} \\
x_{\sigma(2)}^{(n)} \\
\vdots \\
x_{\sigma(d)}^{(n)} \\
\end{array}
\right]
\right)_{\sigma^{-1}}.
\end{equation}
Thus, it is also a kind of symmetry (intuitively, a ``vertical'' one,
as opposed to the componentwise symmetry discussed above).
This easily leads us to the following result concerning componentwise extensions
of unidimensional fusion functions.

\begin{proposition}\label{Prop:rotationinvcomponent}
If $\func{F}_{\func{G}_1,\dots,\func{G}_d}$ is an orthogonal equivariant
componentwise extension of $\func{G}_i: \mathbb{R}^{n}\to\mathbb{R}$, $i\in[d]$, then necessarily
it is a componentwise fusion function: there exists $\func{G}$ such that $\func{CwG}=\func{F}_{\func{G}_1,\dots,\func{G}_d}$.
Moreover, it necessarily holds that $\func{G}(x_1,\dots,x_n)=-\func{G}(-x_1,\dots,-x_n)$.
\end{proposition}

We already noted that $\func{CwAMean}$ is an orthogonal equivariant componentwise fusion function.
However, the above necessary conditions are not sufficient:
it turns out that the componentwise median, $\func{CwMedian}$, is not orthogonal equivariant.

\begin{figure}[htb!]
\centering
\begin{minipage}[t]{6cm}
\centering
\includegraphics[width=5.5cm]{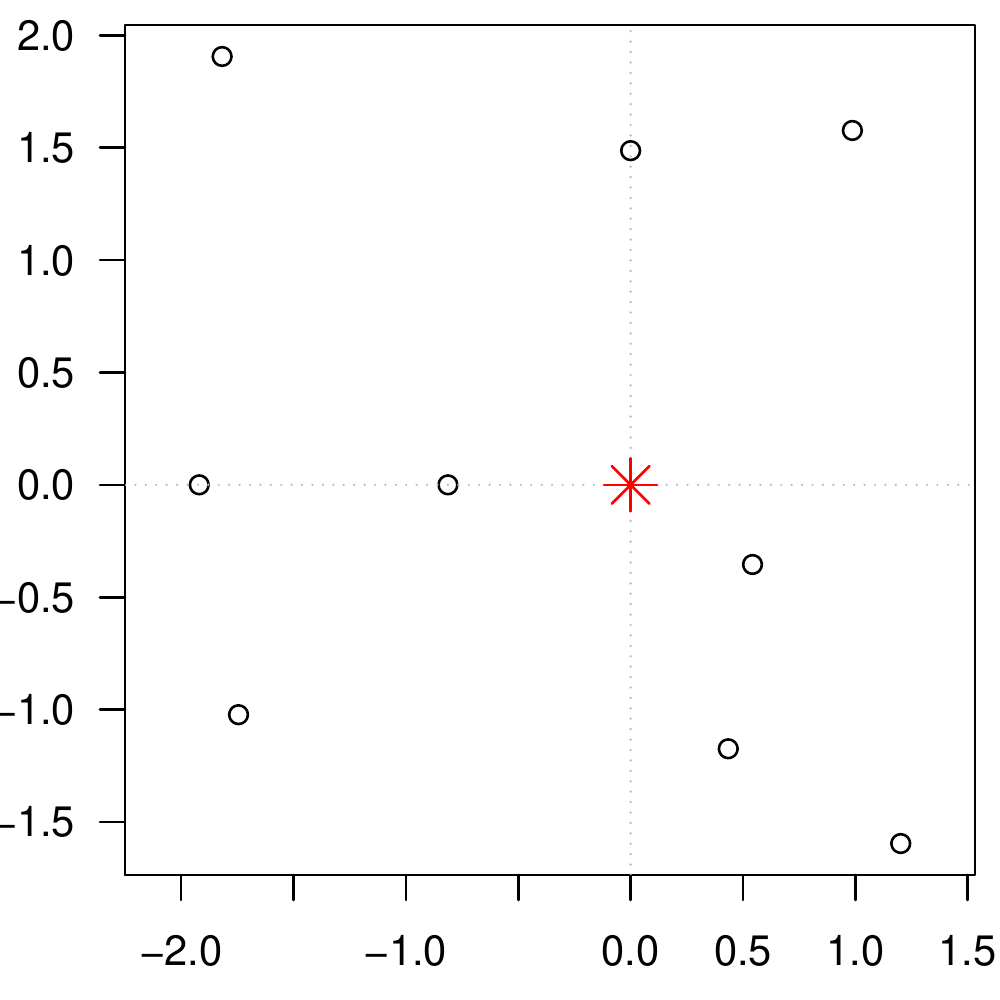}

\centering (a) A sample data set and its componentwise median $\func{CwMedian}(\vect{X})$ at $(0,0)$.
\end{minipage}
\begin{minipage}[t]{6cm}
\centering
\includegraphics[width=5.5cm]{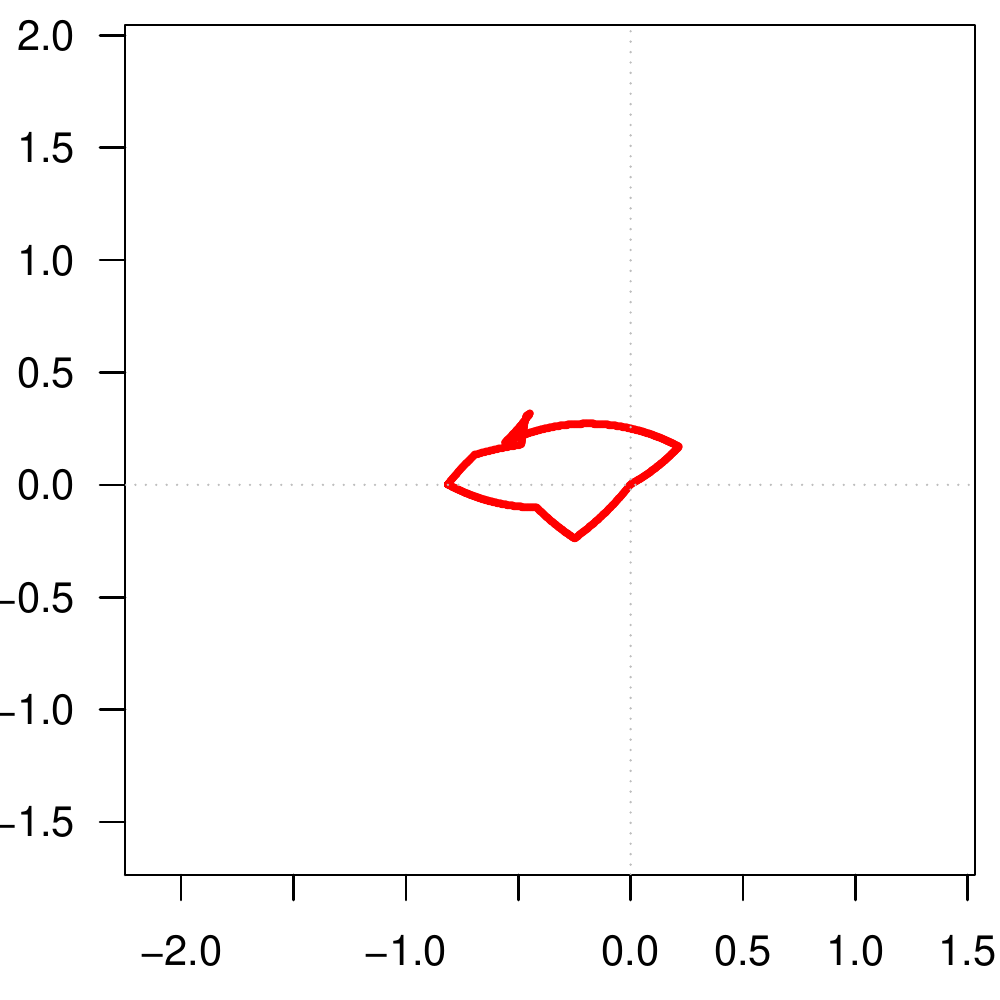}

\centering (b) Componentwise medians' trace over all rotations of the input data set,
$\mathbf{A}^{-1} \func{CwMedian}(\mathbf{A}\vect{X})$.
\end{minipage}

\caption{\label{Fig:component_median_rotation} Componentwise median and its dependence on the choice of a coordinate system.}
\end{figure}

\bigskip
Interestingly, even if we are given a non-orthogonal equivariant
fusion function, we may \index{orthogonalization}\emph{orthogonalize} it.
Below we explain two particularly appealing  orthogonalization methods,
which may be used in the case of, e.g., the componentwise median, $\func{CwMedian}$,
see Figure~\ref{Fig:component_median_rotation}.

\paragraph{Orthomedian.}%
\index{OrMedian@$\mathsf{OrMedian}$|see {orthomedian}}%
\index{orthomedian}%
The orthomedian by Grubel \cite{Grubel1996:orthogonalization} is an interesting instance of
the concept of orthogonalization, originally applied on the componentwise median. Basically, it is the averaged
median of all orthogonally transformed versions of the input data set.
As the group $\mathcal{O}(d)$ of orthogonal $d\times d$ matrices is compact,
we may introduce a fusion function:
\begin{equation}
   \func{OrMedian}(\vect{X}) = \int_{\mathcal{O}(d)} \mathbf{A}^{-1}
   \func{CwMedian}(\mathbf{A}\vect{X})\,d\mathbf{A},
\end{equation}
which is orthogonal equivariant (by construction) and additionally translation
and uniform scale equivariant (but not $d$-scale equivariant).
Interestingly, it is no longer $\le_d$-nondecreasing,
so this new property is introduced at some cost.
The idea behind orthogonalization is quite general and may be applied in the case of other fusion functions as well.

One may (and should) ask how the orthogonal median may be computed.
The above integral may of course be approximated via some Monte Carlo quadrature scheme.
In such a case, random matrices sampled uniformly from ${\mathcal{O}(d)}$ can
be generated (see Algorithm~\ref{Arg:rortho}).
However, this is computationally demanding and we observe a quite
slow rate of convergence (e.g., for $d=2$ we need at least $1000$ MC iterations
to get satisfiable results for a few dozen of points).

Another approach is to consider a set of $N$ points on a unit $d$-hypersphere,
$\vect{a}^{(1)}, \dots, \vect{a}^{(N)}$.
Then the orthomedian may be approximated, see \cite[Section~5]{Grubel1996:orthogonalization},  by:
\[
\func{OrMedian}(\vect{X}) \simeq d\func{CwAMean}(\vect{y}^{(1)},\dots,\vect{y}^{(N)}),
\]
where:
\[
\vect{y}^{(i)} = \func{Median}\left({\vect{a}^{(i)}}^T\vect{x}^{(1)}, \dots, {\vect{a}^{(i)}}^T\vect{x}^{(n)}\right)\vect{a}^{(i)}.
\]
The points on the hypersphere can be sampled randomly, %
but this process has an even slower convergence rate than the above-mentioned one.
It is best to rely on a quasi-Monte Carlo approach and sample the points
uniformly. This is easy for $d=2$. In higher dimensions, however, the problem,
at least for arbitrary $N$, becomes quite difficult. It is because
we have to solve the (hyper)Sphere Packing (L\'{a}szl\'{o} Fejes T\'{o}th's) problem,
see \cite{ConwaySloane1998:spherepackings}, which concerns the task of placing
$N$ points on a $d$-dimensional hypersphere so as
to maximize the minimal distance (or equivalently the minimal angle) between them.
Such a task may be treated with a stochastic optimization routine (e.g., simulated
annealing) or using an algorithm proposed in, e.g., \cite{LovisoloSilva2001:unifdistribhypersphere}.
Note that the probed points may be tabulated and stored for later use.

\medskip
Before moving to the second orthogonalization method,
let us briefly recall a statistical procedure called principal component analysis.

\begin{remark}\index{PCA|see {principal component analysis}}%
\index{principal component analysis}%
\label{Remark:PCA}%
Principal component analysis (PCA)
uses an orthogonal transformation to convert a set of observations of
possibly correlated variables into a set of values of linearly uncorrelated variables,
see \cite[Section~3.4 and Section~14.5]{HastieTibshiraniFriedman2013:esl}.
Let:
\[\vect{X}_c=\vect{X}-\func{CwAMean}(\vect{x}^{(1)},\dots,\vect{x}^{(n)})\]
be a centered version of $\vect{X}$.
Then the sample covariance matrix is given by $\vect{S}=\vect{X}_c\vect{X}_c^T/n\in\mathbb{R}^{d\times d}$.
Let us take the eigendecomposition of:
\[
\vect{X}_c\vect{X}_c^T = \vect{V}\vect{D}^2\vect{V}^T.
\]
\index{SVD|see {singular value decomposition}}\index{singular value decomposition}%
This may be obtained by taking the singular value decomposition (SVD):
\[\vect{X}_c^T = \vect{U}\vect{D}\vect{V}^T,\]
where $\vect{U}$ is an $n\times n$ orthogonal matrix,
$\vect{D}$ is an $n\times d$ diagonal matrix with nonnegative elements,
and $\vect{V}$ is a $d\times d$ orthogonal matrix,
see the \package{LAPACK} \cite{lapack} library routine \texttt{DGESDD}.
The eigenvectors $\vect{v}^{(i)}$ are called \emph{principal component directions} of $\vect{X}_c$.
The first principal component direction $\vect{v}^{(1)}$ has the property that
$\vect{z}^{(1)} = \vect{X}_c^T \vect{v}^{(1)}$ has the largest sample variance, $d^2_1/n$ among
all normalized linear combinations of $\vect{X}_c$'s rows.
Subsequent principal components have maximum variance subject
to being orthogonal to the earlier ones.
\end{remark}

\paragraph{SVD-based orthogonalization.}
Given a unidimensional fusion function, $\func{G}:\mathbb{R}^n\to\mathbb{R}$
such that $(\forall x_i)$ $\func{G}(x_1,\dots,x_n)=-\func{G}(-x_1,\dots,-x_n)$,
here is a simple way to orthogonalize its componentwise extension, $\func{CwG}$.
Assuming that the SVD of $(\vect{X}-\func{CwAMean}(\vect{X}))^T = \vect{U}\vect{D}\vect{V}^T$ and knowing
that $\vect{V}={\vect{V}^T}^{-1}$,
we may set
\begin{equation}
\func{OrG2}(\vect{X}) = \vect{V} \func{CwG}\left(\vect{V}^T(\vect{X}-\func{CwAMean}(\vect{X}))\right)+\func{CwAMean}(\vect{X}).
\end{equation}
It is easily seen that in such a way we obtain not only an orthogonal
equivariant fusion function, but also one that is translation equivariant.
Note that the condition $\func{G}(x_1,\dots,x_n)=-\func{G}(-x_1,\dots,-x_n)$ is crucial,
as the $\vect{U},\vect{V}$ matrices might be ambiguous:
the singular vectors are only defined up to sign;
if we change the sign of a left singular vector,
an equivalent SVD decomposition may be obtained by changing the sign of the
corresponding right vector.

Here, if $\func{G}$ is nondecreasing,
then the resulting fusion function is nondecreasing with respect to the
direction that has the maximal variance
(and other directions that are orthogonal to it and also maximize the remaining variance).

\begin{remark}
Let $\func{OrMedian2}$ be a SVD-orthogonalized version of the componentwise median for the case $d=2$.
Given $\vect{x}^{(1)}=(1,1), \vect{x}^{(2)}=(1,-1), \vect{x}^{(3)}=(-1,-1), \vect{x}^{(4)}=(-1,1)$,
we have $\func{OrMedian2}(\dots)=(0,0)$. Now letting $\vect{x}'^{(1)}=\vect{x}^{(1)}+(0,2)$,
we get $\func{F}(\dots)\approx(-0.25,0.07)$.
Thus, $\func{OrMedian2}$ is not componentwise monotone.
\end{remark}

\subsection{Equivariance to similarity transforms}\label{Sec:equivSimTrans}

The class of similarity transforms includes translation, uniform scaling
in each direction, rotation, and reflection. Equivariance to similarity transforms
can be conceived as a ``lightweight'' version of the corresponding property
with regard to affine transforms.

For any $\|\cdot\|$ matrix norm,
if $\vect{A}\in\mathbb{R}^{d\times d}$ is a nondegenerate matrix,
and $\vect{t}\in\mathbb{R}$,
then $(\vect{A},\vect{t})$ represents a similarity transform,
whenever $\frac{1}{\|\vect{A}\|}\vect{A}$ is orthogonal.

Let us go back to the above-derived SVD-based componentwise fusion function
orthogonalization scheme. For a given $\func{G}:\mathbb{R}^n\to\mathbb{R}$,
under the assumption that $\vect{U}\vect{D}\vect{V}^T$ is the SVD decomposition of $(\vect{X}-\func{CwAMean}(\vect{X}))^T$,
and by noting that for any $s>0$ we have:
\[
   s(\vect{X}-\func{CwAMean}(\vect{X})) = \vect{U}(s\vect{D})\vect{V}^T
\]
we can define a similarity transform-equivariant fusion function as:
\[
   \func{SimG}(\vect{X}) =
   {\|\vect{D}\|} {\vect{V}^{-1}}^T\func{CwG}\left(\frac{1}{\|\vect{D}\|}\vect{V}^T(\vect{X}-\func{CwAMean}(\vect{X}))\right)+\func{CwAMean}(\vect{X}),
\]
or, alternatively:
\[
   \func{SimG}'(\vect{X}) =
   \vect{V} \vect{D}^T \func{CwG}(\vect{U}^T)+\func{CwAMean}(\vect{X}).
\]

\subsection{Affine equivariance}

An affine transformation is a map that preserves
hyperplanes: ratios of Euclidean distances of points lying on a straight line remain the same.
Every linear transformation is affine, but not every affine transformation is linear.

\begin{figure}[htb!]
\centering
\begin{minipage}{6cm}
\centering
\includegraphics[width=5.5cm]{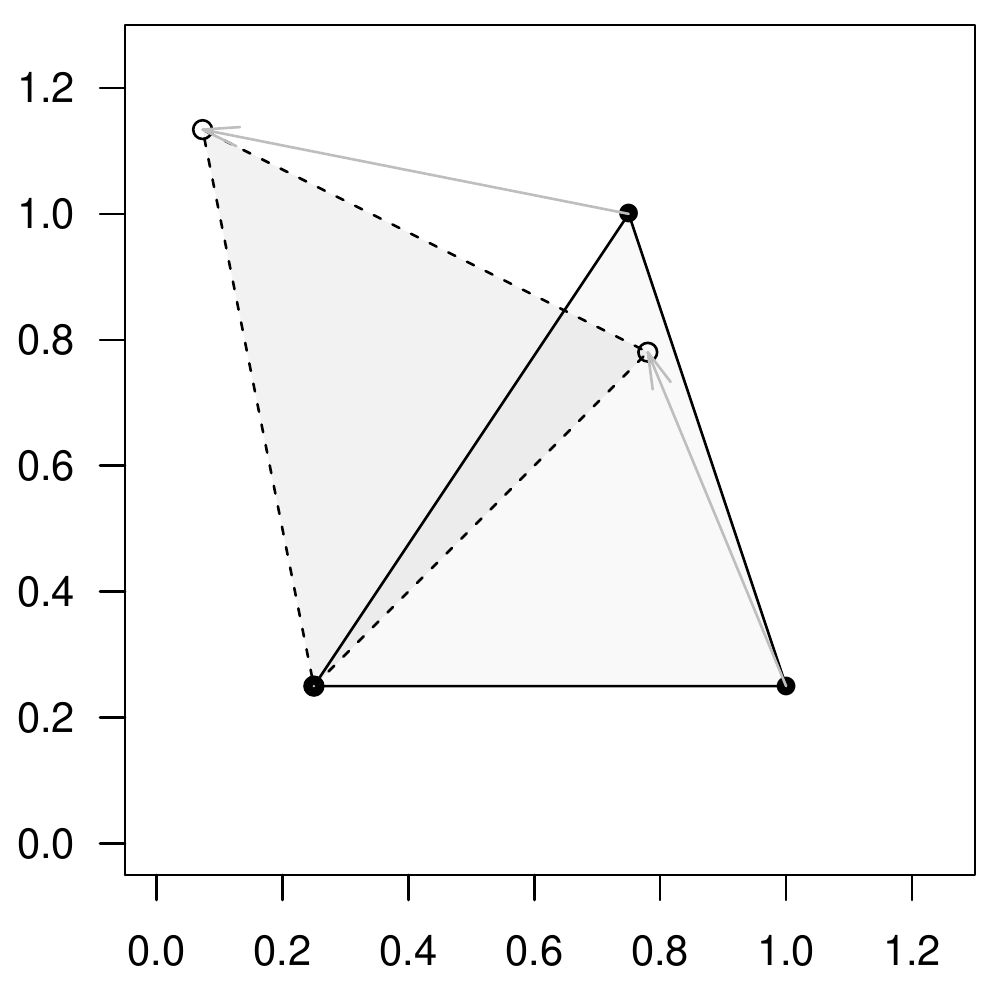}

\centering (a) Translation, rotation, translation.
\end{minipage}
\begin{minipage}{6cm}
\centering
\includegraphics[width=5.5cm]{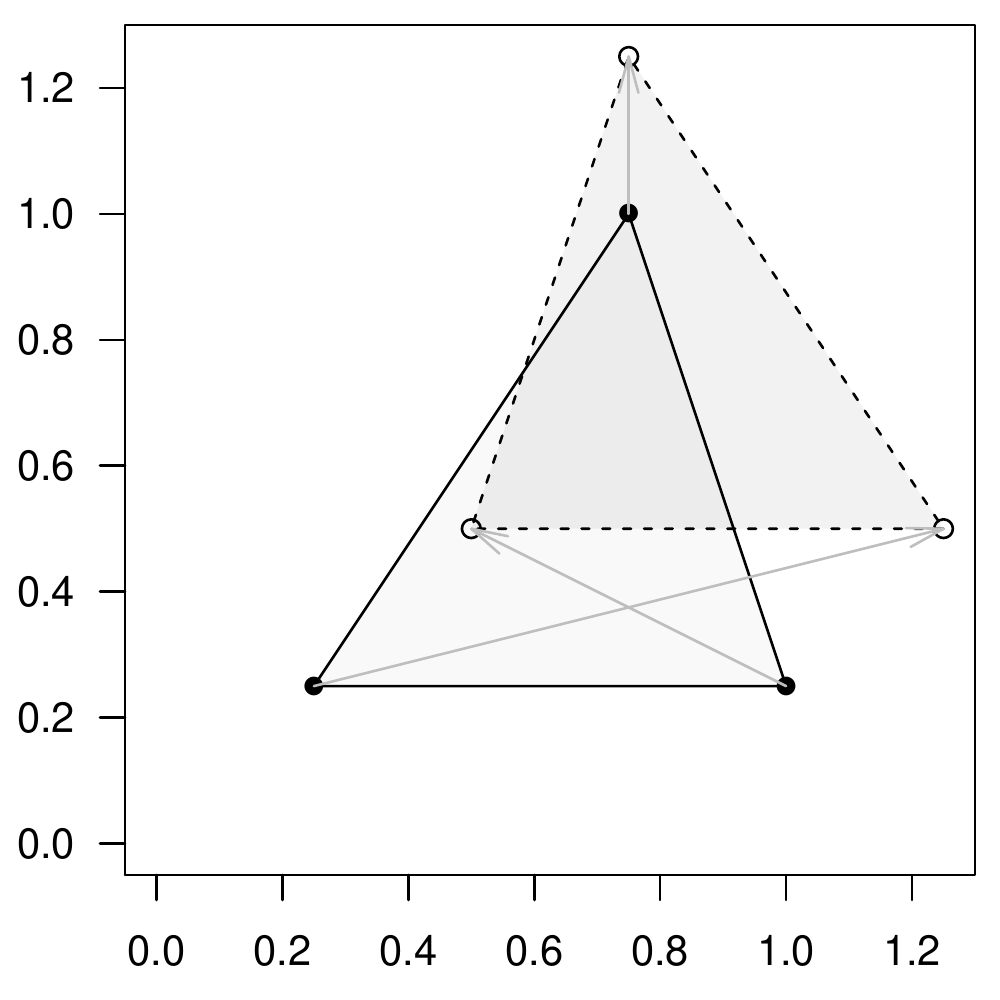}

\centering (b) Reflection against $OY$, translation.
\end{minipage}

\smallskip
\begin{minipage}{6cm}
\centering
\includegraphics[width=5.5cm]{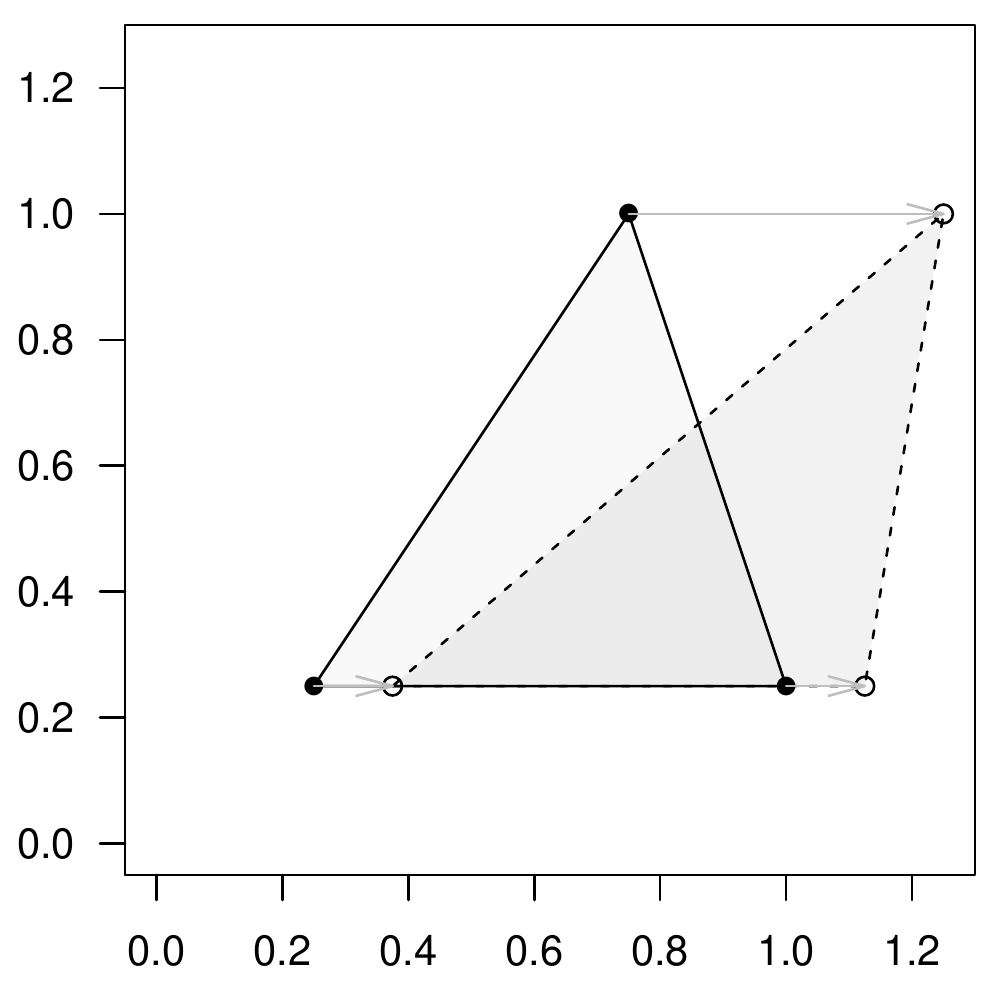}

\centering (c) Horizontal shear.
\end{minipage}
\begin{minipage}{6cm}
\centering
\includegraphics[width=5.5cm]{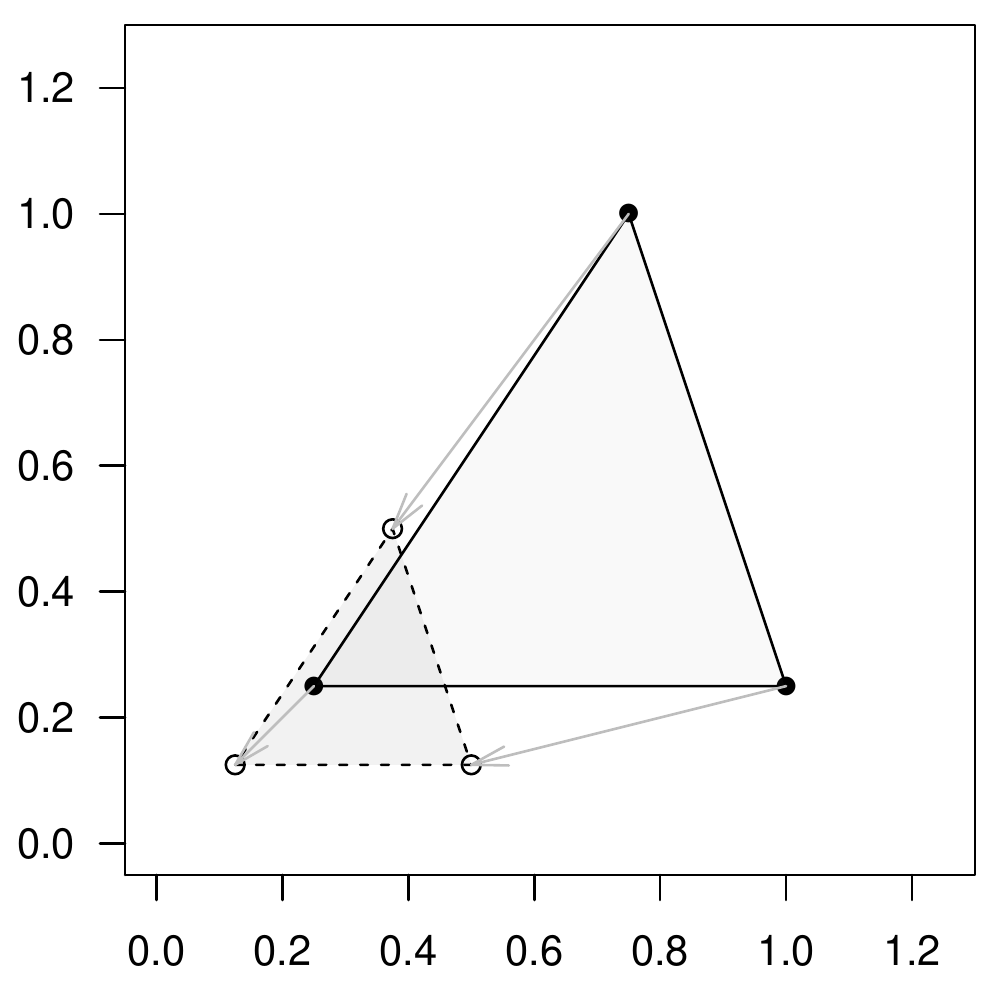}

\centering (d) Scaling.
\end{minipage}

\caption{\label{Fig:affine2dex} Exemplary affine transformations in $\mathbb{R}^2$.}
\end{figure}

\begin{example}
Table \ref{Table:affine2dex} lists exemplary affine transformations
in $\mathbb{R}^2$. Among them we find, e.g., translation, scaling,
rotation, %
shear mapping, reflection, and also any of their compositions.
Figure~\ref{Fig:affine2dex} depicts some affine mappings.
\end{example}

\begin{remark}
Note that we require  $\mathrm{det}\,\vect{A}\neq 0$.
Otherwise, transformations such as projections onto $OX$ and $OY$ axes
would also be included in our discussion. Yet, classically they are omitted
as they lead to ``too drastic'' data loss.
\end{remark}

It turns out that if affine equivariance is important to us,
then we should be interested in rather ``complex'' fusion functions
(see also Proposition~\ref{Prop:rotationinvcomponent}).
This is because of the following fact characterizing
affine equivariant componentwise fusion functions.

\begin{proposition}\label{Prop:affineequicw}
\index{componentwise arithmetic mean}%
A continuous and bounding-box internal fusion function $\func{CwG}$,
being a componentwise extension of $\func{G}:\mathbb{R}^n\to\mathbb{R}$, is
affine equivariant if and only if $\func{G}$ is a weighted arithmetic mean.
\end{proposition}
The above result follows from the fact that $\func{G}$ must necessarily be
additive, compare Theorem~\ref{Thm:CharacterizationAdditive}.
Thus, the only componentwise symmetric, internal, continuous, and affine equivariant
fusion function is formed by extending  the arithmetic mean.

Among non-componentwise fusion functions that are affine equivariant
we find, e.g., the Tukey median, $\func{TkMedian}$.

\begin{remark}
It turns out that an affine transformation $(\vect{A},\vect{t})$
may be expressed using a single $(d+1)\times(d+1)$ square matrix.
\index{homogeneous coordinates}%
For that, the so-called \emph{homogeneous coordinate system}, introduced
by A.F.~M\"{o}bius, is typically used.
In order to do so, we first construct the \emph{augmented matrix}:
\[
   \vect{B}=\left[
\begin{array}{ccc|c}
 & & &  \\
 & \vect{A} & & \vect{t} \\
& & &  \\
\hline
0 & 0 & 0 & 1 \\
\end{array}
\right]
\]
Then, instead of operating on vectors like $\vect{x}\in\mathbb{R}^d$,
we rather consider $[\vect{x}\ 1]^T\in\mathbb{R}^{d+1}$,
i.e., a version of the original inputs with an additional coordinate equal to $1$ added.
In such a way:
\[\vect{y}=\vect{A}\vect{x}+\vect{t}\]
may now be written as:
\[
\left[
\begin{array}{c}
 \\
\vect{y} \\
 \\
1
\end{array}
\right]
=
\left[
\begin{array}{ccc|c}
 & & &  \\
 & \vect{A} & & \vect{t} \\
& & &  \\
\hline
0 & 0 & 0 & 1 \\
\end{array}
\right]
\left[
\begin{array}{c}
 \\
\vect{x} \\
 \\
1
\end{array}
\right].
\]
\end{remark}

\begin{figure}[p!]
   \centering

   \includegraphics[width=4.5in]{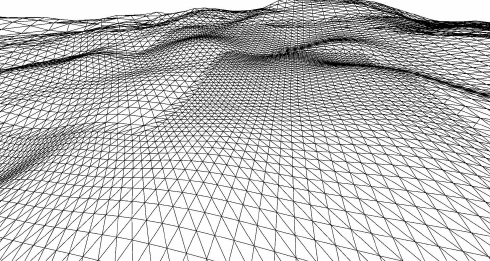}

   \caption{\label{Fig:Naturenv} An exemplary virtual 3D world simulation.}
\end{figure}

\begin{example}
The homogeneous coordinate system
is very common in 3D computer graphics, especially in games.
This is the case of, e.g., First Person Perspective (FPP) shooters
(like \textit{Doom}, \textit{Wolfenstein}, \textit{Duke Nukem 3D}, \textit{Quake}, or \textit{Counter-Strike})
or flight simulators.
Figure~\ref{Fig:Naturenv} gives a screenshot of an untextured terrain mesh in an
exemplary 3D world simulation.

In such a setting, an agent is most often represented as a point $(0,0,0)$ and faces
towards the $(1,0,0)$ vector. Here, the translation vector $\vect{t}$ may designate
the current position of an agent. The affine $\vect{A}$ matrix provides
the direction in which it looks. This is often provided by a composition of 3 rotation matrices given via the Euler angles
-- roll (OX rotation $\gamma$; unused in FPP shooters), pitch (OY  rotation $\beta$, look up/down),
and yaw (OZ rotation $\alpha$, turn left/right):
\[\small
\left[
\begin{array}{ccc}
\cos\alpha \cos\beta & \cos\alpha \sin\beta \sin\gamma - \sin\alpha \cos\gamma & \cos\alpha \sin\beta \cos\gamma + \sin\alpha \sin\gamma \\
\sin\alpha \cos\beta & \sin\alpha \sin\beta \sin\gamma + \cos\alpha \cos\gamma & \sin\alpha \sin\beta \cos\gamma - \cos\alpha \sin\gamma\\
-\sin\beta           & \cos\beta  \sin\gamma                                   & \cos\beta  \cos\gamma \\
\end{array}
\right].
\]

It is worth noting that modern graphics cards take advantage of homogeneous coordinates
when a programmer implements vector and matrix algebra (e.g., via vertex shaders).
\package{OpenGL} and \package{Direct3D} libraries allow for efficient data processing with 4-element (\texttt{float}) registers.
\end{example}

\begin{table}[p!]
\caption{\label{Table:affine2dex} Exemplary affine transformations in $\mathbb{R}^2$.}

\centering
\begin{tabularx}{1.0\linewidth}{p{4.5cm}p{4.5cm}p{3.4cm}}
\toprule
\bf\small description & \multicolumn{2}{l}{\bf\small transformation $\vect{x}\mapsto \vect{A}\vect{x}+\vect{t}$} \\
\midrule
Translation & $ \vect{A}=\left[\begin{array}{cc}
1 & 0 \\
0 & 1 \\
\end{array}\right]
$
&
$
\vect{t} = \left[
\begin{array}{c}
t_1 \\
t_2 \\
\end{array}
\right]
$
\\
\midrule
Rotation by $\vartheta$ &
$
\vect{A}=\left[\begin{array}{cc}
\cos\vartheta & -\sin\vartheta \\
\sin\vartheta & \cos\vartheta \\
\end{array}\right]
$
&
$
\vect{t} = \left[
\begin{array}{c}
0 \\
0 \\
\end{array}
\right]
$
\\
\midrule
Uniform scaling by $s$ &
$
\vect{A}=\left[\begin{array}{cc}
s & 0 \\
0 & s \\
\end{array}\right]
$
&
$
\vect{t} = \left[
\begin{array}{c}
0 \\
0 \\
\end{array}
\right]
$
\\
\midrule
Horizontal shear by $m_x$ &
$
\vect{A}=\left[\begin{array}{cc}
1 & m_x \\
0 & 1 \\
\end{array}\right]
$
&
$
\vect{t} = \left[
\begin{array}{c}
0 \\
0 \\
\end{array}
\right]
$
\\
\midrule
Reflection against $OX$ & $ \vect{A}=\left[\begin{array}{cc}
1 & 0 \\
0 & -1 \\
\end{array}\right]
$
&
$
\vect{t} = \left[
\begin{array}{c}
0 \\
0 \\
\end{array}
\right]
$
\\
\bottomrule
\end{tabularx}
\end{table}

\begin{table}[p!]
\centering
\caption[Exemplary fusion functions and some properties they fulfill.]{\label{Tab:extransmultidim}
Exemplary fusion functions and some properties they fulfill:
M -- componentwise monotonicity, T -- translation,
uS -- uniform scale, $d$S -- $d$-scale, O~--~orthogonal, and A -- affine equivariance.}

\centering
\begin{tabularx}{1.0\linewidth}{Xcccccc}
\toprule
\bf\small function                & \bf\small M & \bf\small T & \bf\small uS & \bf\small $d$S & \bf\small O & \bf\small A \\
\midrule
$\func{CwAMean}$                  & \pt  & \pt  & \pt   & \pt     & \pt  & \pt \\
$\func{CwMedian}$                 & \pt  & \pt  & \pt   & \pt     & \pf  & \pf \\
$\func{1center}_{\mathfrak{d}_2}$ & \pf  & \pt  & \pt   & \pf     & \pt  & \pf \\
$\func{1median}_{\mathfrak{d}_2}$ & \pf  & \pt  & \pt   & \pf     & \pt  & \pf \\
$\func{TkMedian}$                 & \pf  & \pt  & \pt   & \pt     & \pt  & \pt \\
\bottomrule
\end{tabularx}
\end{table}

\clearpage

\paragraph{Affinitization.}\index{affinitization}%
It turns out that each fusion function $\func{F}$ may be easily modi\-fied
so that it fulfills affine equivariance. This may be done via
the \index{transformation-retransformation}\emph{transformation-retransformation} technique, see \cite{ChaudhuriSengupta1993:signtestmultidim},
Let us fix a set of $d$ unique indices, $\mathcal{I}=\{i_0,i_1,i_2,\dots,\allowbreak i_d\}\subseteq[n]$
and take:
\begin{equation}
\mathbf{A}_\mathcal{I} = [ \vect{x}^{(i_1)}-\vect{x}^{(i_0)}\ \dots\ \vect{x}^{(i_d)}-\vect{x}^{(i_0)} ]\in\mathbb{R}^{d\times d}.
\end{equation}
Assuming that the $\mathbf{A}_\mathcal{I}$ matrix is invertible, it can be treated
as the basis matrix for a \index{data-driven coordinate system}\emph{data-driven coordinate system} in which a transformed version of our
input data set is:
\[\vect{Y} = \mathbf{A}_\mathcal{I}^{-1}\, \vect{X}.\]

Thus, a modified version of a componentwise extension of $\func{G}:\mathbb{R}^n\to\mathbb{R}$,
$\func{AffG}$, may be given by:
\begin{equation}
   \func{AffG}(\vect{X}) = \mathbf{A}_\mathcal{I}\,
      \func{CwG}\left(\mathbf{A}_\mathcal{I}^{-1}\, (\vect{X}-\func{CwAMean}(\vect{X}))\right)
      + \func{CwAMean}(\vect{X}).
\end{equation}
Such a construct is general, and its special case for $\func{G}=\func{Median}$
was first proposed by Chakraborty and Chaudhuri \cite{ChakrabortyChaudhuri1996:transretrans}
(see also \cite{MottonenETAL2010:astheospatialmed} for a discussion on
affinitization of the 1-median).
This time, unfortunately, the resulting fusion function is no longer symmetric.

\begin{example}
   As a summary, Table~\ref{Tab:extransmultidim} lists the properties fulfilled
   by idempotent fusion functions discussed so far.
   We see that the componentwise mean meets all of them.
\end{example}

\section{Idempotence, internality, and weak monotonicity}

First of all, let us note that if $\func{G}$ is idempotent, then
its componentwise extension $\func{CwG}$ is also idempotent
in the sense that for any $\vect{x}\in\mathbb{R}^d$
we have $\func{CwG}(n\ast\vect{x})=\vect{x}$.

\index{internality}%
More generally, if $\func{G}$ is internal (recall Proposition~\ref{Proposition:internalVSidempotent}), then
$\func{CwG}(\vect{x}^{(1)},\allowbreak\dots,\allowbreak\vect{x}^{(n)})$ is in the (axis-aligned) bounding box (orthotope, hyperrectangle)
of $\vect{X}$:
\[
\Bigg[\bigwedge_{i=1}^n x^{(i)}_1, \bigvee_{i=1}^n x^{(i)}_1\Bigg]
\times\dots\times
\Bigg[\bigwedge_{i=1}^n x^{(i)}_d, \bigvee_{i=1}^n x^{(i)}_d\Bigg].
\]

\begin{remark}
The above property is not necessarily an attractive generalization of
ordinary internality: it seems to be too weak.
Let $d=2$, $n=3$ and consider
$\vect{x}^{(1)} = (1,0)$, $\vect{x}^{(2)} = (0,0)$,
$\vect{x}^{(3)} = (0,1)$.

For instance, if $\func{G}(\vect{y})=\frac{1}{n}\sum_{i=1}^n y_i$,
then $\func{CwG}(\vect{x}^{(1)}, \vect{x}^{(2)}, \vect{x}^{(3)})=(\frac{1}{3}, \frac{1}{3})$.
However, if $\func{G}(\vect{y})=\bigvee_{i=1}^n y_i$,
then $\func{CwG}(\vect{x}^{(1)}, \vect{x}^{(2)}, \vect{x}^{(3)})=(1, 1)$.
\end{remark}

In  multivariate data analysis  and computational geometry, the use of convex hulls
\index{convex hull}%
is quite natural, see \cite{LiuPareliusSingh1999:multivaranal}.
To recall, the convex hull $\mathrm{CH}(\vect{x}^{(1)},\dots,\allowbreak\vect{x}^{(n)})$
of a finite set of points is the smallest convex set (polytope)
that includes all the provided points. Equivalently, it is the set
\index{convex combination}%
of all convex combinations of
$\vect{x}^{(1)},\dots,\vect{x}^{(n)}$:
\[
    \mathrm{CH}(\vect{x}^{(1)},\dots,\vect{x}^{(n)}) =    \Big\{
   \sum_{i=1}^n w_i\mathbf{x}^{(i)}: \\
   \text{for all vectors }\vect{w}\ge\mathbf{0}\text{ with }\sum_{i=1}^n w_i=1
   \Big\}.
\]
The convex hull may be determined algorithmically.
For example, if $d\in \{2,3\}$, the Chan algorithm \cite{Chan1996:chull}
has $O(n\log h)$ time complexity,
where $h$ is the number of vertices of $\mathrm{CH}(\vect{X})$.
On the other hand, if $d>3$, then an $O(n^{\lfloor d/2\rfloor})$ algorithm
exists \cite{Chazelle1993:chull}.

Let us now introduce a new type of internality.

\begin{definition}
\index{CH-internality|see {convex-hull based internality}}\index{convex-hull based internality}%
A fusion function $\func{F}$ is \emph{CH-internal} if and only if
for all $\vect{X}$  we have that $\func{F}(\vect{X})\in\mathrm{CH}(\vect{X})$.
\end{definition}

Please note that for $d=1$ we have
$\mathrm{CH}(\vect{X})=[\bigwedge_{i=1}^n x^{(i)}_1, \bigvee_{i=1}^n x^{(i)}_1]$,
i.e., it is the smallest real interval containing all the input samples.
Because of that, for univariate fusion functions, the
CH-internality and ordinary internality coincide.

One may wonder about the relationship between the CH- and bounding box-based
internality. Of course, each CH-internal function fulfills the straightforward
extension of ordinary internality. However, e.g., $\func{CwMedian}$
is bounding box- but not CH-internal. The following result
states that these two notions are equivalent when rotation equivariant fusion
functions are concerned.

\begin{proposition}\label{Prop:rotequichull}
Let $\func{F}:(\mathbb{R}^{d})^n\to\mathbb{R}^{d}$ be rotation %
equivariant and such that for any $\mathbf{X}$ it holds that $\func{F}(\mathbf{X})$ is in the bounding box
of $\mathbf{X}$. Then $\func{F}(\mathbf{X})\in\mathrm{CH}(\mathbf{X})$.
\end{proposition}

A simple proof of this proposition is based on the fact that
the convex hull is equivariant to rotations %
and that it is a subset of the bounding box.
Moreover, the convex hull may be expressed as the intersection of appropriate
halfspaces \cite{Edelsbrunner1987:algcomputgeom}.
$\vect{X}$ may always be rotated so that any convex hull's face
is aligned within the axes. Then the hyperplane that includes
such a face coincides with the hyperplane including the bounding box's
face.

\bigskip
As for monotonicity,
we already noted that  componentwise nondecreasingness is problematic.
Instead, however, we may consider a straightforward componentwise extension of
weak monotonicity, compare Definition~\ref{Def:weakmonotonicity}\index{weak monotonicity}.

\begin{definition}
A fusion function $\func{F}:(\mathbb{R}^{d})^n\to\mathbb{R}^d$ is \emph{weakly monotone}
whenever $\func{F}(\vect{X}+\vect{t})\ge_d \func{F}(\vect{X})$
for any $\vect{t}\ge_d (d\ast 0)$ and $\vect{X}\in(\mathbb{R}^{d})^n$.
\end{definition}

Surely, every translation equivariant fusion function is weakly monotone,
but the converse is not necessarily true.

\medskip
Here is a ``multidimensional'' counterpart of
Proposition \ref{Prop:CompositionProps}.

\begin{proposition}\label{Prop:CompositionProps2}
Let $\func{F}:(\mathbb{R}^d)^{k}\to\mathbb{R}^d$ for some $k$, $\func{G}_1,\dots,\func{G}_k:(\mathbb{R}^d)^n\to\mathbb{R}^d$,
and $\func{H}:(\mathbb{R}^d)^{n}\to\mathbb{R}^d$ be given by $\func{H}(\vect{X})=\func{F}(\func{G}_1(\vect{X}),\dots,\func{G}_k(\vect{X}))$
for $\vect{X}\in(\mathbb{R}^d)^n$.
\begin{itemize}
   \item If $\func{F},\func{G}_1,\dots,\func{G}_k$ are idempotent (respectively,
   bounding box-internal, CH-internal, translation, uniform scale,
   $d$-scale, orthogonal, affine equivariant),
   then $\func{H}$ is also idempotent (respectively,
   bounding box-internal, CH-internal, and so forth).

   \item If $\func{F}$ is weakly monotone and
   $\func{G}_1,\dots,\func{G}_k$ are translation equivariant, then
   $\func{H}$ is weakly monotone.
\end{itemize}
\end{proposition}

Note that a different form of monotonicity could also be defined by requiring that
if $(\forall i\in[n])$ $\vect{x}^{(i)}\le_d\vect{y}^{(i)}$,
then $\func{F}(\vect{X})\not>_d\func{F}(\vect{Y})$.
However, it is not even fulfilled by the Euclidean 1-median, compare Example~\ref{Ex:1mediannotmonotone}.

\section[Data depth, corresponding medians, and ordering of inputs]%
{Data depth, corresponding medians,\newline and ordering of inputs}

The purpose of the notion of data depth is to measure  how ``central'' or ``deep'' a point $\vect{y}$
is with respect to a point cloud $\vect{X}$.
It may be used, e.g., to visualize (mostly bivariate) data sets
\cite{LiuPareliusSingh1999:multivaranal},
compute statistical hypothesis tests \cite{ChenouriSmall2012:testdepth,LiLiu2004:nonpartest},
 design control charts, and even support decision making \cite{RousseuwRuts1999:popdistrdepth}.
It has been studied extensively by data analysts and computer scientists.

What is crucial to us in this monograph is that
with any depth notion, its corresponding multidimensional median may be defined,
which may serve as a robust estimator of location, see
\cite{Small1990:medianssurv,RousseeuwStruyf2004:robuststats,Aloupis2006:datadepth}
for some surveys on the topic. A depth-based median is a point of the maximal
depth (or the center of gravity of a set of points
of maximal depth, if there is no single point with such a property).

Let us assume that  $\vect{X}$ is a $d$-dimensional data set in \index{regular position}\emph{regular position},
i.e., with no more than $d$ points lying in a $(d-1)$-dimensional subspace.
In particular, in the bivariate  case, we have that no more than two observations are colinear.
In the following paragraphs we review the most notable data depth notions
(like Tukey's, Liu's, and Oja's) and their corresponding affine equivariant medians.
Later on we shall note that the concept of data depth leads to orderings of
the input points, which will enable us to define new, quite interesting fusion functions.

\index{data depth}%
It is assumed that the depth of a point $\vect{y}\in\mathbb{R}^d$ relative to $\vect{X}\in(\mathbb{R}^{d})^n$
is quantified via a bounded function
$
\func{depth}:\mathbb{R}^d\times(\mathbb{R}^{d})^n\to[0,b]
$
for some~$b$.
Zuo and Serfling in \cite{ZuoSerfling2000:gennotstdepth} list some desirable properties
that this notion should fulfill, namely, for any $\vect{X}$ and~$\vect{y}$ they require:
\begin{itemize}
   \item affine invariance\footnote{Note that in this book we made a clear distinction between
equivariance and invariance to specific transformations.}:
   for all $\vect{A}\in\mathbb{R}^{d\times d}$ of full rank and $\vect{t}\in\mathbb{R}^d$:
   \[\func{depth}(\vect{A}\vect{y}+\vect{t}; \vect{A}\vect{X}+\vect{t})
   =\func{depth}(\vect{y}; \vect{X}),\]
   \item monotonicity relative to the deepest point:
   if $\vect{y}=\sup_\vect{y} \func{depth}(\vect{y}; \vect{X})$,
   then for all $\vect{z}$ and $\alpha\in[0,1]$ it holds:
   \[\func{depth}(\vect{z}; \vect{X}) \le  \func{depth}(\alpha\vect{y} + (1-\alpha)\vect{z}; \vect{X}),\]
   \item vanishing at infinity: $\func{depth}(\vect{y}; \vect{X})\to 0$
   as $\|\vect{y}\|\to \infty$.
\end{itemize}
Note that depth notions are often considered in a statistical environment,
so other properties may additionally be of interest, e.g.,
maximality at center: $\sup_\vect{y} \func{depth}(\vect{y}; \vect{X})=\func{depth}(\boldsymbol\mu; \vect{X})$
where $\boldsymbol\mu$ is the point of symmetry of the empirical distribution of $\vect{X}$ (if it exists), etc.

\subsection{Tukey's halfplane location depth and median}

In 1974, Tukey \cite{Tukey1974:mathpict} introduced
the concept of the \emph{depth}\label{Def:depth} of a value $y$ with respect to
a unidimensional set of points $\vect{x}=(x_1,\dots,x_n)$.
It is defined as the minimum number of data points
from $\vect{x}$ on the left and on the right of~$y$:
\begin{equation}
\func{tdepth}_1(y; x_1,\dots,x_n) = |\{i: x_i\le y\}|\wedge |\{i: x_i\ge y\}|.
\end{equation}
The Tukey depth is related to the observations' ranking.
The sample minimum and maximum are the points of depth 1,
the median is of depth $n/2$ (the ``deepest'' value), and the first and the third quartiles
are of depth $n/4$.
As noted in \cite{DonohoGasko1992:breakdown}, one can define trimmed means
by, say, averaging points of depth $\ge n/10$.
This notion has been used to develop robust regression
techniques, see, e.g., \cite{RousseeuwHubert1999:regressiondepth}.

As a matter of fact, Tukey in the same paper \cite{Tukey1974:mathpict} introduced
a generalization of this idea too. \emph{The halfplane location depth}
of $\vect{y}$ relative to $\vect{X}$ is the smallest number of points in
$\vect{X}$ contained in any closed halfhyperplane
with boundary line through $\vect{y}$. In other words, see also \cite{DonohoGasko1992:breakdown}:

\begin{definition}\index{Tukey depth}%
The \emph{Tukey depth} of $\vect{y}\in\mathbb{R}^d$ relative to $\vect{X}\in(\mathbb{R}^{d})^n$
is an integer such that:
\begin{eqnarray*}
\func{tdepth}_d(\vect{y};\vect{x}^{(1)},\dots,\vect{x}^{(n)})
&=&\min_{\vect{u}\in\mathbb{R}^d, \|\vect{u}\|=1} \func{tdepth}_1\left(\vect{u}^T\vect{y}; \vect{u}^T\vect{X}\right)\\
&=&\min_{\vect{u}\in\mathbb{R}^d, \|\vect{u}\|=1} |\{i: \vect{u}^T\vect{x}^{(i)}\ge \vect{u}^T\vect{y}\}|.
\end{eqnarray*}
\end{definition}

\begin{remark}\index{projection pursuit}%
Multidimensional Tukey depth is defined via projection pursuit, see \cite{Huber1985:projectionpursuit}.
It results in applying all possible one-dimensional projections of the data set to a line
and computing the univariate Tukey depth.
\end{remark}

It is easily seen that a set of all points of depth $\ge \delta$
(a $\delta$-depth contour), for any given $\delta>0$, is either empty or is a convex polytope
(e.g., polygon for $d=2$).

Note that a point outside the convex hull of $\vect{X}$ is
always of depth~$0$ \cite{RousseeuwRuts1998:bivartukmed}.
On the other hand, for all $\vect{y}$ we have
$\func{tdepth}_d(\vect{y};\vect{X}) \le n$.
In fact, we may be slightly more precise about the upper limit for $d=2$.

\begin{proposition}
If $\vect{X}$ is a bivariate data set in {regular position}, then
the maximal Tukey depth, $\delta=\max_{\vect{y}'\in\mathbb{R}^d}\func{tdepth}_d(\vect{y}';\vect{X})$,
fulfills:
\[
   \left\lceil \frac{n}{3} \right\rceil \le \delta \le \left\lfloor\frac{n}{2}\right\rfloor.
\]
\end{proposition}
The upper bound was proved by Rousseeuw and Ruts in \cite{RousseeuwRuts1998:bivartukmed}
while the lower bound was given by Donoho and Gasko \cite{DonohoGasko1992:breakdown}.
More generally, for any $d$, by \cite{DonohoGasko1992:breakdown},
we have that $\left\lceil \frac{n}{d+1} \right\rceil \le \delta \le \left\lceil\frac{n}{2}\right\rceil$.

Note that the deepest point might not be uniquely defined.
In order to overcome this issue, we may consider the following fusion function.

\begin{definition}
\index{Tukey median}%
Let $\mathcal{R}$ be the deepest Tukey depth region with respect to given $\vect{X}\in\mathbb{R}^{d\times n}$, i.e.,
$\mathcal{R}=\left\{\vect{y}\in\mathbb{R}^d:
   \func{tdepth}_d(\vect{y};\vect{X})=\delta
   \right\}$, where $\delta=\max_{\vect{y}'\in\mathbb{R}^d}\func{tdepth}_d(\vect{y}';\vect{X})$.
The center of gravity of such a region:
\begin{equation}
   \func{TkMedian}(\vect{X})=\frac{\int_{\mathbb{R}^d}  x \indicator(x\in\mathcal{R})\,dx}{\int_{\mathbb{R}^d}  \indicator(x\in\mathcal{R})\,dx },
\end{equation}
is called the \emph{Tukey median} of $\vect{X}$.
\end{definition}

For $d=1$, the Tukey median generalizes the concept of a \index{median}median.
Thus, in higher dimensions this fusion function can be thought of as a multidimensional median.

\begin{figure}[t!]
\centering

\includegraphics[width=8.25cm]{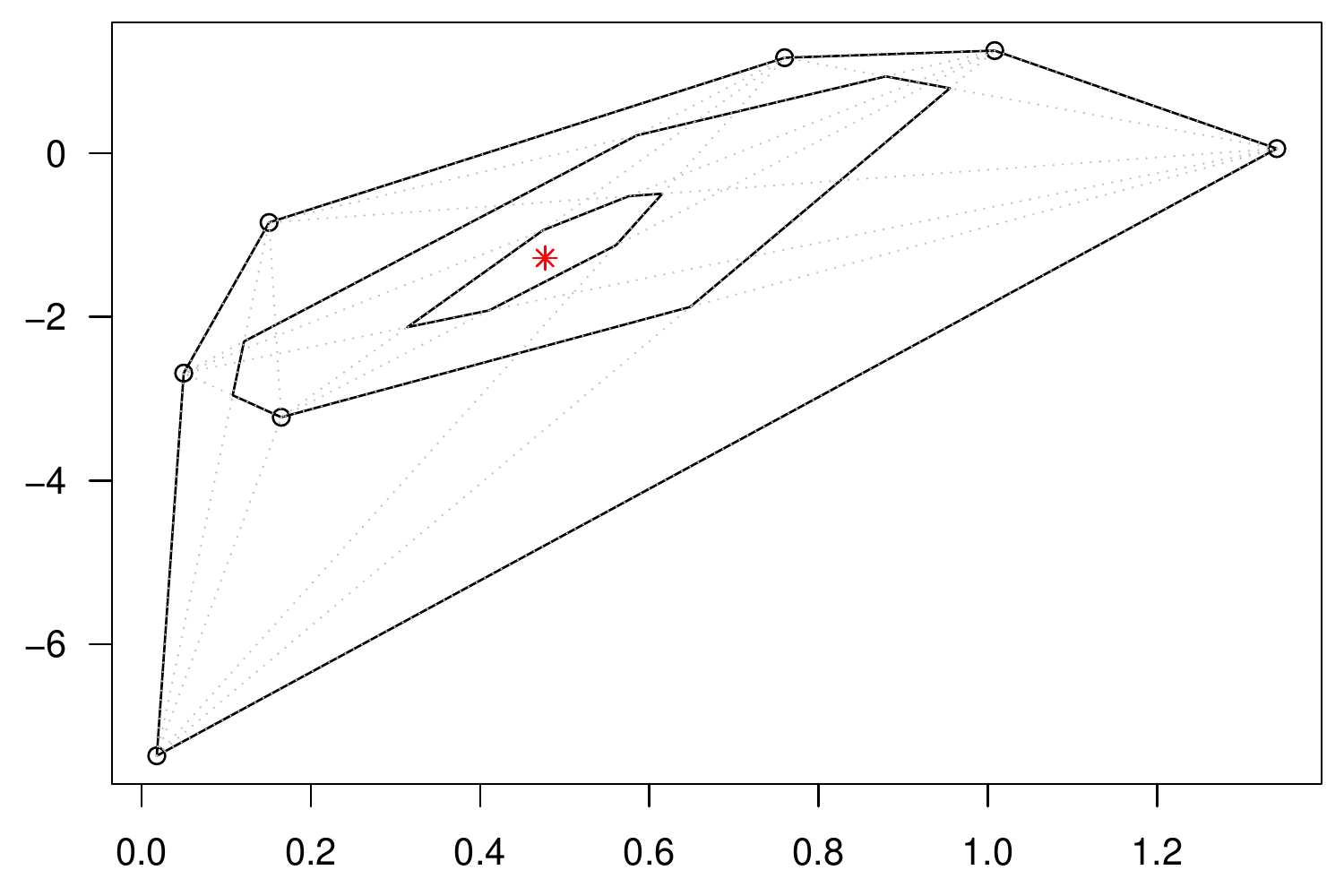}

\caption[Tukey depth contours and Tukey median of a data set.]%
{\label{Fig:tdepth} Tukey depth contours and Tukey median $(\ast)$ of a data set.}
\end{figure}

\begin{example}
Figure \ref{Fig:tdepth} depicts an exemplary data set, the three Tukey depth contours,
and the center of gravity of the deepest Tukey depth region, i.e., the Tukey median.
\end{example}

\begin{remark}\index{center of gravity}%
Let $(x_1,\dots,x_m)$ and $(y_1,\dots,y_m)$ be coordinates of a convex polygon in $\mathbb{R}^2$,
ordered clockwise. Then its center of gravity, $(C_x, C_y)$, is given by:
\begin{eqnarray*}
C_x & = & \frac{\sum_{i=1}^m (x_i+x_{i+1})(x_i y_{i+1}-x_{i+1} y_i)}{3\sum_{i=1}^m (x_i y_{i+1}-x_{i+1} y_i)} , \\
C_y & = & \frac{\sum_{i=1}^m (y_i+y_{i+1})(x_i y_{i+1}-x_{i+1} y_i)}{3\sum_{i=1}^m (x_i y_{i+1}-x_{i+1} y_i)} ,  \\
\end{eqnarray*}
where, for brevity of notation, $x_{m+1}=x_1$ and $y_{m+1}=y_1$.
For $d>2$, e.g., one may perform a Delaunay triangulation of a given convex polytope
and calculate sums of appropriate integrals (for each simplex independently).
\end{remark}

\begin{figure}[t!]
\centering

\includegraphics[width=8.25cm]{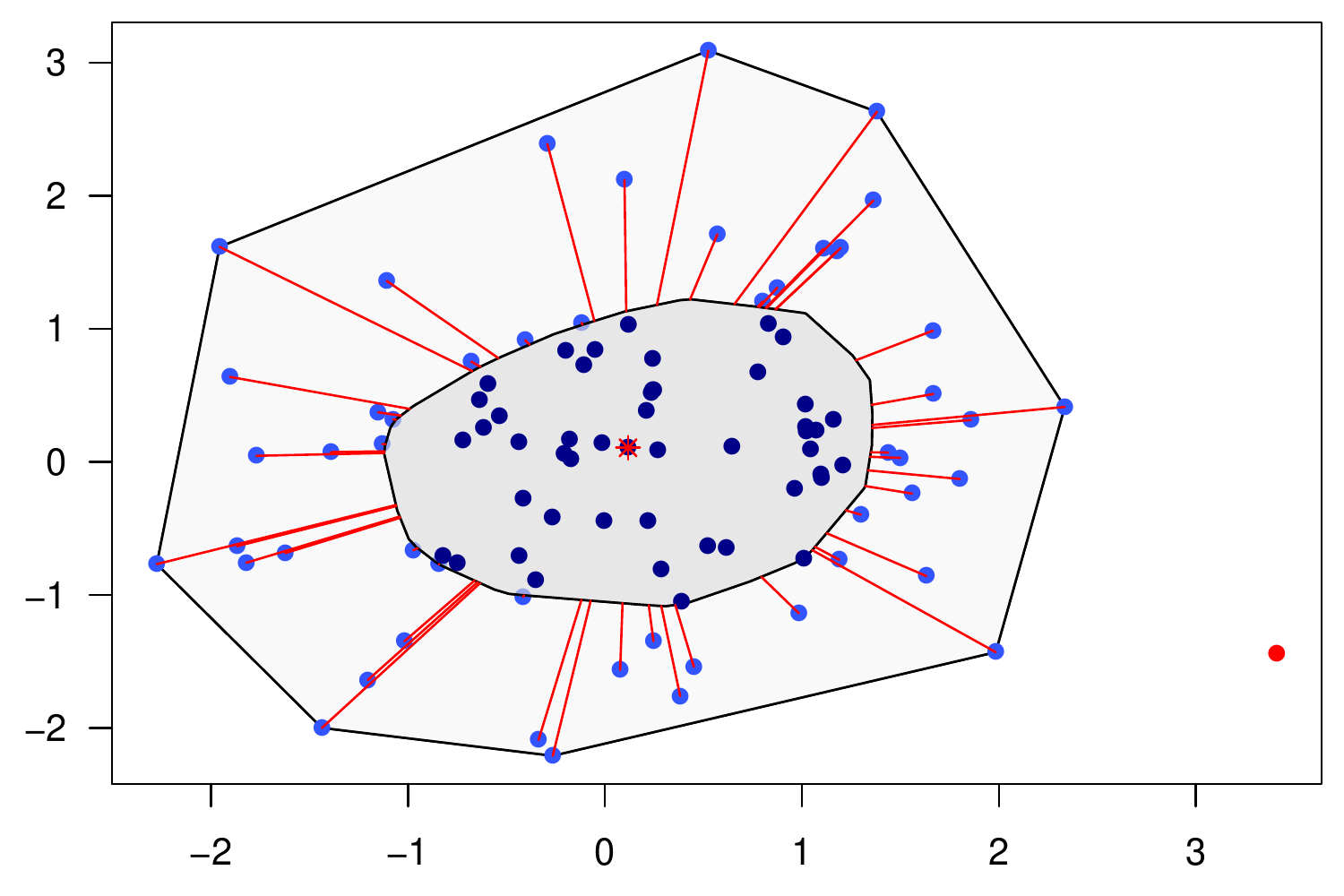}

\includegraphics[width=8.25cm]{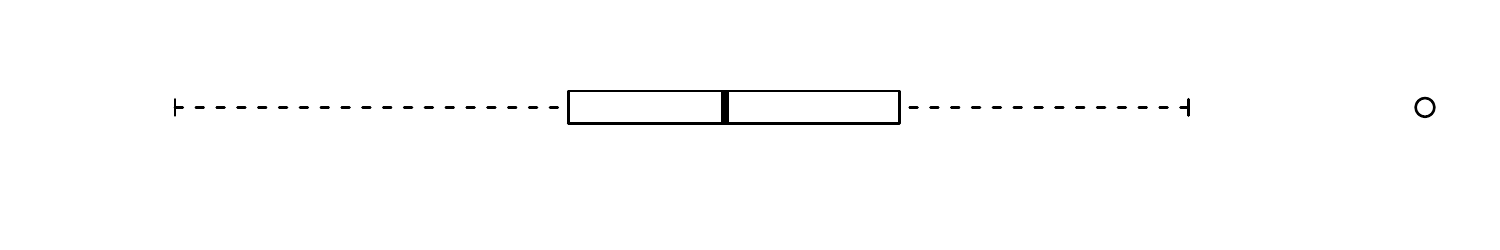}

\caption{\label{Fig:bagplot} A bagplot of a bivariate data set and a boxplot of
its projection onto $OX$ generated with R (\texttt{aplpack::bagplot}).}
\end{figure}

\begin{example}\index{bagplot}%
A \emph{bagplot}, a bivariate version of the box-and-whisker plot, is based on the discussed
notions, see Figure~\ref{Fig:bagplot}.
It consists of the Tukey median, a \textit{bag} that contains 50\% of the data points
(it results in a linear interpolation of two Tukey depth regions),
and a \textit{fence} that separates inliers form outliers (originally,
an inflated version of the bag scaled by a factor of 3).
For more details the reader is referred to \cite{RousseeuwRutsTukey1999:bagplot}.
\end{example}

\begin{remark}
The Tukey depth, as well as its corresponding median,
is affine equivariant, see \cite[Lemma 2.1]{DonohoGasko1992:breakdown}.
Moreover, the Tukey depth is monotonic relative to the deepest point
and vanishes at infinity.
\end{remark}

For $d=2$, a na\"{i}ve algorithm to compute the Tukey depth requires $O(n^2)$ time.
However, in \cite{RousseuwRuts1996:as307} an optimal (see \cite{AloupisETAL2002:lowerboundsdepth})
$O(n\log n)$ algorithm \texttt{LDEPTH} was given.
It is implemented in \R's \texttt{depth} package and available via a call to \texttt{depth(..., method="{}Tukey"{})}.
For $d=3$ there exists a $O(n^2\log n)$ exact algorithm, see \cite{RousseeuwStruyf1998:comptlocd}.
For larger $d$, there is an approximate Monte Carlo-type algorithm,
also provided in \cite{RousseeuwStruyf1998:comptlocd}.

\begin{algorithm}\label{Alg:TukeyDepth}
Here is how we may approximate $\func{tdepth}_d(\vect{y};\vect{x}^{(1)},\dots,\vect{x}^{(n)})$
for arbitrary $d$, see \cite[Section 2.3]{RousseeuwStruyf1998:comptlocd}.
\begin{enumerate}
   \item[1.] Let $D:=n$;
   \item[2.] Repeat $m$ times (for a given $m$):
   \begin{enumerate}
   \item[2.1.] Draw a random sample of size $d$ from $\mathrm{U}\{\vect{x}^{(1)},\dots,\vect{x}^{(n)}\}$;
   \item[2.2.] Determine a direction $\vect{u}$ perpendicular to the above subset;
   \item[2.3.] Project the points in $\vect{X}$ to the line $L$ through $\vect{y}$ with direction~$\vect{u}$;
   \item[2.4.] Compute the univariate Tukey depth $D'$ of $\vect{y}$ on ${L}$;
   \item[2.5.] Set $D:= D\wedge D'$;
   \end{enumerate}
   \item[3.] Return $D$ as result;
\end{enumerate}
\end{algorithm}

A point with the largest Tukey depth
(there may be many such points) may be found in expected
$O(d\log n)$ time  for $d=2$ and in $O(n^{d-1})$ expected time
for $d\ge 3$, see \cite{Chan2004:opttukeydepth}.
For $d=2$ the \package{ISODEPTH} \cite{RutsRousseeuw1996:pointcloud}
algorithm determines vertices of a depth contour in $O(n^2 \log n)$ time
and the \package{HALFMED} \cite{RousseeuwRuts1998:bivartukmed}
algorithm for computing the Tukey median is $O(n^2 \log^2 n)$.
Additionally, in \cite{LangermanSteiger2003:maxdepth}
we may find an algorithm to compute a high depth point (not necessarily
the median) in $O(n\log^2 n)$ and a lower bound for this task $\Omega(n\log n)$.
The interested reader is referred to \cite{Aloupis2006:datadepth,BremnerETAL2008:outputsensitivetukey}
for further discussion on Tukey depth-related algorithms.

\subsection{Liu's simplical depth and median}

Note that for $i_1,\dots,i_{d+1}\in[n]$, if $\vect{x}^{(i_1)},\dots,\vect{x}^{(i_{d+1})}$
are affinely independent, then the convex hull of $d+1$ points,
$\mathrm{CH}\left(\vect{x}^{(i_1)},\dots,\vect{x}^{(i_{d+1})}\right)$, defines a $d$-dimensional simplex.
In particular, for $d=2$, $\mathrm{CH}\left(\vect{x}^{(i_1)},\dots,\vect{x}^{(i_{3})}\right)$
simply designates a triangle. Another notion of data depth of a point $\vect{y}$
relative to $\vect{X}$ is by Liu \cite{Liu1990:simplicesdepth}.
In the bivariate case it is defined as the number of triangles formed by any three elements
in $\vect{X}$ that contain $\vect{y}$.
Intuitively, a ``deep'' or ``central'' point is of large Liu depth.
More generally, we have what follows.

\begin{definition}\index{Liu depth|see {simplical depth}}\index{sdepth@$\mathsf{sdepth}$|see {simplical depth}}%
\index{simplical depth}%
The Liu \emph{simplical depth} of $\vect{y}\in\mathbb{R}^d$
with respect to $\vect{X}\in\mathbb{R}^{d\times n}$ is defined as:
\[
   \func{sdepth}(\vect{y};\vect{x}^{(1)},\dots,\vect{x}^{(n)}) = \frac{\left|\left\{
     \{i_1,\dots,i_{d+1}\}: \vect{y}\in \mathrm{CH}\left(\vect{x}^{(i_1)},\dots,\vect{x}^{(i_{d+1})}\right)
   \right\}\right|}{\displaystyle{n \choose d+1}}.
\]
\end{definition}
It is easily seen that the Liu depth is affine invariant,
that is its result does not change under arbitrary affine transformations.

\begin{remark}
Recall that it is easy to check whether
$\vect{y}\in \mathrm{CH}(\vect{x}^{(i_1)},\dots,\vect{x}^{(i_{d+1})})$.
For a nondegenerate simplex, it suffices to solve $\vect{y}=[\vect{x}^{(i_1)}\,\dots\,\vect{x}^{(i_{d+1})}]\,\boldsymbol\alpha$
for $\boldsymbol\alpha$ under the constraint $\sum_{i=1}^{d+1} \alpha_i=1$
and verify whether $\boldsymbol\alpha\ge_{d+1}\mathbf{0}$.
\end{remark}

\index{SMedian@$\mathsf{SMedian}$|see {simplical median}}%
\index{simplical median}%
The \emph{simplical median}, $\func{SMedian}$, may be defined similarly to the
Tukey median -- as a point with the greatest simplical depth or the center of
gravity of the deepest Liu depth region.
This leads to an affine equivariant fusion function.

For $d=2$, a straightforward algorithm to compute the simplical depth of
a point requires $O(n^3)$ time.
An optimal (see \cite{AloupisETAL2002:lowerboundsdepth}) $O(n\log n)$
algorithm for that very purpose
was proposed in \cite{RousseuwRuts1996:as307}.
It is available in \R{}  via a call to \texttt{depth::depth(..., method="{}Liu"{})}.
Moreover, for $d=2$ there exists an $O(n^4)$ time  algorithm for finding
the simplical median, see \cite{AloupisETAL2003:algbivmed}.

\subsection{Oja's  depth and median}

The {Oja depth} (also known as the simplical volume depth) has been introduced in
\cite{Oja1983:descrstatmultivar}.
Here we provide its slightly transformed version, as given in \cite{LiuPareliusSingh1999:multivaranal},
since the original definition is not compatible with the aforementioned
depth measures: we would like to assure that a central point is of the greatest depth.

\begin{definition}\index{simplical volume depth|see {Oja depth}}\index{Oja depth}%
\index{odepth@$\mathsf{odepth}$|see {Oja depth}}%
The \emph{Oja depth} of $\vect{y}\in\mathbb{R}^n$ with respect to $\vect{X}\in(\mathbb{R}^{d})^n$
is given~by:
\[
   \func{odepth}(\vect{y};\vect{x}^{(1)},\dots,\vect{x}^{(n)}) =
   {n \choose d}^{-1}
   \frac{1}{
     1+
     \displaystyle\sum_{\{i_1,\dots,i_{d}\}} \mathrm{vol}\left(\mathrm{CH}\left(\vect{y}, \vect{x}^{(i_1)},\dots,\vect{x}^{(i_{d})}\right)\right)
   },
\]
where $\mathrm{vol}(\cdot)$ designates the volume of a given simplex.
\end{definition}

In particular, for $d=2$, the Oja depth of a point is the sum of all the areas of triangles
formed by this point and two points in an input data set.

\begin{remark}
The volume of a nondegenerate $d$-dimensional simplex given by vertices
$\vect{x}^{(i_1)},\dots,\vect{x}^{(i_{d+1})}$
equals~to:
\[
\mathrm{vol}\left(\mathrm{CH}\left(\vect{x}^{(i_1)},\dots,\vect{x}^{(i_{d+1})}\right)\right)
=\mathrm{abs}\left(
\frac{1}{d!}\,
\mathrm{det}\left[
\begin{array}{cccc}
1 & 1 & \cdots & 1 \\
x_1^{(i_1)} & x_1^{(i_2)} & \dots & x_1^{(i_{d+1})} \\
\vdots & \vdots & \ddots & \vdots \\
x_d^{(i_1)} & x_d^{(i_2)} & \dots & x_d^{(i_{d+1})} \\
\end{array}
\right]
\right).
\]
\end{remark}

This depth measure is not affine invariant.
It is because for an affine transformation $T(\vect{x})=\vect{A}\vect{x}+\vect{t}$
we have:
\[\mathrm{vol}(T(\vect{x}^{(i_1)}),\dots,T(\vect{x}^{(i_{d+1})}))
=\mathrm{abs}(|\vect{A}|)\,\mathrm{vol}(\vect{x}^{(i_1)},\dots,\vect{x}^{(i_{d+1})}),\]
see \cite[Lemma 2.1]{Oja1983:descrstatmultivar}.
However, its corresponding median, defined as a point with the maximum depth, is affine equivariant.

\begin{definition}
\index{OMedian@$\mathsf{OMedian}$|see {Oja median}}%
\index{Oja median}%
The \emph{Oja median} of $\vect{X}\in\mathbb{R}^{d\times n}$ is given by:
\begin{equation}
   \func{OMedian}(\vect{X})=\argmin_{\vect{y}\in\mathbb{R}^d}
   \displaystyle\sum_{\{i_1,\dots,i_{d}\}} \mathrm{vol}(\mathrm{CH}(\vect{y}, \vect{x}^{(i_1)},\dots,\vect{x}^{(i_{d})})).
\end{equation}
\end{definition}

\begin{remark}\index{median}%
Note that the Oja median generalizes the one-dimensional median,
which is a point $y$ that minimizes $\sum_{i} |x_i-y|$.
\end{remark}

For $d=2$ there exists an $O(n \log^3 n)$ algorithm \cite{AloupisETAL2003:algbivmed} for finding the Oja median
and an $O(n\log n)$ algorithm \cite{AloupisMcleish2005:lowerboundojadepth} for computing the Oja depth of a given point,
see also \texttt{depth::depth(..., method="{}Oja"{})} in \R.
Another algorithm for finding the Oja median for any $d$ was proposed in
\cite{RonkainenETAL2003:compojamed} and runs in $O(dn^d\log n)$ time.

Note that in \cite{NiinimaETAL1990:breadownOja} a generalization of the Oja median
was proposed:
\begin{equation}
\func{OMedian}_\alpha(\vect{X})=\argmin_{\vect{y}\in\mathbb{R}^d}
\displaystyle\sum_{\{i_1,\dots,i_{d}\}} \mathrm{vol}(\mathrm{CH}(\vect{y}, \vect{x}^{(i_1)},\dots,\vect{x}^{(i_{d})}))^\alpha
\end{equation}
for some predefined $\alpha\in[1,2]$.

\subsection{Other depth notions}

Below we list some other approaches for defining data depth.

\paragraph{$L_1$ depth.}\index{L1 depth@$L_1$ depth}
The $L_1$ depth by Vardi and Zhang \cite{VardiZhang2000:multivarL1}
is closely connected to the 1-median given by Equation~\eqref{Eq:1median}.
In particular, $L_1$ depth is maximized at the 1-median.

\paragraph{Projection depth.}\index{projection depth}%
\index{pdepth@$\mathsf{pdepth}$|see {projection depth}}%
The projection depth by Zuo \cite{Zuo2003:projectionbaseddepth}
is a generalization of the concept by Donoho and Gasko \cite{DonohoGasko1992:breakdown}, see also
\cite{RousseeuwStruyf2004:robuststats}.
\index{projection pursuit}%
A measure of outlyingness of a point $\vect{y}$ is given via projection pursuit:
\begin{equation}
   \func{O}(\vect{y}; \vect{X}) = \sup_{\|\vect{u}\|=1} \frac{|\vect{u}^T\vect{y} - \func{F}(\vect{u}^T\vect{y})|}{\func{V}(\vect{u}^T\vect{y})},
\end{equation}
where $\func{F}$ is some unidimensional aggregation function (e.g., median)
and $\func{V}$ is some spread measure (e.g., median absolute deviation, see Section~\ref{Sec:SpreadMeasures}).
This leads to a depth measure:
\begin{equation}
   \func{pdepth}(\vect{y}; \vect{X}) = \frac{1}{1+\func{O}(\vect{y}; \vect{X})}.
\end{equation}

\paragraph{Perihedral depth.}\index{perihedral depth}%
Perihedral depth, see \cite{Durocher2014:combdepth}, is given
by the number of subsets of $\vect{X}$ whose convex hulls
contain $\vect{y}$.

\paragraph{Convex hull peeling depth.}\index{convex hull peeling depth}%
The convex hull peeling depth (see \cite{Eddy1982:chpeel};
according to \cite{Huber1972:lecture} the idea was
proposed by Tukey) is determined by consecutively computing
a convex hull of a set of points and removing values lying outside its boundary.
The corresponding median may be constructed
by computing the center of gravity of the ``last'' convex hull.
According to \cite{Aloupis2006:datadepth},
convex hull peeling may be done in $O(n\log^2 n)$ time for $d=2$.

\paragraph{Delaunay depth.}\index{Delaunay depth}%
Recall that a Delaunay triangulation of $\vect{X}$ is a triangulation (tessellation) $\mathrm{DT}(\vect{X})$
such that all points in $\vect{X}$ are not in circum-hyperspheres of any simplices in $\mathrm{DT}(\vect{X})$.
The Delaunay depth (according to \cite{AbellanasClaverolHurtado2007:delaunaydepth}
introduced by Green in \cite{Green1981:peelingbivariatedata})
of $\vect{y}$ with respect to $\vect{X}$ is the length of the shortest path in $\mathrm{DT}(\vect{X})$
from $\vect{y}$ to the convex hull of $\vect{X}$.

\begin{figure}[htb!]
\centering

\includegraphics[width=8.25cm]{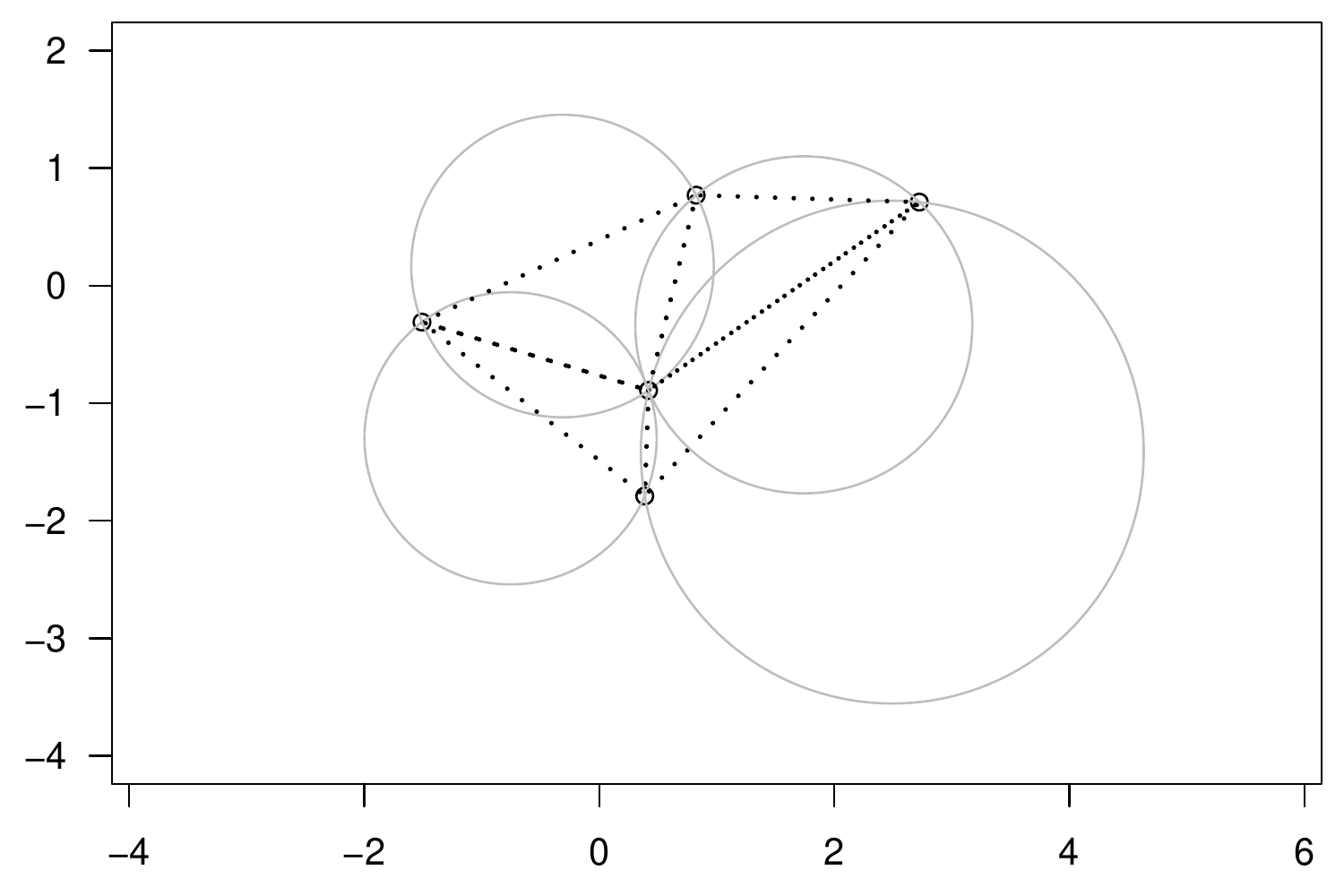}

\caption{\label{Fig:delaunay} Delaunay triangulation of an exemplary bivariate set of points,
together with circumcircles of all the triangles in the given tesselation.}
\end{figure}

\begin{example}
Figure \ref{Fig:delaunay} depicts an exemplary Delaunay triangulation
of a set of 5 points in $\mathbb{R}^2$. For each of the triangles in the triangulation,
its circumcircle is also plotted.
\end{example}

In point of fact, the Delaunay depth is a member of a special class called \textit{proximity depths},
defined as the number of edges in a proximity graph that must be visited
to reach $\mathrm{CH}(\vect{X})$.

\paragraph{Zonoid data depth.}\index{zonoid data depth}%
\index{zdepth@$\mathsf{zdepth}$|see {zonoid data depth}}%
The {zonoid data depth}, see \cite{DyckerhoffETAL1996:zonoiddd},
of $\vect{y}\in\mathbb{R}^d$ with respect to $\vect{X}$ is defined as:
\begin{equation}
   \func{zdepth}(\vect{y};\vect{X}) = \sup\{\alpha\in[0,1]: \vect{y}\in D_\alpha(\vect{X})\},
\end{equation}
with convention $\sup\{\emptyset\}=0$,
where $D_\alpha$ is the \emph{$\alpha$-trimmed region} \cite{KoshevoyMosler1997:zonoidtrimming}
of the empirical distribution generated by $\mathbf{X}$,
i.e.:
\begin{equation}
   D_\alpha(\vect{X})=\left\{
   \sum_{i=1}^n w_i\vect{x}^{(i)}: \sum_{i=1}^n w_i=1,\ (\forall i)\ w_i\ge 0,\ \alpha w_i \le 1/n
   \right\}.
\end{equation}
Note that for $\alpha\in[0,1/n]$ we have $D_\alpha=\mathrm{CH}(\mathbf{X})$.
Moreover, $D_1$ is a singleton containing the centroid of $\vect{X}$
and for $\alpha<\alpha'$ we have $D_{\alpha'}\subset D_\alpha$.
Of course, if $\vect{y}\not\in \mathrm{CH}(\vect{X})$, then
$\func{zdepth}(\vect{y};\vect{X})=0$.
The zonoid data depth fulfills some important properties:
it is affine invariant, continuous with respect to $\vect{y}$ and each $\vect{x}^{(i)}$,
and monotone. The computation of depth of a given point may be reduced to a linear
programming task, see \cite{DyckerhoffETAL1996:zonoiddd}.
However, the point with the greatest zonoid depth corresponds
to $\func{CwAMean}(\vect{X})$.

For a summary of other depth notions, such as the Mahalanobis depth,
majority depth, or the likelihood depth, see
\cite{LiuPareliusSingh1999:multivaranal}.

\begin{remark}
Here is a possible application of the concept of data depth
in regression analysis.
Assume that we are given $\vect{X}\in{(\mathbb{R}^{d})^n}$, $\vect{y}\in\mathbb{R}^n$
and we would like to fit a hyperplane $H_{\boldsymbol\vartheta}$, $\boldsymbol\vartheta\in\mathbb{R}^{d+1}$,
defined by $y=\vartheta_1 x_1+\dots+\vartheta_d x_d+\vartheta_{d+1}$,
such that $H_{\boldsymbol\vartheta}(\vect{X})$ is as close to $\vect{y}$ as possible.

The \index{regression depth}\emph{regression depth}
(introduced by Rousseeuw and Hubert in \cite{RousseeuwHubert1999:regressiondepth})
of $H_{\boldsymbol\vartheta}$ relative
to $\vect{X}$ and $\vect{y}$ is defined as the smallest number of indices like $i$ such that the residual
$r_i = \vartheta_1 x_1^{(i)}+\dots+\vartheta_d x_d^{(i)}+\vartheta_{d+1}-y_i$
needs to change its sign to make $H_{\boldsymbol\vartheta}$ \emph{nonfit},
i.e., there exists a hyperplane $V$ such that no $\vect{x}^{(i)}$
is on $V$, $r_i>0$ for all $\vect{x}^{(i)}$ in one of $V$'s open halfspace
and $r_i<0$ for all $\vect{x}^{(i)}$ in the other halfspace.

Intuitively, it is the smallest number of observations in $\vect{X}$
that would need to be removed in order
to make a computed regression model a \emph{nonfit}. It measures how well a
hyperplane fit represents data: a good fit is of larger depth than a bad one.
Thus, a fit with large depth is well-balanced relative to the input data.

There is an exact  $O(n\log n)$-time algorithm for computing the Tukey-based regression depth
for the case $d=1$, see \cite{RousseeuwHubert1999:regressiondepth}.
It was extended to arbitrary $d$ in \cite{RousseeuwStruyf1998:comptlocd}, but its
time complexity is $O(n^{d} \log n)$; obviously, for large $d$ and $n$ such a routine is practically unusable.
However, an approximate approach, similar to the one in Algorithm \ref{Alg:TukeyDepth},
may be used in such a case, see also \cite{RousseeuwStruyf1998:comptlocd,RousseeuwETAL1999:regressiondepthrejoinder}.
There is also an algorithm to compute hyperplanes with the greatest
depths \cite{vanKreveldETAL2008:effalgomaxregdepth}.
\end{remark}\ignorespaces

\subsection{Symmetrization of fusion functions}

Recall that a fusion function is symmetric, whenever for all permutations
$\sigma$ of $[n]$ it holds $\func{F}(\vect{x}^{(1)},\dots,\vect{x}^{(n)})
=\func{F}(\vect{x}^{\sigma(1)},\dots,\vect{x}^{\sigma(n)})$.

Given a non-symmetric unidimensional function, one may easily symmetrize it
by referring to the notion of an order statistic, i.e., the $i$th smallest
value among a set of input elements.
It is because, by Proposition~\ref{Prop:symmetrization},
$\func{F}:\mathbb{R}^{n}\to\mathbb{R}$ is symmetric if and only if there exists
a function $\func{G}:\mathbb{R}^n\to\mathbb{R}$ such that:
\[\func{F}(\vect{x}^{(1)},\dots,\vect{x}^{(n)})=\func{G}(\vect{x}^{(\sigma(1))},\dots,\vect{x}^{(\sigma(n))}),\]
where $\sigma$ is an ordering permutation of the input values.
In such a way, e.g., a weighted arithmetic mean becomes an OWA operator.
Such a construction is only valid, however, in the $d=1$ case, as here
a natural linear order $\le$ is defined, see \cite{LiuPareliusSingh1999:multivaranal} for discussion.

In other words, if $d>1$, then it is not easy to determine which values are
``small'' or ``large'', especially if we allow a set of points
to be orthogonally transformed.

\begin{figure}[t!]
   \centering
   \includegraphics[width=12cm]{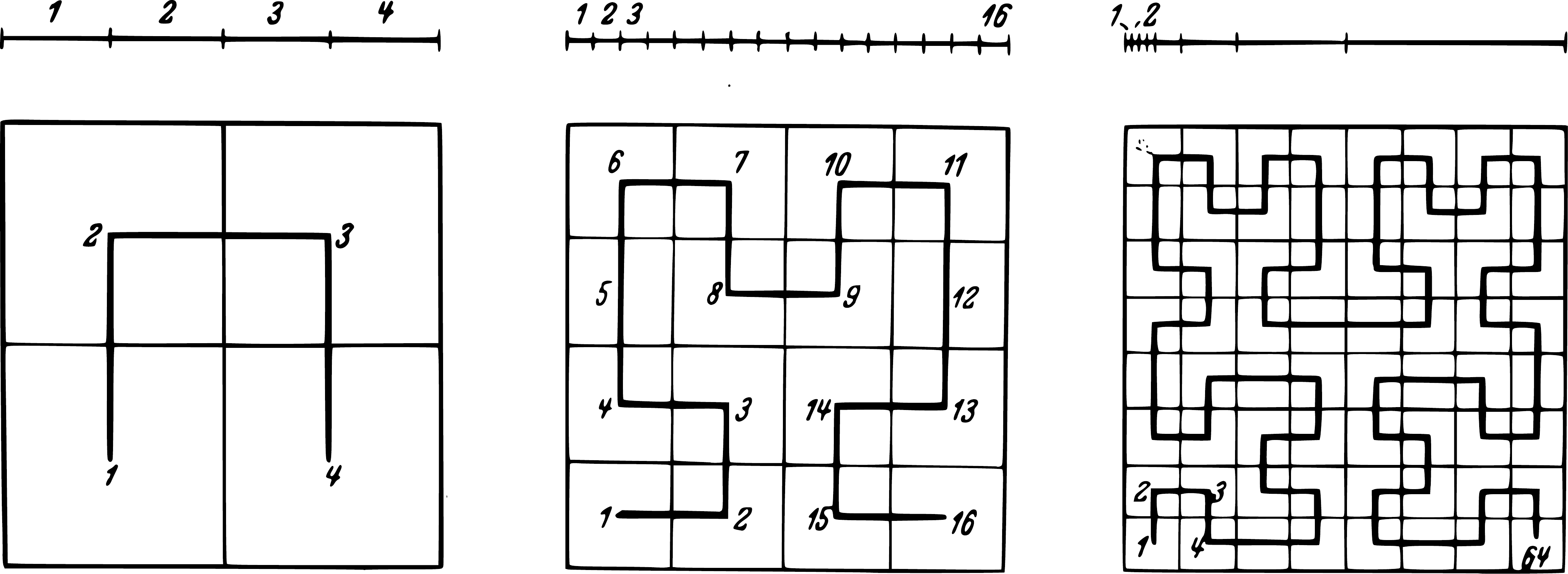}

   \caption{\label{Fig:HilbertCurve} Three iterations of the Hilbert curve creation process as depicted in
   Hilbert's original 1891 paper \cite{Hilbert1891:flaschenstuck}.}
\end{figure}

One possible way to order a set of points in $\mathbb{R}^2$ is to use one of
the so-called \emph{plane-filling curves}.
For instance, let us consider the fractal-like Hilbert curve.
Its building process is recursive and its  first few steps are depicted
in Figure~\ref{Fig:HilbertCurve}.
A set of points may be sorted by considering the order in which they appear on
such a plane-filling curve. Notably, the \package{CGAL} \cite{cgal:eb-14b}
library has effective procedures to do so, also in higher dimensions.
Such a sorting scheme may be used to speed up some geometric algorithms.
Unfortunately, it is easily seen that the resulting ordering is neither
translation nor, e.g., rotation invariant (but it might be made translation
and uniform scale invariant by transforming the input data set).

Another way to sort a multivariate data set is to
order the input values with respect to increasing distances from a fixed point,
e.g., the set's componentwise mean. If the Euclidean distance is used, the introduced
sorting scheme shall be affine equivariant. Yet, it might not be unique for some data sets.
If ties occur, one may first order the observations relatively using the same ordering
as in the input data set (this may be easily done by applying a stable sort algorithm).

More elaborate approaches may be based on the concept of
data depth. With these, the points $\vect{x}^{(i)}$, $i=1,\dots,n$, may be ordered with
respect to their decreasing or increasing depths. In other words, we may make use of a permutation
$\sigma$ of $\{1,\dots,n\}$ such that
$\sigma(i)\le\sigma(j)$ implies that for $i<j$:
\[\func{depth}(\vect{x}^{(\sigma(i))}; \vect{x}^{(1)}, \dots, \vect{x}^{(n)})
\le \func{depth}(\vect{x}^{(\sigma(j))}; \vect{x}^{(1)}, \dots, \vect{x}^{(n)}),\]
where $\func{depth}$ is some data depth measure.
In this way, we get so-called \emph{depth order statistics}, see \cite{LiuPareliusSingh1999:multivaranal}.
Note that, unlike in the univariate case, they are not ordered from the ``smallest'' to the ``largest'',
but from the ``most central'' to the ``least central''.

Having an ordered version of the input set of points,
one may easily define, e.g., multidimensional versions of
trimmed or Winsorized means, see \cite{Masse2009:mtrimeanTukeydepth}.

\section{Penalty-based fusion functions}\label{Sec:PenaltyMultidim}

At the very beginning of this chapter, we introduced
some notable fusion functions: the componentwise mean, 1-median,
and 1-center (with respect to the Euclidean metric).
Let us now discuss them, as well as their generalizations, in greater detail.

\subsection{1-median}\label{Sec:1medianfixedd}

Circa~1650, Evangelista Torricelli proposed a solution to a problem posed by Pierre
de Fermat in the early 17th century: given three points in a plane,
find the fourth point for which the sum of its distances to the three
given points is as small as possible (compare \cite{KrarupVajda1997:Torricelli}).
This task can be formulated for an arbitrary number of points as follows.
Find $\vect{y}$ such that:
\begin{equation}
\func{1median}_{\mathfrak{d}}(\vect{x}^{(1)},\dots,\vect{x}^{(n)})
= \argmin_{\vect{y}\in\mathbb{R}^{d}} \frac{1}{n} \sum_{i=1}^n \mathfrak{d}(\vect{x}^{(i)}, \vect{y}),
\end{equation}
where $\mathfrak{d}$ is a metric (originally the Euclidean one). Such a point, called in the literature
\index{geometric median|see {Euclidean 1-median}}%
\index{spatial median|see {Euclidean 1-median}}%
the \emph{1-median}, geometric median, spatial median, mediancenter, $L_1$-median, Fermat-Weber, or Torricelli point,
generalizes the concept of a one-dimensional median (i.e., for $d=1$ it is equal to $\func{Median}$
for arbitrary $L_p$ metric $\mathfrak{d}_p$ and odd $n$).

\paragraph{Euclidean metric.}
\index{Euclidean 1-median}If $\mathfrak{d}=\mathfrak{d}_2$, the Euclidean 1-median
is slightly less sensitive to outliers than the centroid ($\func{CwAmean}$,
see below), compare Figure~\ref{Fig:FermatWeber}.

\begin{figure}[htb!]
\centering

\includegraphics[width=8.25cm]{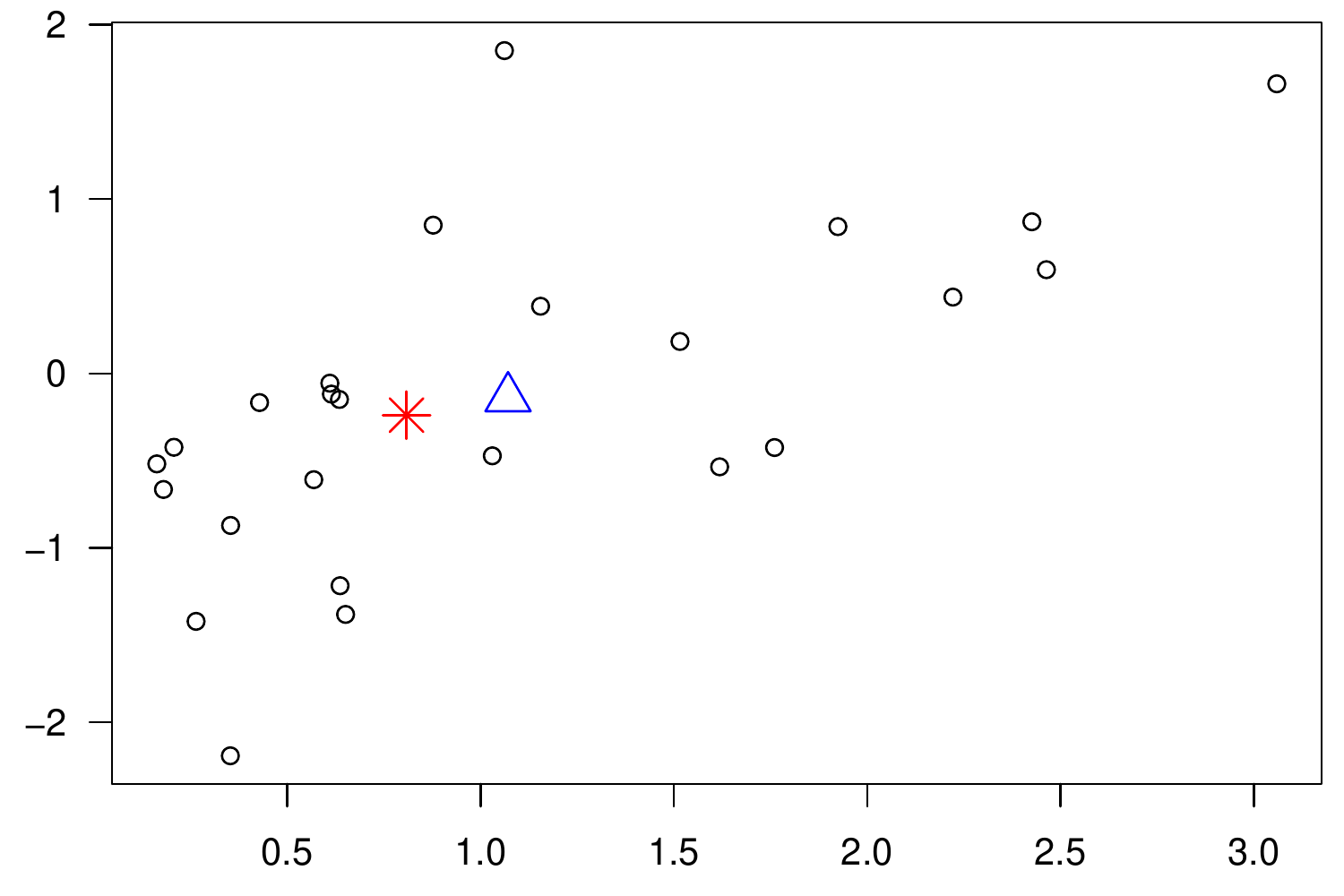}

\caption[1-median and centroid of an exemplary data set.]%
{\label{Fig:FermatWeber} 1-median ({\Large\textasteriskcentered})
and centroid ($\bigtriangleup$) of an exemplary data set.
By definition, 1-median is less sensitive to outliers.}
\end{figure}

In the unidimensional case, as noted above, the solution reduces to the sample median
and thus it might not be unique. However, for $d\ge 2$ and $\vect{X}$ such that
it is not concentrated on a line,
Milasevic and Ducharme showed \cite{MilasevicDucharme1987:spatialunique}
that the spatial median is always well-defined.

Note that Euclidean 1-median is sometimes used as an estimate of the underlying
multidimensional probability distribution's theoretical median.
Moreover, Brown in \cite{Brown1983:statusespatialmed} generalized
the two-sample statistical hypothesis sign  test for the equality of medians
in one dimension by using their spatial analogues (the angle test).

\paragraph{Weighted Euclidean metric.}
Let us consider a more general version of the above-presented case,
closely related to the Fermat-Weber problem, see, e.g., \cite{Brimberg1995:FermatWeberrevisited,Tellier1972:webersolint},
\index{Fermat-Weber problem}%
which aims at finding the location for a new facility that minimizes the sum of transportation costs
to $n$ destination points (e.g., customers), having in mind different
costs per unit distance.

\index{weighted Euclidean 1-median}%
Given a weighting vector $\vect{w}$, the \emph{weighted geometric median} is defined as:
\begin{equation}
\func{1median}_{\mathfrak{d}_2,\vect{w}}(\vect{X}) = \argmin_{\vect{y}\in\mathbb{R}^{d}} \sum_{i=1}^n w_i \mathfrak{d}_2(\vect{x}^{(i)}, \vect{y}),
\end{equation}
Unfortunately, in general, no analytic formula expressing the solution to the above equation exists,
even in the $(\forall i)\ w_i=1/n$ case.
By considering the partial derivatives of the above objective function,
it may be shown, see \cite{VardiZhang2000:multivarL1}, that it is a point $\vect{y}$ such that:
\begin{equation}\label{Eq:SpatialMedianStationaryPoint}
\sum_{i=1}^n \frac{w_i\vect{y}}{\mathfrak{d}_2(\vect{x}^{(i)}, \vect{y})}
= \sum_{i=1}^n \frac{w_i\vect{x}^{(i)}}{\mathfrak{d}_2(\vect{x}^{(i)}, \vect{y})}.
\end{equation}
However, from Equation~\eqref{Eq:SpatialMedianStationaryPoint}, we may derive the following
algorithm to compute the fusion function of  interest.

\begin{algorithm}\label{Alg:Weiszfeld}
\index{Weiszfeld procedure}Weiszfeld procedure \cite{Weiszfeld1937:FermatWeberalgorithm}:
\begin{enumerate}
   \item[1.] Choose a starting point $\vect{y}^{(0)}$ in the convex hull of $\{\vect{x}^{(1)},\dots,\vect{x}^{(n)}\}$;
   \item[2.] For $j=1,2,\dots$ do:
   \begin{enumerate}
      \item[2.1.] If $\vect{y}^{(j-1)} = \vect{x}^{(i)}$ for some $i\in[n]$, then let $\vect{y}^{(j)} := \vect{x}^{(i)}$;
      \item[2.2.] Otherwise, let
      $\vect{y}^{(j)} := \displaystyle\frac{\displaystyle\sum_{i=1}^n \frac{w_i \vect{x}^{(i)}}{\mathfrak{d}_2(\vect{x}^{(i)}-\vect{y}^{(j-1)})}}{%
{\displaystyle\sum_{i=1}^n \frac{w_i}{\mathfrak{d}_2(\vect{x}^{(i)}-\vect{y}^{(j-1)})}}}$;
      \item[2.3.] If $\mathfrak{d}_2(\vect{y}^{(j)}, \vect{y}^{(j-1)})\le \varepsilon$ for some fixed $\varepsilon>0$, then return $\vect{y}^{(i)}$ as result.
   \end{enumerate}
\end{enumerate}
\end{algorithm}
It may be shown, see \cite{Brimberg1995:FermatWeberrevisited}, that
the Weiszfeld algorithm converges to an optimal solution for all but
a countable set of starting points $\vect{y}^{(0)}$.
An exemplary implementation of the above algorithm is given in Figure~\ref{Fig:Weiszfeld},
see also its more robust version called SOR-Weiszfeld introduced in \cite{KarkkainenAyramo2005:compspatmedrobdm}
and the AS78 algorithm \cite{Gower1974:mediancentre} which is based on the steepest descent heuristic.

Note that Equation~\eqref{Eq:SpatialMedianStationaryPoint} implies that:
\begin{equation}\label{Eq:1medianconvexhull}
\vect{y} = \sum_{i=1}^n v_i \vect{x}^{(i)},\quad
v_i = \displaystyle\frac{\displaystyle\frac{w_i}{\mathfrak{d}_2(\vect{x}^{(i)}, \vect{y})}}{\displaystyle\sum_{i=1}^n \frac{w_i}{\mathfrak{d}_2(\vect{x}^{(i)}, \vect{y})}}.
\end{equation}
We see that $(v_1,\dots v_n)$ is a weighting vector. Hence,
the 1-median fulfills the convex hull-based internality.
Moreover, it
is orthogonal, uniform scale, and translation equivariant but not
$d$-scale and thus not affine equivariant, see \cite{MottonenETAL2010:astheospatialmed} for  discussion. %
Also, its symmetry depends solely on the form of the weighting vector $\vect{w}$.

\paragraph{Manhattan distance.}\index{componentwise median}%
Interestingly, it turns out that by setting $\mathfrak{d}$ to be the Manhattan
$\mathfrak{d}_1$ metric, we get the already mentioned
componentwise median, $\func{CwMedian}$,
see \cite{BedallZimmermann1979:mediancentre}.
Recall that this fusion function is nondecreasing, translation and $d$-scale
equivariant, but not rotation equivariant (note how the Manhattan distance
behaves under rotations).
Note that the $k$-medians algorithm (more robust to outliers than $k$-means)
was  originally based on the 1-median with respect to $\mathfrak{d}_1$.

\paragraph{Other Minkowski distances.}
There exists  a Newton-Raphson-like algorithm
\cite{BedallZimmermann1979:mediancentre} which computes the 1-median
in the case $\mathfrak{d}_p$ for arbitrary $p\ge 1$.
As a matter of fact, for moderate values of $p$ and sample sizes, this task may be easily
determined using a generic nonlinear optimization solver, for example:

\begin{lstlisting}[language=R]
one_median_Lp <- function(X, p) {
   optim(rowMeans(X), function(c) {
      sum(colSums(abs(X-c)^p)^(1/p))
   }, method="BFGS", control=list(reltol=1e-16))$par
}
\end{lstlisting}

\begin{figure}[htb!]
\centering

\includegraphics[width=8.25cm]{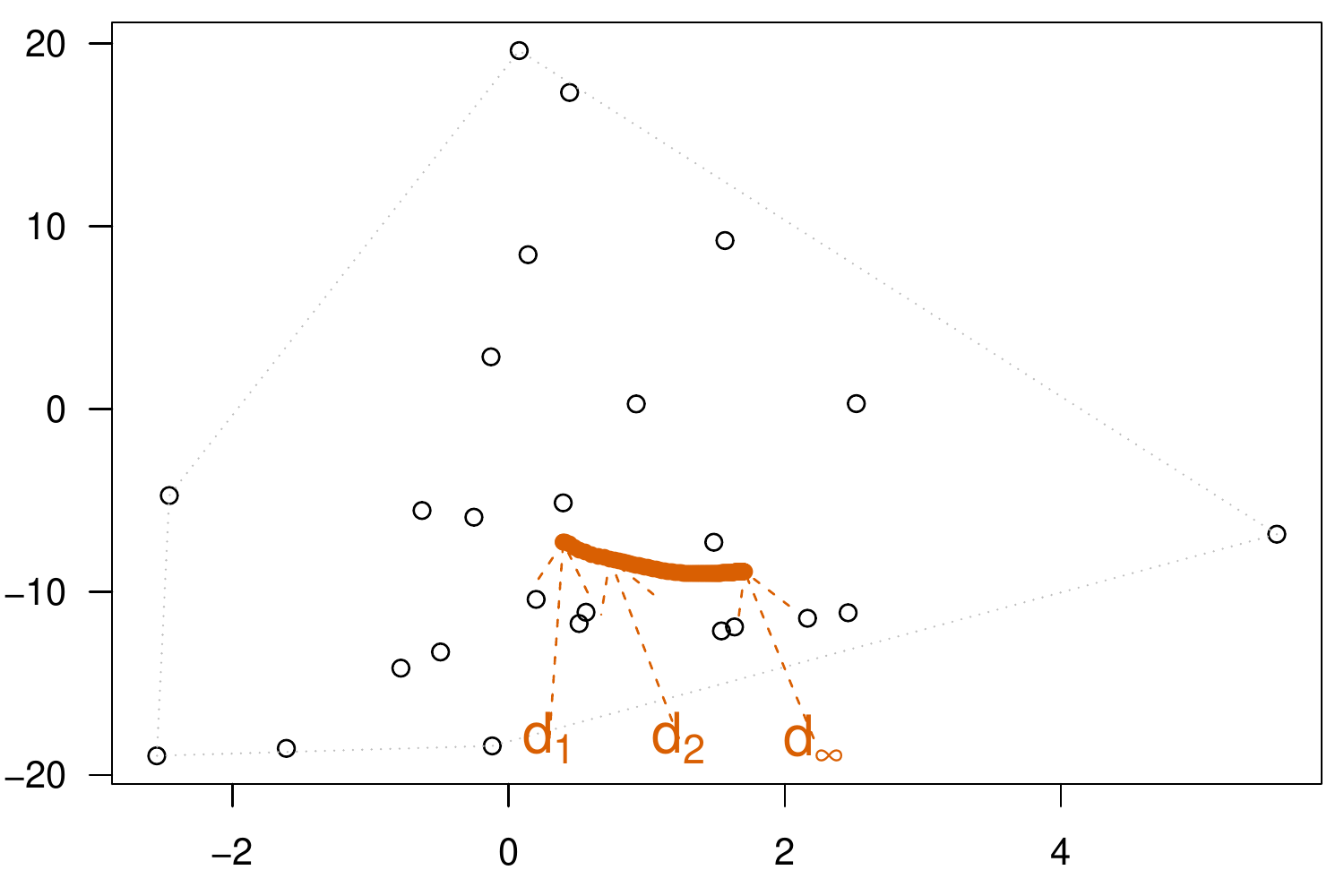}

\caption{\label{Fig:1median_trace} $\func{1median}_{\mathfrak{d}_p}$
trace as a function of $p\in[0,\infty]$.}
\end{figure}

\begin{example}Figure~\ref{Fig:1median_trace} depicts 1-median trace
of an exemplary data set. It is assumed that the 1-median is computed
with respect to Minkowski $\mathfrak{d}_p$ metrics and the trace is
generated by varying $p\in[1,\infty]$.
\end{example}

\subsection{Medoid}\label{Sec:medoid}

The 1-median should not be confused with the concept of a \emph{medoid} or set median,
\index{set median|see {medoid}}%
\index{medoid}%
\index{exemplar}%
which is a point $\vect{y}$ such that:
\begin{equation}\label{Eq:medoid}
\func{Medoid}_\mathfrak{d}(\vect{y}) = \argmin_{\vect{y}\in\{\vect{x}^{(1)},\dots,\vect{x}^{(n)}\}}
\frac{1}{n} \sum_{i=1}^n \mathfrak{d}(\vect{x}^{(i)}, \vect{y}),
\end{equation}
for arbitrary metric $\mathfrak{d}$ (usually Euclidean or Manhattan).
The difference is that we do not look among all the vectors in $\mathbb{R}^d$, but
restrict ourselves to the input data set (hence, the medoid is a kind
of \emph{exemplar}, compare one of possible definitions of internality on page~\pageref{Internality2}).
In other words, a medoid is a point in a given data set, for which average
dissimilarity to all the other objects in the set is minimal.

\begin{remark}
A medoid may be non-uniquely defined. This is the case
for $\mathfrak{d}_2$ and three vertices of an equilateral triangle
(or more generally, $d$ vertices of a regular $d$ simplex).
In such a situation, the computer science perspective is to choose
\emph{any} point that fulfills Equation~\eqref{Eq:medoid}.
Yet, for $\func{Medoid}_\mathfrak{d}$ to be a proper fusion function,
we should choose some method of distinguishing \emph{the} medoid of interest.
In particular, we may assume that we return one that has the smallest index $i$
among $\{i\in[n]: \vect{x}^{(i)}=\vect{y} \}$.
\end{remark}

Medoids are useful, e.g., in clustering problems (the $k$-medoids
algorithm, see \cite{ParkJun2009:simplefastkmedoids})
or as rough estimates of 1-medians.
They are internal as well as translation, uniform scale, and rotation equivariant
for $\mathfrak{d}=\mathfrak{d}_2$.

Note that we shall refer to this concept once again when discussing
aggregation in arbitrary pseudometric spaces, see Section~\ref{Sec:PseudometricSpace}.

\subsection{Centroid}\label{Sec:centroid}

\index{weighted centroid}%
Given a weighting vector $\vect{w}$, the \emph{weighted centroid}
is a point $\vect{y}$ such that:
\begin{equation}
\vect{y} = \argmin_{\vect{y}\in\mathbb{R}^{d}} \sqrt{\sum_{i=1}^n w_i \left(\mathfrak{d}_2(\vect{x}^{(i)}, \vect{y})\right)^2},
\end{equation}
where $\mathfrak{d}_2$ is the Euclidean metric.

Please notice the similarity between the above definition and the definition of
the weighted Euclidean 1-median. ${\mathfrak{d}_2}^2$ is of course no longer a metric, but
a kind of \emph{dissimilarity measure}. Due to this simplification
it turns out that the solution to the above equation is very easy:
it is the componentwise extension of the weighted arithmetic mean.
Thus, it is componentwise monotonic. Moreover, we already noted that
it is an affine invariant fusion function which fulfills convex hull-based
internality. Also note that the centroid minimizes the variance of distances
from the observations to itself.

The centroid is a basis for the $k$-means
clustering algorithm, see \cite{MacQueen1967:kmeans,Forgy1965:clusteranalkmeans}.
On the other hand, its weighted version is used in the fuzzy $c$-means
procedure \cite{Bezdek1981:fcm}.
In physics, the discussed notion reflects the center of mass of a system
of particles.

Notably, the centroid is a special case of the Fr\'{e}chet mean for $\mathfrak{d}=\mathfrak{d}_2$.

\subsection{1-center}

\index{Euclidean 1-center}%
\index{seb|see {smallest enclosing ball}}%
\index{smallest enclosing ball|see {Euclidean 1-center}}%
For a given metric $\mathfrak{d}$ the \emph{1-center}
(\emph{smallest enclosing ball}, seb)  problem aims at finding:
\begin{equation}\label{Eq:1center}
\func{1center}_\mathfrak{d}(\vect{x}^{(1)},\dots,\vect{x}^{(n)})
=\argmin_{\vect{y}\in\mathbb{R}^{d}} \bigvee_{i\in[n]} \mathfrak{d}(\vect{x}^{(i)}, \vect{y}).
\end{equation}
In particular, if $\mathfrak{d}$ is the Euclidean metric $\mathfrak{d}_2$, the above task
is called the Euclidean 1-center problem and was first proposed by
James Sylvester in 1857 \cite{Sylvester1857:questiongeom}.
Note that this task is not the same as finding the center of a circumscribed circle.

\begin{figure}[htb!]
   \centering
   \includegraphics[width=5.5cm]{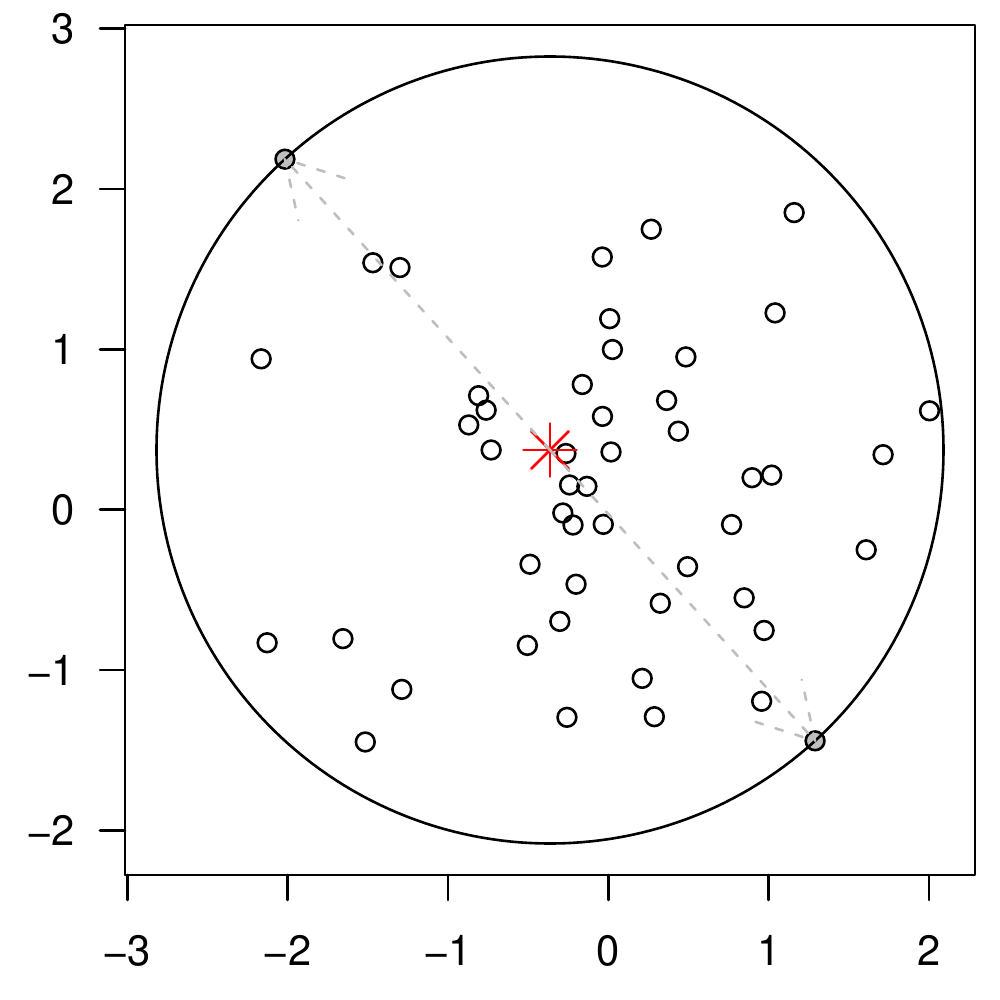}
   \includegraphics[width=5.5cm]{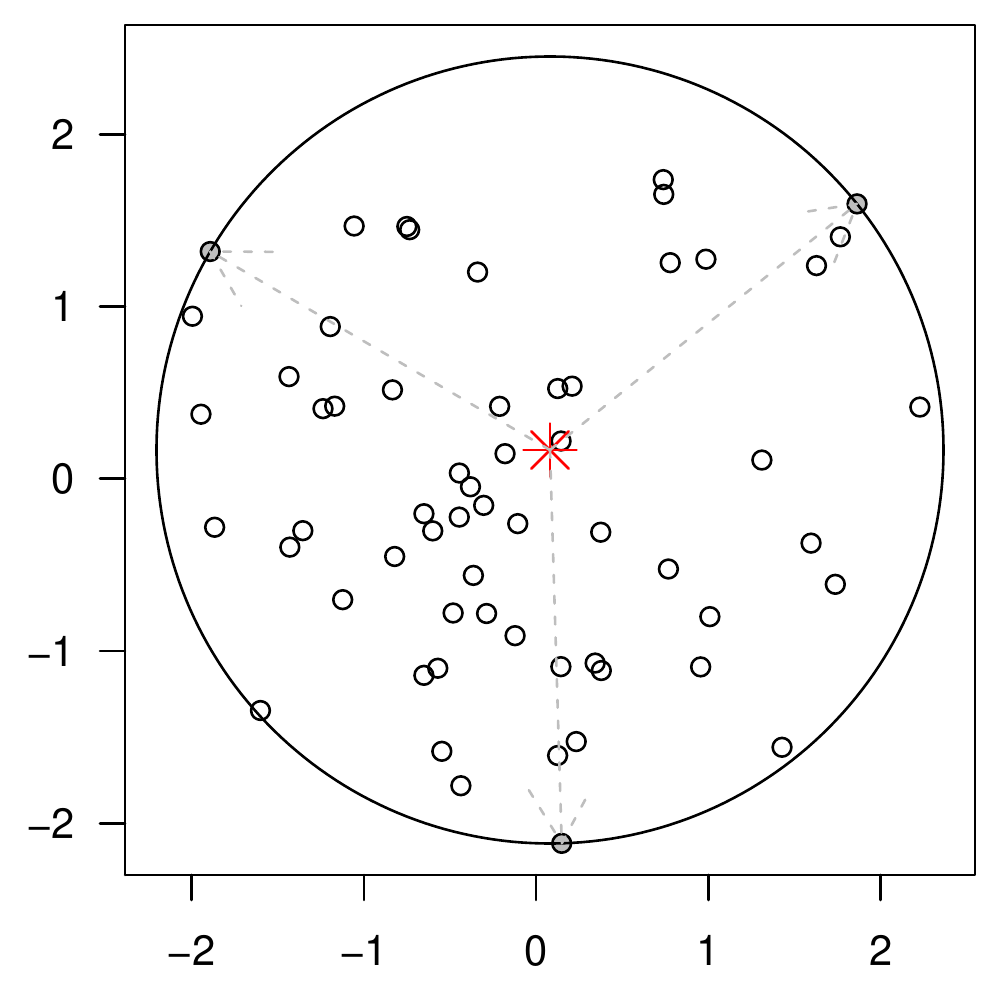}

   \caption{\label{Fig:sebexample} Euclidean 1-centers of two exemplary data sets.}
\end{figure}

Figure \ref{Fig:sebexample} depicts Euclidean 1-centers of two exemplary two-dimen\-sio\-nal
data sets. Such a formulation is used in many real-world
applications, see, e.g., \cite{Gartner1999:fastrobustsmallestencballs},
which include:
pattern recognition (finding reference points), computational biology (protein analysis),
support vector machines -- high-dimensional clustering,
and nearest neighbor search.
In particular, for $d=3$ these may be used in computer graphics, e.g., visibility culling,
ray tracing, and object collision detection.
However, we should be careful when using it in data analysis:
it is extremely sensitive to outliers.
What is more, for $d=2$ we have an important operational research application,
known as the facility location problem, when one aims to seek the location
of the distribution center that minimizes the distance
to a customer that is situated farthest away.

\bigskip
It may be shown, see \cite{GartnerSchonherr2000:qpball}, that the solution
to the Euclidean 1-center problem can be
expressed as:
\begin{equation}\label{Eq:sebform}
   \vect{y} = \sum_{i=1}^n v_i \vect{x}^{(i)} = \vect{v}\vect{X}^T,
\end{equation}
where the weighting vector $\vect{v}$
is computed by solving the quadratic programming (QP) problem:
\[
   \mathrm{minimize}\ \vect{v}^T \vect{X}^T\vect{X} \vect{v} - (\mathrm{diag}(\vect{X}^T\vect{X}))^T \vect{v}
   \quad \text{w.r.t.~}\vect{v}
\]
subject to:
\begin{eqnarray*}
\vect{1}^T\vect{v} & = & 1, \\
\vect{v} & \ge_n & \vect{0}.
\end{eqnarray*}
Note again that special care should be taken while choosing a software library
to compute this QP task, compare Remark~\ref{Remark:CGAL}.
For example, the \textsf{quadprog} package for R, which implements the dual method of Goldfarb
and Idnani \cite{GoldfarbIdnani1983:quadprog},
is only able to find a solution if $\vect{D}$ is positive-definite, which
-- in general -- is not our case.
Instead, for this task we may use a generic QP solver
given in Figures~\ref{Fig:CGAL_quadprog1} and \ref{Fig:CGAL_quadprog2},
which relies on the \package{CGAL} library.
Figure~\ref{Fig:seb_algo} gives an exemplary \package{Rcpp}
implementation of a routine to compute the {smallest enclosing~ball}.

A different, combinatorial algorithm (that resembles the simplex algorithm
for linear programming) has been proposed in
\cite{FischerGartnerKuts2003:fastencball}. Moreover, the \package{CGAL}
\cite{cgal:eb-14b} library includes an implementation of Welzl's
routine \cite{Welzl1991:smdisk}.

\bigskip
From Equation~\eqref{Eq:sebform} it follows that the Euclidean 1-center
is necessarily convex hull internal.
What is more, it is translation, orthogonal, and uniform scale equivariant
(but not $d$-scale equivariant).

\bigskip
On the other hand, the Chebyshev 1-center
is a componentwise extension of $\func{F}(\vect{x})=(\func{Max}(\vect{x})+\func{Min}(\vect{x}))/2$,
i.e., it is the center of the points' bounding rectangle.

Moreover, similarly to the concept of a medoid,
we may define a \index{seboid}\emph{seboid},
\index{exemplar}which is an exemplar
minimizing the function in Equation~\eqref{Eq:1center},
see Section~\ref{Sec:PseudometricSpace} for further information.

\subsection{A more general framework}\label{Sec:genpenaltyoptmultidim}

\index{penalty-based function}%
Similarly as in Definition~\ref{Def:PenaltyFunction}, we may introduce
the notion of a penalty-based fusion function for aggregation
of points in $\mathbb{R}^d$. This time, however, we should rather
assume that the set of minimizers of a penalty function $P:\mathbb{R}^d\times(\mathbb{R}^d)^{n}\to[0,\infty]$
is a convex polytope and that a $P$-based fusion function $\func{F}$
is given as the center of gravity of such a set.
For $d=1$, this setting generalizes the one from the previous chapter.
Surely, each idempotent function $\func{F}$ is a penalty-based one for some $P$.

\index{data depth}Let $\func{depth}$ be some data depth notion which is
bounded from above by a value $m\in[0,\infty]$.
\index{data depth-based penalty function}%
By setting $P(\vect{y};\vect{X})=m-\func{depth}(\vect{y};\vect{X})$
we get that the median corresponding to $\func{depth}$ is a $P$-based fusion function.

All the other fusion functions presented in this section may be generalized as follows, see Table \ref{Tab:PenaltyMultidim}

\begin{definition}
Let $\func{D}:[0,\infty]^n\to[0,\infty]$ be a nondecreasing fusion function
such that $\func{D}(n\ast 0)=0$ and $\mathfrak{d}$ be an arbitrary pseudometric.
\index{distance-based penalty function}%
Then a \emph{distance}-based penalty function is given by:
\begin{equation}
   P(\vect{y};\vect{X}) = \func{D}\left( \mathfrak{d}(\vect{x}^{(1)}, \vect{y}),\dots,\mathfrak{d}(\vect{x}^{(n)}, \vect{y}) \right).
\end{equation}
\end{definition}

Note that not all metrics lead to proper penalty functions, though.
This is the case of the Hamming distance (see below).

\begin{proposition}\label{Prop:DistanceBasedPenaltyMultidim}
If $P$ is a distance-based penalty function generated by $\func{D}$, $\mathfrak{d}$,
and $\func{F}$ is a $P$-based fusion function,
then we observe the following regularities:
\begin{itemize}
   \item $\func{F}$ is idempotent.
   \item If $\mathfrak{d}$ is a norm-generated metric, then $\func{F}$ is translation equivariant. %
   \item If $\mathfrak{d}$ is a norm-generated metric and $\func{D}$ is scale equivariant, then $\func{F}$ is uniform scale equivariant.
   \item If $\mathfrak{d}$ is the Euclidean metric, then $\func{F}$ is  orthogonal equivariant. %
   \item If $\mathfrak{d}$ is the Euclidean metric and $\func{D}$ is strictly increasing, then $\func{F}$ is  CH-internal. %
   \item If $\mathfrak{d}$ is the Manhattan metric and $\func{D}$ is strictly increasing, then $\func{F}$ is  bounding box-internal. %
\end{itemize}
\end{proposition}

\begin{table}[htb!]
\centering
\caption{\label{Tab:PenaltyMultidim} Examples of distance-based penalty functions.}

\begin{tabularx}{1.0\linewidth}{XX}
\toprule
\small\bf $\func{D}$ & \small\bf $\func{D}$ minimizer  \\
\midrule
Arithmetic mean & 1-median \\
\midrule
Weighted arithmetic mean & Weighted 1-median \\
\midrule
Maximum & 1-center  \\
\midrule
Quadratic mean & Centroid \\
\midrule
Weighted quadratic mean & Weighted centroid \\
\bottomrule
\end{tabularx}
\end{table}

Other fusion functions may be used instead of those listed in Table \ref{Tab:PenaltyMultidim},
for example $\func{D}(\vect{d})=d_{(\lfloor n/2\rfloor)}$, will give us the center
of the smallest ball containing approximately half of the input points
(may be useful in the process of constructing metric tree-based data structures,
e.g., vp-trees \cite{Yianilos1993:vptree}).

\begin{figure}[htb!]
\centering

\begin{tabular}{ccc}
\includegraphics[width=3.8cm]{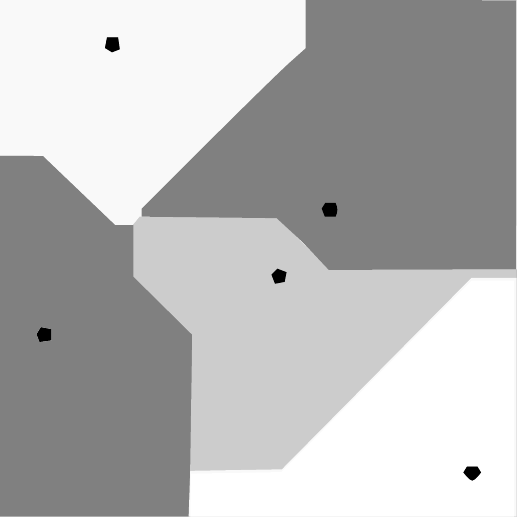} &
\includegraphics[width=3.8cm]{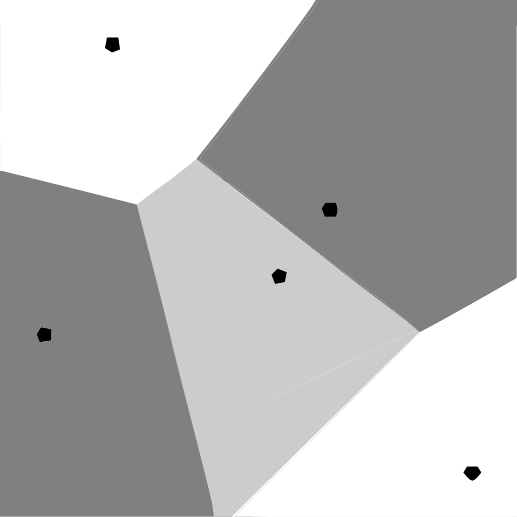} &
\includegraphics[width=3.8cm]{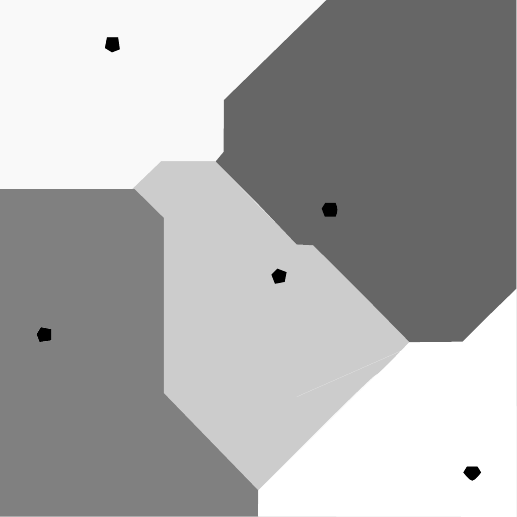} \\
(a) $\mathfrak{d}_1$ &
(b) $\mathfrak{d}_2$ &
(c) $\mathfrak{d}_\infty$ \\
\end{tabular}

\caption{\label{Fig:Voronoi} Dirichlet (Voronoi) regions generated
by 5 points in $\mathbb{R}^2$ and different metrics.}
\end{figure}

\begin{remark}
The idea of incorporating generic penalty minimizers in clustering tasks was discussed
by Leisch in \cite{Leisch2006:kcentroidstoolbox}.
He proposes a generalization of the $k$-means and $k$-medians algorithm
which works for any metric and its minimizer. To recall, the aim of such
algorithms is to find, for a given $k$, the centers of clusters
$\boldsymbol\mu^{(1)},\dots, \boldsymbol\mu^{(k)}$ which partition
input data points into $k$ disjoint groups.
The $i$th point's membership to one of the clusters, $c(i)\in[k]$,
is expressed in terms of its proximity to one of the cluster centers
(cluster centers generate Dirichlet (Voronoi) regions, see Figure \ref{Fig:Voronoi},
\index{k-means clustering}%
\index{Voronoi region}%
which determine a center's ``attraction area'').
This type of clustering algorithms tries to approach a solution such that the total distance
between all input points $\vect{x}^{(i)}$ and their corresponding  clusters' centers $\boldsymbol\mu^{(c(i))}$
is as small as possible, i.e.:
\[
\mathrm{minimize}\ \sum_{i\in n} \mathfrak{d}(\vect{x}^{(i)}, \boldsymbol\mu^{(c(i))})
\quad \text{w.r.t.~}c:[n]\to[k] \text{ (onto)},
\]
where:
\begin{equation}\label{Eq:kmeancentr}
\boldsymbol\mu^{(i)} = \argmin_{\vect{y}\in\mathbb{R}^d} \func{F}\{\vect{x}^{(j)}: j\in[n], c(j)=i\}.
\end{equation}
A $k$-means-like algorithm is a heuristic which aims to solve the above optimization
problem in the following manner:
\begin{enumerate}
   \item[1.] Initialize $c(i), i\in[n]$, e.g., randomly;
   \item[2.] Update the centroids according to Equation~\eqref{Eq:kmeancentr};
   \item[3.] Repeat Step 2. until convergence.
\end{enumerate}

\end{remark}

\begin{remark}
There are various ways that can aid in choosing a fusion function
for practical use. One of them may be based on the set of
useful properties (such as a particular type of equivariance) that an aggregation method fulfills.
The other ones rely on a fusion function's numerical characteristics
(compare the notion of a breakdown point in Section~\ref{Sec:FusionFunctionsCharacteristics}).

Nevertheless, these properties or characteristics are non-probabilistic in their nature.
As multidimensional fusion functions are frequently investigated
by computational statisticians, it is also  interesting to inspect their
behavior on random input data.

For instance, Mass\'{e} and Plante in \cite{MassePlante2003:mcstudybivariate}
perform a {M}onte {C}arlo study on the accuracy and robustness of ten bivariate location estimators:
the centroid, Tukey Median, Liu median, Oja median, depth-based trimmed medians (Liu and Tukey, $\alpha=0.05,0,1$), spatial median,
and componentwise median. They consider 26 random data scenarios for different $n$ and $d=2$, including various types
of samples' contamination, with the point of reference set to the population median (center of symmetry).
It turns out that the best performance is exhibited by the Euclidean 1-median, the Oja and the Tukey median,
as well as the componentwise median.
\end{remark}

\section{Aggregation on product lattices}\label{Sec:ProdLatAg}

In Section~\ref{Sec:ProdLat} we explored the topic of fusion
of data which were objects in some bounded poset. Let us extend the discussion
slightly to the case of information items that are instances of poset sequences.
This is exactly the situation, for example, occurring in a decision making task where $n$
experts express their opinions on $d$ alternatives and there is a need
to obtain their ``averaged'' view on all of the alternatives.

\subsection{Cartesian product}

\index{Cartesian product}%
The Cartesian product (see, e.g., \cite{Birkhoff1967:latticetheory})
of $d$ identical bounded posets
$\mathcal{P}=(P, \sqsubseteq, \underline{0}, \overline{1})$
is the bounded poset $\mathcal{P}^d=(P^d, \sqsubseteq^d, \underline{0}^d, \overline{1}^d)$
with $P^d=P\times\dots\times P$,
$\underline{0}^d=(d\ast\underline{0})$,
$\overline{1}^d=(d\ast\overline{1})$.
Here, the partial ordering relation $\sqsubseteq^d$ is given by:
\begin{equation}
   (x_1,\dots,x_d)\sqsubseteq^d(y_1,\dots,y_d) \Longleftrightarrow
   x_1\sqsubseteq y_1 \text{ and } \dots \text{ and }
   x_d\sqsubseteq y_d.
\end{equation}

\begin{remark}
Most constructions presented in this section
may be quite easily extended to the case of a Cartesian product
of non-identical bounded posets. We do not follow such a route
for better readability of the material.
\end{remark}

Additionally, if we consider a product of $d$ identical bounded lattices
$(P, \sqsubseteq, \sqcap, \sqcup, \underline{0}, \overline{1})$,
then we get a bounded lattice with join $\sqcap^d$ and meet $\sqcup^d$ operations, respectively, given by:
\begin{eqnarray*}
(x_1,\dots,x_d)\sqcap^d(y_1,\dots,y_d) &=& (x_1\sqcap y_1, \dots, x_d\sqcap y_d),\\
(x_1,\dots,x_d)\sqcup^d(y_1,\dots,y_d) &=& (x_1\sqcup y_1, \dots, x_d\sqcup y_d).
\end{eqnarray*}

\begin{remark}
If $\mathcal{P}$ is a bounded chain,
then $\mathcal{P}^d$ is  a bounded lattice
(product of chains is only a chain in trivial cases: $d=1$ or $|P|=1$).
For instance, let $\mathcal{P}=([0,1], \le,\allowbreak \wedge, \vee, 0, 1)$.
Considering $\mathcal{P}^2$, we have $(0,1)\not\le_2\not\ge_2(1,0)$,
hence $\le_2$ is not a linear order.
\end{remark}

Similarly, any function $\func{F}: P^n\to{P}$,
i.e., taking $n$ objects in ${P}$ as input,
may be extended in a componentwise manner.
This way, we obtain $\func{F}:({P}^d)^n\to{P}^d$ as follows:
\begin{equation}
   \func{F}^d(\vect{x}^{(1)},\dots,\vect{x}^{(n)})
   =\left(\func{F}(x_1^{(1)},\dots,x_1^{(n)}), \dots, \func{F}(x_d^{(1)},\dots,x_d^{(n)})\right)
\end{equation}
for all $\vect{x}^{(1)},\dots,\vect{x}^{(n)}\in{P}^d$.

Recalling the discussion on componentwise fusion functions in the case ${P}=\mathbb{R}$,
we have the following result.

\begin{proposition}
If $\func{F}$ is an aggregation function on a bounded poset ${P}$,
see Definition~\ref{Def:AgFunLattice},
then $\func{F}^d$ is an aggregation function on ${P}^d$.
More generally, if $\func{F}_1,\dots,\func{F}_d$ are aggregation functions on ${P}$,
then the componentwise (decomposable, see \cite{KomornikovaMesiar2011:agbndposet}) fusion function
$(\func{F}_1,\dots,\func{F}_d)$ is an aggregation function on ${P}^d$.
\end{proposition}

As we know from previous sections, of course, one does not have to limit
him/herself to such simple extensions of fusion functions. If some kind
of dependency between variables exists in an input data set, more elaborate solutions
may be necessary. For instance, in decision making we may want to introduce
fusion functions that ignore the answers of experts who constantly (for all the attributes)
provide contrasting answers. We may also do so for experts whose answers
are characterized by a very small variability, and so forth.

\subsection{Penalty-based aggregation on product lattices}

Assume that $P=\{p_1,p_2,\dots\}$ is countable
and that $\sqsubseteq$ is a linear order
\index{natural metric on product chains}%
with $p_i \sqsubseteq p_j$ whenever $i\le j$.
We may consider a \emph{natural metric} on ${P}^d$
(see \cite{BustinceETAL2014:cartprodlat})
such that for any $(p_{i_1}, \dots, p_{i_d}), (p_{j_1}, \dots, p_{j_d})\in P^d$
it holds:
\begin{equation}
\mathfrak{d}_\mathrm{N}\left((p_{i_1}, \dots, p_{i_d}), (p_{j_1}, \dots, p_{j_d})\right)=\sum_{u=1}^d |i_u-j_u|.
\end{equation}

With this metric, penalty-based fusion functions such as some of those considered
in Section~\ref{Sec:PenaltyMultidim} may be introduced.
Note that  a form of weighting of different dimensions and relative
elements' order may also be  incorporated here so that we get:
\begin{equation}
\mathfrak{d}_{\vect{w},\boldsymbol\varphi}\left((p_{i_1}, \dots, p_{i_d}), (p_{j_1}, \dots, p_{j_d})\right)=\sum_{u=1}^d w_u|\varphi_u(i_u)-\varphi_u(j_u)|.
\end{equation}
for some increasing and convex $\varphi_1,\dots,\varphi_d:\mathbb{N}\to\mathbb{R}$
and a weighting vector $\vect{w}\in[0,1]^d$.

\subsection{Conjunctive, disjunctive, and averaging functions}

Let us go back to the Komorn\'{i}kov\'{a}-Mesiar classification
of fusion functions, see \cite{KomornikovaMesiar2011:agbndposet} and Section~\ref{Sec:AgLatClass}.

De Baets and Mesiar in \cite{DeBaetsMesiar1999:trinormprodlat} showed that
the componentwise extension of $d$ t-norms is also a t-norm.
On the other hand, as shown by Jenei and De Baets in \cite{JeneiDeBaets2003:tnormprodlat},
there may exist t-norms on product lattices ${P}^d$ that are not
direct products of t-norms on ${P}$.

By \cite[Proposition 5]{KomornikovaMesiar2011:agbndposet},
we have that  $(\func{F}_1,\dots,\func{F}_d)$ is strongly conjunctive (disjunctive)
if and only if each $\func{F}_i$ is strongly conjunctive (disjunctive).

Other properties are not necessarily inherited so easily as indicated in the following example.

\begin{example}[\cite{KomornikovaMesiar2011:agbndposet}]
Consider the bounded chain $([0,1], \le,\allowbreak 0, 1)$.
Here, the sample median $\func{Median}$ is strongly averaging.
But if we act on a product of three such chains
we get that $\func{Median}^3$ is not even weakly averaging.
\end{example}

\subsection{Other orders on product lattices}

It turns out that the product order is only one of many possible extensions
of $\sqsubseteq$ to a product lattice. Other popular choices
include the inf-, sup-based, and lexicographic ordering. In decision making these correspond
to maximin, maximax (see, e.g., \cite{DuboisETAL1996:maximin}) and leximin
(see, e.g., \cite{DuboisETAL2001:leximin,Fishburn1974:lexicographicorders}) approaches, respectively.

\begin{definition}\index{inf-based order}%
Let $(P, \sqsubseteq, \sqcap,\sqcup)$ be a  lattice.
Then the \emph{inf-based ordering} is given for every $\vect{p},\vect{q}\in P^d$
by $\vect{p}\sqsubseteq_\mathrm{min}\vect{q}$ if and only
if $\bigsqcap_{i=1}^d p_i \sqsubseteq \bigsqcap_{i=1}^d q_i$.
\end{definition}

\begin{definition}\index{sup-based order}%
The \emph{sup-based ordering} is given for every $\vect{p},\vect{q}\in P^d$
by $\vect{p}\sqsubseteq_\mathrm{max}\vect{q}$ if and only
if $\bigsqcup_{i=1}^d p_i \sqsubseteq \bigsqcup_{i=1}^d q_i$.
\end{definition}

In other words, the two above orders say that a lattice element $\vect{p}$
is dominated by $\vect{q}$, whenever the satisfaction degree of the least
(respectively, greatest) satisfied constraint  in the first object is
not greater than  the corresponding observation in the second one.

\begin{definition}\index{lexicographic order}%
The \emph{lexicographic ordering} is given for every $\vect{p},\vect{q}\in P^d$ by:
\begin{eqnarray}
\vect{p}\sqsubseteq_\mathrm{lex}\vect{q}&\Longleftrightarrow&
(\exists i\in[d]) (\forall j\in[i-1])\ p_j=q_j\text{ and }\\
&&
 p_i \sqsubset q_i \text{ if $i<d$ and } p_{i}\sqsubseteq q_{i}\text{ otherwise}.\nonumber
\end{eqnarray}
\end{definition}

Note that an extended version of the above order shall be studied in the next Chapter.

\begin{remark}
As we already mentioned, the lexicographic order is particularly
appealing in decision making. Imagine we have a set of criteria, ordered with respect to their importance,
like ``child safety'', ``price'', and ``attractive outlook'' in the case of a decision making task
to determine which car should be bought by an agent.
If a car $A$ is less safe than $B$, no matter what the satisfaction degrees of other
criteria are, $B$ is preferred to $A$. On the other hand, if $A$ is as safe
as $B$, then one should also consider its price and then -- perhaps -- its
general appearance.
\end{remark}

In the three discussed cases if $\sqsubseteq$ is a linear order,
then the above-defined orders are at least total preorders.

\bigskip
Note that some of the results presented in Section~\ref{Sec:ProdLat}
may be utilized in any of these new, ``multidimensional'' settings.
A combination of posets gives us yet another poset
and the methods presented in the previous Chapter are still valid
here. This is because they are very general in their nature.
We therefore decide not to explore them any further in this book.

\section{Aggregation of character sequences}\label{Sec:nominalscaled}

In Section~\ref{Sec:nominalscale1} we noted that aggregation
of $n$ elements on a nominal scale was neither very challenging
nor interesting. Nevertheless, the situation is quite different
in the case of $n$ vectors of length $d$ with elements in some
alphabet $\Sigma$ (and will be even more engaging in the next chapter, where
we deal with character strings).

All the fusion functions considered in this  section are distance penalty-based ones.
Perhaps the most frequently used metric on $\Sigma^d$ is the one introduced by Hamming,
see \cite{Hamming1950:errorcodes}.

\begin{definition}\label{Def:HammingDistance}\index{Hamming distance}%
The \emph{Hamming distance} is defined for $\vect{x},\vect{y}\in\Sigma^d$ as:
\begin{equation}
   \mathfrak{d}_\mathrm{H}(\vect{x}, \vect{y})=\sum_{i=1}^d \indicator(x_i \neq y_i).
\end{equation}
\end{definition}
In other words, it is the total number of indices at which two given vectors~differ.

Exemplary applications of fusion functions based on the Hamming distance minimizers
include finding gene clusters, creating diagnostic probes, or discovering potential drug
targets, see, e.g., \cite{LanctotETAL2003:stringselproblems}, especially if we
compute them over the DNA or protein sequences domain.
Also, they are useful in error correction tasks:
imagine that a few signals were sent with errors, the ``central'' one
(this is particularly the case of the median vector discussed below)
may represent the underlying correct information piece.

Let us briefly review possibly interesting properties of such fusion functions.
Of course, there is no ordering relation on $\Sigma$, thus we cannot
refer to any notion of monotonicity here. Instead, we may consider
if for every $\vect{x},\vect{x}^{(1)},\dots,\vect{x}^{(n)}\in\Sigma^d$ it holds:
\begin{itemize}
   \item $\func{F}(n\ast\vect{x})=\vect{x}$, \hfill(idempotency)
   \item $\func{F}(\vect{x}^{(1)},\dots,\vect{x}^{(n)})=\func{F}(\vect{x}^{(\sigma(1))},\dots,\vect{x}^{(\sigma(n))})$ for all $\sigma\in\mathfrak{S}_{[n]}$,
   \hfill(symmetry)
   \item if $\vect{y}=\func{F}(\vect{x}^{(1)},\dots,\vect{x}^{(n)})$, then
   $y_i\in\{x_i^{(1)},\dots,x_i^{(n)}\}$, \hfill(internality)
   \item $\func{F}(\vect{x}^{(1)},\dots,\vect{x}^{(n)})=\func{F}({\vect{x}'}^{(1)},\dots,{\vect{x}'}^{(n)})=\vect{y}$
   where $\vect{x'}^{(j)}\in\{\vect{x}^{(j)}, \vect{y}\}$,\par\hfill(decomposability)
\end{itemize}
and for extended fusion functions:
\begin{itemize}
   \item $\func{F}(\vect{x}^{(1)},\dots,\vect{x}^{(n)})=\vect{y}$,
then $\func{F}(k\ast\vect{y},\vect{x}^{(1)},\dots,\vect{x}^{(n)})=\vect{y}$ for all $k$,\par\hfill(L-stability)
   \item $\func{F}(\vect{x}^{(1)},\dots,\vect{x}^{(n)})=\vect{y}$,
then $\func{F}(\vect{x}^{(1)},\dots,\vect{x}^{(n)},k\ast\vect{y})=\vect{y}$ for all $k$.\par\hfill(R-stability)
\end{itemize}

\subsection{Median}

Let us first study the problem of finding:
\[
\func{Median}_{\mathfrak{d}_\mathrm{H}}(\vect{x}^{(1)},\dots,\vect{x}^{(n)})
=\argmin_{\vect{x}\in\Sigma^d} \sum_{i\in[n]}
\mathfrak{d}_\mathrm{H}(\vect{x}^{(i)}, \vect{x}).
\]
As the solution might be ambiguous, we may  rather be interested in
determining any $\vect{x}^*\in\Sigma^d$ such that:
\begin{equation}\label{Eq:MedianHamming}
   \sum_{i\in[n]} \mathfrak{d}_\mathrm{H}(\vect{x}^{(i)}, \vect{x}^*) :=
   \min_{\vect{x}\in\Sigma^d} \sum_{i\in[n]} \mathfrak{d}_\mathrm{H}(\vect{x}^{(i)}, \vect{x}).
\end{equation}
It turns out that such a vector can be determined easily. For that,
we may use the following algorithm.

\begin{algorithm}
To determine all solutions to Equation~\eqref{Eq:MedianHamming}, proceed as follows:
\begin{enumerate}
   \item[1.] For $i=1,\dots,d$ do:
   \begin{enumerate}
   \item[1.1.] Let $k_i:=\max_{x\in\{x_i^{(1)},\dots,x_i^{(n)}\}} \sum_{j\in[n]} \indicator(x = x_i^{(j)})$,
   i.e., the number of occurrences of the most frequently occurring character at index $i$;
   \item[1.2.] Let $E_i = \{ x\in\{x_i^{(1)},\dots,x_i^{(n)}\}:
      \sum_{j\in[n]} \indicator(x = x_i^{(j)}) = k_i \}$, i.e., the set of
      all characters that occur exactly $k_i$ times at index~$i$;
   \end{enumerate}
   \item[2.] Return all $\vect{x^*}\in E_1\times\dots\times E_d$.
\end{enumerate}
\end{algorithm}

\begin{remark}
If we are interested in any $\vect{x^*}$ which is a solution to Equation~\eqref{Eq:MedianHamming},
then the above procedure may be implemented in such a way that, e.g.,
(a) it uses $O(|\Sigma|+d)$ additional memory units and $O(d|\Sigma|+dn)$ time (a bucket sort-like algorithm)
or (b) with the usage of $O(n+d)$ additional memory units and $O(dn\log n)$ time,
see also Remark~\ref{Algorithm:mode}.
\end{remark}

Figure~\ref{Fig:medianhamming} gives an exemplary \lang{C++} implementation
which is based on hash tables and has an amortized run time of  $O(d|\Sigma'|+dn)$,
where $\Sigma'\subseteq\Sigma$ consists only of letters used in the input strings.
Note that an input data set is given via a $d\times n$ integer matrix there.

\begin{example}\label{Example:MedianHamming}
Let us set $d=3$, $n=6$, and $\Sigma=\{0,1,2,3\}$. Consider the following data set:
\begin{center}
\begin{tabularx}{1.0\linewidth}{XXXXXXX}
\toprule
\bf\small $j$ & \bf\small 1 & \bf\small 2 & \bf\small 3 & \bf\small 4 & \bf\small 5 & \bf\small 6 \\
\midrule
\bf\small $s_1^{(j)}$ & 2  &  1  &  3  &  1 &   2   & 1\\
\bf\small $s_2^{(j)}$ & 2  &  3  &  1  &  1 &   0   & 2 \\
\bf\small $s_3^{(j)}$ & 3  &  0  &  0  &  2 &   0   & 0 \\
\bottomrule
\end{tabularx}
\end{center}
Noticing that $\min_{\vect{x}\in\Sigma^d} \sum_{i\in[n]} \mathfrak{d}_\mathrm{H}(\vect{x}^{(i)}, \vect{x})=9$,
there are two solutions to Equation~\eqref{Eq:MedianHamming}:
$(1,1,0)$ and $(1,2,0)$.
One of them is among the input vectors (this is not a rule in general),
so it also corresponds to the set's medoid.
\end{example}

The median with respect to the Hamming distance
is definitely symmetric, idempotent, internal, decomposable, and stable.

\begin{remark}
   The above algorithm may easily be extended to find
   a~weighted median, i.e., $\argmin_{\vect{x}\in\Sigma^d} \sum_{i\in[n]}
w_i \mathfrak{d}_\mathrm{H}(\vect{x}^{(i)}, \vect{x})$, where $\vect{w}$ is a weighting vector.
\end{remark}

\subsection{Center}

Now we shall focus on determining:
\[
\func{Center}_{\mathfrak{d}_\mathrm{H}}(\vect{x}^{(1)},\dots,\vect{x}^{(n)})
=\argmin_{\vect{x}\in\Sigma^d} \bigvee_{i\in[n]}
\mathfrak{d}_\mathrm{H}(\vect{x}^{(i)}, \vect{x}).
\]
Again, the solution may be non-unique, therefore we rather shall be aiming at
determining any $\vect{x}^*\in\Sigma^d$ such that:
\begin{equation}\label{Eq:CentroidHamming}
   \bigvee_{i\in[n]} \mathfrak{d}_\mathrm{H}(\vect{x}^{(i)}, \vect{x}^*) :=
   \min_{\vect{x}\in\Sigma^d} \bigvee_{i\in[n]} \mathfrak{d}_\mathrm{H}(\vect{x}^{(i)}, \vect{x}).
\end{equation}
Such a fusion function is of interest in coding theory \cite{FrancesLitman1997:coveringcodes},
gene clustering \cite{DinuIonescu2012:efficientrankcloseststring}, and other bioinformatics
tasks \cite{BoucherMa2011:closeststringoutliers}.

\begin{example}
Let us go back to data in Example~\ref{Example:MedianHamming}.
There are two centers: $(1,2,0)$ and $(2,1,0)$.
We have $\min_{\vect{x}\in\Sigma^d} \bigvee_{i\in[n]} \mathfrak{d}_\mathrm{H}(\vect{x}^{(i)}, \vect{x})=2$.
\end{example}

A center character sequence is at least idempotent, symmetric, and internal
(or more precisely, there is at least one internal solution).%

\bigskip
Unfortunately, there is no polynomial time-algorithm (with respect
to~$n$ -- it can be reduced to 3SAT) for computing it (unless $\mathrm{P}=\mathrm{NP}$) even for $d=2$,
see, e.g., \cite{FrancesLitman1997:coveringcodes,LanctotETAL2003:stringselproblems}.

\index{IP task|see {integer programming task}}\index{integer programming task}%
Among exact algorithms, which aim to find a string
within some maximal distance threshold, $e$,
that is $\vect{x}$ with $\bigvee_{i\in[n]} \mathfrak{d}(\vect{x}^{(i)}, \vect{x})\le e$,
we may list \cite{MenesesETAL2004:optimalcloseststringip,GrammETAL2003:fixedparamclostr,ChimaniETAL2011:closercloseststr},
which are based on integer programming (IP), see \cite{Lenstra1983:integerprogramming}.
One of the simplest formulations of the discussed problem may be written
in terms of an IP task as follows (see \cite{MenesesETAL2004:optimalcloseststringip}):
\begin{equation}
      \mathrm{minimize}\ \delta
      \quad \text{w.r.t.~}\delta\in\mathbb{Z}, \vect{t}\in\mathbb{Z}^d, \vect{z}\in \mathbb{Z}^{n\times d}
\end{equation}
subject to:
\begin{eqnarray*}
\delta - \sum_{j=1}^d z_{i,j} &\ge& 0, \quad i=1,\dots,n\\
t_j - kz_{i,j} & \le & s_j^{(i)}, \quad i=1,\dots,n, j=1,\dots,d\\
kz_{i,j} - t_j & \ge & s_j^{(i)}, \quad i=1,\dots,n, j=1,\dots,d\\
\delta  & \in & \{0,\dots,d\},\\
t_j     & \in & \{1,\dots,k\}, \quad j=1,\dots,d\\
z_{i,j} & \in & \{0, 1\}, \quad i=1,\dots,n, j=1,\dots,d.
\end{eqnarray*}
Here we assume that $\Sigma=\{1,2,\dots,k\}$ (the original $\Sigma$
may always be reencoded in such a way).
The solution is stored in the $\vect{t}$ vector.
This can be solved, e.g., using the \package{COIN-OR SYMPHONY} library
(via the \package{Rsymphony} package in \R).
Please note that the above formulation leads to a practically unusable
implementation (unless $d$, $n$ are small).

Among other exact algorithms we may find the one given in
\cite{ChenWang2011:exactclosesstring}. Here, we start with a string
in the input data set and then consecutively modify no more than $e$
letters in the candidate string at a time.
Another is given in \cite{HufskyETAL2011:swiftlycenterstr}. It
is based on some data reduction techniques and search tree algorithms.
What is more, in \cite{BoucherMa2011:closeststringoutliers} an algorithm
to compute the closest string in the presence of outliers is given, i.e.,
one within a Hamming distance of $\delta$ to at least $n-k$ of the input strings
for some $k$.

There are also polynomial-time approximation schemes,
see, e.g., \cite{LiETAL2002:closeststring,MazumdarETAL2013:chebyshevradius,LanctotETAL2003:stringselproblems}.
For instance, Lanctot et al.~derive a polynomial-time $(4/3+\varepsilon)$-approximation
algorithm for any small $\varepsilon>0$, see \cite{LanctotETAL2003:stringselproblems}.

\begin{remark}\index{1 approximation algorithm@$(1+\varepsilon)$-approximation algorithm}%
We say that a procedure is a $(1+\varepsilon)$-approximation algorithm
whenever the ratio of the \textit{quality} of the result (here, expressed
in terms of the Hamming distance) as compared to the optimal solution is guaranteed to be not
greater than $1+\varepsilon$ for any $\varepsilon>0$.
\end{remark}

As in practice exact algorithms exhibit poor performance, here let us discuss
a so-called evolutionary strategy to approximate the center string.
The first \emph{genetic algorithms} were introduced by Fraser,
\index{genetic algorithm}%
see, e.g., \cite{Fraser1957:simulgensyst,FraserBurnell1970:compmodgene}.
This is a class of adaptive, approximate optimization algorithms inspired by
the biological process of natural selection.
Of course, they do not guarantee that the global maximum
of a fitness function $f$ shall be found.
However, such  techniques are especially useful if the objective
function is defined on a discrete space (and this is our case).

\begin{algorithm}\label{Alg:Genetic}
For a given $k$ (population size), $\eta$ (number of iterations),
and some fit measure $f$:
\begin{enumerate}
   \item[1.] Generate a random initial population,
   i.e., a set of $k$ initial elements (individuals) $P=\{\vect{p}^{(1)},\dots,\vect{p}^{(k)}\}$;
   \item[2.] Determine the fit measure $f_j$ for each individual $\vect{p}^{(j)}$;
   \item[3.] Set $P$ to be the best population considered so far, $P^*$;
   \item[4.] For $i=1,\dots,\eta$ do:
   \begin{enumerate}
   \item[4.1.] Selection: randomly select $k$ pairs of vectors in $P$
   in, e.g., such a way that each vector occurs in the resulting sample
   with probability proportional to some function of $f_j$ (fitness proportio\-nate selection);
   \item[4.2.] Crossover: generate a new population $P'$,
   such that each new individual is obtained by combining elements
   in a pair of vectors selected in the previous step
   (some characters are taken from the first vector in a pair,
   the other ones are taken from the second vector);
   \item[4.3.] Mutation: replace a few randomly chosen elements
   of vectors in $P'$ with some other characters;
   \item[4.4.] Set $P:=P'$ and recompute the fit measures $f_j$, $j=1,\dots,k$;
   \item[4.5.] If the current population includes an individual of
   the best fit so far, i.e., $\max_j f_j > \max_j f_j^*$, set $P^*:=P$;
   \end{enumerate}
   \item[5.] Return the best individual from $P^*$,
   i.e., $\argmin_{{\vect{p}^*}^{(j)}, j\in[k]} f_j^*$.
\end{enumerate}
\end{algorithm}

In our case, the fit measure of an individual $\vect{p}^{(j)}$
is inversely proportional to
$\bigvee_{i\in[n]} \mathfrak{d}_\mathrm{H}(\vect{x}^{(i)}, \vect{p}^{(j)})$.
Figure~\ref{Fig:hamming_closest_ga} gives an exemplary \R{} implementation
of Algorithm~\ref{Alg:Genetic}, which aims to determine an approximate
solution to Equation~\eqref{Eq:CentroidHamming}.
Note that the crossover scheme choice is crucial here: we observe
that the uniform crossover works far better than its one- or two-point version.
There are some other possible options too, e.g., a crossover  based on
three parents or different selection phase schemes.

\begin{table}[htbp!]
\caption[ASCII codes and their corresponding code points.]{\label{Tab:ASCII} ASCII
codes (Unicode block \textit{C0 Controls and Basic Latin}) and their corresponding
code points (\textbf{char}s). \textbf{dec} stands for a decimal code and \textbf{bit} gives its
corresponding 7-bit sequence.}

\centering\footnotesize
\begin{tabularx}{0.3\linewidth}{lll}
\toprule
\small\bf dec & \small\bf bit & \small\bf char \\
\midrule
0 & 0000000 & \it NUL \\
  1 & 0000001 & \it SOH \\
  2 & 0000010 & \it STX \\
  3 & 0000011 & \it ETX \\
  4 & 0000100 & \it EOT \\
  5 & 0000101 & \it ENQ \\
  6 & 0000110 & \it ACK \\
  7 & 0000111 & \it BEL \\
  8 & 0001000 & \it BS \\
  9 & 0001001 & \it HT \\
  10 & 0001010 & \it LF \\
  11 & 0001011 & \it VT \\
  12 & 0001100 & \it FF \\
  13 & 0001101 & \it CR \\
  14 & 0001110 & \it SO \\
  15 & 0001111 & \it SI \\
  16 & 0010000 & \it DLE \\
  17 & 0010001 & \it DC1 \\
  18 & 0010010 & \it DC2 \\
  19 & 0010011 & \it DC3 \\
  20 & 0010100 & \it DC4 \\
  21 & 0010101 & \it NAK \\
  22 & 0010110 & \it SYN \\
  23 & 0010111 & \it ETB \\
  24 & 0011000 & \it CAN \\
  25 & 0011001 & \it EM \\
  26 & 0011010 & \it SUB \\
  27 & 0011011 & \it ESC \\
  28 & 0011100 & \it FS \\
  29 & 0011101 & \it GS \\
  30 & 0011110 & \it RS \\
  31 & 0011111 & \it US \\
  32 & 0100000 & \it (space)  \\
  33 & 0100001 & \tt ! \\
  34 & 0100010 & \tt "{} \\
  35 & 0100011 & \tt \# \\
  36 & 0100100 & \tt \$ \\
  37 & 0100101 & \tt \% \\
  38 & 0100110 & \tt \& \\
  39 & 0100111 & \tt ' \\
  40 & 0101000 & \tt ( \\
  41 & 0101001 & \tt ) \\
  42 & 0101010 & \tt * \\
\bottomrule
\end{tabularx}
\begin{tabularx}{0.3\linewidth}{lll}
 \toprule
\small\bf dec & \small\bf bit & \small\bf char \\
\midrule
  43 & 0101011 & \tt + \\
  44 & 0101100 & \tt , \\
  45 & 0101101 & \tt - \\
  46 & 0101110 & \tt . \\
  47 & 0101111 & \tt / \\
  48 & 0110000 & \tt 0 \\
  49 & 0110001 & \tt 1 \\
  50 & 0110010 & \tt 2 \\
  51 & 0110011 & \tt 3 \\
  52 & 0110100 & \tt 4 \\
  53 & 0110101 & \tt 5 \\
  54 & 0110110 & \tt 6 \\
  55 & 0110111 & \tt 7 \\
  56 & 0111000 & \tt 8 \\
  57 & 0111001 & \tt 9 \\
  58 & 0111010 & \tt : \\
  59 & 0111011 & \tt ; \\
  60 & 0111100 & \tt $<$ \\
  61 & 0111101 & \tt = \\
  62 & 0111110 & \tt $>$ \\
  63 & 0111111 & \tt ? \\
  64 & 1000000 & \tt @ \\
  65 & 1000001 & \tt A \\
  66 & 1000010 & \tt B \\
  67 & 1000011 & \tt C \\
  68 & 1000100 & \tt D \\
  69 & 1000101 & \tt E \\
  70 & 1000110 & \tt F \\
  71 & 1000111 & \tt G \\
  72 & 1001000 & \tt H \\
  73 & 1001001 & \tt I \\
  74 & 1001010 & \tt J \\
  75 & 1001011 & \tt K \\
  76 & 1001100 & \tt L \\
  77 & 1001101 & \tt M \\
  78 & 1001110 & \tt N \\
  79 & 1001111 & \tt O \\
  80 & 1010000 & \tt P \\
  81 & 1010001 & \tt Q \\
  82 & 1010010 & \tt R \\
  83 & 1010011 & \tt S \\
  84 & 1010100 & \tt T \\
  85 & 1010101 & \tt U \\
\bottomrule
\end{tabularx}
\begin{tabularx}{0.3\linewidth}{lll}
\toprule
\small\bf dec & \small\bf bit & \small\bf char \\
\midrule
  86 & 1010110 & \tt V \\
  87 & 1010111 & \tt W \\
  88 & 1011000 & \tt X \\
  89 & 1011001 & \tt Y \\
  90 & 1011010 & \tt Z \\
  91 & 1011011 & \tt [ \\
  92 & 1011100 & \tt $\backslash$ \\
  93 & 1011101 & \tt ] \\
  94 & 1011110 & \verb|^| \\
  95 & 1011111 & \tt \_ \\
  96 & 1100000 & \tt ` \\
  97 & 1100001 & \tt a \\
  98 & 1100010 & \tt b \\
  99 & 1100011 & \tt c \\
  100 & 1100100 & \tt d \\
  101 & 1100101 & \tt e \\
  102 & 1100110 & \tt f \\
  103 & 1100111 & \tt g \\
  104 & 1101000 & \tt h \\
  105 & 1101001 & \tt i \\
  106 & 1101010 & \tt j \\
  107 & 1101011 & \tt k \\
  108 & 1101100 & \tt l \\
  109 & 1101101 & \tt m \\
  110 & 1101110 & \tt n \\
  111 & 1101111 & \tt o \\
  112 & 1110000 & \tt p \\
  113 & 1110001 & \tt q \\
  114 & 1110010 & \tt r \\
  115 & 1110011 & \tt s \\
  116 & 1110100 & \tt t \\
  117 & 1110101 & \tt u \\
  118 & 1110110 & \tt v \\
  119 & 1110111 & \tt w \\
  120 & 1111000 & \tt x \\
  121 & 1111001 & \tt y \\
  122 & 1111010 & \tt z \\
  123 & 1111011 & \tt \{ \\
  124 & 1111100 & \tt $|$ \\
  125 & 1111101 & \tt \} \\
  126 & 1111110 & \tt \~{} \\
  127 & 1111111 & \it DEL \\
  \phantom{127} & \phantom{1111111} & \phantom{DEL} \\
\bottomrule
\end{tabularx}
\end{table}
\clearpage{\pagestyle{empty}\cleardoublepage}
\newcommand{\quot}{\text{\tt{"{}}}}

\chapter{Aggregation of strings}\label{Chap:anydim}

\lettrine[lines=3]{U}{p} to now we have discussed different data fusion frameworks
in the case of numeric (quantitative), ordinal, and nominal data. We started
with a mathematically simple case of unidimensional data
and then considered a more complex setting in which the aggregated objects
were tuples of length $d$. In other words, for a given set $X$ we considered:

\begin{itemize}
   \item $\func{F}:X^{n}\to X$ (univariate fusion functions, see Chapter~\ref{Chapter:onedim}),
   \item $\func{F}:(X^d)^n\to X^d$ ($d$-variate fusion functions, see Chapter~\ref{Chap:multidim}).
\end{itemize}
The above frameworks can be extended so that fusion functions which take an
arbitrary number of elements as input are obtained:
\begin{itemize}
   \item $\func{F}:X^*\to X$ (extended univariate fusion functions),
   \item $\func{F}:(X^d)^*\to X^d$ (extended $d$-variate fusion functions).
\end{itemize}
Surely, each extended fusion function may be conceived of as a family of
fusion functions, each acting on tuples of different lengths.

\begin{remark}
Aggregation theoreticians sometimes also consider fusion functions
which act on infinite sequences of elements. This is useful for studying
asymptotic behavior of fusion functions, see, e.g.,
\cite{MesiarPap2008:aggrinfseq,GhiselliRicci2009:asidempotent}.
As this kind of data does not occur in computational tasks
(given a data set, one may always determine $d^+=\max\{|\vect{x}^{(i)}|, i=1,\dots,n\}$),
we do not discuss such a framework in this monograph.
\end{remark}

It turns out that one more type of extension may be useful.
Namely, we can be interested in aggregating vectors of arbitrary
(nonconforming) lengths. Such a scenario from now on is called
fusion of \emph{strings}, \emph{anyvariate} or \emph{variable length data}. More specifically:
\begin{itemize}
   \item $\func{F}:(X^*)^n\to X^*$ ($n$-ary fusion functions to aggregate vectors of any length),
   \item $\func{F}:(X^*)^*\to X^*$ (extended fusion functions to aggregate
   an arbitrary number of vectors of any length).
\end{itemize}
Depending on the choice of $X$, this situation frequently occurs, e.g.,
in the case of informetric data ($X=\Ival$) or character strings (like DNA
or bit sequences -- nominal scale).

\begin{remark}
A sequence of strings with elements being real numbers
may be represented in \R{}/\package{Rcpp} as a \texttt{List} object
(a vector of elements of any type), which stores \texttt{NumericVectors}
as its elements. In the case of character strings, see Section~\ref{Sec:aginformetricAny},
this corresponds to the \texttt{CharacterVector} type,
whose elements are objects of class \texttt{Rcpp::String}.
In pure \lang{C} we may use such data types as
\texttt{double**} and \texttt{char**}, respectively,
and, when using the \lang{C++} Standard Library (or \package{STL})
objects, we may set \texttt{std::vector< std::vector<double> >}
and \texttt{std::vector< std::string >}, respectively.
\end{remark}

Having in mind that the current space is even more ``complex'' than the previous
one, this time let us begin with a review of different orderings on~$X^*$.

\section{Orders in the space of strings}

Let $X^*=\bigcup_{d=1}^\infty X^d$ (note that this time we include vectors of length one).
In a parallel section from the previous chapter we studied -- among others --
the so-called product order, which was a way to extend a partial or linear
ordering relation  $\sqsubseteq$ on some set $P$ to the case of $P^d$ for
some $d$. Here we are naturally interested in a review of different ways of
extending $\sqsubseteq$ to $\sqsubseteq^*$ in such a way that it may act
on $P^*$. More formally, given a poset $\mathcal{P}=(P, \sqsubseteq)$,
our aim is to construct $\mathcal{P}^*=(P^*, \sqsubseteq^*)$. It turns out
that several interesting ways to do so exist in the aggregation literature.

\subsection{Lexicographic order}

The \index{lexicographic order}\emph{lexicographic order}
is defined for $\vect{p},\vect{q}\in P^*$ as:
\begin{eqnarray}
\mathbf{p}\sqsubseteq^*\mathbf{q}&\Longleftrightarrow&
(\exists i\in[d_p\wedge d_q]) (\forall j\in[i-1])\ p_j=q_j\text{ and }\\
&&
 p_i \sqsubset q_i \text{ if }i<d_p \text{ and } p_{i}\sqsubseteq q_{i}\text{ otherwise},\nonumber
\end{eqnarray}
where $|\vect{p}|=d_p$, $|\vect{q}|=d_q$ and,
as usual, $p\sqsubset q$ whenever $p\sqsubseteq q$ with $p\neq q$.

Note that if $\sqsubseteq$ is a linear order,
then also $\sqsubseteq^*$ is one.
Moreover, if $\mathcal{P}$ is bounded from below
with the least element denoted with $\underline{0}$,
then $(\underline{0})$ is the least element of $\mathcal{P}^*$.

\begin{example}
Let $P=\{\mathtt{a}, \mathtt{b},\dots, \mathtt{z}\}$
and $\mathtt{a}\sqsubset\mathtt{b}\sqsubset\dots\sqsubset\mathtt{z}$.
In such a case, we have
$\mathtt{\quot{}a\quot{}} \sqsubset^*
 \mathtt{\quot{}aa\quot{}} \sqsubset^*
 \mathtt{\quot{}aaa\quot{}} \sqsubset^*
 \dots \sqsubset^*
 \mathtt{\quot{}ab\quot{}} \sqsubset^*
 \mathtt{\quot{}aba\quot{}} \sqsubset^*
 \mathtt{\quot{}abaa\quot{}} \sqsubset^*
 \dots \sqsubset^*
 \mathtt{\quot{}abb\quot{}} \sqsubset^*
 \dots \sqsubset^*
 \mathtt{\quot{}b\quot{}}$ etc.,
 where, e.g., $\mathtt{\quot{}abc\quot{}}=(\mathtt{a}, \mathtt{b}, \mathtt{c})$.
\end{example}

\begin{remark}
Lexicographic order determines exactly how character strings are ordered
in many locales. Yet, in natural language processing tasks there are some
exceptions to this rule. For instance, in the Slovak locale (similar rules
exist for Czech), we have
$\mathtt{\quot{}citlivý\quot{}}  \sqsubset^*
\mathtt{\quot{}hladný\quot{}}  \sqsubset^*
\mathtt{\quot{}chladný\quot{}}$
(delicate / hungry / cool). This is because \textit{ch} is treated as a digraph here
and in fact it should be treated as a distinct, single character.
Moreover, if we compare strings which consist of numerals, one might need
a different order here, e.g., one that gives $\mathtt{\quot{}ID\_69\quot{}}  \sqsubset^*
\mathtt{\quot{}ID\_123\quot{}}$; please refer to the Unicode Technical
Standard on string collation \cite{UTS10} for more information.
\end{remark}

Among modified versions of the lexicographic ordering we find,
among others, the Luzin-Sierpiński (Kleene-Brouwer) order, which gives a greater priority to a string
with prefix $\vect{p}$ than to the sole string $\vect{p}$ in its entirety, see \cite{Kleene1955:kleeneorder}.

\subsection{$\alpha$- and $\beta$-, and informetric orderings}\label{Sec:AlphaBetaOrdering}

The so-called $\alpha$- and $\beta$-orderings were introduced
by Carbonell, Mas, and Mayor in \cite{CarbonellMasMayor1997:monotonicextendedowa}
for the purpose of studying extended classical aggregation functions
and constructing weighting triangles, compare Section~\ref{Sec:WeightingTriangles}.
Moreover, they were considered in a more general (lattice) setting by Calvo and Mayor in \cite{CalvoMayor1999:remarks2eaf}.
Assuming that $(P, \sqsubseteq, \sqcap,\sqcup)$ is a complete lattice,
we have what follows.

\begin{definition}\index{alpha-ordering@$\alpha$-ordering}%
Let $\vect{p}\in P^{d_p}, \vect{q}\in P^{d_q}$.
Then $\vect{p}\sqsubseteq_\alpha\vect{q}$ if and only if $d_p\le d_q$ and
$(\forall i\in[d_p])$ $p_i\sqsubseteq q_i$ and if additionally
$d_p < d_q$, then $\bigsqcup_{i=1}^{d_p} p_i \sqsubseteq \bigsqcap_{i=d_p+1}^{d_q} q_i$.
\end{definition}

\begin{definition}\index{beta-ordering@$\beta$-ordering}%
Let $\vect{p}\in P^{d_p}, \vect{q}\in P^{d_q}$.
Then $\vect{p}\sqsubseteq_\beta\vect{q}$ if and only if $d_p\ge d_q$ and
$(\forall i\in[d_q])$ $p_i\sqsubseteq q_i$ and if additionally
$d_p > d_q$, then $\bigsqcup_{i=d_q+1}^{d_p} p_i \sqsubseteq \bigsqcap_{i=1}^{d_q} q_i$.
\end{definition}

We see that for $d_p=d_q$ both orders coincide with the extension of the product order
to $P^*$, $\sqsubseteq^*$, defined as $\vect{p}\sqsubseteq^*\vect{q}$
whenever $d_p=d_q$ and $\vect{p}\sqsubseteq^d\vect{q}$. Formally,
as each binary relation on $X$ is in fact a subset of $X^2$, we have that
$\sqsubseteq^*\, \subseteq \, \sqsubseteq_\alpha$ and $\sqsubseteq^*\, \subseteq \, \sqsubseteq_\beta$.

If $(P,\sqsubseteq,\sqcap,\sqcup, \underline{0}, \overline{1})$ is a bounded lattice,
then $\underline{0}$ is the least element with respect to $\sqsubseteq_\alpha$
and $\overline{1}$ is the greatest one with respect to  $\sqsubseteq_\beta$.

\bigskip
A somehow more relaxed version of $\sqsubseteq_\alpha$ may be formulated as follows.

\begin{definition}\index{gamma-ordering@$\gamma$-ordering}%
Let $\vect{p}\in P^{d_p}, \vect{q}\in P^{d_q}$.
Then $\vect{p}\sqsubseteq_\gamma\vect{q}$ if and only if $d_p\le d_q$ and
$(\forall i\in[d_p])$ $p_i\sqsubseteq q_i$.
\end{definition}

This type of ordering is useful in informetric tasks, see the next section
for details.
If $(P,\sqsubseteq,\sqcap,\sqcup, \underline{0}, \overline{1})$ is a bounded lattice,
then $\underline{0}$ is the least element with respect to $\sqsubseteq_\gamma$.

\bigskip
\begin{proposition}\label{Prop:OrderGamma}
Given a bounded lattice $(P,\sqsubseteq,\sqcap,\sqcup, \underline{0}, \overline{1})$,
we have what follows for every $\vect{p}\in P^d$ and $d\in\mathbb{N}$:
\begin{itemize}
   \item $(p_1,\dots,p_d, \bigsqcap_{i=1}^{d} p_i)\sqsubseteq_\beta(p_1,\dots,p_d)\sqsubseteq_\alpha(p_1,\dots,p_d, \bigsqcup_{i=1}^{d} p_i)$, see \cite{CalvoMayor1999:remarks2eaf},
   \item $(p_1)\sqsubseteq_\gamma(p_1,p_2)\sqsubseteq_\gamma\dots\sqsubseteq_\gamma(p_1,\dots,p_d)\sqsubseteq_\gamma(p_1,\dots,p_d, \underline{0})$.
\end{itemize}
\end{proposition}
Also, if $(P,\sqsubseteq)$ is a chain, then the above extensions
of $\sqsubseteq$ generate lattices.

\bigskip
\begin{remark}
Regarding classical fusion functions based on
the above orderings, we have what follows.
Let $\func{F}:P^*\to P$ be a fusion function monotonic with respect to $\sqsubseteq^*$. Then:
\begin{itemize}
   \item $\func{F}$ is monotonic with respect to $\sqsubseteq_\alpha$
   if and only if $\func{F}(\vect{p})\sqsubseteq\func{F}(\vect{p}, \bigsqcup_{i=1}^{d_p} p_i)$,
   \item $\func{F}$ is monotonic with respect to $\sqsubseteq_\beta$
   if and only if $\func{F}(\vect{p}, \bigsqcap_{i=1}^{d_p} p_i)\sqsubseteq\func{F}(\vect{p})$
\end{itemize}
for all $\vect{p}\in P^*$, see \cite{CalvoMayor1999:remarks2eaf}.

Notably, in \cite{CalvoMayor1999:remarks2eaf} the concept of an extended aggregation function
on $P^*$ has been defined with the requirement of idempotency, as well as $\alpha$-, and
$\beta$-monotonicity.

Also let $\func{F}:P^*\to P$ be a fusion function monotonic with respect to $\sqsubseteq^*$
and $(P,\sqsubseteq,\underline{0}, \overline{1})$ be a bounded poset.
Then $\func{F}$ is monotonic with respect to $\sqsubseteq_\gamma$
if and only if $\func{F}(\vect{p})\sqsubseteq\func{F}(\vect{p}, \underline{0})$.

Yet, in this chapter we are interested in fusion functions like $\func{F}: (P^*)^n\to P^*$.
\end{remark}

\subsection{Aggregation methods}

As by using the listed extensions of $\sqsubseteq$ we get different lattices,
trivially, methods already discussed (note their great generality)
in Section~\ref{Sec:ProdLat} may be used to aggregate such kinds of data.

\section{Aggregation of informetric data}\label{Sec:aginformetricAny}

\begin{figure}[t!]
\centering

\scalebox{0.8}{
   \begin{tikzpicture}

   \begin{scope}[yshift=0cm]
   \draw[fill=black!50] (0,0)--(1.5,0)--(1.5,0.75)--(0.5,0.75)--(0.5,1.5)--(0,1.5)--(0,0);
   \draw[fill=black!30] (0.5,0.75)--(1,1.05)--(1,0.75);
   \draw[fill=black!20] (1,0.75)--(1.5,1.05)--(1.5,0.75);
   \draw[fill=gray!40,draw,decorate,decoration={bumps}] (0.2,1.5)--(0.1,1.9)--(-0.1,2.3)--(0.25,2.5)--(0.6,2.7)--(0.4,2.3)--(0.3,1.5);

   \node at(0.75,-0.5) {Producer $p_1$};

   \node at (2.5,0.15) {$\longrightarrow$};
   \node at (2.5,0.4) {$\longrightarrow$};

   \draw[fill=blue!5,rounded corners] (3.25,-0.25) rectangle (10.5,1);

   \draw[fill=brown!40,rounded corners] (3.5,0.0) rectangle (4.5,0.75);
   \node at(4,0.375) { $x_{1}^{(1)}$ };
   \draw[fill=brown!40,rounded corners] (4.75,0.0) rectangle (5.75,0.75);
   \node at(5.25,0.375) { $x_{2}^{(1)}$ };
   \node at (6.25,0.375) {\dots};
   \draw[fill=brown!40,rounded corners] (6.75,0.0) rectangle (7.75,0.75);
   \node at(7.25,0.375) { $x_{n_1}^{(1)}$ };

   \node at(7,-0.5) {$n_1$ products};

   \node[right] at(10.5,0.375) {$=\vect{x}^{(1)}\in\mathcal{S}_{\le d}$};

   \end{scope}

   \begin{scope}[yshift=-3.5cm]
   \draw[fill=black!50] (0,0)--(1.5,0)--(1.5,0.75)--(0.5,0.75)--(0.5,1.5)--(0,1.5)--(0,0);
   \draw[fill=black!30] (0.5,0.75)--(1,1.05)--(1,0.75);
   \draw[fill=black!20] (1,0.75)--(1.5,1.05)--(1.5,0.75);
   \draw[fill=gray!40,draw,decorate,decoration={bumps}] (0.2,1.5)--(0.1,1.9)--(-0.1,2.3)--(0.25,2.5)--(0.8,2.5)--(0.4,2.3)--(0.3,1.5);

   \node at(0.75,-0.5) {Producer $p_2$};

   \node at (2.5,0.15) {$\longrightarrow$};
   \node at (2.5,0.4) {$\longrightarrow$};

   \draw[fill=blue!5,rounded corners] (3.25,-0.25) rectangle (10.5,1);

   \draw[fill=brown!40,rounded corners] (3.5,0.0) rectangle (4.5,0.75);
   \node at(4,0.375) { $x_{1}^{(2)}$ };
   \draw[fill=brown!40,rounded corners] (4.75,0.0) rectangle (5.75,0.75);
   \node at(5.25,0.375) { $x_{2}^{(2)}$ };
   \node at (6.25,0.375) {\dots};
   \draw[fill=brown!40,rounded corners] (6.75,0.0) rectangle (7.75,0.75);
   \node at(7.25,0.375) { $x_{n_2-2}^{(2)}$ };
   \draw[fill=brown!40,rounded corners] (8,0.0) rectangle (9,0.75);
   \node at(8.5,0.375) { $x_{n_2-1}^{(2)}$ };
   \draw[fill=brown!40,rounded corners] (9.25,0.0) rectangle (10.25,0.75);
   \node at(9.75,0.375) { $x_{n_2}^{(2)}$ };

   \node at(7,-0.5) {$n_2$ products};

   \node[right] at(10.5,0.375) {$=\vect{x}^{(2)}\in\mathcal{S}_{\le d}$};

   \end{scope}

   \begin{scope}[yshift=-4.25cm]
   \node at(0.75, -0.5) {$\vdots$};
   \node at(4, -0.5) {$\vdots$};
   \node at(7.25, -0.5) {$\vdots$};

   \node at(4, -2) {$\vdots$};
   \node at(7.25, -2) {$\vdots$};
   \end{scope}

   \begin{scope}[yshift=-8cm]
   \draw[fill=black!50] (0,0)--(1.5,0)--(1.5,0.75)--(0.5,0.75)--(0.5,1.5)--(0,1.5)--(0,0);
   \draw[fill=black!30] (0.5,0.75)--(1,1.05)--(1,0.75);
   \draw[fill=black!20] (1,0.75)--(1.5,1.05)--(1.5,0.75);
   \draw[fill=gray!40,draw,decorate,decoration={bumps}] (0.2,1.5)--(-0.1,1.7)--(-0.1,2.3)--(0.25,2.4)--(0.6,2.5)--(0.4,2.3)--(0.3,1.5);

   \node at(0.75,-0.5) {Producer $p_k$};

   \node at (2.5,0.15) {$\longrightarrow$};
   \node at (2.5,0.4) {$\longrightarrow$};

   \draw[fill=blue!5,rounded corners] (3.25,-0.25) rectangle (10.5,1);

   \draw[fill=brown!40,rounded corners] (3.5,0.0) rectangle (4.5,0.75);
   \node at(4,0.375) { $x_{1}^{(k)}$ };
   \draw[fill=brown!40,rounded corners] (4.75,0.0) rectangle (5.75,0.75);
   \node at(5.25,0.375) { $x_{2}^{(k)}$ };
   \node at (6.25,0.375) {\dots};
   \draw[fill=brown!40,rounded corners] (6.75,0.0) rectangle (7.75,0.75);
   \node at(7.25,0.375) { $x_{n_k-1}^{(k)}$ };
   \draw[fill=brown!40,rounded corners] (8,0.0) rectangle (9,0.75);
   \node at(8.5,0.375) { $x_{n_k}^{(k)}$ };

   \node at(7,-0.5) {$n_k$ products};

   \node[right] at(10.5,0.375) {$=\vect{x}^{(k)}\in\mathcal{S}_{\le d}$};

   \end{scope}

   \node[fill=red!10,rounded corners] at(6,2.0) {quality ratings $\in\Ival$};
   \draw[red,->] (4.35,2)--(4,1.5)--(4,0.8);
   \draw[red,->] (5.25,1.75)--(5.25,0.8);
   \draw[red,->] (7.25,1.75)--(7.25,0.8);

   \end{tikzpicture}
}

\caption{\label{Fig:PAP} Producers, products, and their quality ratings, see \cite{CenaGagolewski2015:om3fss}.}
\end{figure}
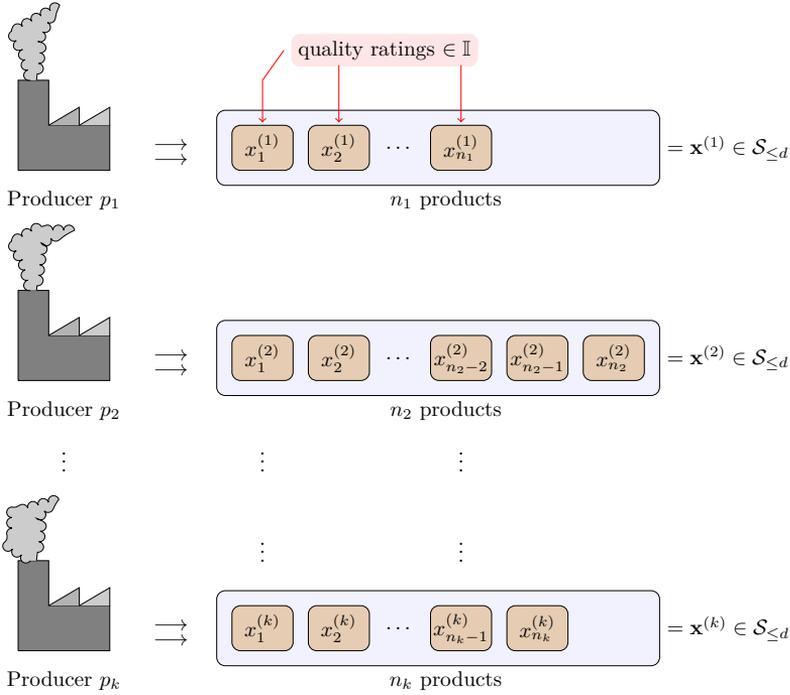

Typical practical situations in which we are faced with the need
to aggregate vectors of any length with elements in some real interval
$\Ival$ (commonly $\Ival=[0,\infty]$ or $\Ival=[-\infty,\infty]$)
include scientometrics, webometrics, marketing,
manufacturing, or quality engineering. Such application domains
are sometimes referred to as \index{informetrics}\emph{informetrics} (information metrics),
and their aim is to deal with quantitative aspects of information processes.
Here we assume that we have a set of abstract \emph{producers} that
output various numbers of \emph{products} and each product is given
a numeric valuation, representing its quality, see Figure~\ref{Fig:PAP},
Table~\ref{Table:InstancesPAP}, and, e.g., \cite{GagolewskiGrzegorzewski2011:ijar,%
CenaGagolewski2015:om3fss,FranceschiniMaisano2009:hmanufacturing,GagolewskiMesiar2012:joi}.

\begin{table}[t!]
\centering \caption[Representative instances of informetric and similar data,
where numeric lists of nonconforming lengths may be encountered.]%
{\label{Table:InstancesPAP}
Representative instances of informetric and similar data,
where numeric lists of nonconforming lengths may be encountered, see, e.g., \cite{CenaGagolewski2015:om3fss}.}

\noindent
\begin{tabularx}{1.0\linewidth}{p{2.9cm}XX}
\toprule
\small\bf producer & \small\bf products & \small\bf rating method  \\
\midrule
\textsf{R} package author & \textsf{R} packages & Number of dependencies \\
\midrule
Developer team & \textsf{Python} packages & Number of namespace imports from other projects \\
\midrule
Web server & Web pages & Number of targeting web-links or Page Rank \\
\midrule
Web service server & JSON/XML-RPC methods  & Number of remote procedure calls \\
\midrule
Developer team & Code repository (git, svn, etc.) & Number of commits
or lines of code\\
\midrule
Publisher & On-line document & Number of downloads  \\
\midrule
Social networking profile & Posts & Number of ``tweets'' or ``likes'' \\
\midrule
StackOverflow users & Answers to other users' questions & Up-votes \\
\midrule
YouTube channels & Videos & Number of views \\
\midrule
Digital library & Subscriber & Number of accesses \\
\midrule
Scientist & Scientific articles & Number of citations  \\
\midrule
Scientific institute & Scientists & The $h$-index \\
\midrule
Factory   & Model-ranges of products  & Sale results \\
\midrule
Factory product  & Supplied lots  & Number of items without defects \\
\midrule
Artist   & Paintings  & Auction price \\
\bottomrule
\end{tabularx}
\end{table}

Most often, the order of elements in input vectors does not matter.
Therefore, we may  assume that the vectors we aggregate are already sorted.
For any $d$ and a fixed $\Ival$, let $\mathcal{S}_d$ designate the set of nonincreasingly ordered
vectors of length $d$, i.e., $\mathcal{S}_d=\{ (x_1, \dots, x_d)\in\IvalPow{d}, x_1\ge\dots\ge x_d\}$.
Moreover, let $\mathcal{S}_{\le d}$ be a set of nonincreasingly ordered
vectors of length at most $d$, that is $\mathcal{S}_{\le d}=\bigcup_{i=1}^d \mathcal{S}_i$.
From now on we also assume that $\mathcal{S}=\mathcal{S}_{\le\infty}$.

Suppose that we are given $n$ producers and that each of them produced no more than $d$ products
for some $d$. Obviously, such $d$ is finite and well defined for each set of producers.
The set of producers may thus be represented as $\mathcal{X} = \{\vect{x}^{(1)},\dots, \vect{x}^{(n)}\}$,
where $\vect{x}^{(i)}=\left(x_1^{(i)},\dots,x_{d_i}^{(i)}\right)\in\mathcal{S}_{\le d}$ for all $i=1,\dots,n$.
For instance, $x_{j}^{(i)}$ may denote the number of citations of the $j$th most cited
paper of the $i$th scholar, or the score of the $j$th best post written
by the $i$th Stack Exchange user.

In this section we are interested in constructing fusion functions like
$\func{F}: \mathcal{S}_{\le d}^n\to \mathcal{S}_{\le d}$
or their extended versions
$\func{F}: \mathcal{S}_{\le d}^*\to \mathcal{S}_{\le d}$.
They aim to  determine the most ``typical'' or ``representative'' output of
a producer in a cluster of producers.
They may be used in, among others, informetric data clustering
tasks, see the papers by Cena and Gagolewski \cite{CenaGagolewski2015:kmeansinformetric,CenaGagolewski2015:fuzzycmeansinformetric}
(for a (fuzzy) $k$-means-like procedure)
and also \cite{OrtegaEtAll:multiinstitutes,ChengLiu:clusteruniversity,Costas2010:bibclass,Ibanez2013:spanish}.
Moreover, note that in Chapter~\ref{Chap:Characteristics} we shall focus on numeric characteristics of informetric data,
which include such tools as the famous Hirsch $h$-index (a particular Sugeno integral).

Possibly desirable properties of fusion functions of our interest here include:
\begin{itemize}
   \item monotonicity with respect to $\sqsubseteq_\gamma$,

   \item symmetry,

   \item $\func{F}(n\ast\vect{x})=\vect{x}$, \hfill(idempotency)

   \item $\func{F}((x_1), (x_1, x_2), \dots, (x_1,x_2,\dots,x_n))=(x_1,x_2,\dots,x_j)$
   for some $j\in[n]$ and all $n$ as well as $x_1, x_2, \dots x_n\in\Ival$, $x_i \ge x_{i+1}$, $i\in[n-1]$,
   \par\hfill(idempotency on common indices)

   \item $d_y\in\left[ d_\mathrm{min}, d_\mathrm{max} \right]$, \hfill(length internality)

   \item $y_j \in [ \bigwedge_{i=1}^n x_j^{(i)}, \bigvee_{i=1}^n x_j^{(i)} ]$ for all $j\in[d_\mathrm{min}]$,
   \par\hfill(componentwise internality on common indices)

   \item $y_j \in [ \bigwedge_{i=1}^n x_{d_i}^{(i)}, \bigvee_{i=1}^n x_1^{(i)} ]$ for every $j\in[d_y]$,
   \hfill(global internality)

   \item $\func{F}(\vect{x}^{(1)},\dots,\vect{x}^{(n)})=\func{F}({\vect{x}'}^{(1)},\dots,{\vect{x}'}^{(n)})=\vect{y}$
   where $\vect{x'}^{(i)}\in\{\vect{x}^{(i)}, \vect{y}\}$ for every $i\in[n]$,

   \item stability,
\end{itemize}
for all $\vect{x},\vect{x}^{(1)},\dots,\vect{x}^{(n)}\in\mathcal{S}_{\le d}$,
where we assume that $\func{F}(\vect{x}^{(1)},\dots,\vect{x}^{(n)})=\vect{y}$,
$d_y = |\vect{y}|$,
$d_i = |\vect{x}^{(i)}|$, $d_\mathrm{min}=\bigwedge_{i=1}^n d_i$, and
$d_\mathrm{max}=\bigvee_{i=1}^n d_i$.

\subsection{Metrics on the space of numeric strings}

In order to construct fusion functions on the considered domain, defined as minimizers of some penalty,
let us study a family of metrics introduced by Cena, Gagolewski, and Mesiar in
\cite{CenaETAL2015:prodclust}.
Recall that among some interesting metrics in the space of vectors of the same lengths $\mathbb{R}^d$,
we have, e.g., the Euclidean $\mathfrak{d}_{2}(\vect{x},\vect{y}) = \sqrt{\sum_{i=1}^d (x_i-y_i)^2}$
or the Manhattan $\mathfrak{d}_{1}(\vect{x}, \vect{y})=\sum_{i=1}^d |x_i-y_i|$ distance.
Metrics on sets of real vectors are most often defined
by considering absolute values of pairwise differences of vectors' elements
(see the notion of a norm-generated metric).
Letting $\sum_{i=u}^v \cdots = 0$ for $u > v$,
one way to redefine, e.g., $\mathfrak{d}_{1}$ so that it acts on elements in $\mathcal{S}_{\le d}$
is to consider:
\begin{equation}\label{Eq:L1primedef}
\mathfrak{d}'_{1}(\vect{x},\vect{y}) = \sum_{i=1}^{d_x}|x_{i} - y_{i}|+
\sum_{i=d_x+1}^{d_y}|y_{i}|,
\end{equation}
where, by symmetry, without loss in generality, we assume that
$d_x\le d_y$.
Note that as $|a-0|=|0-a|=|a|$, we have
$\mathfrak{d}'_{1}(\vect{x},\vect{y})=\mathfrak{d}_{1}(\vect{\tilde{x}}, \vect{\tilde{y}})$,
where, e.g., $\vect{\tilde{x}}=(x_1,\dots,x_{d_x},0,0,\dots,0)\in\mathbb{R}^d$,
i.e., a version of $\vect{x}$ padded with 0s
(a similar idea is reflected in the Hirsch h-index, which in fact is
the Ky Fan metric \cite{Fan1943:metric} applied
to $\vect{x}$ and a $\vect{0}$ vector of the same length as $\vect{x}$).

Unfortunately, $\mathfrak{d}'_{1}$ is only a pseudometric on $\mathcal{S}_{\le d}$:
a vector $(x_1,\dots,\allowbreak x_{d_x})$ is indistinguishable from $(x_1,\dots,x_{d_x},0,0,\dots,0)$.
In other words, nonexistent products are treated in the same way as products of
quality $0$. This setting, however, is not completely valid in our framework.
Thus, an additional penalty for the difference in vectors' lengths may be introduced.

\begin{theorem}\label{Prop:gener}
Let $\mathfrak{d}:\mathcal{S}_{\le d}\times\mathcal{S}_{\le d}\to[0,\infty]$
be such that $\mathfrak{d}(\vect{x},\vect{y})=\mu(\tilde{\vect{x}},\tilde{\vect{y}})+\nu(\vect{x},\vect{y})$,
where $\mu$ is a metric on $\mathbb{R}^d$ and $\nu$ is a pseudometric on
$\mathcal{S}_{\le d}$. Then $\mathfrak{d}$ is a metric on $\mathcal{S}_{\le d}$
if and only if for all $\vect{x},\vect{y}$ such that $\tilde{\vect{x}}=\tilde{\vect{y}}$
it holds $\nu(\vect{x},\vect{y})=0 \Longrightarrow n_x=n_y$.
\end{theorem}

In particular, we may consider $\nu(\vect{x},\vect{y})$ which is just a function
of vector lengths, e.g., $\nu(\vect{x},\vect{y})=p|d_x^r-d_y^r|$
for any $p,r>0$. In such a way, the Manhattan and the Euclidean metric
may be rewritten as:
\begin{equation}\label{Eq:M1def}
\mathfrak{d}_{M_1, p, r}(\vect{x},\vect{y}) = \mathfrak{d}_{1}(\tilde{\vect{x}}, \tilde{\vect{y}})+p|d_x^r-d_y^r|
\end{equation}
and:
\begin{equation}\label{Eq:M2def}
\mathfrak{d}_{M_2, p, r}(\vect{x},\vect{y}) = \mathfrak{d}_{2}(\tilde{\vect{x}}, \tilde{\vect{y}})+p|d_x^r-d_y^r|,
\end{equation}
respectively.
Take any vectors $\vect{x}, \vect{y}$ such that $n_x < n_y$.
Both metric classes possess the important property that
the distance between $\vect{x}$ and $\vect{y}$ is smaller than
the distance between $\vect{x}$ and $(\vect{y},a,a,\dots,a)$,
i.e., $\vect{y}$ padded with at least one value $a\in\Ival$.
In other words, such metrics are able to distinguish vectors of different
lengths from each other.

\subsection{Centroid}

\index{centroid}%
Let us study fusion functions that minimize sums of $\mathfrak{d}_{M_2, p, r}$-based
penalties of the form:
\begin{eqnarray}\label{Eq:SqM2informetricdist}
\mathfrak{d}_{p,r}^2(\vect{x},\vect{y}) &=& \sum_{i=1}^{\min{\{d_x, d_y\}}}(x_{i} - y_{i})^2+
\sum_{i=d_x+1}^{n_y}y_{i}^2+\\\nonumber
&+&\sum_{i=d_y+1}^{n_x}x_{i}^2+p|d_x^r-d_y^r|,
\end{eqnarray}
which lead to centroid-like (see Section~\ref{Sec:centroid}) mappings:
\begin{eqnarray}\label{Eq:SqM2informetricentroid}
   \func{F}(\vect{x}^{(1)},\dots,\vect{x}^{(n)}) = \underset{\vect{y}\in\mathcal{S}} {\operatorname{arg\,min}}
   \sum_{i=1}^n \mathfrak{d}_{p,r}^2(\vect{x}^{(i)}, \vect{y}).
\end{eqnarray}
First of all, let us note that $|\func{F}(\vect{x}^{(1)},\dots,\vect{x}^{(n)})|\le d=\max\{|\vect{x}^{(i)}|: i=1,\dots,n\}$.
It can be shown, see \cite{CenaGagolewski2015:kmeansinformetric}, that the value of $\func{F}$ may be determined
by using the following 2-step procedure.
\begin{enumerate}
\item First of all, for all $j=1,\dots,d$ we compute:
\[
   \vect{y}^{(j)} = \underset{\vect{y}\in\mathcal{S}_d} {\operatorname{arg\,min}}
   \sum_{i=1}^n \mathfrak{d}_{p,r}^2(\vect{x}^{(i)}, \vect{y}).
\]
\item Then we set:
\begin{eqnarray*}
   j^* & = & \underset{j=1,\dots,d} {\operatorname{arg\,min}}
   \sum_{i=1}^n \mathfrak{d}_{p,r}^2(\vect{x}^{(i)}, \vect{y}^{(j)}),\\
   \func{F}(\vect{x}^{(1)},\dots,\vect{x}^{(n)}) &=& \vect{y}^{(j^*)}.
\end{eqnarray*}
\end{enumerate}

It turns out that if we act on $\Ival=[0,\infty]$,
we simply have for all $j\in[d]$ and $i\in[j]$:
\[
y_i^{(j)} = \frac{1}{n} \sum_{k=1}^n \tilde{x}_i^{(k)}.
\]
The obtained fusion function is
symmetric, idempotent, and length internal, among others.

The above formula is unfortunately invalid for arbitrary $\Ival$. This
is due to the fact that the space $\mathcal{S}_{\le d}$ consists of
\emph{ordered} vectors. Thus, in general, the task that aims to determine
a penalty minimizer here is much more difficult.
It may be shown that a procedure given in Figure~\ref{Fig:kmeansprod2} may be used
to compute $\func{F}(\vect{x}^{(1)},\dots,\vect{x}^{(n)})$,
see \cite{CenaGagolewski2015:kmeansinformetric} for more details -- some of the components
have to be averaged. Unfortunately, if we allow negative elements,
the resulting fusion function is definitely not $\sqsubseteq_\gamma$-nondecreasing.

Note that this procedure may relatively easily be generalized
to the case of weighted $\mathfrak{d}_{p,r}^2$ penalty functions,
see \cite{CenaGagolewski2015:fuzzycmeansinformetric}.

\begin{example}[\cite{CenaGagolewski2015:kmeansinformetric}]%
Let:
\begin{equation*}
\mathcal{X}=\left\{
\begin{array}{rrrrrrl}
(&42,&21,&12,&10,&8&),\\
(&1,&0,&-10\phantom{,}&&&),\\
(&0,&-1\phantom{,}&&&&),\\
(&-10,&-13\phantom{,}&&&&)\\
\end{array}
\right\}.
\end{equation*}
Assuming that $p=1, r=1$, we have $\func{F}(\mathcal{X})=(8\frac{1}{4}, 4\frac{1}{4}, 1\frac{2}{3}, 1\frac{2}{3}, 1\frac{2}{3})$.
Here is the output of the procedure given in Figure~\ref{Fig:kmeansprod2} for each $j=1,\dots,d$.

\begin{center}
\begin{tabularx}{1.0\linewidth}{lrrrrrrr}
\toprule
$j$ &    $\sum_{i=1}^n \mathfrak{d}_{1,1}^2(\vect{x}^{(i)}, \vect{y}^{(j)})$
& $y_1^{(j)}$ & $y_2^{(j)}$ & $y_3^{(j)}$ & $y_4^{(j)}$ & $y_5^{(j)}$ & $y_6^{(j)}$ \\
\midrule
1 & 3139.75           & 8.25 \\
2 & 3063.50           & 8.25 &4.25 \\
3 & 3062.50           & 8.25 &4.25 &0.50 \\
4 & 3047.50           & 8.25 &4.25 &1.50 &1.50 \\
5 & \underline{3034.17}  & 8.25 &4.25 &1.67 &1.67 &1.67 \\
6 & 3037.17           & 8.25 &4.25 &1.67 &1.67 &1.67 &0.00 \\
\bottomrule
\end{tabularx}
\end{center}
\end{example}

\begin{example}[\cite{CenaGagolewski2015:kmeansinformetric}]%
Let:
\begin{equation*}
\mathcal{X}=\left\{
\begin{array}{rrrrrrl}
(&-10,&-12,&-14,&-16,&-17&),\\
(&1,&0,&-10\phantom{,}&&&),\\
(&-10,&-15,&-16\phantom{,}&&&),\\
(&-20\phantom{,}&&&&&)\\
\end{array}
\right\}.
\end{equation*}
Then  $\func{F}(\mathcal{X})=(-6.95, -6.95, -6.95, -6.95, -6.95)$
for $p=1, r=1$.

\begin{center}
\begin{tabularx}{1.0\linewidth}{lrrrrrr}
\toprule
$j$ &    $\sum_{i=1}^n \mathfrak{d}_{1,1}^2(\vect{x}^{(i)}, \vect{y}^{(j)})$           & $y_1^{(j)}$ & $y_2^{(j)}$ & $y_3^{(j)}$ & $y_4^{(j)}$ & $y_5^{(j)}$ \\
\midrule
1 & 1694.75           & -9.750 &\\
2 & 1528.50           & -8.250 &-8.250 \\
3 & 1126.50           & -8.250 &-8.250 &-10.000 \\
4 & 1142.75           & -7.625 &-7.625 & -7.625 &-7.625 \\
5 & \underline{1108.95}  & -6.950 &-6.950 & -6.950 &-6.950 & -6.950 \\
\bottomrule
\end{tabularx}
\end{center}
\end{example}

\subsection{1-Median}\label{Sec:1mediannumstr}

\index{Euclidean 1-median}%
Due to the nature of the introduced metrics,
the described 2-step procedure may also be incorporated in the case of
finding the 1-median with respect to
the $\mathfrak{d}_{M_1,p,r}$ and $\mathfrak{d}_{M_2,p,r}$ metrics.

For $\Ival=[0,\infty]$ and $\mathfrak{d}_{M_1,p,r}$, the 1-Median
of course corresponds to the componentwise median (with missing elements
treated as $0$s). That is, for some $j\in[d]$ and $i\in[j]$ we have:
\[
   y_i^{(j)} = \func{Median}(\tilde{x}_i^{(1)}, \dots, \tilde{x}_i^{(n)}).
\]
By the monotonicity of $\func{Median}$ and the fact that $0 \le x_i^{(j)}$
for all $j\in[n]$ and $i\in[d_j]$, we have that if $\vect{x}^{(j)}\in\mathcal{S}$,
then $\vect{y}^{(j)}\in\mathcal{S}$. In other words, the resulting vector is surely
sorted.

\begin{remark}
Inspired by the above derivations, we may introduce the following family
of componentwise fusion functions for numeric strings in the case of $\Ival=[0,\infty]$.
Let $\func{F}:\Ival^n\to\Ival$ be a nondecreasing fusion function.
Note that $\tilde{x}^{(1)},\dots,\tilde{x}^{(n)}\in\IvalPow{d}$ and thus
with data transformed in such a way we obtain a case exactly as in the previous
chapter. Now let $\vect{y}\in\mathcal{S}_d$ be such that for $i\in[d]$ we have:
\[
   y_i = \func{F}(\tilde{x}_i^{(1)}, \dots, \tilde{x}_i^{(n)}).
\]
Then a penalty-based solution may be given as $(y_1,\dots,y_{d'})$
where $d'$ is given by:
\[
   d' = \argmin_{d'\in [d]} P\left((y_1,\dots,y_{d'}); \vect{x}^{(1)},\dots,\vect{x}^{(n)}\right).
\]
For instance, $P$ may be given as
$P\left((y_1,\dots,y_{d'}); \vect{x}^{(1)},\dots,\vect{x}^{(n)}\right)
= \sum_{k=1}^n p|d_k^r-d'^r|$ for some $p,r>0$. Of course, similarly as
in Definition~\ref{Def:PenaltyFunction}, we require
$P:\mathcal{S}\times\mathcal{S}^{n}\to[0,\infty]$ to fulfill:
\begin{itemize}
   \item $P(\vect{y};\vect{x}^{(1)},\dots,\vect{x}^{(n)})=0$
   if $\vect{y}=\vect{x}^{(j)}$ for all $j\in[n]$,
   \item for every fixed $\vect{x}^{(1)},\dots,\vect{x}^{(n)}$, the set of minimizers of $P(\vect{y};\vect{x}^{(1)},\dots,\vect{x}^{(n)})$ is a singleton.
\end{itemize}
Moreover, if $\argmin_{d'\in [d]} P\left((y_1,\dots,y_{d'}); \vect{x}^{(1)},\dots,\vect{x}^{(n)}\right)$ is ambiguous,
then we may choose the smallest one or the largest $d'$ which minimizes the penalty.
\end{remark}

Note that the case of $\mathfrak{d}_{M_2,p,r}$ is slightly more difficult.
First of all, we need the following result.

\begin{proposition}\label{Prop:convexsorted}
For all $d$ and $n$, a convex combination of any $n$ vectors in $\mathcal{S}_d$ is also a vector in $\mathcal{S}_d$.
\end{proposition}

Of course, for any $\vect{x}\in\mathcal{S}_{\le d}$ and $\Ival=[0,\infty]$,
it holds that $\tilde{\vect{x}}\in\mathcal{S}_d$.
Recall that in Section~\ref{Sec:1medianfixedd} we noted that
the 1-median is within the convex hull of a set of input points,
see Equation~\eqref{Eq:1medianconvexhull}. This implies that $\vect{y}^{(j)}$
is sorted. Thus, e.g., the Weiszfeld algorithm may be used in the 1st step
of our 2-step procedure.

\section{Aggregation of character strings}\label{Sec:nominalscaleany}

This time, for a given $X$, let $X^*=\bigcup_{d=0}^\infty X^d$
denote the Kleene closure of $X$.
In particular, an empty vector $\varepsilon\in X^*$.

In Section \ref{Sec:aginformetricAny} we discussed a few methods to aggregate
vectors of nonconforming lengths. Each member of such a vector
was a real number. It was quite a comfortable situation,
as algebraic operations like addition, multiplication, division, and so forth,
were defined there -- we were on an interval scale.
Here we revisit a situation where vectors with elements
on a nominal scale are to be aggregated, see Sections~\ref{Sec:nominalscale1}
and~\ref{Sec:nominalscaled}.

\index{character string}%
A tuple $\vect{x}\in\Sigma^*$ is often called a \emph{character string}
(over a finite set $\Sigma$), and an element $s_i$ -- the $i$th character.
If, say, $\mathtt{a},\mathtt{b}\in\Sigma$,
then we shall sometimes write, e.g., \texttt{\quot{}aba\quot}
instead of $(\mathtt{a},\mathtt{b},\mathtt{a})$.

\begin{example}\index{DNA sequence}%
Going back to Example~\ref{Ex:NominalACGT}, character strings over
$\Sigma=\{\mathtt{A},\mathtt{C},\mathtt{G},\mathtt{T}\}$
may be interpreted as DNA sequences or protein sequences in the case of $|\Sigma|=20$.
On the other hand, referring to Example~\ref{Ex:NominalASCII},
we may also consider ASCII or Unicode character strings.
\end{example}

\begin{example}
Strings over  $\Sigma=\{\mathtt{0},\mathtt{1}\}$ are called bit strings.
In fact, even if generally it is not the most
convenient perspective, each computer file or a digital signal
transmission may be viewed as sequence of bits
(see Table~\ref{Tab:ASCII}).
Interestingly, there are a few different ways to map Unicode
code point sequences to bit sequences: UTF-8, UTF-16LE, UTF-16BE,
UTF-32LE, UTF-32BE, and so on.
\end{example}

Here we are of course  interested in fusion functions
like $\func{F}:(\Sigma^*)^n\to\Sigma^*$.

\begin{remark}
In SQL-oriented relational database management systems,
the term ``string aggregation'' is usually understood as a form
of string concatenation (joining). For instance, in SQLite,
there is an aggregate $\mathtt{GROUP\_CONCAT()}$,
which joins a given list of strings, separating each item with a comma.
We have, e.g.:
\[
\mathtt{GROUP\_CONCAT}(\text{\texttt{"{}a"{}}}, \text{\texttt{"{}bcd"{}}}, \text{\texttt{"{}abc"{}}})
= \text{\texttt{"{}a,bcd,abc"{}}.}
\]
A similar function in Oracle Database
is named $\mathtt{LISTAGG()}$ and in PostgreSQL -- $\mathtt{STRING\_AGG()}$.
Additionally, it is worth noting that it is an associative
string fusion function, see \cite{LehtonenETAL2014:astringfun}.
\end{remark}

As far as desired properties of such fusion functions are concerned,
firstly, let us stress that this time no ordering relation can naturally be taken into account,
so we cannot rely on any type of monotonicity.
On the other hand, at this point we may be expecting:
\begin{itemize}
   \item symmetry,
   \item $\func{F}(n\ast\vect{x})=\vect{x}$,\hfill (idempotency)
   \item $|\func{F}(\vect{x}^{(1)},\dots,\vect{x}^{(n)})|\in
   \left[ \bigwedge_{i=1}^n |\vect{x}^{(i)}|, \bigvee_{i=1}^n |\vect{x}^{(i)}| \right]$,
   \hfill(length internality)
   \item $\func{F}(\vect{x}^{(1)},\dots,\vect{x}^{(n)})\in \Sigma'^*$ with
   $\Sigma' = \{ s_j^{(i)}: i\in[n], j\in[|\vect{x}^{(i)}|] \}$
(the output is only based on characters which are used in the inputs;
this is because we rather do not want the result of aggregating
(\texttt{"{}a"{}}, \texttt{"{}c"{}}, \texttt{"{}c"{}}, \texttt{"{}a"{}})
to be  \texttt{"{}b"{}}),
   \item for some $\mathtt{b_1},\dots,\mathtt{b_j}\in\Sigma$, there exists $i\in[j]$ such that
   $\func{F}(\quot{}\mathtt{b_1}\quot{},\quot{}\mathtt{b_1b_2}\quot{},\allowbreak\dots,\allowbreak
   \quot{}\mathtt{b_1b_2}\dots\mathtt{b_j}\quot{})=\allowbreak
   \quot{}\mathtt{b_1b_2}\dots \mathtt{b_i}\quot{}$,
   \item $\func{F}(\vect{x}^{(1)},\dots,\vect{x}^{(n)})
   =\func{F}({\vect{x}'}^{(1)},\dots,{\vect{x}'}^{(n)})=\vect{y}$
   where $\vect{x'}^{(j)}\in\{\vect{x}^{(j)}, \vect{y}\}$,
   \item stability,
\end{itemize}
for each $\vect{x},\vect{x}^{(1)},\dots,\vect{x}^{(n)}\in\Sigma^*$.

The fusion functions discussed further on are defined as
minimizers of carefully aggregated pseudometrics or metrics.
Thus, in the latter case we naturally expect to obtain idempotent
fusion functions. Nevertheless, we should not go that far now: firstly we shall
define a few types of dissimilarity measures for strings.

\subsection{Dissimilarity measures of character strings}

Let us review the most frequently used dissimilarity measures
of character strings, see also \cite{Kruskal1983:sequencecomparison,%
Navarro2001:stringmatch,Boytsov2011:approxdictsearch,Loo2014:stringdist}.
They are not only used for constructing fusion functions,
but have numerous other applications, e.g., in
spelling correction \cite{Morgan1970:spellcorr,MaysDamerauMercer1991:contextspell},
error-tolerant pattern searching,
fuzzy/approximate pattern matching \cite{WandeltETAL2014:stringsimsearch},
plagiarism detection \cite{BartoszukGagolewski2014:fuzzyrsimilar,BartoszukGagolewski2015:similar2},
text retrieval, optical character recognition,
text clustering \cite{ZadroznyKacprzyk2006:textcat,KacprzykZadrozny2007:textcat},
file revisions comparison (see the UNIX \texttt{diff} utility
or software revision control systems like \texttt{git}, \texttt{svn}, or
\texttt{mercurial}),
and many others.

\index{character string distance}%
Below we discuss the following classes of character string distances:
\begin{itemize}
   \item edit-based distances,
   \item $q$-gram-based distances,
   \item other.
\end{itemize}

\begin{remark}
Let us once again stress  that some of the dissimilarity measures presented below
-- like the Jaccard $q$-gram index -- are not necessarily ``full'' metrics.
A few of them violate the condition ``$\mathfrak{d}(x,y)=0$ if and only if $x=y$''.
They are at least \emph{pseudometrics}: the ``only if'' part of the axiom
might not be met in all the cases.
\end{remark}

\subsubsection{Edit-based distances}

Edit-based distances express the smallest total \textit{cost}
of necessary \textit{changes} that
have to be made to transform one string so as to get another one.

\begin{definition}\index{edit operation}%
An \emph{edit operation} is a pair $(\vect{a},\vect{b})\in\Sigma^*\times\Sigma^*$,
written $\vect{a}\to\vect{b}$ for short.
\end{definition}

For instance, the set $\mathbb{B}\subseteq\Sigma^*\times\Sigma^*$ of admissible edit operations may include:
\begin{itemize}
   \item single character removal ($(\forall\mathtt{a}\in\Sigma)$, $(\quot\mathtt{a}\quot\to\varepsilon) \in \mathbb{B}$),
   \item single character insertion  ($(\forall\mathtt{b}\in\Sigma)$, $(\varepsilon\to\quot\mathtt{b}\quot) \in \mathbb{B}$),
   \item single character substitution ($(\forall\mathtt{a},\mathtt{b}\in\Sigma)$, $(\quot\mathtt{a}\quot\to\quot\mathtt{b}\quot) \in \mathbb{B}$),
   \item transposition of an adjacent pair of characters (swap; $(\forall\mathtt{a},\mathtt{b}\in\Sigma)$, $(\quot\mathtt{ab}\quot\to\quot\mathtt{ba}\quot) \in \mathbb{B}$),
\end{itemize}
compare the historical Damerau paper \cite{Demerau1964:dist} for discussion.

We may apply an edit operation $\vect{a}\to\vect{b}$
at a position $i$ of a string $\vect{u}$, if:
\[ (u_i,u_{i+1},\dots,u_{i+|\vect{a}|-1})=\vect{a}. \]
In such a way, we derive a new string:
\[ \vect{v}=(u_1,\dots,u_{i-1},\vect{b},u_{i+|\vect{a}|},\dots,u_{|\vect{u}|}).\]

Given any two strings $\vect{u},\vect{v}$, we may seek
a \emph{transforming sequence} $(s_1,\dots,s_k)$ of edit operations and positions where they
are applied, $s_i\in\mathbb{B}\times\mathbb{N}$,
such that $\vect{v}$ may be derived step by step from $\vect{u}$.
In order to define an edit distance, we shall assume that
$\mathbb{B}$ is such that for each pair of strings there always exists
at least one such sequence. From now on denote by
$\mathrm{S}(\vect{u},\vect{v})\in(\mathbb{B}\times\mathbb{N})^*$
the set of all transforming sequences
that enable us to derive $\vect{v}$ from $\vect{u}$.

\begin{example}
Assume that the set of admissible transformations
consists only of single character removal and insertion.
In such a setting, at least three steps are needed to
derive \texttt{"{}fiction"{}} from \texttt{"{}function"{}}.

\begin{center}
\begin{tikzpicture}
   \node at (0,0)   {\tt\bfseries f};
   \node at (0.5,0) {\tt\bfseries u};
   \node at (1,0)   {\tt\bfseries n};
   \node at (1.5,0) {\tt\bfseries c};
   \node at (2,0)   {\tt\bfseries t};
   \node at (2.5,0) {\tt\bfseries i};
   \node at (3,0)   {\tt\bfseries o};
   \node at (3.5,0) {\tt\bfseries n};

   \node at (0.5,-0.5) {  \rotatebox{90}{$\looparrowright$} };
   \node[anchor=west] at (4,-0.5) { $\quot\mathtt{u}\quot\to\varepsilon$ (index 2)};
   \draw[dotted] (0.8, -0.55)--(3.9,-0.55);

   \node at (0,-1)   {\tt\bfseries f};
   \node at (0.5,-1) {\tt\bfseries n};
   \node at (1,-1)   {\tt\bfseries c};
   \node at (1.5,-1) {\tt\bfseries t};
   \node at (2,-1)   {\tt\bfseries i};
   \node at (2.5,-1) {\tt\bfseries o};
   \node at (3,-1)   {\tt\bfseries n};

   \node at (0.5,-1.5) {  \rotatebox{90}{$\looparrowright$} };
   \node[anchor=west] at (4,-1.5) { $\quot\mathtt{n}\quot\to\varepsilon$  (index 2)};
   \draw[dotted] (0.8, -1.55)--(3.9,-1.55);

   \node at (0,-2)   {\tt\bfseries f};
   \node at (0.5,-2) {\tt\bfseries c};
   \node at (1,-2)   {\tt\bfseries t};
   \node at (1.5,-2) {\tt\bfseries i};
   \node at (2,-2)   {\tt\bfseries o};
   \node at (2.5,-2) {\tt\bfseries n};

   \node at (0.25,-2.5) {  \rotatebox{90}{$\looparrowright$} };
   \node[anchor=west] at (4,-2.5) { $\varepsilon\to\quot\mathtt{i}\quot$  (index 2)};
   \draw[dotted] (0.55, -2.55)--(3.9,-2.55);

   \node at (0,-3)   {\tt\bfseries f};
   \node at (0.5,-3) {\tt\bfseries i};
   \node at (1,-3)   {\tt\bfseries c};
   \node at (1.5,-3) {\tt\bfseries t};
   \node at (2,-3)   {\tt\bfseries i};
   \node at (2.5,-3) {\tt\bfseries o};
   \node at (3,-3)   {\tt\bfseries n};
\end{tikzpicture}
\end{center}
\end{example}

Also, we might be interested in introducing a function
$\delta:\mathbb{B}\to\mathbb{R}_{0+}$ that measures
the cost of applying any given edit operation.

\begin{definition}\index{edit distance|see {generic edit distance}}\index{generic edit distance}%
A \emph{generic edit distance} relative to a set of edit operations $\mathbb{B}$
and their costs $\delta:\mathbb{B}\to\mathbb{R}_{0+}$
is given for any $\vect{u},\vect{v}\in\Sigma^*$ by:
\begin{equation}
\mathfrak{d}(\vect{u},\vect{v})
=\min_{S\in\mathrm{S}(\vect{u},\vect{v})} \sum_{(b,i)\in S} \delta(b).
\end{equation}
\end{definition}

The following result follows from \cite[Theorem 2.8]{Boytsov2011:approxdictsearch}.

\begin{theorem}\label{Thm:editdstmetric}
Whenever $\mathbb{B}$ and $\delta$ are such that:
\begin{itemize}
   \item if $(\vect{a}\to\vect{b})\in\mathbb{B}$, then also the reverse operation
   $(\vect{b}\to\vect{a})\in\mathbb{B}$; moreover, we have
   $\delta(\vect{a}\to\vect{b})=\delta(\vect{b}\to\vect{a})$,
   \item if $(\vect{a}\to\vect{b})\in\mathbb{B}$,
   then $\delta(\vect{a}\to\vect{b})=0$ implies that
   $\vect{a}=\vect{b}$,
   \item $\mathbb{B}$ is finite,
\end{itemize}
then the generic edit distance relative to $\mathbb{B}$ and $\delta$
is a metric on $\Sigma^*\times\Sigma^*$.
\end{theorem}

Classical edit distances assume that each edit operation has unit cost.

\begin{definition}
\index{LCS distance|see {longest common subsequence distance}}\index{longest common subsequence distance}%
The \emph{longest common subsequence (LCS) distance} \cite{NeedlemanWunsch1970:lcsdist},
$\mathfrak{d}_{\mathrm{LCS}}$, for $\vect{u}, \vect{v}\in\Sigma^*$ is defined as
the minimal number of single character  insertions and deletions
that are used to derive $\vect{v}$ from $\vect{u}$.
In other words, it is an edit distance given by
$\mathbb{B}=\{(\quot\mathtt{a}\quot\to\varepsilon),(\varepsilon\to\quot\mathtt{a}\quot): \mathtt{a}\in\Sigma\}$
and $\delta(b)=1$ for every $b\in\mathbb{B}$.
\end{definition}

If we additionally enable the use of a single character replacement operation,
we might get an extended version of the metric introduced
in Definition~\ref{Def:HammingDistance}.

\begin{remark}
\index{Hamming distance}%
The extended \emph{Hamming  distance}
 $\mathfrak{d}_\mathrm{H}$ of $\vect{u}, \vect{v}\in\Sigma^*$
is given by:
\begin{equation}
   \mathfrak{d}_\mathrm{H}(\vect{u}, \vect{v})=\left\{
   \begin{array}{ll}
   \sum_{i=1}^d \indicator(u_i \neq v_i) & \text{if }|\vect{u}|=|\vect{v}|=d, \\
   \infty & \text{otherwise}.\\
   \end{array}
   \right.
\end{equation}
It may may be conceived as a kind of edit distance,
where single character replacement has unit cost
and character insertion and removal has an infinitely large cost.
\end{remark}

The LCS character modification scheme together with the replacement operation
leads us also to the famous Levenshtein distance.

\begin{definition}
\index{Levenshtein distance}%
The \emph{Levenshtein distance} \cite{Levenshtein1966:binarycodes},
$\mathfrak{d}_{\mathrm{LV}}$, for $\vect{u}, \vect{v}\in\Sigma^*$ is defined as
the minimal number of  single character insertions, deletions, and replacements
that are used to obtain $\vect{u}$ from $\vect{v}$,
i.e., it is an edit distance defined by
$\mathbb{B}=\{(\quot\mathtt{a}\quot\to\varepsilon),(\varepsilon\to\quot\mathtt{a}\quot),
(\quot\mathtt{a}\quot\to\quot\mathtt{b}\quot): \mathtt{a},\mathtt{b}\in\Sigma\}$
and $\delta(b)=1$ for every $b\in\mathbb{B}$.
\end{definition}

\begin{example}
We have $\mathfrak{d}_{\mathrm{LV}}(\text{\texttt{"{}function"{}}},
\text{\texttt{"{}fiction"{}}})=2$.

\begin{center}
\begin{tikzpicture}
   \node at (0,0)   {\tt\bfseries f};
   \node at (0.5,0) {\tt\bfseries u};
   \node at (1,0)   {\tt\bfseries n};
   \node at (1.5,0) {\tt\bfseries c};
   \node at (2,0)   {\tt\bfseries t};
   \node at (2.5,0) {\tt\bfseries i};
   \node at (3,0)   {\tt\bfseries o};
   \node at (3.5,0) {\tt\bfseries n};

   \node at (0.5,-0.5) {  \rotatebox{90}{$\looparrowright$} };
   \node[anchor=west] at (4,-0.5) { $\quot\mathtt{u}\quot\to\varepsilon$ (index 2)};
   \draw[dotted] (0.8, -0.55)--(3.9,-0.55);

   \node at (0,-1)   {\tt\bfseries f};
   \node at (0.5,-1) {\tt\bfseries n};
   \node at (1,-1)   {\tt\bfseries c};
   \node at (1.5,-1) {\tt\bfseries t};
   \node at (2,-1)   {\tt\bfseries i};
   \node at (2.5,-1) {\tt\bfseries o};
   \node at (3,-1)   {\tt\bfseries n};

   \node at (0.5,-1.5) {  \rotatebox{90}{$\looparrowright$} };
   \node[anchor=west] at (4,-1.5) { $\quot\mathtt{n}\quot\to\quot\mathtt{i}\quot$ (index 2)};
   \draw[dotted] (0.8, -1.55)--(3.9,-1.55);

   \node at (0,-2)   {\tt\bfseries f};
   \node at (0.5,-2) {\tt\bfseries i};
   \node at (1,-2)   {\tt\bfseries c};
   \node at (1.5,-2) {\tt\bfseries t};
   \node at (2,-2)   {\tt\bfseries i};
   \node at (2.5,-2) {\tt\bfseries o};
   \node at (3,-2)   {\tt\bfseries n};
\end{tikzpicture}
\end{center}
\end{example}

It turns out that a dynamic programming scheme may be applied to calculate
$\mathfrak{d}_\mathrm{LV}(\vect{u},\vect{v})$, see, e.g.,
\cite{WagnerFischer1974:string2stringcorr},
as well as $\mathfrak{d}_\mathrm{LCS}(\vect{u},\vect{v})$ \cite{NeedlemanWunsch1970:lcsdist},
see also \cite{Vintsyuk1968:speechdiscr}.
For instance, we have that
$\mathfrak{d}_{\mathrm{LV}}(\vect{u}, \vect{v})=d_{|\vect{u}|,|\vect{v}|}$, where
$d_{0,0}=0$, $d_{i,j}=\infty$ if $i\wedge j< 0$ and otherwise:
\begin{equation}\label{Eq:Ldist}
d_{i,j}
=
      \min\left\{\begin{array}{lll}
         d_{i-1, j-1}+1\cdot\indicator(u_i\neq v_j),\\
         d_{i, j-1}+1,\\
         d_{i-1, j}+1.
      \end{array}
      \right\}
\end{equation}
A basic algorithm runs in $O(|\vect{u}|\, |\vect{v}|)$ time
and requires $O(|\vect{u}|\, |\vect{v}|)$ memory, which
makes it practically unusable for long data streams (say, consisting of more than 100,000 characters).
However, its advantage is that we may trace back the changes made in the
first string to get the second string.
If just the value of the edit distance is needed,
only two rows of the $(d_{i,j})$ matrix need be allocated.
This leads to the space complexity of $O(|\vect{u}|\wedge|\vect{v}|)$,
see Figure~\ref{Fig:Levenshtein} for an exemplary implementation in the case
of the Levenshtein distance.
Please note that the proposed implementation acts on integer vectors,
so for character strings it may be computed over Unicode text
(all code points may be converted to UTF-32).

Another algorithm, given by Ukkonen \cite{Ukkonen1983:approxstrmatch}, is able to compute
$\mathfrak{d}_\mathrm{LV}(\vect{u},\vect{v})$ in $O(d\cdot (|\vect{u}|\wedge|\vect{v}|))$ time
and $O(d)$ space, where $d=\mathfrak{d}_\mathrm{LV}(\vect{u},\vect{v})$.
It may be modified to check in $O(t\cdot (|\vect{u}|\wedge|\vect{v}|))$ time whether $d\le t$
for some $t$.
A different algorithm, this time by Masek and Pateson,
can be found in \cite{MasekPateson1980:fastereditdst}:
it leads to $O(n\max\{1, m/\log n\})$ time, where
$n=(|\vect{u}|\vee|\vect{v}|)$ and $m=(|\vect{u}|\wedge|\vect{v}|)$.

\bigskip
If we additionally allow swapping any two adjacent characters,
we get the following dissimilarity measure,
formalized for the first time
by Lowrance and Wagner in \cite{LowranceWagner1975:extstring2string}.

\begin{definition}
\index{Demerau-Levenshtein distance (unrestricted)}%
The \emph{(unrestricted) Damerau–Levenshtein distance}
$\mathfrak{d}_{\mathrm{DL}}$, for $\vect{u}, \vect{v}\in\Sigma^*$ is defined as
the minimal number of single character insertions, deletions, replacements,
and pairwise transpositions that are used to obtain $\vect{u}$ from $\vect{v}$,
i.e., it is an edit distance given by
$\mathbb{B}=\{(\quot\mathtt{a}\quot\to\varepsilon),(\varepsilon\to\quot\mathtt{a}\quot),
(\quot\mathtt{a}\quot\to\quot\mathtt{b}\quot), (\quot\mathtt{ab}\quot\to\quot\mathtt{ba}\quot): \mathtt{a},\mathtt{b}\in\Sigma\}$
and $\delta(b)=1$ for every $b\in\mathbb{B}$.
\end{definition}

Also in this case there exists a dynamic programming approach-based algorithm,
see \cite{LowranceWagner1975:extstring2string}, yet it is more complicated.
We have $\mathfrak{d}_{\mathrm{DL}}(\vect{u}, \vect{v})=d_{|\vect{u}|,|\vect{v}|}$, where
$d_{0,0}=0$, $d_{i,j}=\infty$ if $i\wedge j< 0$ and otherwise:
\begin{equation}\label{Eq:DLdist}
d_{i,j}
=
      \min\left\{\begin{array}{lll}
         d_{i-1, j-1}+1\cdot\indicator(u_i\neq v_j),\\
         d_{i, j-1}+1,\\
         d_{i-1, j}+1, \\
         \displaystyle\bigwedge_{\overset{i'<i,j'<j}{u_i=v_{j'}\text{ and }u_{i'}=v_j}} \left(d_{i'-1, j'-1}+i-i'+j-j'-1\right).
      \end{array}
      \right\}
\end{equation}

\begin{remark}
\index{Demerau-Levenshtein distance (restricted)|see {optimal string alignment distance}}%
\index{optimal string alignment distance}%
As noted by, e.g., Boytsov \cite{Boytsov2011:approxdictsearch} and
Loo \cite{Loo2014:stringdist}, the (unrestricted) Damerau-Levenshtein
distance is very often confused with its restricted version, namely the
\emph{optimal string alignment distance} (OSA).
Informally, in OSA each substring is allowed to be edited only once.
This dissimilarity measure is calculated as in Equation~\eqref{Eq:DLdist},
but the minimum ($\wedge$) loop is computed only in the case of $i'=i-1$ and $j'=j-1$.
For the unrestricted version, we for example have
$\mathfrak{d}_\mathrm{DL}(\mathtt{\quot{}ba\quot},\allowbreak \mathtt{\quot{}acb\quot})=2$.

\medskip
\hspace*{2.5cm}
\begin{tikzpicture}
   \node at (0,0)   {\tt\bfseries b};
   \node at (0.5,0) {\tt\bfseries a};

   \node at (0,-0.5) {  \rotatebox{90}{$\looparrowright$} };
   \node[anchor=west] at (2,-0.5) { $\quot\mathtt{ba}\quot\to\quot\mathtt{ab}\quot$ (index 1)};
   \draw[dotted] (0.3, -0.55)--(1.9,-0.55);

   \node at (0,-1)   {\tt\bfseries a};

   \node at (0.5,-1)   {\tt\bfseries b};

   \node at (0.25,-1.5) {  \rotatebox{90}{$\looparrowright$} };
   \node[anchor=west] at (2,-1.5) { $\varepsilon\to\quot\mathtt{c}\quot$ \it (edits an already modified part)};
   \draw[dotted] (0.55, -1.55)--(1.9,-1.55);

   \node at (0,-2)   {\tt\bfseries a};
   \node at (0.5,-2) {\tt\bfseries c};
   \node at (1,-2)   {\tt\bfseries b};
\end{tikzpicture}

\medskip\noindent
On the other hand, $\mathfrak{d}_\mathrm{OSA}(\mathtt{\quot{}ba\quot}, \mathtt{\quot{}acb\quot})=3$.

\medskip
\hspace*{2.5cm}
\begin{tikzpicture}
   \node at (0,0)   {\tt\bfseries b};
   \node at (0.5,0) {\tt\bfseries a};

   \node at (0,-0.5) {  \rotatebox{90}{$\looparrowright$} };
   \node[anchor=west] at (2,-0.5) { $\quot\mathtt{b}\quot\to\varepsilon$ (index 1)};
   \draw[dotted] (0.3, -0.55)--(1.9,-0.55);

   \node at (0,-1)   {\tt\bfseries a};

   \node at (0.25,-1.5) {  \rotatebox{90}{$\looparrowright$} };
   \node[anchor=west] at (2,-1.5) { $\varepsilon\to\quot\mathtt{c}\quot$ (index 2)};
   \draw[dotted] (0.55, -1.55)--(1.9,-1.55);

   \node at (0,-2)   {\tt\bfseries a};
   \node at (0.5,-2) {\tt\bfseries c};

   \node at (0.75,-2.5) {  \rotatebox{90}{$\looparrowright$} };
   \node[anchor=west] at (2,-2.5) { $\varepsilon\to\quot\mathtt{b}\quot$ (index 3)};
   \draw[dotted] (1.05, -2.55)--(1.9,-2.55);

   \node at (0,-3)   {\tt\bfseries a};
   \node at (0.5,-3) {\tt\bfseries c};
   \node at (1,-3)   {\tt\bfseries b};
\end{tikzpicture}

\noindent
Also note that
$\mathfrak{d}_\mathrm{DL}(\mathtt{\quot{}ba\quot}, \mathtt{\quot{}ab\quot{}})
=\mathfrak{d}_\mathrm{OSA}(\mathtt{\quot{}ba\quot}, \mathtt{\quot{}ab\quot{}})=1$
and
$\mathfrak{d}_\mathrm{DL}(\mathtt{\quot{}ab\quot},\allowbreak \mathtt{\quot{}acb\quot{}})
=\mathfrak{d}_\mathrm{OSA}(\mathtt{\quot{}ab\quot}, \mathtt{\quot{}acb\quot{}})=1$.
We see that OSA does not fulfill the triangle inequality and hence
it is not a metric on $\Sigma^*\times\Sigma^*$.
\end{remark}

As we already mentioned, classical edit distances assume that
each edit operation has unit cost. However, some
software libraries like the \package{stringdist} \cite{Loo2014:stringdist}
package for \R{} allow us to specify costs of each type of edit operation.
For instance, having been given $w_\mathrm{I}, w_\mathrm{D}, w_\mathrm{R}>0$,
i.e., costs of insertion, deletion, and replacement, respectively,
we may set $(\forall \mathtt{a},\mathtt{b}\in\Sigma)$
$\delta(\quot\mathtt{a}\quot\to\varepsilon)=w_\mathrm{D}$,
$\delta(\varepsilon\to\quot\mathtt{B}\quot)=w_\mathrm{I}$,
$\delta(\quot\mathtt{a}\quot\to\quot\mathtt{b}\quot)=w_\mathrm{R}$.
In such a way we get a weighted Levenshtein distance, see \cite{Kohonen1985:medianstrings}.
It is easily seen that this leads to
the following modification of Equation~\eqref{Eq:Ldist}:
\begin{equation*}
d_{i,j}
=
      \min\left\{\begin{array}{lll}
         d_{i-1, j-1}+w_\mathrm{R}\cdot\indicator(u_i\neq v_j),\\
         d_{i, j-1}+w_\mathrm{I},\\
         d_{i-1, j}+w_\mathrm{D}.
      \end{array}
      \right\}
\end{equation*}
According to Theorem~\ref{Thm:editdstmetric},
a weighted Levenshtein distance is a metric
if $w_\mathrm{I}=w_\mathrm{D}$.
Moreover, if $w_\mathrm{R}=w_\mathrm{I}+w_\mathrm{D}$,
then a dissimilarity measure proportional
to the longest common subsequence distance is obtained.

\begin{remark}
More complex edit operations and
cost dispatch schemes are suitable
for natural language processing tasks (e.g.,
automated spell checking).
For instance, in the case of the German language,
we may set $\delta(\quot\texttt{\ss{}}\quot\to\quot\mathtt{ss}\quot)$
to be smaller than the cost of other replacement operations etc.
\end{remark}

It is worth noting that various modifications
of edit distances exist in the literature.
In particular, a constrained version of the Levenshtein distance,
with limits on the number and type of edit operations performed,
was proposed by Oommen in \cite{Oommen1986:constrainededdst}.
Moreover, Marzal and Vidal, see \cite{MarzalVidal1993:normalizededdst},
discuss a normalized edit distance defined as the minimum
with respect to all transforming sequences $\mathrm{S}(\vect{u},\vect{v})$
of $w/p$, where $w$ is the sum of unit costs of
edit operations in a transforming sequence and $p$
stands for the number of such operations.
Additionally, in \cite{RistadYianilos1998:learnstreddst}
an algorithm for edit-distance learning (more precisely, determining
costs of edit operations) is given.

\subsubsection{$Q$-gram-based distances}

\index{q-gram}%
Given a string $\vect{u}$, a \emph{$q$-gram}, $|\vect{u}|\ge q\ge 1$,
is its substring that consists of $q$ consecutive
characters in $\vect{u}$, see \cite{Ukkonen1992:qgrams}.
The concept dates back to Shannon \cite{Shannon1945:theorycomput}
and was used by him to model processes generating discrete
sequences of characters. Nowadays, $q$-grams (at a word level%
\footnote{For instance, Google in 2006 released a 24 GB compressed
data set which consists of 13{,}588{,}391 unique words;
we may find, e.g., 1{,}176{,}470{,}663 5-grams, ranked
by the frequency of their occurrences,
see \url{https://catalog.ldc.upenn.edu/LDC2006T13}.})
are used, among others for automated search engine query completion.

Let $\mathcal{Q}_q(\vect{u})$ denote the set of all
$q$-grams in $\vect{u}$, that is:
\begin{equation}
   \mathcal{Q}_q(\vect{u})=\Big\{
      (u_i, u_{i+1}, \dots, u_{i+q-1})\in\Sigma^q:
      i=1,\dots,|\vect{u}|-q+1
   \Big\}.
\end{equation}
We see that $|\mathcal{Q}_q(\vect{u})|=|\vect{u}|-q+1$.

\begin{example}
Three bigrams can be obtained from \texttt{"{}ACTG{}"}:
it holds $\mathcal{Q}_2(\quot{}\mathtt{ACTG}\quot)
=\{\quot{}\mathtt{AC}\quot,
\quot{}\mathtt{CT}\quot,
\quot{}\mathtt{TG}\quot
\}.$
\end{example}

It turns out that $q$-grams may be used to define dissimilarity measures
for strings.

\begin{definition}
\index{Jaccard q-gram dissimilarity index}%
Given $\vect{u},\vect{v}\in\Sigma^*$
and $q \le |\vect{u}|\wedge|\vect{v}|$,
the \emph{Jaccard $q$-gram dissimilarity index} is given by:
\begin{equation}
   \mathfrak{d}_{\mathrm{J},q}(\vect{u}, \vect{v})=
   1-\frac{|\mathcal{Q}_q(\vect{u})\cap\mathcal{Q}_q(\vect{v})|}%
   {|\mathcal{Q}_q(\vect{u})\cup\mathcal{Q}_q(\vect{v})|}\in[0,1].
\end{equation}
\end{definition}

\begin{remark}
It holds $\mathfrak{d}_{\mathrm{J},2}(\quot\mathtt{aab}\quot, \quot\mathtt{aaab}\quot)=0$,
thus a Jaccard index is not a metric.
The property $\mathfrak{d}_{\mathrm{J},q}(\vect{u},\vect{v})=0
\Longleftrightarrow \vect{u}=\vect{v}$ is violated
for all pairs of strings with non-unique $q$-grams.
However, a Jaccard index is positive, symmetric, and fulfills the triangle
inequality and hence it is a pseudometric.
\end{remark}

Let $c_\vect{q}(\vect{u})$ designate the number of occurrences of
a substring $\vect{q}$ in $\vect{u}$:
\begin{equation}
   c_\vect{q}(\vect{u})=\left|
   \Big\{
      i=1,\dots,|\vect{u}|-|\vect{q}|+1:
      (u_i,\dots,u_{i+|\vect{q}|-1})=\vect{q}
   \Big\}
   \right|.
\end{equation}
Clearly, $c_\vect{q}(\vect{u})>0$ if and only if
$\vect{q}\in\mathcal{Q}_q(\vect{u})$.

\begin{example}
We have $c_\mathtt{\quot{}aa\quot}(\mathtt{\quot{}aaabaa\quot})=3$.
\end{example}

\index{q-gram profile}%
The so-called \emph{$q$-gram profile} allows us to represent a string
as a vector with integer elements of size $|\Sigma|^q$:
$\mathcal{QP}_q(\vect{u})=(c_\vect{q}(\vect{u}))_{\vect{q}\in\Sigma^q}$.

\begin{example}
Let $\Sigma=\{\mathtt{a},\mathtt{b}\}$ and $q=2$.
We have:

\begin{center}
\begin{tabular}{rrccccl}
&& \footnotesize$\quot\mathtt{aa}\quot$ & \footnotesize$\quot\mathtt{ab}\quot$
& \footnotesize$\quot\mathtt{ba}\quot$ & \footnotesize$\quot\mathtt{bb}\quot$ &\\
$\mathcal{QP}_q(\quot\mathtt{abaa}\quot)=$ &(& 1, & 1, & 1, & 0 &), \\
$\mathcal{QP}_q(\quot\mathtt{aabbaa}\quot)=$ &(& 2, & 1, & 1, & 1 &). \\
\end{tabular}
\end{center}
\end{example}

This leads us to the following dissimilarity measure proposed by Ukkonen
in~\cite{Ukkonen1992:qgrams}.

\begin{definition}\index{q-gram distance}%
The \emph{$q$-gram distance}  is defined as:
\begin{equation}\label{Eq:qgramdist}
   \mathfrak{d}_{\mathrm{Q},q}(\vect{u}, \vect{v})=\sum_{\vect{q}\in\Sigma^q}
   |c_\vect{q}(\vect{u})-c_\vect{q}(\vect{v})|.
\end{equation}
\end{definition}
Note that the summation may be made
just over $\vect{q}\in \mathcal{Q}_q(\vect{u})\cup\mathcal{Q}_q(\vect{v})$.

\begin{remark}
According to \cite[Theorem 2.1]{Ukkonen1992:qgrams},
$\mathfrak{d}_{\mathrm{Q},q}$ is a pseudometric for any $q$.
It is not a metric as, for instance, for the
bigram distance we have
$\mathfrak{d}_{\mathrm{Q},2}(\quot\mathtt{abaa}\quot,
\quot\mathtt{aaba}\quot)=0$.
It is because many strings may have the same $q$-gram profiles.
\end{remark}

Ukkonen in \cite{Ukkonen1992:qgrams} provides
an $O(|\vect{u}|+|\vect{v}|)$-time
and $O(|\Sigma|^q+|\vect{u}|+|\vect{v}|)$-space algorithm to compute
the $q$-gram distance.
Also please notice that Equation~\eqref{Eq:qgramdist} is nothing more than the $L_1$ metric
between $\mathcal{QP}_q(\vect{u})$ and $\mathcal{QP}_q(\vect{v})$ and thus
the introduced dissimilarity measure can potentially be easily generalized.

\subsubsection{Other string metrics}

We should point out that many other string distances may be found in the literature.
For instance, the Dinu rank distance $\mathfrak{d}_\mathrm{DR}$ \cite{Dinu2003:rankdiststring},
closely related to the so-called Spearman's footrule, see \cite{DiaconisGraham1977:spearmanfootrule},
has recently been of interest in computational biology.
As for its construction we need some linear order on $\Sigma$ now
(in fact, its nature is irrelevant here), let us without loss of generality
assume that $\Sigma=\{1,\dots,k\}\subseteq\mathbb{N}$ --
any set of characters may be mapped to a set of consecutive integers.

Let us define a mapping $\Sigma\ni u_i \mapsto (u_i, j_i)\in\mathbb{N}^2$, $i\in[d_u]$,
$d_u=|\vect{u}|$,
such that $j_i = \sum_{k=1}^i \indicator(u_k = u_i)$.
In other words, e.g., $(3, 2)$ denotes the 2nd occurrence of character
$3$ in $\vect{u}$. From that we generate
the sequence $\tilde{\vect{u}} = ( (u_{\sigma_u(1)},j_{\sigma_u(1)}),\allowbreak
\dots,\allowbreak (u_{\sigma_u(d_u)}, j_{{\sigma_u(d_u)}}) )$, where
$\sigma_u$ stands for the ordering permutation of $\tilde{\vect{u}}$
with respect to the linear order $\preceq$ such that $(a,j)\preceq(a',j')$
if and only if either $a<a'$, or $a=a'$ and $j\le j'$ (it is a lexicographic
order on $\mathbb{N}^2$). Now for any $(a, j)\in\mathbb{N}^2$ let:
\[
   \mathrm{ord}_\vect{u}(a,j) = \left\{\begin{array}{ll}
      \sigma_u(i) & \text{if }(a,j)=(\tilde{u}_{i}, \tilde{j}_{i})\text{ for some $i$},\\
      0 & \text{if }(a,j)\text{ is not a member of }\tilde{\vect{u}}.
   \end{array}\right.
\]

\begin{example}
For instance, given a string $(2, 1, 1, 3, 3, 4, 1)$, we get:

\begin{center}
\begin{tabularx}{1.0\linewidth}{XXXXlr}
\toprule
\bf\small $i$ & \bf\small $u_i$ & \bf\small  $j_i$ & \bf\small $\tilde{u}_i$ &\bf\small  $\tilde{j}_i$ &
            \bf\small $\mathrm{ord}_\vect{u}(\tilde{u}_i,\tilde{j}_i)$  \\
\midrule
1 & 2 & 1 & 1 & 1 \it(first character $1$) & 2 \\
2 & 1 & 1 & 1 & 2 \it(second character $1$)& 3 \\
3 & 1 & 2 & 1 & 3 \it(third character $1$) & 7 \\
4 & 3 & 1 & 2 & 1 \it(first character $2$) & 1 \\
5 & 3 & 2 & 3 & 1 \it(first character $3$) & 4 \\
6 & 4 & 1 & 3 & 2 \it(second character $3$)& 5 \\
7 & 1 & 3 & 4 & 1 \it(first character $4$) & 6 \\
\bottomrule
\end{tabularx}
\end{center}
\end{example}

\begin{definition}\label{Def:dinurank}\index{Dinu rank distance}%
Let $\vect{u},\vect{v}\in\Sigma^*$.
The \emph{Dinu rank distance} is given by:
\begin{equation}
   \mathfrak{d}_\mathrm{DR}(\vect{u},\vect{v}) = \sum_{(a, j)\in\tilde{\vect{u}}\cup\tilde{\vect{v}}}
   \left|
    \mathrm{ord}_\vect{u}(a,j) -  \mathrm{ord}_\vect{v}(a,j)
   \right|.
\end{equation}
\end{definition}

Thus, it is an $L_1$ distance between the $\mathrm{ord}$ vectors.
It may be shown that $\mathfrak{d}_\mathrm{DR}$ is a metric,
see \cite{Dinu2003:rankdiststring}.

Figure~\ref{Fig:dinurank} gives our own implementation of the algorithm
to compute the Dinu rank distance which operates in $O(d_u\log d_u+d_v\log d_v)$-time.
Note that it is also possible to formulate it in such a way that it runs in $O(d_u|\Sigma|+d_v|\Sigma|)$-time.
The costly step here is to find the stable ordering permutations of
$\vect{u}$ and $\vect{v}$, but if more computations are needed on a set of strings
(e.g., in implementations of hierarchical clustering algorithms
that require roughly $n^2$ distance computations), they may be determined
once in advance, and the time gets reduced to $O(d_u\vee d_v)$.

\medskip
Other string (pseudo)metrics include, for example,
the Jaro or Jaro-Winkler distance (see \cite{Winkler1990:stringcompmetr}),
or the one proposed by Ehrenfeucht and Haussler in
\cite{EhrenfeuchtHaussler1988:newdststring}.
Note that the issue of how to compare DNA sequences
is still in the top of a list of major open problems in
bioinformatics/computational biology, see \cite{Wooley1999:trendscompbiol}.
Yet, the discussed instances are perhaps the most frequently used
and influential ones. Having said that, let us proceed with some seminal
distance-based fusion function construction methods.

\subsection{Median strings and a strings' centroid}

\index{median string}%
The concept of a median string was introduced by Kohonen \cite{Kohonen1985:medianstrings}
for the purpose of smoothing of erroneous versions of strings and
for string classification in, e.g., pattern recognition.
Given $\vect{x}^{(1)},\dots,\vect{x}^{(n)}\in\Sigma^*$
it is a string $\vect{x}^*$ such that:
\[
\vect{x}^*
=\argmin_{\vect{x}\in\Sigma^*} \sum_{i\in[n]}
\mathfrak{d}(\vect{x}^{(i)}, \vect{x}),
\]
where $\mathfrak{d}$ is some string metric, originally
the Levenshtein distance.
Additionally, we may consider a centroid-like search task:
\[
\vect{x}^*
=\argmin_{\vect{x}\in\Sigma^*} \sum_{i\in[n]}
\mathfrak{d}^2(\vect{x}^{(i)}, \vect{x}).
\]
Note that many strings which are solutions to the two above equations may exist.
Thus, the definition of a fusion function to aggregate a set of strings
should be formulated with care.

\begin{remark}
   Note that in the space of character strings, a medoid (set median)
   may in some contexts be more sensible than a median string,
   especially if $n$ is large. This is especially  the case when
   not all the strings in $\Sigma^*$ are ``valid''
   or ``meaningful'' (e.g., when we aggregate words in natural language).
   Recall that in the case of a medoid, we always get a string which is among
   those in the input data set. Another option is to restrict the search
   domain and seek within some \emph{dictionary}.
\end{remark}

\paragraph{The case of two strings.}
Let $n=2$ and $\vect{x}^{(1)},\vect{x}^{(2)}\in\Sigma^*$.
Both of the input objects are within the possible 1-median strings.
In such a case, as a solution one may want to consider a string $\vect{x}$ such that
$\mathfrak{d}(\vect{x}^{(1)},\vect{x})=\lfloor \mathfrak{d}(\vect{x}^{(1)},\vect{x}^{(2)})/2\rfloor$
and $\mathfrak{d}(\vect{x}^{(1)},\vect{x})+\mathfrak{d}^2(\vect{x}^{(2)},\vect{x})
=\mathfrak{d}(\vect{x}^{(1)},\vect{x}^{(2)})$ as a median string,
which is exactly a solution to the string centroid problem.
In other words, here a centroid is always at the same time a medoid.

To compute a Levenshtein metric-based centroid of two strings, we may make use of the fact
that the classical algorithm that determines the value of this distance
(see Equation~\eqref{Eq:Ldist}) also provides us with the information on
how to edit the first string in such a way that the second one may be obtained.
In order to do so, we can apply consecutive edit operations on the first string
until the cumulative cost of edits made so far reaches $\lfloor \mathfrak{d}(\vect{x}^{(1)},\vect{x}^{(2)})/2\rfloor$.
This leads to an algorithm whose source code is given in Figure~\ref{Fig:LevenshteinCentroid2}.
Note that the underlying fusion function is not symmetric. It can be made
such if we first order the two input strings lexicographically.

\begin{remark}
The centroids of $\tt\quot{}1234\quot{}$ and $\tt\quot{}2345\quot{}$
with respect to the Levenshtein, LCS, and Damerau-Levenshtein distances
are exactly $\tt\quot{}234\quot{}$ and $\tt\quot{}12345\quot{}$.
We observe that whichever we choose as a desired output of a centroid-like fusion function,
the length internality property is violated.

However, in order to guarantee length internality, one may always
restrict the search domain and be rather interested in finding, e.g.:
\[
\vect{x}^*
=\argmin_{\vect{x}\in\bigcup_{d=d_\mathrm{min}}^{d_\mathrm{max}} \Sigma^d} \quad\sum_{i\in[n]}
\mathfrak{d}^p(\vect{x}^{(i)}, \vect{x}),
\]
where $p\in\{1,2\}$ and $d_\mathrm{min}=\bigwedge_{i=1}^n |\vect{x}^{(i)}|, d_\mathrm{max}
=\bigvee_{i=1}^n |\vect{x}^{(i)}|$.
\end{remark}

\paragraph{General case.}
There are string distances which guarantee that a median search
is of polynomial-time.  This is in the case of, e.g., the Dinu rank distance,
see \cite{DinuManea2006:efficientrankaggregation}.
However, unlike in the fixed $d$ case and the Hamming distance, it turns out that
finding a median string with respect to the (weighted) Levenshtein
distance is an NP-complete problem as a function of $n$
even if $\Sigma$ is a binary alphabet.
Nicolas and Rivals in \cite{NicolasRivals2005:hardnesscentermedianstring,
NicolasRivals2003:complexitycentermedianstring}
proved that by reduction to the intractable longest common
subsequence problem. An exact algorithm was provided
by Kruskal \cite{Kruskal1983:sequencecomparison}.

In order to find an approximate version of a median string with respect to
the Levenshtein distance, Kohonen \cite{Kohonen1985:medianstrings} suggests
to compute the set median (which may be done easily) and then \textit{to vary
each of the symbol positions of the set median, making ``errors'' of all
three types over the whole alphabet,
and checking whether the sum of distances from the other elements is decreased.}
More elaborate approximate algorithms, in the case of weighted Levenshtein
distances, were given by Martinez et al.~\cite{MartinezETAL2003:medianstrknn}
(together with an application in $k$-nearest neighbor classification)
and Abreu and Juan in \cite{AbreuJuan2014:medianstr} --
yet, they are also based on perturbations over the initial string.
On the other hand, Jiang et al.~\cite{JiangETAL2012:medianstringembed}
incorporate an idea of computing median strings by embedding
them into Euclidean vector spaces.
See also the works by Kohonen and Somervuo
\cite{KohonenSomervuo1998:somsstring,Somervuo2004:onlinesomsstring} for an application in
constructing unsupervised self-organizing maps (SOMs).

\bigskip
Here we shall provide a genetic algorithm (see Algorithm~\ref{Alg:Genetic})
to compute the desired fusion function. Its most interesting facet concerns
a proper crossover and mutation scheme, the selection of which might not be trivial in the space
of vectors of arbitrary lengths. We recommend the following approaches:
\begin{itemize}
   \item a \emph{crossover} between $\vect{u}$ and $\vect{v}$ is set to be
   the centroid of $\{\vect{u}, \vect{v}\}$ (see above),
   \item a single \emph{mutation} operation may consist of a Levenshtein edit:
   with equal probability we choose  to perform at a randomly chosen index
   in a string being mutated, either:
   \begin{itemize}
      \item an insertion of a character,
      \item a removal of a character, or
      \item a replacement of a character with one sampled from $\Sigma$.
   \end{itemize}
\end{itemize}

\begin{remark}
   The discussed crossover scheme is sometimes called
an \index{intermediate recombination}\emph{intermediate recombination}.
We observe that a cut-and-splice crossover (joining a random prefix
of $\vect{u}$ with a random suffix of $\vect{v}$) does not perform well.
Also what does not work best is a scheme used in \cite{DinuIonescu2012:efficientrankcloseststring}
(for center strings with respect to the Dinu rank distance, see below),
which basically is based on joining a random prefix of the first vector
with a permuted version of a sampled suffix of the second vector.
\end{remark}

\begin{remark}
Note that finding the median with respect to the $q$-gram distance
is much more difficult.

Let $P=\{\vect{q}^{(1)},\dots,\vect{q}^{(m)} \}$ be a set of all $q$-grams that appear in at least one of the
input strings, i.e., $P=\bigcup_{i=1}^n \mathcal{Q}_q(\vect{x}^{(i)})$.
We remap each string $\vect{x}^{(i)}$ to a $q$-gram profile
$\vect{c}^{(i)} = (c_{\vect{q}^{(1)}}(\vect{x}^{(i)}), \dots, c_{\vect{q}^{(m)}}(\vect{x}^{(i)}))$,
which gives a vector of nonnegative integers.

Our aim is to find:
\begin{equation}
   \argmin_{\vect{c}\in X} \sum_{i=1}^n \mathfrak{d}_1(\vect{c}, \vect{c}^{(i)}),
\end{equation}
where $X$ is a subset of $\mathbb{N}_0^m$ which denotes a valid $q$-gram
profile, i.e., one from which we may reconstruct a proper character string.

If the search space was just as simple as $\mathbb{N}_0^m$, an integer
programming (IP) solver could be used for determining the 1-median
(perhaps one that is able to iterate through all the optimal solutions).
Yet, for instance, assume that $\tt\quot{}abcb\quot{}$, $\tt\quot{}cbac\quot{}$, $\tt\quot{}acab"$
are three input character strings and $q=2$.
Then the output from an IP solver suggests that the best match consists of
the following bigrams: $\tt\quot{}ac\quot{}$, $\tt\quot{}cb\quot{}$, $\tt\quot{}ab\quot{}$.
It is easily seen that no string can be constructed from such a $q$-gram (multi)set.

\end{remark}

\subsection{Closest strings}

\index{center string|see {closest string}}\index{closest string}%
Recall that the 1-center problem
aims at finding a point which minimizes the maximal
distance towards every point in a given input data set.
What is known in the literature under the name
\emph{closest} or \emph{center} string problem represents exactly
such a type of task, this time however -- in the character string domain.

More precisely, let $\mathfrak{d}$ be a (pseudo)metric on $\Sigma^*$.
Given $\vect{x}^{(1)},\dots,\vect{x}^{(n)}$, we define:
\[
\func{ClosestString}_\mathfrak{d}(\vect{x}^{(1)},\dots,\vect{x}^{(n)})
=\argmin_{\vect{x}\in\Sigma^*} \bigvee_{i\in[n]}
\mathfrak{d}(\vect{x}^{(i)}, \vect{x}).
\]
Note that the solution may be ambiguous.
There are many applications of such a fusion function in computational biology.
Among some instances listed in
\cite{DinuIonescu2012:efficientrankcloseststring}
we find: searching for motifs or common patterns in a set of given DNA sequences
or genetic drugs design with a structure similar to a set of RNA sequences.

\begin{remark}
For $n=2$ a centroid of $\{ \vect{x}^{(1)}, \vect{x}^{(2)}\}$
is also its center string.
\end{remark}

If $\mathfrak{d}$ is the ordinary Levenshtein distance and $|\Sigma|\ge 2$,
then -- as shown by Nicolas and Rivals in
\cite{NicolasRivals2003:complexitycentermedianstring} --
the center string is NP-complete with respect to $n$
(a proof is by reduction to the longest common subsequence problem).
Moreover, the mentioned authors show similar results for the case of the
weighted Levenshtein distance in \cite{NicolasRivals2005:hardnesscentermedianstring}.

Here, also an algorithm for finding the closest string with respect to the Dinu rank distance
is NP-complete, see \cite{DinuPopa2012:closeststring}.
Thus, in \cite{DinuIonescu2012:efficientrankcloseststring} Dinu and Ionescu propose
a genetic algorithm-based approach to approximate a closest string.
Moreover, in \cite{DinuIonescu2012:clusteringcloseststring} they
develop $k$-means-like and hierarchical clustering methods based on closest strings
and the rank distance.

\clearpage{\pagestyle{empty}\cleardoublepage}
\chapter{Aggregation of other data types}\label{Chap:other}

\lettrine[lines=3]{F}{usion} functions defined on more complex domains
than in the previous chapters are the subject of interest in this part
of the monograph. In the consecutive sections we shall assume that we deal with
the following data types:
\begin{enumerate}
   \item directional (e.g., angular) data, %
   \item real intervals,
   \item fuzzy numbers,
   \item random variables, %
   \item trees and other graphs as well as rankings and other relations,
   \item general finite semimetric spaces,
   \item heterogeneous data sets.
\end{enumerate}

\noindent
We already observed that it is possible
to aggregate fusion functions (in particular, regression and classification
models) and metrics themselves. Even though the construction and analysis
of data fusion tools acting on the aforementioned domains may seem much more difficult,
it shall turn out that many of the already known ideas may be easily
extrapolated. For instance, if we deal with a linear space (and thus if we are able
to define  addition and scalar multiplication operations properly), then
a weighted arithmetic mean, i.e., a convex combination, can be defined. If a linear ordering relation
may be introduced quite naturally, then we can refer to the notion of an order statistic.
Moreover, having various metrics or other kinds of dissimilarity
measures, we may consider the notion of a penalty-based fusion function.

\section{Directional data}\label{Sec:Directional}

Let us consider the situation where observations are defined on spheres
$\{\vect{x}\in\mathbb{R}^{d}:\|\vect{x}\|=1\}=\mathbb{S}^{d-1}\subset\mathbb{R}^{d}$,
$d>1$, rather than Euclidean spaces like in Chapter~\ref{Chap:multidim},
see \cite{Mardia1975:dirdata,MardiaJupp1999:dirstatbook,Fisher1993:statanalcirc}.
For instance, this kind of information may occur in:
\begin{itemize}
   \item data on location of earthquake epicenters (as Earth may be modeled by a sphere),
   \item observations of winds, animal migration, paleomagnetism, etc.~(here,
   a natural phenomenon's movement direction is the most relevant),
   \item events occurring periodically, e.g., on a 24-hour clock, yearly calendar
   (whenever there is a cyclic pattern in time),
   \item  handwriting features descriptions (for, e.g., optical character recognition,
   see \cite{Bahlmann2006:directionalhandwrite}),
   \item models of local protein structure, see \cite{BoomsmaETAL2008:proteinstruct},
\end{itemize}
and many others.

Handling \index{directional data}\emph{directional} (e.g, circular/angular for $d=2$ or spherical for $d=3$)
data is quite challenging.
Even if we are on a circle, we do not have a natural ordering of our data.
This is because angles of $-180^\circ$ and $180^\circ$  are equivalent,
as well as $0^\circ$ and $360^\circ$, and so on.
Additionally, observe that the ``average'' of $165^\circ$ and $-165^\circ$
should not be set to $0^\circ$, etc. Our space of discourse here
may be conceived as a ``modulo $2\pi$''-type algebra.

\begin{figure}[htb!]
\centering

\includegraphics[width=8.25cm]{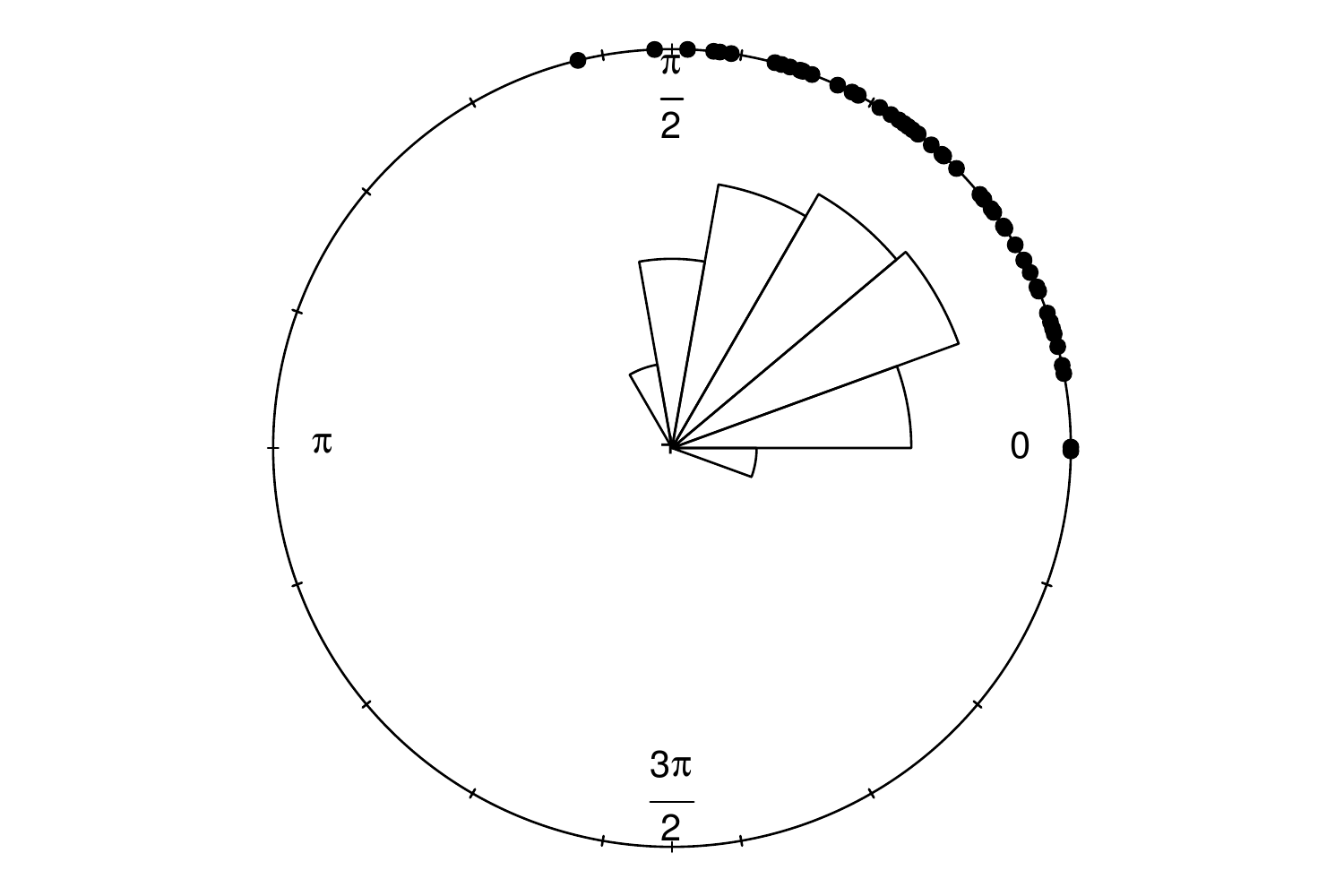}

\caption{\label{Fig:rose} An exemplary rose diagram of a circular data set.}
\end{figure}

\begin{example}
   A \index{rose diagram}\emph{rose diagram} is a modification of an ordinary histogram,
   tailored for depicting circular data.
   Figure \ref{Fig:rose} depicts a rose diagram of an exemplary
   circular data set, being a random sample from a von~Mises distribution
   (the circular counterpart of a normal distribution)
   with expected value of $\pi/4$.
\end{example}

\begin{remark}
What should be the mean of $0, \pi/2, \pi, 3\pi/2$? Of course, there
is no definite answer to this question.
\end{remark}

Let us consider a few fusion functions that may be found in the literature
and which aim to provide information on the \textit{average} value of
a directional data set.
In other words, for $d>1$ and some $n$, this time we are interested in
fusion functions like $\func{F}:(\mathbb{S}^{d-1})^n\to \mathbb{S}^{d-1}$.

Firstly, we should note that in the case of directional data, most researchers discourage
using a stereographic projection of input data
(see \cite{Small1990:medianssurv}), i.e., embedding
$\vect{x}^{(1)},\dots,\vect{x}^{(n)}$ in $\mathbb{R}^d$. Instead, using the Euclidean
space analogues are preferred. Therefore, one may consider:
\begin{itemize}
   \item arcs of a great circle as replacements for straight lines (the ``shortest curve'' joining two points),
   \item arc lengths as replacements for the Euclidean distance,
\end{itemize}
and so on.

\begin{example}
\index{CircMean@$\mathsf{CircMean}$|see {circular mean}}\index{circular mean}%
The mean of a circular data set $\vect{x}^{(1)},\dots,\vect{x}^{(n)}\in\mathbb{S}^1$
is usually defined as follows, see \cite{JammalamadakaSenGupta2001:circstat}.
Let $\vartheta_1,\dots,\vartheta_n\in[-\pi,\pi[$ denote the corresponding angles.
Then:
\begin{eqnarray*}
   && \func{CircMean}(\boldsymbol\vartheta)\\
   &=& \mathrm{atan2}\Big(
      \func{AMean}(\sin\vartheta_1,\dots,\sin\vartheta_n),
      \func{AMean}(\cos\vartheta_1,\dots,\cos\vartheta_n)
   \Big),
\end{eqnarray*}
where for $x,y\in\mathbb{R}\setminus\{0\}$:
\[
   \mathrm{atan2}(y, x) = \left\{
   \begin{array}{ll}
      \arctan \frac{y}{x} & \text{if }x>0,\\
      \arctan \frac{y}{x}+\pi & \text{if }x\le 0\text{ and }y>0,\\
      \arctan \frac{y}{x}-\pi & \text{if }x\le 0\text{ and }y\le 0.\\
   \end{array}\right.
\]
\end{example}

\begin{example}\index{Mardia-type median}%
The Mardia-type (see \cite{MardiaJupp1999:dirstatbook}) median of a circular data
set is a point $\mathbf{y}$ on the unit circle such that:
\begin{itemize}
   \item most of the observations are closer to the median $\mathbf{y}$
   than to the anti-median $\mathbf{y}'$,
   \item the number of observations in each semi-circle determined
   by the diameter $\mathbf{y}\mathbf{y}'$ are equal.
\end{itemize}
Note that for some data sets the Mardia-type median may be ambiguous,
see \cite{Otieno2002:phd} for discussion and some modifications of the above method.
\end{example}

\begin{example}\index{Mardia-Fisher spherical median}%
The extension of the 1-median is called \emph{the Mardia-Fisher spherical median}.
It is a point $\vect{y}$ such that:
\[
   \vect{y}=\argmin_{\vect{y}\in\mathbb{S}^{d-1}} \sum_{i=1}^n \mathfrak{d}(\vect{x}^{(i)},\vect{y}),
\]
where $\mathfrak{d}(\vect{x},\vect{y})=\cos^{-1}(\langle\vect{x},\vect{y}\rangle)$,
i.e., the length of the geodesic arc joining $\vect{x}$ and~$\vect{y}$.
\end{example}

\begin{example}
Data depth notions were also generalized for the case of directional data.
For instance, an Oja-type spherical median was proposed in \cite{Small1987:measurecentrality}
-- instead of simplices we consider the intersection of all closed
hemispheres that contain $d+1$ points.
Moreover, in \cite{LiuSingh1992:orddirectional} the concept of angular simplical depth
and angular Tukey's depth (rotation invariant) is considered and in
\cite{LeySabbahVerdebout2014:quantdir} -- angular Mahalanobis depth.
\end{example}

\section{Aggregation of real intervals}

If input data of numeric type are not precisely given, they sometimes are represented
as real intervals,
see, e.g., \cite{BargielaPedrycz2003:granular,KulczyckiKowalski2011:interval,KlirYuan1995:fuzzybook}.
Let $\mathcal{I}([a,b])$ denote the set of all closed subintervals of $[a,b]$, $a<b$.

\begin{example}
In statistics, data may be provided by means of frequency tables
(which may be much easier to gather manually or using low accuracy measurement devices).
An exemplary grouped (histogram-like) data set is as follows:

\begin{center}
\begin{tabularx}{1.0\linewidth}{XX}
\toprule
\small\bf time & \small\bf frequency \\
\midrule
$[0,5[$  & 5 \\
$[5,10[$ & 9 \\
$[10,15[$ & 6 \\
$[15,20[$ & 3\\
\bottomrule
\end{tabularx}
\end{center}
\end{example}

\begin{example}
Recall from page~\pageref{Def:FP} that most values $x\in\mathbb{R}$
cannot be directly represented in a set of floating-point numbers, $\mathbb{F}$.
Instead of simple rounding, we may model $x$ as the smallest interval
$[\underline{x},\overline{x}]$, where $\underline{x},\overline{x}\in\mathbb{F}$
and $x\in[\underline{x},\overline{x}]$.
Surely, $\underline{x}=\underline{\mathrm{fp}}(x)\le x$ is defined as $x$ rounded towards $-\infty$
and $\overline{x}=\overline{\mathrm{fp}}(x)\ge x$ -- towards $+\infty$.
For instance, the GNU \lang{C} library allows
(on CPU architectures and compilers that support this operation)
to change the rounding mode by a call to the \texttt{int fesetround(int round)} function,
where $\mathtt{round}\in\{\mathtt{FE\_TONEAREST}, \mathtt{FE\_UPWARD}, \mathtt{FE\_DOWNWARD}, \allowbreak\mathtt{FE\_TOWARDZERO}\}$.

\end{example}

In the current setting, a few approaches to interval data fusion are possible.

\paragraph{Intervals as bounded posets.}
On the set of real intervals we may define a partial ordering relation,
e.g., as follows:
\begin{itemize}
   \item $[\underline{x},\overline{x}]\sqsubseteq_I[\underline{y},\overline{y}]$
   if and only if $\overline{x}<\underline{y}$, or $\underline{x}=\underline{y}$ and $\overline{x}=\overline{y}$
   (the so-called \index{interval order}\emph{interval order}),
   \item $[\underline{x},\overline{x}]\le^2 [\underline{y},\overline{y}]$ if and only if  $\underline{x}\le\underline{y}$ and $\overline{x}\le\overline{y}$
   (Cartesian product extension of ordinary $\le$ on the set of real numbers).
\end{itemize}
In both cases we may obtain a bounded lattice, thus aggregation methods
developed already in Section~\ref{Sec:ProdLat} are directly applicable here,
see also \cite{Demirci2006:agposets}.
Of course, it is possible to derive more tailored results as well, such as ones
concerning for instance t-norms on the space of real intervals and the $\le^2$ order, see
\cite{DeBaetsMesiar1999:trinormprodlat,Zhang2005:trinormposet}.
In particular, it can be shown that for every semicontinuous t-norm $\func{T}$ on
$\mathcal{I}([0,1])$ there exists a t-norm $\func{T}'$ on $[0,1]$
such that $\func{T}([\underline{x},\overline{x}], [\underline{y},\overline{y}])
=[\func{T}'(\underline{x},\underline{y}), \func{T}'(\overline{x},\overline{y})]$.
Moreover, L\`{a}zaro and Calvo in \cite{LazaroCalvo2005:xao}
considered aggregation functions monotone with respect to the above orders.

\paragraph{Defuzzification.}
Note that each interval $[\underline{x},\overline{x}]$ may be represented as
$x\pm r$, where $x=(\underline{x}+\overline{x})/2$ is its midpoint
and $r=(\overline{x}-\underline{x})/2\ge 0$ is its halfwidth.
Some statistical data analysis handbooks suggest to ``defuzzify'' interval data
and instead just to consider the corresponding midpoints
(corresponding halfwidths may be aggregated separately to measure
the imprecision of the outcome). Then, classical
fusion functions for unidimensional quantitative data may be used.

\paragraph{Interval arithmetic.}
\index{interval arithmetic}Let us introduce the following extensions of arithmetic operations to the space of real intervals,
see, e.g., \cite{KlirYuan1995:fuzzybook}:
\begin{eqnarray*}
\big[\underline{x},\overline{x}\big]\oplus \big[\underline{y},\overline{y}\big] & = & \big[\underline{x}+\underline{y},\overline{x}+\overline{y}\big], \\
\big[\underline{x},\overline{x}\big]\ominus\big[\underline{y},\overline{y}\big] & = & \big[\underline{x}-\overline{y},\overline{x}-\underline{y}\big], \\
\big[\underline{x},\overline{x}\big]\otimes\big[\underline{y},\overline{y}\big] & = &
   \big[
      \underline{x}\cdot\underline{y}\wedge\overline{x}\cdot\overline{y}\wedge\underline{x}\cdot\overline{y}\wedge\overline{x}\cdot\underline{y},
      \underline{x}\cdot\underline{y}\vee\overline{x}\cdot\overline{y}\vee\underline{x}\cdot\overline{y}\vee\overline{x}\cdot\underline{y}
   \big], \\
\big[\underline{x},\overline{x}\big]\oslash\big[\underline{y},\overline{y}\big] & = &
   \big[
      \underline{x},\overline{x}\big]\otimes\big[1/\overline{y},
      1/\underline{y}
   \big]\text{ whenever }0 \not\in [\underline{y},\overline{y}\big].
\end{eqnarray*}
Functions like $\func{f}:\mathbb{R}\to\mathbb{R}$ may be extended straightforwardly.
For example, if $\func{f}$ is strictly increasing, then let:
\[
   \text{\circled{$\func{f}$}}([\underline{x},\overline{x}]) = [\func{f}(\underline{{x}}), \func{f}(\overline{{x}})].
\]

\begin{remark}
The \package{Interval Arithmetic Library} in \package{Boost} for the \lang{C++} programming language
\cite{BronnimannETAL2006:boost}
is able to programmatically quantify the propagation of rounding errors
in floating point computations by using proper rounding towards $-\infty$ (left) and $+\infty$ (right bound).
\end{remark}

Note that for any $s\ge 0$ it holds $\big[\underline{x},\overline{x}\big]\otimes s
= \big[\underline{x},\overline{x}\big]\otimes [s,s]
= \big[s\underline{x},s\overline{x}\big]$. We see that the set of intervals
is closed under addition and scalar multiplication
and forms a linear space. %
Thus, the notion of a weighted arithmetic mean may easily be introduced.
This leads to an idempotent and $\le^2$-monotone fusion function.
On the other hand, redefining  OWA-like operators is not as trivial,
as the construction of a linear order on $\mathcal{I}([a,b])$ can be
done in many ways (e.g., by considering intervals' midpoints,
halfwidths, etc.).

\paragraph{Penalty-based fusion functions.}
In order to introduce penalty-based functions to aggregate
interval data, let us first recall the most popular interval metrics,
see also  \cite[Chapter~8]{BeliakovETAL2015:practicalbook}.

\begin{definition}
\index{Moore metric}\emph{Moore's interval metric} is given by:
\begin{equation}
\mathfrak{d}_\mathrm{M}\left(\big[\underline{x},\overline{x}\big]\oplus \big[\underline{y},\overline{y}\big]\right)
= |\underline{x}-\underline{y}|\vee|\overline{x}-\overline{y}|.
\end{equation}
\end{definition}

If $[a,b]$ is interpreted as a point in $\mathbb{R}^2$, the Moore metric is exactly
the Chebyshev distance, $\mathfrak{d}_\infty$.
On the other hand, if we rely on the midpoint $\pm$ halfwidth representation,
then this metric is the $\mathfrak{d}_1$ one:
it holds $\mathfrak{d}_\mathrm{M}(x\pm r_x, y\pm r_y) = |x-y|+|r_x-r_y|$.
Therefore, we have what follows (compare also Section~\ref{Sec:PenaltyMultidim}).
\begin{itemize}
   \item The $\mathfrak{d}_\mathrm{M}$-based 1-median of $\vect{x}\in\mathcal{I}([a,b])^n$ is equal to the componentwise
median of the inputs' midpoints and halfwidths, see \cite[Theorem~1]{ChaventSaracco2008:centraltendinterval}.
   \item The $\mathfrak{d}_\mathrm{M}$-based 1-center of $\vect{x}\in\mathcal{I}([a,b])^n$ is given
   as: \begin{equation}\Big[ (\bigvee_i \underline{x}_i+\bigwedge_i \underline{x}_i)/2,
   (\bigvee_i \overline{x}_i+\bigwedge_i \overline{x}_i)/2\Big],\end{equation} see \cite[Theorem 2]{ChaventSaracco2008:centraltendinterval},
\end{itemize}
Moreover, if we assume that $\mathfrak{d}_\mathrm{M_2}(x\pm r_x, y\pm r_y) = \sqrt{|x-y|^2+|r_x-r_y|^2}$,
then:
\begin{itemize}
   \item  $\mathfrak{d}_\mathrm{M_2}$-based centroid  of $\vect{x}\in\mathcal{I}([a,b])^n$  is  equal to the componentwise
arithmetic mean of midpoints and halfwidths, see \cite[Theorem 3]{ChaventSaracco2008:centraltendinterval}.
\end{itemize}
Note that the above results may easily be generalized to hyperrectangles
\index{granular data}%
such that their faces are parallel to axes of a coordinate system.
Such data occur, among others, in the so-called granular
\cite{BargielaPedrycz2003:granular,PedryczETAL2008:granular}
box regression, see \cite{Peters2011:granularregression} and also
\cite{Grzegorzewski2013:granularregression,%
PetersLacic2012:granularregressionoutlier}.

\index{Wasserstein metric}%
\index{Bertoluzza metric}%
Among other interval metrics we find the Wasserstein one,
$\mathfrak{d}_\mathrm{W}(x\pm r_x, y\pm r_y) =
\sqrt{(x-y)^2+(r_x-r_y)/3}$ and the Bertoluzza one
$\mathfrak{d}_\mathrm{B}(x\pm r_x, y\pm r_y) =
\sqrt{(x-y)^2+2(r_x-r_y)/3}$, see, e.g.,
\cite{IrpinoVerde2008:clusterinterval} for a review and a possible application
in data clustering.

\section{Aggregation of fuzzy numbers}\label{Sec:FuzzyNumber}

Fuzzy set theory lets us to quite intuitively represent
imprecise or vague information, see \cite{KlirYuan1995:fuzzybook}.
Fuzzy numbers (FNs), introduced
by Dubois and Prade in \cite{DuboisPrade1978:opfn}, form a particular
subclass of fuzzy sets of the real line.
They play an important role in many practical applications,
e.g., in automation and robotics \cite{KimETAL1994:robotfn},
statistical process control \cite{Hryniewicz2008:fuzzysqc},
survey design in the social sciences \cite{SaaETAL2015:fuzzyquestionnaire}
or decision making \cite{Chen2000:topsisfuzzy},
since we often describe our knowledge about objects
through vague numbers such as ``I'm about 180 cm tall''
or ``The train arrived between 2 and 3 p.m.''.

\begin{definition}\index{fuzzy number}%
A fuzzy set $A$ with membership function $\mu_A:\mathbb{R}\to[0,1]$
is a \emph{fuzzy number}, if it possesses at least the four following properties:
\begin{itemize}
\item[(a)] it is a normalized fuzzy set,
i.e., $\mu_A(x_0)=1$ for some $x_0\in\mathbb{R}$,
\item[(b)] it is fuzzy convex, i.e., for any $x_1,x_2\in\mathbb{R}$
and $\lambda\in[0,1]$ it holds:
\[\mu_A(\lambda x_1 + (1-\lambda) x_2) \ge \mu_A(x_1)\wedge \mu_A(x_2),\]
\item[(c)] the support of $A$ is bounded,
where: \[\mathrm{supp}(A) = \mathrm{cl}(\{x\in\mathbb{R}: \mu_A(x)>0\}),\]
\item[(d)] $\mu_A$ is upper semicontinuous.
\end{itemize}
\end{definition}

\begin{remark}
It may be shown that the membership function of a fuzzy number $A$ is given by:
\begin{equation}
\mu_A(x)=\left\{
\begin{array}{lll}
0 & \text{if} & x<a_{1}, \\
l_{A}(x) & \text{if} & a_{1}\leq x<a_{2}, \\
1 & \text{if} & a_{2}\leq x\leq a_{3}, \\
r_{A}(x) & \text{if} & a_{3}<x\leq a_{4}, \\
0 & \text{if} & x>a_{4},%
\end{array}%
\right.  \label{def fuzzy}
\end{equation}%
where $a_{1},a_{2},a_{3},a_{4}\in \mathbb{R}$,
$l_{A}:[a_{1},a_{2}]\longrightarrow [0,1]$ is a nondecreasing upper
semicontinuous function, $l_{A}(a_{1})=0$, $l_{A}(a_{2})=1$, called the
\textit{left side} of the fuzzy number, and $r_{A}:[a_{3},a_{4}]\longrightarrow [0,1]$
is a nonincreasing upper semicontinuous function,
$r_{A}(a_{3})=1$, $r_{A}(a_{4})=0$, called the \textit{right side}
of the fuzzy number $A$.
\end{remark}

\begin{remark}
A fuzzy number $A$ may also be specified by providing its so-called
$\alpha$-cuts, $\alpha\in[0,1]$. Let:
\begin{equation}
A_\alpha = \{x\in\mathbb{R}: \mu_A(x)\ge \alpha\}
\end{equation}
for $\alpha>0$ and
$   A_0=\mathrm{supp}(A)$.
The $1$-cut is sometimes called the core of $A$.
Every $\alpha$-cut is a closed interval, that is:
\begin{equation}
   A_\alpha=\left[
   A_L(\alpha), A_R(\alpha)
   \right],
\end{equation}
with:
\begin{eqnarray*}
   A_L(\alpha) &=& \inf\{x\in\mathbb{R}: \mu_A(x)\ge \alpha\},\\
   A_R(\alpha) &=& \sup\{x\in\mathbb{R}: \mu_A(x)\ge \alpha\}.
\end{eqnarray*}
Note that if the sides of
the fuzzy number $A$ are strictly monotone, then $A_{L}$ and $A_{U}$ are
inverse functions of $l_{A}$ and $r_{A}$, respectively.
\end{remark}

Let $\mathbb{F}(\mathbb{R)}$ denote the set of all fuzzy
numbers. In practice, e.g., when computations of arithmetic
operations are performed, fuzzy numbers with simple membership functions are
often preferred. A very useful subclass of $\mathbb{F}(\mathbb{R)}$,
especially for computer processing, may be defined by
considering fuzzy numbers with piecewise linear side functions. Thus,
let us consider the following definition,
see \cite{CoroianuETAL2013:piecewise1,CoroianuETALXXXX:piecewise2}.

\begin{definition}
\label{knotfn} Fix $n\in \mathbb{N}_0$.
Given $\boldsymbol{\alpha}\in
\{ (\alpha_0, \alpha_1,\allowbreak \dots, \alpha_{n+1})\in{[0,1]}^{n+2}: 0=\alpha_0 < \alpha_1 < \dots < \alpha_n < \alpha_{n+1}=1 \}$
and
$\mathbf{s}\in\{ (s_1, \dots, s_{2n+4})\in\mathbb{R}^{2n+4}: s_1 \le \dots \le s_{2n+4} \}$,
\index{piecewise linear fuzzy number}%
an $\boldsymbol{\alpha}$-\textit{piecewise linear $n$-knot fuzzy number} $S(\boldsymbol{\alpha}, \mathbf{s})$,
is defined by:
\begin{eqnarray*}
S(\boldsymbol{\alpha}, \mathbf{s})_{L}(\beta) & = &
s_{i+1}+(s_{i+2}-s_{i+1})\, \frac{\beta -\alpha_{i}}{\alpha_{i+1}-\alpha_{i}}%
,  \label{pw 1} \\
S(\boldsymbol{\alpha}, \mathbf{s})_{U}(\beta) & = &
s_{2n+4-i}+(s_{2n+3-i}-s_{2n+4-i})\, \frac{\beta -\alpha_{i}}{%
\alpha_{i+1}-\alpha_{i}},  \label{pw 2}
\end{eqnarray*}%
for some $i\in[0:n]$ such that $\beta \in \left[ \alpha_{i},%
\alpha_{i+1}\right]$.
\end{definition}

\begin{figure}[htb!]
\centering
\par
\par
\includegraphics[width=8.25cm]{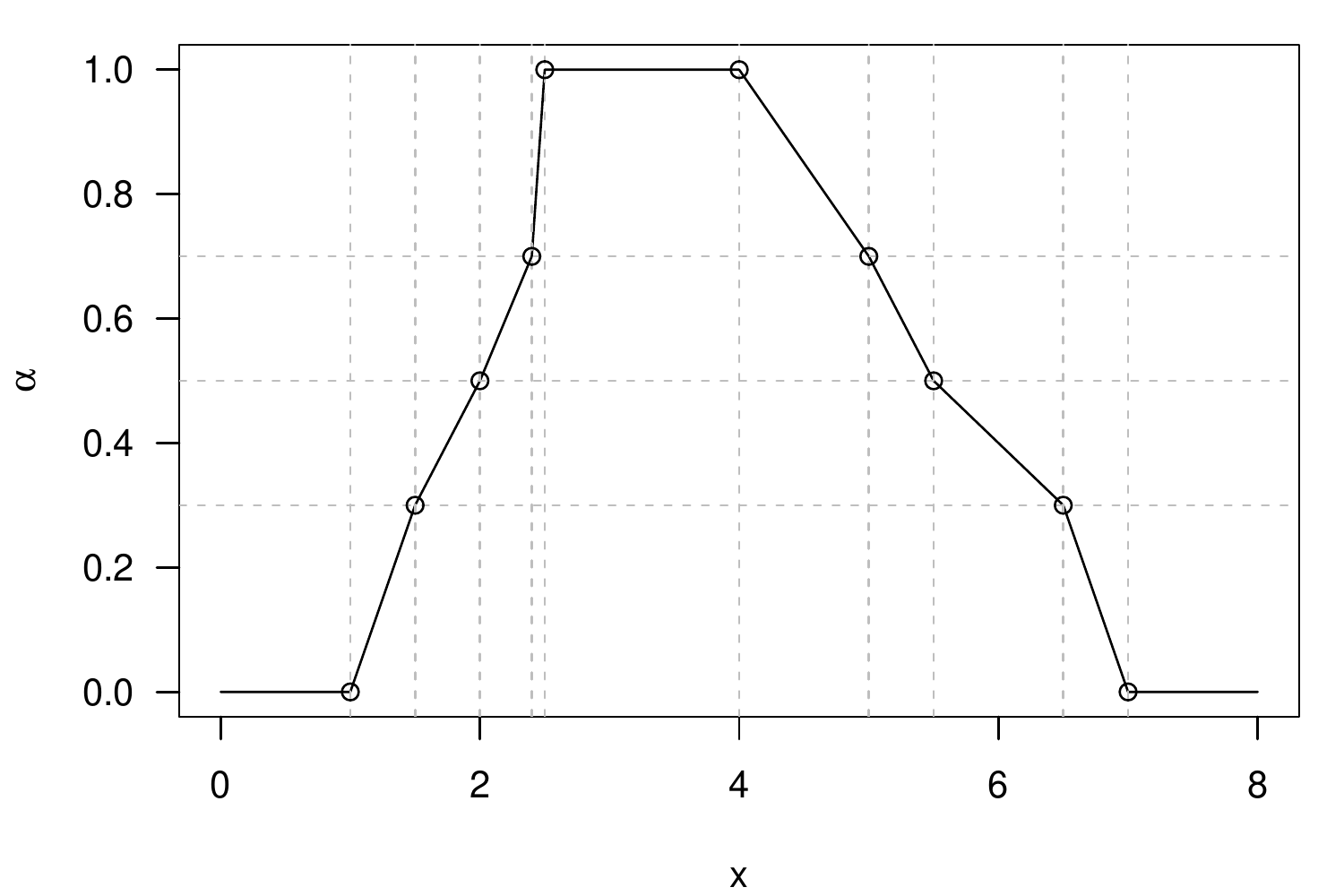}
\caption[Plot of an exemplary $3$-knot piecewise linear fuzzy number.]%
{Plot of an exemplary $3$-knot piecewise linear fuzzy number $S(%
\boldsymbol{\protect\alpha}, \mathbf{s})$, with $\boldsymbol{\protect\alpha}%
=(0, 0.3, 0.5, 0.7, 1)$ and $\mathbf{s}=(1,1.5,2,2.4,2.5,4,5,5.5,6.5,7)$.}
\label{Fig:Example_knot3}
\end{figure}

Please note that
the membership function of $S(\boldsymbol{\alpha}, \mathbf{s})$ is also
piecewise linear in the case when $\mathbf{s}$ is strictly monotone (for an
example see Figure~\ref{Fig:Example_knot3}).

\begin{remark}
It is worth noting that the class of fuzzy sets introduced in Definition
\ref{knotfn} generalizes some well-known subfamilies of
fuzzy numbers. Actually, for $n=0$ and $s_1=s_4$ we get ``crisp'' real
numbers, for $n=0$ and $s_1=s_2, s_3=s_4$ we obtain ``crisp'' real
intervals; if $n=0$ and $s_2=s_3$ we get triangular fuzzy numbers, and by assuming
only $n=0$ we obtain trapezoidal fuzzy numbers.
\end{remark}

Further on we  assume that two fuzzy numbers $A$ and $B$ are equal (denoted $A=B$)
if $A_{L}(\alpha)=B_{L}(\alpha)$ and $A_{U}(\alpha)=B_{U}(\alpha)$ for all $\alpha\in[0,1]$.

\bigskip
Let us consider a fusion function $\func{F}:\mathbb{F}(\mathbb{R})^n\to\mathbb{F}(\mathbb{R})$,
which aims to aggregate $n$ fuzzy numbers so that one fuzzy number is generated
as a result.

\paragraph{Defuzzification methods.}
Concepts such as the expected value \cite{DuboisPrade1987:meanfn} or value \cite{DelgadoETAL1998:canonicalfn}
of a fuzzy number (see Section \ref{Sec:CharacteristicsFuzzyNumbers})
may be used to defuzzify a given fuzzy number.
Together with some measure of nonspecifity, e.g.,
width \cite{Chanas2001:intervapproxfn} or ambiguity \cite{DelgadoETAL1998:canonicalfn},
these may be used to concisely represent $A\in\mathbb{F}(\mathbb{R})$ as $x\pm r$,
i.e., a real interval. If such a level of data loss is accepted, then the aggregation
methods discussed in the previous section may be utilized.

\paragraph{Arithmetic operations.}
In order to extend a binary arithmetic operation $\ast$ (e.g., $+$, $-$, $\times$, $/$)
to the set of fuzzy numbers, most often Zadeh's \index{extension principle}\emph{extension principle},
see \cite{Kerre2011:zadehext,KlirYuan1995:fuzzybook}, is used.
In such a case $C = A\circledast B$ is given via the membership function:
\begin{equation}\label{Eq:ExtensionPrinciple}
   \mu_C(z) = \sup_{z = x\ast y} \left(
      \mu_A(x) \wedge \mu_B(y)
   \right).
\end{equation}
Note that for fuzzy numbers being real intervals,
the extension principle generates exactly the same arithmetic operators
as presented in the previous section.
In particular, in the $\alpha$-cut representation,
the sum $A\oplus B$ and the scalar multiplication $t \otimes A$ (see, e.g., \cite[page~40]{DiamondKloeden1994:metricspacesfs})
for every $\alpha \in [0,1]$ is given by:
\begin{equation}
\left(A\oplus B\right)_{\alpha}=A_{\alpha}\oplus B_{\alpha}= \left[ A_{L}\left(\alpha%
\right)+B_{L}\left(\alpha \right), A_{U}\left(\alpha\right)
+B_{U}\left(\alpha\right) \right]
\end{equation}%
and:
\begin{equation}
\left( t  \otimes A\right)_{\alpha }=t  \otimes A_{\alpha }=\left\{
\begin{array}{ll}
\left[ t\cdot A_{L}\left( \alpha \right), t\cdot A_{U}\left( \alpha \right) %
\right] & \text{if }t \geq 0, \\
\left[ t\cdot A_{U}\left( \alpha \right), t\cdot A_{L}\left( \alpha \right) %
\right] & \text{if }t <0.%
\end{array}%
\right.
\end{equation}

\begin{remark}
   Note that a set of piecewise linear fuzzy numbers with fixed knot configuration
   is closed under addition and scalar multiplication, but not, e.g.,
   multiplication of two arbitrary members of this class. Nevertheless, this suffices to
   introduce the notion of a weighted arithmetic mean.
\end{remark}

Basic fuzzy number arithmetic operations are available in \R{} via the \package{FuzzyNumbers}
package \cite{GagolewskiCaha:fuzzynumberspackage}.
For practical reasons, each arbitrary fuzzy number should be approximated by
a piecewise linear one (using a considerable number of knots),
see \cite{CoroianuETALXXXX:piecewise2} for discussion.

\begin{lstlisting}[language=R]
library("FuzzyNumbers")
A <- TrapezoidalFuzzyNumber(1, 2, 3, 4)
B <- TriangularFuzzyNumber(3, 5.5, 6)
C <- as.PiecewiseLinearFuzzyNumber(A, knot.n=100) *
     as.PiecewiseLinearFuzzyNumber(B, knot.n=100)
alphacut(C, c(0, 1)) # support and core
##       L      U
## 0     3   24.0
## 1    11   16.5
\end{lstlisting}

Please note that the extension principle is based on the $\wedge$ operation.
It turns out that this way of extending arithmetic operations to the set
of fuzzy numbers can be generalized by replacing $\wedge$ in Equation~\eqref{Eq:ExtensionPrinciple}
with, e.g., an arbitrary triangular norm. This leads to the notion
of the so-called interactive fuzzy numbers, where  one is able to take
into account a kind of mutual interdependency between such types of objects
(compare the role of copulas in probability theory).
This idea has been investigated by Full\'{e}r and other researchers,
see, e.g., \cite{CarlssonFullerMajlender2004:addcorrfn,FullerMajlender2003:interactivefn,%
CoroianuFuller2013:multintfn,Coroianu2016:necsufintfn,DuboisETAL2000:fuzzintanal}.

\paragraph{Orders in the space of fuzzy numbers.}
The space of fuzzy numbers, just like its subclass -- real intervals,
has no natural linear order. A relation $A\sqsubseteq_S B$ whenever, e.g.,
$\sup\mathrm{supp}\,A \le \inf\mathrm{supp}\,B$ or $A=B$
is merely a partial ordering.

Nevertheless, in the literature many authors have considered different
ways to construct so-called ranking indices, i.e., functions of the
$r:\mathbb{F}(\mathbb{R})\to\mathbb{R}$ kind, which can be used to
construct a total preorder on the set of fuzzy numbers. Such tools may be useful for symmetrizing
weighted arithmetic means in order to define OWA-like operations.

In particular, Ban and Coroianu in \cite{BanCoroianu2015:simplsearchrank}
characterized all the ranking indices for trapezoidal fuzzy numbers
that fulfill -- among others -- the set of famous Wang and Kerre
\cite{WangKerre2001:reasonable1} axioms, including translation and scale invariance.
Denoting a trapezoidal fuzzy number as $\mathrm{T}(s_1, s_2, s_3, s_4)$,
this very strong result indicates that the only reasonable ranking index
may be a kind of linear combination of $s_1,\dots,s_4$,
given by:
\[r(\mathrm{T})=cs_1+(0.5-c)s_2+(0.5-c)s_3+cs_4\]
for some $c\in[0,1]$.

\paragraph{Metrics in the space of fuzzy numbers and penalty-based fusion functions.}
Perhaps the most often considered metric in the space of fuzzy numbers
is an extension of the Euclidean distance defined by the equation:
\begin{equation}
\mathfrak{d}_2(A,B) = \sqrt{\int_0^1 \left(A_L(\alpha)-B_L(\alpha)\right)^2\,d\alpha
         + \int_0^1 \left(A_U(\alpha)-B_U(\alpha)\right)^2\,d\alpha}.
\end{equation}
In a very similar manner, arbitrary weighted Minkowski distances may be introduced,
see \cite{Grzegorzewski1998:metricsordersfn}.

Ban, Coroianu, and Grzegorzewski in \cite{BanCoroianuGrzegorzewski2011:trapapproxaggr}
considered a trapezoidal fuzzy number fusion problem.
They derived an algorithm for determining a $\mathfrak{d}_2$-based
centroid of $n$ such fuzzy sets using the Karush-Kuhn-Tucker theorem,
which is expressed as:
\[
\func{F}(\vect{t}^{(1)}, \dots, \vect{t}^{(n)})
= \mathrm{T}\left(
\func{AMean}(s_1^{(1)},\dots,s_1^{(n)}),
\dots,
\func{AMean}(s_4^{(1)},\dots,s_4^{(n)})
\right),
\]
where $\vect{t}^{(i)}=\mathrm{T}(s_1^{(i)},\dots,s_4^{(i)})$.

Additionally, the same authors  in \cite{BanCoroianuGrzegorzewski2013:fmed}
studied the conditions for which, given a metric $\mathfrak{d}$ in the space
of fuzzy numbers, the corresponding 1-median exists and is unique.
It is worth noting that their results are based on the R\aa{}dstr\"{o}m embedding theorem
and can be quite easily generalized to some other linear spaces equipped with a norm-generated metric.

\paragraph{Some notes on aggregation of other types of fuzzy quantities.}
There are various possible ways to generalize and/or extend the theory
of classical fuzzy sets. One of them includes the class of
Atanassov's so-called intuitionistic fuzzy sets (AIFS, see, e.g., \cite{Atanassov1986:ifs,Atanassov1999:aifs}).
Here, the degrees of ``belongingness'' and ``nonbelongingness''
of an observation to an AIFS are modeled separately.
Notably, AIFS  are equivalent to interval-valued fuzzy sets
(see \cite{DeschrijverKerre2003:aifsintervalfscorrespondence} for discussion),
so we may rather just model the degree of belongingness using a real interval.

In particular, e.g., Szmidt and Kacprzyk in \cite{SzmidtKacprzyk2000:distaifs}
as well as Grzegorzewski in \cite{Grzegorzewski2004:distaifs}
review possible ways to define metrics in the space of AIFS.
Moreover, Deschrijver in \cite{Deschrijver2011:owaaifs} defines OWA operators
together with quasi-arithmetic means,
and Beliakov, Bustince, James, Calvo, and Fernandez \cite{BeliakovETAL2011:medianaifs}
define median-like fusion functions.
The reader is referred to \cite{BeliakovETAL2015:practicalbook} for a
comprehensive review of these concepts and a list of practical applications
of AIFS, e.g., in image processing.

\section{Aggregation of random variables}\label{Sec:RandomVariables}

Let us assume that we are given $n$ random variables which
are independent and identically distributed \index{i.i.d.}(i.i.d.) --
following a common cumulative distribution function $F$.
That is, let $\vect{X}=(X_1,\dots,X_n)$ i.i.d.~$F$.
Moreover, let $F$ be continuous with support $\Ival=[a,b]$.
Note that basically in aggregation theory we only consider observed values
such as $\vect{x}=(x_1,\dots,x_n)$, i.e., particular realizations of $\vect{X}$.
Probabilistic models provide us with yet another way for dealing with input
data imprecision (in fact, one that is accepted by most of the practitioners).

\begin{definition}
A \index{statistic}\emph{statistic} is any function of random variables.
\end{definition}

Thus, each fusion function defined on a sequence of random variables
is a statistic in the probabilistic sense.
Note that $F(X_1,\dots,X_n)=Y$ is per se another random variable which follows
its own distribution function.

\smallskip
Studies of probabilistic properties of particular classes
of fusion functions appear significantly less frequently in the literature
than research dealing with constructions of particular functions fulfilling
desired statistical properties. Despite this, let us now review a few
fundamental results on general properties of some fusion functions discussed so far.

\paragraph{Order statistics.}
Here are some basic facts on \index{order statistic}order statistics in an i.i.d.~mo\-del,
see \cite[Chapter~2]{DavidNagaraja2003:orderstatistics}.
The cumulative distribution function of the $i$th order statistic
is given by:
\begin{equation}
F_{(i)}(x) = \sum_{j=i}^n {n \choose j} F^j(x)(1-F(x))^{n-j} = I_{F(x)}(i, n-i+1)
\end{equation}
and -- assuming that $f$ is the common density of each $X_i$ --
the probability density function is given by:
\begin{equation}
f_{(i)}(x) =  \frac{F^{i-1}(x)(1-F(x))^{n-j} f(x)}{B(i, n-i+1)},
\end{equation}
where $B(x,y)=\int_0^1 t^{x-1}(1-t)^{y-1}\,dt$
is the \index{Beta function}Beta function and $I_p(x,y)=\int_0^p t^{x-1}(1-t)^{y-1}\,dt / B(x,y)$
is the \index{regularized incomplete Beta function}regularized incomplete Beta function,
$x,y>0$, $p\in[0,1]$.

In particular, the \index{median}sample median for even $n$ follows the c.d.f.:
\begin{eqnarray*}
&&F_\func{Median}(x) = \frac{2}{B(n/2, n/2)}\cdot\\
&\cdot&\int_{-\infty}^x F(y)^{0.5n-1} \left(
(1-F(y))^{0.5n} - (1-F(2x-y))^{0.5n}
\right) \,f(y)\,dy,
\end{eqnarray*}
see \cite{DesuRodine1969:popmedian} for a proof.

\begin{example}
The $i$th order statistic of a sample of i.i.d.~random variables
uniformly distributed on $[0,1]$ has a Beta distribution with parameters $i$ and $n+1-i$.
In any case, generally we can observe that deriving exact yet user-friendly
forms of order statistics' distributions is a difficult task.
\end{example}

We already mentioned that a statistic is a random variable itself.
For large $n$ and arbitrary $p\in]0,1]$,
the $\lceil np\rceil$th order statistic is approximately
normally distributed. More precisely:
\[
   X_{(\lceil np\rceil)} \sim \mathrm{AN}\left(
   \func{F}^{-1}(p), \frac{\sqrt{p(1-p)}}{\sqrt{n} f(F^{-1}(p))}
   \right),
\]
where $\mathrm{AN}(\mu,\sigma)$ denotes that the distribution
is asymptotically normal with expected value $\mu$ and standard deviation $\sigma$.

Of course, do notice that
$X_{(i)}$ and $X_{(j)}$ are no more independent random variables.
However, in the literature we may find general equations giving joint
distributions of pairs, triples, etc., of order statistics.
For more details on stochastic properties of order statistics and
functions of order statistics the reader is referred to the seminal
monograph by David and Nagaraja \cite{DavidNagaraja2003:orderstatistics}.

\begin{remark}
Provided that $F$ is symmetric, the sample median is one of the
possible \emph{estimators} of the expected value of $X$,
that is -- intuitively -- fusion functions which aim to \textit{guess}
one of the parameters or characteristics of the unknown probability distribution
-- solely based on the observed sample. What is worth noting,
``good statistical properties'', such as unbiasedness, consistency, efficiency,
and so forth,
may suggest which fusion function shall be chosen for use in particular applications
(see Section~\ref{Sec:CharacteristicsProb}).
For instance, it is known that the arithmetic mean is an unbiased, minimal variance estimator
of the expected value provided that $X_i$ has finite variance, which in simple
statistical models (e.g., not contaminated by outliers) is a much better choice
than the median.
\end{remark}

\paragraph{Weighted arithmetic means and ordered weighted averages.}
By the famous central limit theorem we know that the arithmetic mean
is asymptotically normally distributed (under certain conditions on $F$).
For arbitrary weighted means, if the expected value of $X$ is finite, then
the expected value of a weighted mean is equal to the expected value of $X$,
because $\mathbb{E}\,\func{WMean}_\vect{w}(\vect{X})=\sum_{i=1}^n w_i\mathbb{E}\,X_i=\mathbb{E}\,X$.

Interestingly, in probability and statistics, OWA operators are special cases of
the so-called L-statistics, i.e., linear combinations of order statistics.
Their properties are quite well-known already, compare \cite{Borovskikh1981:Lstatistics}.
More generally, Kojadinovic and Marichal in \cite{KojadinovicMarichal2009:distribChoquet}
studied the moments and distributions of arbitrary Choquet discrete integrals.

Extended versions of functions from both of the above classes have been considered.
In \cite{Stigler1969:lincombos} the conditions on the triangle of coefficients
choice for which a corresponding $L$-statistic has a limiting normal distribution
is studied. For that we must assume certain weight generating schemes, compare Section~\ref{Sec:WeightingTriangles}.
For instance, in \cite{JamisonETAL1965:convwave} the convergence of weighted averages
is studied, where there is one weight sequence
$(c_1,c_2,\dots)$ and the statistic is of the form
$\func{F}(x_1,\dots,x_n) = \sum_{i=1}^n c_ix_i/\sum_{j=1}^n c_j$.
On the other hand, like, e.g.,  in \cite{Borovskikh1981:Lstatistics,Yager1991:connectives},
we may also assume that
$\func{F}(x_1,\dots,x_n) = \sum_{i=1}^n c_{i,n} x_i/\sum_{j=1}^n c_{j,n}$,
where $c_{i,n}=C(i/{n+1})$ for some coefficient generating function
$C:]0,1[\to\mathbb{R}_{0+}$.

\paragraph{Discrete Sugeno integrals and other lattice polynomial functions.}
Marichal in \cite{Marichal2006:cdflatpoly} derived formulas for cumulative
distribution functions and moments of lattice polynomial functions
in the case of independent (but not necessarily identically distributed)
random variables (real-valued ones). Note again that symmetric lattice polynomial
functions are equivalent to sample quantiles.
The case in which random variables are not necessarily independent
was studied by Dukhovny in \cite{Dukhovny2007:latpolyrv}.
Also, Marichal and Kojadinovic studied the behavior of linear
combinations of lattice polynomial functions in the case of uniformly distributed
input data, see \cite{MarichalKojadinovic2008:distrfunclinlatpoly}.
The i.i.d.~case for weighted lattice polynomial functions
was studied in \cite{Marichal2008:wlatpolyiid}.
Asymptotic behavior of the discrete Sugeno integral
was studied by Gagolewski and Grzegorzewski
in \cite{GagolewskiGrzegorzewski2010:smps}. In particular, they showed
the asymptotic normality of this fusion function and that it is a consistent
estimator of some underlying probability distribution's characteristic of location.

\begin{remark}
Knowing the probabilistic behavior of fusion functions enables us
to construct tools, e.g., aiming at statistical inference
or decision making. For instance, a two-sample statistical hypothesis test
for equality of Pareto distribution parameters based on the differences in the outputs
of a particular Sugeno integral (the Hirsch index, see Section~\ref{Sec:ImpactFunctions})
was proposed by Gagolewski in \cite{Gagolewski2012:smps}.
\end{remark}

\paragraph{Operations on random variables.}
Taking into account the above and other facts from probability theory,
we may infer some new results concerning other classes of fusion functions.
For instance, let $\varphi$ be a strictly increasing and continuous function
and assume that $Y=\varphi(X)$. Knowing that:
\begin{equation}
   F_Y(u) = F_X(\varphi^{-1}(u)),
\end{equation}
we may easily deduce the form of the cumulative distribution function
of a quasi-arithmetic mean from the form of the c.d.f.~of the arithmetic mean etc.
What is more, note that under the current assumptions the density function is given by:
\begin{equation}
   f_Y(u) = f_X(\varphi^{-1}(u)) \frac{d}{du} \varphi^{-1}(u).
\end{equation}

Basic arithmetic operations on independent random variables,
see, e.g., \cite{Springer1979:algebrarv}, are given by:
\begin{eqnarray}
f_{X+Y}(u) = (f_X\oplus  f_Y)(u)       & = & \int_{-\infty}^{+\infty} f_X(t) f_Y(u-t)\, dt, \\
f_{X-Y}(u) = (f_X\ominus f_Y)(u)       & = & \int_{-\infty}^{+\infty} f_X(t) f_Y(t-u)\, dt, \\
f_{X\times Y}(u) = (f_X\otimes f_Y)(u) & = & \int_{-\infty}^{+\infty} \frac{f_X(t) f_Y(u/t)}{|t|} \, dt, \\
f_{X/Y}(u) = (f_X\oslash f_Y)(u)       & = & \int_{-\infty}^{+\infty} \frac{f_X(t) f_Y(t/u) |t|}{u^2} \, dt.
\end{eqnarray}
Jaroszewicz and Korzeń in \cite{JaroszewiczKorzen2012:arithrv}
study families of probability distributions closed under the above operations.
Moreover, they develop a methodology for approximating arbitrary densities
using piecewise Chebyshev interpolation. The \package{PaCAL} (probabilistic calculator)
package for \lang{Python} \cite{KorzenJaroszewicz2014:pacal}
is based on these results.

\paragraph{Orders in the space of random variables.}
There are many possible ways to introduce a partial order on
the family of random variables, see, e.g., \cite{Lehmann1955:ordprob}.
In particular, \index{first order stochastic dominance}first order stochastic dominance
\index{0st@$\preceq_\mathrm{st}$|see {first order stochastic dominance}}%
is defined as:
\begin{equation}
   X \preceq_\mathrm{st} Y \text{ if and only if } (\forall x)\ F_X(x)\ \ge F_Y(x),
\end{equation}
and the \index{likelihood ratio order}likelihood ratio order as:
\begin{equation}
   X \preceq_\mathrm{lr} Y \text{ if and only if } g(u)=\frac{f_Y(u)}{f_X(u)} \text{ is an increasing
   function of $u$}.
\end{equation}
Linear orders may be introduced by considering, e.g.,  numerical characteristics
of probability distributions such as the expected value (see Section~\ref{Sec:CharacteristicsProb}).
Moreover, it is not uncommon to consider various dissimilarity measures
(which might not fulfill the triangle inequality),
like the Kullback-Leibler divergence \cite{KullbackLeibler1951:divergence}
and the Kolmogorov-Smirnov, Cramer-von Mises, or Anderson-Darling statistics which
appear in the corresponding goodness-of-fit tests, see \cite{Stephens1974:edfgof}.

\paragraph{Randomness and fuzzy numbers.}
On a side note, we may also consider randomness and fuzziness together.
In particular, Puri and Ralescu in \cite{PuriRalescu1986:fuzzyrandomvariables}
defined the concept of a random fuzzy variable as a mapping
from a sample space $\Omega$ to the set of fuzzy numbers
(see, e.g., \cite{Kwakernaak1978:fuzzyrandomvariables1} for one of the possible
alternative approaches).
In such a framework, e.g.,
Sinova and others \cite{SinovaETAL2012:medianrandomfn,%
SinovaETAL2010:medianrandomint,SinovaETAL2013:medianrandomint,SinovaETAL2015:medianrandomfn}
considered various types of median-like fusion functions
for random intervals and random fuzzy numbers.

\section{Aggregation of graphs and relations}

Recall (compare Remark~\ref{Ex:HasseBeautiful})
that, at least for the purpose of this book,
we may assume that there is a one-to-one correspondence between
graphs and binary relations. Nevertheless, data fusion methods for
the two classes of objects differ from each other as they most often
serve much different practical purposes.

\paragraph{Aggregation of rankings and other relations.}
Suppose that $P=\{p_1,\dots,p_k\}$ and let
$\mathcal{L}(P)$ denote the set of all linear strict ordering relations on $P$.
Our aim is to construct a fusion function $\func{F}: \mathcal{L}(P)^n\to\mathcal{L}(P)$
that aggregates $n$ linear ordering relations
into one that is as much ``concordant'' with the inputs as possible.
From now on we assume that the set of input orders $\vect{x}=(\sqsubset^{(1)},\dots,\sqsubset^{(n)})$ is fixed.

The construction of fair election methods continues to be of interest to many
researchers since the 18th century.
For instance, the famous \index{Borda count}\emph{Borda count} assigns to each
$p_j$, $j\in[k]$, a particular number of points relative to $p_j$'s
position in a ranking $\sqsubset^{(i)}$, $i\in[n]$, namely:
\begin{equation}
b_{i,j} = 1+\left|\left\{ l\in[k]: p_l \sqsubset^{(i)} p_j \right\}\right|.
\end{equation}
Then the position of $p_j$ in the aggregated ranking is
a function of the total number of points, $\bar{b}_j = \sum_{i=1}^n b_{i,j}$.
The reader is referred to the extensive
literature on the subject for a treatment of those kinds of data fusion
methods at an appropriate level of detail, e.g.,
\cite{Arrow1963:socialchoice,Colomer2004:handbookelect,%
Fishburn1977:condorcet,%
Lin2010:rankaggregation,BouyssouPerny1992:valprefrel}.

Nevertheless, we shall at least sketch two noteworthy classes of rank
\index{Kemeny aggregation scheme}%
aggregation method construction. A \emph{Kemeny-like optimal aggregation scheme},
compare \cite{Kemeny1959:mathnonum},
aims at finding a ranking $\sqsubset^*\ \in\mathcal{L}(P)$
which for some dissimilarity measure $\mathfrak{d}$ (e.g., a metric)
has the property that:
\begin{equation}
   \sum_{i=1}^n \mathfrak{d}(\sqsubset^*, \sqsubset^{(i)})
   \le \sum_{i=1}^n \mathfrak{d}(\sqsubset, \sqsubset^{(i)}),
\end{equation}
for all $\sqsubset\ \in\mathcal{L}(P)$.
For instance, $\mathfrak{d}$ may be the already mentioned Dinu
\index{Dinu rank distance}\cite{DinuManea2006:efficientrankaggregation}
rank distance or a function of the \index{Kendall correlation coefficient}Kendall correlation coefficient $\tau$ \cite{ChinETAL2004:approxdynrankagg},
see also \cite{DidehvarEslahchi2007:algrankagg} for a different choice.
This approach to rank fusion is in fact an instance of a penalty-based scheme.
As most often exact algorithms for determining a Kemeny optimal solution
are computationally intractable, various approximate methods are used
in practice, e.g., ones that are based on evolutionary strategies
(compare Algorithm~\ref{Alg:Genetic}).

A second approach is based on a notion of monotonicity.
For instance, Rademaker and De~Baets in \cite{RademakerDeBaets2014:rankingmono} %
considered the following measure $\func{S}_\mathbf{x}: \mathcal{L}(P)^2\to[0:n]$
of strength of support for a pair $(p_i, p_j)$:
\begin{equation}
   \func{S}_\mathbf{x}(p_i, p_j) = \sum_{l=1}^n \indicator(p_i \sqsupset^{(l)} p_j),
\end{equation}
see also \cite{RademakerDeBaets2011:agmonrecip}.
For $p_i\neq  p_j$ it holds that $\func{S}_\mathbf{x}(p_i, p_j) + \func{S}_\mathbf{x}(p_j, p_i) = n$.
The authors suggest that the aggregated ranking $\sqsubset^*$ should fulfill the following monotonicity condition
for all $p_i, p_i', p_j, p_j'\in P$:
\begin{multline*}
   (p_i \sqsupseteq^* p_i') \text{ and } (p_j \sqsupseteq^* p_j') \text{ and }
   \left(
   (p_i \sqsupset^* p_i')
   \text{ or }
   (p_j \sqsupset^* p_j')
   \right)\\
   \Longrightarrow
   \func{S}_\mathbf{x}(p_i, p_j) \ge \func{S}_\mathbf{x}(p_i', p_j').
\end{multline*}
If the construction of such a ranking is not possible,
it is allowed to ``slightly modify'' the values returned by $\func{S}_\mathbf{x}$
so that a concordance becomes possible (note that the result might not be unique).
Unfortunately, the procedure proposed in \cite{RademakerDeBaets2014:rankingmono}
requires that all the possible rankings in $\mathcal{L}(P)$ shall be considered.
Nevertheless, the reader is already aware that, e.g., a genetic algorithm
 may quite easily be constructed to approximate
the desired solution.

For an approach to aggregating arbitrary partial ordering relations,
see, e.g., \cite{RademakerDeBaets2010:thresholdaggpord},
in which pairwise preferences are learned through a majority-based
voting process computed iteratively (using the notion of transitive closure)
in such a way that cyclical and contradictory
preferences are avoided.

Another interesting problem considers an aggregation of equivalence relations.
For instance, Gionis, Mannila, and Tsaparas in \cite{GionisETAL2007:clustagr}
discuss the issue of \index{clustering aggregation}\emph{clustering
aggregation}. Given $n$ partitions $C_1,\dots,C_n$
of the same data set $X$ into an equal number of subsets, $d$,
they review different methods to find a single $d$-partition
that minimizes the total disagreement with the $n$ input clusterings.

\paragraph{Aggregation of trees and other graphs.}
Graph representation of objects is particularly useful in
various pattern recognition tasks, see, e.g.,
\cite{Bille2005:treeedit,BunkeRiesen2011:graphpattern}.
We may be faced with a need to aggregate a set of graphs
(possibly with labeled edges or nodes) when
we need to determine a prototypical object in a set of similar
glyphs in an optical character recognition task or to construct
a $k$-means-like procedure for structurally described molecules.
In such a case, we may rely on the notion of a penalty-based fusion function.

The following two classes of dissimilarity measures for graphs
are most often referred to in the literature:
\index{graph metric}\index{graph edit distance}\index{maximal common subgraph}%
\begin{itemize}
   \item metrics based on maximal common subgraphs
or minimal common supergraphs, see, e.g., \cite{FernandezValiente2001:graphdist,%
BunkeShearer1998:graphdistmetric,WallisETAL2001:graphdist}, for instance:
\begin{equation}
   \mathfrak{d}(G_1, G_2) = 1 - \frac{\left|\mathrm{mcs}(G_1, G_2)\right|}{|G_1|\vee|G_2|},
\end{equation}
where $\mathrm{mcs}(G_1,G_2)$ denotes the maximal common subgraph of two input graphs
and $|G|$ gives the number of vertices in a graph $G$,

\item
edit distance-based metrics, see, e.g., \cite{Tai1979:tree2treecorr,ZhangShasha1989:edittree,%
Bille2005:treeedit,GaoETAL2010:survged}, defined by considering
the minimal number of node/edge relabeling, deletions, insertions (and possibly other
types of edit operations) that are needed to transform a given graph to another one.
\end{itemize}
Interestingly, as shown by Bunke in \cite{Bunke1997:grapheditmcs}, there exist graph-based
and $\mathrm{mcs}$-based distances which are equivalent to each other.

\section{Aggregation in finite semimetric spaces}\label{Sec:PseudometricSpace}

Let $X=(\vect{x}^{(1)},\dots,\vect{x}^{(n)})$ be a finite set
and $(X,\mathfrak{d})$ denote a space equipped with a dissimilarity measure
(a \index{semimetric}\emph{semimetric}) $\mathfrak{d}:X\times X\to[0,\infty]$, i.e., one that fulfills:
\begin{itemize}
   \item symmetry, i.e., $\mathfrak{d}(\vect{x}^{(i)}, \vect{x}^{(j)})=\mathfrak{d}(\vect{x}^{(j)}, \vect{x}^{(i)})$ for all $i,j\in[n]$,
   \item $\mathfrak{d}(\vect{x}^{(i)},\vect{x}^{(i)})=0$ for all $i\in[n]$.
\end{itemize}
We would like to construct a fusion function $\func{F}$ which aggregates
all elements in $X$. Clearly, the output should be an element in $X$ as well.
Due to the high generality of the assumed model (which as a matter of fact is quite
realistic), the set of possible operations that may be involved in
the fusion process is limited:
practically, we may only be looking for a penalty-based \index{exemplar}exemplar,
see Section~\ref{Sec:medoid}.

Let $\func{D}:[0,\infty]^n\to[0,\infty]$ be a nondecreasing and idempotent fusion function
such that $\func{D}(n\ast 0)=0$.
\index{distance-based penalty function}%
We consider a fusion function like:
\begin{equation}
   \func{F}(\vect{x}^{(1)},\dots,\vect{x}^{(n)}) = \argmin_{\vect{y}\in X} \func{D}\left( \mathfrak{d}(\vect{x}^{(1)}, \vect{y}),\dots,\mathfrak{d}(\vect{x}^{(n)}, \vect{y}) \right).
\end{equation}
If $\func{D}(d_1,\dots,d_n) = \sum_{i=1}^n d_i$, then we get a \index{medoid}medoid,
and if $\func{D}(d_1,\dots,d_n) = \bigvee_{i=1}^n d_i$,
a \index{seboid}seboid is obtained. Clearly, other choices of $\func{D}$ are also
possible and potentially useful.

A fast way to compute exemplars is crucial in, for instance, clustering
large data sets. In practice, the costly part of all the algorithms
to compute $\func{F}$ involves the computation of $\mathfrak{d}$.
This is the case of, e.g., long DNA sequences and the Levenshtein distance.
Thus, our aim here is to discuss some possible approaches which
keep the number of total calls to $\mathfrak{d}$ as small as possible.

Let us suppose that $\func{D}$ is associative with neutral element $e$.
The simplest approach to computing $\func{F}$ is as follows.

\begin{algorithm}\label{Alg:exemplar1}
To compute $\argmin_{\vect{y}\in X} \func{D}\left( \mathfrak{d}(\vect{x}^{(1)},
\vect{y}),\dots,\mathfrak{d}(\vect{x}^{(n)}, \vect{y}) \right)$ in the case of associative $\func{D}$
with neutral element $e$, proceed as follows:
\begin{enumerate}
   \item[1.] Let $\vect{d} = (n\ast e)$;
   \item[2.] For $i = 1,2,\dots,n-1$ do:
   \begin{enumerate}
      \item[2.1.] For $j = i+1, i+2, \dots, n$ do:
      \begin{enumerate}
         \item[2.1.1.] Let $d' = \mathfrak{d}(\vect{x}^{(i)}, \vect{x}^{(j)})$;
         \item[2.1.2.] $d_i = \func{D}(d_i, d')$;
         \item[2.1.3.] $d_j = \func{D}(d_j, d')$;
      \end{enumerate}
   \end{enumerate}
   \item[3.] Return $\vect{x}^{(i)}$ as result, where $i=\argmin_{i\in[n]} d_i$.
\end{enumerate}
\end{algorithm}

It is easily seen that the above algorithm requires
exactly $n(n-1)/2$ calls to $\mathfrak{d}$, thus, it does not adapt to
input data at all. Therefore, we may consider the following algorithm.

\begin{algorithm}\label{Alg:exemplar2}
To compute $\argmin_{\vect{y}\in X} \func{D}\left( \mathfrak{d}(\vect{x}^{(1)},
\vect{y}),\dots,\mathfrak{d}(\vect{x}^{(n)}, \vect{y}) \right)$ in the case of associative $\func{D}$
with neutral element $e$, proceed as follows:
\begin{enumerate}
   \item[1.] Let $b_d = \infty$;
   \item[2.] Let $b_i = -1$;
   \item[3.] For $i = 1,2,\dots,n$ do:
   \begin{enumerate}
      \item[3.1.] $c_d = e$;
      \item[3.2.] For $j = 1,2, \dots, n$ do:
      \begin{enumerate}
         \item[3.1.1.] Let $d = \mathfrak{d}(\vect{x}^{(i)}, \vect{x}^{(j)})$;
         \item[3.1.2.] $c_d = \func{D}(c_d, d)$;
         \item[3.1.3.] If $c_d \ge b_d$ then break (go to step 3.3);
      \end{enumerate}
      \item[3.3.] If $c_d < b_d$ then:
      \begin{enumerate}
         \item[3.3.1.] $b_d = c_d$;
         \item[3.3.2.] $b_i = i$;
      \end{enumerate}
   \end{enumerate}
   \item[4.] Return $\vect{x}^{(b_i)}$ as result.
\end{enumerate}
\end{algorithm}

This algorithm requires at least $2n$ but no more than $n^2$ calls to $\mathfrak{d}$.
Its performance is thus strongly dependent on the type of input data, form of $\func{D}$,
as well as the order of input elements.
Note that it does not take into account the symmetry of $\mathfrak{d}$.
Some savings would be possible at the cost of utilizing additional $O(n^2)$
memory, but for large $n$ the use of such a cache is highly discouraged:
if $n=100{,}000$ and values of $\mathfrak{d}$ are stored as 8-byte \texttt{double}
type, we would need 40 GB of RAM, which is way beyond memory limits of popular
desktop PCs nowadays.

\begin{example}\label{Ex:medoidalg2}
Let us compare an average speedup (or slowdown)
in terms of number of calls to $\mathfrak{d}$ of the second algorithm
as compared to Algorithm~\ref{Alg:exemplar1}.
The averages are based on $M=10$ Monte Carlo samples and in each considered
scenario $n=10{,}000$ input data items were aggregated.

\begin{center}\small
\begin{tabularx}{1.0\linewidth}{lXp{1.8cm}p{1.8cm}}
\toprule
\small\bf metric $\mathfrak{d}$ & \small\bf type of data in $X$
& \small\bf Alg.~\ref{Alg:exemplar2}\newline speedup\newline (medoid)
& \small\bf Alg.~\ref{Alg:exemplar2}\newline speedup\newline (seboid) \\
\midrule
Euclidean & normal distribution, $d=200$               & 0.54 & 1796.1 \\
\midrule
Manhattan & normal distribution, $d=200$               & 0.54 & 1519.6 \\
\midrule
Maximum & normal distribution, $d=200$                 & 0.55 & 1277.0 \\
\midrule
Dinu & ACTG sequences, $d=200$                         & 0.65 & 1681.8\\
\midrule
Levenshtein & \package{ispell} Polish words dictionary & 0.60 & 1646.9\\
\bottomrule
\end{tabularx}
\end{center}
It turns out
that the use of the second procedure is particularly appealing
when computing a seboid.
\end{example}

In certain applications (such as clustering), for large data sets
we might be interested in a rough estimate of a set exemplar
which is quite close to the true one but  can be computed much faster.
For that we propose the following procedure.

\begin{algorithm}\label{Alg:exemplarApprox}
To approximate $\argmin_{\vect{y}\in X} \func{D}\left( \mathfrak{d}(\vect{x}^{(1)},
\vect{y}),\dots,\mathfrak{d}(\vect{x}^{(n)}, \vect{y}) \right)$ in the case of
associative $\func{D}$ with neutral element $e$, proceed as follows:
\begin{enumerate}
   \item[1.] Let $c_i =$ some random index in $[n]$;
   \item[2.] Let $\vect{v}=(n\ast 0)$;
   \item[3.] Let $b_i = c_i$; \hfill\textit{(current candidate)}
   \item[4.] Let $b_d = \func{D}\left( \mathfrak{d}(\vect{x}^{(1)}, \vect{x}^{(c_i)}),\dots,\mathfrak{d}(\vect{x}^{(n)}, \vect{x}^{(c_i)}) \right)$;
   \item[5.] $v_{c_i} = 1$; \hfill\textit{(mark as visited)}
   \item[6.] Do:
   \begin{enumerate}
      \item[6.1.] Let $\mathit{change} = 0$;
      \item[6.1.] For each $u_i$ in (indices of $k$-nearest neighbors of $c_i$) do:
      \begin{enumerate}
         \item[6.1.1.] If $v_{u_i} = 1$ then continue to step 6.1;
         \item[6.1.2.] Let $u_d = \func{D}\left( \mathfrak{d}(\vect{x}^{(1)}, \vect{x}^{(u_i)}),\dots,\mathfrak{d}(\vect{x}^{(n)}, \vect{x}^{(u_i)}) \right)$;
         \item[6.1.3.] $v_{u_i} = 1$; \hfill\textit{(mark as visited)}
         \item[6.1.4.] If $u_d < b_d$ then: \hfill\textit{(better candidate was found)}
         \begin{enumerate}
            \item[6.1.4.1.] $b_i = u_i$;
            \item[6.1.4.2.] $b_d = u_d$;
            \item[6.1.4.3.] $\mathit{change} = 1$;
         \end{enumerate}
      \end{enumerate}
      \item[6.2.] $c_i = b_i$;
   \end{enumerate}
   while $\textit{change} = 1$;
   \item[7.] Return $\vect{x}^{(b_i)}$ as result.
\end{enumerate}
where $k$ is some fixed but small integer.
\end{algorithm}

This algorithm has been inspired by the steepest-descent optimization technique.
Instead of computing the gradient (which in an arbitrary semimetric space
is of course unavailable), an element's $k$ nearest neighbors are taken into account.
Starting from a randomly chosen point, we proceed in the direction which
gives the best fit (inversely proportional to the value of $\func{D}$)
until we find a local minimum. In order to increase the quality of the result,
it is suggested to run the procedure a few times (note that the $\vect{v}$ vector
should not be overwritten).

Figures~\ref{Fig:MedoidApprox1} and \ref{Fig:MedoidApprox2} present
a possible implementation of the algorithm which assumes by default
$k=5$ and $15$ restarts.
Numerical studies indicate that the procedure works reasonably well in the case
of, e.g., $D(\vect{d})=\sum_{i=1}^n d_i$, and $D(\vect{d})=\sum_{i=1}^n d_i^2$,
but is far from perfect in the case of a seboid search (which anyway can be
performed very efficiently with Algorithm~\ref{Ex:medoidalg2}).

\begin{example}
Let us compute the speedup of an approximate medoid search
in terms of the number of $\mathfrak{d}$ calls in Algorithm~\ref{Alg:exemplar2}.
Scenarios and experiment setup are identical to the those used in Example~\ref{Ex:medoidalg2},
we used $k=5$ and $15$ restarts.

\newcolumntype{H}{>{\setbox0=\hbox\bgroup}c<{\egroup}@{}}
\begin{center}\small
\begin{tabularx}{1.0\linewidth}{lXp{1.8cm}p{1.8cm}HH}
\toprule
\small\bf metric $\mathfrak{d}$ & \small\bf type of data in $X$
& \small\bf Alg.~\ref{Alg:exemplarApprox}\newline speedup\newline (medoid)
& \small\bf Alg.~\ref{Alg:exemplarApprox}\newline rel. err.\newline (medoid)
& \small\bf Alg.~\ref{Alg:exemplarApprox}\newline speedup\newline (seboid)
& \small\bf Alg.~\ref{Alg:exemplarApprox}\newline rel. err.\newline (seboid) \\
\midrule
Euclidean & normal distribution, $d=200$               & 26.82  & 0.0001  & 29.40  & 0.1104    \\
\midrule
Manhattan & normal distribution, $d=200$               & 25.70  & 0.0003  & 29.80  & 0.0927    \\
\midrule
Maximum & normal distribution, $d=200$                 & 25.47  & 0.0052  & 28.06  & 0.4087    \\
\midrule
Dinu & ACTG sequences, $d=200$                         & 15.24  & 0.0065  & 16.01  & 1.1097    \\
\midrule
Levenshtein & \package{ispell} Polish words dictionary & 24.28  & 0.0134  & 34.93  & 1.1713    \\
\bottomrule
\end{tabularx}
\end{center}
We observe that the speedup is considerable while the relative error is kept small.
\end{example}

A different approximate algorithm -- characterized by a competitive performance,
but one which only works in the case of a medoid search task --
is given by Mic\'{o} and Oncina in \cite{MicoOncina2001:mediannonmetric}
(note that our approach may also be used to improve the quality of its output).
There is also an exact algorithm proposed by Juan and Vidal in \cite{JuanVidal1998:fastmediansearch}.
It may be applied in the cases when $\mathfrak{d}$ is a metric. Nevertheless,
it does not perform well for high-dimensional data (due to the so-called
\index{curse of dimensionality}\emph{curse of dimensionality}, compare, e.g., \cite{AggarwalETAL2001:curse}).

\section{Aggregation of heterogeneous data}

Aggregation of complex data sets consisting of heterogeneous variables
(like those representing information coming from different types
of sources and/or having incompatible representations) faces us with
new challenges that perhaps were not present before. Nevertheless,
it turns out that many of the methods we have already discussed are still
valid in such a setting, compare \cite{Bloch1996:combinationoperators,Torra2010:fusionhandbook}.
Other ones need to be adjusted accordingly
or combined with other data fusion and data mining tools which start to
serve their purpose when they are considered as a whole system.

Let us, however, note that the data fusion algorithms already presented are very general in their
nature and thus can be described and examined via a plethora of formal methods
and approaches. On the other hand, data science practitioners
are aware of the fact that dealing with complex data sets sometimes
may appear to be more art than science: each database often needs a customized
treatment and it is not trivial to find frequently occurring, common patterns.
Nevertheless, we shall at least try to explore limitations of the data
fusion methods reviewed so far, suggest some heuristics to overcome them,
as well as indicate few new types of data mining tasks, where their usage
may be advantageous.

In the commencing sections of the second Chapter we discussed in detail
data fusion methods to aggregate points in a $d$-dimensional space.
We relied on an implicit assumption that the combined variables
were homogeneous. In such a scenario, operations like rotations were
fully justified. However, in a heterogeneous setting, this might not be the case.
The easiest way to deal with data in complex domains is to apply simple
componentwise fusion functions, that is, treat each of the variables
independently. This is an imperfect solution, as any interactions
between features cannot be taken into account in this way.
Therefore, we may try to  group variables of similar type and apply data fusion
methods separately for each block. We can do the same
with respect to clustered records that denote similar entities.

Penalty-based approaches may be quite powerful here too, especially
if we have variables of mixed types, like categorical, ordinal, and numerical
 in one data set. Once a set of variables is partitioned, various
dissimilarity measures may be introduced on each group, and then such
measures may be aggregated (for instance, it is known that a conical combination
of different metrics generates a new metric, etc.).

As usual, proper data wrangling -- that is preprocessing, remapping,
and reencoding -- is a crucial initial stage of the data analysis process.
An important step often consists of decorrelation of variables,
e.g., via  principal component analysis, correspondence analysis,
or manifold learning procedures, see \cite{HastieTibshiraniFriedman2009:esl}.
This can also lead to reduction in data dimensionality.

Fusion functions are also required in the process of improving data quality.
For example, in the case of missing observations, it is customary to group
(cluster) observations which represent similar entities and fill
information that is not available with ``averaged'' results, compare \cite{RubinLittle2002:statanalNA}.

Another area where they may be found useful deals with data deduplication
and consolidation. That is, when  removal of similar entities is needed.
In such a case, we merge multiple
redundant entries and replace them with aggregated ones, those that
minimize information loss. For instance, Bronselaer, Szymczak, Zadrożny,
and De~Tr\'{e} \cite{BronselaerETAL2016:datafusion} develop a framework
that during such a process takes into account a natural ordering relation
that is learned dynamically from a data set.
Moreover, in \cite{BronselaerDeTre2010:aspectsmerge} a theoretical
model for automated coreferent object detection and processing is
proposed.

Let us also mention the record linkage task, see
\cite{Winkler2006:recordlinkage,BilenkoETAL2003:namematching,DomingoTorra2003:recordlink},
which aim is to combine a set $\{D_1,\dots,D_n\}$ of
different, inconsistent, or non-unified databanks (e.g., \lang{SQL} tables)
into a single new database somehow. Typically, this is done by
identifying all records in a database $D_j$ that correspond to a record $r$
in a databank $D_i$, $i\neq j$, either exactly (this may be done by simple
join-like operations) or approximately in cases where data are contaminated
by errors. In the latter setup, fast fuzzy
matching algorithms are needed, including
distance- and clustering-based ones, compare \cite{Torra2010:fusionhandbook}.
This typically involves the use of complex data structures
such as vp- and kd-trees, GNAT, or similar
\cite{Yianilos1993:vptree,Brin1995:gnat,LessmannWurtz2012:fastnnsearch}, which speed up
searching for similar objects.

\clearpage{\pagestyle{empty}\cleardoublepage}
\chapter{Numerical characteristics of objects}\label{Chap:Characteristics}

\lettrine[lines=3]{S}{ynthetic} measures of diverse characteristics of objects
are useful whenever there is a need to quantify how similar to or different from
each other are given entities in terms of some carefully distinguished features.
One  such notion discussed already is a vector norm (see Definition~\ref{Def:norm}).
We shall see that in order to capture exactly a type of behavior or property that is
of interest to a practitioner in a particular setting,
we should rely on its proper mathematical axiomatization.
In this chapter we are interested in exploring various ways to numerically
characterize probability distributions, spread of numerical sequences
(in their entirety), the degree of decision makers'
consensus, economic inequality or poverty, entropy, empirical distribution shape,
fuzzy numbers, as well as fusion functions themselves.
We end the discussion with the notion of a checksum function,
which shall appear different in its very nature from all the other
measures.

\section{Characteristics of probability distributions}\label{Sec:CharacteristicsProb}

The development of currently widely used
measures of data central tendency and variability is inevitably connected
with the history of probability and statistics.
Perhaps the first official (published) use of the term
\index{standard deviation}\emph{standard deviation}
(in the context of probability) is due to Pearson
\cite[Part~I]{Pearson1894:stdev}:
\begin{quote}\it
Let the equation to the probability-curve be $y'=\frac{1}{\sigma \sqrt{2\pi}}
e^{-x^2 / (2\sigma^2)}.$
Then $\sigma$ will be termed its {\normalfont standard-deviation}
(error of mean square).
\cite[page~80]{Pearson1894:stdev}
\end{quote}
However, it is known that the terms ``root mean squared error''
and ``mean error'' were already used by Gauss.

On the other hand, the term \index{variance}\emph{variance} was probably
first defined in the paper by R.A.~Fisher \cite{Fisher1918:variance}:
\begin{quote}\it
It is therefore desirable in analyzing the causes of variability
to deal with the square of the standard deviation
as the measure of variability.
We shall term this quantity the {\normalfont Variance} (\dots).
\cite[page~399]{Fisher1918:variance}
\end{quote}

Please note that both quotations discuss in fact
the underlying probability distribution characteristic, and not
the (observed) sample-based estimates. The debate leading to the
acceptance of a proper distinction between the objects
being of interest of probability theory, on the one hand,
and statistics, on the other, engaged many leading researchers
for many years in the first decades of the 20th century%
\footnote{According to the on-line encyclopedia ``Earliest Known Uses
of Some of the Words of Mathematics'' (maintained by J.~Aldrich,
see http://jeff560.tripod.com/mathword.html, see also
``The Oxford Dictionary of Statistical Terms''):
{\it Although Student (1908) had used the phrases,
``mean of the population'' and ``mean of the sample'',
it was not until the 1930s that such terms
as sample mean or population standard deviation became prominent.}}.
The need to discriminate between a (probabilistic) population
with its characteristics (such as expected value $\mu$ or variance $\sigma^2$),
and a (statistical, observed) sample
from which we may calculate the characteristics' estimates
(like mean $\bar{\vect{x}}$ or \textit{sample} variance $s^2$)
is explained, e.g., in  Fisher's paper
\cite{Fisher1922:mathfound}:
\begin{quote}\it
(\dots) it has happened that in statistics a purely verbal confusion
has hindered the distinct formulation of statistical problems;
for it is customary to apply the same name, {\normalfont mean, standard
deviation, correlation coefficient}, etc., both to the true value
which we should like to know, but can only estimate, and to
the particular value at which we happen to arrive by our methods of
estimation; so also in applying the term probable error, writers sometimes
would appear to suggest that the former quantity, and not merely the
latter, is subject to error.
\end{quote}

Moreover, Fisher in the same paper \cite{Fisher1922:mathfound}
considered two estimates of the (population's)
standard deviation $\sigma$
(in some statistical model), namely the \index{ME@$\mathsf{ME}$|see {mean error}}\index{mean error}\emph{mean error}:
\begin{equation}
\func{ME}(\vect{x}) = \frac{1}{n} \sqrt{\frac{\pi}{2}} \sum_{i=1}^n |x_i - \func{AMean}(\vect{x})|,
\end{equation}
and the \index{mean squared error}\index{MSE@$\mathsf{MSE}$|see {mean squared error}}\emph{mean squared error} defined as:
\begin{equation}
\func{MSE}(\vect{x}) = \sqrt{\frac{1}{n} \sum_{i=1}^n (x_i - \func{AMean}(\vect{x}))^2}.
\end{equation}
This paper is a beautiful early example of a comparative study
concerning the usage of two different sample statistics that shall measure
the same quantity.

\bigskip
In this section our main focus is on various methods that can be used to
measure central tendency or dispersion of probability distributions.
Note that in Section~\ref{Sec:RandomVariables} we considered fusion functions
that act on \emph{random data} and this time we are interested in functions
that are used to numerically summarize some aspects of the underlying probability
distribution's behavior.

Then, we shall consider the link between the two approaches.
More precisely, we concentrate on the properties that a statistic
(a fusion function acting on random data) should fulfill in order
to call it an \textit{estimator} of a probability distribution
characteristic.

\subsection{Measures of location}

Denote by $\mathcal{D}_d$ the set of probability distributions
in $\mathbb{R}^d$, $d\ge 1$. Note that if a random variable $X$ is $f$-distributed,
that is $X\sim f\in\mathcal{D}_d$, then we shall also denote this
fact by $X\in\mathcal{D}_d$ for brevity.
Moreover, if $\varphi:\mathcal{D}_d\to Z$ for some set $Z$,
then instead of writing $\varphi(f)$ we shall also use the notation $\varphi(X)$.

\index{measure of location}%
Oja in \cite{Oja1983:descrstatmultivar} considered the following
axiomatization of measures of location, which is a multivariate generalization
of a model introduced by Bickel and Lehmann in \cite{BickelLehmann1975:descrstat12}.

\begin{definition}
We call $\func{L}:\mathcal{D}_d\to\mathbb{R}^d$ a measure of location in the Oja sense,
whenever:
\begin{enumerate}
   \item[(a)] for any $X,Y\in\mathcal{D}_d$ if
   $X \preceq_\mathrm{st} Y$, then $\func{L}(X)\le\func{L}(Y)$,
   \item[(b)] for all matrices $\vect{A}\in\mathbb{R}^{d\times d}$ of full rank,
   all $\vect{t}\in\mathbb{R}^d$, and $X\in\mathcal{D}_d$ such that
   $\vect{A}X+\vect{t}\in\mathcal{D}_d$ it holds that
   $\func{L}(\vect{A}X+\vect{t}) = \vect{A}\func{L}(X)+\vect{t}$.
\end{enumerate}
\end{definition}
In other words, a measure of location is first order stochastic dominance-monotone
and affine equivariant.

Note that if the distribution of $X$ is symmetric about $\boldsymbol\mu$,
that is $\boldsymbol\mu - X$ has the same distribution as $X-\boldsymbol\mu$,
then $\func{L}(X)=\boldsymbol\mu$. In other words, if $\mathcal{D}_d$
is a class consisting solely of symmetrical probability distributions,
then all measures of location coincide.

Apart from the expected value, $\mathbb{E}\,X$, also, e.g., the population version
of the Oja median is an example of a location measure.

\subsection{Measures of dispersion}\label{Sec:DispersionProb}

\index{measure of dispersion}%
Bickel and Lehmann in \cite{BickelLehmann1976:descrstat3}
considered measures of dispersion for a family
of symmetric univariate probability distributions $\mathcal{D}_1$.
We consider $X$ less dispersed than $Y$, denoted $X \preceq_\mathrm{d} Y$,
whenever $|X-\mu_X|\preceq_\mathrm{st} |Y-\mu_Y|$,
where $\mu_X$ and $\mu_Y$ denote the points of symmetry of $X$ and $Y$, respectively.
Then $\func{S}: \mathcal{D}_1\to[0,\infty]$ is called a dispersion (scatter)
measure, whenever:
\begin{enumerate}
   \item[(a)] for all $X,Y\in\mathcal{D}_1$, if $X\preceq_\mathrm{d} Y$,
   then $\func{S}(X)\le\func{S}(Y)$,
   \item[(b)] for all $s,t\in\mathbb{R}$, $X\in\mathcal{D}_1$,
   if $sX+t\in\mathcal{D}_1$, then $\func{S}(sX+t) = |s|\func{S}(X)$.
\end{enumerate}
Equivalently, dispersion measures are $\preceq_\mathrm{d}$-monotone,
scale equivariant, and translation invariant.

The presented notion has been generalized by Oja in \cite{Oja1983:descrstatmultivar}.
He defined $\func{S}: \mathcal{D}_d\to[0,\infty]$ to be a scatter measure,
if it fulfills generalized $\preceq_\mathrm{d}$-monotonicity
(the concept is based on areas of appropriate multidimensional simplices),
as well as such that $\func{S}(\mathbf{A}X+\mathbf{t})=|\mathrm{det}(\vect{A})|\func{S}(X)$.
In particular, this class includes the following measures for any $p>0$:
\begin{itemize}
   \item generalizations of the unidimensional standard deviation:
   \[ \mathrm{Sd}\,X = \sqrt{\mathrm{Var}\,X} = \sqrt{\mathbb{E}\,X^2-(\mathbb{E}\,X)^2} = \sqrt{(\mathbb{E}(\mathbb{E}\,X - X)^2)},\]
   such as:
   \[ \func{S}(X) = \sqrt[p]{ \mathbb{E}\big( \mathrm{vol}(\mathrm{CH}(\mathbb{E}X, X_1,\dots,X_d))\big)^p}, \]
   \item generalizations of the Gini mean difference such as:
   \[ \func{S}(X) = \sqrt[p]{ \mathbb{E}\big( \mathrm{vol}(\mathrm{CH}(X_1,\dots,X_{d+1}))\big)^p}, \]
\end{itemize}
where $f\in\mathcal{D}_d$ and $X, X_1,\dots,X_{d+1}$ i.i.d.~$f$.

Additionally, Bickel and Lehmann in \cite{BickelLehmann1979:descrstat4}
considered measures of spread for univariate but not necessarily symmetric
distributions.

\subsection{Point estimation}

\index{estimator}Suppose that $\func{S}$ is a statistic and that $F_\vartheta$ is a probability
distribution characterized by some parameter $\vartheta$.
Assume that $\vect{X}=(X_1,\dots,X_n)$ is a sequence of random variables following $F_\vartheta$
(most often they are considered to be independent).
The aim of point estimation, see, e.g., \cite{Shao2007:mathematicalstat,LehmannCasella1998:pointest},
is to determine whether $\func{S}(X_1,\dots,X_n)$ may be used somehow to \textit{guess}
-- in an educated manner -- the value of $\vartheta$.
As $\vartheta$ is a fixed value and $\func{S}(X_1,\dots,X_n)$ is a random variable,
there are many possible ways to relate these two objects.
In particular, we may be interested in measuring an estimator's:
\begin{itemize}
   \item \index{bias}\emph{Bias} or expected systematic error, i.e.:
   \begin{equation}
      \func{Bias}_\vartheta(\func{S}) = \mathbb{E}\left( \func{S}(\vect{X})-\vartheta \right).
   \end{equation}
   Note that if $\func{Bias}_\vartheta(\func{S})=0$, we call $\func{S}$ an \index{unbiased estimator}\emph{unbiased estimator}
   of $\vartheta$.
   \item \index{mean squared error}\emph{Mean squared error}, i.e.:
   \begin{equation}
      \func{MSE}_\vartheta(\func{S}) = \mathbb{E}\left( \func{S}(\vect{X})-\vartheta \right)^2.
   \end{equation}
   It is well-known that $\func{MSE}_\vartheta(\func{S})=\mathrm{Var}\, \func{S}(\vect{X}) + (\func{Bias}_\vartheta(\func{S}))^2$.
   \item \index{efficiency}\emph{Efficiency}, which is equal to $1$
   whenever $\func{S}$ is unbiased
   and has the smallest possible mean squared error (and hence variance)
   among all unbiased estimators of $\vartheta$.
\end{itemize}
For instance, it may be shown that if $(X_1,\dots,X_n)$ is a sample
of i.i.d.~random variables with expectation of $\mu$ and variance of $\sigma^2$,
then the sample variance given by:
\begin{equation}
   \func{Var}(\vect{X}) = \frac{1}{n-1} \sum_{i=1}^n (X_i-\func{AMean}(\vect{X}))^2
\end{equation}
is an unbiased estimator of $\sigma^2$.
Moreover, the arithmetic mean is an unbiased estimator of $\mu$.
If, additionally, the random variables are normally distributed, then $\func{AMean}$
is of efficiency $1$. In such a case, the sample median is an unbiased estimator
of $\mu$ too, but yet not as efficient.

Additionally, asymptotic properties (for arbitrarily large $n$) may also be studied.
If $(X_1,X_2,\dots)$ is a sequence of $F_\vartheta$-distributed random variables, these include:
\begin{itemize}
   \item asymptotic unbiasedness,
   \item asymptotic efficiency,
   \item asymptotic normality,
   \item \index{consistent estimator}consistency,
   which holds if $\lim_{n\to\infty} \Pr_\vartheta( |\func{S}(X_1,\dots,X_n) - \vartheta| < \varepsilon ) = 1$
   for all $\varepsilon>0$,
\end{itemize}
and so on.
For instance, the sample standard deviation, $\func{SD}$, is only an asymptotically
unbiased estimator of the population standard deviation, $\sigma$.

From the described perspective, it is not unusual to take a unidimensional fusion
function $\func{F}$, treat it as a statistic, and answer questions such as:
\begin{itemize}
   \item What does $\func{F}$ estimate?
   \item How well does it perform in doing so?
\end{itemize}
in particular probability models.
Such an approach may provide a new insight into the existing fusion functions
(compare Section~\ref{Sec:RandomVariables} too).

\section{Spread measures}\label{Sec:SpreadMeasures}

Many introductory textbooks on applied statistics and academic lectures on
the subject include a review of the so-called descriptive statistics,
i.e., methods for summarizing quantitative unidimensional data
for performing exploratory data analysis.
Most often such methods are divided into at least two classes
(see \cite[Chapter~1]{Aczel1996:businessstat} and, e.g., \cite{Cramer1946:mathmethstat}):
\begin{enumerate}
    \item[1.] \emph{Measures of central tendency}
    (also known as measures of location
    or centrality of observations);
    e.g., sample quantiles (including median, min, and max),
    arithmetic mean, mode,
    trimmed and Winsorized mean etc.
    \item[2.] \emph{Measures of variability} (or data spread),
    e.g., range, interquartile range, variance, standard deviation.
\end{enumerate}
As we noted in the first chapter,
aggregation theory classically focuses on (among others) the broadly-conceived means.
However, we often need a very different kind of a proper
synthesis of multidimensional numeric data into a single number -- the one that
falls into the second category above.

It turns out that popular measures of data variability may be divided further
into the following subclasses:
\begin{enumerate}
   \item[2a.] \textit{Measures of absolute data spread},
   e.g., standard deviation, interquartile range, median absolute deviation.
   In this case, an absolute spread measure $\func{V}$ may
   accompany an aggregation function $\func{A}$ in order to state
   that a numeric list $\vect{x}$ is concisely described
   as $\func{A}(\vect{x})\pm\func{V}(\vect{x})$.

   \item[2b.] \textit{Measures of relative data spread}
   (e.g., Gini coefficient, coefficient of variation),
   which are dependent on the order of magnitude of a numeric list's elements.
   For instance, imagine that we have two groups of people.
   The first group consists of $(1, 2, 3)$-year-olds
   and the second one of $(101, 102, 103)$-year-olds.
   Intuitively, the relative spread of age in the first group
   is greater than that of the second group.
\end{enumerate}

In this section we would like to focus on measures of absolute data spread
from the perspective of aggregation theory. For that, we shall properly axiomatize
this class so that we establish exactly our universe of discourse.

\begin{remark}
Pitman in  \cite{Pitman1939:estlocscale} claimed that the function $\func{C}$,
used to estimate the scale parameter $c$ in his simple translate-scale model
(see Remark~\ref{Remark:MeanPitman}), should satisfy the conditions:
\begin{enumerate}
   \item[(c1)] $\func{C}(x_1,\dots,x_n)\ge 0$, \hfill(nonnegativity)
   \item[(c2)] $\func{C}\left( \frac{x_1+\lambda}{\mu}, \dots,
   \frac{x_n+\lambda}{\mu}\right) = \frac{\func{C}(x_1,\dots,x_n)}{\mu}$,
   for all $\lambda\in[-\infty, \infty]$ and $\mu>0$.\par
   \hfill(scale equivariance and translation invariance)
\end{enumerate}
Unfortunately, his setting does not serve our purposes:
it is too weak. Let:
\[
\func{S}(\vect{x})=\left\{
\begin{array}{ll}
0 & \text{for }\vect{x}=(n\ast c)\text{ for some }c,\\
\dfrac{\sum_{i=1}^n (x_i - \vect{\bar{x}})^2}{\sum_{i=1}^n (x_i - x_{(1)})} & \text{otherwise}.
\end{array}
\right.
\]
It is easily seen that $\func{S}$ is nonnegative,
translation invariant,
and $\func{S}(s\vect{x})=s\func{S}(\vect{x})$ for all $s\ge 1$.
However, it holds that $\func{S}(0,4,100)\simeq 61.64 > \func{S}(0,10,107)\simeq 59.71$,
which is counter-intuitive.
Moreover, it may be easily seen that
some ``classical'' spread measures, e.g., the sample
variance, do not fulfill (c2).
\end{remark}

\subsection{Measures of absolute spread for unidimensional data}

Here we shall rather rely on the axiomatization of measures of absolute data spread
which was proposed by Gagolewski in \cite{Gagolewski2015:spread}.

\begin{definition}\index{relation preccurlyeq@relation $\preccurlyeq_n$}%
For some $\Ival=[a,b]$, given $\vect{x},\vect{x}'\in\IvalPow{n}$,
we write $\vect{x}\preccurlyeq_n\vect{x}'$
and say that $\vect{x}$ \emph{has not greater absolute spread than} $\vect{x}'$,
if and only if for all $i,j\in[n]$ it holds:
\begin{equation}\label{Eq:spreaddef_orig}
(x_i-x_j)(x_i'-x_j')\ge 0 \text{ and } |x_i-x_j|\le |x_i'-x_j'|.
\end{equation}
\end{definition}

Please note that $\preccurlyeq_n$ is a preorder on $\IvalPow{n}$,
that is, a relation that is reflexive and transitive.
What is more, it is not necessarily total, i.e., not all vectors
are comparable with each other.

Additionally, whether $\preccurlyeq_n$ holds for given $\vect{x},\vect{x}'$
depends on how the elements in both vectors are jointly ordered.
The left side of \eqref{Eq:spreaddef_orig} implies that
if $\vect{x}\preccurlyeq_n\vect{x}'$, then $\vect{x},\vect{x}'$ are {comonotonic}.

\begin{figure}[t!]
\centering
\includegraphics{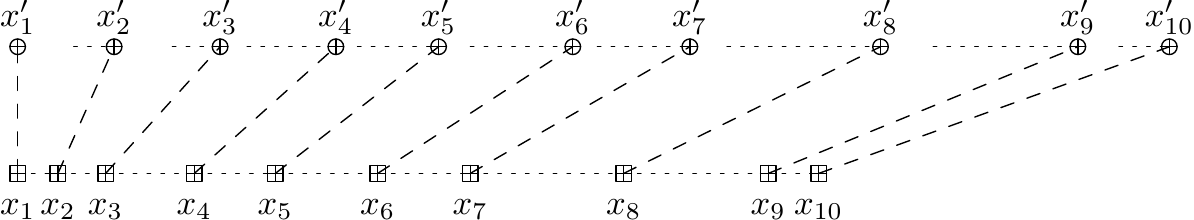}

\caption[Two exemplary numeric lists with
different spreads.]{\label{Fig:SpreadExample} Two exemplary numeric lists with
different spreads: $\vect{x}\preccurlyeq_n\vect{x}'$.}
\end{figure}

Figure~\ref{Fig:SpreadExample} illustrates two vectors:
$\vect{x}$ and its modified version $\vect{x}'$ with increased distances between
consecutive elements.

\begin{remark}
Let us study how $\preccurlyeq_n$ behaves
under scaling and translation of elements in a given vector.

   It is easily seen that for all $s\ge 1$
   and $\vect{x}\in\IvalPow{n}$ such that $s\vect{x}\in\IvalPow{n}$
   we have    $\vect{x}\preccurlyeq_n s\vect{x}$.
   Additionally, for all $t\in\mathbb{R}$
   for which $t+\vect{x}\in\IvalPow{n}$
   it holds $\vect{x}\preccurlyeq_n t+\vect{x}$
   and, at the same time, $t+\vect{x}\preccurlyeq_n\vect{x}$.
   Thus, $\preccurlyeq_n$ is not antisymmetric.

What is more,  for all $c\in\Ival$,
$(n\ast c)$ is a minimal element of $(\IvalPow{n}, \preccurlyeq_n)$,
i.e., for any $\vect{x}$ we have $(n\ast c)\preccurlyeq_n\vect{x}$.
This relation is also convex:
for all $\vect{x},\vect{x}', \alpha\in[0,1]$
it holds $\vect{x}\preccurlyeq_n \alpha\vect{x}+(1-\alpha)\vect{x}'\preccurlyeq_n\vect{x}'$
whenever $\vect{x}\preccurlyeq_n\vect{x}'$.
\end{remark}

Let us proceed with the definition of objects in which
we have a special interest in this section.

\begin{definition}[\cite{Gagolewski2015:spread}]\label{Def:SpreadMeasure}
A \index{spread measure}\emph{spread measure} is a mapping
$\func{V}:\IvalPow{n}\to[0,\infty]$ such that:
\begin{enumerate}
 \item[(v1)] for each $\vect{x}\preccurlyeq_n\vect{x}'$
 it holds $\func{V}(\vect{x})\le\func{V}(\vect{x}')$,
 \item[(v2)] for any $c\in\Ival$
 it holds $\func{V}(n\ast c)=0$.
\end{enumerate}
\end{definition}

Note that the first characteristic property implies that each spread measure
is {translation invariant}. Moreover, for all $s\ge 1$ and $\vect{x}\in\IvalPow{n}$ such that $s\vect{x}\in\IvalPow{n}$
it holds $\func{V}(\vect{x})\le \func{V}(s\vect{x})$.

In \cite{Gagolewski2015:spread} it has been shown that this class includes,
among others, the following spread measures:
\begin{itemize}
   \item $\func{Var}(\vect{x})=\frac{1}{n-1} \sum_{i=1}^n \left( x_i - \func{AMean}(\vect{x}) \right)^2$,
   \hfill\index{Var@$\mathsf{Var}$|see {variance}}\index{variance}(\emph{sample variance})
   \item $\func{SD}(\vect{x})=\sqrt{\func{Var}(\vect{x})}$,
   \hfill\index{standard deviation}(\emph{standard deviation})
   \item $\func{Range}(\vect{x})=\func{Max}(\vect{x})-\func{Min}(\vect{x})$,
   \hfill\index{range}(\emph{range})
   \item $\func{IQR}(\vect{x})=\func{Q}_{0.75}(\vect{x})-\func{Q}_{0.25}(\vect{x})$,
   \hfill\index{IQR@$\mathsf{IQR}$|see {interquartile range}}\index{interquartile range}(\emph{interquartile range})
   \item $\func{MAD}(\vect{x})=1.4826\func{Median}(|\vect{x}-\func{Median}(\vect{x})|)$,\par
   \hfill\index{median absolute deviation}(\emph{median absolute deviation})
   \item $\func{ME}(\vect{x})=\frac{1}{n} \sqrt{\frac{\pi}{2}} \sum_{i=1}^n |x_i - \func{AMean}(\vect{x})|$,
   \hfill(\emph{Fisher's mean error})
\end{itemize}
that is functions widely used in exploratory data analysis (all of them are symmetric).
Note that the sample variance, standard deviation, mean error, and range are 3-incremental fusion functions.

\begin{proposition}
Let $\func{V}$ be a spread measure such that $\sup_{\vect{x}\in\IvalPow{n}} \func{V}(\vect{x}) = u$.
Then for each nondecreasing function $\varphi:[0,u]\to[0,(b-a)]$
such that $\varphi(0)=0$, $\varphi\circ\func{V}$ is a spread measure too.
\end{proposition}

Taking the above into account, the following further classes of fusion functions
(together with their monotone transforms) may be distinguished:
\begin{itemize}
   \item $\func{V}(\vect{x})=\sum_{i=1}^n \sum_{k=1}^n |x_i-x_k|^p$ for some $p\ge 1$,
   in particular, the sample variance:
   \begin{equation}
   \func{Var}(\vect{x})=\frac{1}{2n(n-1)} \sum_{i=1}^n \sum_{k=1}^n \left(x_i-x_k\right)^2
   \end{equation}
   and the \index{MD@$\mathsf{MD}$|see {mean difference}}\index{mean difference}\emph{Gini mean difference}:
   \begin{equation}
   \func{MD}(\vect{x})=\frac{1}{n\,(n-1)} \sum_{i=1}^n \sum_{k=1}^n |x_i-x_k|,
   \end{equation}
   \item $\func{V}(\vect{x})={\func{A}}\left( \left|x_{1}-\func{Q}_\alpha(\vect{x})\right|,\dots,
   \left|x_{n}-\func{Q}_\alpha(\vect{x})\right|\right)$ for some $n$-ary classical aggregation function $\func{A}$
   and $\alpha\in[0,1]$, e.g., $\func{MAD}$, $\func{IQR}$, and $\func{Range}$,
   \item \index{WDpWAM@$\mathsf{WD}_p\mathsf{WAM}$ operators}\index{WDpWAM@$\mathsf{WD}_p\mathsf{OWA}$ operators}%
   WD${}_2$WAM spread measures of the form:
   \begin{equation}
   \func{V}(\vect{x})=\sum_{i=1}^n w_i\left(x_i-\sum_{j=1}^n w_j x_j\right)^2
   \end{equation}
   for some weighting vector $\vect{w}$
   as well as their symmetrized counterparts (WD${}_2$OWA operators),
   e.g., the sample variance,
   \item WD${}_1$WAM spread measures $\func{V}(\vect{x})=\sum_{i=1}^n w_i\left|x_i-\sum_{j=1}^n w_i x_j\right|$
   and the corresponding WD${}_1$OWA operators,
   e.g., the Fisher mean error,
   \item WD${}_\infty$WAM operators $\func{V}(\vect{x})=\max_{i=1,\dots,n} \left|x_i-\sum_{j=1}^n w_j x_j\right|$
   as well as the corresponding WD${}_\infty$OWA operators.
\end{itemize}
Note that in \cite{Gagolewski2015:normalizedspread} Gagolewski considered
normalized versions of some of the above spread measures classes
which can be used in decision making. They attain the greatest possible value
equal to $(b-a)$ (in particular, $1$ if $\Ival=[0,1]$). For example:
\begin{equation}
\func{NWD}_2\func{WAM}_\vect{w}(\vect{x}) =
\frac{\sqrt{\sum_{i=1}^n w_i \left(x_i-\sum_{j=1}^n w_jx_j\right)^2}}{(b-a)\sqrt{p(1-p)}},
\end{equation}
where $p=\max_{A\subseteq [n], \sum_{i\in A} w_i\le 0.5} \sum_{i\in A} w_i$,
and:
\begin{equation}
\func{NWD}_1\func{WAM}_\vect{w}(\vect{x}) = \frac{\sum_{i=1}^n w_i
\left|x_i-\sum_{j=1}^n w_jx_j\right|}{2p(1-p)(b-a)},
\end{equation}
where $p=\max_{A\subseteq [n], \sum_{i\in A} w_i\le 0.5} \sum_{i\in A} w_i$.

\bigskip
Let us proceed with an appealing characterization of measures of absolute spread.
For any given $\vect{x}\in\IvalPow{n}$, let
$\func{diff}(\vect{x})=(x_{(2)}-x_{(1)},\dots,x_{(n)}-x_{(n-1)})\in[0,(b-a)]^{n-1}$
denote the \index{iterated difference}\index{diff@$\mathsf{diff}$|see {iterated difference}}%
\emph{iterated difference} between consecutive ordered components
of a given vector.
Please note that such a function is available in some programming languages:
in particular, it may be computed by calling \texttt{diff(sort(x))} in \textsf{R}.
We see that if $\boldsymbol\delta=\func{diff}(\vect{x})$, then
$0\le \delta_i\le b-a$ and $\sum_{i=1}^{n-1} \delta_i\le b-a$.
Intuitively, if $\vect{x}$ is already ordered, then this operation
may be viewed as a kind of ``vector differentiation''.
On the other hand,
for $\tilde{\vect{x}}=\func{cumsum}(x_{(1)}, \boldsymbol\delta)
=(x_{(1)},x_{(1)}+\delta_1,x_{(1)}+\delta_1+\delta_2,\dots,x_{(1)}+\delta_1+\dots+\delta_n)$
denoting the  \index{iterated difference}\index{cumsum@$\mathsf{cumsum}$|see {cumulative sum}}\emph{cumulative sum}
of $\hat{\boldsymbol\delta}=(x_{(1)}, \boldsymbol\delta)$
we have $x_{(i)}=\tilde{x}_{i}$, $x_i=\tilde{x}_{\sigma^{-1}(i)}$, where
$\sigma$ is such that $\vect{x}\in\IvalPow{n}_\sigma$.
Thus, $\vect{x}$ may be reconstructed from $x_{(1)}, \boldsymbol\delta,$ and $\sigma$.

We are now in a position to provide an equivalent definition
of the relation defined by Equation~\eqref{Eq:spreaddef_orig}.

\begin{lemma}[\cite{Gagolewski2015:spread}]\label{Lemma:SpreadAlternative}
For any $\vect{x},\vect{x}'\in\IvalPow{n}$
it holds $\vect{x}\preccurlyeq_n\vect{x}'$
if and only if $\vect{x},\vect{x}'$ are comonotonic
and $\func{diff}(\vect{x})\le_{n-1}\func{diff}(\vect{x}')$.
\end{lemma}

Therefore, we have what follows.

\begin{theorem}[\cite{Gagolewski2015:spread}]\label{Thm:diff}
$\func{V}:\Ival^n\to[0,\infty]$ is a spread measure if and only if
the following conditions are valid:
\begin{enumerate}
 \item[(v1')] for each comonotonic $\vect{x},\vect{x}'$
 such that $\func{diff}(\vect{x})\le_{n-1}\func{diff}(\vect{x}')$
 we have $\func{V}(\vect{x})\le\func{V}(\vect{x}')$,
 \item[(v2')] $\inf_{\vect{x}\in\IvalPow{n}} \func{V}(\vect{x})=0$.
\end{enumerate}
\end{theorem}

\begin{corollary}
For any $\func{V}:\Ival^n\to[0,\infty]$,
$\func{V}|_{\sigma}$ fulfills (v1) and (v2)
if and only if there exists $\tilde{\func{A}}:[0,b-a]^{n-1}\to[0,\infty]$
such that $\func{V}|_\sigma(\vect{x})=\tilde{\func{A}}(\func{diff}(\vect{x}))$
is nondecreasing and lower endpoint-preserving.
\end{corollary}

We see that symmetric absolute spread measures are nothing more than aggregation
functions computed on iterated differences of an input vector.

\subsection{Measures of relative spread}

As indicated in \cite{Gagolewski2015:spread},
some ``normalized'' measures of \textit{relative} spread may also be considered.
At the most general level, these are functions of the form:
\begin{equation}
\func{S}(\vect{x}) = \frac{\func{V}(\vect{x})}{\func{A}(\vect{x})},
\end{equation}
where $\func{V}$ is an absolute spread measure, and $\func{A}$ is an
aggregation function.

For instance, the well known (unit-free) \index{Gini coefficient}\emph{Gini coefficient}, defined as:
\begin{equation}
\func{Gini}(\vect{x})=\frac{\func{MD}(\vect{x})}{2\func{AMean}(\vect{x})}
\end{equation}
is definitely not a measure of absolute spread.
This is because it is not even translation invariant: we have
$\func{Gini}(0,2,4)=2/3$, and $\func{Gini}(2,4,6)=1/3$.
Moreover, even though $(0,2,4)\preccurlyeq_n(0,3,5)$,
we have $\func{G}(0,3,5)=5/8<2/3=\func{G}(0,2,4)$.
A similar observation may be made about the so-called \index{coefficient of variation}\emph{coefficient of variation}:
\begin{equation}
\func{CV}(\vect{x})=\frac{{\func{SD}(\vect{x})}}{\func{AMean}(\vect{x})}.
\end{equation}
Both functions take into account the order of magnitude of the observations,
and are ratio scale invariant (i.e., $\func{S}(s\vect{x})=\func{S}(\vect{x})$ for all $s>0$)
as well as continuous but not translation invariant.

\subsection{Spread measures for multidimensional data}\label{Sec:SpreadMultidimSample}

Similarly as in Chapter~\ref{Chap:multidim}, let us again assume that we are
given $\vect{X}\in(\mathbb{R}^d)^n$.
As noted, e.g., in \cite{LiuPareliusSingh1999:multivaranal}, there are two
ways to quantify dispersion of a multivariate data set:
as a matrix or as a scalar. The latter is of course much easier to construct
and fits the overall setting established in this chapter.
Nevertheless, let us at least mention that, e.g., the sample
\index{covariance matrix}\emph{covariance matrix}, given  by:
\begin{equation*}
   \func{Cov}(\vect{X}) = \frac{1}{n-1} \left(\vect{X}-\func{CwAMean}(\vect{X})\right) \left(\vect{X}-\func{CwAMean}(\vect{X})\right)^T
   \in\mathbb{R}^{d\times d},
\end{equation*}
can reveal other useful information on a dataset, such as the orientation of the
empirical probability mass distribution and the dispersion of individual
variates or covariates.

In a recent contribution, Kołacz and Grzegorzewski \cite{KolaczGrzegorzewskiXXXX:dispersionmultidim}
considered multidimensional spread measures defined as functions
$\func{V}:(\mathbb{R}^d)^n\to[0,\infty]$ that are:
\begin{itemize}
   \item symmetric,
   \item translation and rotation invariant,
   \item homogeneous, i.e., there exists a nondecreasing
   function $\varphi:[0,\infty[\to [0,\infty[$
   such that $\func{V}(s\vect{X}) = \varphi(s) \func{V}(\vect{X})$
   for all $s>0$ and  $\vect{X}\in(\mathbb{R}^d)^n$.
   \item such that $\func{V}(n\ast\vect{x}) = 0$ for all $\vect{x}\in\mathbb{R}^d$.
\end{itemize}
Similarly as the Pitman \cite{Pitman1939:estlocscale} axiomatization
of a scale parameter estimate, their setting seems to be too mild,
as it only concerns uniform scaling in each direction, translation, and rotation
transforms. It would be informative, if it referred to some ordering relation.
Nevertheless, this axiomatization is a good starting point for future
research on the topic -- to our best knowledge there are no alternatives
to this proposal in the literature yet. Interestingly, the authors explore
the relationships between spread measures for vectors of different arities
and functions that are generated via particular so-called multidistances
\cite{MartinMayor2011:multiargdist,MartinMayor2010:propmultidist}.

\begin{remark}\label{Remark:spreadrelgeneralization}
The Oja simplex volume-approach
(see \cite{Oja1983:descrstatmultivar} and Section~\ref{Sec:DispersionProb})
gives one possibility for generalizing the $\preccurlyeq_n$ relation
defined in \cite{Gagolewski2015:spread}.
Note that for $d=1$
the condition $|\vect{x}^{(i)}-\vect{x}^{(j)}|\le |\vect{x}'^{(i)}-\vect{x}'^{(j)}|$
presented in Equation~\eqref{Eq:spreaddef_orig}
may be written also as:
\[
   \mathrm{vol}\left(\mathrm{CH}(\vect{x}^{(i)},\vect{x}^{(j)})\right)
   \le
   \mathrm{vol}\left(\mathrm{CH}(\vect{x}'^{(i)},\vect{x}'^{(j)})\right),
\]
which now can be generalized for any $d$ as:
\[
   \mathrm{vol}\left(\mathrm{CH}(\vect{x}^{(i_1)},\dots,\vect{x}^{(i_{d+1})})\right)
   \le
   \mathrm{vol}\left(\mathrm{CH}(\vect{x}'^{(i_1)},\dots,\vect{x}'^{(i_{d+1})})\right),
\]
where $i_1, \dots,i_{d+1}\in[n]$.
Regardless of some problems with defining comonotonicity
(which can be quite easily bypassed),
a fusion function monotone with respect to the above partial ordering is automatically
translation and rotation invariant. Moreover, uniform scaling in each direction
shall never lead to a decrease in its output.
\end{remark}

\begin{figure}[htb!]
\centering
\begin{minipage}[t]{6cm}
\centering
\includegraphics[width=5.5cm]{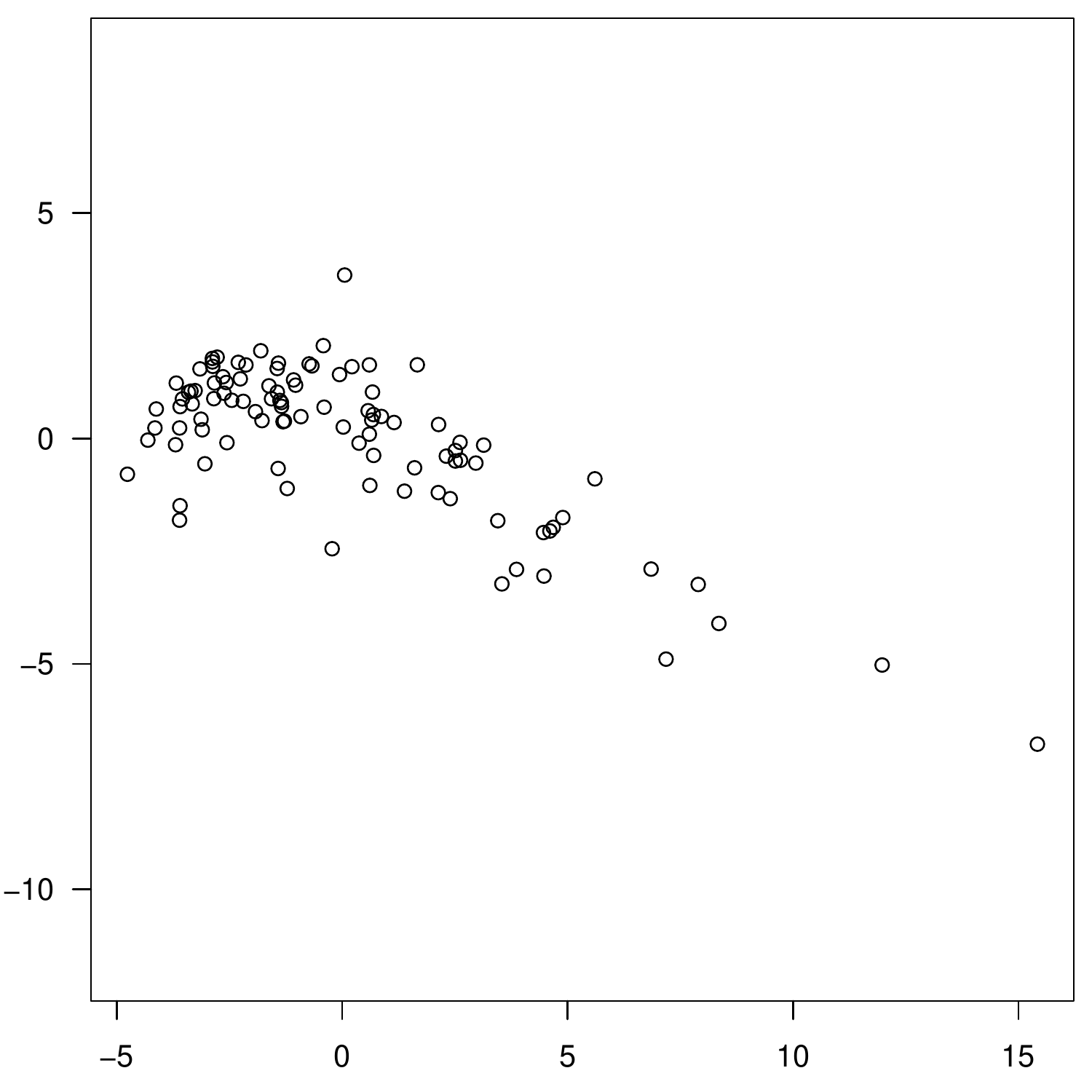}

\centering (a) A sample data set.
\end{minipage}
\begin{minipage}[t]{6cm}
\centering
\includegraphics[width=5.5cm]{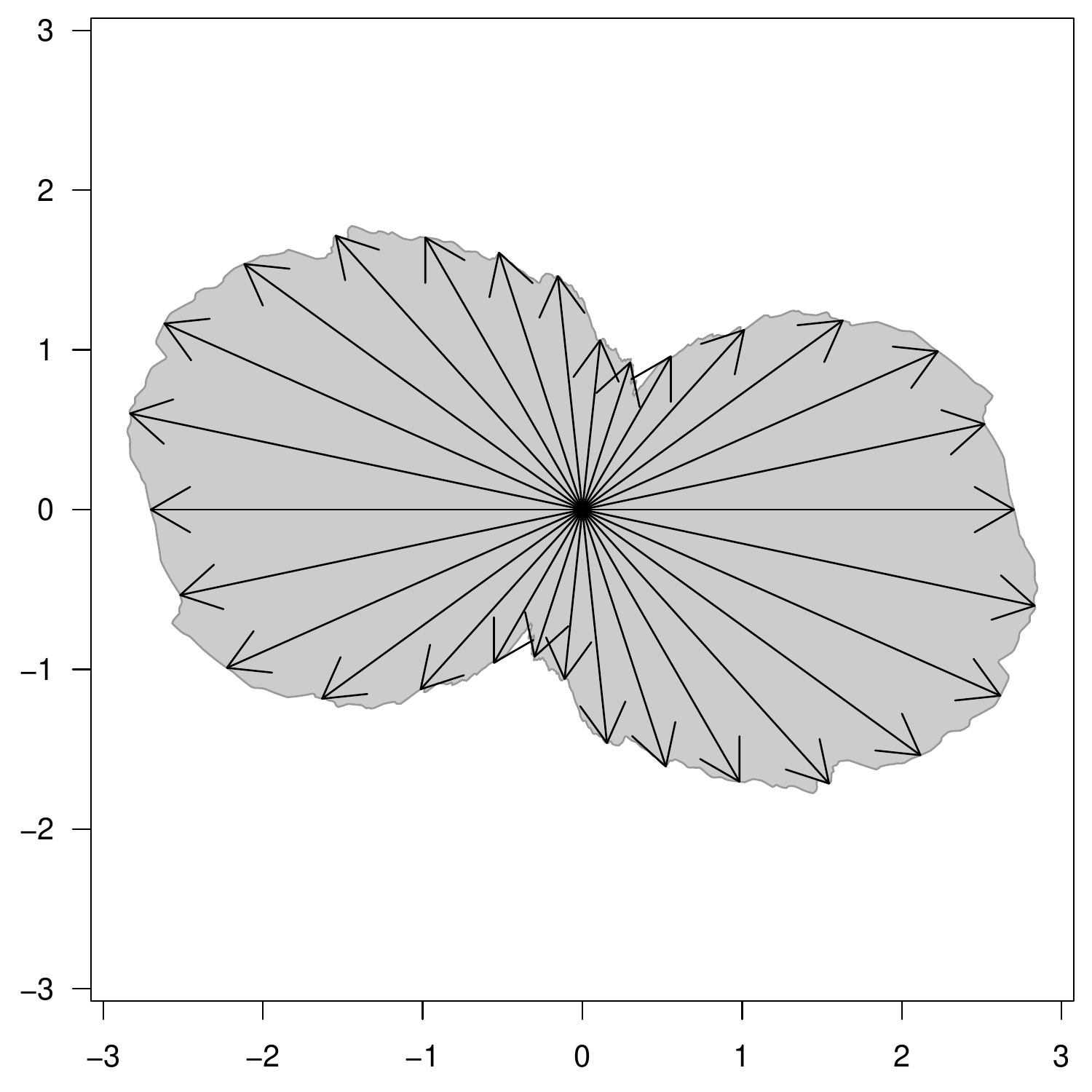}

\centering (b) Median absolute deviation.
\end{minipage}

\bigskip
\begin{minipage}[t]{6cm}
\centering
\includegraphics[width=5.5cm]{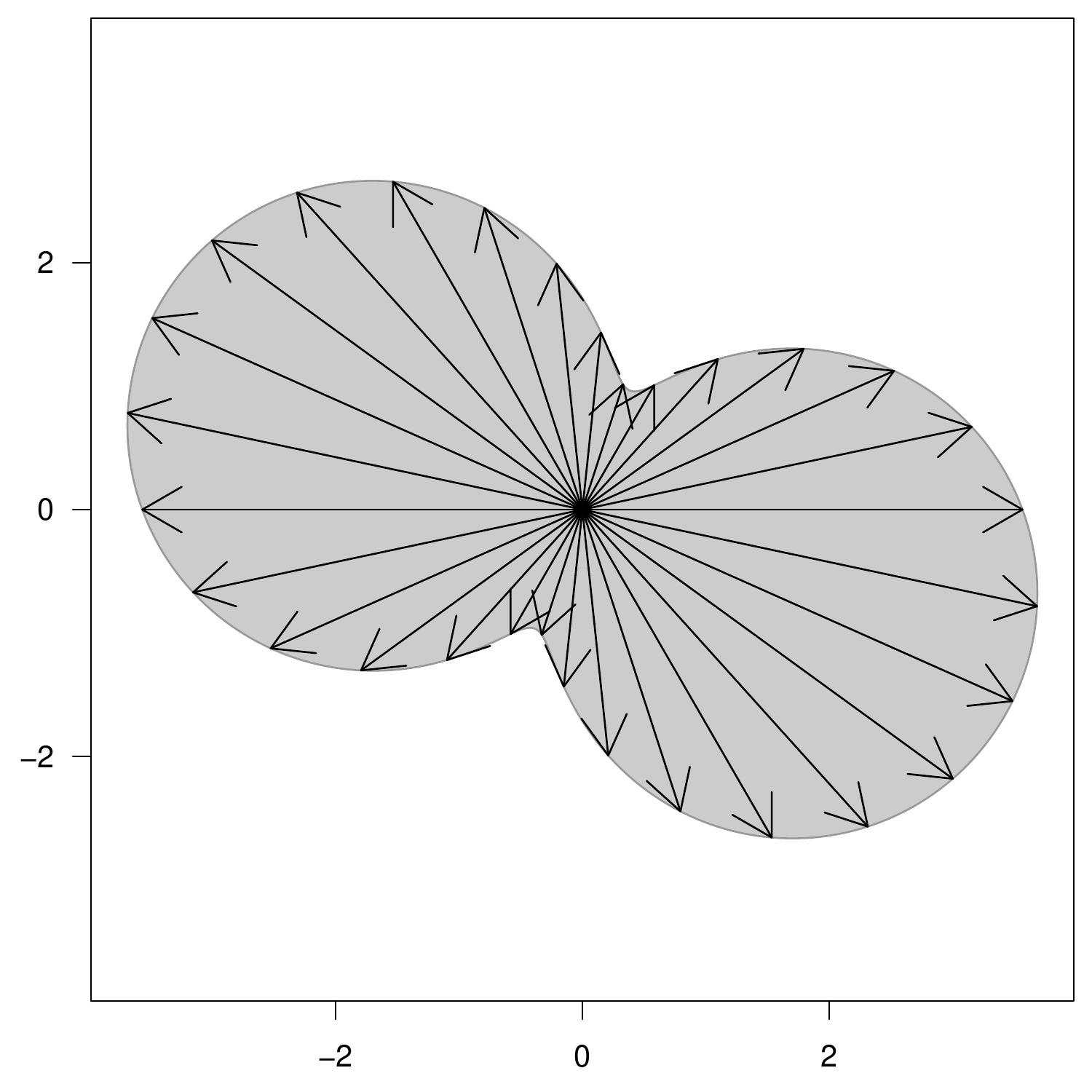}

\centering (c) Standard deviation.
\end{minipage}
\begin{minipage}[t]{6cm}
\centering
\includegraphics[width=5.5cm]{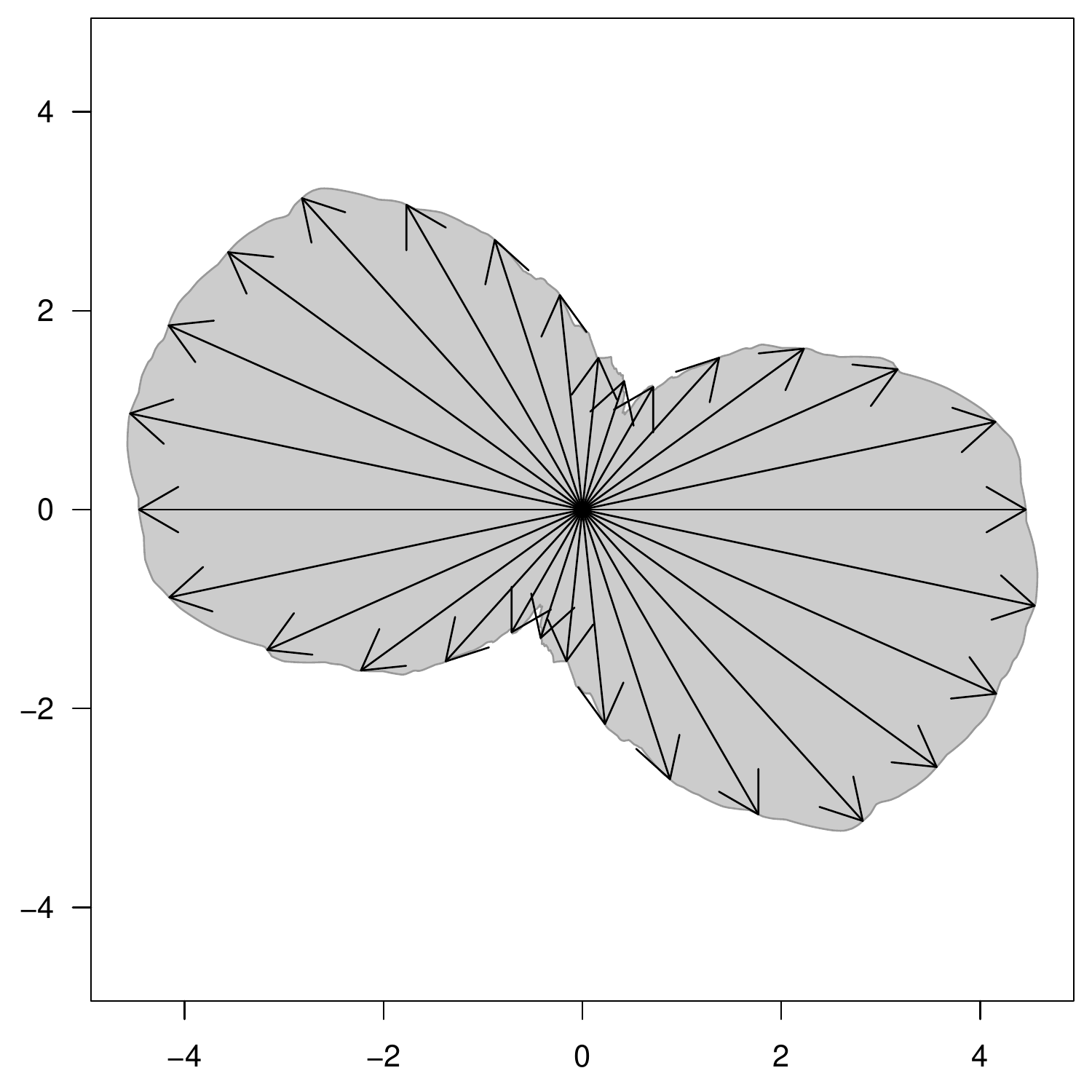}

\centering (d) Interquartile range.
\end{minipage}

\caption[An exemplary 2D data set together with one dimensional dispersion measures computed for its projections in every direction.]%
{\label{Fig:dispersion_pursuit} An exemplary 2D data set together
with one dimensional dispersion measures computed for its projections in every direction
(represented as arrow lengths).
Note that $\func{MAD}$ and $\func{IQR}$ give nonsmooth shapes.}
\end{figure}

Here are a few particular classes and/or general construction methods
for dispersion measures.
\begin{itemize}
   \item Unidimensional spread measures may be generalized to any $d$
   via \index{projection pursuit}projection pursuit, see \cite{Huber1985:projectionpursuit}.
   This is because we may apply all possible one-dimensional projections
   of the data set and compute the univariate~$\func{V}$.
   For instance:
   \[
      \func{V}'(\vect{X}) = \sup_{\|\vect{u}\|=1} \func{V}(\vect{u}^T\vect{X}),
   \]
   which gives the maximal possible directional variance (compare the Principal
   Component Analysis method), or:
   \[
      \func{V}''(\vect{X}) = \int_{\|\vect{u}\|=1} \func{V}(\vect{u}^T\vect{X})\,d\vect{u},
   \]
   which gives the averaged dispersion, see Figure~\ref{Fig:dispersion_pursuit}
   for a graphical illustration.

\item Given a semimetric $\mathfrak{d}$ on $\mathbb{R}^d$
and a fusion function $\func{F}: [0,\infty]^{n(n-1)/2}\to[0,\infty]$,
compute:
\[
   \func{V}'(\vect{X}) = \func{F}\left(
   \mathfrak{d}(\vect{x}^{(1)}, \vect{x}^{(2)}), \mathfrak{d}(\vect{x}^{(1)}, \vect{x}^{(3)}),
   \dots, \mathfrak{d}(\vect{x}^{(n-1)}, \vect{x}^{(n)})
   \right),
\]
in particular $\func{F}$ can be the arithmetic mean.

\item Given a semimetric $\mathfrak{d}$ on $\mathbb{R}^d$,
a fusion function $\func{A}:(\mathbb{R}^d)^n\to\mathbb{R}^d$,
and a fusion function $\func{F}: [0,\infty]^{n}\to[0,\infty]$,
compute:
\[
   \func{V}'(\vect{X}) = \func{F}\left(
   \mathfrak{d}(\vect{x}^{(1)}, \func{A}(\vect{X})), \dots, \mathfrak{d}(\vect{x}^{(n-1)}, \func{A}(\vect{X}))
   \right),
\]
in particular $\func{F}$ can be the quadratic mean or the $\func{Max}$ function
and $\func{A}$ -- the componentwise arithmetic mean (centroid).
\end{itemize}
Moreover, Liu, Parelius, and Singh in \cite{LiuPareliusSingh1999:multivaranal}
consider a few data depth-based dispersion measures.

\begin{remark}
Various methods for measuring dispersion of directional data
exist as well. Given a circular data sample $\vartheta_1,\dots,\vartheta_n\in[-\pi,\pi[$, for instance:
\[
\func{V}(\vartheta_1,\dots,\vartheta_n) =
\sqrt{\func{AMean}(\sin\vartheta_1,\dots,\sin\vartheta_n)^2+\func{AMean}(\cos\vartheta_1,\dots,\cos\vartheta_n)^2}
\]
is quite often used in practice. The interested reader is referred to \cite{Otieno2002:phd}
for further references.
\end{remark}

\section{Consensus, inequality, and other measures}

Somehow related to spread measures are numerical characteristics
that originate from decision making, ecology, and economics.

\paragraph{Measures of consensus and ecological evenness.}
Recently, Beliakov, Calvo, and James in \cite{BeliakovCalvoJames2014:consensusmeasures}
studied measures of decision makers' \index{consensus measure}\emph{consensus} that are based on Bonferroni means
and fuzzy implications. They postulate that these should be functions like
$\func{C}:[0,1]^n\to[0,1]$ which fulfill at least the following properties:
\begin{itemize}
   \item symmetry (unanimity),
   \item for all $x\in[0,1]$ it holds $\func{C}(n\ast x)=1$ (maximal consensus),
   \item $\func{C}(\lfloor n/2\rfloor\ast 0, \lceil n/2\rceil\ast 1)=0$
   and $\func{C}(\lfloor n/2\rfloor\ast 1, \lceil n/2\rceil\ast 0)=0$
   (minimal consensus),
   \item monotonicity with respect to the majority,
   i.e., for each $c\in[0,1]$ and $\vect{x},\vect{y}\in[0,1]^{\lfloor n/2\rfloor}$,
   if $|c-\vect{x}|\le_{\lfloor n/2\rfloor}|c-\vect{y}|$, then
   $\func{C}(\lceil n/2\rceil\ast c, \vect{x})\ge \func{C}(\lceil n/2\rceil\ast c, \vect{y})$.
\end{itemize}

\paragraph{Indices of social inequality and poverty.}
Economists find their interests in measures of social \index{inequality}\emph{inequality} (unevenness, poverty, etc.)
For instance, Marques Pereira and others \cite{GarciaLaprestaETAL2015:fuzzypoverty,BortotMarquesPereira2015:povertyemean,AristondoETAL2013:inequality}
(see, e.g., \cite{KobusMilos2012:inequalitydecomp,Kobus2012:inequalitydecomp} for a different setting)
study poverty measures for nonnegative vectors defined as functions -- among others --
monotone with respect to the Lorenz majorization relation $\sqsubseteq_\mathrm{L}$,
which is defined as $\vect{x}\sqsubseteq_\mathrm{L}\vect{y}$
if and only if $\func{AMean}(\vect{x})=\func{AMean}(\vect{y})$
and
$\func{cumsum}(x_{(n)},\dots,x_{(1)}) \le_n \func{cumsum}(y_{(n)},\dots,y_{(1)})$.
In particular, it is easily seen that $(n\ast\func{AMean}(\vect{x}))\sqsubseteq_\mathrm{L}\vect{x}$
for all $\vect{x}$.
On a side note, recall that absolute spread measures
are given via a $\func{diff}$-based relation $\preccurlyeq_n$
and that $\func{cumsum}$ can be conceived as a dual operation to $\func{diff}$.

\begin{remark}
Monotonicity with respect to $\sqsubseteq_\mathrm{L}$ is also called
\index{Schur-convexity}\emph{Schur-convexity}
in the literature. Interestingly, if $\vect{w},\vect{v}$ are weighting vectors of the same lengths,
then, see, e.g., \cite{BortotPereira2014:gini}:
\begin{itemize}
   \item for all $\vect{x}\in\IvalPow{n}$,
   $\func{OWA}_\vect{w}(\vect{x})\le\func{OWA}_\vect{v}(\vect{x})$
   if and only if $\func{cumsum}(\vect{w}) \ge_n \func{cumsum}(\vect{v})$,
   \item $\func{OWA}_\vect{w}$ is Schur-convex if and only if
   $\vect{w}$ is ordered nondecreasingly.
\end{itemize}
Moreover, an exponential mean $\mathsf{EMean}_\gamma$ is Schur-convex,
whenever $\gamma\ge 0$, see \cite{BortotMarquesPereira2015:povertyemean}.
\end{remark}

It turns out, see \cite{BeliakovJames2015:unifyconsensus,BeliakovJamesNimmo2014:ecologicalconsensus},
that social inequality measures can be related to ecological indices of
\index{ecological evenness indices}\emph{evenness} \cite{Pielou1969:introecology},
which aim to capture how evenly species' populations are distributed over a geographical
region, compare \cite{Camargo1993:dominance,Heip1974:evenness,Pielou1975:ecodiversity}.

\paragraph{Entropy of discrete probability mass functions.}
A noteworthy characterization of measures of entropy or uncertainty
of discrete probability mass functions (represented
as numeric vectors in $[0,1]^n$ with elements summing up to 1) was proposed
by Mart\'{i}n, Mayor, and Su{\~n}er in \cite{MartinMayorSuner2001:dispersionmeasures},
compare also \cite{SimoviciJaroszewicz2002:axiomentropy,Renyi1959:entropy}
for axiomatizations on different kinds of domains.
Such a class includes the Shannon \index{entropy}entropy,
$\func{Entropy}(\vect{w}) = -\sum_{i=1}^n w_i \log w_i$,
and alike, see also~\cite{KostalLanskyPokora2013:statisticaldispersionShannon}.
Here, monotonicity with respect to a partial order $\sqsubseteq_\mathrm{D}$
such that $\vect{w}\sqsubseteq_\mathrm{D}\vect{v}$ if and only if
for all $i\in[n]$ $w_i\le v_i\le 1/n$ or $w_i\ge v_i\ge 1/n$
is considered useful.

\paragraph{Measures of shape of empirical distributions.}
We strongly believe that this short overview would have left the reader with
a feeling of dissatisfaction if the two following measures of an input
data vector's empirical distribution shape had not been considered.

A measure of \index{skewness}\emph{skewness} quantifies the degree of non-symmetry of an empirical distribution.
A negative or positive skew is observed if the mass of the distribution
is concentrated on the right or, respectively, left of the corresponding data histogram.
In this case, we may consider, e.g.:
\begin{equation}
   \mathrm{skewness}(\vect{x})=\frac{\tfrac{1}{n} \sum_{i=1}^n (x_i-\func{AMean}(\vect{x}))^3}%
   {\left(\tfrac{1}{n-1} \sum_{i=1}^n (x_i-\func{AMean}(\vect{x}))^2\right)^{3/2}}.
\end{equation}
Notably, Liu, Parelius, and Singh in \cite{LiuPareliusSingh1999:multivaranal}, compare also
the work of Oja \cite{Oja1983:descrstatmultivar}, study different types of
symmetry of multidimensional data samples, such as spherical, elliptical,
antipodal, or angular ones.

On the other hand, a measure of \index{kurtosis}\emph{kurtosis}
(peakedness/flatness) shall be sensitive to the movement of the probability mass from the shoulders
of a distribution to its center or tails, e.g.:
\begin{equation}
   \mathrm{kurtosis}(\vect{x})=\frac{\tfrac{1}{n} \sum_{i=1}^n (x_i - \func{AMean}(\vect{x}))^4}%
   {\left(\tfrac{1}{n} \sum_{i=1}^n (x_i - \func{AMean}(\vect{x}))^2\right)^2} - 3.
\end{equation}
Please note that for samples following a normal distribution
skewness and kurtosis are -- on average -- equal to 0.

\section{Impact functions for informetric data}\label{Sec:ImpactFunctions}

Let us assume that $\Ival=[0,\infty]$ and that our universe of discourse
consists of informetric strings as in Section~\ref{Sec:aginformetricAny}.
This time, however, we would like to compute a numerical characteristic
of a given $\vect{x}\in\mathcal{S}=\{(x_1,x_2,\dots,x_d): d\in\mathbb{N}, (\forall i\in[d])\ x_i\in\Ival, x_1\ge x_2\ge\dots\ge x_d\}$
such that it reflects both:
\begin{itemize}
   \item the number of items (e.g., scientific articles, posts, software packages)
   produced by an abstract information resources producer
   (e.g., a~scientist, StackOverflow user, software engineer) and
   \item the quality of individual products.
\end{itemize}
In the informetric (in particular, scientometric) literature
it is widely accepted, see, e.g., \cite{Woeginger2008:axiomatich,
Woeginger2008:axiomaticg,
Woeginger2008:symmetryaxiom,Rousseau2008:woegingerax,Quesada2009:monotonicityh,
Quesada2010:moreaxiomatics,GagolewskiGrzegorzewski2011:ijar,FranceschiniMaisano2011:structevalh},
that such an \index{impact function}\emph{impact function}
$\func{F}:\mathcal{S}\to[0,\infty]$ to be applied in
the so-called Producers Assessment Problem (PAP) should at least be:
\begin{itemize}
   \item $\sqsubseteq_\gamma$-nondecreasing (compare Section~\ref{Sec:AlphaBetaOrdering}) and
   \item such that $\func{F}(0)=0$.
\end{itemize}
Note that $\sqsubseteq_\gamma$-nondecreasingness implies both
monotonicity with respect to each component as well as the vector's size (arity),
see \cite{GagolewskiGrzegorzewski2011:ijar} for a proof.

\begin{remark}
   Note that, originally, many proposals for bibliometric
indices assumed that we aggregate the number of papers' citations,
i.e., sequences with elements in $\Naturals_0$.
Generally, however, the paper quality measures may be arbitrary
real numbers, for example when citations are normalized
according to the number of coauthors,
paper's time of publication, quality of a journal,
and so forth, see, e.g., \cite{GagolewskiMesiar2012:joi}.
\end{remark}

Some of the notable examples of impact functions are as follows:
\begin{itemize}
\item Total number of product qualities:
\begin{equation}\label{Eq:IndexSum}
\func{Sum}(x_1,\dots,x_n)=\sum_{i=1}^{n} x_{i},
\end{equation}
or, more generally, a weighted sum of elements of $\vect{x}\in\mathcal{S}$.
This includes, e.g., ``the total number of citations of the five most cited
papers''.

\item The Hirsch $h$-index \index{h-index}\cite{Hirsch2005:hindex}:
\begin{equation}%
   \func{H}(x_1,\dots,x_n) = \max\left\{h\in[0:n]: {x}_{h} \ge h\right\}, %
\end{equation}
with convention $x_0=x_1$.

\item The Kosmulski MaxProd-index \index{MaxProd-index}\cite{Kosmulski2007:maxprod}:
\begin{equation}%
   \func{MP}(x_1,\dots,x_n) = \max\left\{i\cdot{{x}}_{i}: i\in[n] \right\}.
\end{equation}
This index is a particular case of the (projected) $l_p$-indices, $p\ge 1$,
see \cite{GagolewskiGrzegorzewski2009:geometricapproach}.

\item The Egghe $g$-index \index{g-index}\cite{Egghe2006:g}:
\begin{equation}%
   \func{G}(x_1,\dots,x_n) = \max\left\{g\in[0:n]: \sum_{i=1}^g {{x}}_{i} \ge g^2\right\},
\end{equation}
with convention $\sum_{i=1}^0 \cdots = 0$ and $x_{n+1}=x_{n+2}=\dots=0$.

\item The Woeginger $w$-index \index{w-index}\cite{Woeginger2008:axiomatich}:
\begin{equation}%
\func{W}(x_1,\dots,x_n) = \max\left\{w\in[0:n]: {x}_{i}\ge w-i+1\text{ for all } i\le w\right\}.
\end{equation}
The $h$- and $w$-index are generalized by, e.g., the class of $r_p$-indices, $p\ge 1$,
see \cite{GagolewskiGrzegorzewski2009:geometricapproach}.

\item The $h(2)$-index \cite{Kosmulski2006:h2}:
\begin{equation}\label{Eq:IndexH2}
\func{H2}(x_1,\dots,x_n) = \max\left\{h\in[0:n]: {x}_{h} \ge h^2\right\}. %
\end{equation}
Note that the $h(2)$-index is one of the many examples of very
simple, direct modifications of the $h$-index.
Many authors considered  settings other than ``$h^2$'' on the right
side of Equation~\eqref{Eq:IndexH2}, e.g., ``$\alpha h$'' for some
$\alpha > 0$ or ``$h^\beta$'', $\beta\ge 1$, see \cite{AlonsoETAL2009:hreview}.

\end{itemize}
All the introduced impact functions are \index{zero-insensitivity}\emph{zero-insensitive},
that is, for all $\vect{x}\in\mathcal{S}$ it holds $\func{F}(\vect{x})=\func{F}(\vect{x},0)$.
Moreover, the $h$-, $w$-, and $h(2)$-indices are symmetric minitive,
see \cite{Gagolewski2012:ipmu}, and additionally the $h$-index is also maxitive and modular.

\subsection{Impact functions generated by universal integrals}

\newcommand{\IvalInfNonIncr}{\mathcal{S}}
Let us study the connection between zero-insensitive impact functions
and universal integrals, see Section~\ref{Sec:Integrals}. In this setting, with no loss in generality,
we may assume that the vectors we characterize are padded with $0$s
and that they are elements in $\mathcal{S}_d=\{(x_1,x_2,\dots,x_d)\in\IvalPow{d}:  x_1\ge x_2\ge\dots\ge x_d\}$ for some fixed $d$.
We shall need a transformation from the vector
space $\mathcal{S}_d$ into the space $\mathcal{R}^{(\Omega,\mathcal{F})}$
for some $(\Omega,\mathcal{F})$. Although the most straightforward
choice is of course the measurable space $(\Naturals, 2^\Naturals)$,
it is not necessarily the most convenient one. Thus, we fix
the space to $(\Ival,\mathcal{B}(\Ival))$.

Given $\vect{x}\in\mathcal{S}_d$, let
$\langle\vect{x}\rangle\in\mathcal{R}^{(\Ival,\mathcal{B}(\Ival))}$ such that:
\[
   \langle\vect{x}\rangle(t) = x_{\lfloor t+1\rfloor}, \quad t\in\Ival.
\]
It is easily seen that $\langle\vect{x}\rangle$ is a nonincreasing step
function with steps possible only in points from $\Naturals$.
As a matter of fact, $\langle\vect{x}\rangle$ is often called by bibliometricians
the \emph{citation function} for the vector $\vect{x}$.

Let us consider the family $\Phi$ of functions $\func{F}:\mathcal{S}_d\to\Ival$
given by the equation:
\begin{equation}\label{Eq:model}
   \func{F}(\vect{x}) = \eta\Big( \mathcal{I}\big(\mu, \langle\varphi(\vect{x})\rangle\big) \Big)
\end{equation}
where:
\begin{itemize}
   \item $\varphi:\mathcal{S}_d\to\mathcal{S}_d$ -- a function nondecreasing in
   each variable,
   $\varphi(0,0,\dots,0)=(0,0,\dots,0)$,

   \item $\mu:\mathcal{B}(\Ival)\to[0,\infty]$ -- a monotone measure,

   \item $\mathcal{I}$ -- a universal integral on
   $\mathcal{M}^{(\Ival,\mathcal{B}(\Ival))}
   \times
   \mathcal{R}^{(\Ival,\mathcal{B}(\Ival))}$,

   \item $\eta:\Ival\to\Ival$ --  an increasing function,
   $\eta(0)=0$.
\end{itemize}
Noteworthily, Gagolewski and Mesiar in \cite{GagolewskiMesiar2014:integrals}
provide an easy-to-use algorithm that may be used to compute the function given by Equation~\eqref{Eq:model}.

We have what follows, see \cite{GagolewskiMesiar2014:integrals}.

\begin{theorem}
Each function $\func{F}$ given by
Equation~\eqref{Eq:model} is a zero-insensitive impact function.
\end{theorem}

It is important to discuss  the implications of choosing different
$\varphi$, $\mu$, $\mathcal{I}$, and $\eta$
on the aggregation process.
Please note that the $\varphi$ function may be used,  e.g., to
normalize citation records, and often will be set by extending a
function of one variable $\varphi'$ to $\mathcal{S}$,
that is $\varphi(\vect{x})=(\varphi'(x_1),\varphi'(x_2),\dots)$.
Many classical (citation-based) bibliometric indices assume that
$\varphi'(x)=\lfloor x\rfloor$ or $\varphi'(x)=x$.
The $\eta$ function may be used to ``calibrate'' the output values,
especially if we would like to compare the values of different
impact functions.
On the other hand, the monotone measure $\mu$ shall in turn often be set to be
the Lebesgue measure $\lambda$ or some monotone transformation of $\lambda$.

\begin{example}
It is easily seen that:
\begin{itemize}
\item $\func{Sum}(\vect{x})=\mathrm{Ch}(\lambda,\langle\vect{x}\rangle)$,
i.e., a Choquet integral, see Equation~\eqref{Eq:IntChoquet},
\item $\func{H}(\vect{x})=\mathrm{Su}(\lambda,\langle\lfloor\vect{x}\rfloor\rangle)
= \lfloor\mathrm{Su}(\lambda,\langle\vect{x}\rangle)\rfloor$
(Sugeno integral, Equation~\eqref{Eq:IntSugeno1}),
see also \cite{TorraNarukawa2008:h2fuzzyintegrals},
\item $\func{MP}(\vect{x})=\mathrm{Sh}(\lambda,\langle\vect{x}\rangle)$
(Shilkret integral, Equation~\eqref{Eq:IntShilkret}).
\end{itemize}
\end{example}

\begin{example}\label{Ex:differentMu}
Let $\varphi=\mathrm{id}$,
$\mathcal{I}=\mathrm{Ch}$, and $\eta=\mathrm{id}$.
\begin{itemize}
\item If $\mu=\lambda$, then we get of course $\mathcal{I}(\lambda, \langle\vect{x}\rangle)=
\sum_{i} x_i$.

\item For $\mu(A)=\lambda(A)^2$ (a convex transformation),
we obtain $\mathcal{I}(\lambda^2, \langle\vect{x}\rangle)=
\sum_i (i^2-(i-1)^2)\cdot x_i=
1x_1+3x_2+5x_3+7x_5+9x_6+\dots$.
Thus, we put higher weight for productivity here.

\item If $\mu(A)=\sqrt{\lambda(A)}$ (a concave transformation),
then $\mathcal{I}(\sqrt{\lambda}, \langle\vect{x}\rangle)=
\sum_i (\sqrt{i}-\sqrt{i-1})\cdot x_i\simeq
1.00x_1+ 0.41x_2+ 0.32x_3+ 0.27x_4+ 0.24x_5+ 0.21x_6+\dots$.
In consequence, the top-cited papers are of greater significance.
\end{itemize}

For instance, consider two vectors
$\vect{y}=(60,30,10,4,0,0,\dots)$ (higher quality) and
$\vect{z}=(15,13,11,11,9,8,7,7,6,5,3,3,2,1,1,1,1,0,0,\dots)$ (higher productivity).
We have $\mathcal{I}(\lambda, \langle\vect{y}\rangle)=\mathcal{I}(\lambda, \langle\vect{z}\rangle)=104$,
$\mathcal{I}(\lambda^2, \langle\vect{y}\rangle)\simeq 228 <
\mathcal{I}(\lambda^2, \langle\vect{z}\rangle)\simeq 1050$,
and
$\mathcal{I}(\sqrt{\lambda}, \langle\vect{y}\rangle)\simeq 76.7 >
\mathcal{I}(\sqrt{\lambda}, \langle\vect{z}\rangle)\simeq 36.9$
\end{example}

\begin{example}
   Let $\mathcal{I}=\mathrm{Su}$, $\mu=\lambda$, $\eta=\mathrm{id}$.
   We know that by choosing $\varphi(\vect{x})=\lfloor \vect{x}\rfloor$
   we obtain the $h$-index, $\func{H}$.
   It is easily seen that, e.g., $\mathrm{Su}(\lambda, \langle\lfloor\sqrt{\vect{x}}\rfloor\rangle)
   =\func{H2}(\vect{x})$.
   As we already indicated,
   many other Hirsch-based indices actually use simple transformations
   of the input vector, such as the one above.
   Moreover, by dropping the floor function we  obtain
   the generalization of the $h$-index that is real-valued.

   The $\varphi$ function may be used, e.g., to change the impact
   of extremely high-cited publications, like when we choose
   $\varphi(\vect{x})=\log(\vect{x}+1)$.
\end{example}

\begin{example}
Consideration of more complex $\varphi:\mathcal{S}\to\mathcal{S}$ functions
may lead us to other notable numerical characteristics. For example,
the $g$- and $w$-index.
Let $\func{cummin},\func{cumsum}:\IvalPow{d}\to\IvalPow{d}$
\index{cumulative minimum}\index{cumulative sum}denote the cumulative minimum and sum, respectively,
i.e.:
\begin{eqnarray*}
\func{cummin}(\vect{x})&=&(x_1,x_1\wedge x_2,x_1\wedge x_2\wedge x_3,\dots),\\
\func{cumsum}(\vect{x})&=&(x_1,x_1+x_2,x_1+x_2+x_3,\dots).
\end{eqnarray*}
Given $\vect{x}\in\mathcal{S}_d$ it holds:
\begin{eqnarray*}
\func{G}(\vect{x}) &=& \mathrm{Su}\Bigg(\lambda, \Big\langle\Big\lfloor 0\vee \func{cummin}\big(\func{cumsum}(\vect{x})-(1^2,2^2,\dots)+(1,2,\dots)\big)\Big\rfloor\Big\rangle\Bigg),\\
\end{eqnarray*}
and:
\begin{eqnarray*}
\func{W}(\vect{x}) &=& \mathrm{Su}\Bigg(\lambda, \Big\langle\Big\lfloor\mathtt{cummin}\big(\vect{x}+(1,2,\dots)-1\big)\Big\rfloor\Big\rangle\Bigg).
\end{eqnarray*}
\end{example}

\begin{example}
   Let $\mathcal{I}=\mathrm{Sh}$, $\mu=\lambda$, $\varphi=\mathrm{id}$.
   By setting $\eta=\mathrm{id}$ we of course get the MaxProd-index,
   $\func{MP}$.
   We may note, however, that the valuations generated by
   this index cannot be easily compared
   to that of the $h$-index.
   For example, we get $\func{H}(n\ast n, 0, 0, \dots)=n$ and $\func{MP}(n\ast n, 0, 0, \dots) = n^2$.
   Thus, by setting $\eta(x)=\sqrt{x}$ we may obtain
   the ``calibrated'' version of the MaxProd index.
\end{example}

Of course, integrals other than the classical Choquet,
Sugeno, or Shilkret, may also lead to interesting indices.

\subsection{Properties of impact functions}

Apart from zero-insensitivity, here are some other
properties of impact functions that can  be useful in practice
while aggregating vectors of varying lengths, see \cite{CenaGagolewski2015:om3fss}:
\begin{itemize}
   \item \index{F-insensitivity}\emph{$F$-insensitivity}, see \cite{GagolewskiGrzegorzewski2010:ipmu,Woeginger2008:axiomatich},
see also ``conservative productivity increment'' in \cite{Miroiu2013:axiomathirschqualquant}
and the notion of ``stability'' in \cite{BeliakovJames2013:stable},
which holds if for all $\vect{x}\in\mathcal{S}$ and $0\le y\le \func{F}(\vect{x})$
we have $\func{F}(\vect{x},y)=\func{F}(\vect{x})$,

\item \index{F+sensitivity}\emph{$F+$sensitivity}, see \cite{GagolewskiGrzegorzewski2010:ipmu,Woeginger2008:axiomatich},
see also ``productivity responsiveness'' in \cite{Miroiu2013:axiomathirschqualquant},
that is for all $\vect{x}\in\mathcal{S}$ and $y>\func{F}(\vect{x})$
we have $\func{F}(\vect{x},y)>\func{F}(\vect{x})$,

\item \index{multiplicative coherence}\emph{multiplicative coherence}, compare \cite{WaltmanEck2012:inconsistencyh},
i.e., for all $\vect{x},\vect{y}\in\mathcal{S}$ and $d\ge 1$
if $\func{F}(\vect{x})\le\func{F}(\vect{y})$,
then $\func{F}(d\vect{x})
\le\func{F}(d\vect{y})$,

\item \index{additive coherence}\emph{additive coherence},
i.e., for all $\vect{x},\vect{y}\in\mathcal{S}$ and $e\ge 0$
if $\func{F}(\vect{x})\le\func{F}(\vect{y})$,
then $\func{F}(\vect{x}+e)
\le\func{F}(\vect{y}+e)$,

\item \index{independence}\emph{independence}, which was considered in
\cite{BouyssouMarchant2011:rankingconsistent}, and
states that the relative ranking of two producers
should not change after an addition of  products
of the same quality; in other words, for all $\vect{x},\vect{y}\in\mathcal{S}$
and $z\in\Ival$ it holds
$\func{F}(\vect{x})\le\func{F}(\vect{y})$ $\Rightarrow$
$\func{F}(\vect{x},z)\le\func{F}(\vect{y},z)$,

\item \index{consistency}\emph{consistency}, see \cite{BouyssouMarchant2011:rankingconsistent},
which considers joint output of consortia of producers:
if a producer A is dominated by producer B, and C is dominated by D,
then it is reasonable that A and C together (i.e., their combined outputs)
shall be dominated by B and D; in other words, whenever for all $\vect{x},\vect{x}',\vect{y},\vect{y}'$
such that
$\func{F}(\vect{x})\le\func{F}(\vect{y})$ and
$\func{F}(\vect{x}')\le\func{F}(\vect{y}')$ it holds
$\func{F}(\vect{x},\vect{x}')\le\func{F}(\vect{y},\vect{y}')$.
\end{itemize}

\begin{example}
Let us consider the following impact functions:
\begin{itemize}
   \item $\func{Max}(\vect{x})=x_1$ (sample maximum),
   \item $\func{MaxN}(\vect{x})=x_1\wedge n$,
   \item $\func{Q5}(\vect{x})=x_{5}$ if $n\ge 5$ and $0$ otherwise ($\sim$ the fifth quantile),
   \item $\func{H}(\vect{x})=\bigvee_{i=1}^n \lfloor{x_{i}}\rfloor\wedge i$ (the Hirsch index),
   \item $\widetilde{\func{H}}(\vect{x})=\bigvee_{i=1}^n {x_{i}}\wedge i$ (a real-valued Hirsch index),
   \item $\func{H2}(\vect{x})=\bigvee_{i=1}^n \lfloor \sqrt{x_{i}}\rfloor\wedge i$ (the $h^{(2)}$-index),
   \item $\widetilde{\func{H2}}(\vect{x})=\bigvee_{i=1}^n \sqrt{x_{i}}\wedge i$ (a real-valued $h^{(2)}$ index),
   \item $\func{N}(\vect{x})=n$ (sample length),
   \item $\func{NP}(\vect{x})=\sum_{i=1}^n \indicator(x_i > 0)=\bigvee_{i=1}^n \indicator(x_{i}>0)b\wedge i$ (number of elements with non-zero quality).
\end{itemize}
All of these are symmetric minitive, maxitive, as well as modular.
Table~\ref{Tab:propertieswhich} summarizes which of the properties
discussed in this section are fulfilled by the above functions.
The function that obeys the greatest number
of properties is the $\func{Max}$ function.
\end{example}

\begin{table}[htb!]
\caption[Exemplary impact functions and some properties they fulfill.]%
{\label{Tab:propertieswhich} Exemplary impact functions and some properties they fulfill, see~\cite{CenaGagolewski2015:om3fss}.}

\newcolumntype{H}{>{\setbox0=\hbox\bgroup}c<{\egroup}@{}}

\smallskip\centering\noindent\small
\begin{tabularx}{1.0\linewidth}{XcHccHccccccc}
\toprule
&&&&&&&&&&&&\\[-6pt]
\bf\small property name  & $\func{Max}$ & $\func{Max2}$ & $\func{MaxN}$ & $\func{Q5}$ & $\func{Q15}$
& $\func{H}$ & $\widetilde{\func{H}}$ & $\func{H2}$ & $\widetilde{\func{H2}}$
& $\func{N}$ & $\func{NP}$ & $\sum$ \\
&&&&&&&&&&&&\\[-6pt]
\midrule
arity-monotonicity       & \pt & \pt  & \pt  & \pt  & \pf  & \pt  & \pt & \pt & \pt & \pt   & \pt & 9 \\
\midrule
continuity               & \pt & \pt  & \pt  & \pt  & \pt  & \pf  & \pt & \pf & \pt & \pt   & \pf & 6 \\
\midrule
zero-insensitivity       & \pt & \pt  & \pf  & \pt  & \pf  & \pt  & \pt & \pt & \pt & \pf   & \pt & 7 \\
\midrule
F-insensitivity          & \pt & \pf  & \pf  & \pt  & \pf  & \pt  & \pt & \pt & \pf & \pf   & \pf & 5 \\
\midrule
F+sensitivity            & \pt & \pt  & \pt  & \pf  & \pf  & \pf  & \pf & \pf & \pf & \pt   & \pt & 4 \\
\midrule
multiplicative coh.  & \pt & \pt  & \pf  & \pt  & \pt  & \pf  & \pf & \pf & \pf & \pt   & \pt & 4 \\
\midrule
additive coherent        & \pt & \pt  & \pf  & \pf  & \pt  & \pf  & \pf & \pf & \pf & \pt   & \pf & 2 \\
\midrule
independence             & \pt & \pt  & \pf  & \pf  & \pf  & \pf  & \pf & \pf & \pf & \pt   & \pt & 3 \\
\midrule
consistency              & \pt & \pt  & \pt  & \pf  & \pf  & \pf  & \pf & \pf & \pf & \pt   & \pt & 4 \\
\midrule
\hfill$\sum$             & 9  &      &  4   & 5    &      &   3  & 4   & 3   & 3   & 7     & 6   & $\sum$ \\
\bottomrule
\end{tabularx}

\end{table}

\section{Characteristics of fusion functions}\label{Sec:FusionFunctionsCharacteristics}

Numerical characteristics of fusion functions may give us a better insight
into particular aspects of their behavior. Moreover, they can aid in
selecting an aggregation function that best fits a practitioner's needs.

\subsection{Orness and related measures}

First we shall focus on idempotent fusion functions like $\func{F}:\IvalPow{n}\to\Ival$,
see Chapter~\ref{Chapter:onedim}. Let us incorporate the traditional assumption
that $\Ival=[0,1]$. However, please note that the measures introduced
in this section may be easily generalized to any $\Ival$ of finite width.

First let us note that the average value of $\func{F}^{(n)}$ as defined in
\cite[Chapter~10]{GrabischETAL2009:aggregationfunctions}, that is:
\begin{equation}
   \mathrm{average}(\func{F}^{(n)}) = \int_{\Ival^n} \func{F}^{(n)}(\vect{x})\,d\vect{x}
\end{equation}
is nothing more than the expected value of $\func{F}^{(n)}$ under the assumption
that it is applied on a random vector uniformly distributed on $\IvalPow{n}$.

Moreover, recall that in the class of idempotent aggregation functions,
$\func{Min}$ and $\func{Max}$ are the least and the greatest fusion
tools, respectively.

\begin{lemma}[\cite{Dujmovic1973:orness}]
For any $n$ it holds:
\begin{itemize}
   \item $\mathrm{average}(\func{Min}^{(n)}) = \int_{\Ival^n} \func{Min}^{(n)}(\vect{x})\,d\vect{x} = \frac{1}{n+1}$,
   \item $\mathrm{average}(\func{Max}^{(n)}) = \int_{\Ival^n} \func{Max}^{(n)}(\vect{x})\,d\vect{x} = \frac{n}{n+1}$.
\end{itemize}
\end{lemma}

The orness measure for averaging functions was
introduced by Dujmović in \cite{Dujmovic1974:orness} under the name
\textit{disjunction degree}.
This numerical characteristic aims to quantify how far -- on average --
a fusion function's value is from the least and the greatest
averaging functions.

\begin{definition}
Let $\func{F}^{(n)}$ be an idempotent aggregation function.
\index{orness}Its degree of \emph{orness} is given by:
\begin{equation}
\mathrm{orness}(\func{F}^{(n)}) =
   \frac{\int_{\Ival^n} \func{F}^{(n)}(\vect{x})\,d\vect{x} - \int_{\Ival^n} \func{Min}^{(n)}(\vect{x})\,d\vect{x}}%
   {\int_{\Ival^n} \func{Max}^{(n)}(\vect{x})\,d\vect{x} - \int_{\Ival^n} \func{Min}^{(n)}(\vect{x})\,d\vect{x}} \in[0,1].
\end{equation}
\end{definition}

It is easily seen that $\mathrm{orness}(\func{Min}^{(n)})=0$ and $\mathrm{orness}(\func{Max}^{(n)})=1$.
Additionally, $\mathrm{orness}(\func{AMean}^{(n)})=0.5$.
As noted in \cite{BeliakovETAL2007:aggregationpractitioners}, exact orness values
are known only for a few of the most notable fusion functions and arities, in particular:
\begin{itemize}
   \item $\mathrm{orness}(\func{QMean}^{(2)}) = \frac{1}{3}\left(1 + \frac{\log (1+\sqrt{2})}{\sqrt{2}}\right)$,
   \item $\mathrm{orness}(\func{HMean}^{(2)}) = \frac{4}{3} \left(1-\log 2\right)$,
   \item $\mathrm{orness}(\func{GMean}^{(n)}) = \frac{n+1}{n-1} \left(\frac{n}{n+1}\right)^n - \frac{1}{n-1}$.
\end{itemize}
In other cases, computations may be performed numerically (via, e.g.,
numerical cubatures \cite{GenzMalik1980:cubatures,BerntsenETAL1991:cubatures}
or Monte Carlo integration, especially if $n$ is large).
For instance:

\begin{lstlisting}[language=R]
qmean <- function(x) sqrt(mean(x^2)) # quadratic mean
# Compute orness measures for 1,2,3,4-ary quadratic means:
sapply(1:4, function(n) {
   cubature::adaptIntegrate(qmean,
      lowerLimit=rep(0, n),
      upperLimit=rep(1, n),
      tol=1e-12
   )
})

##              [,1]       [,2]       [,3]       [,4]
## integral     0.5        0.5410751  0.554598   0.5609498
## error        5.5511e-15 5.4056e-13 5.5459e-13 5.609497e-13
## fEvaluations 15         62509      2872617    103726263
\end{lstlisting}
Please note that the number of function evaluations increases
drastically as $n$ gets larger.

\begin{remark}
We can also consider the \index{andness}\emph{andness}
measure defined as:
\begin{equation}
   \mathrm{andness}(\func{F}^{(n)}) = 1-\mathrm{orness}(\func{F}^{(n)}).
\end{equation}
\end{remark}

A slightly different measure was introduced by Fern\'{a}ndez Salido and
Murakami in \cite{FernandezMurakami2003:aveorness}.

\begin{definition}
The \index{average orness}\emph{average orness} of an idempotent aggregation function
$\func{F}^{(n)}$ is given by:
\begin{equation}
\mathrm{aveorness}(\func{F}^{(n)}) = \displaystyle\int_{\Ival^n}
\frac{\func{F}^{(n)}(\vect{x})-\func{Min}^{(n)}(\vect{x})}{\func{Max}^{(n)}(\vect{x})-\func{Min}^{(n)}(\vect{x})}\,d\vect{x},
\end{equation}
under the assumption that $0/0=0$.
\end{definition}

Again, $\mathrm{aveorness}(\func{Min}^{(n)})=0$ and $\mathrm{aveorness}(\func{Max}^{(n)})=1$.
Please note that the nature of the introduced measures of disjunctivity
is such that they are based on the fact that $\func{F}$ outputs a single numeric value.
Therefore, they cannot be easily generalized to fusion functions like
$\func{F}^{(n)}:(\IvalPow{d})^n\to\IvalPow{d}$. A possible idea to overcome this
limitation would be to consider, e.g., a quite different measure based
on a properly normalized expected Euclidean distance between the
outputs generated by $\func{F}^{(n)}$ and the boundary of the $\IvalPow{d}$ set
under the assumption that input data are independent and uniformly distributed
on $\IvalPow{d}$. Alternatively, taking into account the fact that
if $X\sim\mathrm{U}(\IvalPow{d})$, then $\mathbb{E}\,X=(d\ast 0.5)$, we
may consider simply a mean squared error-like measure
$\mathbb{E}\, \mathfrak{d}_2(\func{F}^{(n)}(X_1,\dots,X_n), (d\ast 0.5))$.

\bigskip
What is more, if $\func{F}^{(n)}$ is a Lipschitz function,
then its corresponding Lipschitz constant may also be used as its numerical characteristic.

\subsection{Weighting vector's entropy}

For classes of aggregation operators that are parametrized via
a weighting vector -- such as weighted arithmetic means or OWA operators
-- their orness measures are dependent solely on the distribution of weights.

\begin{remark}
The orness and average orness measures coincide in the case of OWA operators,
see \cite{FernandezMurakami2003:aveorness}.
For any weighting vector $\vect{w}$ it holds that:
\[
   \mathrm{orness}(\func{OWA}_\vect{w}) = \mathrm{aveorness}(\func{OWA}_\vect{w})= \sum_{i=1}^n  \frac{n-i}{n-1}w_i.
\]
Nevertheless, this is not the case for arbitrary fusion functions.
\end{remark}

Therefore, such types of numerical characteristics may aid in choosing a particular
fusion function, see, e.g., \cite{FilevYager1998:issueobtainowaweights}
for an example of fitting OWA operators to empirical data under the
constraint that the orness measure is fixed to some pre-established value.

In the current setting, a weighting vector's \index{entropy}\emph{entropy}:
\begin{equation}
   \func{Entropy}(\vect{w}) = -\sum_{i=1}^n w_i \log w_i
\end{equation}
can be conceived as a measure of the degree to which all the input data are used
in the aggregation process, compare \cite{BeliakovETAL2007:aggregationpractitioners}.
It is easily seen that the entropy is maximized
for $\vect{w}=(n\ast (1/n))$. During the choice of a suitable OWA operator,
it is not unusual to fix the desired orness and then maximize the weights' entropy.
Moreover, we can note that, e.g., spread or related measures introduced above may
be used for this purpose as well.

\subsection{Breakdown points and values}\label{Sec:BreakdownPoint}

An \index{outlier}\emph{outlier} is most often defined as an observation
that is too distant from other data points and thus it is in some way suspicious.
It may be present due to a measurement or data input error,
or simply because we are analyzing a sample following a heavy-tailed distribution.
Beckman and Cook in \cite{BeckmanCook1983:outliers} note what follows:

\begin{quote}\it\small
The concern over outliers is old and undoubtedly dates back to the first
attempt to base conclusions on a set of statistical data.
Comments by Bernoulli (1777) indicate that the practice of discarding
discordant observations was commonplace 200 years ago.
\end{quote}

The notion of a breakdown point of a fusion function as discussed in this monograph
has been introduced by Donoho in \cite{Donoho1982:phd}.
It is meant to serve as a measure of a function's robustness to the presence of potential outliers.
Its aim is to express ``the smallest amount of contamination which can cause
the estimator to give an arbitrarily bad answer''.

\begin{definition}[\cite{Donoho1982:phd}]
The \index{breakdown point}\emph{breakdown point} of a fusion function
$\func{F}^{(n)}:(\mathbb{R}^d)^n\to\mathbb{R}^d$, $d\ge 1$,
at $\vect{X}\in(\mathbb{R}^d)^n$ is given by:
\[
   \varepsilon(\func{F}^{(n)},\vect{X})=\min_{m\in[n]}\left\{
   \frac{m}{n}: \sup_{\vect{Y}_m\in(\bar{\mathbb{R}}^d)^n} \| \func{F}^{(n)}(\vect{X})-\func{F}^{(n)}(\vect{Y}_m) \| = \infty
   \right\},
\]
where the supremum is over all possible data sets $\vect{Y}_m$ obtained from
$\vect{X}$ in such a way that exactly $m$ points are replaced with arbitrary values.
\end{definition}

Clearly, the higher the breakdown point, the more insensitive
to outliers a fusion function is at a given point. As it is noted by Lopuha\"{a} and Rousseeuw
in \cite{LopuhaaRousseeuw1991:breakdown}, for most of the functions
studied in the literature, $\varepsilon$ does not depend on $\vect{X}$.
In any case, one might be interested in quantifying
the ``global'' \index{breakdown value}\emph{breakdown value}:
\begin{equation}
   \mathrm{breakval}(\func{F}^{(n)})=\inf_{\vect{X}\in(\mathbb{R}^d)^n} \varepsilon(\func{F}^{(n)},\vect{X}) \in\left[\frac{1}{n},1\right].
\end{equation}
Please note that for translation equivariant fusion functions we have that
$\mathrm{breakval}(\func{F}^{(n)})\le 0.5$, see \cite{Donoho1982:phd}.
It turns out that we have what follows.

\begin{theorem}[\cite{Bassett1991:equimonbreak}]
\index{componentwise median}The componentwise median, $\func{CwMedian}^{(n)}$,
is the only fusion function that is componentwise nondecreasing,
translation and uniform-scale equivariant that has a 50\% breakdown value.
\end{theorem}

On a side note, it is worth stressing that this result is derived from
the notion of monotonicity which is generally very rare in the computational
statistics literature.

\begin{example}Table \ref{Tab:breakdown} gives breakdown points for
some of the fusion functions studied in this book, see also
\cite{DonohoGasko1992:breakdown,Aloupis2006:datadepth}.
We may note that, e.g., the componentwise arithmetic mean (centroid)
is extremely sensitive to outliers.
\end{example}

\begin{table}[t!]
\caption{\label{Tab:breakdown} Breakdown values of exemplary fusion functions.}
\centering
\begin{tabularx}{1.0\linewidth}{Xll}
\toprule
\small\bf fusion function & \small\bf breakdown point & \small\bf reference \\
\midrule
centroid & $\frac{1}{n}$ & --- \\
\midrule
componentwise median & $\frac{1}{2}$ & \cite{LopuhaaRousseeuw1991:breakdown}, see also \cite{Bassett1991:equimonbreak} \\
\midrule
1-median & $\frac{1}{2}$ & \cite{LopuhaaRousseeuw1991:breakdown} \\ %
\midrule
Tukey median & $\in\left[\frac{1}{d+1}, \frac{1}{3}\right]$ & \cite{Aloupis2006:datadepth} \\
\midrule
Oja median & $\frac{2}{n+2}$ for $d=2$ & \cite{NiinimaETAL1990:breadownOja} \\ %
\midrule
Convex hull peeling median & $\le \frac{n+d+1}{(n+2)(d+1)}$ & \cite{Donoho1982:phd,Small1990:medianssurv} \\
\bottomrule
\end{tabularx}

\end{table}

\begin{remark}
   Outliers are not the only types of data ``contamination'' that
   can be considered. Among others we have \index{missing value}\emph{missing values},
   see \cite{RubinLittle2002:statanalNA}, and \index{censored observation}\emph{censored observations},
   see \cite[Section~8.5]{DavidNagaraja2003:orderstatistics}.
\end{remark}

\section{Characteristics of fuzzy numbers}\label{Sec:CharacteristicsFuzzyNumbers}

Recall from Section~\ref{Sec:FuzzyNumber} that a fuzzy number
is a kind of  fuzzy subset of the real line. Let us briefly review the
numerical characteristics of such objects.

\paragraph{Measures of central tendency (defuzzifiers).}
Let us first mention the notion of the \emph{expected interval}
of a fuzzy number $A$, proposed by Dubois and Prade in \cite{DuboisPrade1987:meanfn}:
\begin{equation}
\mathrm{EI}(A) = \left[ \int_0^1 A_L(\alpha)\,d\alpha, \int_0^1 A_U(\alpha)\,d\alpha \right].
\end{equation}
The midpoint of the expected interval is called the \index{expected value of a fuzzy number}\emph{expected value}
of a fuzzy number. It is given by:
\begin{equation}
\mathrm{EV}(A) = \frac{\int_0^1 A_L(\alpha)\,d\alpha + \int_0^1 A_U(\alpha)\,d\alpha}{2}.
\end{equation}

Sometimes a generalization of the expected value,
called \emph{weighted expected value}, is useful.
For a given $w\in[0,1]$, it is defined as:
\begin{equation}
\mathrm{EV}_w(A) = (1-w)\int_0^1 A_L(\alpha)\,d\alpha + w\int_0^1 A_U(\alpha)\,d\alpha.
\end{equation}
It is easily seen that $\mathrm{EV}_{0.5}(A)=\mathrm{EV}(A)$.

\medskip
On the other hand, the \index{value of a fuzzy number}\emph{value} of $A$ was defined by
Delgado, Vila, and Voxman in \cite{DelgadoETAL1998:canonicalfn}, see also \cite{CarlssonFuller2001:possibilisticmeanfn}, as:
\begin{equation}
\mathrm{val}(A) = \int_0^1 \alpha\left(A_L(\alpha)+A_U(\alpha)\right)\,d\alpha.
\end{equation}
The $\alpha$ term may be replaced with a generic weighting function
$\func{w}(\alpha)$, most often such that $\int_0^1 \func{w}(\alpha)\,d\alpha = 0.5$.

Please note that the expected value or  value may be used  to ``defuzzify'' $A$.
The introduced measures are translation and scale equivariant.

\begin{example}
If $A$ denotes a trapezoidal fuzzy number, $A=\mathrm{T}(s_1, s_2, s_3, s_4)$, then
$\mathrm{EV}(A) = (s_1+s_2+s_3+s_4)/4$
and $\mathrm{val}(A) = (s_1+s_4)/6 + (s_2+s_3)/3$.
\end{example}

\paragraph{Measures of nonspecifity.}
Among notions of ``nonspecifity'' of a fuzzy number we find, among others, what follows.
The \index{width of a fuzzy number}\emph{width} of $A$ \cite{Chanas2001:intervapproxfn} is defined as
the width of its expected interval, that is:
\begin{equation}
\mathrm{width}(A) = \int_0^1 A_U(\alpha)\,d\alpha - \int_0^1 A_L(\alpha)\,d\alpha.
\end{equation}
The \index{ambiguity of a fuzzy number}\textit{ambiguity} of $A$ \cite{DelgadoETAL1998:canonicalfn} is defined as:
\begin{equation}
\mathrm{amb}(A) = \int_0^1 \alpha\left(A_U(\alpha)-A_L(\alpha)\right)\,d\alpha.
\end{equation}
Moreover, the \index{standard deviation of a fuzzy number}\emph{standard deviation} of $A$,
as introduced by Carlsson and Full\'{e}r in \cite{CarlssonFuller2001:possibilisticmeanfn},
is given by:
\begin{equation}
\mathrm{sd}(A) = \sqrt{\frac{1}{2} \int_0^1 \alpha\left(A_U(\alpha)-A_L(\alpha)\right)^2\,d\alpha}.
\end{equation}
The three measures are translation invariant and scale equivariant.
A defuzzifier together with a nonspecifity measure may be used to project
a fuzzy number to a real interval.

\begin{example}
If $A$ denotes a trapezoidal fuzzy number, $A=\mathrm{T}(s_1, s_2, s_3, s_4)$, then
$\mathrm{width}(A) = (s_3+s_4)/2-(s_1+s_2)/2$ and $\mathrm{amb}(A) = (s_3-s_2)/3+(s_4-s_1)/6$.
If $A$ is a triangular fuzzy number (i.e., a trapezoidal one with $s_2=s_3$) we additionally have $\mathrm{sd}(A) = (s_4-s_1)/\sqrt{24}$.
\end{example}

\paragraph{Measures of fuzziness.}\index{fuzziness of a fuzzy set}%
Closely related to nonspecifity characteristics are measures of \emph{fuzziness} of
arbitrary fuzzy sets, see, e.g., \cite{SanchezTrillas2012:fuzziness,Weber1984:fuzziness,ZengLi2006:inclusionsimilarityfuzziness},
which are often axiomatized as functions: (a) outputting value of $0$ if and only if
they are applied on a crisp set and (b) monotone with respect to a partial ordering relation $\sqsubseteq_\mathrm{F}$
such that $A \sqsubseteq_\mathrm{F} B$ ($A$ is less fuzzy than $B$) if and only if
$(0.5\wedge \mu_A(x)) \le (0.5\wedge \mu_B(x))$ and $(0.5\vee \mu_A(x)) \ge (0.5\vee \mu_B(x))$ for all $x$.
We see that the membership degree of $0.5$ is considered as the most vague.

\section{Checksums}

Let us end this chapter -- as well as the whole monograph --
with a class of numerical characteristics
which is quite different from the above measures: those which are supposed to be
very difficult to study analytically.

A \index{checksum}\emph{checksum} function
provides us with a way to verify data integrity that can be broken due to:
\begin{itemize}
   \item errors in data transmission,
   \item cryptographic attacks,
   \item malware (malicious software, e.g., viruses, Trojan horses, backdoors) injection,
\end{itemize}
and so on. Such tools are related to hash functions and fingerprint algorithms,
which are also used to map (perhaps uniquely) a (possibly) large and complex data
set into a much simpler domain. However, their purpose is quite different than that
of checksums -- they aim to aid in efficient object dictionary look-up.

Most of the checksum functions studied in the literature assume that an
input data stream consists of chunks of bit sequences of a fixed length,
$\Sigma=\{0,1\}^d$ for, e.g., $d=8$, $16$, $32$, or $64$.
They are incremental (compare Definition~\ref{Def:incrementality})
functions: to compute their value only a single pass through a data stream is required.
Checksum functions map the data into a set of binary sequences of fixed length $d'$, e.g.:
\begin{itemize}
   \item $d'=32$ for the CRC-32 algorithm (which is based on cyclic codes introduced by Prange \cite{Prange1957:crc},
   see also \cite{PetersonBrown1961:cycliccodes,CastagnoliETAL1993:crc}),
   \item $d'=128$ for the MD5 checksum (introduced in RFC1321\footnote{See \url{https://tools.ietf.org/html/rfc1321}.})
   by R.~Rivest, compare also, e.g., \cite{Berson1993:md5}),
   \item $d'=256$ in the case of the SHA-256 algorithm (which was developed by the National Security Agency (NSA),
   see, e.g., \cite{MendelETAL2013:sha256}).
\end{itemize}
Typically, we represent checksum routine outputs as character strings
consisting of hexadecimal digits ($\mathtt{0},\dots,\mathtt{9},\mathtt{a},\dots,\mathtt{f}$).
However, please note that each bit sequence may be mapped to an unsigned binary
(base-2) number:
\begin{equation}
   (b_{d-1} b_{d-2} \cdots b_0)_2 = \sum_{i=0}^{d-1} b_i 2^i.
\end{equation}
Because of this, each checksum is an integer number
and thus a function that computes it may be conceived as a kind of  bit
sequence numerical characteristic, $\func{F}:\Sigma^*\to[0:2^{d'}-1]\subseteq\mathbb{N}_0$.

\begin{example}
We have:
\begin{lstlisting}[language=R]
x <- "fusion functions" # input string
paste(charToRaw(x), collapse="") # hex sequence
## [1] "667573696f6e2066756e6374696f6e73"
digest::digest(x, "crc32")  # CRC-32 checksum
## [1] "659348f9"
digest::digest(x, "md5")    # MD5 checksum
## [1] "ecb11c3a8afdbfdec37956b4997fcf55"
\end{lstlisting}
\end{example}

By default, checksum algorithms involve the following operations on data chunks:
\begin{itemize}
   \item bitwise NOT, AND, OR, and XOR (exclusive OR),
   \item rotate-no-carry (e.g., $00010111\to 10001011$),
   \item right-logical-shift with 0-padding (e.g., $00010111\to \textit{0}0001011$),
   \item addition (modulo $2^{d'}$).
\end{itemize}

\begin{algorithm}
To get a general intuition about how checksum algorithms look,
here is a fragment of \lang{C++} code to compute CRC-32.
\begin{lstlisting}
/* COPYRIGHT (C) 1986 Gary S. Brown.  You may use
 * this program, or code or tables extracted from it,
 * as desired without restriction. */

static uint32_t crc32_tab[256] = {
   0x00000000, 0x77073096, 0xee0e612c, 0x990951ba,
   // ......., .........., .........., ..........,
   0xb40bbe37, 0xc30c8ea1, 0x5a05df1b, 0x2d02ef8d
};

uint32_t crc32(const uint8_t* buf, size_t n)
{
   uint32_t crc = 0 ^ ~0U;
   while (n--)
      crc = crc32_tab[(crc ^ *buf++) & 0xFF] ^ (crc >> 8);
   return crc ^ ~0U;
}
\end{lstlisting}
Here \verb@^@ stands for bitwise XOR,
\verb@&@ for AND, \verb@~@ for NOT, and \verb@>>@ for right-logical-shift.
The data stream is read byte by byte.
\end{algorithm}

Desired properties of checksum functions like $\func{F}$,
especially in cryptographic tasks, include (compare \cite{ContiniETAL2007:criticalhash}):
\begin{itemize}
   \item \textit{preimage resistance} -- given a checksum $\vect{y}$,
   it should be computationally infeasible to find a data stream $\vect{x}$ such that
   $\func{F}(\vect{x})=\vect{y}$,
   \item \textit{collision resistance} -- given a data stream $\vect{x}$,
   it should be computationally infeasible to find a data stream $\vect{x}'\neq\vect{x}$ such that
   $\func{F}(\vect{x})=\func{F}(\vect{x}')$.
\end{itemize}
Generally, $\func{F}$ itself should be relatively easy to compute
but difficult to analyze and thus break (invert).

Another useful feature that $\func{F}$ should possess
is in full opposition to the Lipschitz continuity property:
we would like a checksum to change drastically even for a very small
perturbation in input streams. The dissimilarity degree can be expressed as, for instance:
\begin{itemize}
   \item the Hamming distance in the case of base-2 representation of outputs,
   \item the Levenshtein distance for the character string (hexadecimal) form,
   \item absolute difference in the case of the numeric representation.
\end{itemize}

\begin{example}
Let us slightly modify a string from the previous example:
\begin{lstlisting}[language=R]
x <- "Fusion functions" # input string
paste(charToRaw(x), collapse="") # hex sequence
## [1] "467573696f6e2066756e6374696f6e73"
digest::digest(x, "crc32")  # CRC-32 checksum
## [1] "72ac3384"
digest::digest(x, "md5")    # MD5 checksum
## [1] "5c303067a54b75760940ba0d20ad01b7"
\end{lstlisting}
Please note that the checksums are very different.
For instance, if CRC-32 checksums are interpreted as unsigned integers,
the corresponding decimal numbers are equal to
1704151289 and 1923888004, respectively.
\end{example}

\clearpage{\pagestyle{empty}\cleardoublepage}
\appendix

\titleformat
{\chapter} %
[display] %
{\sf\bfseries\Large\itshape} %
{Appendix\ \thechapter} %
{0.5ex} %
{
    \vspace{1ex}
    \huge\sffamily
} %
[
\textcolor{gray}{\rule{\textwidth}{5pt}}\vspace{1em}
] %

\fancyfoot{}
\fancyhf{}
\fancyhead[LO]{\sffamily\footnotesize Listings\hfill \sffamily\footnotesize\thepage}
\fancyhead[RO]{}
\fancyhead[LE]{}
\fancyhead[RE]{\flushleft\sffamily\footnotesize\thepage\hfill\sffamily\footnotesize Listings}

\chapter{Listings}

\lettrine[lines=3]{S}{ource} codes of scripts or programs included in this book
are licensed under the MIT license. The license permits code reuse within
proprietary software provided that all copies of the software include the
license terms and the copyright notice.

\bigskip
\noindent
\begin{footnotesize}%
\begin{quote}\sffamily\itshape
Copyright \copyright{} 2015 Marek Gagolewski

\medskip
Permission is hereby granted, free of charge, to any person obtaining a copy
of this software and associated documentation files (the ``Soft\-ware''), to deal
in the Software without restriction, including without limitation the rights
to use, copy, modify, merge, publish, distribute, sublicense, and/or sell
copies of the Software, and to permit persons to whom the Soft\-ware is
furnished to do so, subject to the following conditions:

\medskip
The above copyright notice and this permission notice shall be included in
all copies or substantial portions of the Software.

\medskip
THE SOFTWARE IS PROVIDED ``AS IS'', WITHOUT WARRANTY OF ANY KIND, EXPRESS OR
IMPLIED, INCLUDING BUT NOT LIMITED TO THE WARRANTIES OF MERCHANTA\-BI\-LITY,
FITNESS FOR A PARTICULAR PURPOSE AND NONINFRINGEMENT. IN NO EVENT SHALL THE
AUTHORS OR COPYRIGHT HOLDERS BE LIABLE FOR ANY CLAIM, DAMAGES OR OTHER
LIABILITY, WHETHER IN AN ACTION OF CONTRACT, TORT OR OTHERWISE, ARISING FROM,
OUT OF OR IN CONNECTION WITH THE SOFTWARE OR THE USE OR OTHER DEALINGS IN
THE SOFTWARE.
\end{quote}%
\end{footnotesize}

\clearpage

\begin{figure}[p!]
\begin{lstlisting}
#include <algorithm>
// [[Rcpp::plugins("cpp11")]]

struct Comparer {
   const double* v;
   Comparer(const double* _v) { v = _v; }
   bool operator()(const int& i, const int& j) {
      // returns true if the first argument is less than
      // (i.e. is ordered before) the second.
      return (v[i] < v[j]);
   }
};

bool is_comonotonic(NumericVector x, NumericVector y) {
   int n = x.size();
   if (y.size() != n) stop("lengths of x and y differ");

   // recall that array elements in C++ are numbered from 0
   // let s = (0,1,...,n-1)
   std::vector<int> s(n); for (int i=0; i<n; ++i) s[i] = i;

   Comparer ltx(REAL(x));
   std::sort(s.begin(), s.end(), ltx);
   // now s is an ordering permutation of x

   Comparer lty(REAL(y));
   int i1 = 0;
   while (i1 < n) { /* now search for the longest subsequence
                       consisting of equal x's */
      int i2 = i1+1;
      while (i2 < n && x[s[i1]] == x[s[i2]]) ++i2;
      // sort the subsequence if necessary:
      if (i2-i1 > 1) std::sort(s.begin()+i1, s.begin()+i2, lty);
      // y[s[i1-1]]>y[s[i1]] => x and y are not comonotonic:
      if (i1 > 0 && y[s[i1-1]] > y[s[i1]]) return false;
      i1 = i2;
   }

   // as a by-product, (s[0]+1, s[1]+1, ..., s[n-1]+1)
   // is a permutation that orders both x and y
   return true;
}
\end{lstlisting}
\caption{\label{Fig:comonotonic} A \lang{C++} implementation of an $O(n\log n)$ algorithm
to determine if two vectors of length $n$ are comonotonic, see \cite{Gagolewski2015:normalizedspread}.}
\end{figure}

\begin{figure}[p!]
\centering
\begin{lstlisting}[language=R]
#' @param D a symmetric positive-semidefinite n*n matrix
#' @param c a numeric vector of length n
#' @param A an m*n numeric matrix
#' @param b a numeric vector of length m
#' @param r a character vector of length m
#'        with elements like <=, ==, or >=;
#'        specifies types of linear constraints
#' @param l a numeric vector of length n which gives
#'        lower bounds for corresponding x variables,
#'        -Inf gives no bound
#' @param u a numeric vector of length n which gives
#'        upper bounds for corresponding x variables,
#'        Inf gives no bound
#' @param c0 a single numeric value
#'
#' @return
#' A list with the following components:
#'  par: The best set of parameters, x, found;
#'  value: The value of the objective function at par;
#'  counts: The number of iterations that it took
#'         to solve the program;
#'  status: Solution status - 0 for optimal
cgal_qp_solver <- function(D, c, A, b, r=rep(">=", length(b)),
      l=rep(-Inf, length(c)), u=rep(Inf, length(c)), c0=0.0)
{
   stopifnot(is.numeric(D), is.finite(D), is.matrix(D))
   stopifnot(is.numeric(A), is.finite(A), is.matrix(A))
   stopifnot(is.numeric(c), is.finite(c))
   stopifnot(is.numeric(b), is.finite(b))
   stopifnot(is.character(r), r %
   stopifnot(is.numeric(l), !is.na(l) & !is.nan(l))
   stopifnot(is.numeric(u), !is.na(u) & !is.nan(u))
   stopifnot(is.numeric(c0), is.finite(c0))
   stopifnot(length(b) == nrow(A), length(r) == nrow(A))
   stopifnot(isSymmetric(D), ncol(D) == ncol(A))
   stopifnot(length(c) == nrow(D), length(c0) == 1)
   stopifnot(length(l) == nrow(D), length(u) == nrow(D))

   r <- match(r, c("<=", "==", ">="))-2 # values in {-1, 0, 1}
   fl <- is.finite(l) # which lower bounds for x are active
   fu <- is.finite(u) # which upper bounds for x are active

   .cgal_qp_solver(length(c), length(b), A, b, r,
      fl, l, fu, u, D, c, c0)
}
\end{lstlisting}
\caption{\label{Fig:CGAL_quadprog1} An \R{} interface to the \package{CGAL} \cite{cgal:eb-14b} library quadratic programming solver, part I.}
\end{figure}

\begin{figure}[p!]
\centering%
\begin{lstlisting}
#include <CGAL/QP_functions.h>
#include <CGAL/MP_Float.h>
typedef CGAL::MP_Float ET;
typedef CGAL::Quadratic_program_solution<ET> Solution;
typedef CGAL::Quadratic_program_from_iterators<double**,
   double*, CGAL::Comparison_result*,
   int*, double*, int*, double*, double**, double*> Program;
// [[Rcpp::plugins("cpp11")]]
// [[Rcpp::export(".cgal_qp_solver")]]
List cgal_qp_solver(int n, int m, NumericMatrix A,
   NumericVector b, IntegerVector r, LogicalVector fl,
   NumericVector l, LogicalVector fu, NumericVector u,
   NumericMatrix D, NumericVector c, double c0)
{
  double *Aptr = REAL((SEXP)(A)), *Dptr = REAL((SEXP)(D));
  double **_A = new double*[n], **_D = new double*[n];
  for (int i=0; i<n; ++i) // ith column
  {   _A[i] = Aptr+i*m; _D[i] = Dptr+i*n; }
  CGAL::Comparison_result* _r = new CGAL::Comparison_result[m];
  for (int j=0; j<m; ++j)
     _r[j] = (r[j] < 0 ? CGAL::SMALLER
                 : (r[j] > 0 ? CGAL::LARGER : CGAL::EQUAL));
  List retval;
  Program qp(n, m, _A, REAL((SEXP)(b)), _r,
     (int*)LOGICAL((SEXP)(fl)), REAL((SEXP)(l)),
     (int*)LOGICAL((SEXP)(fu)), REAL((SEXP)(u)),
     _D, REAL((SEXP)(c)), c0);
  Solution s(CGAL::solve_quadratic_program(qp, ET()));
  // generate output solution:
  NumericVector solution(n);
  int i=0;
  for (auto it = s.variable_values_begin();
        it != s.variable_values_end(); ++it)
     solution[i++] = to_double(*it);
  retval = List::create(
     _("par") = solution,
     _("value") = to_double(s.objective_value()),
     _("counts") = s.number_of_iterations(),
     _("status") = (s.status() == CGAL::QP_OPTIMAL ? 0
                 : (s.status() == CGAL::QP_INFEASIBLE ? 1
                 : (s.status() == CGAL::QP_UNBOUNDED ? 2
                 : -1)))
  );
  delete [] _r; delete [] _A; delete [] _D;
  return retval;
}
\end{lstlisting}
\caption{\label{Fig:CGAL_quadprog2} An \R{} interface to the \package{CGAL} \cite{cgal:eb-14b} library quadratic programming solver, part II.}
\end{figure}

\clearpage

\begin{figure}[p!]
\centering
\begin{lstlisting}[language=R]
fit_wam_L2_quadprog <- function(X, Y) {
   stopifnot(is.matrix(X), is.matrix(Y))
   n <- nrow(X); m <- ncol(X)
   stopifnot(1 == nrow(Y), m == ncol(Y))

   # Linear constraint (sum(w) == 1):
   A <- matrix(1, ncol=n, nrow=1)
   B <- 1

   # Objective function definition:
   D <- tcrossprod(X)     #    X %
   C <- -tcrossprod(X, Y) # - (X %

   res <- cgal_qp_solver(D, C, A, B, r="==", l=rep(0, n))
   stopifnot(res$status == 0)
   res$par # return value
}
\end{lstlisting}
\caption{\label{Fig:fit_wam_L2_quadprog} \R{} code for least squares fitting of weighted arithmetic mean's weights.}
\end{figure}

\begin{figure}[p!]
\centering
\begin{lstlisting}[language=R]
fit_wam_L1_linprog <- function(X, Y) {
   stopifnot(is.matrix(X), is.matrix(Y))
   n <- nrow(X); m <- ncol(X)
   stopifnot(1 == nrow(Y), m == ncol(Y))

   A <- rbind(
      cbind(t(X), -diag(m), diag(m)),
      c(rep(1, n), rep(0, 2*m))
   )
   B <- c(Y, 1)
   C <- c(rep(0, n), rep(1, 2*m))
   D <- matrix(0, nrow=n+2*m, ncol=n+2*m) # an LP problem
   res <- cgal_qp_solver(D, C, A, B, r=rep("==", nrow(A)),
                                     l=rep(0, n+2*m))
   stopifnot(res$status == 0)
   stopifnot(0 == max(apply(MARGIN=2, FUN=min,
      X=matrix(res$par[-(1:n)], nrow=2, byrow=TRUE))))
   res$par[1:n] # return value: first n parameters
}
\end{lstlisting}
\caption{\label{Fig:fit_wam_L1_linprog} \R{} code for least absolute deviation fitting of a weighted arithmetic mean's weights.}
\end{figure}

\begin{figure}[p!]
\centering
\begin{lstlisting}[language=R]
fit_wam_LInf_linprog <- function(X, Y) {
   stopifnot(is.matrix(X), is.matrix(Y))
   n <- nrow(X); m <- ncol(X)
   stopifnot(1 == nrow(Y), m == ncol(Y))
   A <- rbind(
      cbind(t(X), -1),
      cbind(t(X), +1),
      c(rep(1, n), 0)
   )
   B <- c(Y, Y, 1)
   C <- c(rep(0, n), 1)
   D <- matrix(0, nrow=n+1, ncol=n+1) # an LP problem
   res <- cgal_qp_solver(D, C, A, B,
      r=c(rep("<=", m), rep(">=", m), "=="),
      l=rep(0, n+1))
   stopifnot(res$status == 0)
   res$par[1:n] # return value: first n parameters
}
\end{lstlisting}
\caption{\label{Fig:fit_wam_LInf_linprog} \R{} code for least Chebyshev metric fitting of a weighted arithmetic mean's weights.}
\end{figure}

\begin{figure}[p!]
\centering
\begin{lstlisting}[language=R]
fit_wqam_L2_optim <- function(X, Y, phi, phiInv, phiInvPrime) {
   stopifnot(is.matrix(X), is.matrix(Y))
   n <- nrow(X); m <- ncol(X)
   stopifnot(1 == nrow(Y), m == ncol(Y))
   stopifnot(is.function(phi),  # generator function
      is.function(phiInv),      # its inverse
      is.function(phiInvPrime)) # derivative of inverse
   phiX <- phi(X)
   w0 <- runif(n); w0 <- w0/sum(w0)
   lambda0 <- log(w0) # initial parameters
   E <- function(lambda) { # goodness-of-fit measure
      w <- exp(lambda)/sum(exp(lambda))
      sum((phiInv(t(w) %
   }
   gradE <- function(lambda) { # its gradient
      w <- exp(lambda)/sum(exp(lambda))
      Z <- as.numeric(t(w) %
      2*w*(((phiInv(Z)-Y)*phiInvPrime(Z)) %
   }
   res <- optim(lambda0, E, gradE, method="BFGS")
   stopifnot(res$convergence == 0)
   exp(res$par)/sum(exp(res$par)) # return value
}
\end{lstlisting}
\caption{\label{Fig:fit_wqam_L2_optim} \R{} code for least squares fitting of a weighted quasi-arithmetic mean's weights.}
\end{figure}

\clearpage

\begin{figure}[p!]
\centering
\begin{lstlisting}[language=R]
fit_wqam_L1_optim_approx <- function(X, Y,
      phi, phiInv, phiInvPrime, eps=1e-12) {
   stopifnot(is.matrix(X), is.matrix(Y))
   n <- nrow(X); m <- ncol(X)
   stopifnot(1 == nrow(Y), m == ncol(Y))
   stopifnot(is.function(phi),  # generator function
      is.function(phiInv),      # its inverse
      is.function(phiInvPrime)) # derivative of inverse
   stopifnot(is.numeric(eps), length(eps) == 1, eps > 0)

   phiX <- phi(X)
   w0 <- runif(n); w0 <- w0/sum(w0)
   lambda0 <- log(w0) # initial parameters

   E <- function(lambda) { # goodness-of-fit measure
      w <- exp(lambda)/sum(exp(lambda))
      e <- phiInv(t(w) %
      sum(sqrt(e^2+eps^2))
   }

   gradE <- function(lambda) { # its gradient
      w <- exp(lambda)/sum(exp(lambda))
      Z <- as.numeric(t(w) %
      w*(
        ((phiInv(Z)-Y)*phiInvPrime(Z)/
           sqrt((phiInv(Z)-Y)^2 + eps^2)) %
        (t(phiX)-Z)
      )
   }

   res <- optim(lambda0, E, gradE, method="BFGS")
   stopifnot(res$convergence == 0)
   exp(res$par)/sum(exp(res$par)) # return value
}
\end{lstlisting}
\caption{\label{Fig:fit_wqam_L1_optim} \R{} code for approximate least absolute
deviation fitting of a weighted quasi-arithmetic mean's weights.}
\end{figure}

\begin{figure}[p!]
\centering
\begin{lstlisting}[language=R]
fit_powmean_L2_optim <- function(X, Y, pmin=0.1, pmax=10) {
   stopifnot(is.matrix(X), is.matrix(Y))
   n <- nrow(X); m <- ncol(X)
   stopifnot(1 == nrow(Y), m == ncol(Y))

   p <- 1 # this will be a parameter shared by the 3 functions:
   phi         <- function(x)
      x^p
   phiInv      <- function(x)
      exp(log(x)/p) # x^(1/p)
   phiInvPrime <- function(x)
      exp((1-p)*log(x)/p)/p # (x^(1/p-1))/p
   envir_p <- new.env()
   envir_p[["p"]] <- p
   environment(phi) <- envir_p
   environment(phiInv) <- envir_p
   environment(phiInvPrime) <- envir_p

   E <- function(p) {
      assign("p", p, environment(phi)) # affects 3 functions
      w <- fit_wqam_L2_optim(X, Y, phi, phiInv, phiInvPrime)
      sum((as.numeric((t(X^p) %
   }

   optimize(E, c(pmin, pmax))$minimum
}
\end{lstlisting}
\caption{\label{Fig:fit_powmean_L2_optim} \R{} code for determining best exponent $p$
in a least squares error power mean fitting task;
calls a function given in Figure~\ref{Fig:fit_wqam_L2_optim}.}
\end{figure}

\begin{figure}[p!]
\begin{lstlisting}[language=R]
seb <- function(X) {
   stopifnot(is.numeric(X), is.matrix(X))
   n <- ncol(X)
   # QP solver in CGAL determines argmin_v 0.5 v^T D v + c^T v
   XtX <- crossprod(X) # (t(X) %
   D <- 2.0*XtX
   C <- -diag(XtX)
   A <- matrix(rep(1, n), ncol=n)
   B <- 1
   res <- cgal_qp_solver(D, C, A, B, r="==", l=rep(0, n))
   stopifnot(res$convergence == 0)
   v <- res$par
   as.numeric(tcrossprod(v, X)) # v %
}
\end{lstlisting}
\caption{\label{Fig:seb_algo} An \R{} implementation of a QP-based \cite{GartnerSchonherr2000:qpball} Euclidean 1-center finder.}
\end{figure}

\begin{figure}[p!]
\begin{lstlisting}
NumericMatrix rortho(int d) {
  if (d < 1) stop("d < 1");
  NumericMatrix A(d, d); // resulting matrix
  NumericVector x(d);    // auxiliary vector

  /* --- Step 1. --- */
  double theta = Rf_runif(0.0, 2.0*M_PI); // U[0,2pi]
  double b = double(Rf_runif(0.0, 1.0)<0.5)*2.0-1.0; // U{-1,1}
  A(0,0) =    cos(theta);    A(0,1) =    sin(theta);
  A(1,0) = -b*sin(theta);    A(1,1) =  b*cos(theta);

  /* --- Step 2.--- */
  for (int i=3; i<=d; ++i) {
     /* --- Steps 2.1. and 2.2. together --- */
     double xnorm = 0.0;
     for (int j=0; j<i; ++j) {
        x[j] = Rf_rnorm(0.0, 1.0);
        xnorm += x[j]*x[j];
     } // non-normalized z
     xnorm = sqrt(xnorm);
     x[0] = 1.0-x[0]/xnorm;
     double xnorm2 = x[0]*x[0];
     for (int j=1; j<i; ++j) {
        x[j] = -x[j]/xnorm;
        xnorm2 += x[j]*x[j];
     } // non-normalized x
     xnorm2 = sqrt(xnorm2);
     for (int j=0; j<i; ++j) x[j] /= xnorm2;
     /* --- Step 2.3. --- */
     for (int k=i-1; k>0; --k)
        for (int j=i-1; j>0; --j)
           A(j,k) = A(j-1, k-1);
     for (int j=1; j<i; ++j) A(0,j) = A(j,0) = 0.0;
     A(0,0) = 1.0; // now previous A is extended
      for (int k=0; k<i; ++k) {
        double x2 = 0.0;
        for (int j=0; j<i; ++j)
           x2 += x[j]*A(j,k);   // t(x)*extA(.,k)
        for (int j=0; j<i; ++j)
           A(j,k) -= 2*x[j]*x2; // extA-2*x*above
     }
  }
  return A; // Step 3.
}
\end{lstlisting}
\caption{\label{Fig:rortho} A \lang{C++} implementation of Algorithm~\ref{Arg:rortho}:
Generation of a random orthogonal $d\times d$ matrix.}
\end{figure}

\begin{figure}[p!]
\begin{lstlisting}
#include <algorithm>
#include <utility>
// [[Rcpp::plugins("cpp11")]]
NumericVector Weiszfeld1median(NumericMatrix X,
      NumericVector w, NumericVector y0, double eps=1.0e-9) {
   int d = X.nrow();
   int n = X.ncol();
   if (w.length()  != n) stop("w.length()  != n");
   if (y0.length() != d) stop("y0.length() != d");

   NumericVector y_last(d); // a new vector
   NumericVector y_cur(Rcpp::clone(y0)); // a deep copy of y0

   double lasterr;
   do {
      std::swap(y_last, y_cur); // swaps underlying pointers
      for (int j=0; j<d; ++j)
         y_cur[j] = 0.0;
      double w_over_d_x_y = 0.0;
      for (int i=0; i<n; ++i) {
         double d_xi_y = 0.0;
         for (int j=0; j<d; ++j)
            d_xi_y += (X(j, i)-y_last[j])*(X(j, i)-y_last[j]);
         d_xi_y = sqrt(d_xi_y);
         if (d_xi_y <= eps) return y_last; /* Step 2.1. */
         double w_over_d_xi_y = w[i]/d_xi_y;
         w_over_d_x_y += w_over_d_xi_y;
         for (int j=0; j<d; ++j)
            y_cur[j] += w_over_d_xi_y*X(j, i);
      }

      lasterr = 0.0;
      for (int j=0; j<d; ++j) {
         y_cur[j] /= w_over_d_x_y;
         lasterr += (y_cur[j]-y_last[j])*(y_cur[j]-y_last[j]);
      }
   } while (lasterr > eps*eps); /* Step 2.3. */
   return y_cur;
}
\end{lstlisting}
\caption{\label{Fig:Weiszfeld} A \lang{C++} implementation of the Weiszfeld procedure,
see Algorithm \ref{Alg:Weiszfeld}, for determining the weighted 1-median.}
\end{figure}

\begin{figure}[p!]
\begin{lstlisting}
#include <unordered_map>
// [[Rcpp::plugins("cpp11")]]
IntegerVector median_hamming(IntegerMatrix X) {
   int n = X.ncol();
   int d = X.nrow();
   IntegerVector out(d);
   for (int i=0; i<d; ++i) {
      std::unordered_map<int, int> ht; // hasthable
      for (int j=0; j<n; ++j)
         ht[X(i,j)]++; /* count occurrences of each
            letter; ints are default-constructed as 0 */

      int max = 0, argmax = -1;
      for (auto it=ht.cbegin(); it != ht.cend(); ++it)
         if (max < (*it).second) { // find a most frequently
            max = (*it).second;    // occurring letter
            argmax = (*it).first;
         }
      out[i] = argmax;
   }
   return out;
}
\end{lstlisting}
\caption{\label{Fig:medianhamming} A \lang{C++} implementation of a procedure to determine a solution
to Equation~\eqref{Eq:MedianHamming} -- a median with respect to the Hamming distance.}
\end{figure}

\begin{figure}[p!]
\begin{lstlisting}
// [[Rcpp::export]]
IntegerVector hamming_dist_max(IntegerMatrix Y,
      IntegerMatrix X) {
   int nx = X.ncol(), ny = Y.ncol(), d = Y.nrow();
   if (X.nrow() != d) stop("X.nrow() != Y.nrow()");
   IntegerVector out(ny);
   for (int i=0; i<ny; ++i) {
      int max_hamming = 0;
      for (int j=0; j<nx; ++j) {
         // Hamming distance between Y[,i] and X[,j]
         int h = 0;
         for (int k=0; k<d; ++k) h += (int)(Y(k,i) != X(k,j));
         if (h > max_hamming) max_hamming = h;
      }
      out[i] = max_hamming;
   }
   return out;
}
\end{lstlisting}
\caption{\label{Fig:haming_dist_max} A helper function used in Figure~\ref{Fig:hamming_closest_ga};
determines the maximal Hamming distance between each vector in \texttt{Y} and all vectors in \texttt{X}.}
\end{figure}

\clearpage

\begin{figure}[htb!]
\begin{lstlisting}[language=R]
hamming_closest_ga <- function(X, k=length(X)*8, niter=2000,
      lambdaMutMult=0.001) {
   n <- ncol(X)
   d <- nrow(X)
   S <- unique(as.integer(X))
   # expected value of number of bits to mutate per iteration:
   lambdaMut <- max(1, k*d*lambdaMutMult)

   selection <- function(P, f) {
      p <- (d-f+1)^3 # max(f) == d
      p <- p/sum(p)
      P[,sample(k, replace=TRUE, size=2*k, prob=p)]
   }

   crossover <- function(P2) { # uniform crossover
      P <- P2[,1:k]
      for (i in 1:k) {
         b <- sample(d, d/2)
         P[b, i] <- P2[b, i+k]
      }
      P
   }

   mutation <- function(P) {
      m <- sample(length(P), min(k*d, rpois(1, lambdaMut)))
      P[m] <- sample(S, length(m), replace=TRUE)
      P
   }

   # initial population: points in X and random ones (mixed):
   P <- matrix(nrow=d, sample(S, k*d, replace=TRUE))
   P[,sample(k, min(n, k))] <- X[,sample(n, min(n, k))]

   # store the best solution so far:
   f <- hamming_dist_max(P, X)
   bestP <- t(unique(t(P[,f==min(f)])))
   bestF <- min(f)

   for (i in 1:niter) {
      P <- mutation(crossover(selection(P, f)))
      f <- hamming_dist_max(P, X)
      if (bestF > min(f)) { # we got a better solution
         bestP <- t(unique(t(P[,f==min(f)])))
         bestF <- min(f)
      }
   }
   bestP # return value
}
\end{lstlisting}
\caption{\label{Fig:hamming_closest_ga} An \R{} implementation of a genetic algorithm-based
approximate solution to the closest vector with respect to the Hamming distance finding problem.}
\end{figure}

\begin{figure}[p!]
\begin{lstlisting}
double dpr2_dist(List X, NumericVector y,
                int dy, double p, double r) {
   int n = X.size();
   double dist = 0.0;

   for (int i=0; i<n; ++i) {
      NumericVector x(X[i]);
      int dx = x.size();
      int min_dx_dy = std::min(dx, dy);

      for (int j=0; j<min_dx_dy;  ++j)
         dist += (x[j]-y[j])*(x[j]-y[j]);

      for (int j=min_dx_dy; j<dx; ++j)
         dist += x[j]*x[j];

      for (int j=min_dx_dy; j<dy; ++j)
         dist += y[j]*y[j];

      dist += p*abs(pow(dx, r)-pow(dy, r));
   }

   return dist;
}
\end{lstlisting}
\caption{\label{Fig:kmeansprod_dist2} A \lang{C++} implementation of
to compute the sum of $\mathfrak{d}_{p,r}^2$ penalty functions,
see Equation~\eqref{Eq:SqM2informetricdist}, between the first \texttt{dy} observations
in a vector \texttt{y} and each vector in \texttt{X}.}
\end{figure}

\begin{figure}[p!]
\begin{lstlisting}
#include <deque>
// [[Rcpp::plugins("cpp11")]]
NumericVector dpr2_centroid(List X, double p, double r) {
   int l=X.size();
   int d=calc_max_vector_length(X);

   NumericVector xtilde(d);
   for (auto it=X.begin(); it != X.end(); ++it) {
      NumericVector x(*it);
      int dx = x.size();
      for (int j=0; j<dx; ++j) xtilde[j] += x[j];
   }

   // a linked list (a stack):
   std::deque< std::pair<int, int> > part;
   NumericVector y(d);
   NumericVector best_y=NumericVector(0);
   double best_dist=INFINITY;

   for (int n=1; n<=d; ++n) {
      // C++ arrays use 0-based indices
      part.push_front( std::pair<int, int>(n-1, n-1) );
      y[n-1]=xtilde[n-1]/l;
      auto it=part.begin();
      while (it+1!=part.end() &&
            y[(*it).first] > y[(*(it+1)).second]) {
         // merge:
         int p1=(*it).second-(*it).first+1;
         int p2=(*(it+1)).second-(*(it+1)).first+1;
         y[(*it).second] = (y[(*it).second]*p1+
            y[(*(it+1)).second]*p2)/(p1+p2);
         for (int j=(*it).second-1; j>=(*(it+1)).first; --j)
            y[j]=y[(*it).second];
         (*(it+1)).second=(*it).second;
         // erase current it and move forward (pop stack)
         it=part.erase(it);
      }
      double cur_dist=dpr2_dist(X, y, n, p, r);
      if (cur_dist<best_dist) {
         best_dist=cur_dist;
         best_y=NumericVector(y.begin(), y.begin()+n);
      }
   }
   return best_y;
}
\end{lstlisting}
\caption{\label{Fig:kmeansprod2} A \lang{C++}  implementation of a function to
compute centroid-like fusion function for informetric data given by Equation~\eqref{Eq:SqM2informetricentroid};
a function from Figure~\ref{Fig:kmeansprod_dist2} is called; see \cite{CenaGagolewski2015:kmeansinformetric}.}
\end{figure}

\clearpage

\begin{figure}[p!]
\begin{lstlisting}
int levenshtein_smallmem(int* s1, int* s2, int n1, int n2) {
   if (n1 < n2) {
      std::swap(s1, s2); // pointer swap
      std::swap(n1, n2);
   }

   int* v_cur = new int[n2+1];
   int* v_last = new int[n2+1]; // n2 <= n1
   for (int j=0; j<=n2; ++j)
      v_cur[j] = j;

   for (int i=1; i<=n1; ++i) {
      std::swap(v_last, v_cur); // pointer swap
      v_cur[0] = i;
      for (int j=1; j<=n2; ++j) {
         v_cur[j] = std::min(std::min(
               v_last[j-1]+(int)(s1[i-1]!=s2[j-1]),
               v_cur[j-1]+1),
               v_last[j]+1);
      }
   }

   int ret = v_cur[n2];
   delete [] v_cur;
   delete [] v_last;
   return ret;
}


// [[Rcpp::export]]
int levenshtein_smallmem(IntegerVector s1, IntegerVector s2) {
   // Rcpp interface to the above function
   return levenshtein_smallmem(INTEGER(s1), INTEGER(s2),
      LENGTH(s1), LENGTH(s2));
}

\end{lstlisting}
\caption{\label{Fig:Levenshtein} A memory-efficient \lang{C++} implementation of a
Wagner-Fisher version \cite{WagnerFischer1974:string2stringcorr} of the
Levenshtein distance computation algorithm.}
\end{figure}

\begin{figure}[p!]
\begin{lstlisting}
// [[Rcpp::export]]
IntegerVector Levenshtein_centroid2(IntegerVector s1,
      IntegerVector s2) {
   int n1 = s1.size(), n2 = s2.size();
   NumericMatrix D(n1+1, n2+1);
   IntegerMatrix T(n1+1, n2+1);
   for (int i=1; i<=n1; ++i) { // deletion
      D(i,0) = D(i-1,0)+1; T(i,0) = 4;
   }
   for (int j=1; j<=n2; ++j) { // insertion
      D(0,j) = D(0,j-1)+1; T(0,j) = 2;
   }
   for (int i=1; i<=n1; ++i)
      for (int j=1; j<=n2; ++j) {
         T(i,j) = 0;
         if (s1[i-1]==s2[j-1])
            D(i,j) = D(i-1,j-1);      // no change
         else {
            double m1 = D(i-1,j-1)+1;       // sub
            double m2 = D(i,j-1)+1;         // ins
            double m3 = D(i-1,j)+1;         // del
            D(i,j) = std::min(std::min(m1, m2), m3);
            if (D(i,j) == m1) T(i,j) |= 1;
            if (D(i,j) == m2) T(i,j) |= 2;
            if (D(i,j) == m3) T(i,j) |= 4;
         }
      }
   int maxd = (int)(D(n1, n2)*0.5);
   if (maxd <= 0) return s1;
   std::list<int> l1(s1.begin(), s1.end());
   auto it1 = l1.end(); --it1;
   auto it2 = s2.end(); --it2;
   int x = n1, y = n2;
   for (int curd=0; curd < maxd; ) {
      curd += (int)(T(x,y) != 0);
      if (T(x,y) == 0) {                         // no change
         x--; y--; --it1; --it2;
      } else if (T(x,y) & 1) {                         // sub
         x--; y--; (*(it1--)) = (*(it2--));
      } else if ((T(x, y) & 2) && ((!(T(x,y)&4))
            || ((int)l1.size() < std::max(n1,n2)))) {  // ins
         y--; it1 = l1.insert(++it1, *(it2--)); --it1;
      } else {                                         // del
         x--; it1 = l1.erase(it1); --it1;
      }
   }
   return IntegerVector(l1.begin(), l1.end());
}


\end{lstlisting}
\caption{\label{Fig:LevenshteinCentroid2} A \lang{C++} implementation
of an algorithm to compute the Levenshtein distance-based centroid of two strings.}
\end{figure}

\clearpage

\begin{figure}[p!]
\begin{lstlisting}
#include <algorithm>

struct Comparer {
   // just like the comparer for the comonotonicity algorithm
   const int* v;
   Comparer(const int* _v) { v = _v; }
   bool operator()(const int& i, const int& j) const {
      return v[i] < v[j];
   }
};


// [[Rcpp::export]]
double dinudist(IntegerVector x, IntegerVector y) {
   int nx = x.size(), ny = y.size();

   // ordering permutation of x:
   std::vector<int> ox(nx);
   for (int i=0; i<nx; ++i) ox[i] = i;
   std::stable_sort(ox.begin(), ox.end(),
      Comparer(INTEGER(x)));

   // ordering permutation of y:
   std::vector<int> oy(ny);
   for (int i=0; i<ny; ++i) oy[i] = i;
   std::stable_sort(oy.begin(), oy.end(),
      Comparer(INTEGER(y)));

   double d = 0.0;
   int ix = 0, iy = 0;
   while (ix < nx && iy < ny) {
      if (x[ox[ix]] == y[oy[iy]])
         d += std::abs((ox[ix++]+1) - (oy[iy++]+1));
      else if (x[ox[ix]] < y[oy[iy]])
         d += std::abs((ox[ix++]+1) - 0);
      else
         d += std::abs(0 - (oy[iy++]+1));
   }
   while (ix < nx)
      d += std::abs((ox[ix++]+1) - 0);
   while (iy < ny)
      d += std::abs(0 - (oy[iy++]+1));

   return d;
}
\end{lstlisting}
\caption{\label{Fig:dinurank} A \lang{C++} implementation of an algorithm
to compute the Dinu rank distance.}
\end{figure}

\clearpage

\begin{figure}[p!]
\begin{lstlisting}
struct NNItem {
   size_t index;
   double dist;

   NNItem(size_t index, double dist) :
      index(index), dist(dist) {}

   NNItem() :
      index(SIZE_MAX), dist(-INFINITY) {}

   inline bool operator<( const NNItem& o ) const {
      return dist < o.dist;
   }
};

class Distance { // abstract class, must be overloaded
private:
   size_t n;
   virtual double compute(size_t v1, size_t v2) = 0;

public:
   Distance(Function distance, RObject objects);
   inline size_t getObjectCount() { return n; }
   inline double operator()(size_t v1, size_t v2) {
      return compute(v1, v2);
   }
};

double sumd_nn(Distance* dist, size_t i, size_t nntry,
    std::priority_queue<NNItem>& queue, double limit=INFINITY)
{
  double totd = 0.0;
  for (size_t j=0; j<dist->getObjectCount(); ++j) {
    if (i == j) continue;
    double curd = (*dist)(i, j);
    if (queue.empty())
       queue.push( NNItem(j, curd) );
    else if (curd <= queue.top().dist) {
       if (queue.size() >= nntry && curd < queue.top().dist) {
          double oldtop = queue.top().dist;
          while (!queue.empty() && queue.top().dist == oldtop)
             queue.pop();
       }
       queue.push( NNItem(j, curd) );
    }
    totd += curd;
    if (totd >= limit) return INFINITY;
  }
  return totd;
}
\end{lstlisting}
\caption{\label{Fig:MedoidApprox1} Approximate medoid search in an arbitrary finite semimetric space, part I.}
\end{figure}

\clearpage

\begin{figure}[p!]
\centering%
\begin{lstlisting}
// [[Rcpp::export]]
RObject medoid_approx(Function distance, RObject objects,
      int iters=15, int nntry=5)
{
  RObject result(R_NilValue);
  Distance* dist = new Distance(distance, objects);
  size_t n = dist->getObjectCount();
  std::vector<bool> active(n, true);
  size_t besti_overall = -1;
  double bestd_overall = INFINITY;
  for (size_t r=0; r<(size_t)iters; ++r) {
      std::priority_queue<NNItem> queue;
      size_t besti = (size_t)(unif_rand()*n);
      if (!active[besti]) continue;
      active[besti] = false;
      double bestd = sumd_nn(dist, besti, nntry, queue);
      std::priority_queue<NNItem> bestqueue;
      bool change = true;
      while (change) {
         change = false;
         while (!queue.empty()) {
            NNItem nncur = queue.top();
            queue.pop();
            size_t curi = nncur.index;
            if (!active[curi]) continue;
            active[curi] = false;
            std::priority_queue<NNItem> curqueue;
            double curd = sumd_nn(dist, curi, nntry,
                                 curqueue, bestd);
            if (curd < bestd) {
              change = true;
              bestd = curd;
              besti = curi;
              bestqueue = curqueue;
            }
         }
         queue = bestqueue;
      }
      if (bestd < bestd_overall) {
         bestd_overall = bestd;
         besti_overall = besti;
      }
  }
  result = besti_overall+1;
  if (dist) delete dist;
  return result;
}
\end{lstlisting}
\caption{\label{Fig:MedoidApprox2} Approximate medoid search in an arbitrary finite semimetric space, part II.}
\end{figure}

\clearpage{\pagestyle{empty}\cleardoublepage}

\renewcommand{\chaptermark}[1]{\markboth{{#1}}{}}
\renewcommand{\sectionmark}[1]{\markright{{#1}}{}}

\fancyfoot{}
\fancyhf{}
\fancyhead[LO]{\sffamily\footnotesize References\hfill \sffamily\footnotesize\thepage}
\fancyhead[RO]{}
\fancyhead[LE]{}
\fancyhead[RE]{\flushleft\sffamily\footnotesize\thepage\hfill\sffamily\footnotesize References}

\bibliographystyle{acm}

\clearpage{\pagestyle{empty}\cleardoublepage}
\fancyfoot{}
\fancyhf{}
\fancyhead[LO]{\sffamily\footnotesize Index\hfill \sffamily\footnotesize\thepage}
\fancyhead[RO]{}
\fancyhead[LE]{}
\fancyhead[RE]{\flushleft\sffamily\footnotesize\thepage\hfill\sffamily\footnotesize Index}

\printindex

\end{document}